\documentclass[aps,prd,onecolumn,groupedaddress,showpacs,nofootinbib,amssymb]{revtex4}
\usepackage[dvips]{graphicx}
\usepackage{amssymb}
\usepackage{amsmath}
\usepackage{graphicx,,color}
\usepackage{amsfonts}
\usepackage{bm}
\usepackage{cancel}
\usepackage{comment}
\usepackage{floatflt}
\usepackage{slashed}
\usepackage{appendix}
\usepackage{array}
\usepackage{tabularx}
\usepackage[a4paper,margin=2.5cm]{geometry}
\usepackage{enumitem}
\usepackage{xcolor}
\newcommand\be{\begin{equation}}
\newcommand\ee{\end{equation}}

\allowdisplaybreaks[4]

\begin{document}

\title{An Analytic Scale-dependent Dark Matter Profile and the Baryonic Tully-Fisher Relation}
\author{V.K. Oikonomou$^{1,2}$}\email{voikonomou@gapps.auth.gr,v.k.oikonomou1979@gmail.com}
\affiliation{$^{1)}$Department of Physics, Aristotle University of
Thessaloniki, Thessaloniki 54124, Greece\\
$^{2)}$Center for Theoretical Physics, Khazar University, 41
Mehseti Str., Baku, AZ-1096, Azerbaijan}

\tolerance=5000

\begin{abstract}
In this work we use the recently introduced concept of
self-interacting dark matter with scale-dependent equation of
state, and we provide an analytic model of dark matter that can
produce viable rotation curves even for low-surface-brightness
galaxies, irregular galaxies, low-luminosity spirals and dwarf
galaxies, all known to challenge the cold dark matter description.
The radius dependent effective equation of state of the
self-interacting dark matter model we shall introduce is assumed
to be an isothermal equation of state of the form
$P(r)=K(r)\left(\frac{\rho(r)}{\rho_{\star}}\right)$, where the
energy density will have the form $\rho(r)=\frac{\rho_0}{\left(
1+\frac{r^2}{\alpha^2}\right)^{5/2}}$, while the entropy function
$K(r)$ is $K(r)=\frac{K_0}{\left(
1+\frac{r^2}{\alpha^2}\right)^{1/2}}$. The resulting model is
confronted in detail with the SPARC galaxy data and 175 galaxies
are used and tested. It proves that the analytic model can
successfully produce the rotation curves of 116 galaxies, most of
which are small mass spirals, irregular galaxies,
low-surface-brightness and low-luminosity spirals and dwarf
galaxies. For some of these, especially the intermediate mass
spirals, we used the rotation velocity of the other components of
the galaxy, namely the gas, baryon and disk. On the other hand, 59
galaxies cannot be successfully described by our analytic model.
We tested statistically the correlation between the parameter
$K_0$ of the entropy function corresponding to the viable
galaxies, and the flat rotation velocity $V_{flat}$ and the
maximum rotation velocity $V_{max}$ of the galaxies from the SPARC
data. We also examined the baryon mass $M_b$-$K_0$ relation and
the luminosity $L$-$K_0$ relation. We have been able to produce
the baryonic Tully-Fisher relation for the viable galaxies,
directly from the correlation $K_0$-$M_b$ and $K_0$-$V_{flat}$,
with the resulting relation being $M_b\sim V_{flat}^{4.026 \pm
0.371}$, however we failed to produce the canonical Tully-Fisher
relation. We thoroughly discuss this outcome, which must be a
model-dependent result, since near isothermal equation-of-state
scale-dependent models of the form
$P(r)=K(r)\left(\frac{\rho(r)}{\rho_{\star}}\right)^{\gamma(r)}$,
succeed in providing a semi-empirical proof of both the canonical
and baryonic Tully-Fisher relations.
\end{abstract}

\pacs{04.50.Kd, 95.36.+x, 98.80.-k, 98.80.Cq,11.25.-w}

\maketitle

\section{Introduction}

The large scale structure of the Universe is mysterious and many
questions related to it are still unanswered in a definitive way.
Although there is a rather clear picture on facts like, how the
quantum fluctuations of the Universe created the instabilities on
which the dark matter (DM) started to built up gravitational
potentials, on which sequentially, baryons have built up to create
proto-galactic structure, which resulted eventually to the
galaxies. Much of this complex process is still unknown since many
non-linear relaxation regimes of both baryonic matter and DM are
still vague to our understanding. The galactic generation process
must certainly be hierarchical, and some parts of the galaxy must
have been formed earlier. Most of the galaxies are speculated to
contain a DM  halo. This DM halo was hypothesized by Zwicky a
hundred years ago, after comparing the mass-to-light ratio in the
Coma cluster and compared it to the mass-to-light ratio of spirals
based on the rotation curves of their visible parts. Zwicky
concluded that there is 400 times dark mass in the Coma cluster. A
DM halo can explain the uniformity of the behavior of the rotation
curves of spirals and irregulars and even dwarfs, which either
increases and remains flat in the outer skirts of the galaxies,
for spirals, or behaves linearly as in irregulars and dwarfs.
Especially in dwarfs, DM is expected to contribute to the dynamics
of the galaxy at a percentage $90\%$. To date, the
$\Lambda$-Cold-Dark-Matter ($\Lambda$CDM) model is the most
consistent and successful model cosmologically and at galactic
scales, but it faces challenges. Specifically, the problems it
faces regarding galactic dynamics are the cusp-core problem, the
too-big-to-fail problem and the diversity problem
\cite{Tulin:2017ara}. Self-interacting DM offers an interesting
and consistent solution to these problems.

Building up and continuing a previous work
\cite{Oikonomou:2025bsi}, in this work we shall assume that the DM
is self-interacting in its nature, without assuming a specific
theoretical model, but the main assumption will be on its
effective equation-of-state (EoS). Specifically we assume that it
is radius-dependent, we will refer to it as scale-dependent, and
it is an isothermal EoS of the form,
$P(r)=K(r)\left(\frac{\rho}{\rho_{\star}}\right)$, with the energy
density $\rho(r)$ and the entropy function $K(r)$ being analytic
functions, of the form,
\begin{align}\label{iniDMpres}
    \rho(r) &= \frac{\rho_0}{\left(
1+\frac{r^2}{\alpha^2}\right)^{\frac{5}{2}}}, \\
    K(r) &= \frac{K_0}{\left(
1+\frac{r^2}{\alpha^2}\right)^{\frac{1}{2}}}\, .
\end{align}
This will prove to be a two parameter model, since only $\rho_0$
and $K_0$ will be the parameters determining the dynamics of the
model, and the transition radius $\alpha$ can be evaluated from
these two. This radius-dependent isothermal EoS is an analytic
solution to the hydrodynamic equilibrium equation, thus the
rotation curves of galaxies having such a DM component can be
evaluated fully analytically. For more general discussions on
self-interacting DM and for DM with polytropic EoS, see
\cite{Slepian:2011ev,Saxton:2014swa,Alonso-Alvarez:2024gdz,Kaplinghat:2015aga,Ahn:2004xt,Saxton:2012ja,Saxton:2010jk,Saxton:2016ozz,Arbey:2003sj,
Heikinheimo:2015kra,Wandelt:2000ad,Spergel:1999mh,Loeb:2010gj,Ackerman:2008kmp,Goodman:2000tg,Arbey:2006it,Yang:2025ume,Abedin:2025dis}.
Although in principle, many self-interacting DM models can
accommodate such a scale-dependent EoS, we find the mirror DM
perspective rather fascinating
\cite{Kobzarev:1966qya,Hodges:1993yb,Foot:2004pa,Berezhiani:2003wj,Silagadze:2008fa,Foot:2000tp,Chacko:2005pe,Berezhiani:2000gw,Blinnikov:2009nn,Tulin:2017ara,Mohapatra:2001sx,
Blinnikov:1982eh,Blinnikov:1983gh,Foot:2016wvj,Foot:2014osa,Foot:2014uba,
Foot:2004pq,Foot:2001ft,Foot:2004dh,Foot:1999hm,Foot:2001pv,Foot:2001ne,Foot:2000iu,Pavsic:1974rq,Foot:1993yp,Ignatiev:2000yy,Ignatiev:2003js,
Ciarcelluti:2004ik,Ciarcelluti:2004ip,Ciarcelluti:2010zz,Dvali:2009fw,Foot:2013msa,Foot:2013vna,Cui:2011wk,Foot:2015mqa,Foot:2014mia,Cline:2013zca,Ibe:2019ena,
Foot:2018qpw,Howe:2021neq,Cyr-Racine:2021oal,Armstrong:2023cis,Ritter:2024sqv,Mohapatra:1996yy,Mohapatra:2000qx,Goldman:2013qla,Berezhiani:1995am,Oikonomou:2024geq},
since the mirror DM can describe such scale-dependent behaviors
due to the fact that it allows pressure gradients changes. It is
also important to note that the Tully-Fisher relation finds a nice
explanation in the context of mirror DM \cite{Foot:2016wvj}.

In this article we shall investigate the phenomenological
implications of the isothermal scale-dependent EoS, focusing on
the rotation curves of the SPARC data \cite{Lelli:2016zqa}. We
shall examine 175 galaxies from the SPARC data, and as we shall
demonstrate, a total number of 116 galaxies can be perfectly
described by our model and the rotation curves are perfectly
fitted. The rest 59 galaxies are not well described by our model,
regarding their rotation curves. Our two parameter model
(\ref{iniDMpres}) seems to describe accurately dwarfs, irregular
galaxies, low-surface-brightness spirals and low-luminosity
spirals. The parameters that determine the dynamics are $K_0$, the
entropy parameter, and also $\rho_0$ the central density
parameter. We examined the correlation of $K_0$, the entropy
parameter corresponding to the viable galaxies, and the $V_{flat}$
and $V_{max}$ of the galaxies from the SPARC data. We also
examined the baryon mass $M_b$-$K_0$ relation and in addition, the
luminosity $L$-$K_0$ relation. We produced the baryonic
Tully-Fisher relation for the viable galaxies, directly from the
correlations $K_0$-$M_b$ and the $K_0$-$V_{flat}$, with the
resulting baryonic Tully-Fisher relation being $M_b\sim
V_{flat}^{4.026 \pm 0.371}$. On the other hand we failed to
produce the canonical Tully-Fisher relation from the relations
$L-K_0$ and $K_0-V_{max}$. We discuss this outcome in some detail,
since near isothermal equation-of-state scale-dependent models of
the form
$P(r)=K(r)\left(\frac{\rho(r)}{\rho_{\star}}\right)^{\gamma(r)}$,
succeed in providing a semi-empirical proof of both the canonical
and baryonic Tully-Fisher relations, as was demonstrated in Ref.
\cite{Oikonomou:2025bsi}.

Regarding the presentation of the galactic rotation curves, we
chose a rather small number of 10 viable and non-viable galaxies
from the total of 175 galaxies of the SPARC data. We include the
complete study of the rest 165 galaxies in the Appendix of the
arXiv version of this article and not the journal version of this
article.

\section{An Analytic Isothermal Scale-dependent DM Profile}

As we mentioned in the introduction, in this work we shall assume
that DM is self interacting and also that it has a scale-dependent
(radius dependent) EoS of the form
$P(r)=K(r)\left(\frac{\rho}{\rho_{\star}}\right)$, with the energy
density $\rho(r)$ and the entropy function $K(r)$ being analytic
functions of the form,
\begin{align}\label{ScaledependentEoSDM}
      \rho(r) &=\frac{\rho_0}{\left(
1+\frac{r^2}{\alpha^2}\right)^{\frac{5}{2}}}, \\
    K(r) &= \frac{K_0}{\left(
1+\frac{r^2}{\alpha^2}\right)^{\frac{1}{2}}}\, ,
\end{align}
and we shall assume that $\rho_\star=1$$M_\odot\,{\rm Kpc}^{-3}$
for simplicity. Note that the model is essentially two parameter,
because only $K_0$ and $\rho_0$ are free parameters, and the
parameter $\alpha$ can be obtained by the relation,
\begin{equation}\label{parameteralpha}
K_0= \frac{2 \pi  \alpha ^2 G \rho _0}{9 \rho _*}\, .
\end{equation}
The hydrostatic equilibrium in a spherical symmetry setting reads,
\begin{equation}
\frac{dP}{dr} \;=\; -\,\frac{G\,M(r)\,\rho(r)}{r^{2}},
\end{equation}
and the enclosed mass is,
\begin{equation}
M(r)\;=\;\int_0^r 4\pi r'^{2}\,\rho(r')\,dr'.
\end{equation}
The corresponding differential equation obeyed by the total mass
is the standard relation,
\begin{equation}
\frac{dM}{dr} = 4\pi r^{2}\,\rho(r).
\end{equation}
Therefore, the complete set of differential equations governing
the hydrodynamic equilibrium is the following,
\begin{align}\label{differentialequationsmainset}
\frac{dP}{dr} &= -\rho(r)\frac{G M(r)}{r^2}, \\
P(r) &= K(r) \left(\frac{\rho(r)}{\rho_{\star}}\right)^{\gamma(r)}, \\
\frac{dM}{dr} &= 4\pi r^2 \rho(r).
\end{align}
The functions $\rho(r)$ and $K(r)$ from Eq.
(\ref{ScaledependentEoSDM}) with an isothermal relation between
them $P(r)=K(r)\left(\frac{\rho}{\rho_{\star}}\right)$,
analytically satisfy the set of differential equations
(\ref{differentialequationsmainset}). For these functions, we
shall derive the rotation curves and fit 175 galaxies from the
SPARC data. When necessary we shall invoke the other galactic
parts that contribute to the rotation curves, namely, the gas, the
bulge and disk components. In this case, the total rotation curve
velocity will be,
\[
V_{\text{total}}^2(r) = V_{\text{disk}}^2(r) +
V_{\text{bulge}}^2(r) + V_{\text{gas}}^2(r) + V_{\text{DM}}^2(r)\,
.
\]
Regarding the physical units, the pressure has units,
\begin{equation}
[P] = [\rho][v^2] = M_\odot\,{\rm Kpc}^{-3} \, ({\rm km/s})^2\, ,
\end{equation}
and also,
\[
P(r) = K(r)\frac{\rho(r)}{\rho_\star},
\]
thus,
\begin{equation}
[K(r)] = [P] = M_\odot\,{\rm Kpc}^{-3} \, ({\rm km/s})^2\, .
\end{equation}

\section{Simulations of the Analytic Model of Scale-dependent EoS DM with SPARC Galaxies}

In this section we shall present simulations of the rotation
curves for some characteristic galaxies, using the model of Eq.
(\ref{ScaledependentEoSDM}). The model is essentially a two
parameter model, with only $K_0$ and $\rho_0$ varying. We shall
run some optimization codes for the model, trying to find the
optimal values of $K_0$ and $\rho_0$ which provide perfect fits
for the rotation curves of the 175 galaxies from the SPARC data
\cite{Lelli:2016zqa}. For some galaxies, the model
(\ref{ScaledependentEoSDM}) suffices to provide a perfect fit for
the rotation curves, but for some other galaxies we shall include
the other components of the galaxy, namely the disk, bulge and
gas, and then the desirable result is obtained. From the 175
galaxies, 116 are found to be viable, meaning that the model can
mimic their rotation curves optimally, using simply the model or
the combination of the model in the presence of gas, bulge and
disk components. On the other hand, 59 galaxies are found to be
non-viable. Specifically, the following galaxies are found to be
viable, \textbf{CamB, D512-2, D564-8, D631-7, DDO064, DDO154,
DDO161, ESO079-G014, ESO116-G012, F565-V2, F568-3, F568-V1,
F571-8, F571-V1, F574-1, F583-1, F583-4, KK98-251, NGC0055,
NGC0100, NGC0300, NGC0801, NGC0891, NGC1003, NGC1090, NGC2403,
NGC2683, NGC2841, NGC2903, NGC2915, NGC2955, NGC2976, NGC2998,
NGC3109, NGC3198, NGC3521, NGC3726, NGC3741, NGC3769, NGC3877,
NGC3893, NGC3917, NGC3949, NGC3953, NGC3972, NGC3992, NGC4010,
NGC4013, NGC4051, NGC4068, NGC4085, NGC4088, NGC4100, NGC4138,
NGC4157, NGC4183, NGC4217, NGC4559, NGC5005, NGC5033, NGC5055,
NGC5371, NGC5585, NGC5907, NGC5985, NGC6015, NGC6195, NGC6503,
NGC6674, NGC6789, NGC6946, NGC7331, NGC7793, NGC7814, UGC00128,
UGC00191, UGC00634, UGC00731, UGC00891, UGC01281, UGC02259,
UGC02487, UGC02885, UGC02916, UGC02953, UGC03205, UGC03546,
UGC03580, UGC04278, UGC04325, UGC04483, UGC04499, UGC05005,
UGC05253, UGC05414, UGC05716, UGC05721, UGC05750, UGC05764,
UGC05829, UGC05918, UGC05986, UGC06399, UGC06446, UGC06614,
UGC06667, UGC06786, UGC06787, UGC06818, UGC06917, UGC06923,
UGC06930, UGC06983, UGC07089, UGC07125, UGC07151, UGC07232,
UGC07261, UGC07323, UGC07399, UGC07524, UGC07559, UGC07577,
UGC07603, UGC07690, UGC07866, UGC08286, UGC08490, UGC08550,
UGC08699, UGC08837, UGC09037, UGC09133, UGC09992, UGC10310,
UGC11455, UGC11557, UGC11820, UGC11914, UGC12506, UGC12632,
UGC12732, UGCA442, UGCA444, F561-1, F563-1, F563-V1, F563-V2,
F567-2, F568-1, F574-2, F579-V1, NGC1705, NGC2366, NGC4214,
NGC4389, NGC6946, PGC51017, UGC01230, UGC02023, UGC04305,
UGC05999, UGC06628, UGC06973, UGC07608, UGCA281}.

On the other hand the non-viable galaxies are, \textbf{DDO168,
DDO170, ESO079-G014, ESO116-G012, ESO444-G084, ESO563-G021,
F565-V2, F568-3, F568-V1, F571-8, F571-V1, F574-1, F583-1, F583-4,
IC2574, IC4202, KK98-251, NGC0024, NGC0055, NGC0100, NGC0247,
NGC0289, NGC0300, NGC0801, NGC0891, NGC1003, NGC1090, NGC2403,
NGC2683, NGC2841, NGC2903, NGC2915, NGC2955, NGC2976, NGC2998,
NGC3109, NGC3198, NGC3521, NGC3726, NGC3741, NGC3769, NGC3877,
NGC3893, NGC3917, NGC3949, NGC3953, NGC3972, NGC3992, NGC4010,
NGC4013, NGC4051, NGC4068, NGC4085, NGC4088, NGC4100, NGC4138,
NGC4157, NGC4183, NGC4217, NGC4559, NGC5005, NGC5033, NGC5055,
NGC5371, NGC5585, NGC5907, NGC5985, NGC6015, NGC6195, NGC6503,
NGC6674, NGC6789, NGC6946, NGC7331, NGC7793, NGC7814, UGC00128,
UGC00191, UGC00634, UGC00731, UGC00891, UGC01281, UGC02259,
UGC02487, UGC02885, UGC02916, UGC02953, UGC03205, UGC03546,
UGC03580, UGC04278, UGC04325, UGC04483, UGC04499, UGC05005,
UGC05253, UGC05414, UGC05716, UGC05721, UGC05750, UGC05764,
UGC05829, UGC05918, UGC05986, UGC06399, UGC06446, UGC06614,
UGC06667, UGC06786, UGC06787, UGC06818, UGC06917, UGC06923,
UGC06930, UGC06983, UGC07089, UGC07125, UGC07151, UGC07232,
UGC07261, UGC07323, UGC07399, UGC07524, UGC07559, UGC07577,
UGC07603, UGC07690, UGC07866, UGC08286, UGC08490, UGC08550,
UGC08699, UGC08837, UGC09037, UGC09133, UGC09992, UGC10310,
UGC11455, UGC11557, UGC11820, UGC11914, UGC12506, UGC12632,
UGC12732, UGCA442, UGCA444, F561-1, F563-1, F563-V1, F563-V2,
F567-2, F568-1, F574-2, F579-V1, NGC1705, NGC2366, NGC4214,
NGC4389, NGC6946, PGC51017, UGC01230, UGC02023, UGC04305,
UGC05999, UGC06628, UGC06973}.

In the following subsections we present five viable and five
non-viable examples from the total of the 175 galaxies. The rest
of the study is presented in the Appendix of the arXiv version of
this article.

\subsection{Sample of Viable Galaxies}

\subsubsection{The Galaxy CamB}

For this galaxy, the optimization method we used, ensures maximum
compatibility of the analytic SIDM model of Eq.
(\ref{ScaledependentEoSDM}) with the SPARC data, if we choose
$\rho_0=1.20484\times 10^7$$M_{\odot}/\mathrm{Kpc}^{3}$ and
$K_0=206.963
$$M_{\odot} \, \mathrm{Kpc}^{-3} \, (\mathrm{km/s})^{2}$, in which case the reduced $\chi^2_{red}$ value is
$\chi^2_{red}=0.493168$. Also the parameter $\alpha$ in this case
is $\alpha=2.51$Kpc.

\begin{figure}[h!]
\centering
\includegraphics[width=20pc]{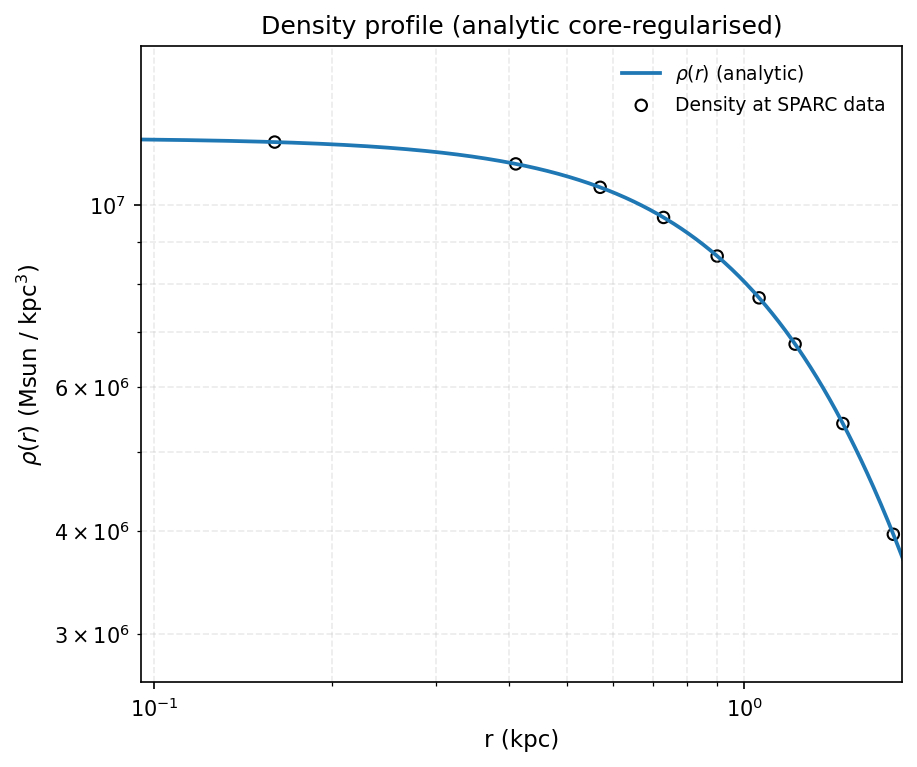}
\caption{The density of the SIDM model of Eq.
(\ref{ScaledependentEoSDM}) for the galaxy CamB, versus the
radius.} \label{CamBdens}
\end{figure}

\begin{figure}[h!]
\centering
\includegraphics[width=35pc]{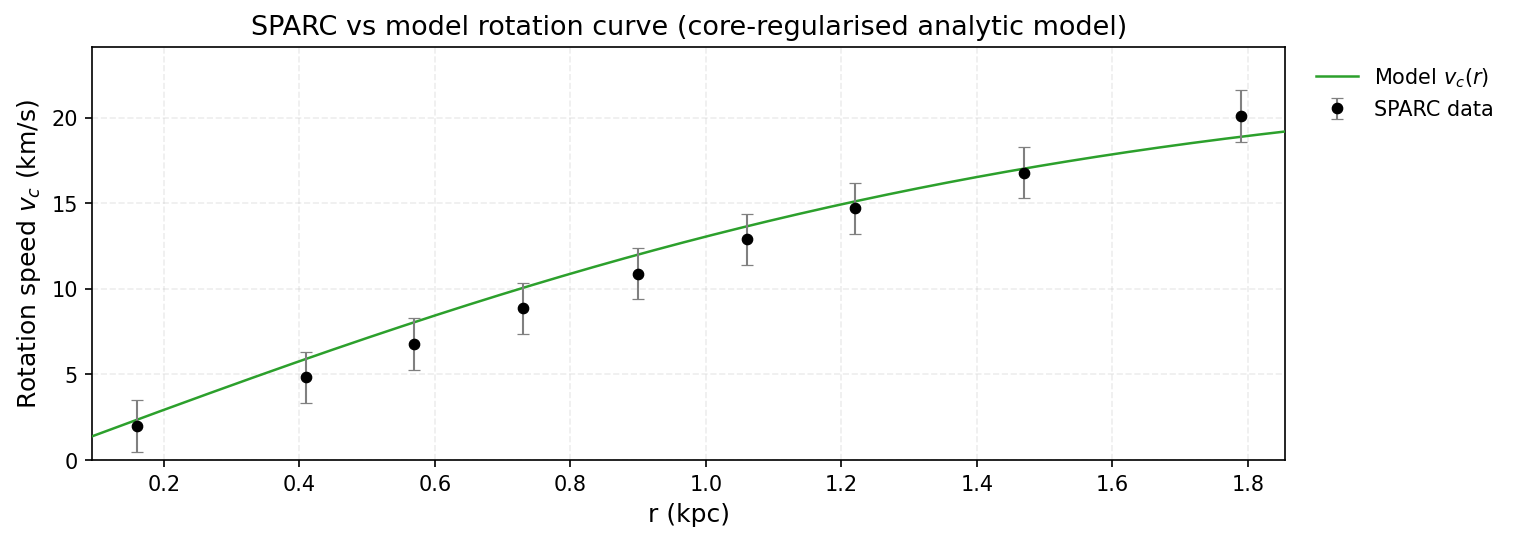}
\caption{The predicted rotation curves for the optimized SIDM
model of Eq. (\ref{ScaledependentEoSDM}), versus the SPARC
observational data for the galaxy CamB.} \label{CamB}
\end{figure}

In Table \ref{collCamB} we present the optimized values of $K_0$
and $\rho_0$ for the analytic SIDM model of Eq.
(\ref{ScaledependentEoSDM}) for which the maximum compatibility
with the SPARC data is achieved.
\begin{table}[h!]
  \begin{center}
    \caption{SIDM Optimization Values for the galaxy CamB}
    \label{collCamB}
     \begin{tabular}{|r|r|}
     \hline
      \textbf{Parameter}   & \textbf{Optimization Values}
      \\  \hline
     $\rho_0 $  ($M_{\odot}/\mathrm{Kpc}^{3}$) & $1.20484\times
     10^7$
\\  \hline $K_0$ ($M_{\odot} \,
\mathrm{Kpc}^{-3} \, (\mathrm{km/s})^{2}$)& 206.963
\\  \hline
    \end{tabular}
  \end{center}
\end{table}
In Figs. \ref{CamBdens}, \ref{CamB} we present the density of the
analytic SIDM model, the predicted rotation curves for the SIDM
model (\ref{ScaledependentEoSDM}), versus the SPARC observational
data and the sound speed, as a function of the radius
respectively. As it can be seen, for this galaxy, the SIDM model
produces viable rotation curves which are compatible with the
SPARC data.

\subsubsection{The Galaxy D512-2}

For this galaxy, the optimization method we used, ensures maximum
compatibility of the analytic SIDM model of Eq.
(\ref{ScaledependentEoSDM}) with the SPARC data, if we choose
$\rho_0=4.00327\times 10^7$$M_{\odot}/\mathrm{Kpc}^{3}$ and
$K_0=552.819
$$M_{\odot} \, \mathrm{Kpc}^{-3} \, (\mathrm{km/s})^{2}$, in which
case the reduced $\chi^2_{red}$ value is $\chi^2_{red}=0.0925236$.
Also the parameter $\alpha$ in this case is $\alpha=2.14454$Kpc.

\begin{figure}[h!]
\centering
\includegraphics[width=20pc]{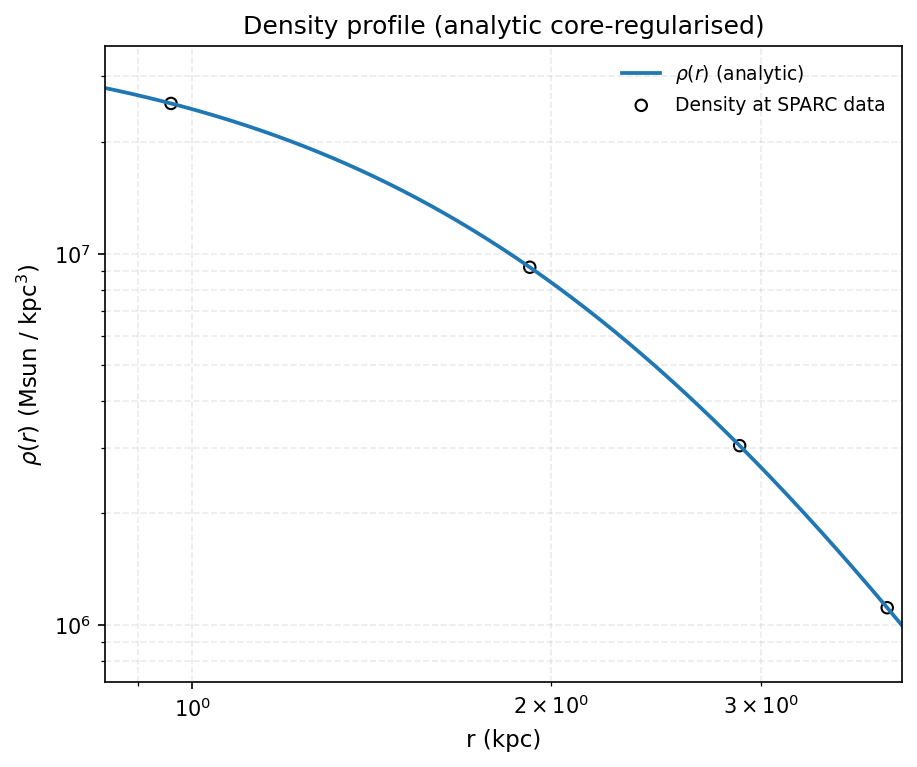}
\caption{The density of the SIDM model of Eq.
(\ref{ScaledependentEoSDM}) for the galaxy D512-2, versus the
radius.} \label{D512-2dens}
\end{figure}

\begin{figure}[h!]
\centering
\includegraphics[width=35pc]{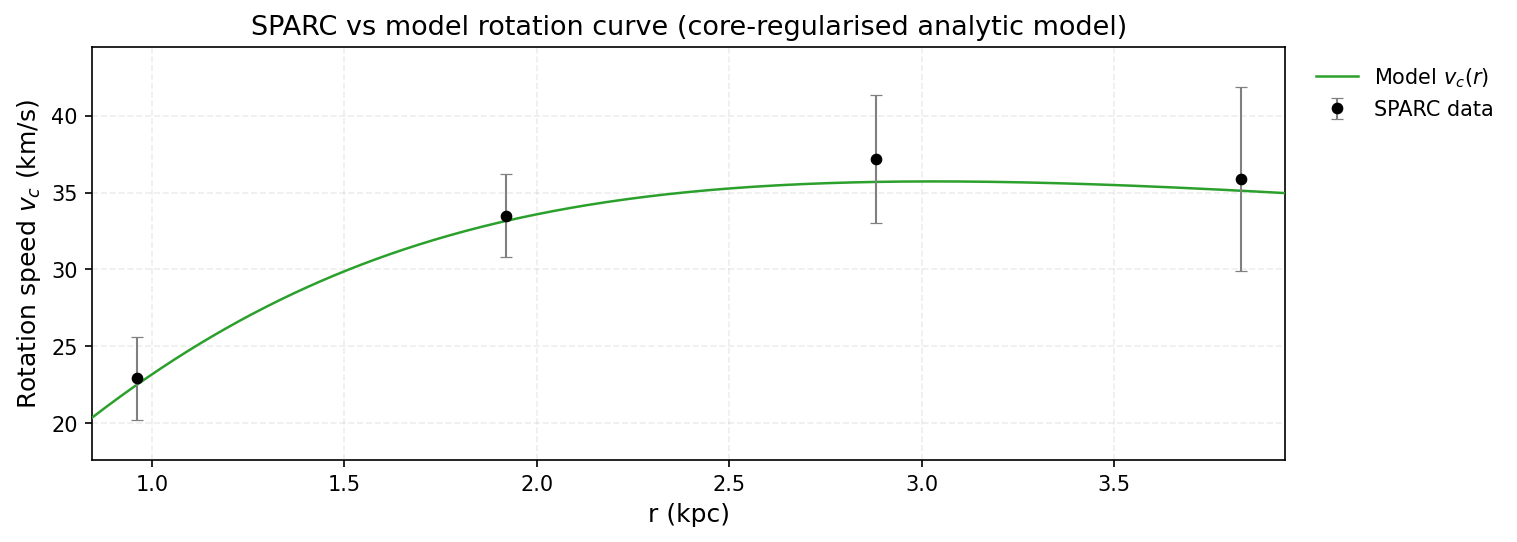}
\caption{The predicted rotation curves for the optimized SIDM
model of Eq. (\ref{ScaledependentEoSDM}), versus the SPARC
observational data for the galaxy D512-2.} \label{D512-2}
\end{figure}

In Table \ref{collD512-2} we present the optimized values of $K_0$
and $\rho_0$ for the analytic SIDM model of Eq.
(\ref{ScaledependentEoSDM}) for which the maximum compatibility
with the SPARC data is achieved.
\begin{table}[h!]
  \begin{center}
    \caption{SIDM Optimization Values for the galaxy D512-2}
    \label{collD512-2}
     \begin{tabular}{|r|r|}
     \hline
      \textbf{Parameter}   & \textbf{Optimization Values}
      \\  \hline
     $\rho_0 $  ($M_{\odot}/\mathrm{Kpc}^{3}$) & $4.00327\times 10^7$
\\  \hline $K_0$ ($M_{\odot} \,
\mathrm{Kpc}^{-3} \, (\mathrm{km/s})^{2}$)& 552.819
\\  \hline
    \end{tabular}
  \end{center}
\end{table}
In Figs. \ref{D512-2dens}, \ref{D512-2} we present the density of
the analytic SIDM model, the predicted rotation curves for the
SIDM model (\ref{ScaledependentEoSDM}), versus the SPARC
observational data and the sound speed, as a function of the
radius respectively. As it can be seen, for this galaxy, the SIDM
model produces viable rotation curves which are compatible with
the SPARC data.

\subsubsection{The Galaxy D564-8}

For this galaxy, the optimization method we used, ensures maximum
compatibility of the analytic SIDM model of Eq.
(\ref{ScaledependentEoSDM}) with the SPARC data, if we choose
$\rho_0=1.46575\times 10^7$$M_{\odot}/\mathrm{Kpc}^{3}$ and
$K_0=242.656
$$M_{\odot} \, \mathrm{Kpc}^{-3} \, (\mathrm{km/s})^{2}$, in which
case the reduced $\chi^2_{red}$ value is $\chi^2_{red}=0.277623$.
Also the parameter $\alpha$ in this case is $\alpha=2.3481 $Kpc.

\begin{figure}[h!]
\centering
\includegraphics[width=20pc]{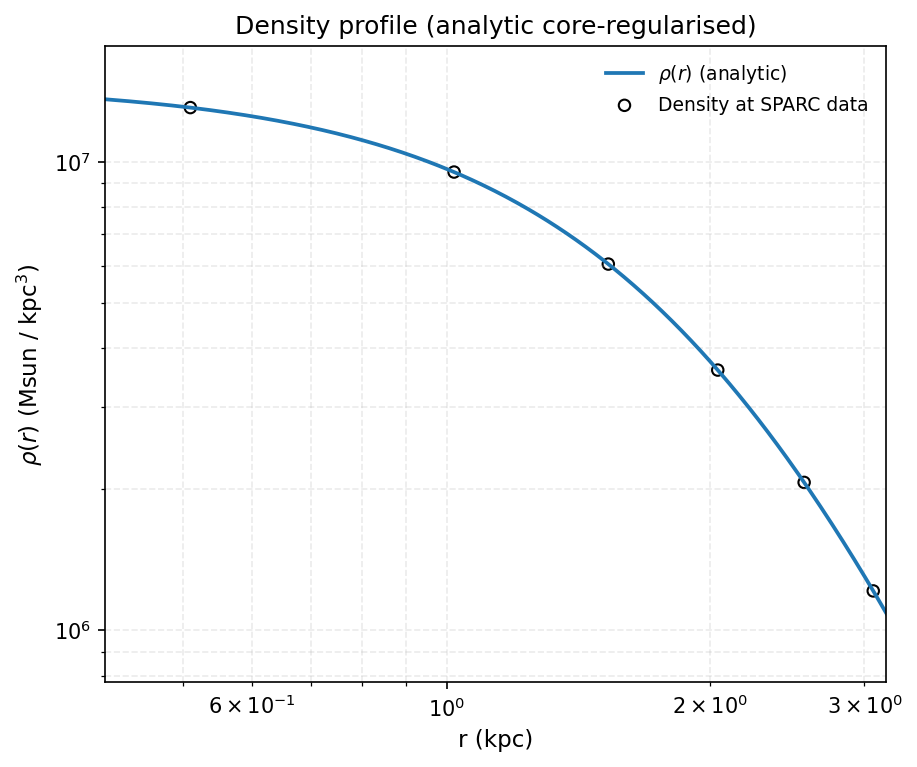}
\caption{The density of the SIDM model of Eq.
(\ref{ScaledependentEoSDM}) for the galaxy D564-8, versus the
radius.} \label{D564-8dens}
\end{figure}

\begin{figure}[h!]
\centering
\includegraphics[width=35pc]{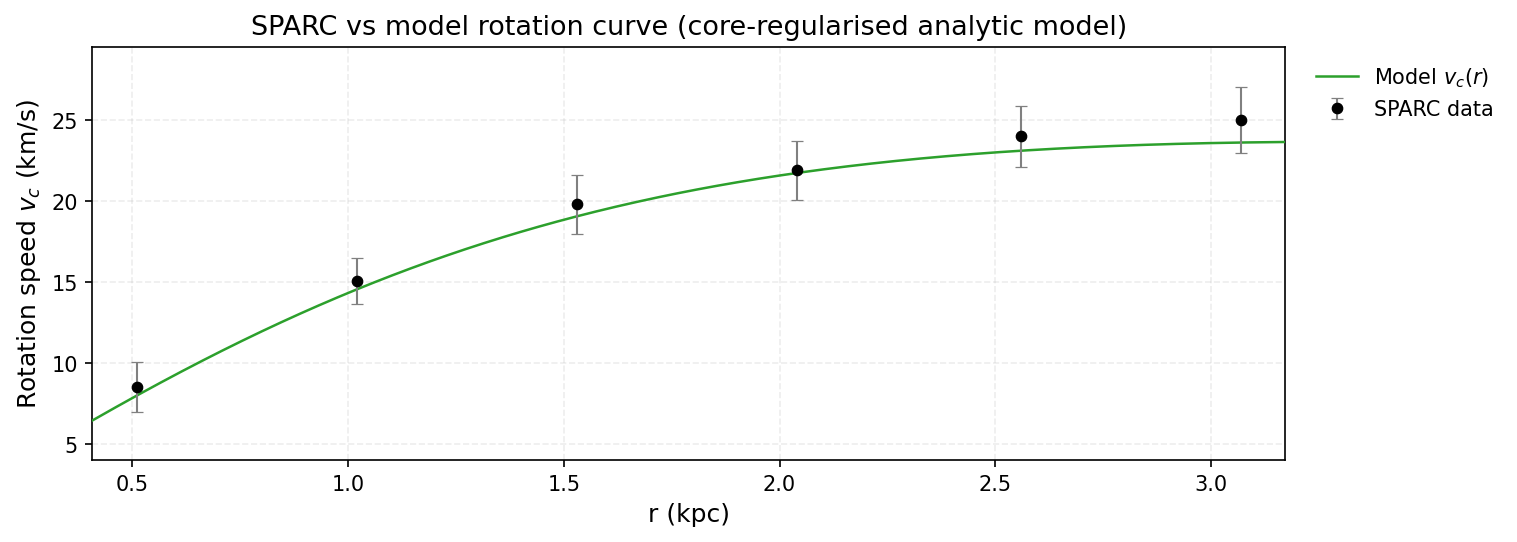}
\caption{The predicted rotation curves for the optimized SIDM
model of Eq. (\ref{ScaledependentEoSDM}), versus the SPARC
observational data for the galaxy D564-8.} \label{D564-8}
\end{figure}

In Table \ref{collD564-8} we present the optimized values of $K_0$
and $\rho_0$ for the analytic SIDM model of Eq.
(\ref{ScaledependentEoSDM}) for which the maximum compatibility
with the SPARC data is achieved.
\begin{table}[h!]
  \begin{center}
    \caption{SIDM Optimization Values for the galaxy D564-8}
    \label{collD564-8}
     \begin{tabular}{|r|r|}
     \hline
      \textbf{Parameter}   & \textbf{Optimization Values}
      \\  \hline
     $\rho_0 $  ($M_{\odot}/\mathrm{Kpc}^{3}$) & $1.46575\times 10^7$
\\  \hline $K_0$ ($M_{\odot} \,
\mathrm{Kpc}^{-3} \, (\mathrm{km/s})^{2}$)& 242.656
\\  \hline
    \end{tabular}
  \end{center}
\end{table}
In Figs. \ref{D564-8dens}, \ref{D564-8} we present the density of
the analytic SIDM model, the predicted rotation curves for the
SIDM model (\ref{ScaledependentEoSDM}), versus the SPARC
observational data and the sound speed, as a function of the
radius respectively. As it can be seen, for this galaxy, the SIDM
model produces viable rotation curves which are compatible with
the SPARC data.

\subsubsection{The Galaxy DDO154, Marginally Viable, Extended Viable by One Miss}

For this galaxy, the optimization method we used, ensures maximum
compatibility of the analytic SIDM model of Eq.
(\ref{ScaledependentEoSDM}) with the SPARC data, if we choose
$\rho_0=2.566\times 10^7$$M_{\odot}/\mathrm{Kpc}^{3}$ and
$K_0=965.604
$$M_{\odot} \, \mathrm{Kpc}^{-3} \, (\mathrm{km/s})^{2}$, in which
case the reduced $\chi^2_{red}$ value is $\chi^2_{red}=4.31061$.
Also the parameter $\alpha$ in this case is $\alpha=3.54015 $Kpc.

\begin{figure}[h!]
\centering
\includegraphics[width=20pc]{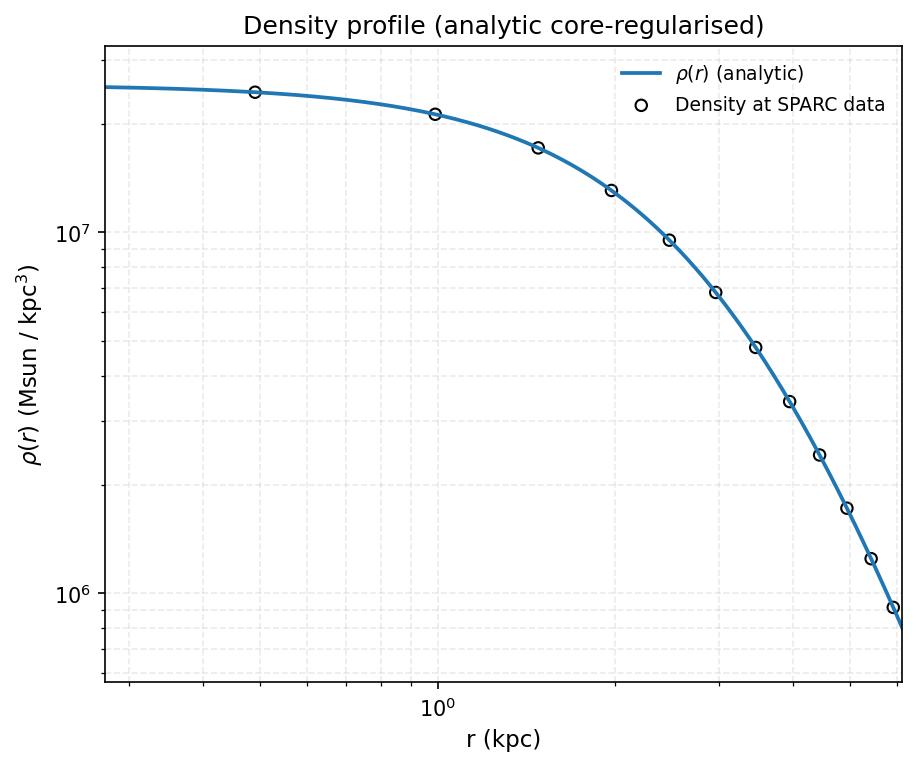}
\caption{The density of the SIDM model of Eq.
(\ref{ScaledependentEoSDM}) for the galaxy DDO154, versus the
radius.} \label{DDO154dens}
\end{figure}

\begin{figure}[h!]
\centering
\includegraphics[width=35pc]{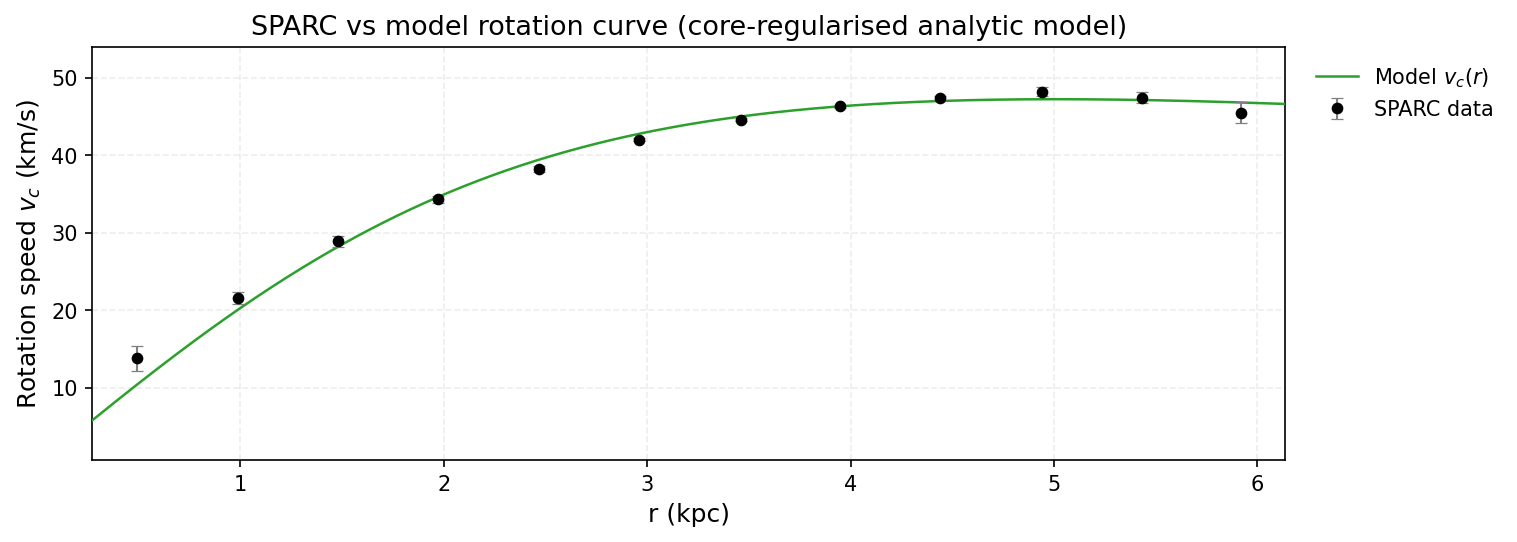}
\caption{The predicted rotation curves for the optimized SIDM
model of Eq. (\ref{ScaledependentEoSDM}), versus the SPARC
observational data for the galaxy DDO154.} \label{DDO154}
\end{figure}

In Table \ref{collDDO154} we present the optimized values of $K_0$
and $\rho_0$ for the analytic SIDM model of Eq.
(\ref{ScaledependentEoSDM}) for which the maximum compatibility
with the SPARC data is achieved.
\begin{table}[h!]
  \begin{center}
    \caption{SIDM Optimization Values for the galaxy DDO154}
    \label{collDDO154}
     \begin{tabular}{|r|r|}
     \hline
      \textbf{Parameter}   & \textbf{Optimization Values}
      \\  \hline
     $\rho_0 $  ($M_{\odot}/\mathrm{Kpc}^{3}$) & $2.566\times 10^7$
\\  \hline $K_0$ ($M_{\odot} \,
\mathrm{Kpc}^{-3} \, (\mathrm{km/s})^{2}$)& 965.604
\\  \hline
    \end{tabular}
  \end{center}
\end{table}
In Figs. \ref{DDO154dens}, \ref{DDO154} we present the density of
the analytic SIDM model, the predicted rotation curves for the
SIDM model (\ref{ScaledependentEoSDM}), versus the SPARC
observational data and the sound speed, as a function of the
radius respectively. As it can be seen, for this galaxy, the SIDM
model produces marginally viable rotation curves which are
marginally compatible with the SPARC data.

\begin{figure}[h!]
\centering
\includegraphics[width=20pc]{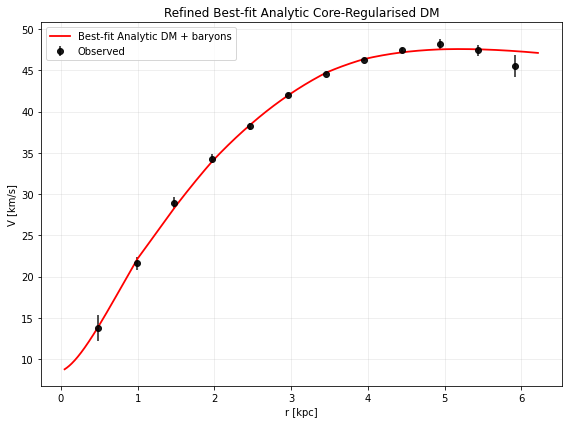}
\caption{The predicted rotation curves after using an optimization
for the SIDM model (\ref{ScaledependentEoSDM}), and the extended
SPARC data for the galaxy DDO154. We included the rotation curves
of the gas, the disk velocities, the bulge (where present) along
with the SIDM model.} \label{extendedDDO154}
\end{figure}

Now we shall include contributions to the rotation velocity from
the other components of the galaxy, namely the disk, the gas, and
the bulge if present. In Fig. \ref{extendedDDO154} we present the
combined rotation curves including all the components of the
galaxy along with the SIDM. As it can be seen, the extended
collisional DM model is almost viable (missed one point).
Also in Table \ref{evaluationextendedDDO154} we present the
optimized values of the free parameters of the SIDM model for
which  we achieve the maximum compatibility with the SPARC data,
for the galaxy DDO154, and also the resulting reduced
$\chi^2_{red}$ value.
\begin{table}[h!]
\centering \caption{Optimized Parameter Values of the Extended
SIDM model for the Galaxy DDO154.}
\begin{tabular}{lc}
\hline
Parameter & Value  \\
\hline
$\rho_0 $ ($M_{\odot}/\mathrm{Kpc}^{3}$) & $1.73089\times 10^7$   \\
$K_0$ ($M_{\odot} \,
\mathrm{Kpc}^{-3} \, (\mathrm{km/s})^{2}$) & 852.337   \\
$ml_{\text{disk}}$ & 0.8442 \\
$ml_{\text{bulge}}$ & 0.9721 \\
$\alpha$ (Kpc) & 4.04919 \\
$\chi^2_{red}$ & 0.765527 \\
\hline
\end{tabular}
\label{evaluationextendedDDO154}
\end{table}

\subsubsection{The Galaxy UGC00634}

For this galaxy, the optimization method we used, ensures maximum
compatibility of the analytic SIDM model of Eq.
(\ref{ScaledependentEoSDM}) with the SPARC data, if we choose
$\rho_0=1.51522\times 10^7$$M_{\odot}/\mathrm{Kpc}^{3}$ and
$K_0=5162.06
$$M_{\odot} \, \mathrm{Kpc}^{-3} \, (\mathrm{km/s})^{2}$, in which
case the reduced $\chi^2_{red}$ value is $\chi^2_{red}=1.24441$.
Also the parameter $\alpha$ in this case is $\alpha=10.6518 $Kpc.

\begin{figure}[h!]
\centering
\includegraphics[width=20pc]{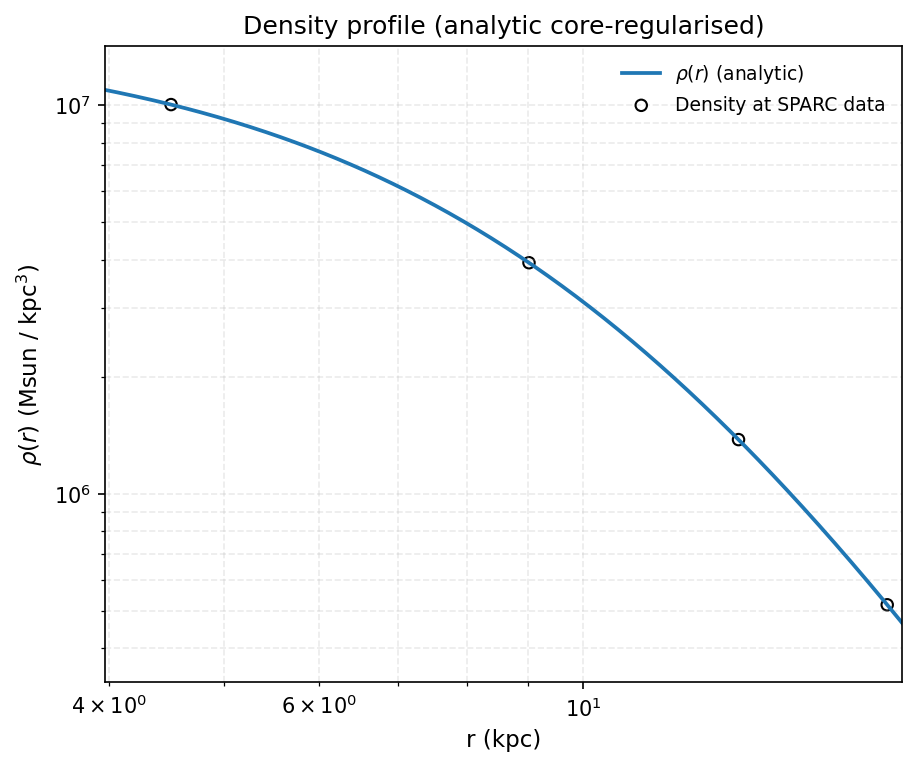}
\caption{The density of the SIDM model of Eq.
(\ref{ScaledependentEoSDM}) for the galaxy UGC00634, versus the
radius.} \label{UGC00634dens}
\end{figure}

\begin{figure}[h!]
\centering
\includegraphics[width=35pc]{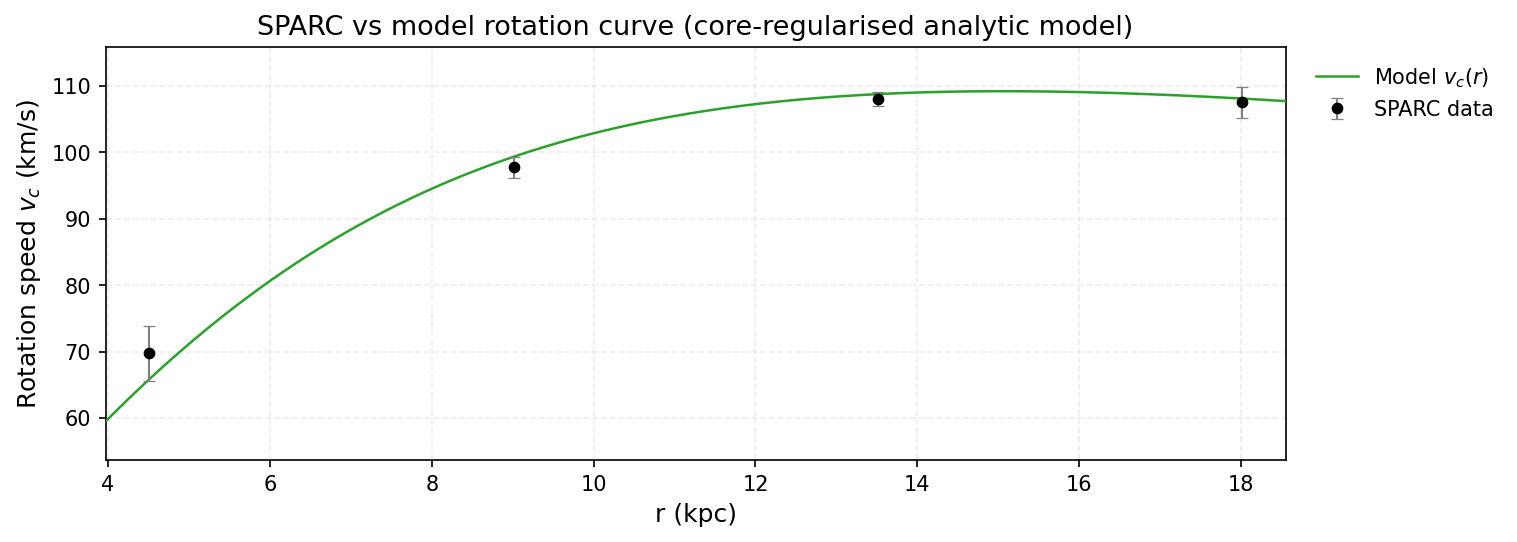}
\caption{The predicted rotation curves for the optimized SIDM
model of Eq. (\ref{ScaledependentEoSDM}), versus the SPARC
observational data for the galaxy UGC00634.} \label{UGC00634}
\end{figure}

In Table \ref{collUGC00634} we present the optimized values of
$K_0$ and $\rho_0$ for the analytic SIDM model of Eq.
(\ref{ScaledependentEoSDM}) for which the maximum compatibility
with the SPARC data is achieved.
\begin{table}[h!]
  \begin{center}
    \caption{SIDM Optimization Values for the galaxy UGC00634}
    \label{collUGC00634}
     \begin{tabular}{|r|r|}
     \hline
      \textbf{Parameter}   & \textbf{Optimization Values}
      \\  \hline
     $\rho_0 $  ($M_{\odot}/\mathrm{Kpc}^{3}$) & $1.51522\times 10^7$
\\  \hline $K_0$ ($M_{\odot} \,
\mathrm{Kpc}^{-3} \, (\mathrm{km/s})^{2}$)& 5162.06
\\  \hline
    \end{tabular}
  \end{center}
\end{table}
In Figs. \ref{UGC00634dens}, \ref{UGC00634} we present the density
of the analytic SIDM model, the predicted rotation curves for the
SIDM model (\ref{ScaledependentEoSDM}), versus the SPARC
observational data and the sound speed, as a function of the
radius respectively. As it can be seen, for this galaxy, the SIDM
model produces viable rotation curves which are compatible with
the SPARC data.

\subsection{Sample of Non-Viable Galaxies}

\subsubsection{The Galaxy ESO444-G084, Non-viable}

For this galaxy, the optimization method we used, ensures maximum
compatibility of the analytic SIDM model of Eq.
(\ref{ScaledependentEoSDM}) with the SPARC data, if we choose
$\rho_0=1.16532\times 10^8$$M_{\odot}/\mathrm{Kpc}^{3}$ and
$K_0=1660.95
$$M_{\odot} \, \mathrm{Kpc}^{-3} \, (\mathrm{km/s})^{2}$, in which
case the reduced $\chi^2_{red}$ value is $\chi^2_{red}=4.40899$.
Also the parameter $\alpha$ in this case is $\alpha=2.17875 $Kpc.

\begin{figure}[h!]
\centering
\includegraphics[width=20pc]{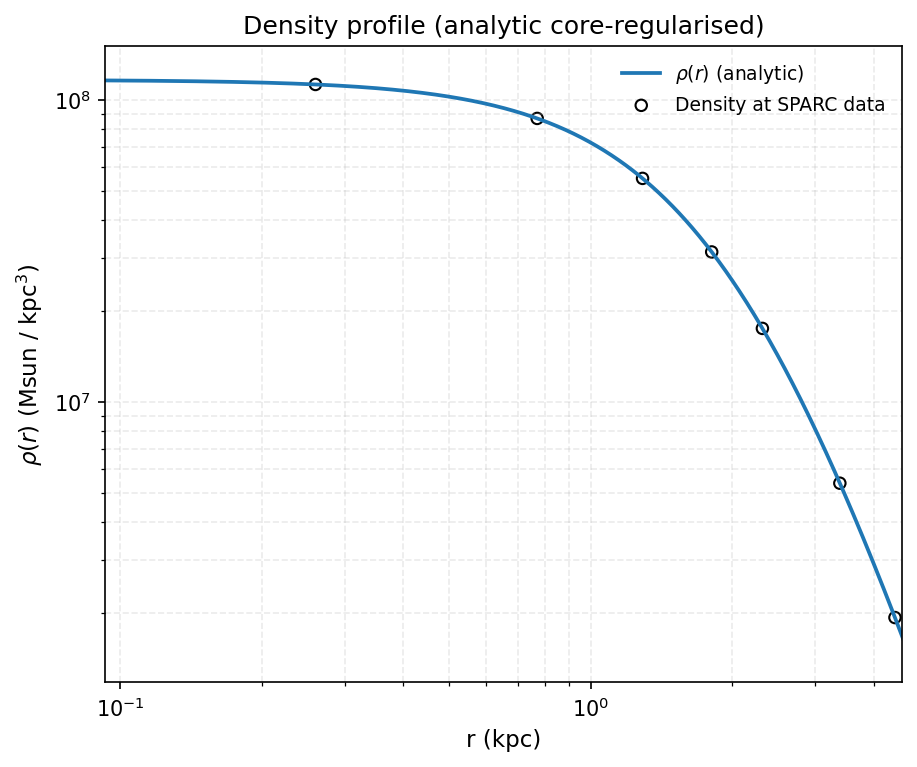}
\caption{The density of the SIDM model of Eq.
(\ref{ScaledependentEoSDM}) for the galaxy ESO444-G084, versus the
radius.} \label{ESO444-G084dens}
\end{figure}

\begin{figure}[h!]
\centering
\includegraphics[width=35pc]{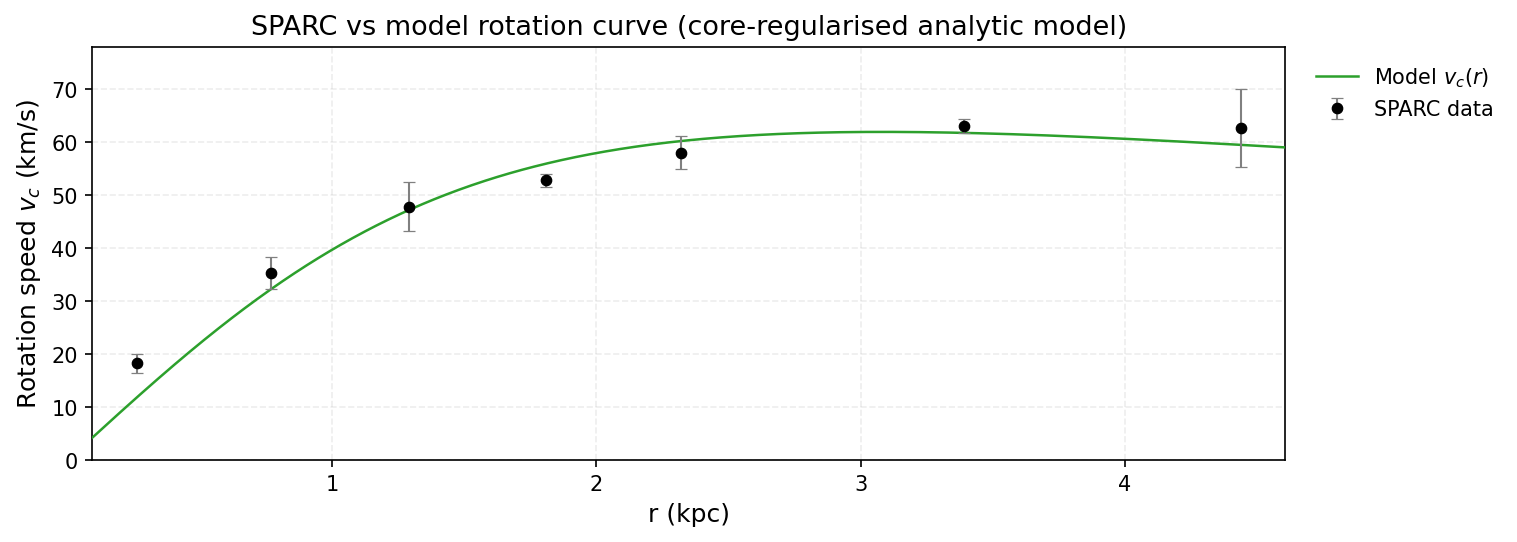}
\caption{The predicted rotation curves for the optimized SIDM
model of Eq. (\ref{ScaledependentEoSDM}), versus the SPARC
observational data for the galaxy ESO444-G084.}
\label{ESO444-G084}
\end{figure}

In Table \ref{collESO444-G084} we present the optimized values of
$K_0$ and $\rho_0$ for the analytic SIDM model of Eq.
(\ref{ScaledependentEoSDM}) for which the maximum compatibility
with the SPARC data is achieved.
\begin{table}[h!]
  \begin{center}
    \caption{SIDM Optimization Values for the galaxy ESO444-G084}
    \label{collESO444-G084}
     \begin{tabular}{|r|r|}
     \hline
      \textbf{Parameter}   & \textbf{Optimization Values}
      \\  \hline
     $\rho_0 $  ($M_{\odot}/\mathrm{Kpc}^{3}$) & $1.16532\times 10^8$
\\  \hline $K_0$ ($M_{\odot} \,
\mathrm{Kpc}^{-3} \, (\mathrm{km/s})^{2}$)& 1660.95
\\  \hline
    \end{tabular}
  \end{center}
\end{table}
In Figs. \ref{ESO444-G084dens}, \ref{ESO444-G084} we present the
density of the analytic SIDM model, the predicted rotation curves
for the SIDM model (\ref{ScaledependentEoSDM}), versus the SPARC
observational data and the sound speed, as a function of the
radius respectively. As it can be seen, for this galaxy, the SIDM
model produces marginally viable rotation curves which are
marginally compatible with the SPARC data.

\begin{figure}[h!]
\centering
\includegraphics[width=20pc]{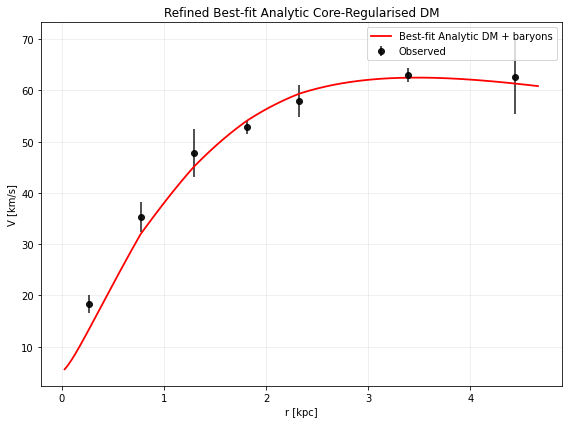}
\caption{The predicted rotation curves after using an optimization
for the SIDM model (\ref{ScaledependentEoSDM}), and the extended
SPARC data for the galaxy ESO444-G084. We included the rotation
curves of the gas, the disk velocities, the bulge (where present)
along with the SIDM model.} \label{extendedESO444-G084}
\end{figure}

Now we shall include contributions to the rotation velocity from
the other components of the galaxy, namely the disk, the gas, and
the bulge if present. In Fig. \ref{extendedESO444-G084} we present
the combined rotation curves including all the components of the
galaxy along with the SIDM. As it can be seen, the extended
collisional DM model is marginally viable.
Also in Table \ref{evaluationextendedESO444-G084} we present the
optimized values of the free parameters of the SIDM model for
which  we achieve the maximum compatibility with the SPARC data,
for the galaxy ESO444-G084, and also the resulting reduced
$\chi^2_{red}$ value.
\begin{table}[h!]
\centering \caption{Optimized Parameter Values of the Extended
SIDM model for the Galaxy ESO444-G084.}
\begin{tabular}{lc}
\hline
Parameter & Value  \\
\hline
$\rho_0 $ ($M_{\odot}/\mathrm{Kpc}^{3}$) & $7.0461\times 10^7$   \\
$K_0$ ($M_{\odot} \,
\mathrm{Kpc}^{-3} \, (\mathrm{km/s})^{2}$) & 1530.99   \\
$ml_{\text{disk}}$ & 1 \\
$ml_{\text{bulge}}$ & 0.2225 \\
$\alpha$ (Kpc) & 2.68972\\
$\chi^2_{red}$ & 3.66799 \\
\hline
\end{tabular}
\label{evaluationextendedESO444-G084}
\end{table}

\subsubsection{The Galaxy IC4202, Non-viable}

For this galaxy, the optimization method we used, ensures maximum
compatibility of the analytic SIDM model of Eq.
(\ref{ScaledependentEoSDM}) with the SPARC data, if we choose
$\rho_0=1.36561\times 10^8$$M_{\odot}/\mathrm{Kpc}^{3}$ and
$K_0=27737.5
$$M_{\odot} \, \mathrm{Kpc}^{-3} \, (\mathrm{km/s})^{2}$, in which
case the reduced $\chi^2_{red}$ value is $\chi^2_{red}=6.79581$.
Also the parameter $\alpha$ in this case is $\alpha=8.22471 $Kpc.

\begin{figure}[h!]
\centering
\includegraphics[width=20pc]{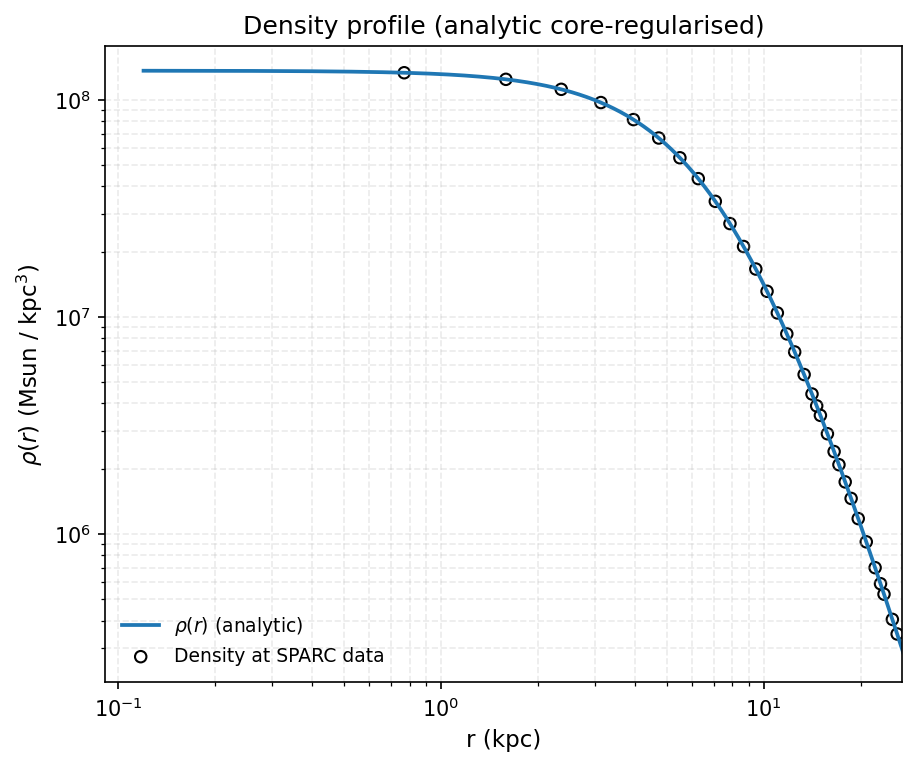}
\caption{The density of the SIDM model of Eq.
(\ref{ScaledependentEoSDM}) for the galaxy IC4202, versus the
radius.} \label{IC4202dens}
\end{figure}

\begin{figure}[h!]
\centering
\includegraphics[width=35pc]{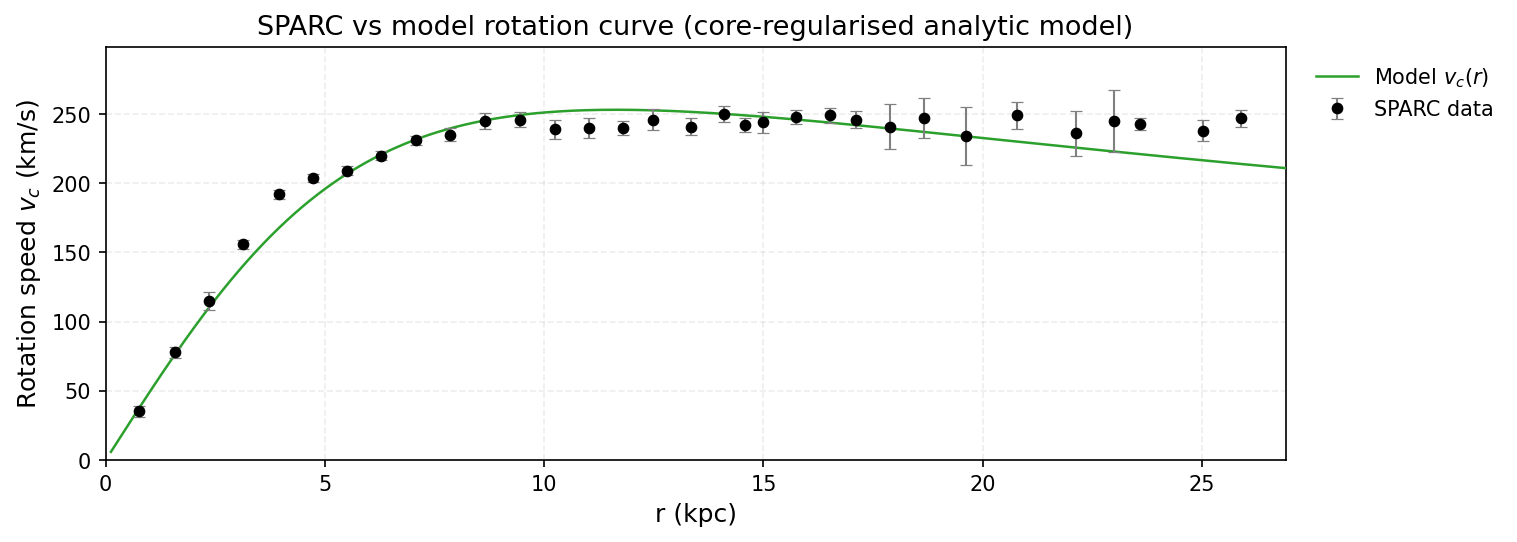}
\caption{The predicted rotation curves for the optimized SIDM
model of Eq. (\ref{ScaledependentEoSDM}), versus the SPARC
observational data for the galaxy IC4202.} \label{IC4202}
\end{figure}

In Table \ref{collIC4202} we present the optimized values of $K_0$
and $\rho_0$ for the analytic SIDM model of Eq.
(\ref{ScaledependentEoSDM}) for which the maximum compatibility
with the SPARC data is achieved.
\begin{table}[h!]
  \begin{center}
    \caption{SIDM Optimization Values for the galaxy IC4202}
    \label{collIC4202}
     \begin{tabular}{|r|r|}
     \hline
      \textbf{Parameter}   & \textbf{Optimization Values}
      \\  \hline
     $\rho_0 $  ($M_{\odot}/\mathrm{Kpc}^{3}$) & $1.36561\times 10^8$
\\  \hline $K_0$ ($M_{\odot} \,
\mathrm{Kpc}^{-3} \, (\mathrm{km/s})^{2}$)& 27737.5
\\  \hline
    \end{tabular}
  \end{center}
\end{table}
In Figs. \ref{IC4202dens}, \ref{IC4202} we present the density of
the analytic SIDM model, the predicted rotation curves for the
SIDM model (\ref{ScaledependentEoSDM}), versus the SPARC
observational data and the sound speed, as a function of the
radius respectively. As it can be seen, for this galaxy, the SIDM
model produces non-viable rotation curves which are incompatible
with the SPARC data.

\begin{figure}[h!]
\centering
\includegraphics[width=20pc]{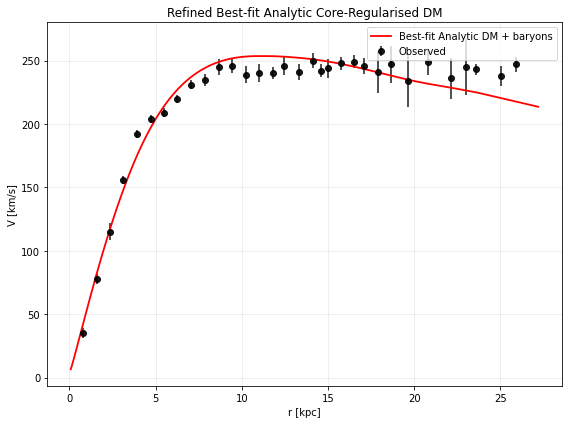}
\caption{The predicted rotation curves after using an optimization
for the SIDM model (\ref{ScaledependentEoSDM}), and the extended
SPARC data for the galaxy IC4202. We included the rotation curves
of the gas, the disk velocities, the bulge (where present) along
with the SIDM model.} \label{extendedIC4202}
\end{figure}

Now we shall include contributions to the rotation velocity from
the other components of the galaxy, namely the disk, the gas, and
the bulge if present. In Fig. \ref{extendedIC4202} we present the
combined rotation curves including all the components of the
galaxy along with the SIDM. As it can be seen, the extended
collisional DM model is non-viable.
Also in Table \ref{evaluationextendedIC4202} we present the
optimized values of the free parameters of the SIDM model for
which  we achieve the maximum compatibility with the SPARC data,
for the galaxy IC4202, and also the resulting reduced
$\chi^2_{red}$ value.
\begin{table}[h!]
\centering \caption{Optimized Parameter Values of the Extended
SIDM model for the Galaxy IC4202.}
\begin{tabular}{lc}
\hline
Parameter & Value  \\
\hline
$\rho_0 $ ($M_{\odot}/\mathrm{Kpc}^{3}$) & $1.55781\times 10^8$   \\
$K_0$ ($M_{\odot} \,
\mathrm{Kpc}^{-3} \, (\mathrm{km/s})^{2}$) & 27821.3   \\
$ml_{\text{disk}}$ & 0.7 \\
$ml_{\text{bulge}}$ & 0.0523 \\
$\alpha$ (Kpc) & 7.7113\\
$\chi^2_{red}$ & 4.67298 \\
\hline
\end{tabular}
\label{evaluationextendedIC4202}
\end{table}

\subsubsection{The Galaxy UGC00128, Non-viable}

For this galaxy, the optimization method we used, ensures maximum
compatibility of the analytic SIDM model of Eq.
(\ref{ScaledependentEoSDM}) with the SPARC data, if we choose
$\rho_0=7.17146\times 10^6$$M_{\odot}/\mathrm{Kpc}^{3}$ and
$K_0=7652.87
$$M_{\odot} \, \mathrm{Kpc}^{-3} \, (\mathrm{km/s})^{2}$, in which
case the reduced $\chi^2_{red}$ value is $\chi^2_{red}=40.4415$.
Also the parameter $\alpha$ in this case is $\alpha=18.8521 $Kpc.

\begin{figure}[h!]
\centering
\includegraphics[width=20pc]{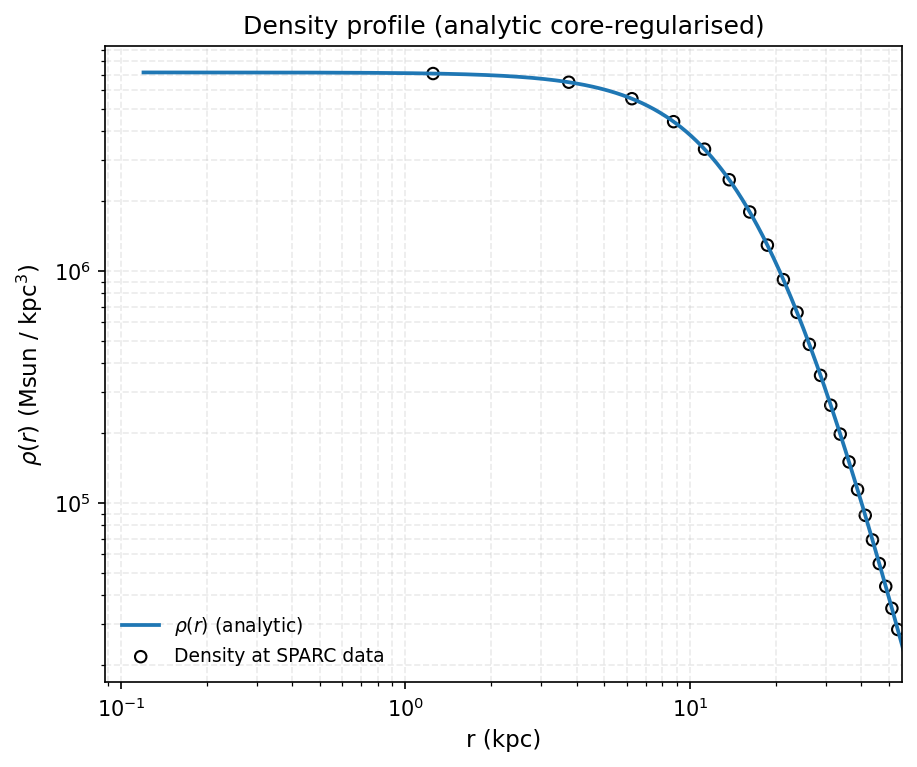}
\caption{The density of the SIDM model of Eq.
(\ref{ScaledependentEoSDM}) for the galaxy UGC00128, versus the
radius.} \label{UGC00128dens}
\end{figure}

\begin{figure}[h!]
\centering
\includegraphics[width=35pc]{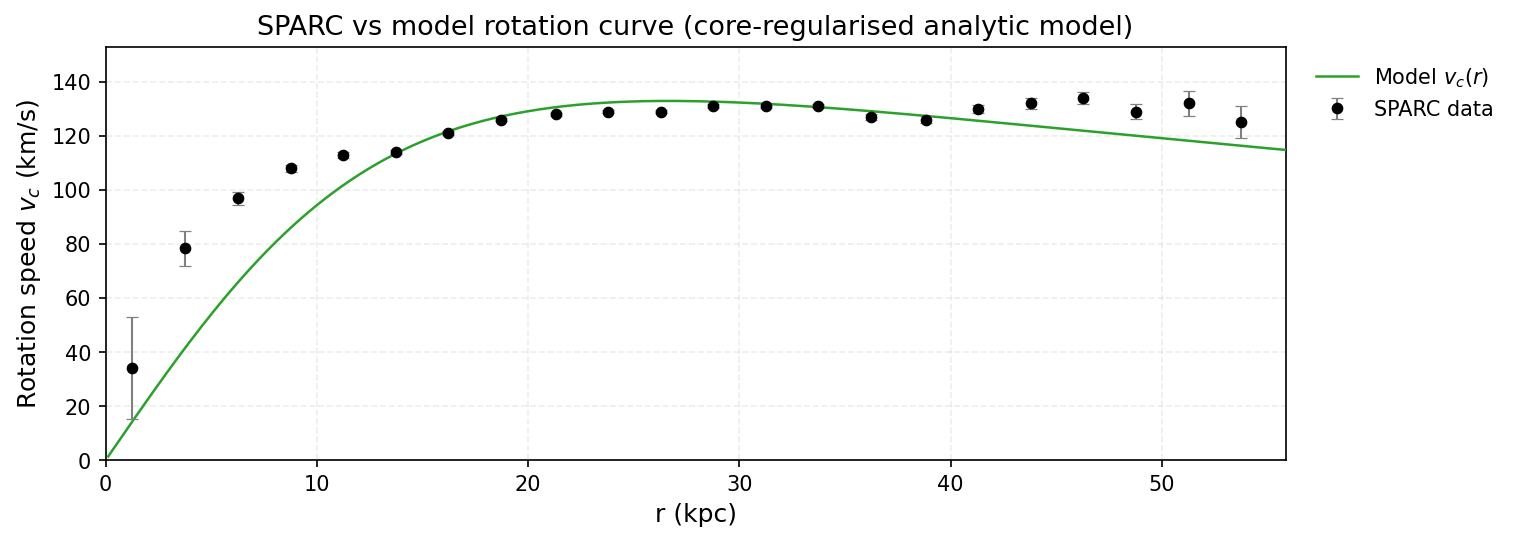}
\caption{The predicted rotation curves for the optimized SIDM
model of Eq. (\ref{ScaledependentEoSDM}), versus the SPARC
observational data for the galaxy UGC00128.} \label{UGC00128}
\end{figure}

In Table \ref{collUGC00128} we present the optimized values of
$K_0$ and $\rho_0$ for the analytic SIDM model of Eq.
(\ref{ScaledependentEoSDM}) for which the maximum compatibility
with the SPARC data is achieved.
\begin{table}[h!]
  \begin{center}
    \caption{SIDM Optimization Values for the galaxy UGC00128}
    \label{collUGC00128}
     \begin{tabular}{|r|r|}
     \hline
      \textbf{Parameter}   & \textbf{Optimization Values}
      \\  \hline
     $\rho_0 $  ($M_{\odot}/\mathrm{Kpc}^{3}$) & $7.17146\times 10^6$
\\  \hline $K_0$ ($M_{\odot} \,
\mathrm{Kpc}^{-3} \, (\mathrm{km/s})^{2}$)& 7652.87
\\  \hline
    \end{tabular}
  \end{center}
\end{table}
In Figs. \ref{UGC00128dens}, \ref{UGC00128}  we present the
density of the analytic SIDM model, the predicted rotation curves
for the SIDM model (\ref{ScaledependentEoSDM}), versus the SPARC
observational data and the sound speed, as a function of the
radius respectively. As it can be seen, for this galaxy, the SIDM
model produces non-viable rotation curves which are incompatible
with the SPARC data.

\begin{figure}[h!]
\centering
\includegraphics[width=20pc]{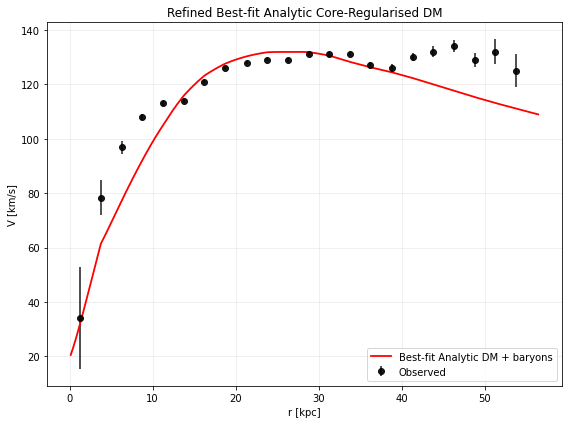}
\caption{The predicted rotation curves after using an optimization
for the SIDM model (\ref{ScaledependentEoSDM}), and the extended
SPARC data for the galaxy UGC00128. We included the rotation
curves of the gas, the disk velocities, the bulge (where present)
along with the SIDM model.} \label{extendedUGC00128}
\end{figure}

Now we shall include contributions to the rotation velocity from
the other components of the galaxy, namely the disk, the gas, and
the bulge if present. In Fig. \ref{extendedUGC00128} we present
the combined rotation curves including all the components of the
galaxy along with the SIDM. As it can be seen, the extended
collisional DM model is non-viable.
Also in Table \ref{evaluationextendedUGC00128} we present the
optimized values of the free parameters of the SIDM model for
which  we achieve the maximum compatibility with the SPARC data,
for the galaxy UGC00128, and also the resulting reduced
$\chi^2_{red}$ value.
\begin{table}[h!]
\centering \caption{Optimized Parameter Values of the Extended
SIDM model for the Galaxy UGC00128.}
\begin{tabular}{lc}
\hline
Parameter & Value  \\
\hline
$\rho_0 $ ($M_{\odot}/\mathrm{Kpc}^{3}$) & $5.5792\times 10^6$   \\
$K_0$ ($M_{\odot} \,
\mathrm{Kpc}^{-3} \, (\mathrm{km/s})^{2}$) & 5775.12   \\
$ml_{\text{disk}}$ & 1 \\
$ml_{\text{bulge}}$ & 0.3491 \\
$\alpha$ (Kpc) & 18.5648\\
$\chi^2_{red}$ & 30.8249 \\
\hline
\end{tabular}
\label{evaluationextendedUGC00128}
\end{table}

\subsubsection{The Galaxy UGC00191, Non-viable}

For this galaxy, the optimization method we used, ensures maximum
compatibility of the analytic SIDM model of Eq.
(\ref{ScaledependentEoSDM}) with the SPARC data, if we choose
$\rho_0=1.02285\times 10^8$$M_{\odot}/\mathrm{Kpc}^{3}$ and
$K_0=3293.19
$$M_{\odot} \, \mathrm{Kpc}^{-3} \, (\mathrm{km/s})^{2}$, in which
case the reduced $\chi^2_{red}$ value is $\chi^2_{red}=26.1093$.
Also the parameter $\alpha$ in this case is $\alpha=3.27457 $Kpc.

\begin{figure}[h!]
\centering
\includegraphics[width=20pc]{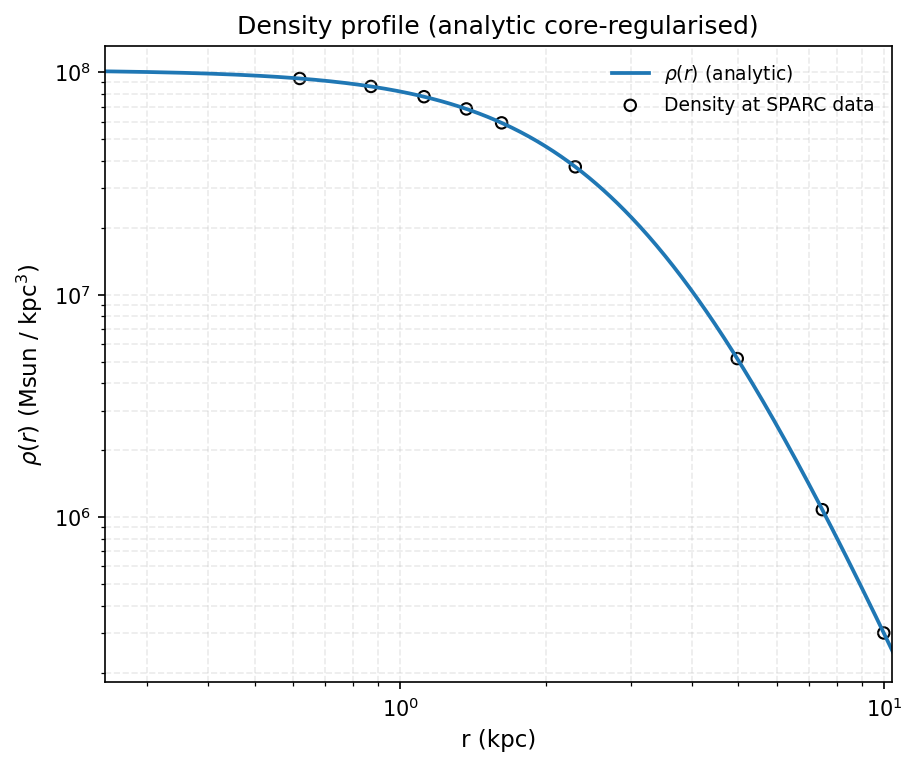}
\caption{The density of the SIDM model of Eq.
(\ref{ScaledependentEoSDM}) for the galaxy UGC00191, versus the
radius.} \label{UGC00191dens}
\end{figure}

\begin{figure}[h!]
\centering
\includegraphics[width=35pc]{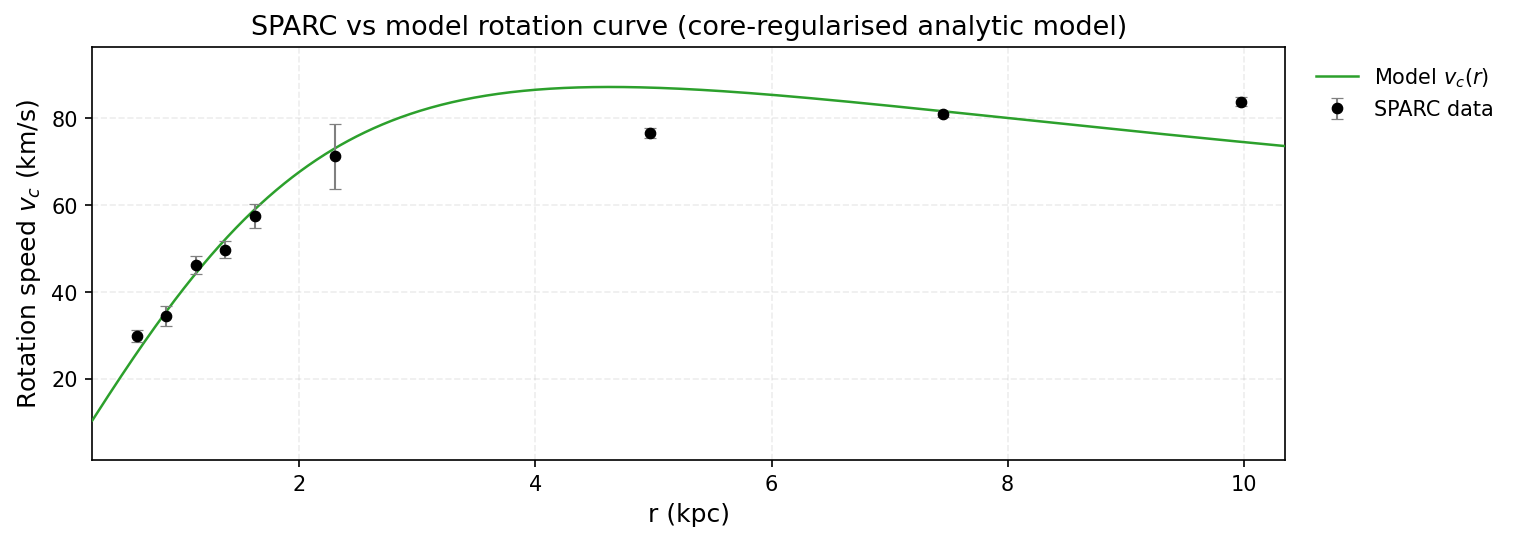}
\caption{The predicted rotation curves for the optimized SIDM
model of Eq. (\ref{ScaledependentEoSDM}), versus the SPARC
observational data for the galaxy UGC00191.} \label{UGC00191}
\end{figure}

In Table \ref{collUGC00191} we present the optimized values of
$K_0$ and $\rho_0$ for the analytic SIDM model of Eq.
(\ref{ScaledependentEoSDM}) for which the maximum compatibility
with the SPARC data is achieved.
\begin{table}[h!]
  \begin{center}
    \caption{SIDM Optimization Values for the galaxy UGC00191}
    \label{collUGC00191}
     \begin{tabular}{|r|r|}
     \hline
      \textbf{Parameter}   & \textbf{Optimization Values}
      \\  \hline
     $\rho_0 $  ($M_{\odot}/\mathrm{Kpc}^{3}$) & $1.02285\times 10^8$
\\  \hline $K_0$ ($M_{\odot} \,
\mathrm{Kpc}^{-3} \, (\mathrm{km/s})^{2}$)& 3293.19
\\  \hline
    \end{tabular}
  \end{center}
\end{table}
In Figs. \ref{UGC00191dens}, \ref{UGC00191} we present the density
of the analytic SIDM model, the predicted rotation curves for the
SIDM model (\ref{ScaledependentEoSDM}), versus the SPARC
observational data and the sound speed, as a function of the
radius respectively. As it can be seen, for this galaxy, the SIDM
model produces non-viable rotation curves which are incompatible
with the SPARC data.

\begin{figure}[h!]
\centering
\includegraphics[width=20pc]{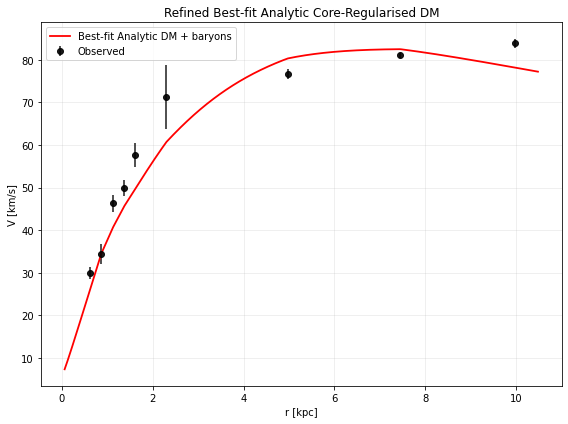}
\caption{The predicted rotation curves after using an optimization
for the SIDM model (\ref{ScaledependentEoSDM}), and the extended
SPARC data for the galaxy UGC00191. We included the rotation
curves of the gas, the disk velocities, the bulge (where present)
along with the SIDM model.} \label{extendedUGC00191}
\end{figure}

Now we shall include contributions to the rotation velocity from
the other components of the galaxy, namely the disk, the gas, and
the bulge if present. In Fig. \ref{extendedUGC00191} we present
the combined rotation curves including all the components of the
galaxy along with the SIDM. As it can be seen, the extended
collisional DM model is non-viable.
Also in Table \ref{evaluationextendedUGC00191} we present the
optimized values of the free parameters of the SIDM model for
which  we achieve the maximum compatibility with the SPARC data,
for the galaxy UGC00191, and also the resulting reduced
$\chi^2_{red}$ value.
\begin{table}[h!]
\centering \caption{Optimized Parameter Values of the Extended
SIDM model for the Galaxy UGC00191.}
\begin{tabular}{lc}
\hline
Parameter & Value  \\
\hline
$\rho_0 $ ($M_{\odot}/\mathrm{Kpc}^{3}$) & $2.92141\times 10^7$   \\
$K_0$ ($M_{\odot} \,
\mathrm{Kpc}^{-3} \, (\mathrm{km/s})^{2}$) & 1897.33   \\
$ml_{\text{disk}}$ & 1 \\
$ml_{\text{bulge}}$ & 0.2668 \\
$\alpha$ (Kpc) & 4.6502\\
$\chi^2_{red}$ & 15.5091 \\
\hline
\end{tabular}
\label{evaluationextendedUGC00191}
\end{table}

\subsubsection{The Galaxy UGC00731, Non-viable}

For this galaxy, the optimization method we used, ensures maximum
compatibility of the analytic SIDM model of Eq.
(\ref{ScaledependentEoSDM}) with the SPARC data, if we choose
$\rho_0=2.29854\times 10^7$$M_{\odot}/\mathrm{Kpc}^{3}$ and
$K_0=2228.14
$$M_{\odot} \, \mathrm{Kpc}^{-3} \, (\mathrm{km/s})^{2}$, in which
case the reduced $\chi^2_{red}$ value is $\chi^2_{red}=5.02761$.
Also the parameter $\alpha$ in this case is $\alpha=5.68194 $Kpc.

\begin{figure}[h!]
\centering
\includegraphics[width=20pc]{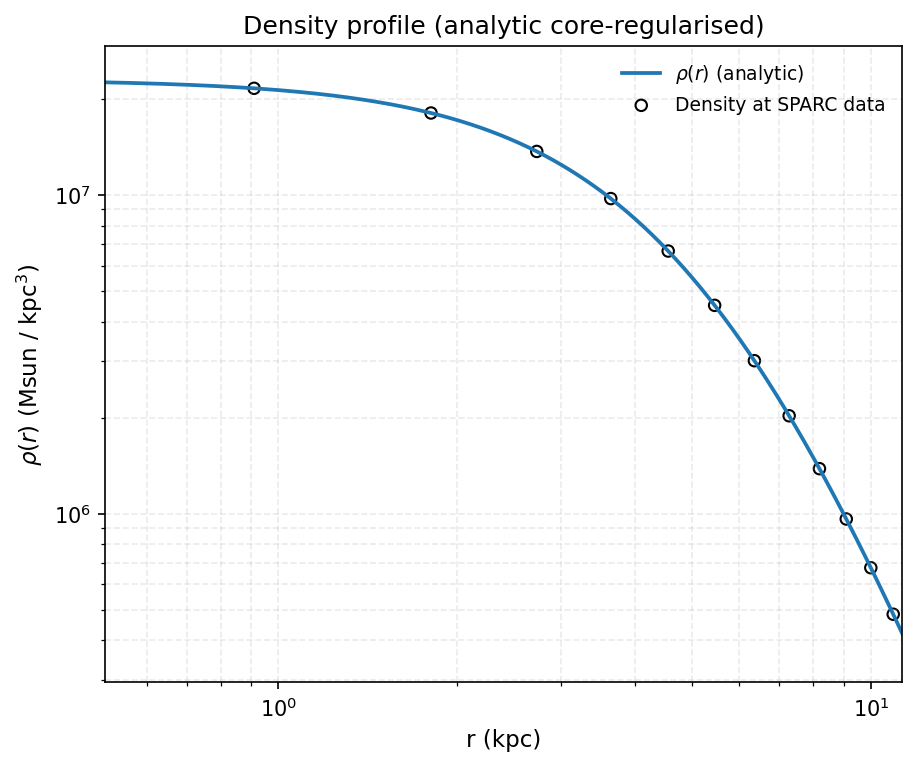}
\caption{The density of the SIDM model of Eq.
(\ref{ScaledependentEoSDM}) for the galaxy UGC00731, versus the
radius.} \label{UGC00731dens}
\end{figure}

\begin{figure}[h!]
\centering
\includegraphics[width=35pc]{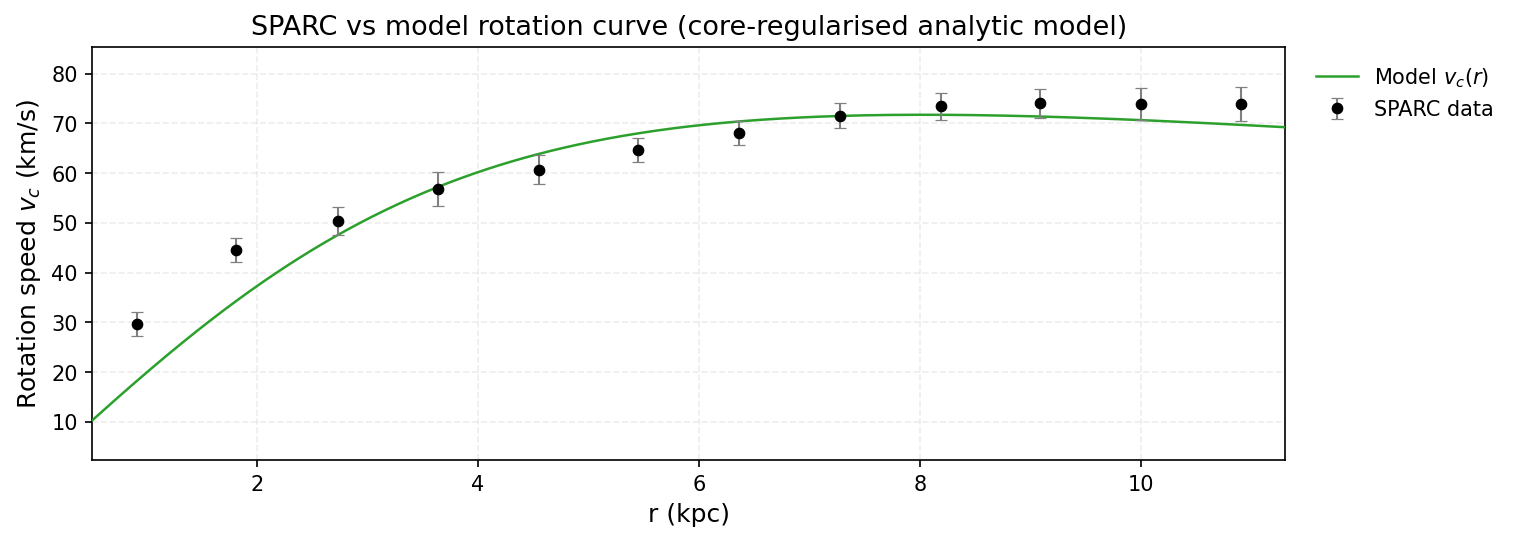}
\caption{The predicted rotation curves for the optimized SIDM
model of Eq. (\ref{ScaledependentEoSDM}), versus the SPARC
observational data for the galaxy UGC00731.} \label{UGC00731}
\end{figure}

In Table \ref{collUGC00731} we present the optimized values of
$K_0$ and $\rho_0$ for the analytic SIDM model of Eq.
(\ref{ScaledependentEoSDM}) for which the maximum compatibility
with the SPARC data is achieved.
\begin{table}[h!]
  \begin{center}
    \caption{SIDM Optimization Values for the galaxy UGC00731}
    \label{collUGC00731}
     \begin{tabular}{|r|r|}
     \hline
      \textbf{Parameter}   & \textbf{Optimization Values}
      \\  \hline
     $\rho_0 $  ($M_{\odot}/\mathrm{Kpc}^{3}$) & $2.29854\times 10^7$
\\  \hline $K_0$ ($M_{\odot} \,
\mathrm{Kpc}^{-3} \, (\mathrm{km/s})^{2}$)& 2228.14
\\  \hline
    \end{tabular}
  \end{center}
\end{table}
In Figs. \ref{UGC00731dens}, \ref{UGC00731} we present the density
of the analytic SIDM model, the predicted rotation curves for the
SIDM model (\ref{ScaledependentEoSDM}), versus the SPARC
observational data and the sound speed, as a function of the
radius respectively. As it can be seen, for this galaxy, the SIDM
model produces non-viable rotation curves which are incompatible
with the SPARC data.

\begin{figure}[h!]
\centering
\includegraphics[width=20pc]{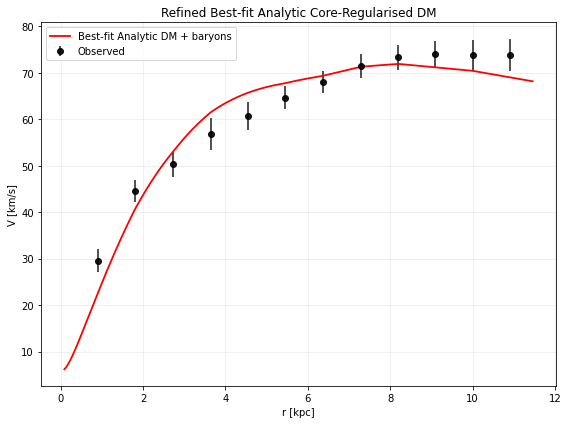}
\caption{The predicted rotation curves after using an optimization
for the SIDM model (\ref{ScaledependentEoSDM}), and the extended
SPARC data for the galaxy UGC00731. We included the rotation
curves of the gas, the disk velocities, the bulge (where present)
along with the SIDM model.} \label{extendedUGC00731}
\end{figure}

Now we shall include contributions to the rotation velocity from
the other components of the galaxy, namely the disk, the gas, and
the bulge if present. In Fig. \ref{extendedUGC00731} we present
the combined rotation curves including all the components of the
galaxy along with the SIDM. As it can be seen, the extended
collisional DM model is non-viable.
Also in Table \ref{evaluationextendedUGC00731} we present the
optimized values of the free parameters of the SIDM model for
which  we achieve the maximum compatibility with the SPARC data,
for the galaxy UGC00731, and also the resulting reduced
$\chi^2_{red}$ value.
\begin{table}[h!]
\centering \caption{Optimized Parameter Values of the Extended
SIDM model for the Galaxy UGC00731.}
\begin{tabular}{lc}
\hline
Parameter & Value  \\
\hline
$\rho_0 $ ($M_{\odot}/\mathrm{Kpc}^{3}$) & $3.25951\times 10^7$   \\
$K_0$ ($M_{\odot} \,
\mathrm{Kpc}^{-3} \, (\mathrm{km/s})^{2}$) & 1759.7   \\
$ml_{\text{disk}}$ & 1 \\
$ml_{\text{bulge}}$ & 0.5214 \\
$\alpha$ (Kpc) & 4.23974\\
$\chi^2_{red}$ & 2.85359 \\
\hline
\end{tabular}
\label{evaluationextendedUGC00731}
\end{table}

\subsection{Viable Galaxies and Their Characteristics}

A central outcome of the present work is the clear distinction of
the SPARC galaxy sample into two main classes: firstly systems for
which our analytic isothermal-like halo model fits well the
rotation curves, and systems for which the rotation curves are not
fitted optimally by the analytic model. In this section we shall
discuss the common properties of these two classes and we shall
attempt to provide a physical interpretation of why the model
succeeds in the former while fails in the latter class of
galaxies.

The viable galaxies share several qualitative properties, first,
viable galaxies are mostly late-type systems, typically classified
in the literature as irregulars, dwarfs, or late-type (small size)
spirals. These galaxies are generally bulge-less, or have only
negligible bulge components. Their stellar mass distributions are
therefore dynamically subdominant over the radii of the galaxies,
and especially in the outer skirts of the galaxies, where the
rotation curves approach a quasi-flat regime, at least for the
spirals. Second, the viable galaxies tend to be gas-rich systems
with extended rotation curves. Observationally, these galaxies
exhibit well-defined measurements of the flat rotation velocity
$V_{flat}$ which is different from the maximum inner velocity
$V_{max}$. Thirdly, the viable galaxies typically show smooth and
slowly rising rotation curves, with a small-scale structure. This
behavior is consistent with systems that are close to a
hydrodynamical equilibrium and are not strongly perturbed by bar
structures. In these galaxies, the assumption of a hydrostatic
equilibrium for the DM halo, which is the basis of our analytic
model, is physically well motivated.

In contrast, a significant fraction of the non-viable systems are
early-type spirals or galaxies with prominent bulge components. In
these objects, the inner gravitational potential is mostly
dominated by baryons, and the rotation curve often reaches its
maximum value well inside the optical radius. Non-viable galaxies
also seem to show more complex rotation curve morphologies,
including  inner steep rises, strongly declining outer profiles,
and strong features that must be associated with the complex
physics of the spiral arms, the bars, or even warps. This
complexity indicates distinctive departures from the hydrodynamic
(and thermodynamic) equilibrium conditions assumed for our
analytic DM halo model. Thus, in the case of non-viable galaxies,
a single and smooth, mostly pressure-supported halo, cannot simply
reproduce the observed kinematics of the galaxies. In addition,
several non-viable galaxies are relatively compact systems for
which the baryonic mass remains dynamically important up to very
large radii. In these galaxies, the dark halo does not dominate
the gravitational potential regions where the model attempts to
achieve a quasi-flat velocity plateau. Hence, the interplay
between halo thermodynamics and the observed kinematics breaks
down. Our model basically describes a finite-mass, isothermal dark
halo, with a radius dependent pressure profile,
\[
P(r) = K(r)\rho(r),
\]
with the parameter $K_0$, the entropy parameter, encoding the
global thermodynamic state of the halo. In late-type, DM-dominated
galaxies, the outer rotation curve is mainly governed by the DM
halo, and thus the baryonic components sub-dominantly affect the
dynamics. Under these conditions, the hydrostatic equilibrium
provides a correlation between the entropy parameter $K_0$ and the
characteristic flat circular velocity.

Thus concluding, our results indicate that the analytic
isothermal, scale-dependent EoS DM halo model, is compatible with
galaxies that are DM dominated and structurally simple. For these
systems, the model links the halo thermodynamics to the rotation
curve kinematics and the baryonic mass.

\section{The Model Fails to Produce the Canonical Tully-Fisher Law}

One of the central goals of this work is to assess whether the
fully analytic, scale-dependent isothermal EoS DM model can
reproduce the observed baryonic scaling relations of disk
galaxies, and specifically the baryonic Tully-Fisher relation.
Firstly however, let us investigate the canonical Tully-Fisher
relation using an empirical, bottom-up approach. We shall
investigate whether the observed luminosity-velocity scaling
$L_0\sim V_{max}^4$ can emerge naturally in our model, from the
following relations, the $L_0-K_0$ relation and the $K_0-V_{max}$
relation. If both relations are well described by power-law
behavior, then the model should in principle predict an effective
Tully-Fisher relation based on the role of $K_0$, without any
direct coupling between the baryons and the kinematics. For our
analysis, we shall restrict ourselves to the subset of viable
galaxies, which are those for which the fully analytic,
scale-dependent isothermal EoS DM model can optimally reproduce
their rotation curves. We first fit the luminosity-halo relation
$L_0-K_0$, using a power-law ansatz of the form,
\begin{equation}
L_0 = A_L\,K_0^{\alpha}\, .
\end{equation}
Using the SPARC data for the viable galaxies, and the values of
$K_0$ we found, we obtained,
\begin{align}
A_L &= (1.63 \pm 2.91)\times 10^{-5}, \\
\alpha &= 1.62 \pm 0.19.
\end{align}
This result indicates a moderately strong scaling between the
luminosity and the thermodynamic normalization $K_0$ of the halo.
The slope $\alpha > 1$ implies that more luminous systems reside
in the halos with a substantially larger effective pressure
support.

\begin{figure}[h!]
\centering
\includegraphics[width=20pc]{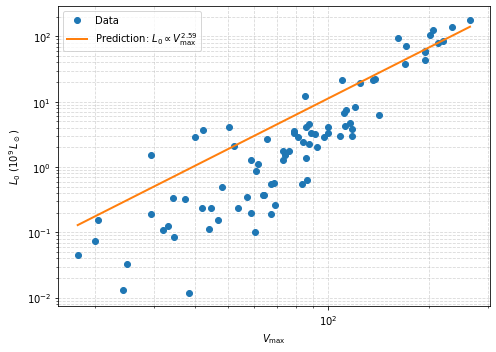}
\caption{The plot $L-V_{max}$ for the 116 viable galaxies we found
from the SPARC data and the optimal fit of it with the orange
line.} \label{plot1}
\end{figure}

Next, we fitted the relation between the halo normalization $K_0$
and the maximum circular velocity $V_{max}$,
\begin{equation}
K_0 = A_K\,V_{max}^{\beta},
\end{equation}
which yielded the following results,
\begin{align}
A_K &= 2.57 \pm 0.96, \\
\beta &= 1.60 \pm 0.07.
\end{align}
This relation suggests that $K_0$ is strongly correlated to the
depth of the gravitational potential, and this result may be
interpreted as an effective thermodynamic characteristic of the
halo mass and velocity scale. By combining the two relations we
found above, the model predicts the following effective
luminosity-velocity $L_0-V_{max}$ relation,
\begin{equation}
L_0 \propto V_{\max}^{\gamma}, \qquad \gamma = \alpha\,\beta.
\end{equation}
Using the best-fit values from our results we find that the
analytic halo model predicts,
\begin{equation}
L_0 \propto V_{\max}^{2.6},
\end{equation}
which is significantly shallower than the well-known canonical
Tully-Fisher relation $L_0\sim V_{max}^4$. In Fig. \ref{plot1} we
present the $L_0-V_{max}$ relation for the the fully analytic,
scale-dependent isothermal EoS DM model. The discrepancy between
the predicted slope $\gamma \simeq 2.6$ and the observed value
near $4$ clearly demonstrates that the analytic isothermal
scale-dependent EoS DM halo model fails to reproduce the canonical
Tully-Fisher law.
This is in contrast to the nearly isothermal model we developed in
Ref. \cite{Oikonomou:2025bsi}, and it seems that the presence of a
polytropic exponent, even if it is nearly isothermal, plays a
role. We do not fully understand this result, since we expected
that the isothermal EoS will reproduce the exact canonical
Tully-Fisher relation. But is seems that this result might be
model-dependent, and for the analytic model, the canonical
Tully-Fisher relation is not reproduced. A correct and definitive
answer to this can be obtained if the model
\cite{Oikonomou:2025bsi} is analyzed through the prism of a purely
isothermal EoS. Then a direct comparison can be made, and we aim
to perform this analysis in a forthcoming article.

\section{The Baryonic Tully-Fisher Relation for the Viable Galaxies and its Derivation from $K_0-M_b$ and $K_0-V_{flat}$ Relations}

While the failure of the luminosity-based Tully-Fisher relation
highlights the limitations of the analytic scale-dependent EoS DM
model, the physically more fundamental relation related to purely
DM dominated galaxies, is the baryonic Tully-Fisher relation,
which links the total baryonic mass $M_b$ to the asymptotic flat
rotation velocity $V_{flat}$. Unlike the luminosity, the baryonic
mass is a conserved quantity and is not a quantity which is
directly affected by the star-formation history or from stellar
mass-to-light ratio variations. Consequently, the baryonic
Tully-Fisher provides a more clear probe of the underlying
dynamical equilibrium of galaxies. And also, since the flat
velocity is used, it is more related to the strength of the DM
gravitational potential.

\begin{figure}[h!]
\centering
\includegraphics[width=20pc]{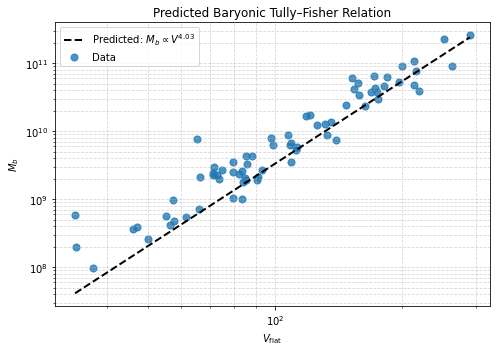}
\caption{The plot $K_0-V_{flat}$ for the viable galaxies we found
from the SPARC data, for which a $V_{flat}$ is given. The orange
line is the best fit.} \label{plot2}
\end{figure}

In this section, we shall investigate whether the analytic
isothermal scale-dependent EoS DM model, is capable of reproducing
the baryonic Tully-Fisher relation, when the asymptotic velocity
$V_{flat}$ is used as the relevant kinematic variable and the
baryonic mass is used instead of the luminosity. We proceed in
direct analogy with the previous luminosity-based analysis.
Firstly, let us examine the correlation between the halo
thermodynamic normalization parameter $K_0$ and the flat rotation
velocity $V_{flat}$, fitting a power-law relation of the form
\begin{equation}
K_0 = A\,V_{flat}^{n}.
\end{equation}
The best-fit parameters we found are,
\begin{align}
A &= 2.62 \pm 1.21, \\
n &= 1.61 \pm 0.09.
\end{align}
Next, we fit the baryonic mass as a function of the halo
normalization parameter $K_0$,
\begin{equation}
M_b = A\,K_0^{n},
\end{equation}
and in this case we found,
\begin{align}
A &= 2.68 \pm 4.91, \\
n &= 2.49 \pm 0.19.
\end{align}
Combining the two empirical relations we found above, the model
predicts a baryonic Tully-Fisher relation of the form,
\begin{equation}
M_b \propto V_{flat}^{\gamma}, \qquad \gamma = n_{(M_b\!-\!K_0)}
\times n_{(K_0\!-\!V_{flat})}.
\end{equation}
Using the best-fit values, we get,
\begin{equation}
\gamma = 4.03 \pm 0.37,
\end{equation}
which is fully consistent with the observed baryonic Tully-Fisher
relation,
\begin{equation}
M_b \propto V_{\mathrm{flat}}^{4}.
\end{equation}
Also in Fig. \ref{plot2} we present the resulting $M_b-V_{flat}$
relation using the intermediate relations $M_b-K_0$ and
$K_0-V_{flat}$. The orange line is the optimal fitting.
Thus the analytic isothermal scale-dependent EoS DM model is
capable of producing the baryonic Tully-Fisher relation. The key
distinction relative to the failed luminosity-based canonical
Tully-Fisher relation lies, probably, in using of $V_{flat}$
instead of $V_{max}$. The flat velocity probes directly the outer
halo, which is dominated by DM, and where the galactic system
reaches the hydrostatic equilibrium. There, the analytic DM
density profile provides an accurate description of the DM
gravitational potential. Hence, the effective pressure radial
support embodied in the entropy parameter $K_0$ directly controls
the flat circular velocity. However, in Ref.
\cite{Oikonomou:2025bsi}, both the canonical and the baryonic
Tully-Fisher relation were produced in a semi-empirical and
semi-theoretical way. In the present framework only the baryonic
Tully-Fisher relation was derived by the $M_b-K_0$ and
$K_0-V_{flat}$ relations. The reason for this difference has to be
model-dependent, and we will be certain on this, after further
work is done on the model developed \cite{Oikonomou:2025bsi} in
its purely isothermal limit.

\section{Conclusions}

In this work we extended the scale-dependent EoS self-interacting
DM framework developed in Ref. \cite{Oikonomou:2025bsi}, and
provided an analytic framework of scale-dependent EoS
self-interacting DM with an isothermal EoS. Our model can provide
perfect fit for the rotation curves of 116 galaxies taken from the
SPARC data \cite{Lelli:2016zqa}, however 59 galaxies cannot be
fitted optimally. The rotation curves of the 116 viable galaxies
are either described solely by the analytic scale-dependent EoS
self-interacting DM, or by the combination of the latter with the
other galaxy components such as the bulge (where present), the
disk and the gas. The viable galaxies consist mostly of late-time
galaxies, which are low-luminosity spirals, low-mass spirals, low
surface brightness spirals, irregular galaxies and dwarfs. On the
other hand, the non-viable galaxies consist of early-time objects,
such as massive spirals with prominent bulge and bar structures.
For the viable galaxies, we also thoroughly sought for
correlations of the entropy parameter $K_0$ with the maximum
velocity of each galaxy $V_{max}$, the flat velocity $V_{flat}$,
the luminosity $L_0$ and the baryonic mass $M_b$. In this way the
correlation between $K_0-V_{flat}$ and $M_b-K_0$ for the viable
galaxies, yielded naturally the baryonic Tully-Fisher relation
$M_b\sim V_{flat}^4$. However, this model failed to produce the
canonical Tully-Fisher relation via the correlations $L_0-K_0$ and
$K_0-V_{max}$. On the contrary the model used in Ref.
\cite{Oikonomou:2025bsi} was able to reproduce both the baryonic
Tully-Fisher relation and the canonical Tully-Fisher relation
almost naturally. Honestly we cannot explain this feature of the
present work. Intuitively, one expects that the findings of
\cite{Oikonomou:2025bsi} would apply in the present context too,
because the model \cite{Oikonomou:2025bsi} yields viable results
for a near isothermal scale-dependent EoS. And one naturally
expects the correlations $K_0\sim V_{max}^2$ and $K_0-V_{flat}^2$
do hold true in nearly isothermal halos. We did not find such
evidence for the present model and we do not fully understand why.
Perhaps it is a model dependent feature, and the only way to be
certain is to take the direct isothermal limit of Ref.
\cite{Oikonomou:2025bsi} and compare the results. If the findings
of the new research directive do not yield the correlations
$K_0\sim V_{max}^2$ and $K_0-V_{flat}^2$, this will be a clear
indication that the presence of an exponent in the
radius-dependent EoS
$P(r)=K(r)\left(\frac{\rho(r)}{\rho_{\star}}\right)^{\gamma(r)}$,
even if it is nearly isothermal, eventually plays a role. If the
correlations $K_0\sim V_{max}^2$ and $K_0-V_{flat}^2$ are
validated even for the isothermal version of Ref.
\cite{Oikonomou:2025bsi}, then this would imply that the
correlations $K_0\sim V_{max}^2$ and $K_0-V_{flat}^2$ are somehow
a model dependent feature. We suspect that the presence of a
non-trivial exponent, even if it is nearly isothermal, stabilizes
the halo and provides more concrete radial pressure support that
eventually realizes the correlations $K_0\sim V_{max}^2$ and
$K_0-V_{flat}^2$ in the halo. Work and research is planned towards
this problem and we expect to report on this issue soon.

Of course, there is much work to be done even for the analytic
model we introduced, this was an introductory article. Having a
scale-dependent EoS self-interacting DM, this would mean that DM
is dissipative, so there is always the possibility of having a
dark disk \cite{Foot:2014uba,Fan:2013yva,McCullough:2013jma}. This
must be discussed also in view of the maximum disk hypothesis. The
gravothermal stability in these models must also be studied, but
these studies stretch far beyond the aims and scopes of this
article.

In addition, an interesting class of galaxies in the Universe that
might be perfectly theoretically harbored by the scale-dependent
EoS self-interacting DM framework are the dark galaxies. Recent
evidence from the Cloud-9 galaxy indicated that it is gas
dominated and DM dominated but contains no stars
\cite{Anand:2025czy,Benitez-Llambay:2024mcp,Benitez-Llambay:2023kyu}.
So it is a dark galaxy, not perfectly described (if described at
all) by the $\Lambda$CDM. Let us discuss this issue, because it is
rather physically interesting. The Cloud-9 system, is
characterized by extreme DM dominance, and also negligible stellar
luminosity, and challenges directly standard galaxy formation
scenarios which are implied by the $\Lambda$CDM model.

In the conventional $\Lambda$CDM models, such objects would
require a combination of inefficient star formation, strong
feedback, and a truly fine-tuned halo, in order to suppress the
baryonic collapse and in parallel, maintaining a dynamically cold
and extended halo. However, even under these assumptions,
reproducing the observed smoothness and equilibrium remains
problematic, as collisionless DM halos do not naturally thermalize
without invoking an additional baryonic regulation.

In contrast, the analytic model we developed here, based on a
radius-dependent isothermal EoS, provides a natural and
self-consistent framework for the Cloud-9 galaxy. In our
framework, the DM halo is an effective fluid with a pressure
proportional to density, which allows the system to reach a
hydrostatic equilibrium. The absence of a luminous disk is thus
not a pathology but a rather expected outcome of the DM halo,
without requiring baryon feedback to stabilize the galaxy. Hence,
the existence of Cloud-9 types of dark galaxies, supports the
notion that DM halos may possess an intrinsic, approximately
isothermal structure on galactic scales, with a pressure support
and the thermalization playing a central dynamical role.
$\Lambda$CDM cannot easily harbor such pressure supported
baryon-less galactic structures. On the other hand, DM emulator
theoretical constructions, such as MOND theories, cannot describe
Cloud-9 types of galaxies. Specifically, the Cloud-9 system itself
by observation, imposes fundamental problems for the MOND
description, precisely because its observed kinematics are not
those of a rotating disk galaxy. Cloud-9 is basically a starless
neutral atomic hydrogen (HI) cloud, the dynamical state of which
is inferred from the narrow $21\,$cm line widths, and the
associated spatially resolved gas morphology, and not from an
extended rotation curve, like in baryon existing spirals. The
observations indicate that the Cloud-9 system is dynamically cold,
it lacks rotation, and it is consistent with a gas cloud embedded
in a DM halo.

In the MOND theoretical construction, the gravitational dynamics
are determined entirely by the total distribution of the baryonic
matter, and thus MOND aims to explain the rotationally supported
systems where circular velocities trace the baryonic mass
distribution. For the Cloud-9, however, the baryonic content is
small-if present at all. The observed HI line width cannot be
straightforwardly interpreted as circular velocity, and even if it
was, the baryonic mass is very small to provide dynamical support
in the context of MOND without perhaps invoking unrealistically
strong external field effects. Moreover, MOND lacks a natural
framework to describe a pressure-supported, starless equilibrium
systems. The very own existence of a long-lived, and dynamically
coherent HI cloud with a negligible stellar mass and also no
rotational support, apparently contradicts the MOND central theme,
that the baryons by themselves can determine galactic dynamics. In
this sense, the Cloud--9 is a sound example of objects that
directly contradict the core axioms of MOND, and regardless if
fine-tunings are used, MOND cannot describe dark galaxies at all.
The scale-dependent EoS self-interacting DM provides a natural
theoretical framework that can harbor pressure supported hydrogen
clouds embedded in a pressure self-regulated DM halo in
hydrodynamic and thermodynamic equilibrium.

\appendix

\section*{Appendix: Complete List of Galactic Rotation Curves Simulations for All the SPARC Galaxies}

In this appendix we present the full analysis of the fitting of
the analytic model of scale-dependent EoS DM with the SPARC
galaxies .


\subsection{The Galaxy D631-7}

For this galaxy, the optimization method we used, ensures maximum
compatibility of the analytic SIDM model of Eq.
(\ref{ScaledependentEoSDM}) with the SPARC data, if we choose
$\rho_0=1.80239\times 10^7$$M_{\odot}/\mathrm{Kpc}^{3}$ and
$K_0=1349.32
$$M_{\odot} \, \mathrm{Kpc}^{-3} \, (\mathrm{km/s})^{2}$, in which
case the reduced $\chi^2_{red}$ value is $\chi^2_{red}=0.631699$.
Also the parameter $\alpha$ in this case is $\alpha=4.99326 $Kpc.

\begin{figure}[h!]
\centering
\includegraphics[width=20pc]{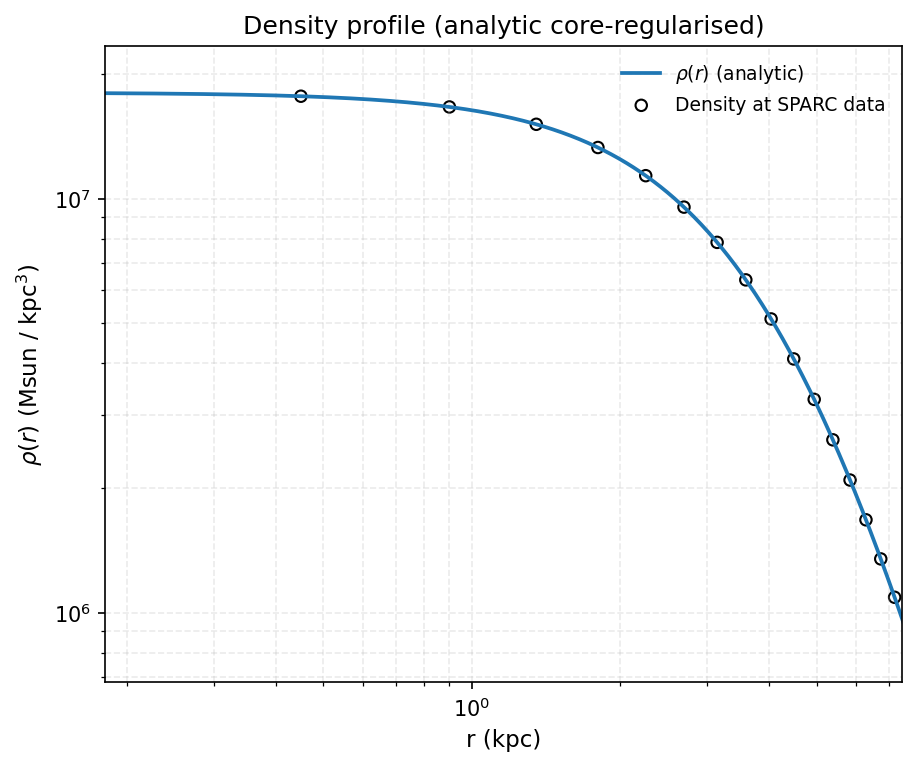}
\caption{The density of the SIDM model of Eq.
(\ref{ScaledependentEoSDM}) for the galaxy D631-7, versus the
radius.} \label{D631-7dens}
\end{figure}

\begin{figure}[h!]
\centering
\includegraphics[width=35pc]{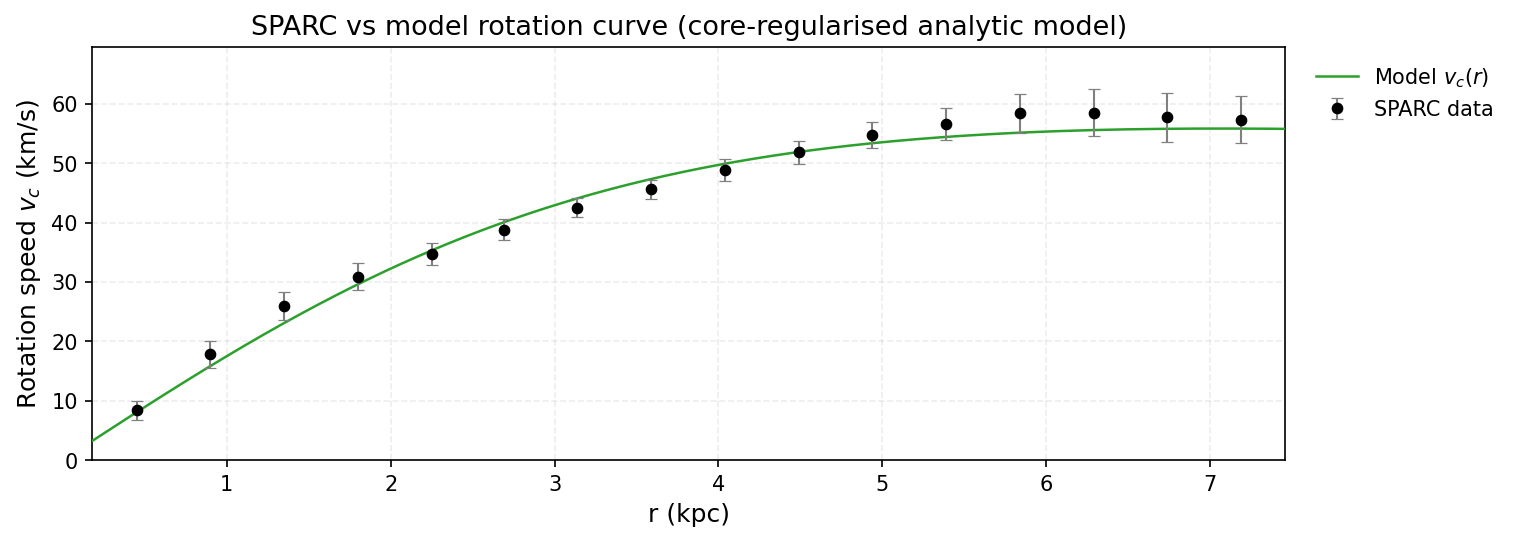}
\caption{The predicted rotation curves for the optimized SIDM
model of Eq. (\ref{ScaledependentEoSDM}), versus the SPARC
observational data for the galaxy D631-7.} \label{D631-7}
\end{figure}

In Table \ref{collD631-7} we present the optimized values of $K_0$
and $\rho_0$ for the analytic SIDM model of Eq.
(\ref{ScaledependentEoSDM}) for which the maximum compatibility
with the SPARC data is achieved.
\begin{table}[h!]
  \begin{center}
    \caption{SIDM Optimization Values for the galaxy D631-7}
    \label{collD631-7}
     \begin{tabular}{|r|r|}
     \hline
      \textbf{Parameter}   & \textbf{Optimization Values}
      \\  \hline
     $\rho_0 $  ($M_{\odot}/\mathrm{Kpc}^{3}$) & $1.80239\times 10^7$
\\  \hline $K_0$ ($M_{\odot} \,
\mathrm{Kpc}^{-3} \, (\mathrm{km/s})^{2}$)& 1349.32
\\  \hline
    \end{tabular}
  \end{center}
\end{table}
In Figs. \ref{D631-7dens}, \ref{D631-7}  we present the density of
the analytic SIDM model, the predicted rotation curves for the
SIDM model (\ref{ScaledependentEoSDM}), versus the SPARC
observational data and the sound speed, as a function of the
radius respectively. As it can be seen, for this galaxy, the SIDM
model produces viable rotation curves which are compatible with
the SPARC data.

\subsection{The Galaxy DDO064}

For this galaxy, the optimization method we used, ensures maximum
compatibility of the analytic SIDM model of Eq.
(\ref{ScaledependentEoSDM}) with the SPARC data, if we choose
$\rho_0=5.49286\times 10^7$$M_{\odot}/\mathrm{Kpc}^{3}$ and
$K_0=1385.86
$$M_{\odot} \, \mathrm{Kpc}^{-3} \, (\mathrm{km/s})^{2}$, in which
case the reduced $\chi^2_{red}$ value is $\chi^2_{red}=0.456646$.
Also the parameter $\alpha$ in this case is $\alpha=2.89875 $Kpc.

\begin{figure}[h!]
\centering
\includegraphics[width=20pc]{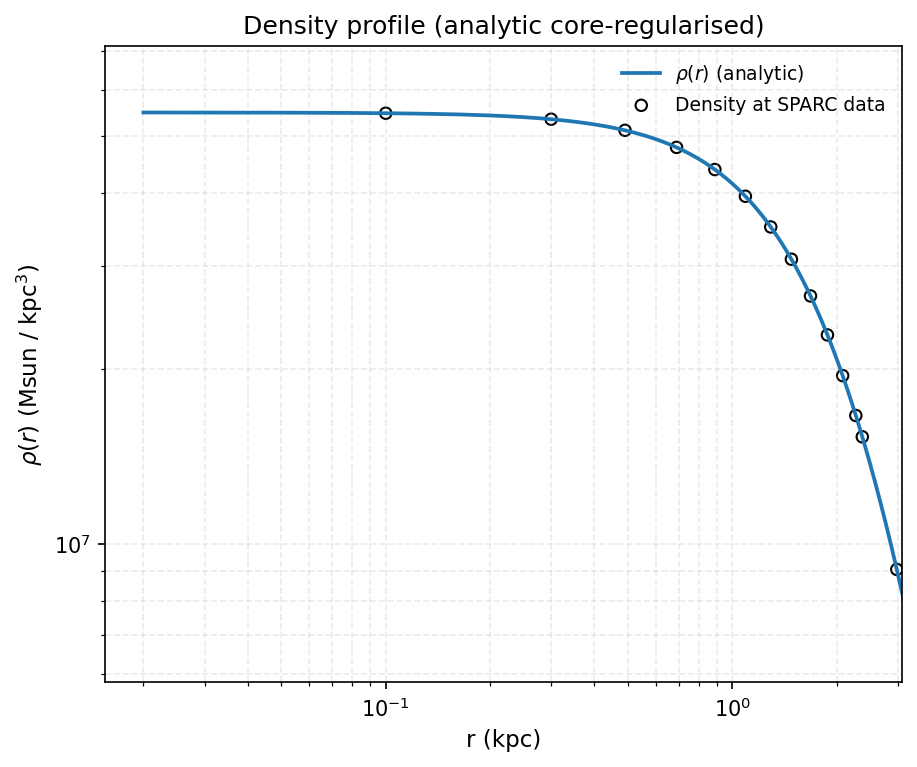}
\caption{The density of the SIDM model of Eq.
(\ref{ScaledependentEoSDM}) for the galaxy DDO064, versus the
radius.} \label{DDO064dens}
\end{figure}

\begin{figure}[h!]
\centering
\includegraphics[width=35pc]{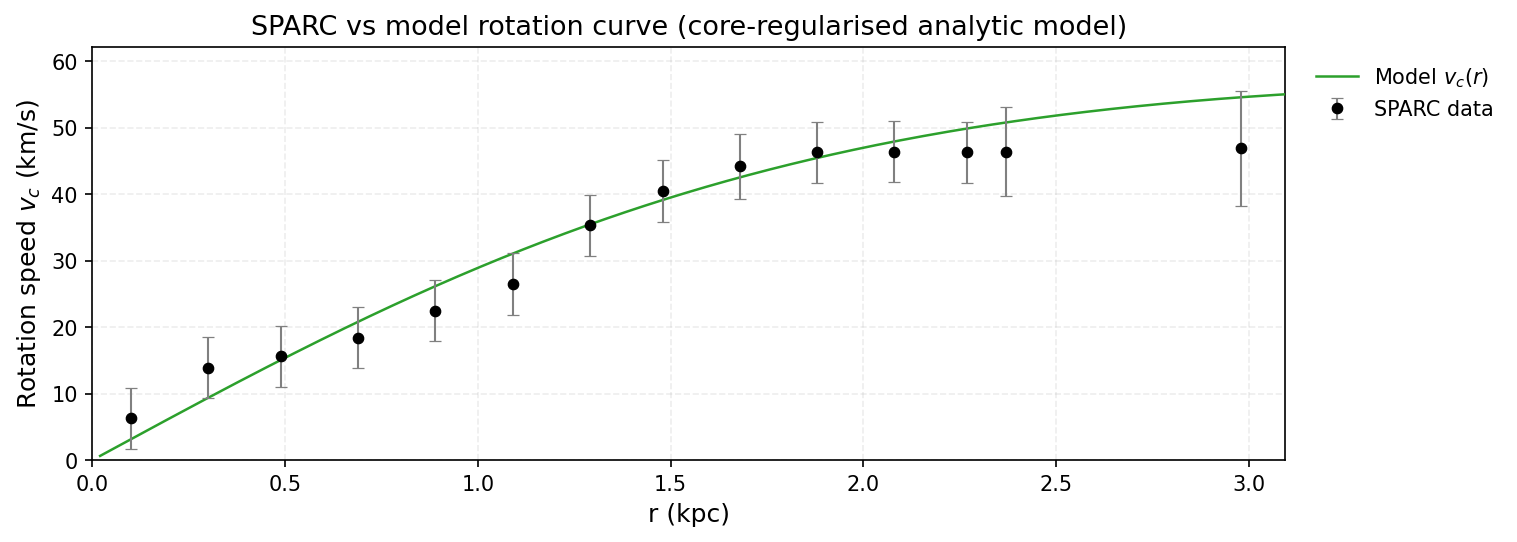}
\caption{The predicted rotation curves for the optimized SIDM
model of Eq. (\ref{ScaledependentEoSDM}), versus the SPARC
observational data for the galaxy DDO064.} \label{DDO064}
\end{figure}

In Table \ref{collDDO064} we present the optimized values of $K_0$
and $\rho_0$ for the analytic SIDM model of Eq.
(\ref{ScaledependentEoSDM}) for which the maximum compatibility
with the SPARC data is achieved.
\begin{table}[h!]
  \begin{center}
    \caption{SIDM Optimization Values for the galaxy DDO064}
    \label{collDDO064}
     \begin{tabular}{|r|r|}
     \hline
      \textbf{Parameter}   & \textbf{Optimization Values}
      \\  \hline
     $\rho_0 $  ($M_{\odot}/\mathrm{Kpc}^{3}$) & $5.49286\times 10^7$
\\  \hline $K_0$ ($M_{\odot} \,
\mathrm{Kpc}^{-3} \, (\mathrm{km/s})^{2}$)& 1385.86
\\  \hline
    \end{tabular}
  \end{center}
\end{table}
In Figs. \ref{DDO064dens}, \ref{DDO064} we present the density of
the analytic SIDM model, the predicted rotation curves for the
SIDM model (\ref{ScaledependentEoSDM}), versus the SPARC
observational data and the sound speed, as a function of the
radius respectively. As it can be seen, for this galaxy, the SIDM
model produces viable rotation curves which are compatible with
the SPARC data.


\subsection{The Galaxy DDO161, Non-viable, Extended Viable}

For this galaxy, the optimization method we used, ensures maximum
compatibility of the analytic SIDM model of Eq.
(\ref{ScaledependentEoSDM}) with the SPARC data, if we choose
$\rho_0=9.31455\times 10^6$$M_{\odot}/\mathrm{Kpc}^{3}$ and
$K_0=1875.28
$$M_{\odot} \, \mathrm{Kpc}^{-3} \, (\mathrm{km/s})^{2}$, in which
case the reduced $\chi^2_{red}$ value is $\chi^2_{red}=2.22523$.
Also the parameter $\alpha$ in this case is $\alpha=8.18847 $Kpc.

\begin{figure}[h!]
\centering
\includegraphics[width=20pc]{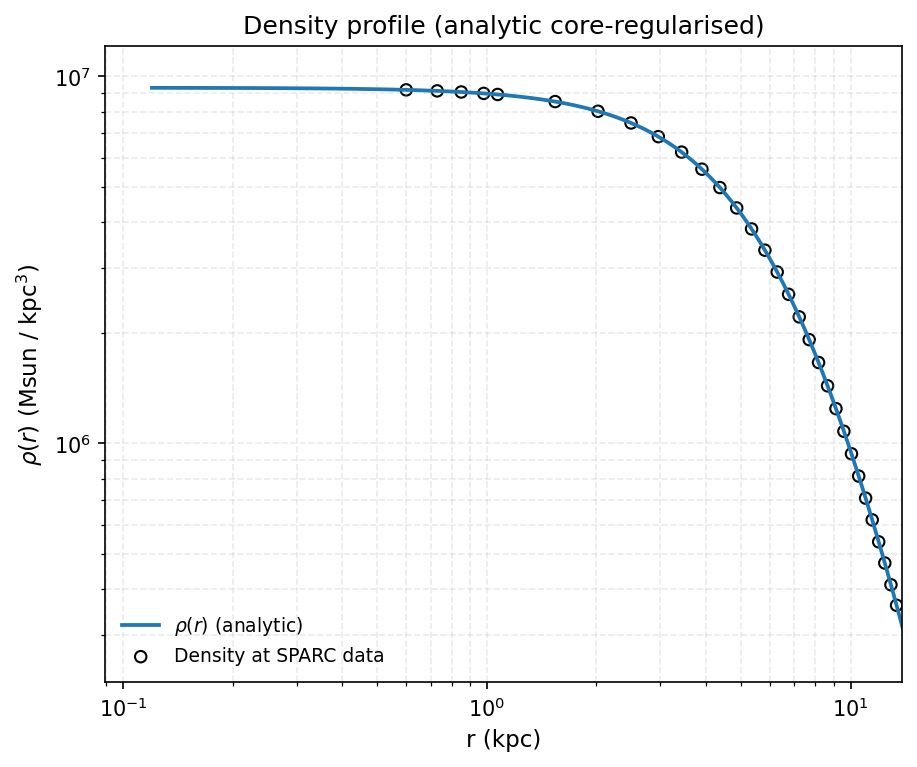}
\caption{The density of the SIDM model of Eq.
(\ref{ScaledependentEoSDM}) for the galaxy DDO161, versus the
radius.} \label{DDO161dens}
\end{figure}

\begin{figure}[h!]
\centering
\includegraphics[width=35pc]{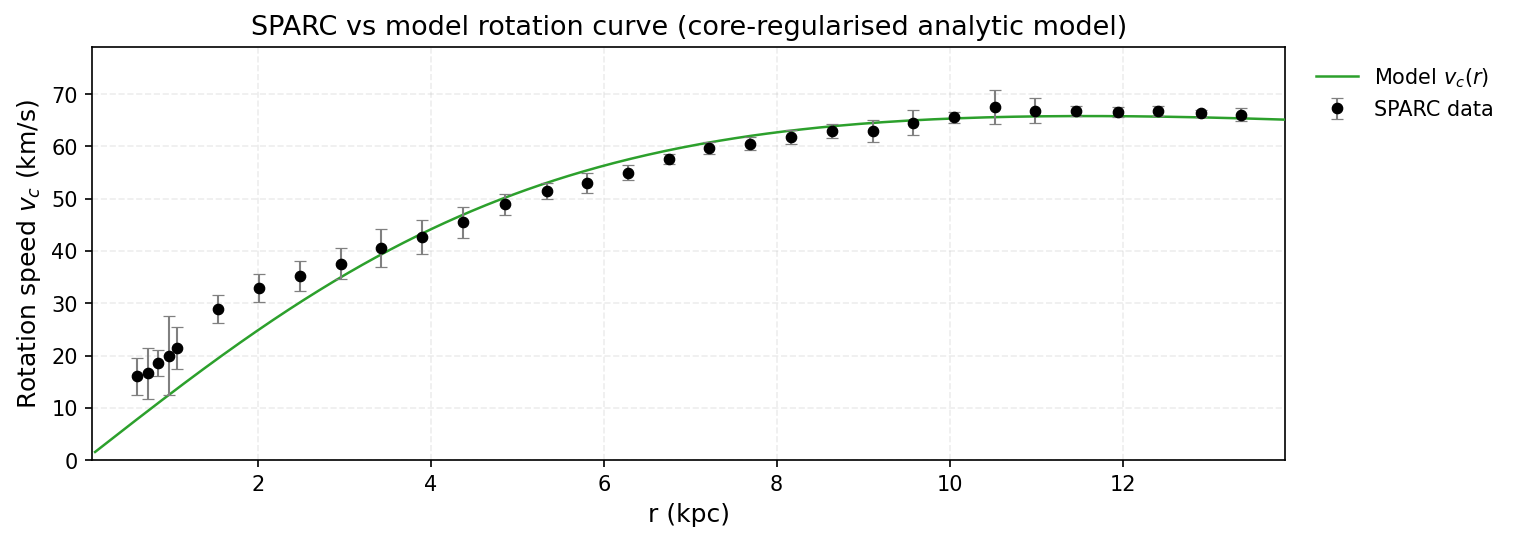}
\caption{The predicted rotation curves for the optimized SIDM
model of Eq. (\ref{ScaledependentEoSDM}), versus the SPARC
observational data for the galaxy DDO161.} \label{DDO161}
\end{figure}

In Table \ref{collDDO161} we present the optimized values of $K_0$
and $\rho_0$ for the analytic SIDM model of Eq.
(\ref{ScaledependentEoSDM}) for which the maximum compatibility
with the SPARC data is achieved.
\begin{table}[h!]
  \begin{center}
    \caption{SIDM Optimization Values for the galaxy DDO161}
    \label{collDDO161}
     \begin{tabular}{|r|r|}
     \hline
      \textbf{Parameter}   & \textbf{Optimization Values}
      \\  \hline
     $\rho_0 $  ($M_{\odot}/\mathrm{Kpc}^{3}$) & $9.31455\times 10^6$
\\  \hline $K_0$ ($M_{\odot} \,
\mathrm{Kpc}^{-3} \, (\mathrm{km/s})^{2}$)& 1875.28
\\  \hline
    \end{tabular}
  \end{center}
\end{table}
In Figs. \ref{DDO161dens}, \ref{DDO161} we present the density of
the analytic SIDM model, the predicted rotation curves for the
SIDM model (\ref{ScaledependentEoSDM}), versus the SPARC
observational data and the sound speed, as a function of the
radius respectively. As it can be seen, for this galaxy, the SIDM
model produces non-viable rotation curves which are incompatible
with the SPARC data.

\begin{figure}[h!]
\centering
\includegraphics[width=20pc]{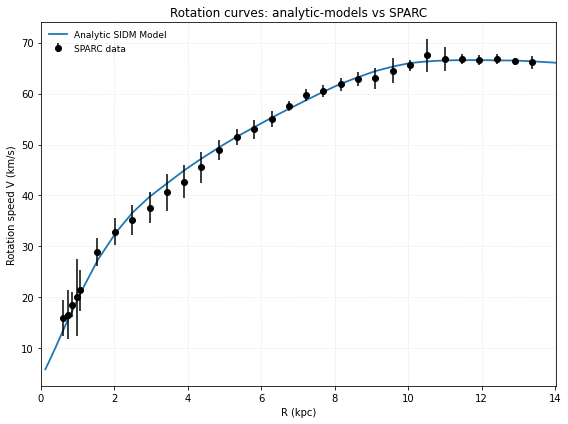}
\caption{The predicted rotation curves after using an optimization
for the SIDM model (\ref{ScaledependentEoSDM}), and the extended
SPARC data for the galaxy DDO161. We included the rotation curves
of the gas, the disk velocities, the bulge (where present) along
with the SIDM model.} \label{extendedDDO161}
\end{figure}

Now we shall include contributions to the rotation velocity from
the other components of the galaxy, namely the disk, the gas, and
the bulge if present. In Fig. \ref{extendedDDO161} we present the
combined rotation curves including all the components of the
galaxy along with the SIDM. As it can be seen, the extended
collisional DM model is viable.
Also in Table \ref{evaluationextendedDDO161} we present the
optimized values of the free parameters of the SIDM model for
which  we achieve the maximum compatibility with the SPARC data,
for the galaxy DDO161, and also the resulting reduced
$\chi^2_{red}$ value.
\begin{table}[h!]
\centering \caption{Optimized Parameter Values of the Extended
SIDM model for the Galaxy DDO161.}
\begin{tabular}{lc}
\hline
Parameter & Value  \\
\hline
$\rho_0 $ ($M_{\odot}/\mathrm{Kpc}^{3}$) & $4.03068 \times 10^6$   \\
$K_0$ ($M_{\odot} \,
\mathrm{Kpc}^{-3} \, (\mathrm{km/s})^{2}$) & 1549.15   \\
$ml_{\text{disk}}$ & 0.8271 \\
$ml_{\text{bulge}}$ & 0.5529 \\
$\alpha$ (Kpc) & \\
$\chi^2_{red}$ & 0.202594 \\
\hline
\end{tabular}
\label{evaluationextendedDDO161}
\end{table}

\subsection{The Galaxy DDO168, Non-viable}

For this galaxy, the optimization method we used, ensures maximum
compatibility of the analytic SIDM model of Eq.
(\ref{ScaledependentEoSDM}) with the SPARC data, if we choose
$\rho_0=4.32263\times 10^7$$M_{\odot}/\mathrm{Kpc}^{3}$ and
$K_0=1301.75
$$M_{\odot} \, \mathrm{Kpc}^{-3} \, (\mathrm{km/s})^{2}$, in which
case the reduced $\chi^2_{red}$ value is $\chi^2_{red}=2.5666$.
Also the parameter $\alpha$ in this case is $\alpha=3.16695 $Kpc.

\begin{figure}[h!]
\centering
\includegraphics[width=20pc]{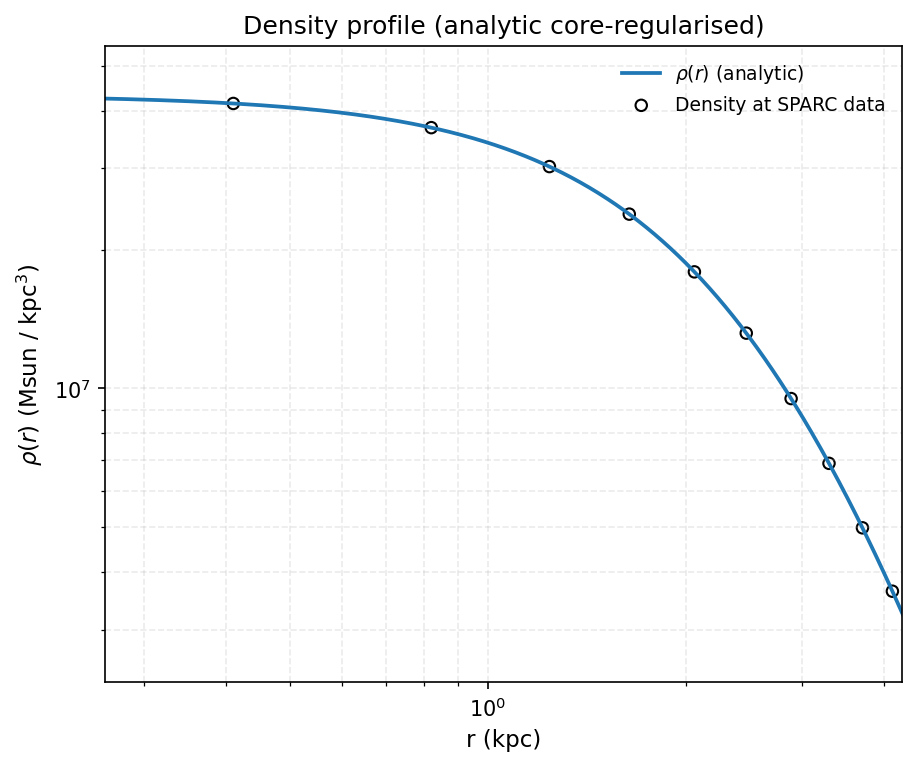}
\caption{The density of the SIDM model of Eq.
(\ref{ScaledependentEoSDM}) for the galaxy DDO168, versus the
radius.} \label{DDO168dens}
\end{figure}

\begin{figure}[h!]
\centering
\includegraphics[width=35pc]{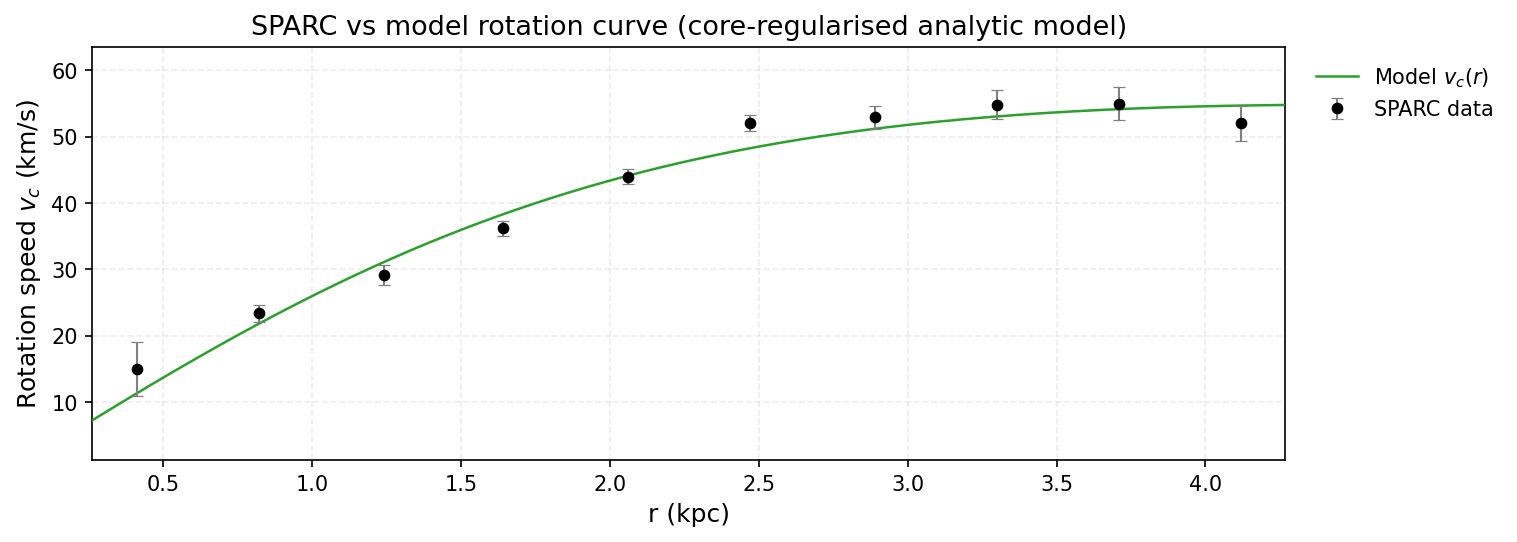}
\caption{The predicted rotation curves for the optimized SIDM
model of Eq. (\ref{ScaledependentEoSDM}), versus the SPARC
observational data for the galaxy DDO168.} \label{DDO168}
\end{figure}

In Table \ref{collDDO168} we present the optimized values of $K_0$
and $\rho_0$ for the analytic SIDM model of Eq.
(\ref{ScaledependentEoSDM}) for which the maximum compatibility
with the SPARC data is achieved.
\begin{table}[h!]
  \begin{center}
    \caption{SIDM Optimization Values for the galaxy DDO168}
    \label{collDDO168}
     \begin{tabular}{|r|r|}
     \hline
      \textbf{Parameter}   & \textbf{Optimization Values}
      \\  \hline
     $\rho_0 $  ($M_{\odot}/\mathrm{Kpc}^{3}$) & $4.32263\times 10^7$
\\  \hline $K_0$ ($M_{\odot} \,
\mathrm{Kpc}^{-3} \, (\mathrm{km/s})^{2}$)& 1301.75
\\  \hline
    \end{tabular}
  \end{center}
\end{table}
In Figs. \ref{DDO168dens}, \ref{DDO168} we present the density of
the analytic SIDM model, the predicted rotation curves for the
SIDM model (\ref{ScaledependentEoSDM}), versus the SPARC
observational data and the sound speed, as a function of the
radius respectively. As it can be seen, for this galaxy, the SIDM
model produces non-viable rotation curves which are incompatible
with the SPARC data.

\begin{figure}[h!]
\centering
\includegraphics[width=20pc]{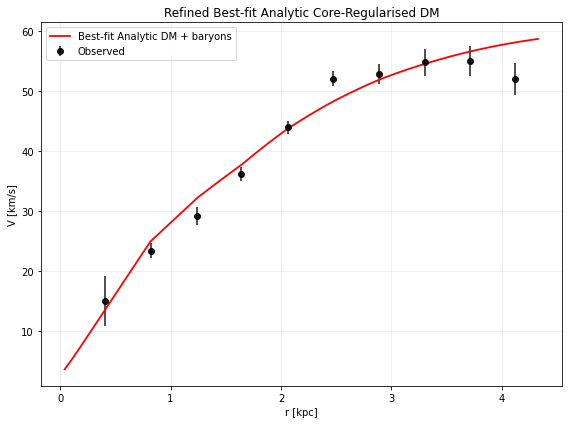}
\caption{The predicted rotation curves after using an optimization
for the SIDM model (\ref{ScaledependentEoSDM}), and the extended
SPARC data for the galaxy DDO168. We included the rotation curves
of the gas, the disk velocities, the bulge (where present) along
with the SIDM model.} \label{extendedDDO168}
\end{figure}

Now we shall include contributions to the rotation velocity from
the other components of the galaxy, namely the disk, the gas, and
the bulge if present. In Fig. \ref{extendedDDO168} we present the
combined rotation curves including all the components of the
galaxy along with the SIDM. As it can be seen, the extended
collisional DM model is non-viable.
Also in Table \ref{evaluationextendedDDO168} we present the
optimized values of the free parameters of the SIDM model for
which  we achieve the maximum compatibility with the SPARC data,
for the galaxy DDO168, and also the resulting reduced
$\chi^2_{red}$ value.
\begin{table}[h!]
\centering \caption{Optimized Parameter Values of the Extended
SIDM model for the Galaxy DDO168.}
\begin{tabular}{lc}
\hline
Parameter & Value  \\
\hline
$\rho_0 $ ($M_{\odot}/\mathrm{Kpc}^{3}$) & $2.59085\times 10^7$   \\
$K_0$ ($M_{\odot} \,
\mathrm{Kpc}^{-3} \, (\mathrm{km/s})^{2}$) & 1359.19   \\
$ml_{\text{disk}}$ & 0.0000 \\
$ml_{\text{bulge}}$ & 0.0738 \\
$\alpha$ (Kpc) & 4.17941 \\
$\chi^2_{red}$ & 3.92914 \\
\hline
\end{tabular}
\label{evaluationextendedDDO168}
\end{table}

\subsection{The Galaxy DDO170 Marginally}

For this galaxy, the optimization method we used, ensures maximum
compatibility of the analytic SIDM model of Eq.
(\ref{ScaledependentEoSDM}) with the SPARC data, if we choose
$\rho_0=1.14592\times 10^7$$M_{\odot}/\mathrm{Kpc}^{3}$ and
$K_0=1569.79
$$M_{\odot} \, \mathrm{Kpc}^{-3} \, (\mathrm{km/s})^{2}$, in which
case the reduced $\chi^2_{red}$ value is $\chi^2_{red}=2.79013$.
Also the parameter $\alpha$ in this case is $\alpha=6.75451 $Kpc.

\begin{figure}[h!]
\centering
\includegraphics[width=20pc]{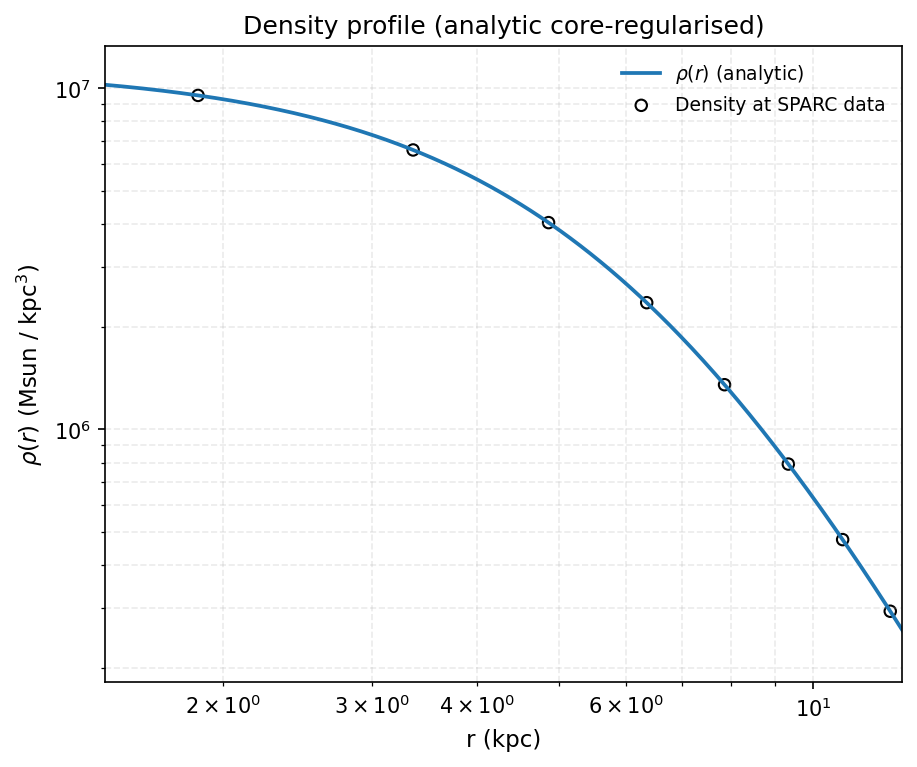}
\caption{The density of the SIDM model of Eq.
(\ref{ScaledependentEoSDM}) for the galaxy DDO170, versus the
radius.} \label{DDO170dens}
\end{figure}

\begin{figure}[h!]
\centering
\includegraphics[width=35pc]{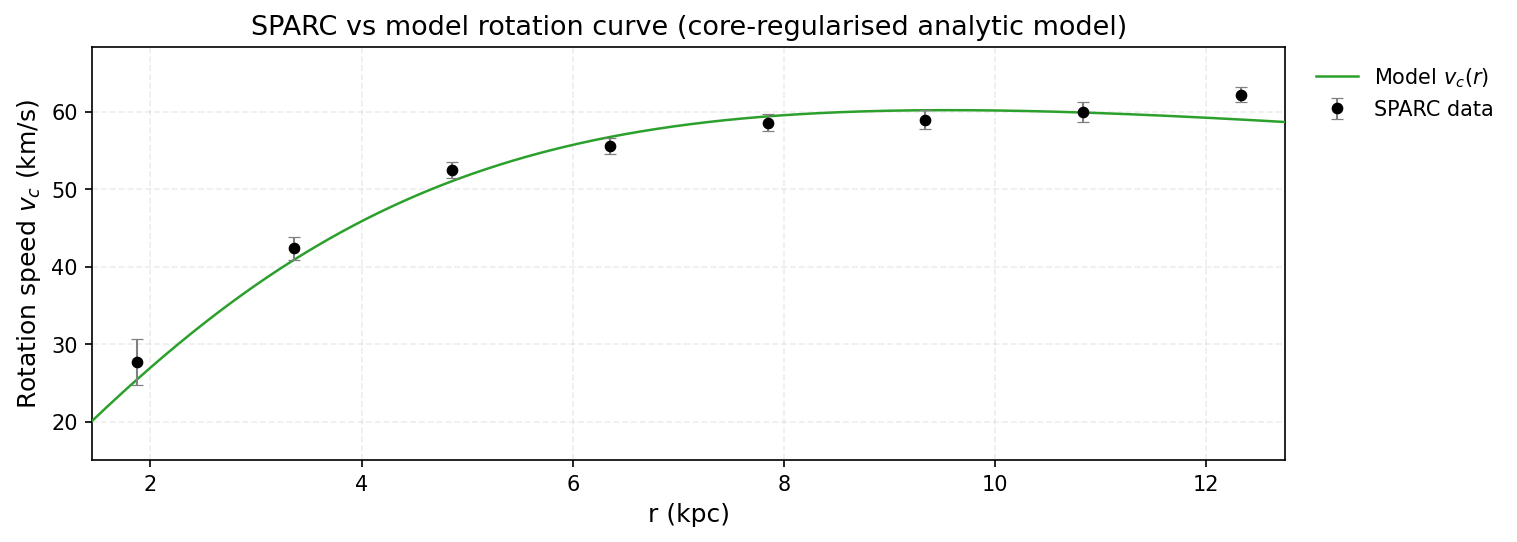}
\caption{The predicted rotation curves for the optimized SIDM
model of Eq. (\ref{ScaledependentEoSDM}), versus the SPARC
observational data for the galaxy DDO170.} \label{DDO170}
\end{figure}

In Table \ref{collDDO170} we present the optimized values of $K_0$
and $\rho_0$ for the analytic SIDM model of Eq.
(\ref{ScaledependentEoSDM}) for which the maximum compatibility
with the SPARC data is achieved.
\begin{table}[h!]
  \begin{center}
    \caption{SIDM Optimization Values for the galaxy DDO170}
    \label{collDDO170}
     \begin{tabular}{|r|r|}
     \hline
      \textbf{Parameter}   & \textbf{Optimization Values}
      \\  \hline
     $\rho_0 $  ($M_{\odot}/\mathrm{Kpc}^{3}$) & $1.14592\times 10^7$
\\  \hline $K_0$ ($M_{\odot} \,
\mathrm{Kpc}^{-3} \, (\mathrm{km/s})^{2}$)& 1569.79
\\  \hline
    \end{tabular}
  \end{center}
\end{table}
In Figs. \ref{DDO170dens}, \ref{DDO170} we present the density of
the analytic SIDM model, the predicted rotation curves for the
SIDM model (\ref{ScaledependentEoSDM}), versus the SPARC
observational data and the sound speed, as a function of the
radius respectively. As it can be seen, for this galaxy, the SIDM
model produces marginally viable rotation curves which are
marginally compatible with the SPARC data.

\begin{figure}[h!]
\centering
\includegraphics[width=20pc]{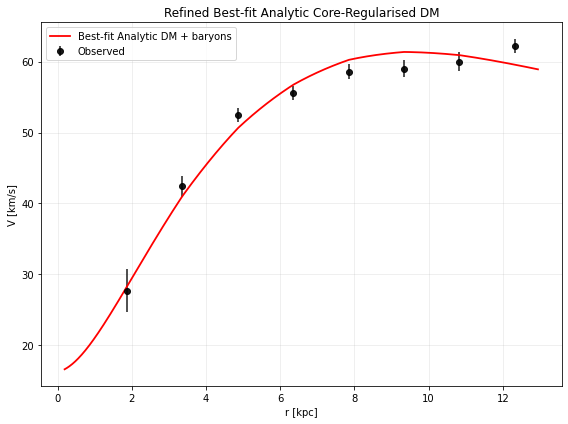}
\caption{The predicted rotation curves after using an optimization
for the SIDM model (\ref{ScaledependentEoSDM}), and the extended
SPARC data for the galaxy DDO170. We included the rotation curves
of the gas, the disk velocities, the bulge (where present) along
with the SIDM model.} \label{extendedDDO170}
\end{figure}

Now we shall include contributions to the rotation velocity from
the other components of the galaxy, namely the disk, the gas, and
the bulge if present. In Fig. \ref{extendedDDO170} we present the
combined rotation curves including all the components of the
galaxy along with the SIDM. As it can be seen, the extended
collisional DM model is marginally viable.
Also in Table \ref{evaluationextendedDDO170} we present the
optimized values of the free parameters of the SIDM model for
which  we achieve the maximum compatibility with the SPARC data,
for the galaxy DDO170, and also the resulting reduced
$\chi^2_{red}$ value.
\begin{table}[h!]
\centering \caption{Optimized Parameter Values of the Extended
SIDM model for the Galaxy DDO170.}
\begin{tabular}{lc}
\hline
Parameter & Value  \\
\hline
$\rho_0 $ ($M_{\odot}/\mathrm{Kpc}^{3}$) & $6.85225\times 10^7$   \\
$K_0$ ($M_{\odot} \,
\mathrm{Kpc}^{-3} \, (\mathrm{km/s})^{2}$) & 1184.82   \\
$ml_{\text{disk}}$ & 1 \\
$ml_{\text{bulge}}$ & 0.8749 \\
$\alpha$ (Kpc) & 7.58762 \\
$\chi^2_{red}$ & 4.81285 \\
\hline
\end{tabular}
\label{evaluationextendedDDO170}
\end{table}

\subsection{The Galaxy ESO079-G014, Marginally Viable, Extended Viable}

For this galaxy, the optimization method we used, ensures maximum
compatibility of the analytic SIDM model of Eq.
(\ref{ScaledependentEoSDM}) with the SPARC data, if we choose
$\rho_0=3.70427\times 10^7$$M_{\odot}/\mathrm{Kpc}^{3}$ and
$K_0=13450.9
$$M_{\odot} \, \mathrm{Kpc}^{-3} \, (\mathrm{km/s})^{2}$, in which
case the reduced $\chi^2_{red}$ value is $\chi^2_{red}=2.79013 $.
Also the parameter $\alpha$ in this case is $\alpha=10.997 $Kpc.

\begin{figure}[h!]
\centering
\includegraphics[width=20pc]{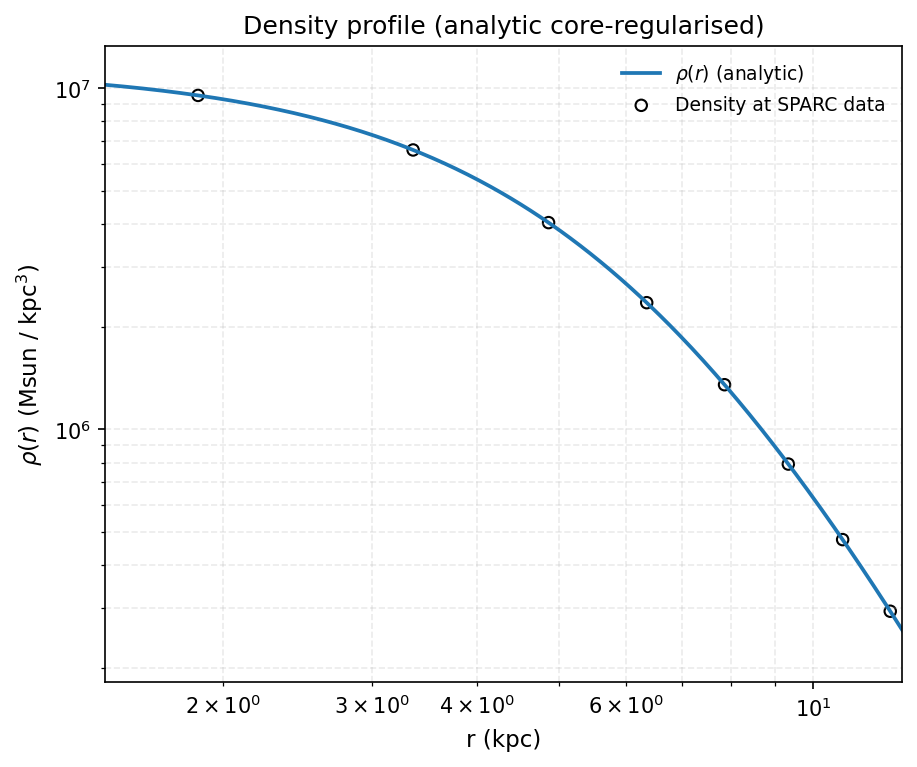}
\caption{The density of the SIDM model of Eq.
(\ref{ScaledependentEoSDM}) for the galaxy ESO079-G014, versus the
radius.} \label{ESO079-G014dens}
\end{figure}

\begin{figure}[h!]
\centering
\includegraphics[width=35pc]{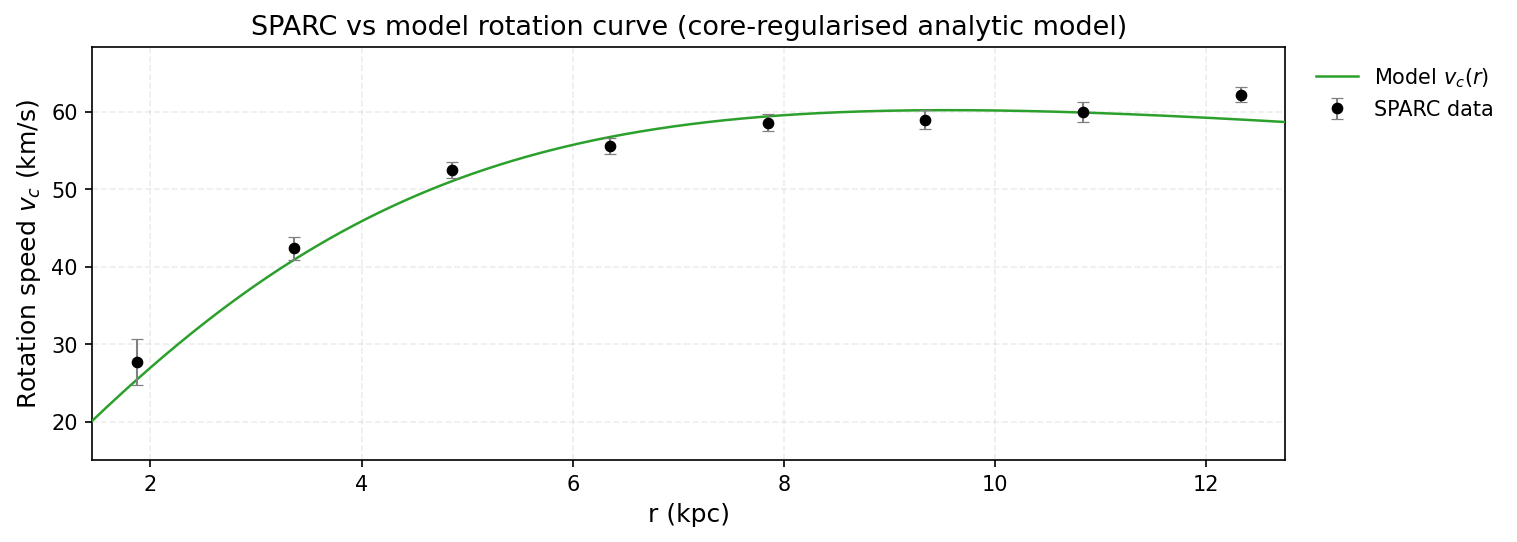}
\caption{The predicted rotation curves for the optimized SIDM
model of Eq. (\ref{ScaledependentEoSDM}), versus the SPARC
observational data for the galaxy ESO079-G014.}
\label{ESO079-G014}
\end{figure}

In Table \ref{collESO079-G014} we present the optimized values of
$K_0$ and $\rho_0$ for the analytic SIDM model of Eq.
(\ref{ScaledependentEoSDM}) for which the maximum compatibility
with the SPARC data is achieved.
\begin{table}[h!]
  \begin{center}
    \caption{SIDM Optimization Values for the galaxy ESO079-G014}
    \label{collESO079-G014}
     \begin{tabular}{|r|r|}
     \hline
      \textbf{Parameter}   & \textbf{Optimization Values}
      \\  \hline
     $\rho_0 $  ($M_{\odot}/\mathrm{Kpc}^{3}$) & $3.70427\times 10^7$
\\  \hline $K_0$ ($M_{\odot} \,
\mathrm{Kpc}^{-3} \, (\mathrm{km/s})^{2}$)& 13450.9
\\  \hline
    \end{tabular}
  \end{center}
\end{table}
In Figs. \ref{ESO079-G014dens}, \ref{ESO079-G014} we present the
density of the analytic SIDM model, the predicted rotation curves
for the SIDM model (\ref{ScaledependentEoSDM}), versus the SPARC
observational data and the sound speed, as a function of the
radius respectively. As it can be seen, for this galaxy, the SIDM
model produces marginally viable rotation curves which are
marginally compatible with the SPARC data.

\begin{figure}[h!]
\centering
\includegraphics[width=20pc]{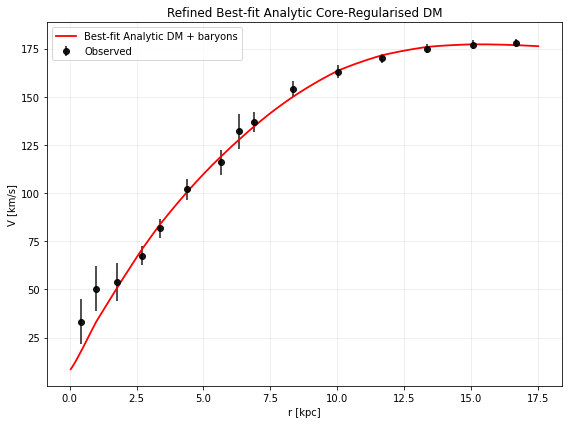}
\caption{The predicted rotation curves after using an optimization
for the SIDM model (\ref{ScaledependentEoSDM}), and the extended
SPARC data for the galaxy ESO079-G014. We included the rotation
curves of the gas, the disk velocities, the bulge (where present)
along with the SIDM model.} \label{extendedESO079-G014}
\end{figure}

Now we shall include contributions to the rotation velocity from
the other components of the galaxy, namely the disk, the gas, and
the bulge if present. In Fig. \ref{extendedESO079-G014} we present
the combined rotation curves including all the components of the
galaxy along with the SIDM. As it can be seen, the extended
collisional DM model is marginally viable.
Also in Table \ref{evaluationextendedESO079-G014} we present the
optimized values of the free parameters of the SIDM model for
which  we achieve the maximum compatibility with the SPARC data,
for the galaxy ESO079-G014, and also the resulting reduced
$\chi^2_{red}$ value.
\begin{table}[h!]
\centering \caption{Optimized Parameter Values of the Extended
SIDM model for the Galaxy ESO079-G014.}
\begin{tabular}{lc}
\hline
Parameter & Value  \\
\hline
$\rho_0 $ ($M_{\odot}/\mathrm{Kpc}^{3}$) & $2.88609\times 10^7$   \\
$K_0$ ($M_{\odot} \,
\mathrm{Kpc}^{-3} \, (\mathrm{km/s})^{2}$) & 11146.3  \\
$ml_{\text{disk}}$ & 0.5030 \\
$ml_{\text{bulge}}$ & 0.6404 \\
$\alpha$ (Kpc) & 11.3399\\
$\chi^2_{red}$ & 0.649054 \\
\hline
\end{tabular}
\label{evaluationextendedESO079-G014}
\end{table}

\subsection{The Galaxy ESO116-G012, Non-viable, Extended Viable}

For this galaxy, the optimization method we used, ensures maximum
compatibility of the analytic SIDM model of Eq.
(\ref{ScaledependentEoSDM}) with the SPARC data, if we choose
$\rho_0=7.7304\times 10^7$$M_{\odot}/\mathrm{Kpc}^{3}$ and
$K_0=5382.81
$$M_{\odot} \, \mathrm{Kpc}^{-3} \, (\mathrm{km/s})^{2}$, in which
case the reduced $\chi^2_{red}$ value is $\chi^2_{red}=3.2133$.
Also the parameter $\alpha$ in this case is $\alpha=4.81564 $Kpc.

\begin{figure}[h!]
\centering
\includegraphics[width=20pc]{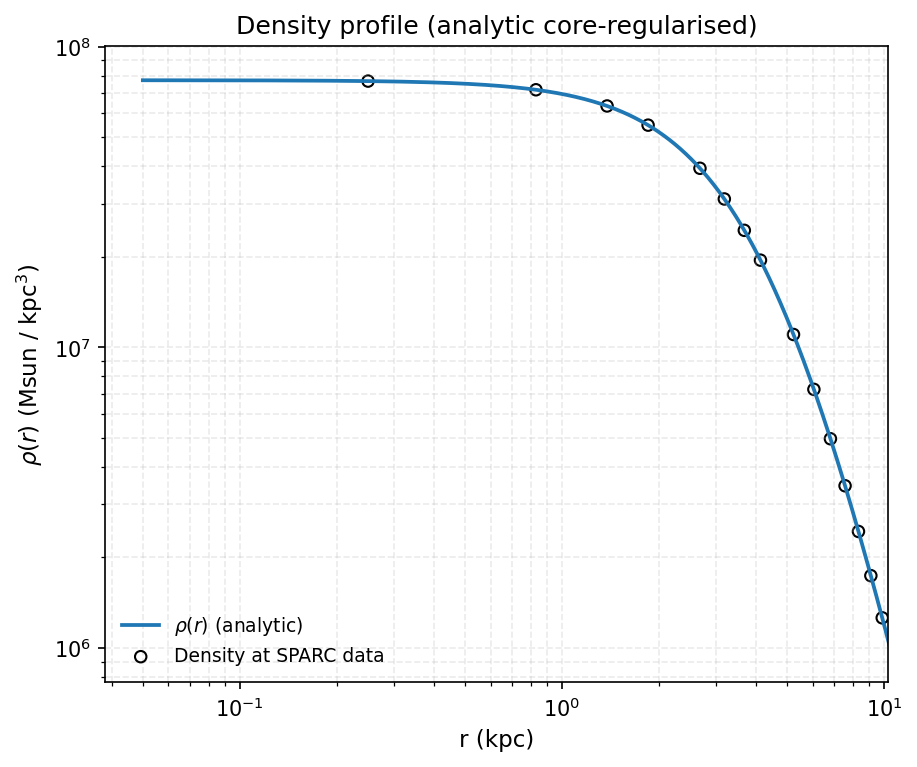}
\caption{The density of the SIDM model of Eq.
(\ref{ScaledependentEoSDM}) for the galaxy ESO116-G012, versus the
radius.} \label{ESO116-G012dens}
\end{figure}

\begin{figure}[h!]
\centering
\includegraphics[width=35pc]{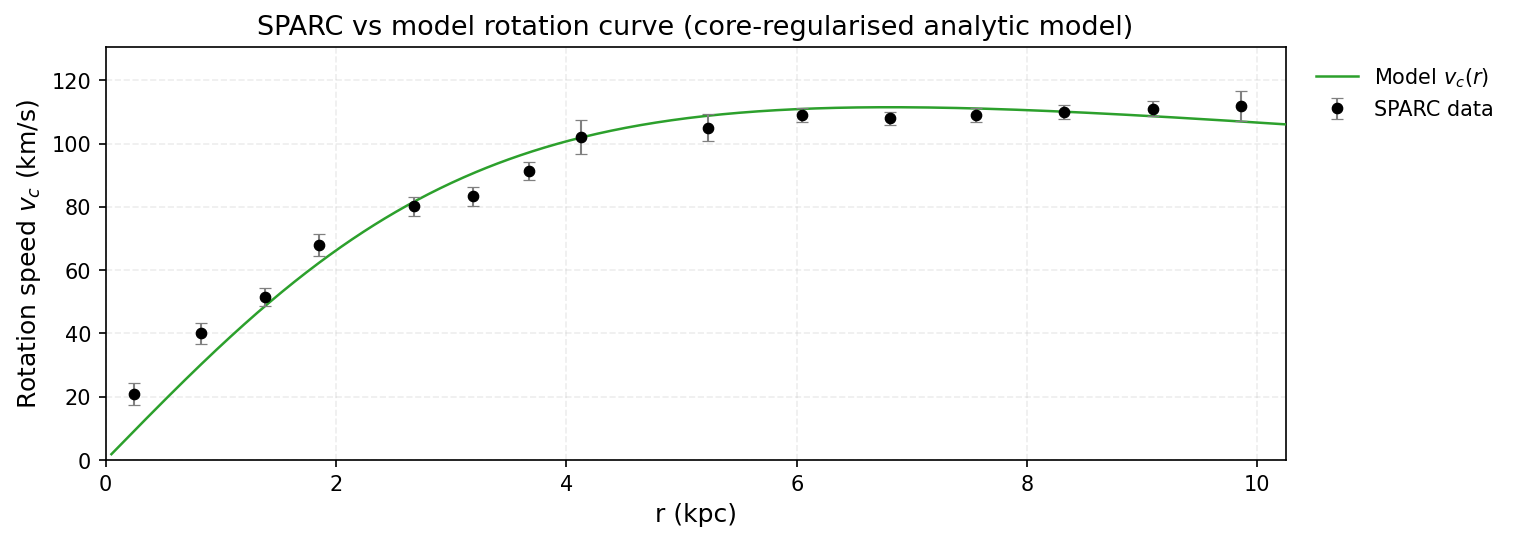}
\caption{The predicted rotation curves for the optimized SIDM
model of Eq. (\ref{ScaledependentEoSDM}), versus the SPARC
observational data for the galaxy ESO116-G012.}
\label{ESO116-G012}
\end{figure}

In Table \ref{collESO116-G012} we present the optimized values of
$K_0$ and $\rho_0$ for the analytic SIDM model of Eq.
(\ref{ScaledependentEoSDM}) for which the maximum compatibility
with the SPARC data is achieved.
\begin{table}[h!]
  \begin{center}
    \caption{SIDM Optimization Values for the galaxy ESO116-G012}
    \label{collESO116-G012}
     \begin{tabular}{|r|r|}
     \hline
      \textbf{Parameter}   & \textbf{Optimization Values}
      \\  \hline
     $\rho_0 $  ($M_{\odot}/\mathrm{Kpc}^{3}$) & $7.7304\times 10^7$
\\  \hline $K_0$ ($M_{\odot} \,
\mathrm{Kpc}^{-3} \, (\mathrm{km/s})^{2}$)& 5382.81
\\  \hline
    \end{tabular}
  \end{center}
\end{table}
In Figs. \ref{ESO116-G012dens}, \ref{ESO116-G012} we present the
density of the analytic SIDM model, the predicted rotation curves
for the SIDM model (\ref{ScaledependentEoSDM}), versus the SPARC
observational data and the sound speed, as a function of the
radius respectively. As it can be seen, for this galaxy, the SIDM
model produces non-viable rotation curves which are incompatible
with the SPARC data.

\begin{figure}[h!]
\centering
\includegraphics[width=20pc]{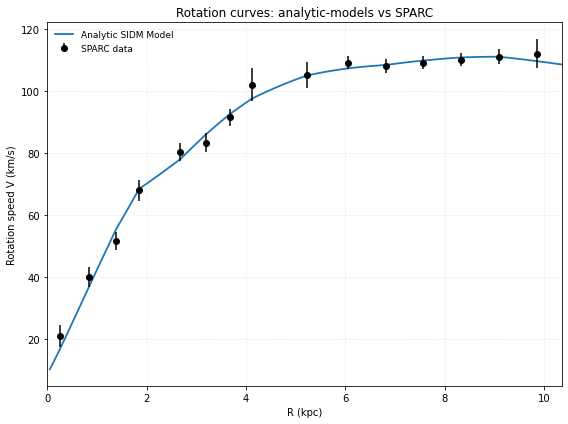}
\caption{The predicted rotation curves after using an optimization
for the SIDM model (\ref{ScaledependentEoSDM}), and the extended
SPARC data for the galaxy ESO116-G012. We included the rotation
curves of the gas, the disk velocities, the bulge (where present)
along with the SIDM model.} \label{extendedESO116-G012}
\end{figure}

Now we shall include contributions to the rotation velocity from
the other components of the galaxy, namely the disk, the gas, and
the bulge if present. In Fig. \ref{extendedESO116-G012} we present
the combined rotation curves including all the components of the
galaxy along with the SIDM. As it can be seen, the extended
collisional DM model is non-viable.
Also in Table \ref{evaluationextendedESO116-G012} we present the
optimized values of the free parameters of the SIDM model for
which  we achieve the maximum compatibility with the SPARC data,
for the galaxy ESO116-G012, and also the resulting reduced
$\chi^2_{red}$ value.
\begin{table}[h!]
\centering \caption{Optimized Parameter Values of the Extended
SIDM model for the Galaxy ESO116-G012.}
\begin{tabular}{lc}
\hline
Parameter & Value  \\
\hline
$\rho_0 $ ($M_{\odot}/\mathrm{Kpc}^{3}$) & $2.86227\times 10^7$   \\
$K_0$ ($M_{\odot} \,
\mathrm{Kpc}^{-3} \, (\mathrm{km/s})^{2}$) & 3959.08   \\
$ml_{\text{disk}}$ & 0.8969 \\
$ml_{\text{bulge}}$ & 0.5851 \\
$\alpha$ (Kpc) & 6.78637 \\
$\chi^2_{red}$ & 0.700004 \\
\hline
\end{tabular}
\label{evaluationextendedESO116-G012}
\end{table}


\subsection{The Galaxy ESO563-G021}

For this galaxy, the optimization method we used, ensures maximum
compatibility of the analytic SIDM model of Eq.
(\ref{ScaledependentEoSDM}) with the SPARC data, if we choose
$\rho_0=9.63711\times 10^7$$M_{\odot}/\mathrm{Kpc}^{3}$ and
$K_0=48595.1
$$M_{\odot} \, \mathrm{Kpc}^{-3} \, (\mathrm{km/s})^{2}$, in which
case the reduced $\chi^2_{red}$ value is $\chi^2_{red}=5.60671$.
Also the parameter $\alpha$ in this case is $\alpha=12.9591 $Kpc.

\begin{figure}[h!]
\centering
\includegraphics[width=20pc]{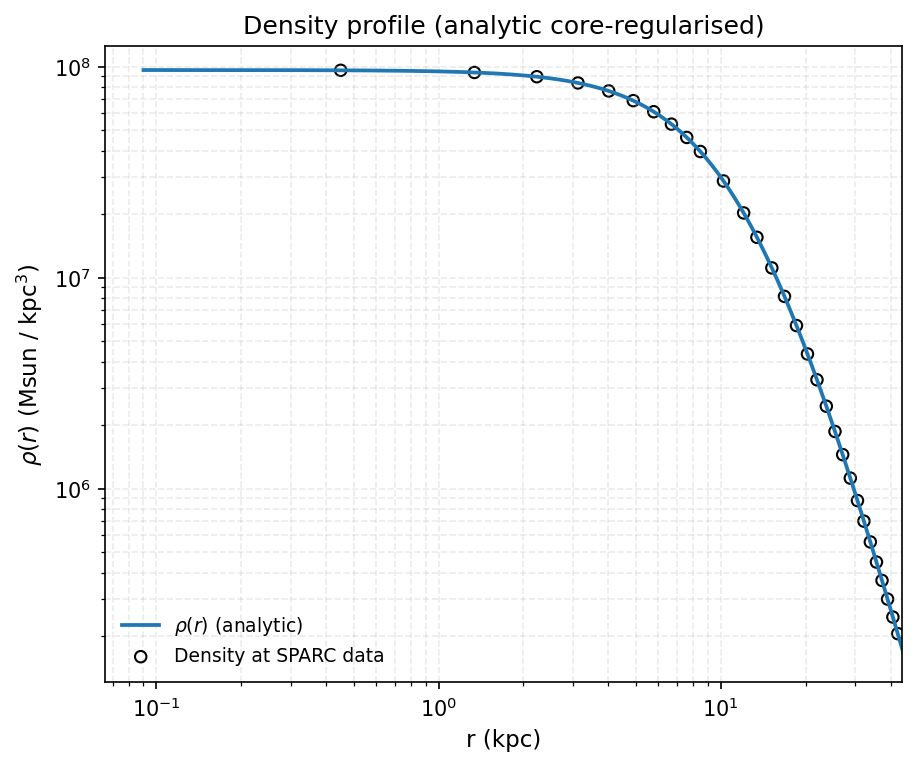}
\caption{The density of the SIDM model of Eq.
(\ref{ScaledependentEoSDM}) for the galaxy ESO563-G021, versus the
radius.} \label{ESO563-G021dens}
\end{figure}

\begin{figure}[h!]
\centering
\includegraphics[width=35pc]{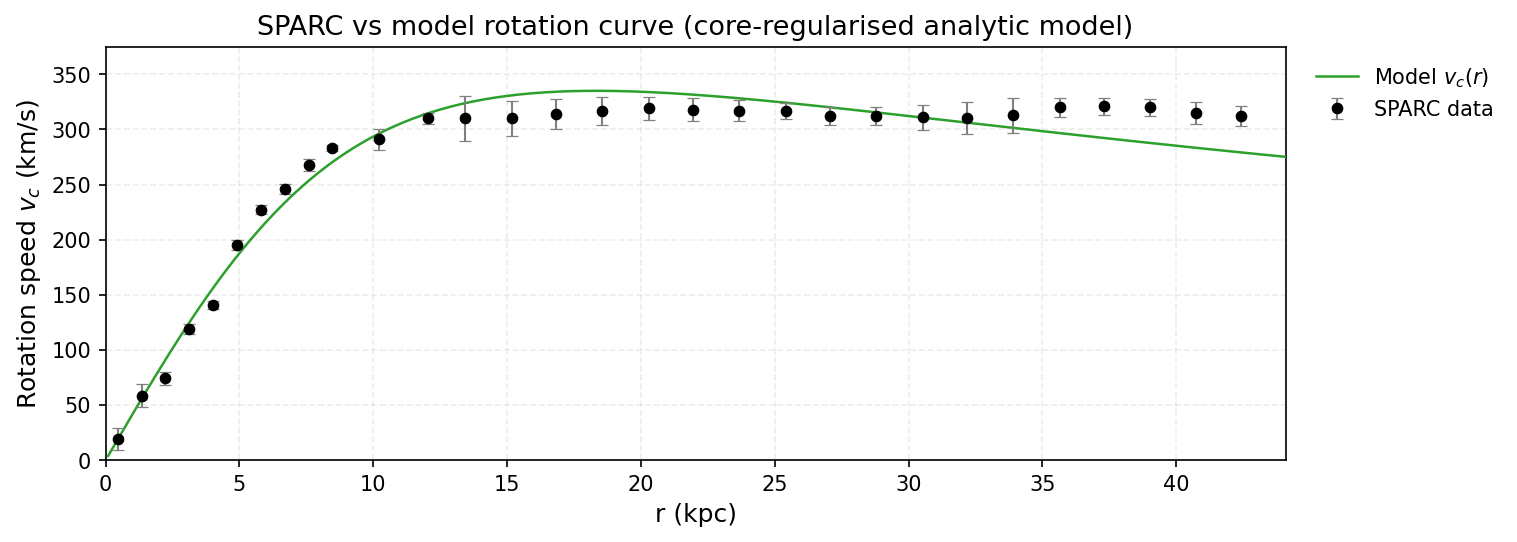}
\caption{The predicted rotation curves for the optimized SIDM
model of Eq. (\ref{ScaledependentEoSDM}), versus the SPARC
observational data for the galaxy ESO563-G021.}
\label{ESO563-G021}
\end{figure}

In Table \ref{collESO563-G021} we present the optimized values of
$K_0$ and $\rho_0$ for the analytic SIDM model of Eq.
(\ref{ScaledependentEoSDM}) for which the maximum compatibility
with the SPARC data is achieved.
\begin{table}[h!]
  \begin{center}
    \caption{SIDM Optimization Values for the galaxy ESO563-G021}
    \label{collESO563-G021}
     \begin{tabular}{|r|r|}
     \hline
      \textbf{Parameter}   & \textbf{Optimization Values}
      \\  \hline
     $\rho_0 $  ($M_{\odot}/\mathrm{Kpc}^{3}$) & $9.63711\times 10^7$
\\  \hline $K_0$ ($M_{\odot} \,
\mathrm{Kpc}^{-3} \, (\mathrm{km/s})^{2}$)& 48595.1
\\  \hline
    \end{tabular}
  \end{center}
\end{table}
In Figs. \ref{ESO563-G021dens}, \ref{ESO563-G021} we present the
density of the analytic SIDM model, the predicted rotation curves
for the SIDM model (\ref{ScaledependentEoSDM}), versus the SPARC
observational data and the sound speed, as a function of the
radius respectively. As it can be seen, for this galaxy, the SIDM
model produces non-viable rotation curves which are incompatible
with the SPARC data.

\begin{figure}[h!]
\centering
\includegraphics[width=20pc]{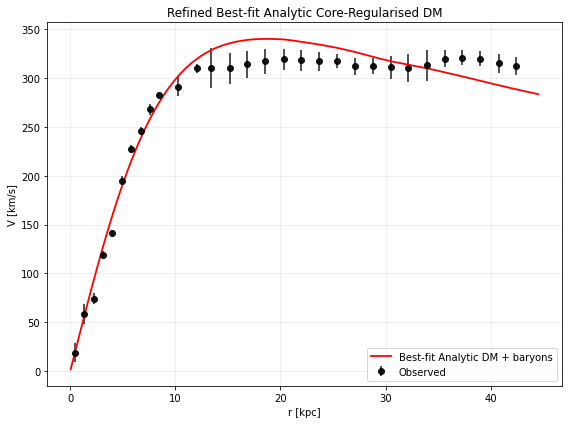}
\caption{The predicted rotation curves after using an optimization
for the SIDM model (\ref{ScaledependentEoSDM}), and the extended
SPARC data for the galaxy ESO563-G021. We included the rotation
curves of the gas, the disk velocities, the bulge (where present)
along with the SIDM model.} \label{extendedESO563-G021}
\end{figure}

Now we shall include contributions to the rotation velocity from
the other components of the galaxy, namely the disk, the gas, and
the bulge if present. In Fig. \ref{extendedESO563-G021} we present
the combined rotation curves including all the components of the
galaxy along with the SIDM. As it can be seen, the extended
collisional DM model is non-viable.
Also in Table \ref{evaluationextendedESO563-G021} we present the
optimized values of the free parameters of the SIDM model for
which  we achieve the maximum compatibility with the SPARC data,
for the galaxy ESO563-G021, and also the resulting reduced
$\chi^2_{red}$ value.
\begin{table}[h!]
\centering \caption{Optimized Parameter Values of the Extended
SIDM model for the Galaxy ESO563-G021.}
\begin{tabular}{lc}
\hline
Parameter & Value  \\
\hline
$\rho_0 $ ($M_{\odot}/\mathrm{Kpc}^{3}$) & $1.00755\times 10^7$   \\
$K_0$ ($M_{\odot} \,
\mathrm{Kpc}^{-3} \, (\mathrm{km/s})^{2}$) & 49854.6   \\
$ml_{\text{disk}}$ & 1 \\
$ml_{\text{bulge}}$ & 0.0253 \\
$\alpha$ (Kpc) & 12.8356\\
$\chi^2_{red}$ & 4.84033 \\
\hline
\end{tabular}
\label{evaluationextendedESO563-G021}
\end{table}

\subsection{The Galaxy F565-V2}

For this galaxy, the optimization method we used, ensures maximum
compatibility of the analytic SIDM model of Eq.
(\ref{ScaledependentEoSDM}) with the SPARC data, if we choose
$\rho_0=1.64594\times 10^7$$M_{\odot}/\mathrm{Kpc}^{3}$ and
$K_0=2804.34
$$M_{\odot} \, \mathrm{Kpc}^{-3} \, (\mathrm{km/s})^{2}$, in which
case the reduced $\chi^2_{red}$ value is $\chi^2_{red}=0.152429$.
Also the parameter $\alpha$ in this case is $\alpha=7.53286 $Kpc.

\begin{figure}[h!]
\centering
\includegraphics[width=20pc]{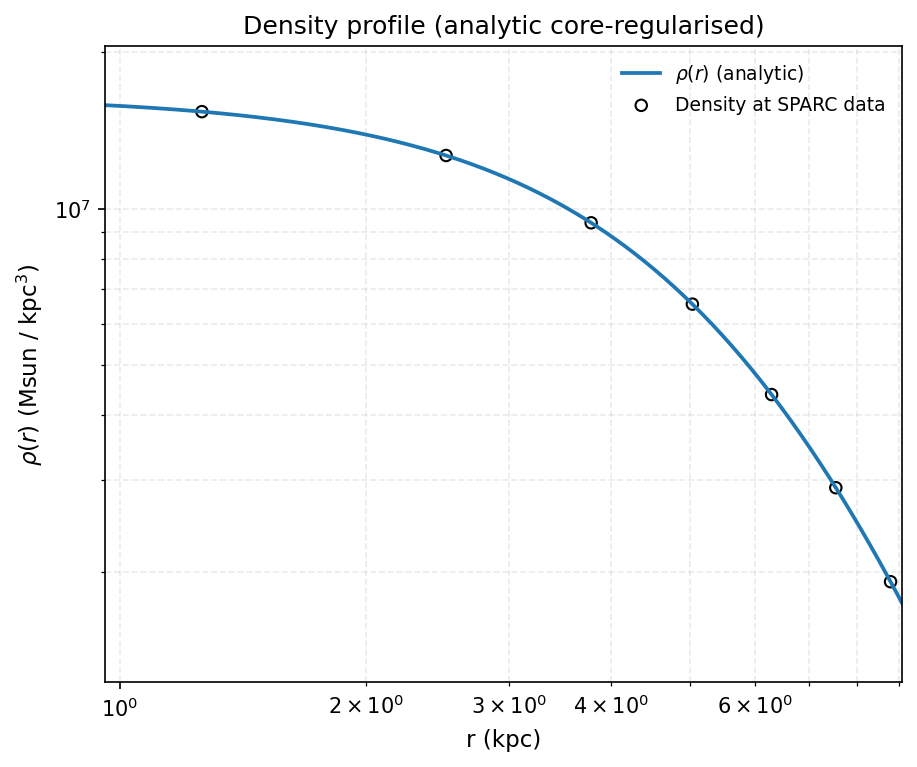}
\caption{The density of the SIDM model of Eq.
(\ref{ScaledependentEoSDM}) for the galaxy F565-V2, versus the
radius.} \label{F565-V2dens}
\end{figure}

\begin{figure}[h!]
\centering
\includegraphics[width=35pc]{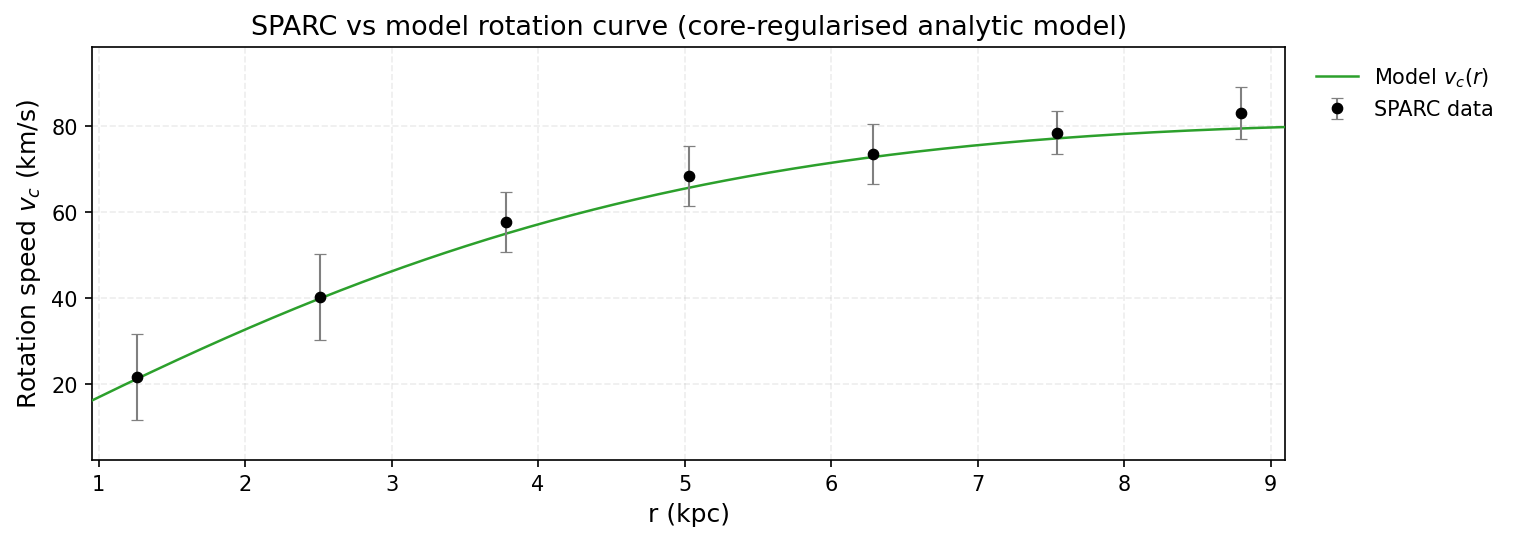}
\caption{The predicted rotation curves for the optimized SIDM
model of Eq. (\ref{ScaledependentEoSDM}), versus the SPARC
observational data for the galaxy F565-V2.} \label{F565-V2}
\end{figure}

In Table \ref{collF565-V2} we present the optimized values of
$K_0$ and $\rho_0$ for the analytic SIDM model of Eq.
(\ref{ScaledependentEoSDM}) for which the maximum compatibility
with the SPARC data is achieved.
\begin{table}[h!]
  \begin{center}
    \caption{SIDM Optimization Values for the galaxy F565-V2}
    \label{collF565-V2}
     \begin{tabular}{|r|r|}
     \hline
      \textbf{Parameter}   & \textbf{Optimization Values}
      \\  \hline
     $\rho_0 $  ($M_{\odot}/\mathrm{Kpc}^{3}$) & $1.64594\times 10^7$
\\  \hline $K_0$ ($M_{\odot} \,
\mathrm{Kpc}^{-3} \, (\mathrm{km/s})^{2}$)& 2804.34
\\  \hline
    \end{tabular}
  \end{center}
\end{table}
In Figs. \ref{F565-V2dens}, \ref{F565-V2} we present the density
of the analytic SIDM model, the predicted rotation curves for the
SIDM model (\ref{ScaledependentEoSDM}), versus the SPARC
observational data and the sound speed, as a function of the
radius respectively. As it can be seen, for this galaxy, the SIDM
model produces viable rotation curves which are compatible with
the SPARC data.

\subsection{The Galaxy F568-3}

For this galaxy, the optimization method we used, ensures maximum
compatibility of the analytic SIDM model of Eq.
(\ref{ScaledependentEoSDM}) with the SPARC data, if we choose
$\rho_0=2.11755\times 10^7$$M_{\odot}/\mathrm{Kpc}^{3}$ and
$K_0=5378.52
$$M_{\odot} \, \mathrm{Kpc}^{-3} \, (\mathrm{km/s})^{2}$, in which
case the reduced $\chi^2_{red}$ value is $\chi^2_{red}=0.680511$.
Also the parameter $\alpha$ in this case is $\alpha=9.1974 $Kpc.

\begin{figure}[h!]
\centering
\includegraphics[width=20pc]{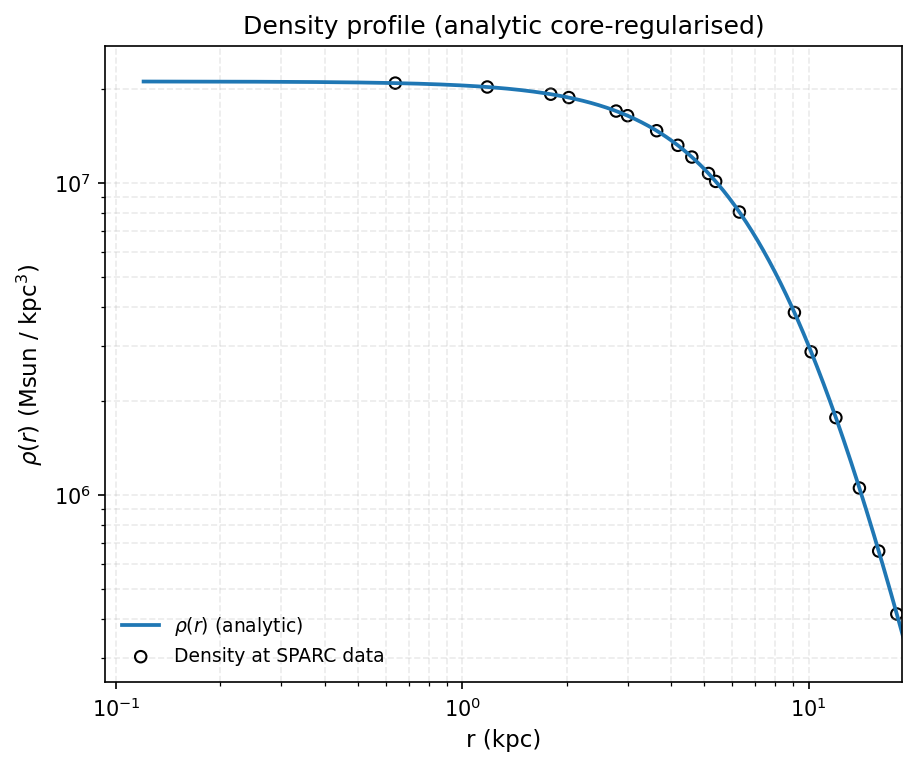}
\caption{The density of the SIDM model of Eq.
(\ref{ScaledependentEoSDM}) for the galaxy F568-3, versus the
radius.} \label{F568-3dens}
\end{figure}

\begin{figure}[h!]
\centering
\includegraphics[width=35pc]{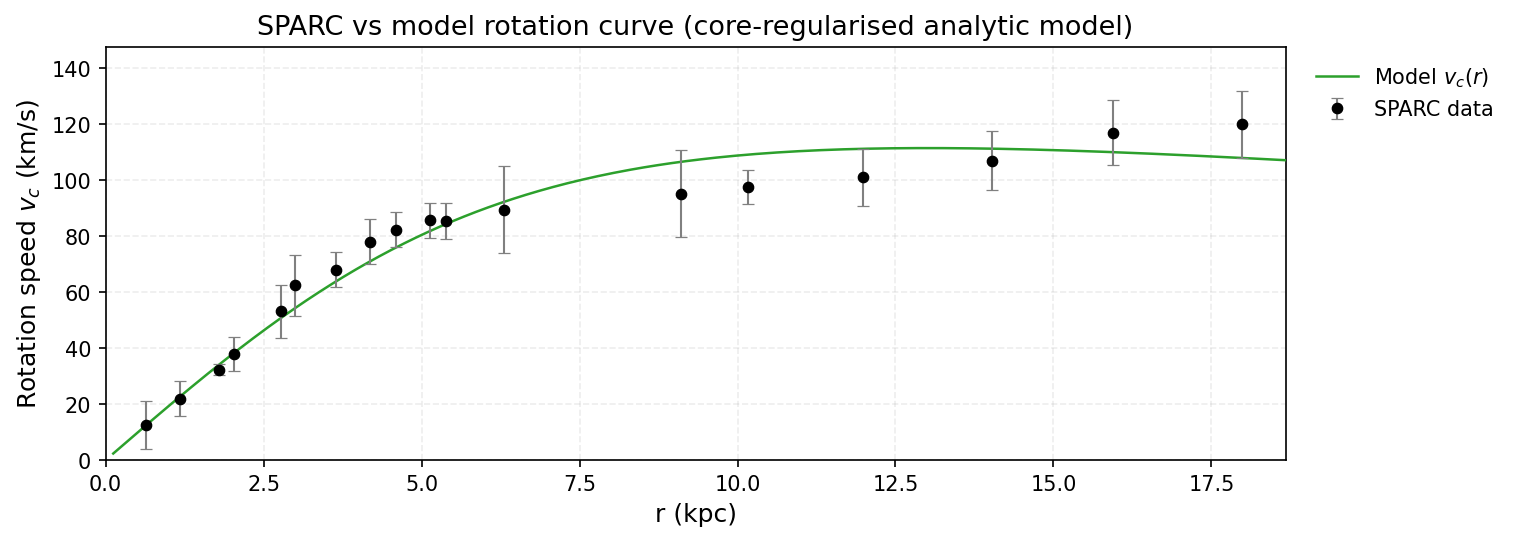}
\caption{The predicted rotation curves for the optimized SIDM
model of Eq. (\ref{ScaledependentEoSDM}), versus the SPARC
observational data for the galaxy F568-3.} \label{F568-3}
\end{figure}

In Table \ref{collF568-3} we present the optimized values of $K_0$
and $\rho_0$ for the analytic SIDM model of Eq.
(\ref{ScaledependentEoSDM}) for which the maximum compatibility
with the SPARC data is achieved.
\begin{table}[h!]
  \begin{center}
    \caption{SIDM Optimization Values for the galaxy F568-3}
    \label{collF568-3}
     \begin{tabular}{|r|r|}
     \hline
      \textbf{Parameter}   & \textbf{Optimization Values}
      \\  \hline
     $\rho_0 $  ($M_{\odot}/\mathrm{Kpc}^{3}$) & $2.11755\times 10^7$
\\  \hline $K_0$ ($M_{\odot} \,
\mathrm{Kpc}^{-3} \, (\mathrm{km/s})^{2}$)& 5378.52
\\  \hline
    \end{tabular}
  \end{center}
\end{table}
In Figs. \ref{F568-3dens}, \ref{F568-3} we present the density of
the analytic SIDM model, the predicted rotation curves for the
SIDM model (\ref{ScaledependentEoSDM}), versus the SPARC
observational data and the sound speed, as a function of the
radius respectively. As it can be seen, for this galaxy, the SIDM
model produces viable rotation curves which are compatible with
the SPARC data.

\subsection{The Galaxy F568-V1}

For this galaxy, the optimization method we used, ensures maximum
compatibility of the analytic SIDM model of Eq.
(\ref{ScaledependentEoSDM}) with the SPARC data, if we choose
$\rho_0=6.25735\times 10^7$$M_{\odot}/\mathrm{Kpc}^{3}$ and
$K_0=6374.56
$$M_{\odot} \, \mathrm{Kpc}^{-3} \, (\mathrm{km/s})^{2}$, in which
case the reduced $\chi^2_{red}$ value is $\chi^2_{red}=0.3568$.
Also the parameter $\alpha$ in this case is $\alpha=5.82479 $Kpc.

\begin{figure}[h!]
\centering
\includegraphics[width=20pc]{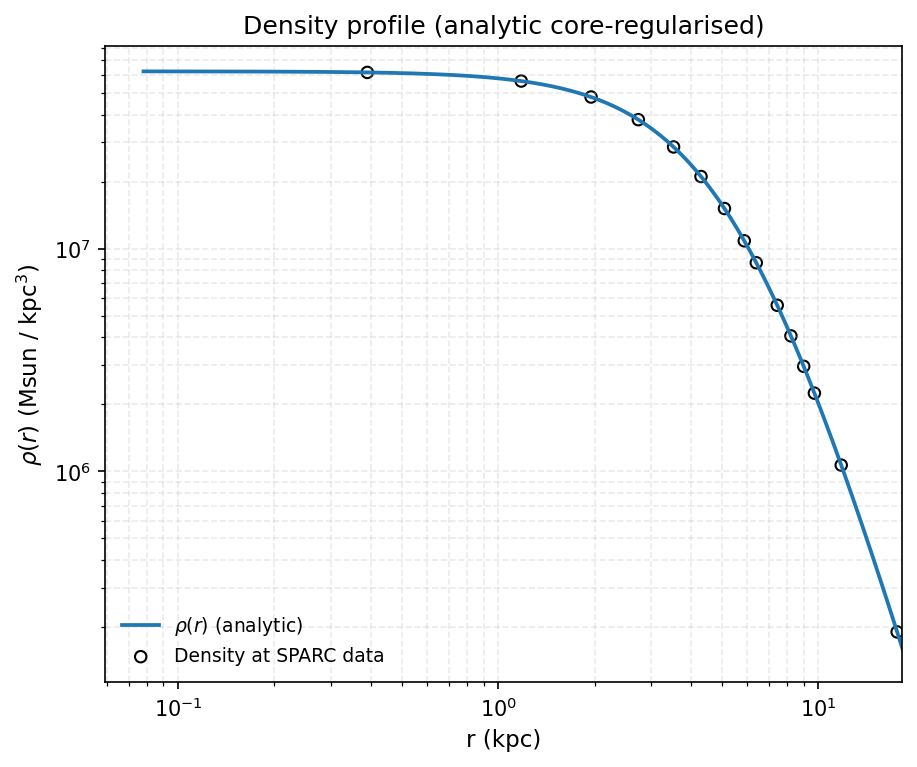}
\caption{The density of the SIDM model of Eq.
(\ref{ScaledependentEoSDM}) for the galaxy F568-V1, versus the
radius.} \label{F568-V1dens}
\end{figure}

\begin{figure}[h!]
\centering
\includegraphics[width=35pc]{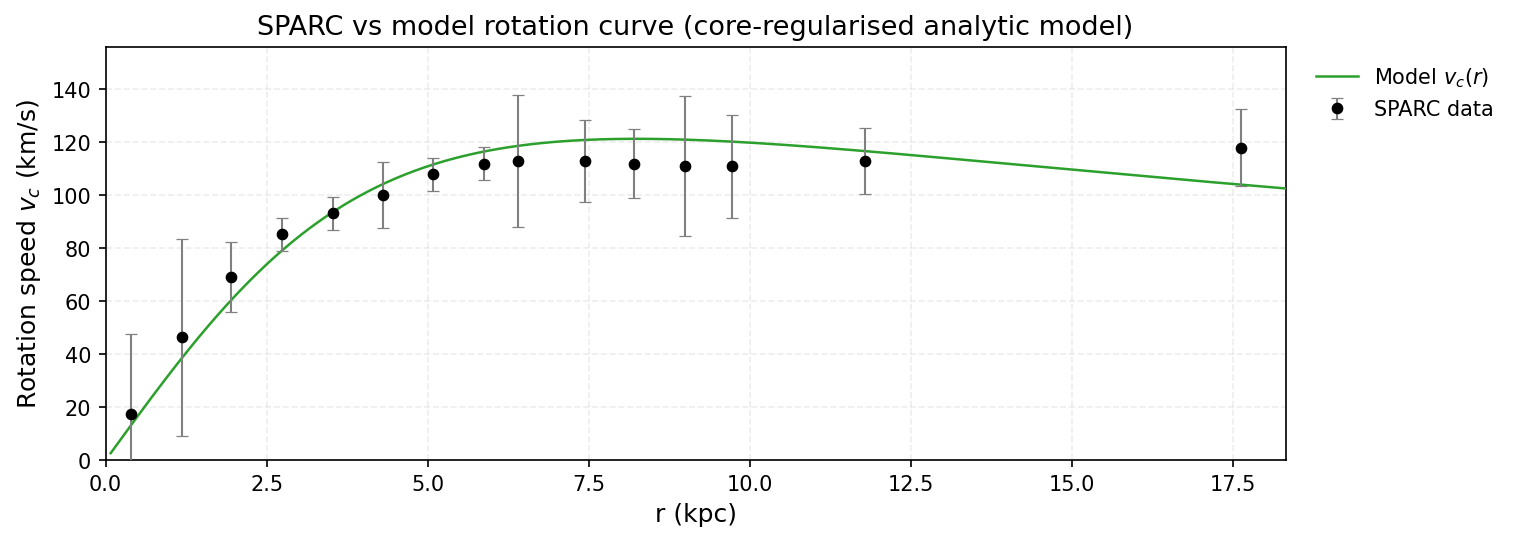}
\caption{The predicted rotation curves for the optimized SIDM
model of Eq. (\ref{ScaledependentEoSDM}), versus the SPARC
observational data for the galaxy F568-V1.} \label{F568-V1}
\end{figure}

In Table \ref{collF568-V1} we present the optimized values of
$K_0$ and $\rho_0$ for the analytic SIDM model of Eq.
(\ref{ScaledependentEoSDM}) for which the maximum compatibility
with the SPARC data is achieved.
\begin{table}[h!]
  \begin{center}
    \caption{SIDM Optimization Values for the galaxy F568-V1}
    \label{collF568-V1}
     \begin{tabular}{|r|r|}
     \hline
      \textbf{Parameter}   & \textbf{Optimization Values}
      \\  \hline
     $\rho_0 $  ($M_{\odot}/\mathrm{Kpc}^{3}$) & $6.25735\times 10^7$
\\  \hline $K_0$ ($M_{\odot} \,
\mathrm{Kpc}^{-3} \, (\mathrm{km/s})^{2}$)& 6374.56
\\  \hline
    \end{tabular}
  \end{center}
\end{table}
In Figs. \ref{F568-V1dens}, \ref{F568-V1} we present the density
of the analytic SIDM model, the predicted rotation curves for the
SIDM model (\ref{ScaledependentEoSDM}), versus the SPARC
observational data and the sound speed, as a function of the
radius respectively. As it can be seen, for this galaxy, the SIDM
model produces viable rotation curves which are compatible with
the SPARC data.

\subsection{The Galaxy F571-V1}

For this galaxy, the optimization method we used, ensures maximum
compatibility of the analytic SIDM model of Eq.
(\ref{ScaledependentEoSDM}) with the SPARC data, if we choose
$\rho_0=1.39508\times 10^7$$M_{\odot}/\mathrm{Kpc}^{3}$ and
$K_0=3247.46
$$M_{\odot} \, \mathrm{Kpc}^{-3} \, (\mathrm{km/s})^{2}$, in which
case the reduced $\chi^2_{red}$ value is $\chi^2_{red}=0.178266$.
Also the parameter $\alpha$ in this case is $\alpha=8.80488 $Kpc.

\begin{figure}[h!]
\centering
\includegraphics[width=20pc]{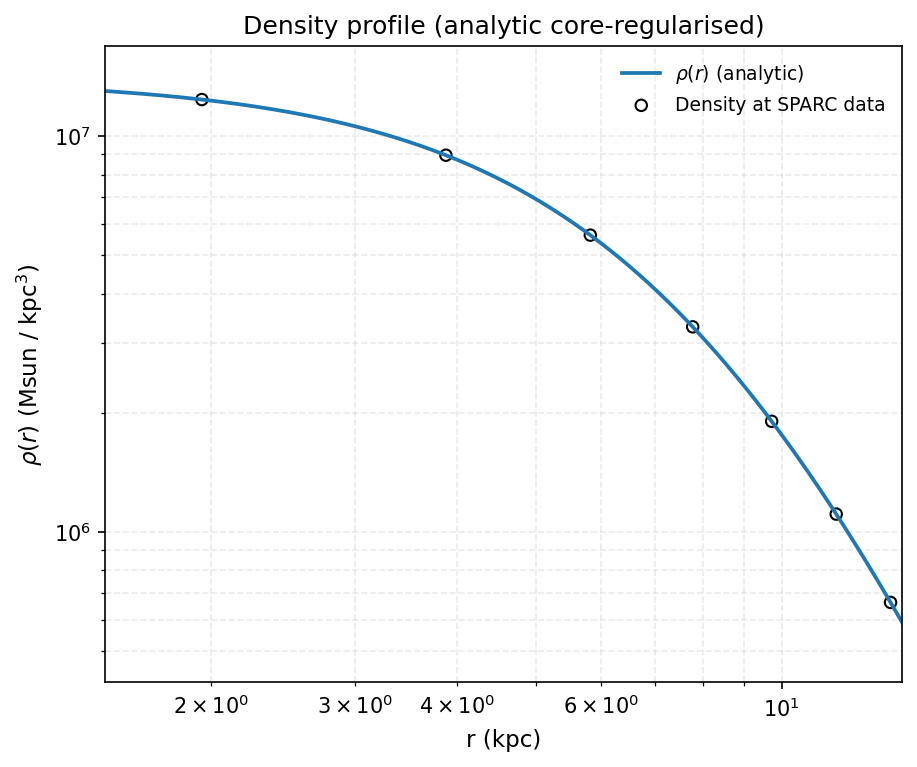}
\caption{The density of the SIDM model of Eq.
(\ref{ScaledependentEoSDM}) for the galaxy F571-V1, versus the
radius.} \label{F571-V1dens}
\end{figure}

\begin{figure}[h!]
\centering
\includegraphics[width=35pc]{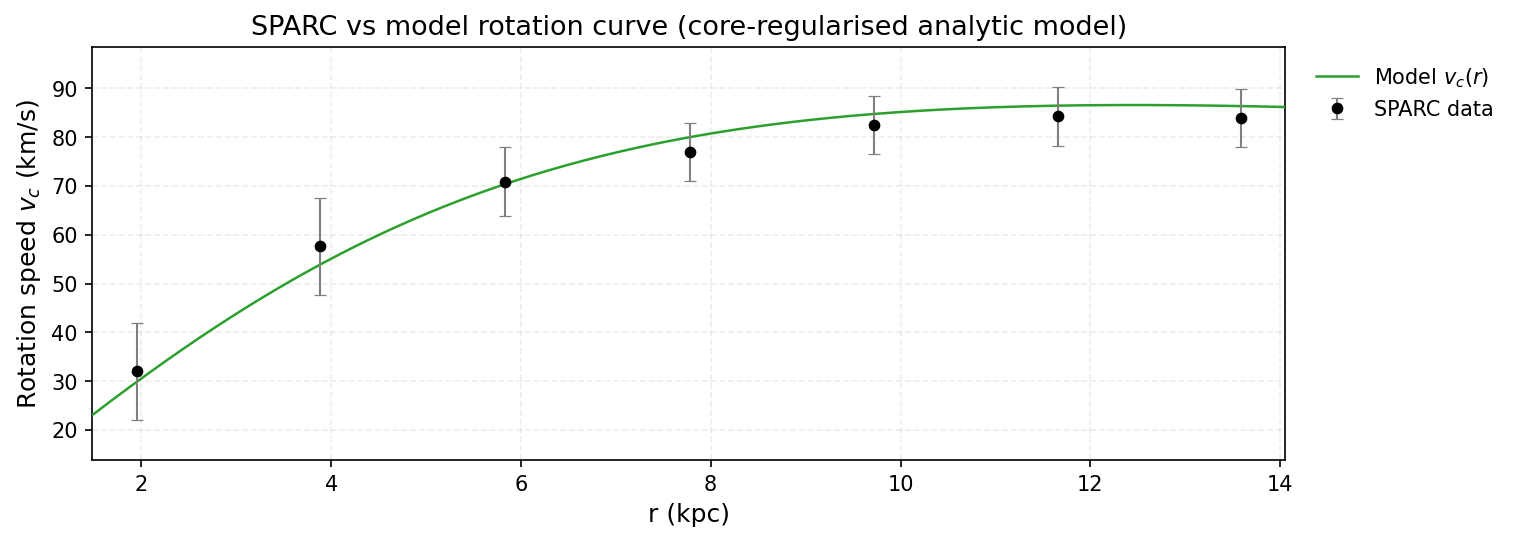}
\caption{The predicted rotation curves for the optimized SIDM
model of Eq. (\ref{ScaledependentEoSDM}), versus the SPARC
observational data for the galaxy F571-V1.} \label{F571-V1}
\end{figure}

In Table \ref{collF571-V1} we present the optimized values of
$K_0$ and $\rho_0$ for the analytic SIDM model of Eq.
(\ref{ScaledependentEoSDM}) for which the maximum compatibility
with the SPARC data is achieved.
\begin{table}[h!]
  \begin{center}
    \caption{SIDM Optimization Values for the galaxy F571-V1}
    \label{collF571-V1}
     \begin{tabular}{|r|r|}
     \hline
      \textbf{Parameter}   & \textbf{Optimization Values}
      \\  \hline
     $\rho_0 $  ($M_{\odot}/\mathrm{Kpc}^{3}$) & $1.39508\times 10^7$
\\  \hline $K_0$ ($M_{\odot} \,
\mathrm{Kpc}^{-3} \, (\mathrm{km/s})^{2}$)& 3247.46
\\  \hline
    \end{tabular}
  \end{center}
\end{table}
In Figs. \ref{F571-V1dens}, \ref{F571-V1} we present the density
of the analytic SIDM model, the predicted rotation curves for the
SIDM model (\ref{ScaledependentEoSDM}), versus the SPARC
observational data and the sound speed, as a function of the
radius respectively. As it can be seen, for this galaxy, the SIDM
model produces viable rotation curves which are compatible with
the SPARC data.

\subsection{The Galaxy F571-8, Non-viable, Extended Viable}

For this galaxy, the optimization method we used, ensures maximum
compatibility of the analytic SIDM model of Eq.
(\ref{ScaledependentEoSDM}) with the SPARC data, if we choose
$\rho_0=3.98342\times 10^7$$M_{\odot}/\mathrm{Kpc}^{3}$ and
$K_0=8709.25
$$M_{\odot} \, \mathrm{Kpc}^{-3} \, (\mathrm{km/s})^{2}$, in which
case the reduced $\chi^2_{red}$ value is $\chi^2_{red}=5.29502$.
Also the parameter $\alpha$ in this case is $\alpha=8.53322 $Kpc.

\begin{figure}[h!]
\centering
\includegraphics[width=20pc]{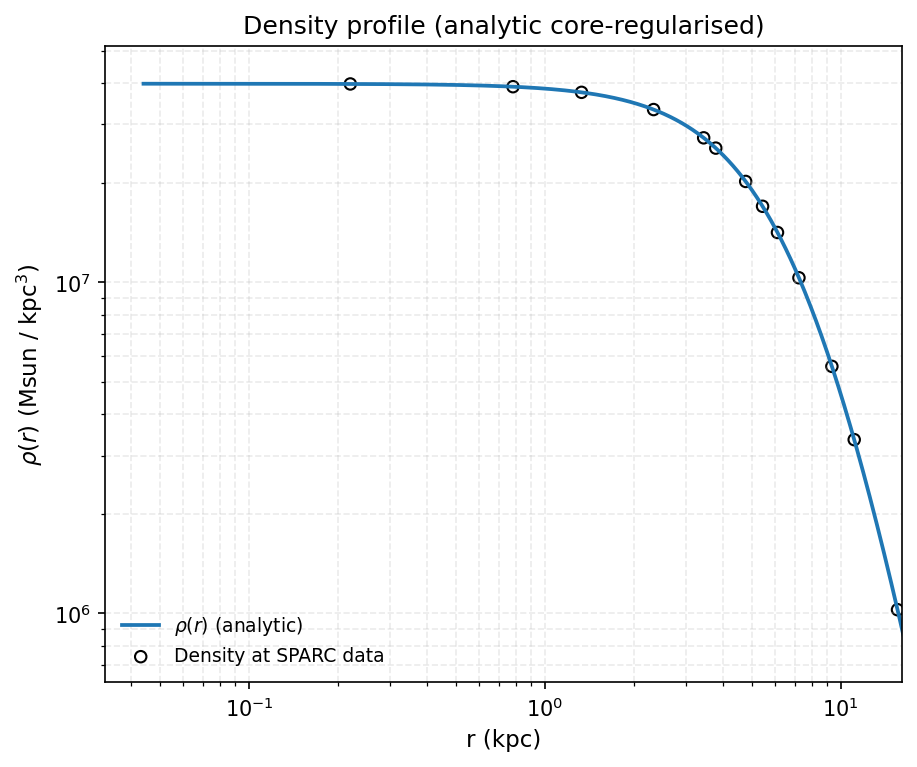}
\caption{The density of the SIDM model of Eq.
(\ref{ScaledependentEoSDM}) for the galaxy F571-8, versus the
radius.} \label{F571-8dens}
\end{figure}

\begin{figure}[h!]
\centering
\includegraphics[width=35pc]{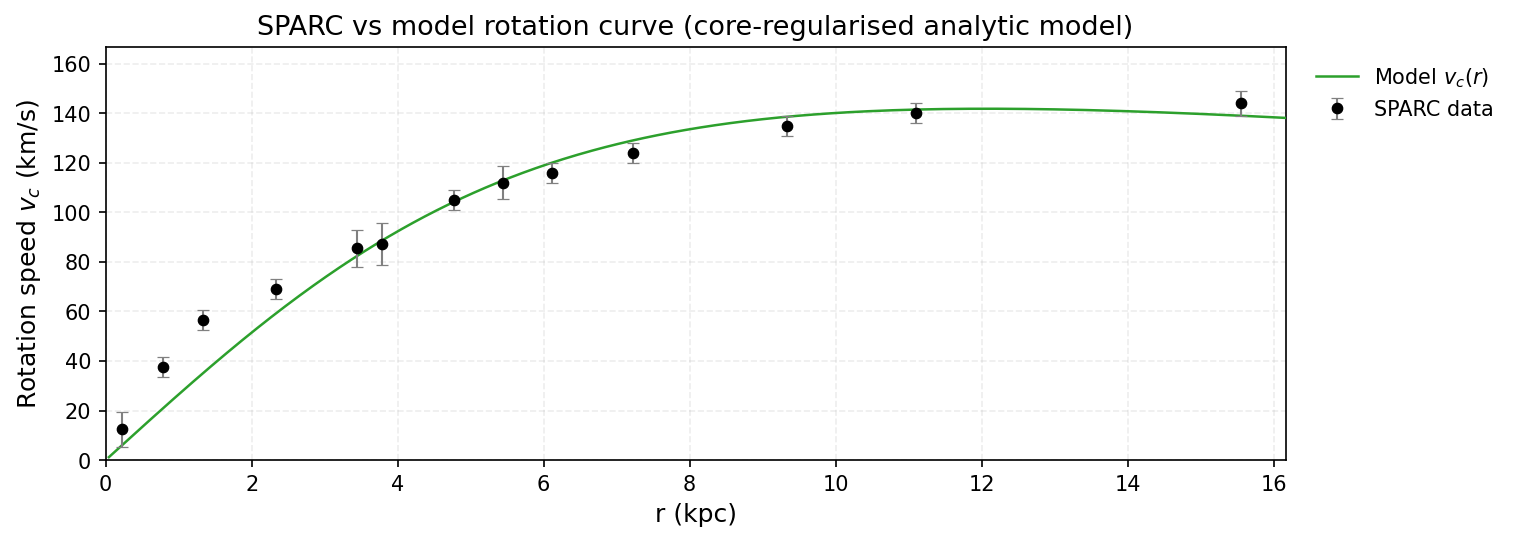}
\caption{The predicted rotation curves for the optimized SIDM
model of Eq. (\ref{ScaledependentEoSDM}), versus the SPARC
observational data for the galaxy F571-8.} \label{F571-8}
\end{figure}

In Table \ref{collF571-8} we present the optimized values of $K_0$
and $\rho_0$ for the analytic SIDM model of Eq.
(\ref{ScaledependentEoSDM}) for which the maximum compatibility
with the SPARC data is achieved.
\begin{table}[h!]
  \begin{center}
    \caption{SIDM Optimization Values for the galaxy F571-8}
    \label{collF571-8}
     \begin{tabular}{|r|r|}
     \hline
      \textbf{Parameter}   & \textbf{Optimization Values}
      \\  \hline
     $\rho_0 $  ($M_{\odot}/\mathrm{Kpc}^{3}$) & $3.98342\times 10^7$
\\  \hline $K_0$ ($M_{\odot} \,
\mathrm{Kpc}^{-3} \, (\mathrm{km/s})^{2}$)& 8709.25
\\  \hline
    \end{tabular}
  \end{center}
\end{table}
In Figs. \ref{F571-8dens}, \ref{F571-8}  we present the density of
the analytic SIDM model, the predicted rotation curves for the
SIDM model (\ref{ScaledependentEoSDM}), versus the SPARC
observational data and the sound speed, as a function of the
radius respectively. As it can be seen, for this galaxy, the SIDM
model produces non-viable rotation curves which are incompatible
with the SPARC data.

\begin{figure}[h!]
\centering
\includegraphics[width=20pc]{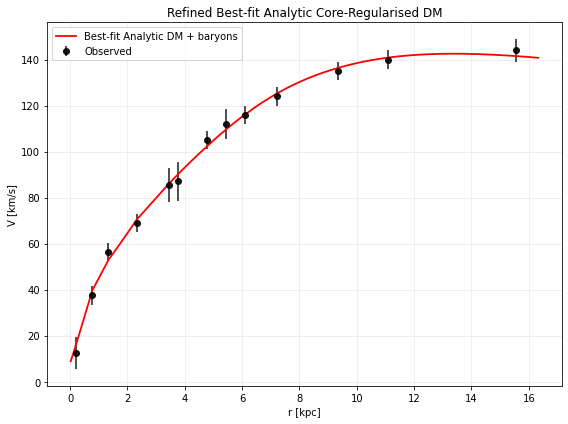}
\caption{The predicted rotation curves after using an optimization
for the SIDM model (\ref{ScaledependentEoSDM}), and the extended
SPARC data for the galaxy F571-8. We included the rotation curves
of the gas, the disk velocities, the bulge (where present) along
with the SIDM model.} \label{extendedF571-8}
\end{figure}

Now we shall include contributions to the rotation velocity from
the other components of the galaxy, namely the disk, the gas, and
the bulge if present. In Fig. \ref{extendedF571-8} we present the
combined rotation curves including all the components of the
galaxy along with the SIDM. As it can be seen, the extended
collisional DM model is viable.
Also in Table \ref{evaluationextendedF571-8} we present the
optimized values of the free parameters of the SIDM model for
which  we achieve the maximum compatibility with the SPARC data,
for the galaxy F571-8, and also the resulting reduced
$\chi^2_{red}$ value.
\begin{table}[h!]
\centering \caption{Optimized Parameter Values of the Extended
SIDM model for the Galaxy F571-8.}
\begin{tabular}{lc}
\hline
Parameter & Value  \\
\hline
$\rho_0 $ ($M_{\odot}/\mathrm{Kpc}^{3}$) & $3.02249\times 10^7$   \\
$K_0$ ($M_{\odot} \,
\mathrm{Kpc}^{-3} \, (\mathrm{km/s})^{2}$) & 8483.67   \\
$ml_{\text{disk}}$ & 0.4126 \\
$ml_{\text{bulge}}$ & 0.2105 \\
$\alpha$ (Kpc) & 9.6673\\
$\chi^2_{red}$ & 0.326601 \\
\hline
\end{tabular}
\label{evaluationextendedF571-8}
\end{table}

\subsection{The Galaxy F574-1, Marginally viable, Extended Viable}

For this galaxy, the optimization method we used, ensures maximum
compatibility of the analytic SIDM model of Eq.
(\ref{ScaledependentEoSDM}) with the SPARC data, if we choose
$\rho_0=3.96165\times 10^7$$M_{\odot}/\mathrm{Kpc}^{3}$ and
$K_0=4201.07
$$M_{\odot} \, \mathrm{Kpc}^{-3} \, (\mathrm{km/s})^{2}$, in which
case the reduced $\chi^2_{red}$ value is $\chi^2_{red}=0.850431$.
Also the parameter $\alpha$ in this case is $\alpha=5.94282 $Kpc.

\begin{figure}[h!]
\centering
\includegraphics[width=20pc]{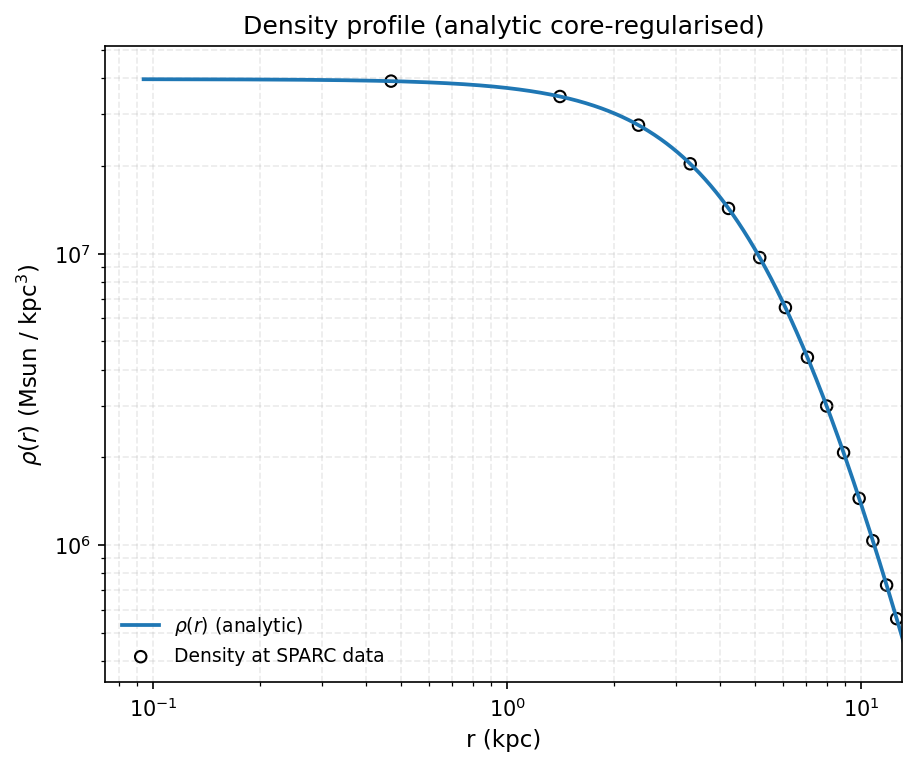}
\caption{The density of the SIDM model of Eq.
(\ref{ScaledependentEoSDM}) for the galaxy F574-1, versus the
radius.} \label{F574-1dens}
\end{figure}

\begin{figure}[h!]
\centering
\includegraphics[width=35pc]{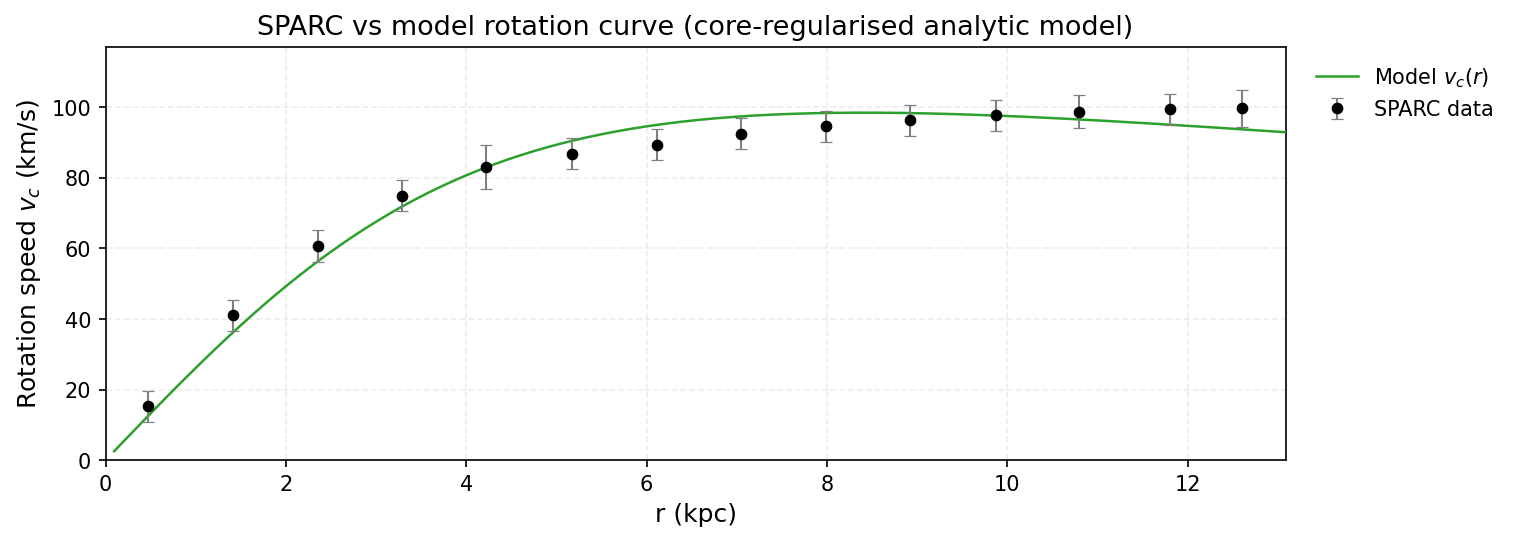}
\caption{The predicted rotation curves for the optimized SIDM
model of Eq. (\ref{ScaledependentEoSDM}), versus the SPARC
observational data for the galaxy F574-1.} \label{F574-1}
\end{figure}

In Table \ref{collF574-1} we present the optimized values of $K_0$
and $\rho_0$ for the analytic SIDM model of Eq.
(\ref{ScaledependentEoSDM}) for which the maximum compatibility
with the SPARC data is achieved.
\begin{table}[h!]
  \begin{center}
    \caption{SIDM Optimization Values for the galaxy F574-1}
    \label{collF574-1}
     \begin{tabular}{|r|r|}
     \hline
      \textbf{Parameter}   & \textbf{Optimization Values}
      \\  \hline
     $\rho_0 $  ($M_{\odot}/\mathrm{Kpc}^{3}$) & $3.96165\times 10^7$
\\  \hline $K_0$ ($M_{\odot} \,
\mathrm{Kpc}^{-3} \, (\mathrm{km/s})^{2}$)& 4201.07
\\  \hline
    \end{tabular}
  \end{center}
\end{table}
In Figs. \ref{F574-1dens}, \ref{F574-1}  we present the density of
the analytic SIDM model, the predicted rotation curves for the
SIDM model (\ref{ScaledependentEoSDM}), versus the SPARC
observational data and the sound speed, as a function of the
radius respectively. As it can be seen, for this galaxy, the SIDM
model produces marginally viable rotation curves which are
incompatible with the SPARC data.

\begin{figure}[h!]
\centering
\includegraphics[width=20pc]{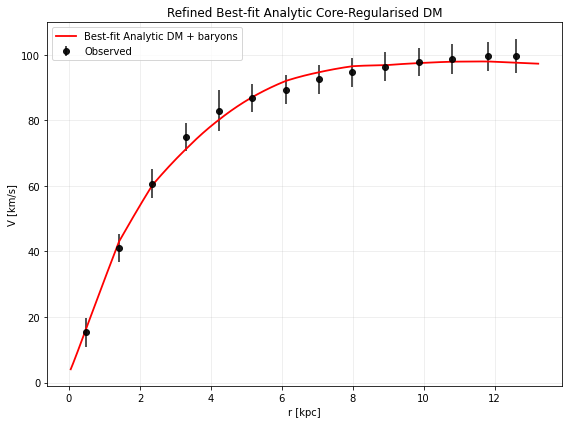}
\caption{The predicted rotation curves after using an optimization
for the SIDM model (\ref{ScaledependentEoSDM}), and the extended
SPARC data for the galaxy F574-1. We included the rotation curves
of the gas, the disk velocities, the bulge (where present) along
with the SIDM model.} \label{extendedF574-1}
\end{figure}

Now we shall include contributions to the rotation velocity from
the other components of the galaxy, namely the disk, the gas, and
the bulge if present. In Fig. \ref{extendedF574-1} we present the
combined rotation curves including all the components of the
galaxy along with the SIDM. As it can be seen, the extended
collisional DM model is viable.
Also in Table \ref{evaluationextendedF574-1} we present the
optimized values of the free parameters of the SIDM model for
which  we achieve the maximum compatibility with the SPARC data,
for the galaxy F574-1, and also the resulting reduced
$\chi^2_{red}$ value.
\begin{table}[h!]
\centering \caption{Optimized Parameter Values of the Extended
SIDM model for the Galaxy F574-1.}
\begin{tabular}{lc}
\hline
Parameter & Value  \\
\hline
$\rho_0 $ ($M_{\odot}/\mathrm{Kpc}^{3}$) & $2.46509\times 10^7$   \\
$K_0$ ($M_{\odot} \,
\mathrm{Kpc}^{-3} \, (\mathrm{km/s})^{2}$) & 2904.38   \\
$ml_{\text{disk}}$ & 0.9765 \\
$ml_{\text{bulge}}$ & 0.6321 \\
$\alpha$ (Kpc) & 6.26334\\
$\chi^2_{red}$ & 0.230896 \\
\hline
\end{tabular}
\label{evaluationextendedF574-1}
\end{table}

\subsection{The Galaxy F583-1}

For this galaxy, the optimization method we used, ensures maximum
compatibility of the analytic SIDM model of Eq.
(\ref{ScaledependentEoSDM}) with the SPARC data, if we choose
$\rho_0=1.88326\times 10^7$$M_{\odot}/\mathrm{Kpc}^{3}$ and
$K_0=3160.22
$$M_{\odot} \, \mathrm{Kpc}^{-3} \, (\mathrm{km/s})^{2}$, in which
case the reduced $\chi^2_{red}$ value is $\chi^2_{red}=0.249447$.
Also the parameter $\alpha$ in this case is $\alpha=7.47573 $Kpc.

\begin{figure}[h!]
\centering
\includegraphics[width=35pc]{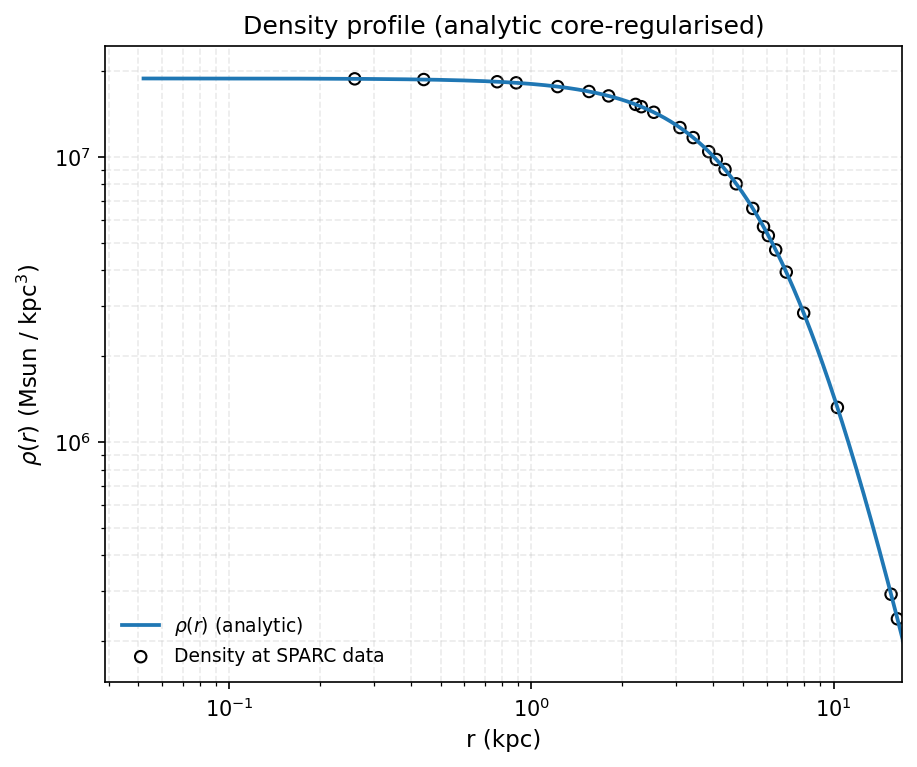}
\caption{The density of the SIDM model of Eq.
(\ref{ScaledependentEoSDM}) for the galaxy F583-1, versus the
radius.} \label{F583-1dens}
\end{figure}

\begin{figure}[h!]
\centering
\includegraphics[width=35pc]{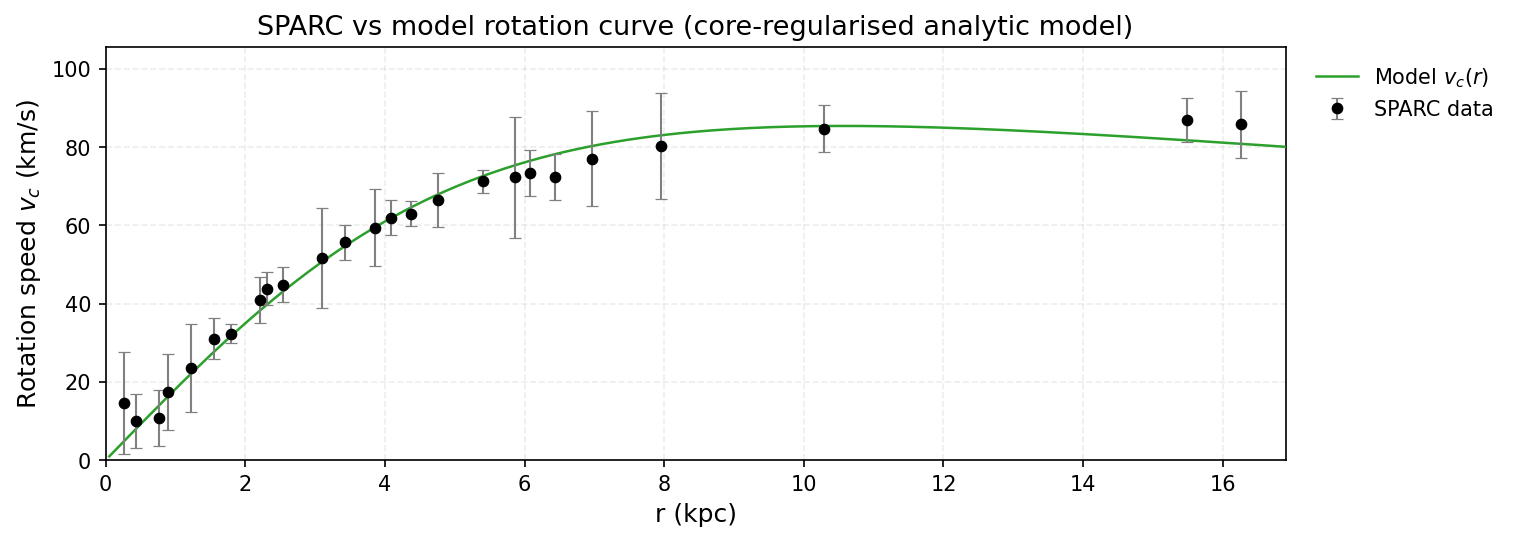}
\caption{The predicted rotation curves for the optimized SIDM
model of Eq. (\ref{ScaledependentEoSDM}), versus the SPARC
observational data for the galaxy F583-1.} \label{F583-1}
\end{figure}

In Table \ref{collF583-1} we present the optimized values of $K_0$
and $\rho_0$ for the analytic SIDM model of Eq.
(\ref{ScaledependentEoSDM}) for which the maximum compatibility
with the SPARC data is achieved.
\begin{table}[h!]
  \begin{center}
    \caption{SIDM Optimization Values for the galaxy F583-1}
    \label{collF583-1}
     \begin{tabular}{|r|r|}
     \hline
      \textbf{Parameter}   & \textbf{Optimization Values}
      \\  \hline
     $\rho_0 $  ($M_{\odot}/\mathrm{Kpc}^{3}$) & $1.88326\times 10^7$
\\  \hline $K_0$ ($M_{\odot} \,
\mathrm{Kpc}^{-3} \, (\mathrm{km/s})^{2}$)& 3160.22
\\  \hline
    \end{tabular}
  \end{center}
\end{table}
In Figs. \ref{F583-1dens}, \ref{F583-1}  we present the density of
the analytic SIDM model, the predicted rotation curves for the
SIDM model (\ref{ScaledependentEoSDM}), versus the SPARC
observational data and the sound speed, as a function of the
radius respectively. As it can be seen, for this galaxy, the SIDM
model produces viable rotation curves which are marginally
compatible with the SPARC data.

\begin{figure}[h!]
\centering
\includegraphics[width=20pc]{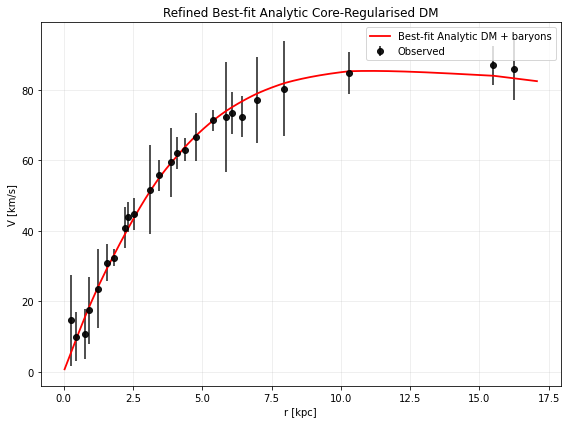}
\caption{The predicted rotation curves after using an optimization
for the SIDM model (\ref{ScaledependentEoSDM}), and the extended
SPARC data for the galaxy F583-1. We included the rotation curves
of the gas, the disk velocities, the bulge (where present) along
with the SIDM model.} \label{extendedF583-1}
\end{figure}

Now we shall include contributions to the rotation velocity from
the other components of the galaxy, namely the disk, the gas, and
the bulge if present. In Fig. \ref{extendedF583-1} we present the
combined rotation curves including all the components of the
galaxy along with the SIDM. As it can be seen, the extended
collisional DM model is viable.
Also in Table \ref{evaluationextendedF583-1} we present the
optimized values of the free parameters of the SIDM model for
which  we achieve the maximum compatibility with the SPARC data,
for the galaxy F583-1, and also the resulting reduced
$\chi^2_{red}$ value.
\begin{table}[h!]
\centering \caption{Optimized Parameter Values of the Extended
SIDM model for the Galaxy F583-1.}
\begin{tabular}{lc}
\hline
Parameter & Value  \\
\hline
$\rho_0 $ ($M_{\odot}/\mathrm{Kpc}^{3}$) & $1.8539\times 10^7$   \\
$K_0$ ($M_{\odot} \,
\mathrm{Kpc}^{-3} \, (\mathrm{km/s})^{2}$) & 2654.75   \\
$ml_{\text{disk}}$ & 0.5424 \\
$ml_{\text{bulge}}$ & 0.5184 \\
$\alpha$ (Kpc) & 6.90502\\
$\chi^2_{red}$ & 0.147575 \\
\hline
\end{tabular}
\label{evaluationextendedF583-1}
\end{table}

\subsection{The Galaxy F583-4, Marginally Viable, Extended Viable}

For this galaxy, the optimization method we used, ensures maximum
compatibility of the analytic SIDM model of Eq.
(\ref{ScaledependentEoSDM}) with the SPARC data, if we choose
$\rho_0=3.99539\times 10^7$$M_{\odot}/\mathrm{Kpc}^{3}$ and
$K_0=1735.41
$$M_{\odot} \, \mathrm{Kpc}^{-3} \, (\mathrm{km/s})^{2}$, in which
case the reduced $\chi^2_{red}$ value is $\chi^2_{red}=1.10839$.
Also the parameter $\alpha$ in this case is $\alpha=3.8034 $Kpc.

\begin{figure}[h!]
\centering
\includegraphics[width=20pc]{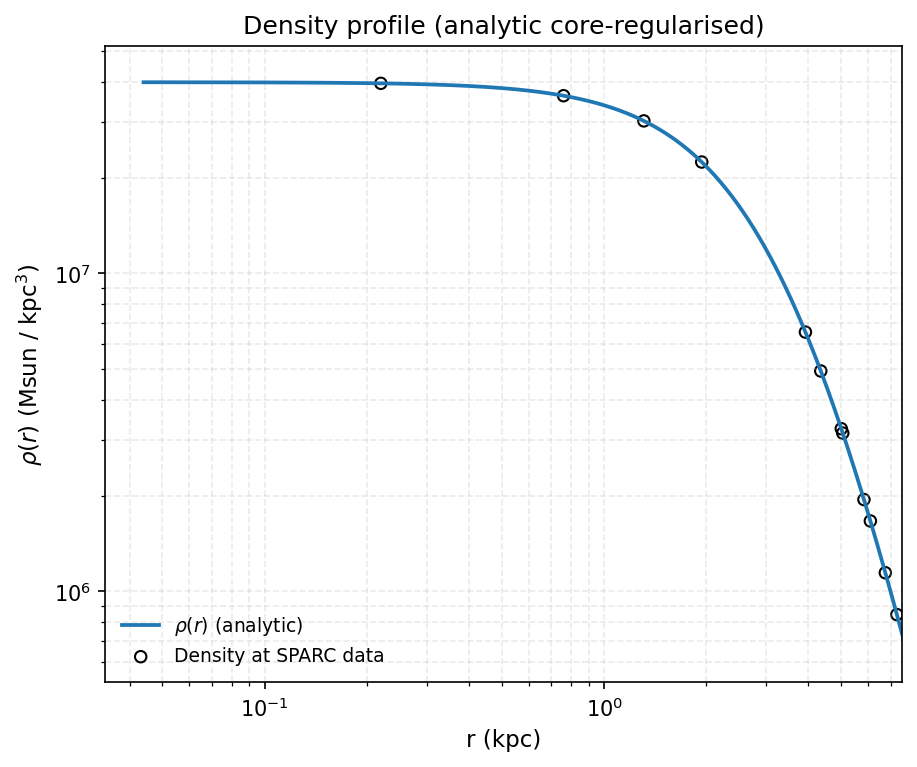}
\caption{The density of the SIDM model of Eq.
(\ref{ScaledependentEoSDM}) for the galaxy F583-4, versus the
radius.} \label{F583-4dens}
\end{figure}

\begin{figure}[h!]
\centering
\includegraphics[width=35pc]{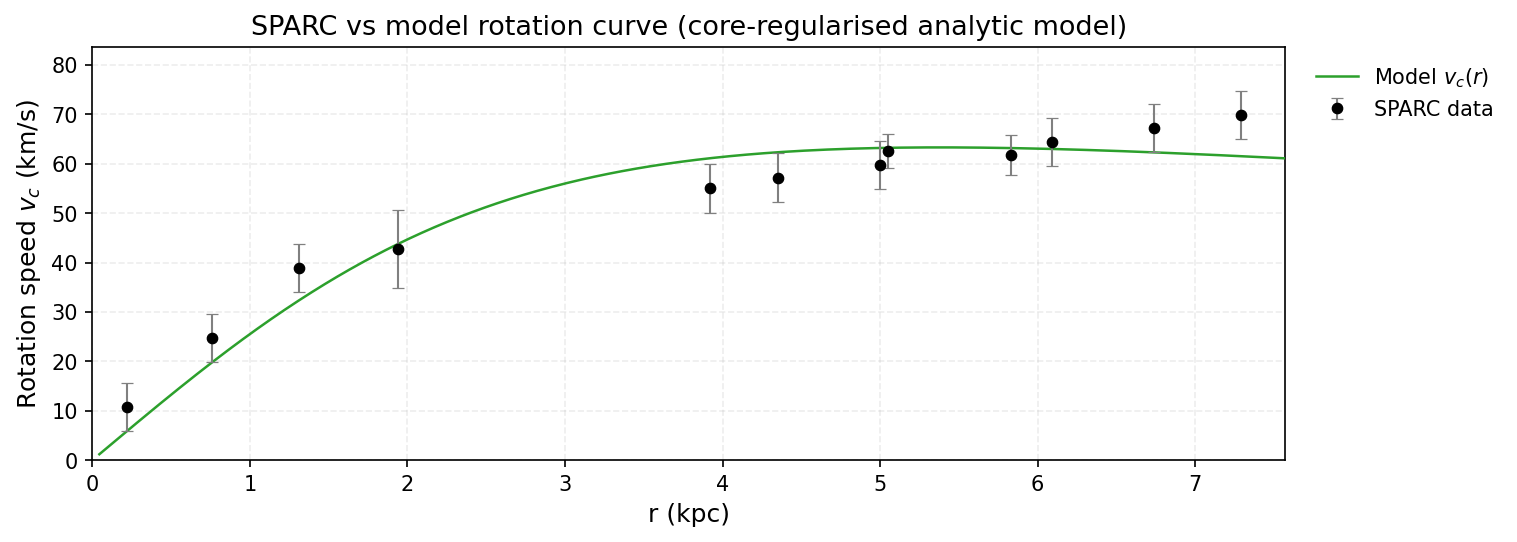}
\caption{The predicted rotation curves for the optimized SIDM
model of Eq. (\ref{ScaledependentEoSDM}), versus the SPARC
observational data for the galaxy F583-4.} \label{F583-4}
\end{figure}

In Table \ref{collF583-4} we present the optimized values of $K_0$
and $\rho_0$ for the analytic SIDM model of Eq.
(\ref{ScaledependentEoSDM}) for which the maximum compatibility
with the SPARC data is achieved.
\begin{table}[h!]
  \begin{center}
    \caption{SIDM Optimization Values for the galaxy F583-4}
    \label{collF583-4}
     \begin{tabular}{|r|r|}
     \hline
      \textbf{Parameter}   & \textbf{Optimization Values}
      \\  \hline
     $\rho_0 $  ($M_{\odot}/\mathrm{Kpc}^{3}$) & $3.99539\times 10^7$
\\  \hline $K_0$ ($M_{\odot} \,
\mathrm{Kpc}^{-3} \, (\mathrm{km/s})^{2}$)& 1735.41
\\  \hline
    \end{tabular}
  \end{center}
\end{table}
In Figs. \ref{F583-4dens}, \ref{F583-4} we present the density of
the analytic SIDM model, the predicted rotation curves for the
SIDM model (\ref{ScaledependentEoSDM}), versus the SPARC
observational data and the sound speed, as a function of the
radius respectively. As it can be seen, for this galaxy, the SIDM
model produces marginally viable rotation curves which are
marginally compatible with the SPARC data.

\begin{figure}[h!]
\centering
\includegraphics[width=20pc]{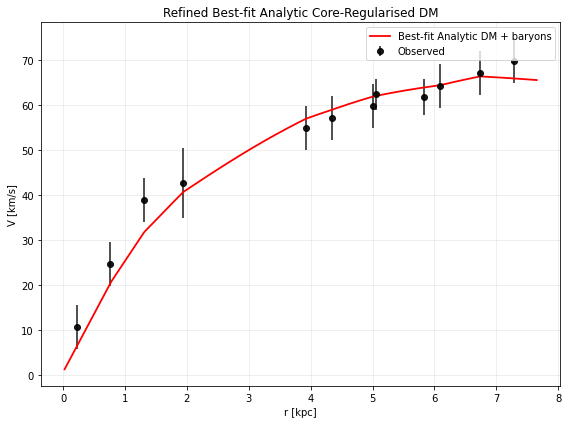}
\caption{The predicted rotation curves after using an optimization
for the SIDM model (\ref{ScaledependentEoSDM}), and the extended
SPARC data for the galaxy F583-4. We included the rotation curves
of the gas, the disk velocities, the bulge (where present) along
with the SIDM model.} \label{extendedF583-4}
\end{figure}

Now we shall include contributions to the rotation velocity from
the other components of the galaxy, namely the disk, the gas, and
the bulge if present. In Fig. \ref{extendedF583-4} we present the
combined rotation curves including all the components of the
galaxy along with the SIDM. As it can be seen, the extended
collisional DM model is marginally viable.
Also in Table \ref{evaluationextendedF583-4} we present the
optimized values of the free parameters of the SIDM model for
which  we achieve the maximum compatibility with the SPARC data,
for the galaxy F583-4, and also the resulting reduced
$\chi^2_{red}$ value.
\begin{table}[h!]
\centering \caption{Optimized Parameter Values of the Extended
SIDM model for the Galaxy F583-4.}
\begin{tabular}{lc}
\hline
Parameter & Value  \\
\hline
$\rho_0 $ ($M_{\odot}/\mathrm{Kpc}^{3}$) & $9.93357\times 10^6$   \\
$K_0$ ($M_{\odot} \,
\mathrm{Kpc}^{-3} \, (\mathrm{km/s})^{2}$) & 1233.91   \\
$ml_{\text{disk}}$ & 1 \\
$ml_{\text{bulge}}$ & 0.3 \\
$\alpha$ (Kpc) & 6.43111\\
$\chi^2_{red}$ & 0.652711 \\
\hline
\end{tabular}
\label{evaluationextendedF583-4}
\end{table}

\subsection{The Galaxy IC2574 Non-viable}

For this galaxy, the optimization method we used, ensures maximum
compatibility of the analytic SIDM model of Eq.
(\ref{ScaledependentEoSDM}) with the SPARC data, if we choose
$\rho_0=5.59403\times 10^6$$M_{\odot}/\mathrm{Kpc}^{3}$ and
$K_0=1904.15
$$M_{\odot} \, \mathrm{Kpc}^{-3} \, (\mathrm{km/s})^{2}$, in which
case the reduced $\chi^2_{red}$ value is $\chi^2_{red}=7.51187$.
Also the parameter $\alpha$ in this case is $\alpha=10.6473 $Kpc.

\begin{figure}[h!]
\centering
\includegraphics[width=20pc]{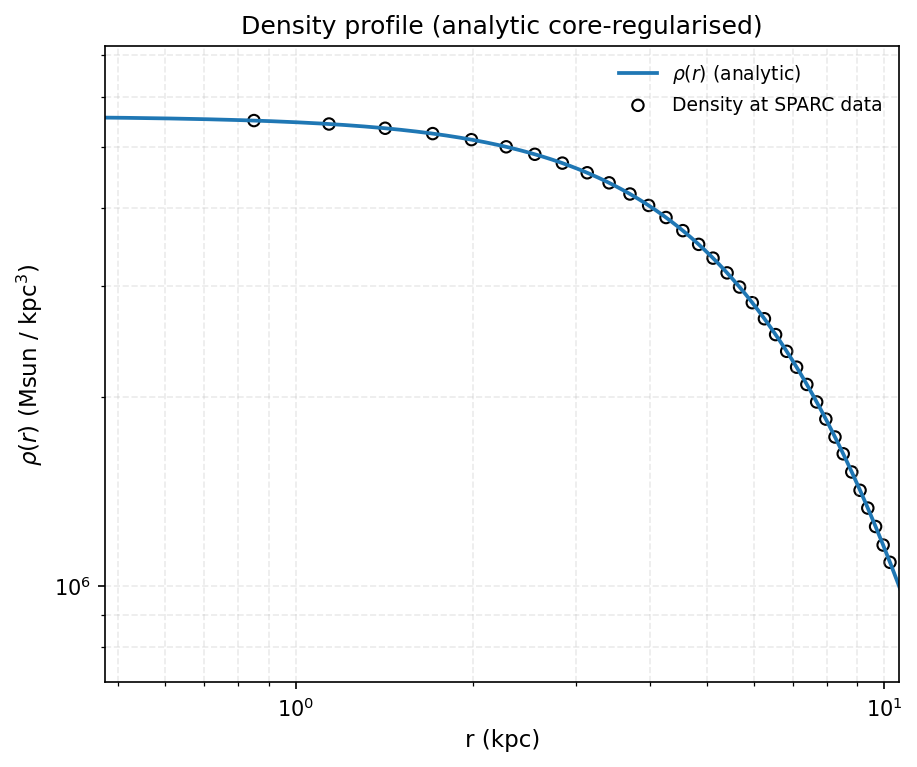}
\caption{The density of the SIDM model of Eq.
(\ref{ScaledependentEoSDM}) for the galaxy IC2574, versus the
radius.} \label{IC2574dens}
\end{figure}

\begin{figure}[h!]
\centering
\includegraphics[width=35pc]{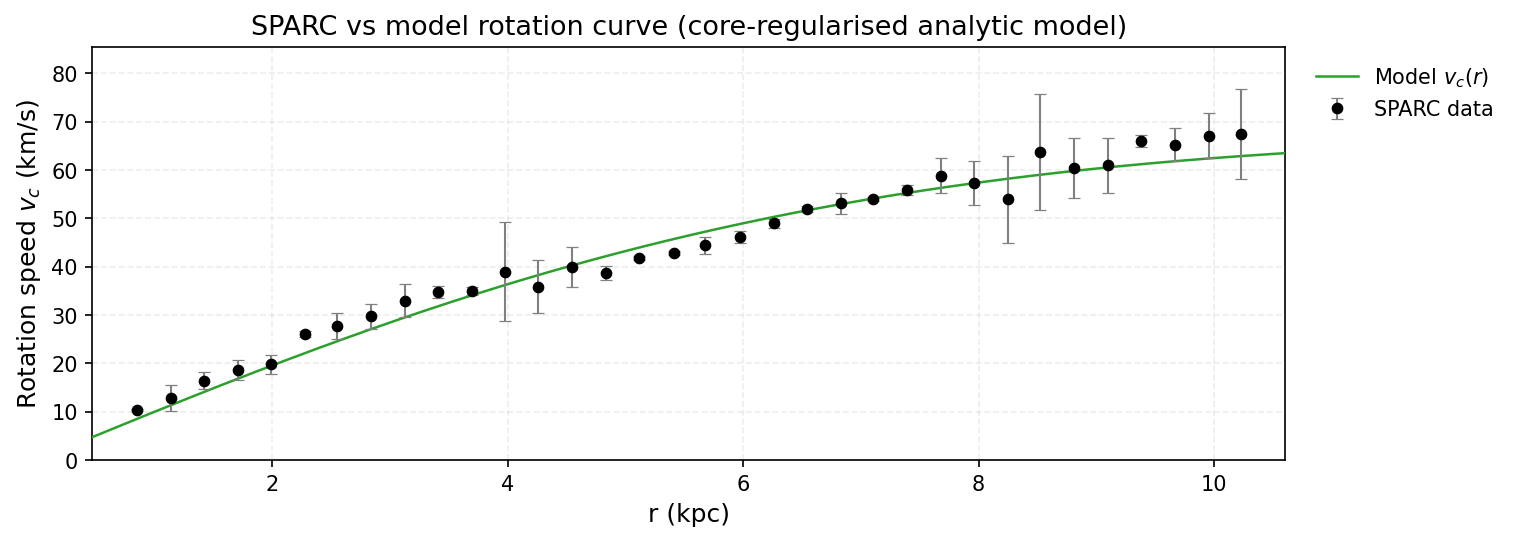}
\caption{The predicted rotation curves for the optimized SIDM
model of Eq. (\ref{ScaledependentEoSDM}), versus the SPARC
observational data for the galaxy IC2574.} \label{IC2574}
\end{figure}

In Table \ref{collIC2574} we present the optimized values of $K_0$
and $\rho_0$ for the analytic SIDM model of Eq.
(\ref{ScaledependentEoSDM}) for which the maximum compatibility
with the SPARC data is achieved.
\begin{table}[h!]
  \begin{center}
    \caption{SIDM Optimization Values for the galaxy IC2574}
    \label{collIC2574}
     \begin{tabular}{|r|r|}
     \hline
      \textbf{Parameter}   & \textbf{Optimization Values}
      \\  \hline
     $\rho_0 $  ($M_{\odot}/\mathrm{Kpc}^{3}$) & $5.59403\times 10^6$
\\  \hline $K_0$ ($M_{\odot} \,
\mathrm{Kpc}^{-3} \, (\mathrm{km/s})^{2}$)& 1904.15
\\  \hline
    \end{tabular}
  \end{center}
\end{table}
In Figs. \ref{IC2574dens}, \ref{IC2574}  we present the density of
the analytic SIDM model, the predicted rotation curves for the
SIDM model (\ref{ScaledependentEoSDM}), versus the SPARC
observational data and the sound speed, as a function of the
radius respectively. As it can be seen, for this galaxy, the SIDM
model produces non-viable rotation curves which are incompatible
with the SPARC data.

\begin{figure}[h!]
\centering
\includegraphics[width=20pc]{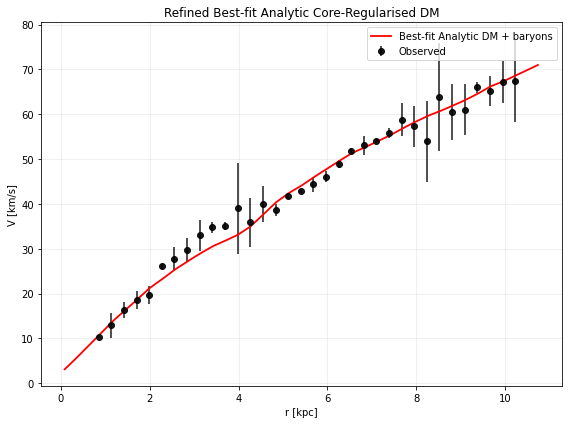}
\caption{The predicted rotation curves after using an optimization
for the SIDM model (\ref{ScaledependentEoSDM}), and the extended
SPARC data for the galaxy IC2574. We included the rotation curves
of the gas, the disk velocities, the bulge (where present) along
with the SIDM model.} \label{extendedIC2574}
\end{figure}

Now we shall include contributions to the rotation velocity from
the other components of the galaxy, namely the disk, the gas, and
the bulge if present. In Fig. \ref{extendedIC2574} we present the
combined rotation curves including all the components of the
galaxy along with the SIDM. As it can be seen, the extended
collisional DM model is non-viable.
Also in Table \ref{evaluationextendedIC2574} we present the
optimized values of the free parameters of the SIDM model for
which  we achieve the maximum compatibility with the SPARC data,
for the galaxy IC2574, and also the resulting reduced
$\chi^2_{red}$ value.
\begin{table}[h!]
\centering \caption{Optimized Parameter Values of the Extended
SIDM model for the Galaxy IC2574.}
\begin{tabular}{lc}
\hline
Parameter & Value  \\
\hline
$\rho_0 $ ($M_{\odot}/\mathrm{Kpc}^{3}$) & $2.41695\times 10^6$   \\
$K_0$ ($M_{\odot} \,
\mathrm{Kpc}^{-3} \, (\mathrm{km/s})^{2}$) & 3730.76   \\
$ml_{\text{disk}}$ & 0.9459 \\
$ml_{\text{bulge}}$ & 0.1211 \\
$\alpha$ (Kpc) & 22.6705 $ ^{**}$\\
$\chi^2_{red}$ & 2.44598 \\
\hline
\end{tabular}
\label{evaluationextendedIC2574}
\end{table}

\subsection{The Galaxy KK98-251}

For this galaxy, the optimization method we used, ensures maximum
compatibility of the analytic SIDM model of Eq.
(\ref{ScaledependentEoSDM}) with the SPARC data, if we choose
$\rho_0=2.10907\times 10^7$$M_{\odot}/\mathrm{Kpc}^{3}$ and
$K_0=616.254
$$M_{\odot} \, \mathrm{Kpc}^{-3} \, (\mathrm{km/s})^{2}$, in which
case the reduced $\chi^2_{red}$ value is $\chi^2_{red}=0.569321$.
Also the parameter $\alpha$ in this case is $\alpha=3.1195 $Kpc.

\begin{figure}[h!]
\centering
\includegraphics[width=20pc]{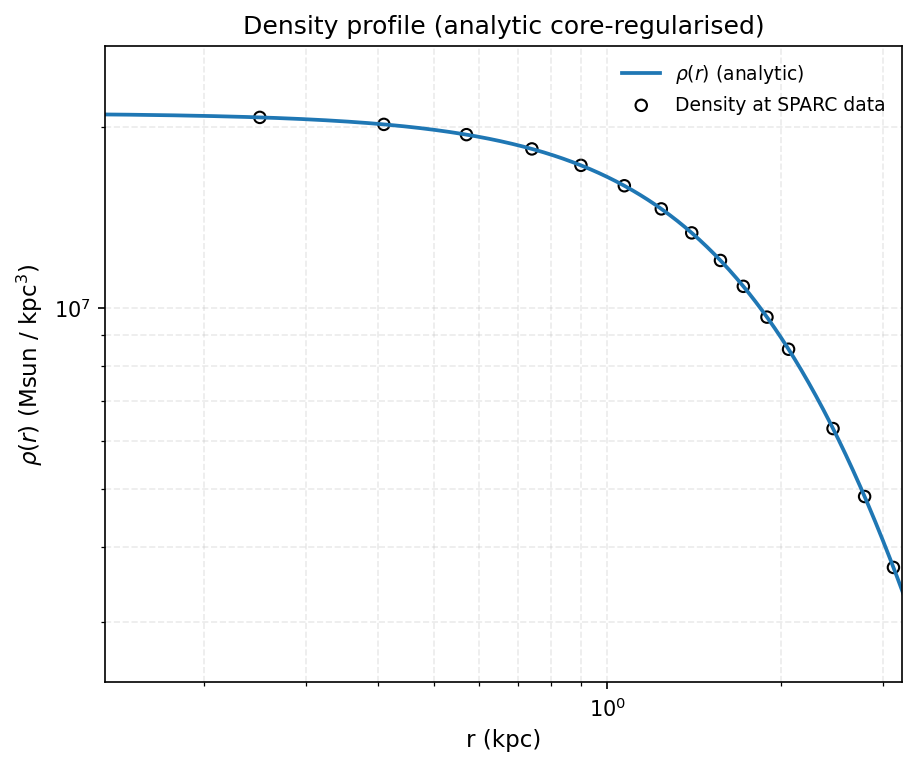}
\caption{The density of the SIDM model of Eq.
(\ref{ScaledependentEoSDM}) for the galaxy KK98-251, versus the
radius.} \label{KK98-251dens}
\end{figure}

\begin{figure}[h!]
\centering
\includegraphics[width=35pc]{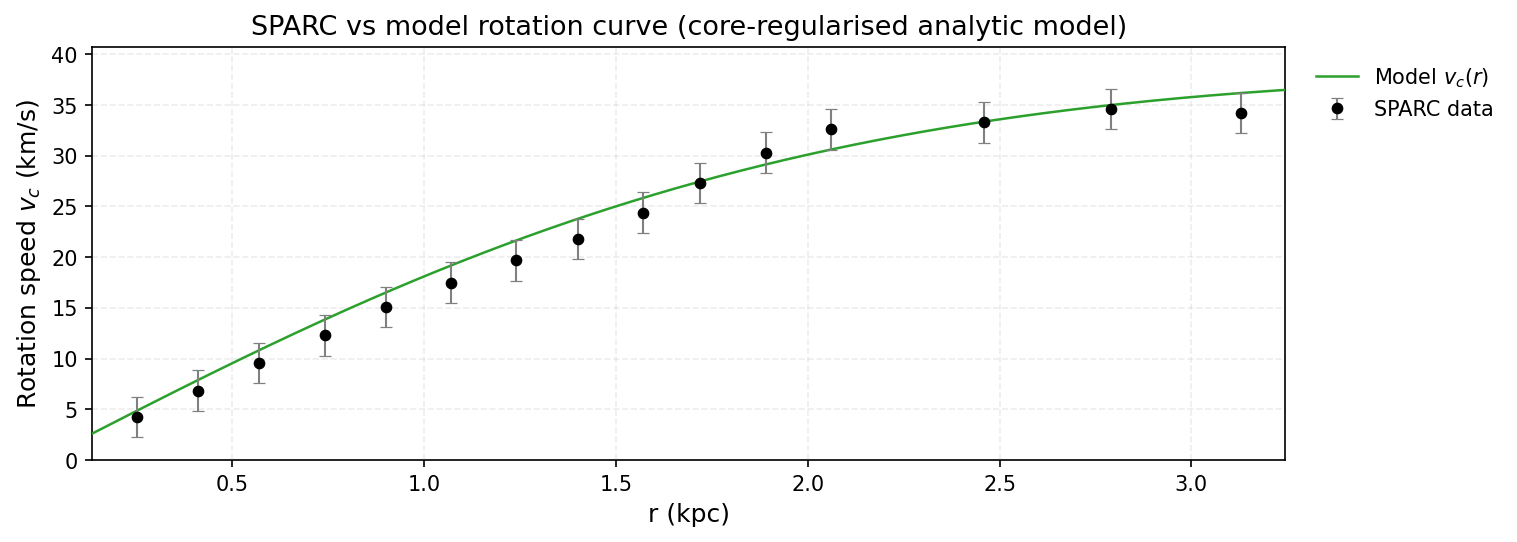}
\caption{The predicted rotation curves for the optimized SIDM
model of Eq. (\ref{ScaledependentEoSDM}), versus the SPARC
observational data for the galaxy KK98-251.} \label{KK98-251}
\end{figure}

In Table \ref{collKK98-251} we present the optimized values of
$K_0$ and $\rho_0$ for the analytic SIDM model of Eq.
(\ref{ScaledependentEoSDM}) for which the maximum compatibility
with the SPARC data is achieved.
\begin{table}[h!]
  \begin{center}
    \caption{SIDM Optimization Values for the galaxy KK98-251}
    \label{collKK98-251}
     \begin{tabular}{|r|r|}
     \hline
      \textbf{Parameter}   & \textbf{Optimization Values}
      \\  \hline
     $\rho_0 $  ($M_{\odot}/\mathrm{Kpc}^{3}$) & $2.10907\times 10^7$
\\  \hline $K_0$ ($M_{\odot} \,
\mathrm{Kpc}^{-3} \, (\mathrm{km/s})^{2}$)& 616.254
\\  \hline
    \end{tabular}
  \end{center}
\end{table}
In Figs. \ref{KK98-251dens}, \ref{KK98-251}  we present the
density of the analytic SIDM model, the predicted rotation curves
for the SIDM model (\ref{ScaledependentEoSDM}), versus the SPARC
observational data and the sound speed, as a function of the
radius respectively. As it can be seen, for this galaxy, the SIDM
model produces viable rotation curves which are compatible with
the SPARC data.

\subsection{The Galaxy NGC0024, Non-viable}

For this galaxy, the optimization method we used, ensures maximum
compatibility of the analytic SIDM model of Eq.
(\ref{ScaledependentEoSDM}) with the SPARC data, if we choose
$\rho_0=3.1262\times 10^8$$M_{\odot}/\mathrm{Kpc}^{3}$ and
$K_0=6437.6
$$M_{\odot} \, \mathrm{Kpc}^{-3} \, (\mathrm{km/s})^{2}$, in which
case the reduced $\chi^2_{red}$ value is $\chi^2_{red}=2.86598$.
Also the parameter $\alpha$ in this case is $\alpha=2.61881 $Kpc.

\begin{figure}[h!]
\centering
\includegraphics[width=20pc]{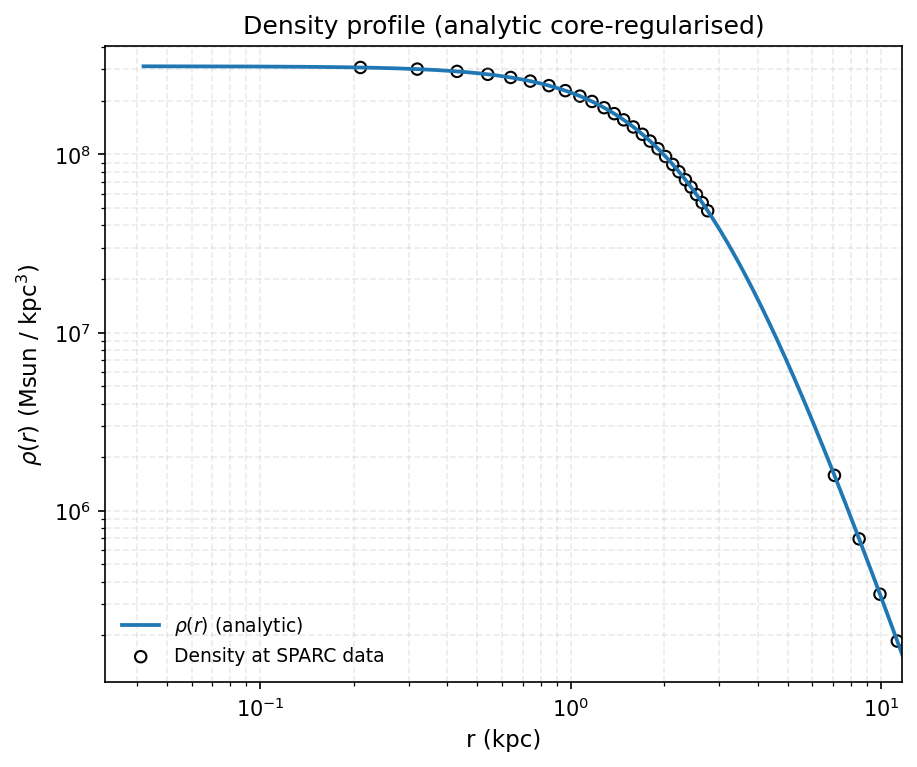}
\caption{The density of the SIDM model of Eq.
(\ref{ScaledependentEoSDM}) for the galaxy NGC0024, versus the
radius.} \label{NGC0024dens}
\end{figure}

\begin{figure}[h!]
\centering
\includegraphics[width=35pc]{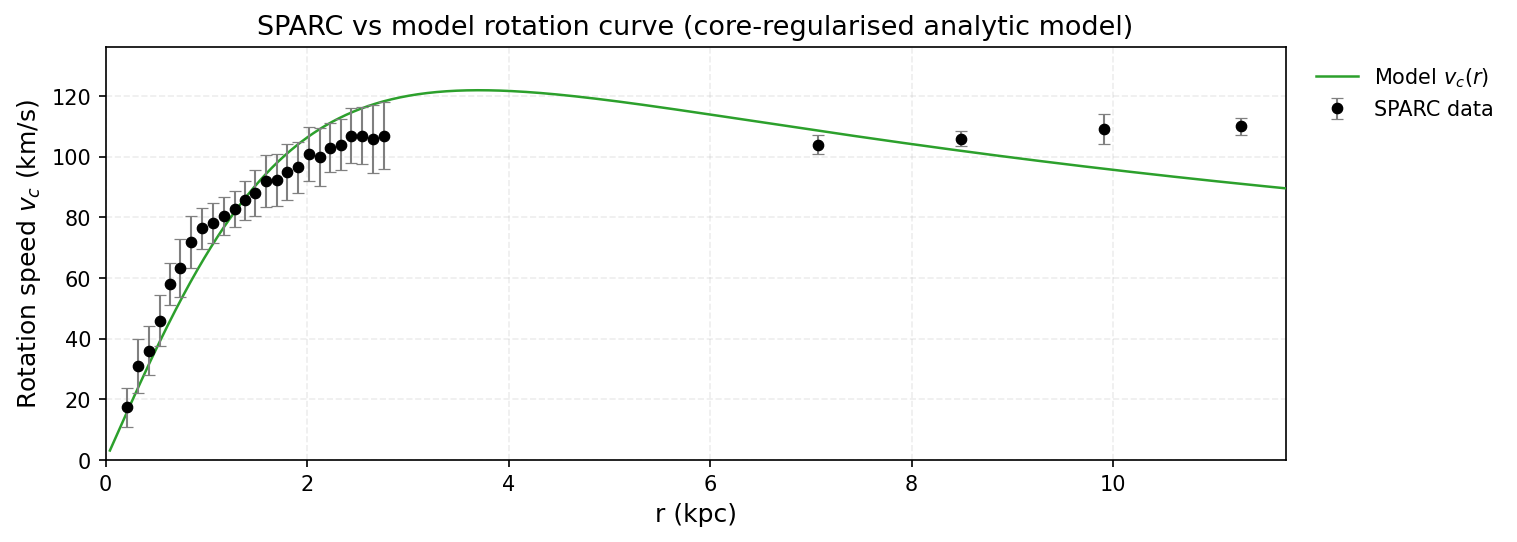}
\caption{The predicted rotation curves for the optimized SIDM
model of Eq. (\ref{ScaledependentEoSDM}), versus the SPARC
observational data for the galaxy NGC0024.} \label{NGC0024}
\end{figure}

In Table \ref{collNGC0024} we present the optimized values of
$K_0$ and $\rho_0$ for the analytic SIDM model of Eq.
(\ref{ScaledependentEoSDM}) for which the maximum compatibility
with the SPARC data is achieved.
\begin{table}[h!]
  \begin{center}
    \caption{SIDM Optimization Values for the galaxy NGC0024}
    \label{collNGC0024}
     \begin{tabular}{|r|r|}
     \hline
      \textbf{Parameter}   & \textbf{Optimization Values}
      \\  \hline
     $\rho_0 $  ($M_{\odot}/\mathrm{Kpc}^{3}$) & $3.1262\times 10^8$
\\  \hline $K_0$ ($M_{\odot} \,
\mathrm{Kpc}^{-3} \, (\mathrm{km/s})^{2}$)& 3.1262
\\  \hline
    \end{tabular}
  \end{center}
\end{table}
In Figs. \ref{NGC0024dens}, \ref{NGC0024} we present the density
of the analytic SIDM model, the predicted rotation curves for the
SIDM model (\ref{ScaledependentEoSDM}), versus the SPARC
observational data and the sound speed, as a function of the
radius respectively. As it can be seen, for this galaxy, the SIDM
model produces non-viable rotation curves which are incompatible
with the SPARC data.

\begin{figure}[h!]
\centering
\includegraphics[width=20pc]{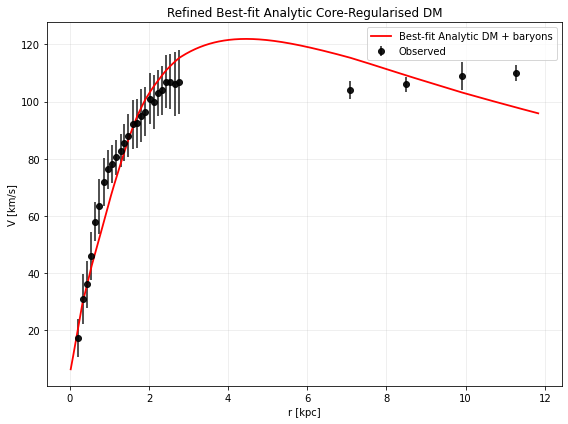}
\caption{The predicted rotation curves after using an optimization
for the SIDM model (\ref{ScaledependentEoSDM}), and the extended
SPARC data for the galaxy NGC0024. We included the rotation curves
of the gas, the disk velocities, the bulge (where present) along
with the SIDM model.} \label{extendedNGC0024}
\end{figure}

Now we shall include contributions to the rotation velocity from
the other components of the galaxy, namely the disk, the gas, and
the bulge if present. In Fig. \ref{extendedNGC0024} we present the
combined rotation curves including all the components of the
galaxy along with the SIDM. As it can be seen, the extended
collisional DM model is non-viable.
Also in Table \ref{evaluationextendedNGC0024} we present the
optimized values of the free parameters of the SIDM model for
which  we achieve the maximum compatibility with the SPARC data,
for the galaxy NGC0024, and also the resulting reduced
$\chi^2_{red}$ value.
\begin{table}[h!]
\centering \caption{Optimized Parameter Values of the Extended
SIDM model for the Galaxy NGC0024.}
\begin{tabular}{lc}
\hline
Parameter & Value  \\
\hline
$\rho_0 $ ($M_{\odot}/\mathrm{Kpc}^{3}$) & $1.18583\times 10^8$   \\
$K_0$ ($M_{\odot} \,
\mathrm{Kpc}^{-3} \, (\mathrm{km/s})^{2}$) & 4583.52   \\
$ml_{\text{disk}}$ & 1 \\
$ml_{\text{bulge}}$ & 0.5 \\
$\alpha$ (Kpc) & 3.58743\\
$\chi^2_{red}$ & 2.11333 \\
\hline
\end{tabular}
\label{evaluationextendedNGC0024}
\end{table}

\subsection{The Galaxy NGC0055, Marginally Viable, Extended Viable}

For this galaxy, the optimization method we used, ensures maximum
compatibility of the analytic SIDM model of Eq.
(\ref{ScaledependentEoSDM}) with the SPARC data, if we choose
$\rho_0=1.87465\times 10^7$$M_{\odot}/\mathrm{Kpc}^{3}$ and
$K_0=3194.12
$$M_{\odot} \, \mathrm{Kpc}^{-3} \, (\mathrm{km/s})^{2}$, in which
case the reduced $\chi^2_{red}$ value is $\chi^2_{red}=0.731137$.
Also the parameter $\alpha$ in this case is $\alpha=7.53296 $Kpc.

\begin{figure}[h!]
\centering
\includegraphics[width=20pc]{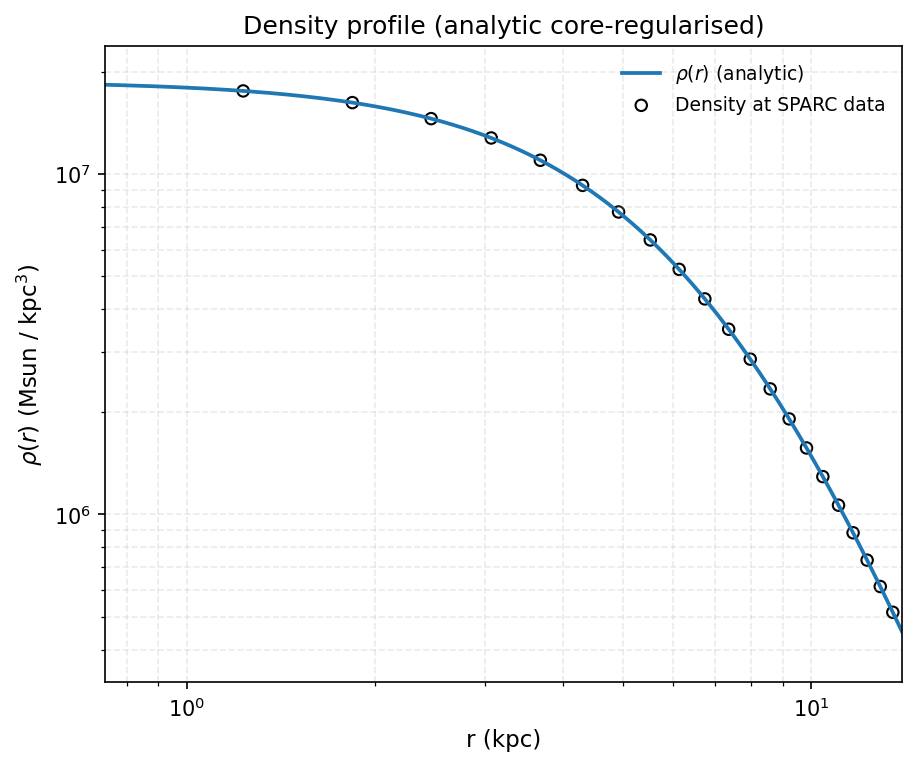}
\caption{The density of the SIDM model of Eq.
(\ref{ScaledependentEoSDM}) for the galaxy NGC0055, versus the
radius.} \label{NGC0055dens}
\end{figure}

\begin{figure}[h!]
\centering
\includegraphics[width=35pc]{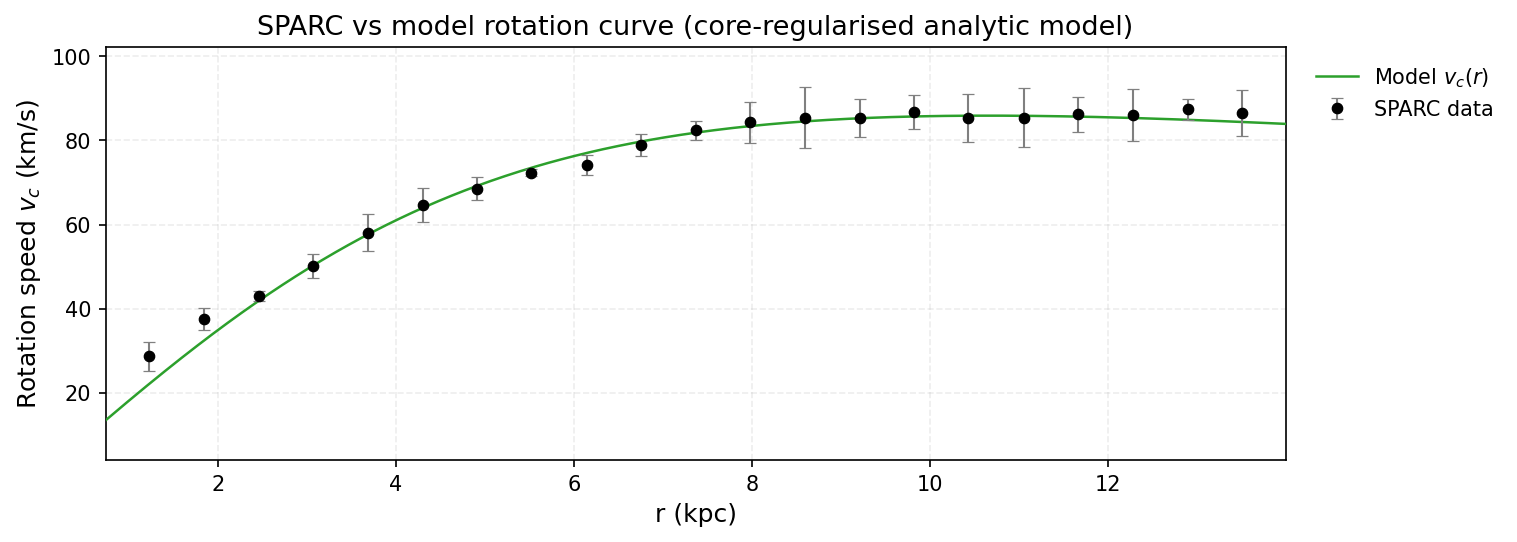}
\caption{The predicted rotation curves for the optimized SIDM
model of Eq. (\ref{ScaledependentEoSDM}), versus the SPARC
observational data for the galaxy NGC0055.} \label{NGC0055}
\end{figure}

In Table \ref{collNGC0055} we present the optimized values of
$K_0$ and $\rho_0$ for the analytic SIDM model of Eq.
(\ref{ScaledependentEoSDM}) for which the maximum compatibility
with the SPARC data is achieved.
\begin{table}[h!]
  \begin{center}
    \caption{SIDM Optimization Values for the galaxy NGC0055}
    \label{collNGC0055}
     \begin{tabular}{|r|r|}
     \hline
      \textbf{Parameter}   & \textbf{Optimization Values}
      \\  \hline
     $\rho_0 $  ($M_{\odot}/\mathrm{Kpc}^{3}$) & $1.87465\times 10^7$
\\  \hline $K_0$ ($M_{\odot} \,
\mathrm{Kpc}^{-3} \, (\mathrm{km/s})^{2}$)& 3194.12
\\  \hline
    \end{tabular}
  \end{center}
\end{table}
In Figs. \ref{NGC0055dens}, \ref{NGC0055} we present the density
of the analytic SIDM model, the predicted rotation curves for the
SIDM model (\ref{ScaledependentEoSDM}), versus the SPARC
observational data and the sound speed, as a function of the
radius respectively. As it can be seen, for this galaxy, the SIDM
model produces marginally viable rotation curves which are
marginally compatible with the SPARC data.

\begin{figure}[h!]
\centering
\includegraphics[width=20pc]{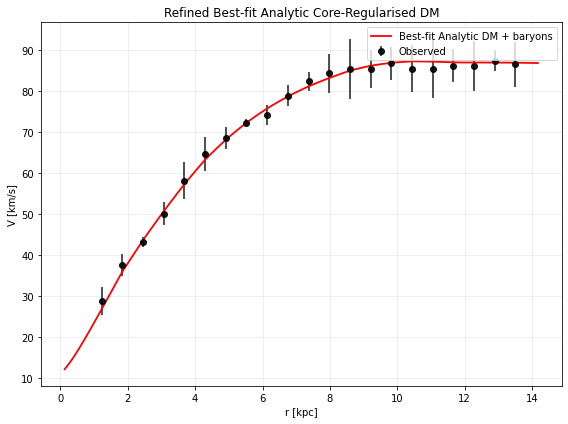}
\caption{The predicted rotation curves after using an optimization
for the SIDM model (\ref{ScaledependentEoSDM}), and the extended
SPARC data for the galaxy NGC0055. We included the rotation curves
of the gas, the disk velocities, the bulge (where present) along
with the SIDM model.} \label{extendedNGC0055}
\end{figure}

Now we shall include contributions to the rotation velocity from
the other components of the galaxy, namely the disk, the gas, and
the bulge if present. In Fig. \ref{extendedNGC0055} we present the
combined rotation curves including all the components of the
galaxy along with the SIDM. As it can be seen, the extended
collisional DM model is viable.
Also in Table \ref{evaluationextendedNGC0055} we present the
optimized values of the free parameters of the SIDM model for
which  we achieve the maximum compatibility with the SPARC data,
for the galaxy NGC0055, and also the resulting reduced
$\chi^2_{red}$ value.
\begin{table}[h!]
\centering \caption{Optimized Parameter Values of the Extended
SIDM model for the Galaxy NGC0055.}
\begin{tabular}{lc}
\hline
Parameter & Value  \\
\hline
$\rho_0 $ ($M_{\odot}/\mathrm{Kpc}^{3}$) & $9.57524\times 10^6$   \\
$K_0$ ($M_{\odot} \,
\mathrm{Kpc}^{-3} \, (\mathrm{km/s})^{2}$) & 2410.56   \\
$ml_{\text{disk}}$ & 0.6115 \\
$ml_{\text{bulge}}$ & 0.2901 \\
$\alpha$ (Kpc) & 9.15547\\
$\chi^2_{red}$ & 0.131558 \\
\hline
\end{tabular}
\label{evaluationextendedNGC0055}
\end{table}

\subsection{The Galaxy NGC0100}

For this galaxy, the optimization method we used, ensures maximum
compatibility of the analytic SIDM model of Eq.
(\ref{ScaledependentEoSDM}) with the SPARC data, if we choose
$\rho_0=3.77865\times 10^7$$M_{\odot}/\mathrm{Kpc}^{3}$ and
$K_0=3237.93
$$M_{\odot} \, \mathrm{Kpc}^{-3} \, (\mathrm{km/s})^{2}$, in which
case the reduced $\chi^2_{red}$ value is $\chi^2_{red}=0.386342$.
Also the parameter $\alpha$ in this case is $\alpha=5.34215 $Kpc.

\begin{figure}[h!]
\centering
\includegraphics[width=20pc]{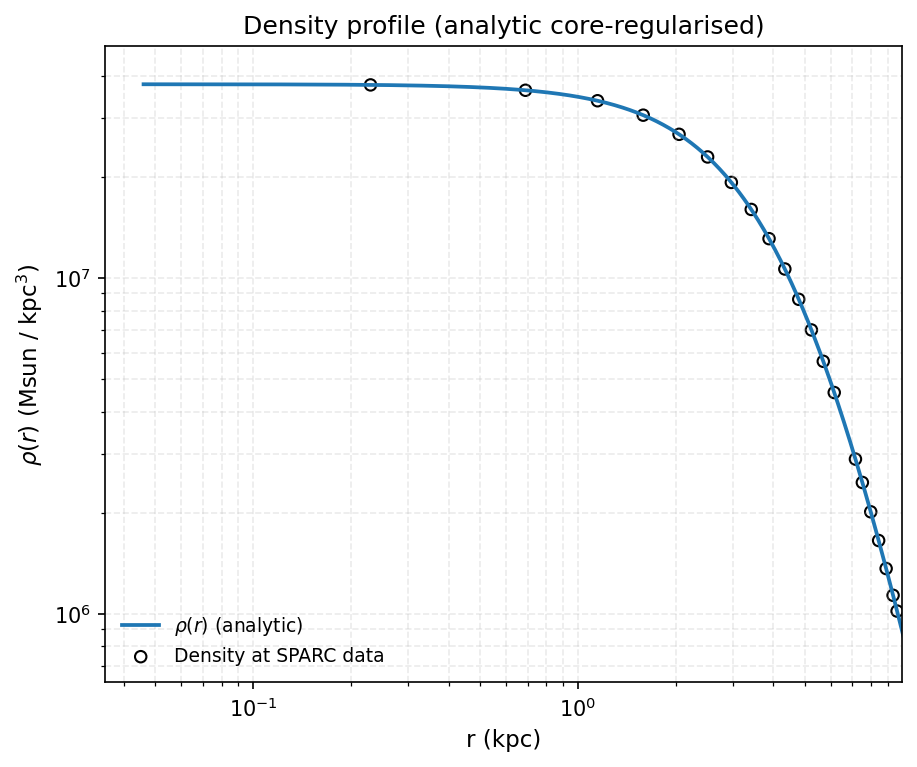}
\caption{The density of the SIDM model of Eq.
(\ref{ScaledependentEoSDM}) for the galaxy NGC0100, versus the
radius.} \label{NGC0100dens}
\end{figure}

\begin{figure}[h!]
\centering
\includegraphics[width=35pc]{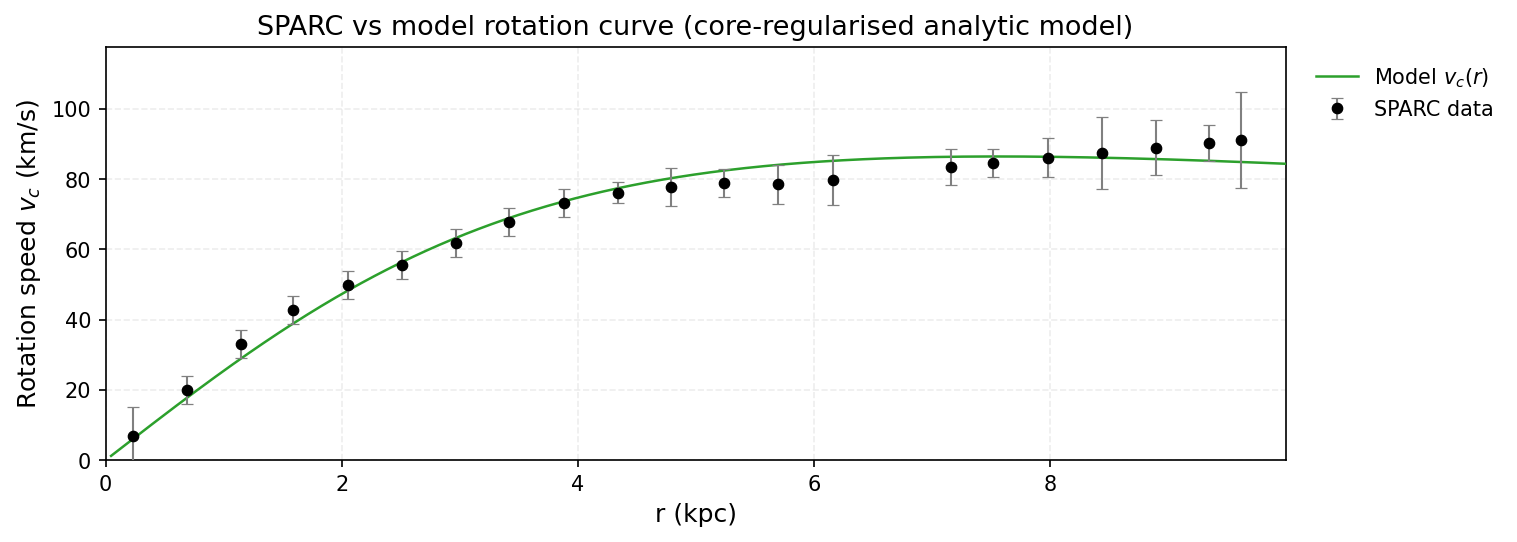}
\caption{The predicted rotation curves for the optimized SIDM
model of Eq. (\ref{ScaledependentEoSDM}), versus the SPARC
observational data for the galaxy NGC0100.} \label{NGC0100}
\end{figure}

In Table \ref{collNGC0100} we present the optimized values of
$K_0$ and $\rho_0$ for the analytic SIDM model of Eq.
(\ref{ScaledependentEoSDM}) for which the maximum compatibility
with the SPARC data is achieved.
\begin{table}[h!]
  \begin{center}
    \caption{SIDM Optimization Values for the galaxy NGC0100}
    \label{collNGC0100}
     \begin{tabular}{|r|r|}
     \hline
      \textbf{Parameter}   & \textbf{Optimization Values}
      \\  \hline
     $\rho_0 $  ($M_{\odot}/\mathrm{Kpc}^{3}$) & $3.77865\times 10^7$
\\  \hline $K_0$ ($M_{\odot} \,
\mathrm{Kpc}^{-3} \, (\mathrm{km/s})^{2}$)& 3237.93
\\  \hline
    \end{tabular}
  \end{center}
\end{table}
In Figs. \ref{NGC0100dens}, \ref{NGC0100} we present the density
of the analytic SIDM model, the predicted rotation curves for the
SIDM model (\ref{ScaledependentEoSDM}), versus the SPARC
observational data and the sound speed, as a function of the
radius respectively. As it can be seen, for this galaxy, the SIDM
model produces viable rotation curves which are compatible with
the SPARC data.

\subsection{The Galaxy NGC0247, Non-viable}

For this galaxy, the optimization method we used, ensures maximum
compatibility of the analytic SIDM model of Eq.
(\ref{ScaledependentEoSDM}) with the SPARC data, if we choose
$\rho_0=2.1361\times 10^7$$M_{\odot}/\mathrm{Kpc}^{3}$ and
$K_0=4233.3
$$M_{\odot} \, \mathrm{Kpc}^{-3} \, (\mathrm{km/s})^{2}$, in which
case the reduced $\chi^2_{red}$ value is $\chi^2_{red}=38.458$.
Also the parameter $\alpha$ in this case is $\alpha=8.12419 $Kpc.

\begin{figure}[h!]
\centering
\includegraphics[width=20pc]{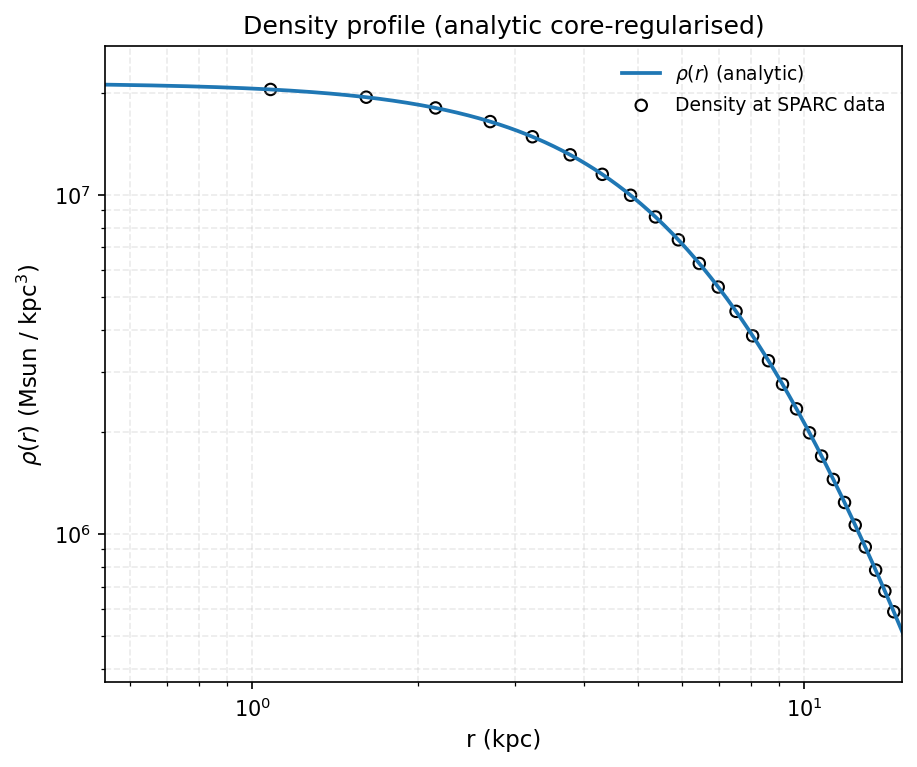}
\caption{The density of the SIDM model of Eq.
(\ref{ScaledependentEoSDM}) for the galaxy NGC0247, versus the
radius.} \label{NGC0247dens}
\end{figure}

\begin{figure}[h!]
\centering
\includegraphics[width=35pc]{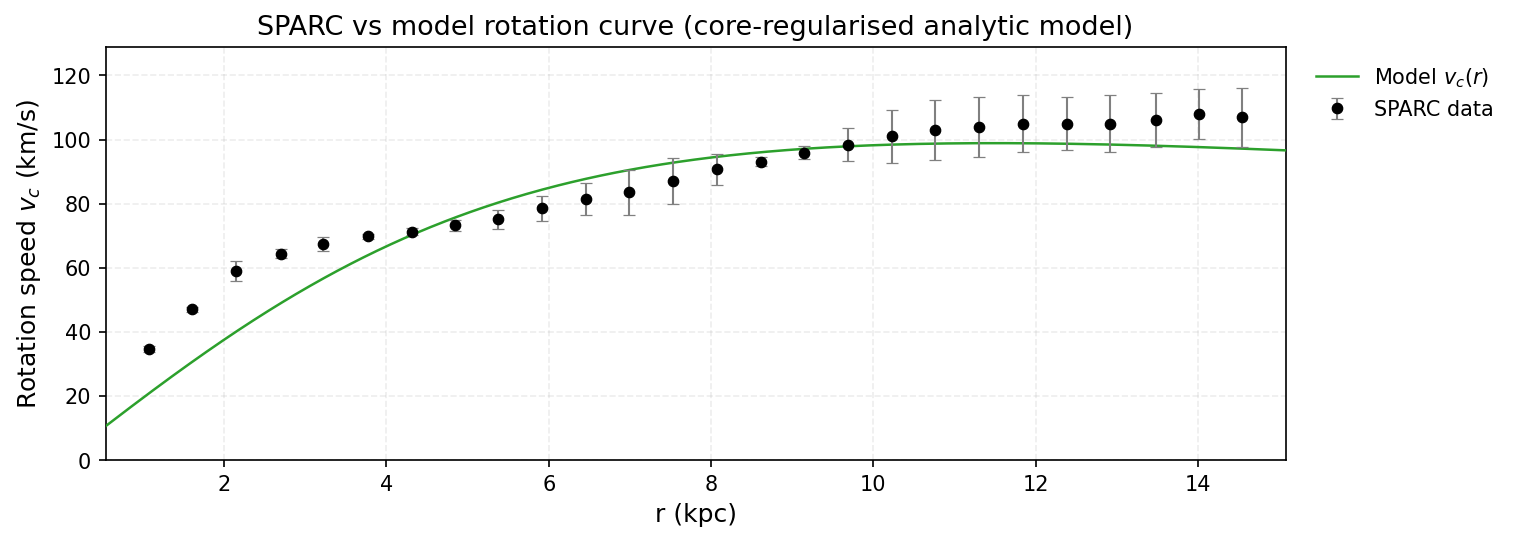}
\caption{The predicted rotation curves for the optimized SIDM
model of Eq. (\ref{ScaledependentEoSDM}), versus the SPARC
observational data for the galaxy NGC0247.} \label{NGC0247}
\end{figure}

In Table \ref{collNGC0247} we present the optimized values of
$K_0$ and $\rho_0$ for the analytic SIDM model of Eq.
(\ref{ScaledependentEoSDM}) for which the maximum compatibility
with the SPARC data is achieved.
\begin{table}[h!]
  \begin{center}
    \caption{SIDM Optimization Values for the galaxy NGC0247}
    \label{collNGC0247}
     \begin{tabular}{|r|r|}
     \hline
      \textbf{Parameter}   & \textbf{Optimization Values}
      \\  \hline
     $\rho_0 $  ($M_{\odot}/\mathrm{Kpc}^{3}$) & $2.1361\times 10^7$
\\  \hline $K_0$ ($M_{\odot} \,
\mathrm{Kpc}^{-3} \, (\mathrm{km/s})^{2}$)& 4233.3
\\  \hline
    \end{tabular}
  \end{center}
\end{table}
In Figs. \ref{NGC0247dens}, \ref{NGC0247} we present the density
of the analytic SIDM model, the predicted rotation curves for the
SIDM model (\ref{ScaledependentEoSDM}), versus the SPARC
observational data and the sound speed, as a function of the
radius respectively. As it can be seen, for this galaxy, the SIDM
model produces non-viable rotation curves which are incompatible
with the SPARC data.

\begin{figure}[h!]
\centering
\includegraphics[width=20pc]{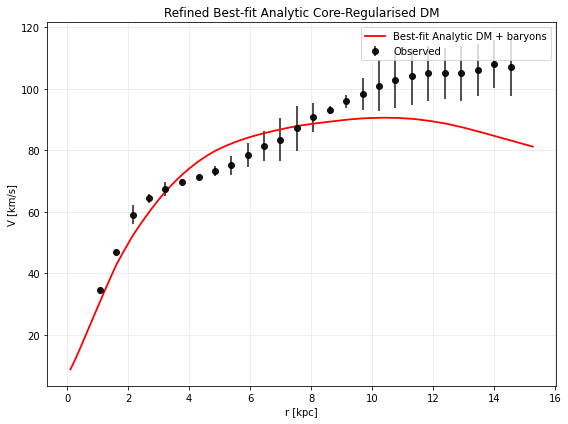}
\caption{The predicted rotation curves after using an optimization
for the SIDM model (\ref{ScaledependentEoSDM}), and the extended
SPARC data for the galaxy NGC0247. We included the rotation curves
of the gas, the disk velocities, the bulge (where present) along
with the SIDM model.} \label{extendedNGC0247}
\end{figure}

Now we shall include contributions to the rotation velocity from
the other components of the galaxy, namely the disk, the gas, and
the bulge if present. In Fig. \ref{extendedNGC0247} we present the
combined rotation curves including all the components of the
galaxy along with the SIDM. As it can be seen, the extended
collisional DM model is non-viable.
Also in Table \ref{evaluationextendedNGC0247} we present the
optimized values of the free parameters of the SIDM model for
which  we achieve the maximum compatibility with the SPARC data,
for the galaxy NGC0247, and also the resulting reduced
$\chi^2_{red}$ value.
\begin{table}[h!]
\centering \caption{Optimized Parameter Values of the Extended
SIDM model for the Galaxy NGC0247.}
\begin{tabular}{lc}
\hline
Parameter & Value  \\
\hline
$\rho_0 $ ($M_{\odot}/\mathrm{Kpc}^{3}$) & $2.28958\times 10^7$   \\
$K_0$ ($M_{\odot} \,
\mathrm{Kpc}^{-3} \, (\mathrm{km/s})^{2}$) & 1861.68   \\
$ml_{\text{disk}}$ & 1 \\
$ml_{\text{bulge}}$ & 0.2749 \\
$\alpha$ (Kpc) & 5.2032\\
$\chi^2_{red}$ & 7.7912 \\
\hline
\end{tabular}
\label{evaluationextendedNGC0247}
\end{table}

\subsection{The Galaxy NGC0289, Non-viable, Extended Marginally Viable}

For this galaxy, the optimization method we used, ensures maximum
compatibility of the analytic SIDM model of Eq.
(\ref{ScaledependentEoSDM}) with the SPARC data, if we choose
$\rho_0=3.07333\times 10^7$$M_{\odot}/\mathrm{Kpc}^{3}$ and
$K_0=18670.3
$$M_{\odot} \, \mathrm{Kpc}^{-3} \, (\mathrm{km/s})^{2}$, in which
case the reduced $\chi^2_{red}$ value is $\chi^2_{red}=17.9695$.
Also the parameter $\alpha$ in this case is $\alpha=14.224 $Kpc.

\begin{figure}[h!]
\centering
\includegraphics[width=20pc]{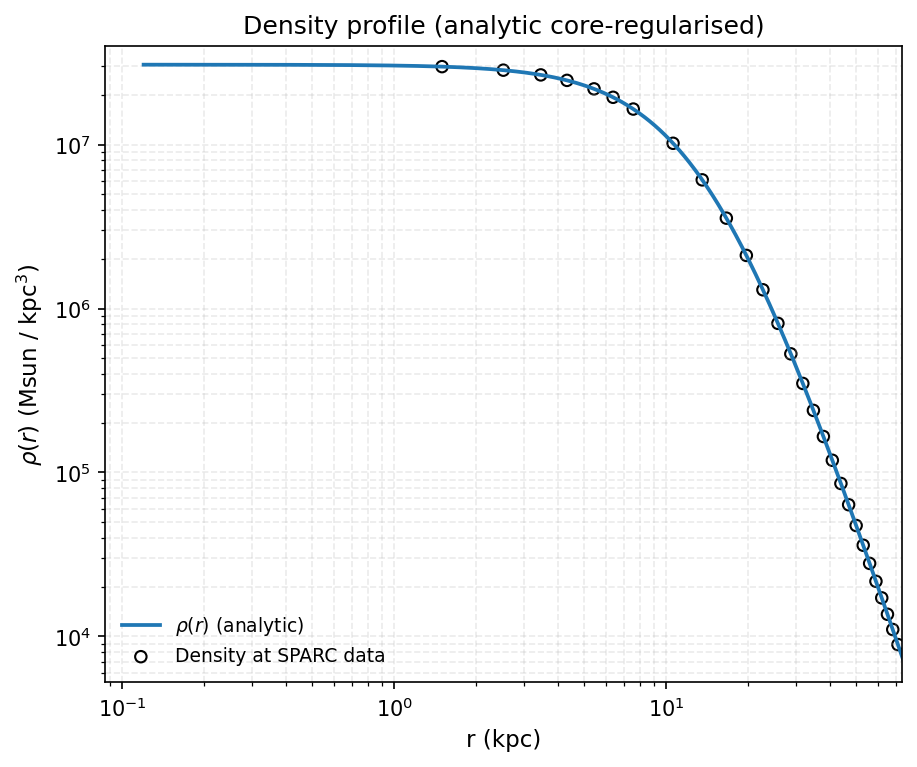}
\caption{The density of the SIDM model of Eq.
(\ref{ScaledependentEoSDM}) for the galaxy NGC0289, versus the
radius.} \label{NGC0289dens}
\end{figure}

\begin{figure}[h!]
\centering
\includegraphics[width=35pc]{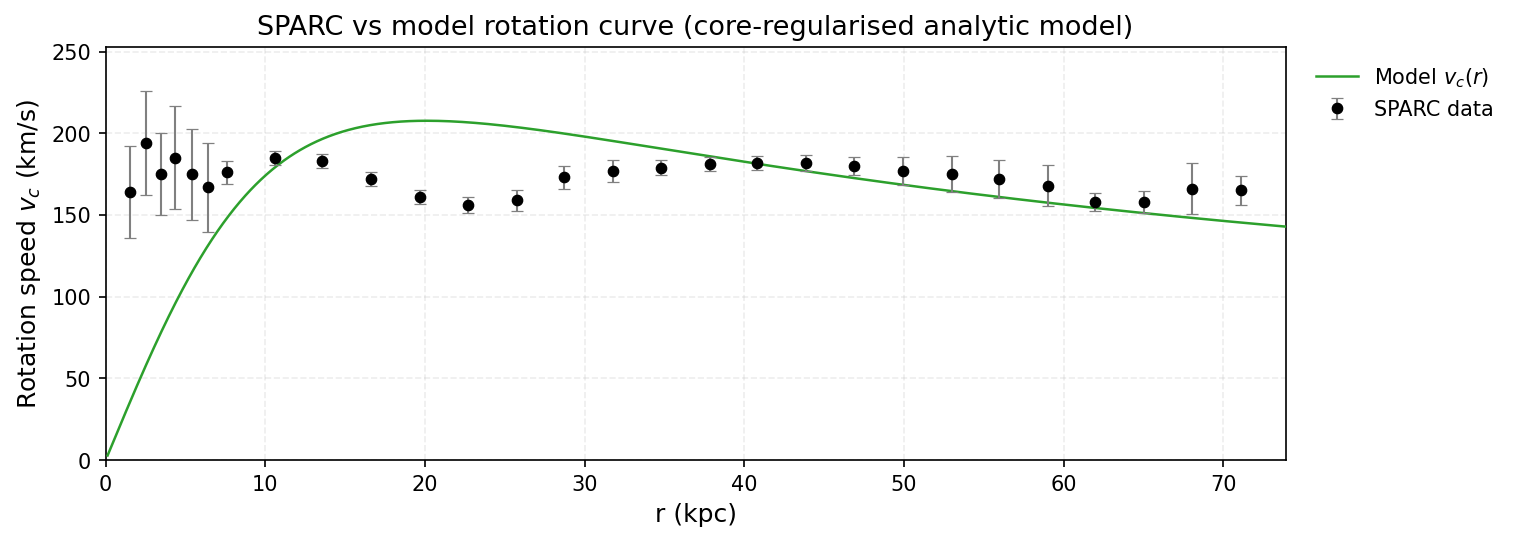}
\caption{The predicted rotation curves for the optimized SIDM
model of Eq. (\ref{ScaledependentEoSDM}), versus the SPARC
observational data for the galaxy NGC0289.} \label{NGC0289}
\end{figure}

In Table \ref{collNGC0289} we present the optimized values of
$K_0$ and $\rho_0$ for the analytic SIDM model of Eq.
(\ref{ScaledependentEoSDM}) for which the maximum compatibility
with the SPARC data is achieved.
\begin{table}[h!]
  \begin{center}
    \caption{SIDM Optimization Values for the galaxy NGC0289}
    \label{collNGC0289}
     \begin{tabular}{|r|r|}
     \hline
      \textbf{Parameter}   & \textbf{Optimization Values}
      \\  \hline
     $\rho_0 $  ($M_{\odot}/\mathrm{Kpc}^{3}$) & $3.07333\times 10^7$
\\  \hline $K_0$ ($M_{\odot} \,
\mathrm{Kpc}^{-3} \, (\mathrm{km/s})^{2}$)& 18670.3
\\  \hline
    \end{tabular}
  \end{center}
\end{table}
In Figs. \ref{NGC0289dens}, \ref{NGC0289} we present the density
of the analytic SIDM model, the predicted rotation curves for the
SIDM model (\ref{ScaledependentEoSDM}), versus the SPARC
observational data and the sound speed, as a function of the
radius respectively. As it can be seen, for this galaxy, the SIDM
model produces non-viable rotation curves which are incompatible
with the SPARC data.

\begin{figure}[h!]
\centering
\includegraphics[width=20pc]{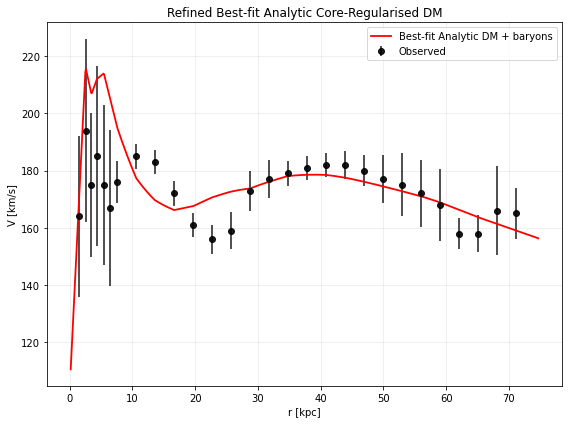}
\caption{The predicted rotation curves after using an optimization
for the SIDM model (\ref{ScaledependentEoSDM}), and the extended
SPARC data for the galaxy NGC0289. We included the rotation curves
of the gas, the disk velocities, the bulge (where present) along
with the SIDM model.} \label{extendedNGC0289}
\end{figure}

Now we shall include contributions to the rotation velocity from
the other components of the galaxy, namely the disk, the gas, and
the bulge if present. In Fig. \ref{extendedNGC0289} we present the
combined rotation curves including all the components of the
galaxy along with the SIDM. As it can be seen, the extended
collisional DM model is non-viable.
Also in Table \ref{evaluationextendedNGC0289} we present the
optimized values of the free parameters of the SIDM model for
which  we achieve the maximum compatibility with the SPARC data,
for the galaxy NGC0289, and also the resulting reduced
$\chi^2_{red}$ value.
\begin{table}[h!]
\centering \caption{Optimized Parameter Values of the Extended
SIDM model for the Galaxy NGC0289.}
\begin{tabular}{lc}
\hline
Parameter & Value  \\
\hline
$\rho_0 $ ($M_{\odot}/\mathrm{Kpc}^{3}$) & $2.60432\times 10^7$   \\
$K_0$ ($M_{\odot} \,
\mathrm{Kpc}^{-3} \, (\mathrm{km/s})^{2}$) & 8825.33   \\
$ml_{\text{disk}}$ & 0.9509 \\
$ml_{\text{bulge}}$ & 0.4 \\
$\alpha$ (Kpc) & 33.5903\\
$\chi^2_{red}$ & 2.09307 \\
\hline
\end{tabular}
\label{evaluationextendedNGC0289}
\end{table}

\subsection{The Galaxy NGC0300, Non-viable, Extended Viable}

For this galaxy, the optimization method we used, ensures maximum
compatibility of the analytic SIDM model of Eq.
(\ref{ScaledependentEoSDM}) with the SPARC data, if we choose
$\rho_0=4.85635\times 10^7$$M_{\odot}/\mathrm{Kpc}^{3}$ and
$K_0=3788.38
$$M_{\odot} \, \mathrm{Kpc}^{-3} \, (\mathrm{km/s})^{2}$, in which
case the reduced $\chi^2_{red}$ value is $\chi^2_{red}=2.38282$.
Also the parameter $\alpha$ in this case is $\alpha=5.09709 $Kpc.

\begin{figure}[h!]
\centering
\includegraphics[width=20pc]{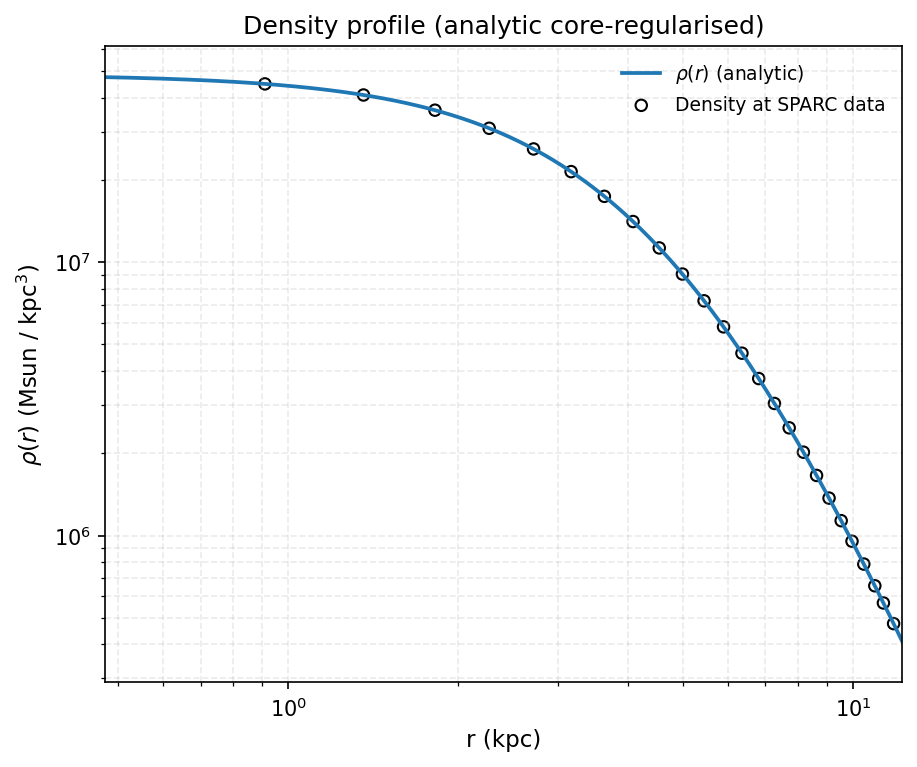}
\caption{The density of the SIDM model of Eq.
(\ref{ScaledependentEoSDM}) for the galaxy NGC0300, versus the
radius.} \label{NGC0300dens}
\end{figure}

\begin{figure}[h!]
\centering
\includegraphics[width=35pc]{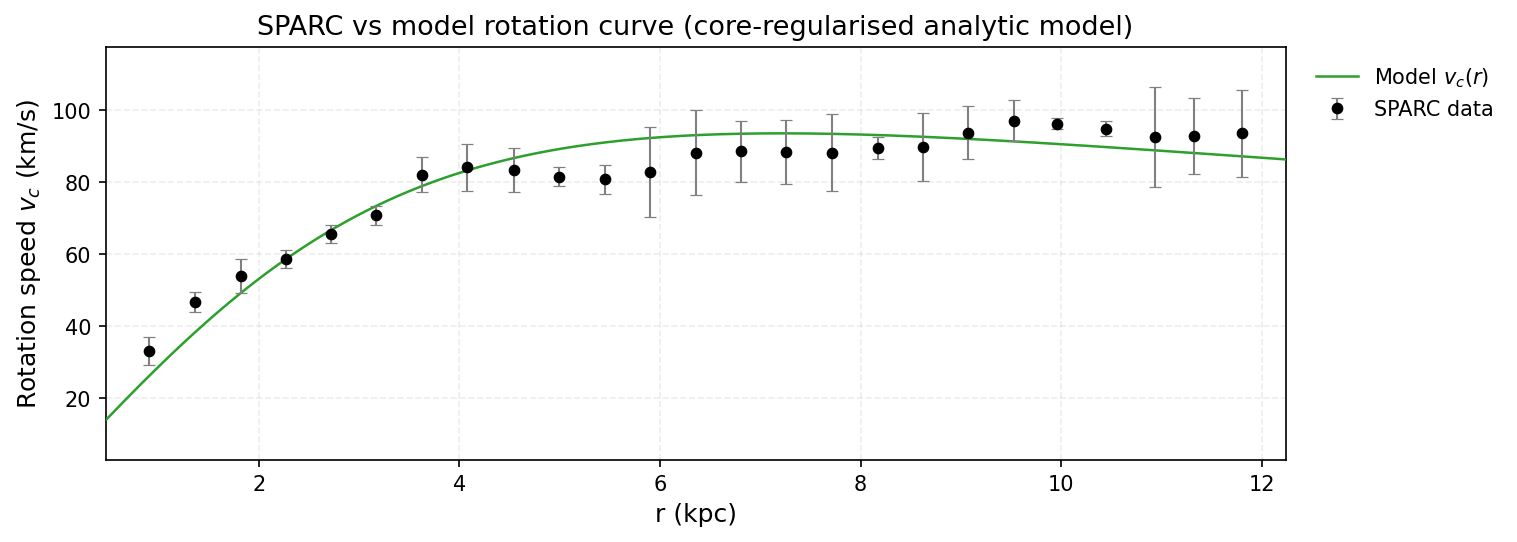}
\caption{The predicted rotation curves for the optimized SIDM
model of Eq. (\ref{ScaledependentEoSDM}), versus the SPARC
observational data for the galaxy NGC0300.} \label{NGC0300}
\end{figure}

In Table \ref{collNGC0300} we present the optimized values of
$K_0$ and $\rho_0$ for the analytic SIDM model of Eq.
(\ref{ScaledependentEoSDM}) for which the maximum compatibility
with the SPARC data is achieved.
\begin{table}[h!]
  \begin{center}
    \caption{SIDM Optimization Values for the galaxy NGC0300}
    \label{collNGC0300}
     \begin{tabular}{|r|r|}
     \hline
      \textbf{Parameter}   & \textbf{Optimization Values}
      \\  \hline
     $\rho_0 $  ($M_{\odot}/\mathrm{Kpc}^{3}$) & $4.85635\times 10^7$
\\  \hline $K_0$ ($M_{\odot} \,
\mathrm{Kpc}^{-3} \, (\mathrm{km/s})^{2}$)& 3788.38
\\  \hline
    \end{tabular}
  \end{center}
\end{table}
In Figs. \ref{NGC0300dens}, \ref{NGC0300} we present the density
of the analytic SIDM model, the predicted rotation curves for the
SIDM model (\ref{ScaledependentEoSDM}), versus the SPARC
observational data and the sound speed, as a function of the
radius respectively. As it can be seen, for this galaxy, the SIDM
model produces non-viable rotation curves which are incompatible
with the SPARC data.

\begin{figure}[h!]
\centering
\includegraphics[width=20pc]{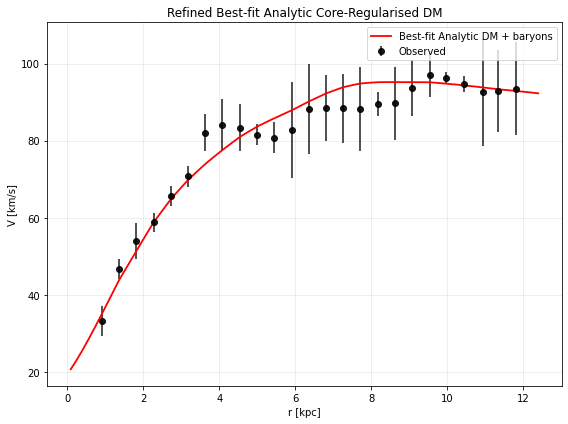}
\caption{The predicted rotation curves after using an optimization
for the SIDM model (\ref{ScaledependentEoSDM}), and the extended
SPARC data for the galaxy NGC0300. We included the rotation curves
of the gas, the disk velocities, the bulge (where present) along
with the SIDM model.} \label{extendedNGC0300}
\end{figure}

Now we shall include contributions to the rotation velocity from
the other components of the galaxy, namely the disk, the gas, and
the bulge if present. In Fig. \ref{extendedNGC0300} we present the
combined rotation curves including all the components of the
galaxy along with the SIDM. As it can be seen, the extended
collisional DM model is viable.
Also in Table \ref{evaluationextendedNGC0300} we present the
optimized values of the free parameters of the SIDM model for
which  we achieve the maximum compatibility with the SPARC data,
for the galaxy NGC0300, and also the resulting reduced
$\chi^2_{red}$ value.
\begin{table}[h!]
\centering \caption{Optimized Parameter Values of the Extended
SIDM model for the Galaxy NGC0300.}
\begin{tabular}{lc}
\hline
Parameter & Value  \\
\hline
$\rho_0 $ ($M_{\odot}/\mathrm{Kpc}^{3}$) & $1.59765\times 10^7$   \\
$K_0$ ($M_{\odot} \,
\mathrm{Kpc}^{-3} \, (\mathrm{km/s})^{2}$) & 3002.46   \\
$ml_{\text{disk}}$ & 1 \\
$ml_{\text{bulge}}$ & 0.1373 \\
$\alpha$ (Kpc) & 7.91032\\
$\chi^2_{red}$ & 0.664034 \\
\hline
\end{tabular}
\label{evaluationextendedNGC0300}
\end{table}

\subsection{The Galaxy NGC0801, Non-viable, Extended Marginally Viable}

For this galaxy, the optimization method we used, ensures maximum
compatibility of the analytic SIDM model of Eq.
(\ref{ScaledependentEoSDM}) with the SPARC data, if we choose
$\rho_0=6.64818\times 10^7$$M_{\odot}/\mathrm{Kpc}^{3}$ and
$K_0=28572.9
$$M_{\odot} \, \mathrm{Kpc}^{-3} \, (\mathrm{km/s})^{2}$, in which
case the reduced $\chi^2_{red}$ value is $\chi^2_{red}=179.525$.
Also the parameter $\alpha$ in this case is $\alpha=11.964 $Kpc.

\begin{figure}[h!]
\centering
\includegraphics[width=20pc]{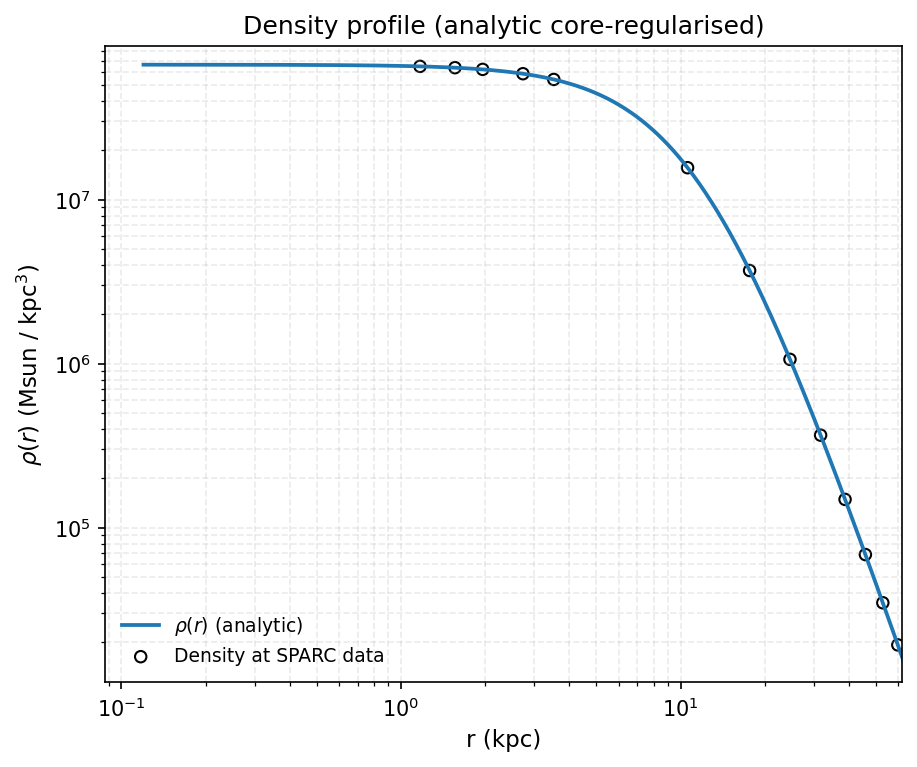}
\caption{The density of the SIDM model of Eq.
(\ref{ScaledependentEoSDM}) for the galaxy NGC0801, versus the
radius.} \label{NGC0801dens}
\end{figure}

\begin{figure}[h!]
\centering
\includegraphics[width=35pc]{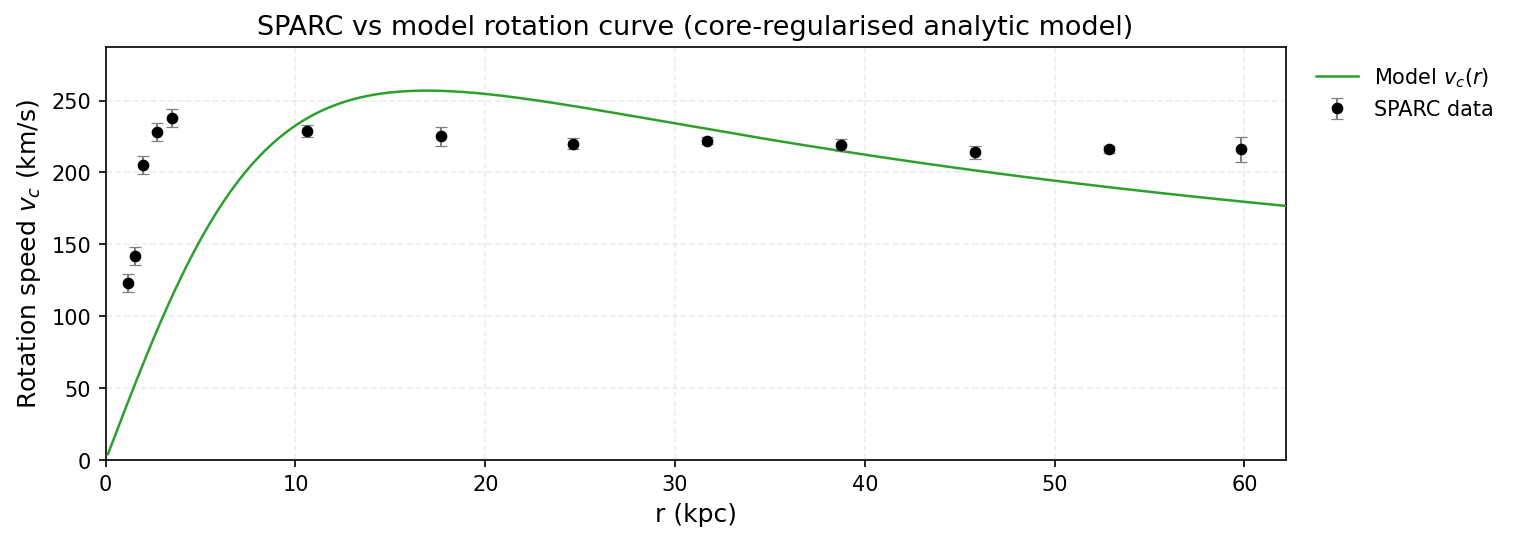}
\caption{The predicted rotation curves for the optimized SIDM
model of Eq. (\ref{ScaledependentEoSDM}), versus the SPARC
observational data for the galaxy NGC0801.} \label{NGC0801}
\end{figure}

In Table \ref{collNGC0801} we present the optimized values of
$K_0$ and $\rho_0$ for the analytic SIDM model of Eq.
(\ref{ScaledependentEoSDM}) for which the maximum compatibility
with the SPARC data is achieved.
\begin{table}[h!]
  \begin{center}
    \caption{SIDM Optimization Values for the galaxy NGC0801}
    \label{collNGC0801}
     \begin{tabular}{|r|r|}
     \hline
      \textbf{Parameter}   & \textbf{Optimization Values}
      \\  \hline
     $\rho_0 $  ($M_{\odot}/\mathrm{Kpc}^{3}$) & $6.64818\times 10^7$
\\  \hline $K_0$ ($M_{\odot} \,
\mathrm{Kpc}^{-3} \, (\mathrm{km/s})^{2}$)& 28572.9
\\  \hline
    \end{tabular}
  \end{center}
\end{table}
In Figs. \ref{NGC0801dens}, \ref{NGC0801} we present the density
of the analytic SIDM model, the predicted rotation curves for the
SIDM model (\ref{ScaledependentEoSDM}), versus the SPARC
observational data and the sound speed, as a function of the
radius respectively. As it can be seen, for this galaxy, the SIDM
model produces non-viable rotation curves which are incompatible
with the SPARC data.

\begin{figure}[h!]
\centering
\includegraphics[width=20pc]{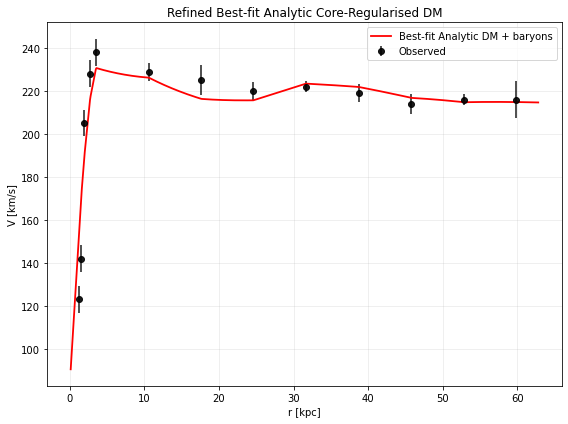}
\caption{The predicted rotation curves after using an optimization
for the SIDM model (\ref{ScaledependentEoSDM}), and the extended
SPARC data for the galaxy NGC0801. We included the rotation curves
of the gas, the disk velocities, the bulge (where present) along
with the SIDM model.} \label{extendedNGC0801}
\end{figure}

Now we shall include contributions to the rotation velocity from
the other components of the galaxy, namely the disk, the gas, and
the bulge if present. In Fig. \ref{extendedNGC0801} we present the
combined rotation curves including all the components of the
galaxy along with the SIDM. As it can be seen, the extended
collisional DM model is non-viable.
Also in Table \ref{evaluationextendedNGC0801} we present the
optimized values of the free parameters of the SIDM model for
which  we achieve the maximum compatibility with the SPARC data,
for the galaxy NGC0801, and also the resulting reduced
$\chi^2_{red}$ value.
\begin{table}[h!]
\centering \caption{Optimized Parameter Values of the Extended
SIDM model for the Galaxy NGC0801.}
\begin{tabular}{lc}
\hline
Parameter & Value  \\
\hline
$\rho_0 $ ($M_{\odot}/\mathrm{Kpc}^{3}$) & $1.18198\times 10^6$   \\
$K_0$ ($M_{\odot} \,
\mathrm{Kpc}^{-3} \, (\mathrm{km/s})^{2}$) & 11504.5   \\
$ml_{\text{disk}}$ & 0.8169 \\
$ml_{\text{bulge}}$ & 0.2987 \\
$\alpha$ (Kpc) & 56.9279\\
$\chi^2_{red}$ & 6.46751 \\
\hline
\end{tabular}
\label{evaluationextendedNGC0801}
\end{table}

\subsection{The Galaxy NGC0891, Non-viable, Extended Viable}

For this galaxy, the optimization method we used, ensures maximum
compatibility of the analytic SIDM model of Eq.
(\ref{ScaledependentEoSDM}) with the SPARC data, if we choose
$\rho_0=3.37381\times 10^8$$M_{\odot}/\mathrm{Kpc}^{3}$ and
$K_0=25248.6
$$M_{\odot} \, \mathrm{Kpc}^{-3} \, (\mathrm{km/s})^{2}$, in which
case the reduced $\chi^2_{red}$ value is $\chi^2_{red}=19.8621$.
Also the parameter $\alpha$ in this case is $\alpha=4.9924 $Kpc.

\begin{figure}[h!]
\centering
\includegraphics[width=20pc]{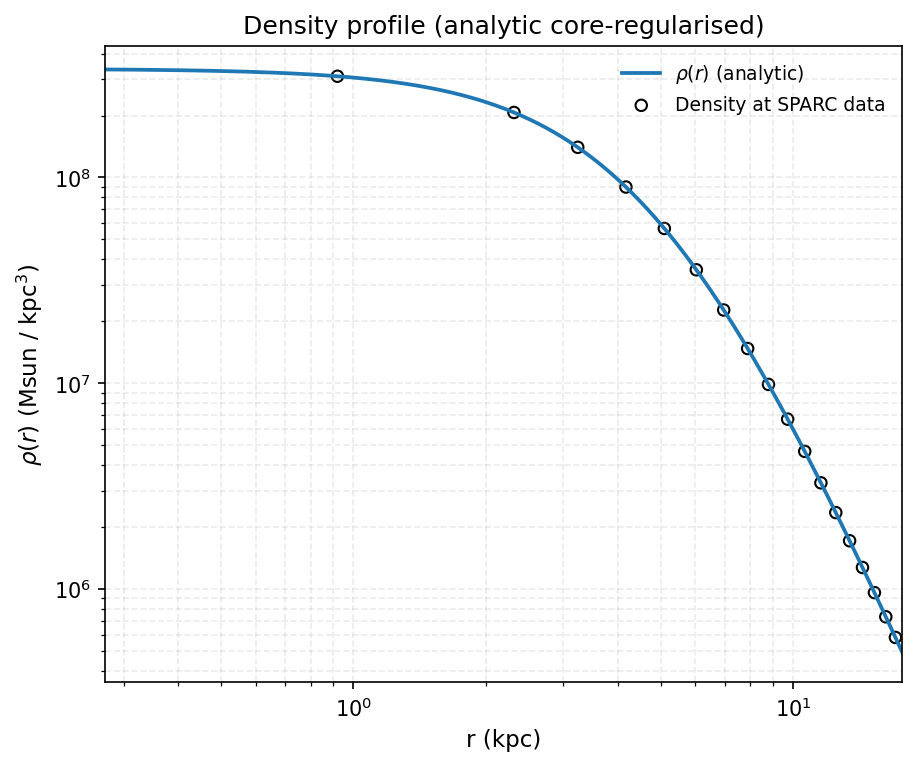}
\caption{The density of the SIDM model of Eq.
(\ref{ScaledependentEoSDM}) for the galaxy NGC0891, versus the
radius.} \label{NGC0891dens}
\end{figure}

\begin{figure}[h!]
\centering
\includegraphics[width=35pc]{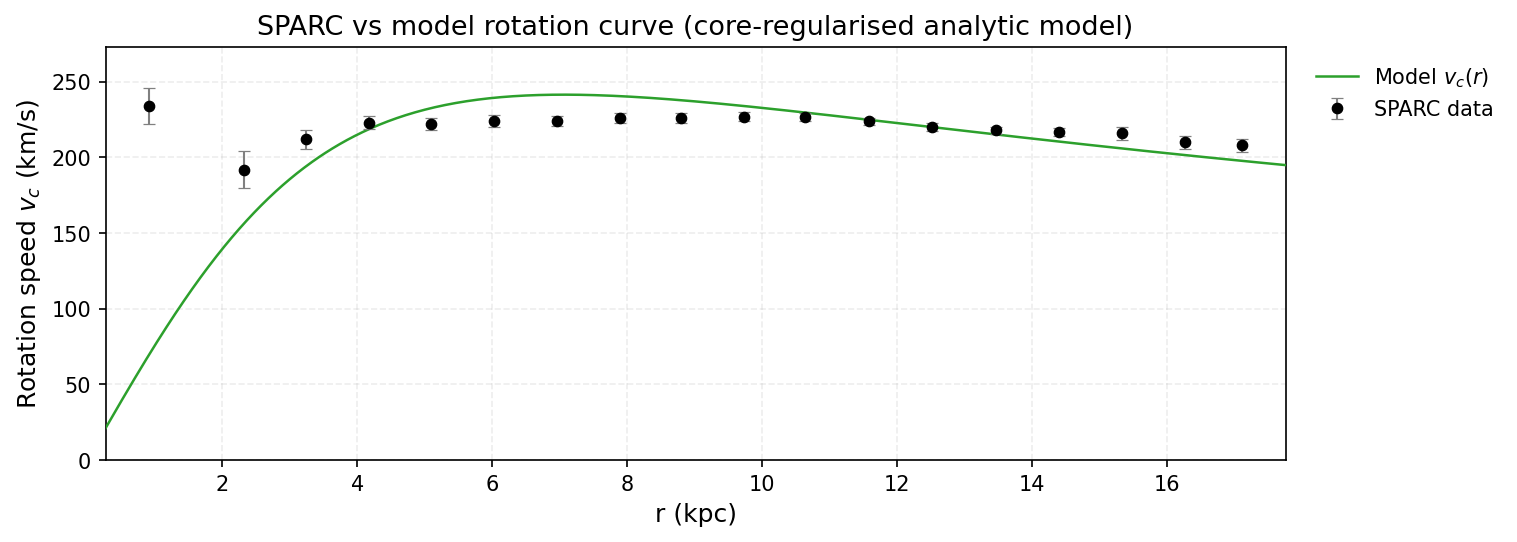}
\caption{The predicted rotation curves for the optimized SIDM
model of Eq. (\ref{ScaledependentEoSDM}), versus the SPARC
observational data for the galaxy NGC0891.} \label{NGC0891}
\end{figure}

In Table \ref{collNGC0891} we present the optimized values of
$K_0$ and $\rho_0$ for the analytic SIDM model of Eq.
(\ref{ScaledependentEoSDM}) for which the maximum compatibility
with the SPARC data is achieved.
\begin{table}[h!]
  \begin{center}
    \caption{SIDM Optimization Values for the galaxy NGC0891}
    \label{collNGC0891}
     \begin{tabular}{|r|r|}
     \hline
      \textbf{Parameter}   & \textbf{Optimization Values}
      \\  \hline
     $\rho_0 $  ($M_{\odot}/\mathrm{Kpc}^{3}$) & $3.37381\times 10^7$
\\  \hline $K_0$ ($M_{\odot} \,
\mathrm{Kpc}^{-3} \, (\mathrm{km/s})^{2}$)& 25248.6
\\  \hline
    \end{tabular}
  \end{center}
\end{table}
In Figs. \ref{NGC0891dens}, \ref{NGC0891} we present the density
of the analytic SIDM model, the predicted rotation curves for the
SIDM model (\ref{ScaledependentEoSDM}), versus the SPARC
observational data and the sound speed, as a function of the
radius respectively. As it can be seen, for this galaxy, the SIDM
model produces non-viable rotation curves which are incompatible
with the SPARC data.

\begin{figure}[h!]
\centering
\includegraphics[width=20pc]{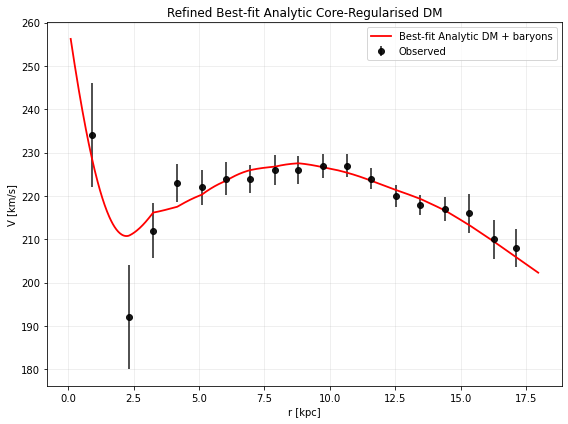}
\caption{The predicted rotation curves after using an optimization
for the SIDM model (\ref{ScaledependentEoSDM}), and the extended
SPARC data for the galaxy NGC0891. We included the rotation curves
of the gas, the disk velocities, the bulge (where present) along
with the SIDM model.} \label{extendedNGC0891}
\end{figure}

Now we shall include contributions to the rotation velocity from
the other components of the galaxy, namely the disk, the gas, and
the bulge if present. In Fig. \ref{extendedNGC0891} we present the
combined rotation curves including all the components of the
galaxy along with the SIDM. As it can be seen, the extended
collisional DM model is non-viable.
Also in Table \ref{evaluationextendedNGC0891} we present the
optimized values of the free parameters of the SIDM model for
which  we achieve the maximum compatibility with the SPARC data,
for the galaxy NGC0891, and also the resulting reduced
$\chi^2_{red}$ value.
\begin{table}[h!]
\centering \caption{Optimized Parameter Values of the Extended
SIDM model for the Galaxy NGC0891.}
\begin{tabular}{lc}
\hline
Parameter & Value  \\
\hline
$\rho_0 $ ($M_{\odot}/\mathrm{Kpc}^{3}$) & $9.35254\times 10^7$   \\
$K_0$ ($M_{\odot} \,
\mathrm{Kpc}^{-3} \, (\mathrm{km/s})^{2}$) & 15975.8   \\
$ml_{\text{disk}}$ & 0.876 \\
$ml_{\text{bulge}}$ & 0.8656 \\
$\alpha$ (Kpc) & 7.54159\\
$\chi^2_{red}$ & 0.51822 \\
\hline
\end{tabular}
\label{evaluationextendedNGC0891}
\end{table}

\subsection{The Galaxy NGC1003, Non-viable}

For this galaxy, the optimization method we used, ensures maximum
compatibility of the analytic SIDM model of Eq.
(\ref{ScaledependentEoSDM}) with the SPARC data, if we choose
$\rho_0=1.89193\times 10^7$$M_{\odot}/\mathrm{Kpc}^{3}$ and
$K_0=5656.14
$$M_{\odot} \, \mathrm{Kpc}^{-3} \, (\mathrm{km/s})^{2}$, in which
case the reduced $\chi^2_{red}$ value is $\chi^2_{red}=33.3223$.
Also the parameter $\alpha$ in this case is $\alpha=9.97834 $Kpc.

\begin{figure}[h!]
\centering
\includegraphics[width=20pc]{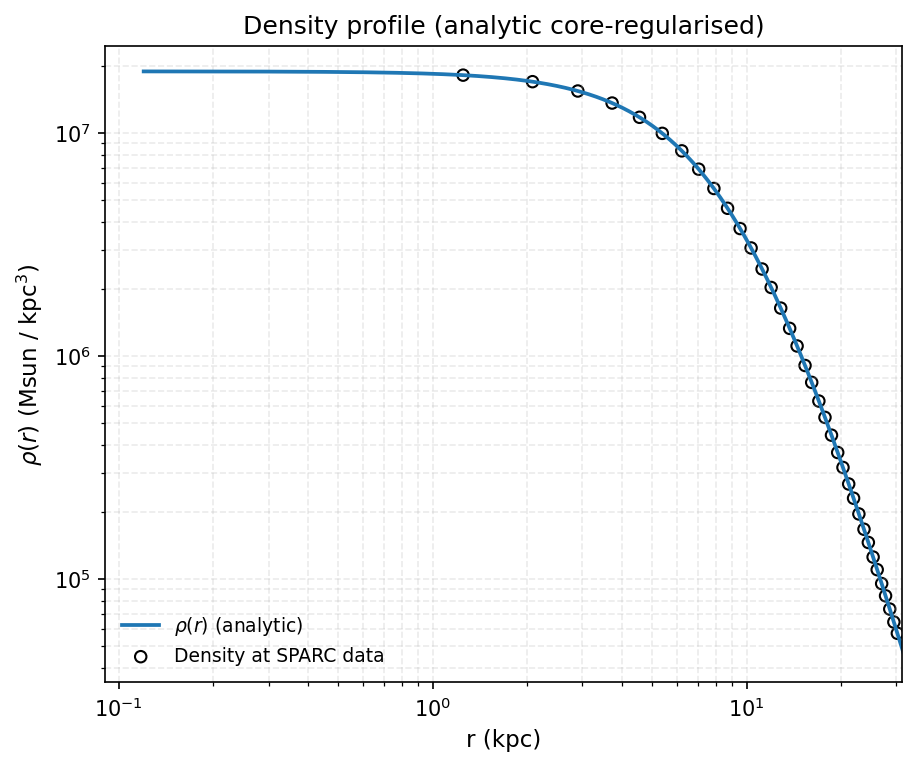}
\caption{The density of the SIDM model of Eq.
(\ref{ScaledependentEoSDM}) for the galaxy NGC1003, versus the
radius.} \label{NGC1003dens}
\end{figure}

\begin{figure}[h!]
\centering
\includegraphics[width=35pc]{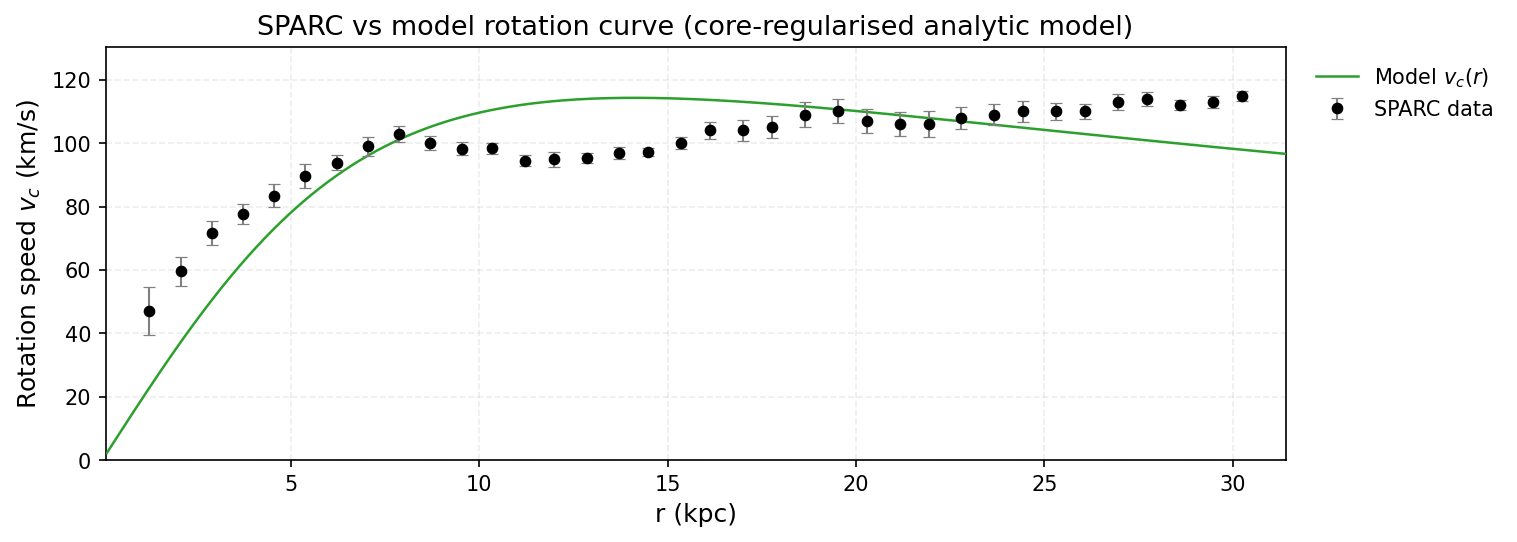}
\caption{The predicted rotation curves for the optimized SIDM
model of Eq. (\ref{ScaledependentEoSDM}), versus the SPARC
observational data for the galaxy NGC1003.} \label{NGC1003}
\end{figure}

In Table \ref{collNGC1003} we present the optimized values of
$K_0$ and $\rho_0$ for the analytic SIDM model of Eq.
(\ref{ScaledependentEoSDM}) for which the maximum compatibility
with the SPARC data is achieved.
\begin{table}[h!]
  \begin{center}
    \caption{SIDM Optimization Values for the galaxy NGC1003}
    \label{collNGC1003}
     \begin{tabular}{|r|r|}
     \hline
      \textbf{Parameter}   & \textbf{Optimization Values}
      \\  \hline
     $\rho_0 $  ($M_{\odot}/\mathrm{Kpc}^{3}$) & $1.89193\times 10^7$
\\  \hline $K_0$ ($M_{\odot} \,
\mathrm{Kpc}^{-3} \, (\mathrm{km/s})^{2}$)& 5656.14
\\  \hline
    \end{tabular}
  \end{center}
\end{table}
In Figs. \ref{NGC1003dens}, \ref{NGC1003} we present the density
of the analytic SIDM model, the predicted rotation curves for the
SIDM model (\ref{ScaledependentEoSDM}), versus the SPARC
observational data and the sound speed, as a function of the
radius respectively. As it can be seen, for this galaxy, the SIDM
model produces non-viable rotation curves which are incompatible
with the SPARC data.

\begin{figure}[h!]
\centering
\includegraphics[width=20pc]{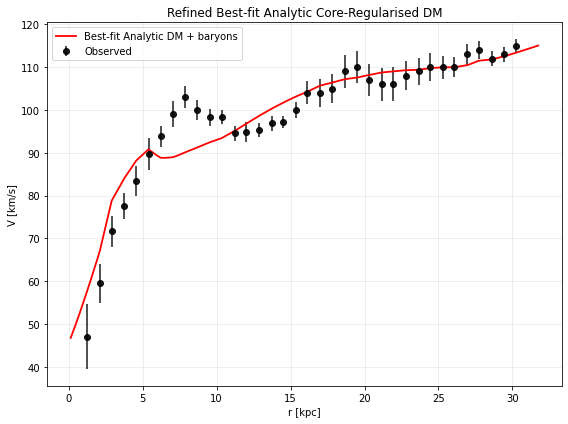}
\caption{The predicted rotation curves after using an optimization
for the SIDM model (\ref{ScaledependentEoSDM}), and the extended
SPARC data for the galaxy NGC1003. We included the rotation curves
of the gas, the disk velocities, the bulge (where present) along
with the SIDM model.} \label{extendedNGC1003}
\end{figure}

Now we shall include contributions to the rotation velocity from
the other components of the galaxy, namely the disk, the gas, and
the bulge if present. In Fig. \ref{extendedNGC1003} we present the
combined rotation curves including all the components of the
galaxy along with the SIDM. As it can be seen, the extended
collisional DM model is non-viable.
Also in Table \ref{evaluationextendedNGC1003} we present the
optimized values of the free parameters of the SIDM model for
which  we achieve the maximum compatibility with the SPARC data,
for the galaxy NGC1003, and also the resulting reduced
$\chi^2_{red}$ value.
\begin{table}[h!]
\centering \caption{Optimized Parameter Values of the Extended
SIDM model for the Galaxy NGC1003.}
\begin{tabular}{lc}
\hline
Parameter & Value  \\
\hline
$\rho_0 $ ($M_{\odot}/\mathrm{Kpc}^{3}$) & $3.29588\times 10^6$   \\
$K_0$ ($M_{\odot} \,
\mathrm{Kpc}^{-3} \, (\mathrm{km/s})^{2}$) & 4532.84   \\
$ml_{\text{disk}}$ & 1 \\
$ml_{\text{bulge}}$ & 0.3459 \\
$\alpha$ (Kpc) & 21.3991\\
$\chi^2_{red}$ & 3.6682 \\
\hline
\end{tabular}
\label{evaluationextendedNGC1003}
\end{table}

\subsection{The Galaxy NGC1090, Non-viable, Extended Marginally Viable}

For this galaxy, the optimization method we used, ensures maximum
compatibility of the analytic SIDM model of Eq.
(\ref{ScaledependentEoSDM}) with the SPARC data, if we choose
$\rho_0=7.47031\times 10^7$$M_{\odot}/\mathrm{Kpc}^{3}$ and
$K_0=14724.3
$$M_{\odot} \, \mathrm{Kpc}^{-3} \, (\mathrm{km/s})^{2}$, in which
case the reduced $\chi^2_{red}$ value is $\chi^2_{red}=11.4881$.
Also the parameter $\alpha$ in this case is $\alpha=8.10212 $Kpc.

\begin{figure}[h!]
\centering
\includegraphics[width=20pc]{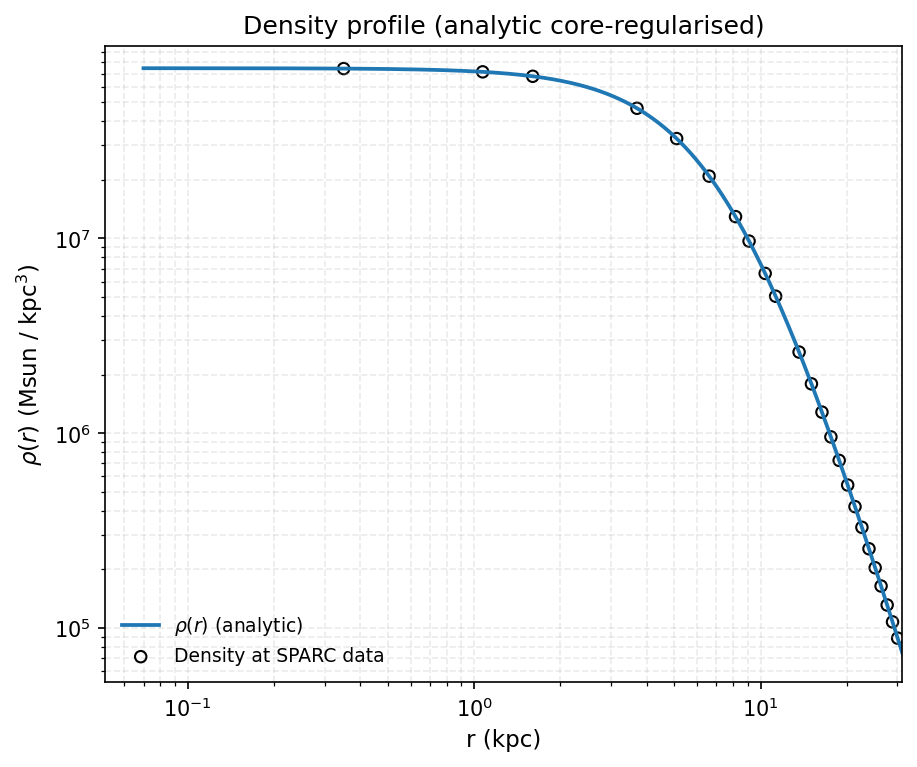}
\caption{The density of the SIDM model of Eq.
(\ref{ScaledependentEoSDM}) for the galaxy NGC1090, versus the
radius.} \label{NGC1090dens}
\end{figure}

\begin{figure}[h!]
\centering
\includegraphics[width=35pc]{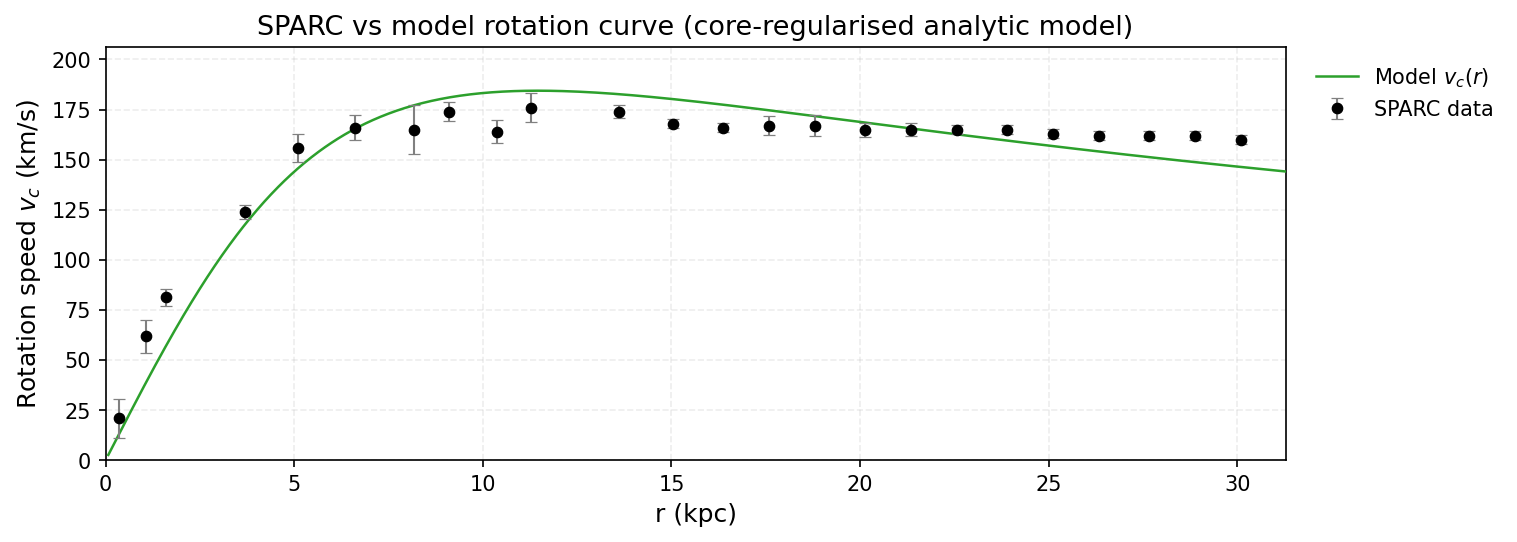}
\caption{The predicted rotation curves for the optimized SIDM
model of Eq. (\ref{ScaledependentEoSDM}), versus the SPARC
observational data for the galaxy NGC1090.} \label{NGC1090}
\end{figure}

In Table \ref{collNGC1090} we present the optimized values of
$K_0$ and $\rho_0$ for the analytic SIDM model of Eq.
(\ref{ScaledependentEoSDM}) for which the maximum compatibility
with the SPARC data is achieved.
\begin{table}[h!]
  \begin{center}
    \caption{SIDM Optimization Values for the galaxy NGC1090}
    \label{collNGC1090}
     \begin{tabular}{|r|r|}
     \hline
      \textbf{Parameter}   & \textbf{Optimization Values}
      \\  \hline
     $\rho_0 $  ($M_{\odot}/\mathrm{Kpc}^{3}$) & $7.47031\times 10^7$
\\  \hline $K_0$ ($M_{\odot} \,
\mathrm{Kpc}^{-3} \, (\mathrm{km/s})^{2}$)& 14724.3
\\  \hline
    \end{tabular}
  \end{center}
\end{table}
In Figs. \ref{NGC1090dens}, \ref{NGC1090} we present the density
of the analytic SIDM model, the predicted rotation curves for the
SIDM model (\ref{ScaledependentEoSDM}), versus the SPARC
observational data and the sound speed, as a function of the
radius respectively. As it can be seen, for this galaxy, the SIDM
model produces non-viable rotation curves which are incompatible
with the SPARC data.

\begin{figure}[h!]
\centering
\includegraphics[width=20pc]{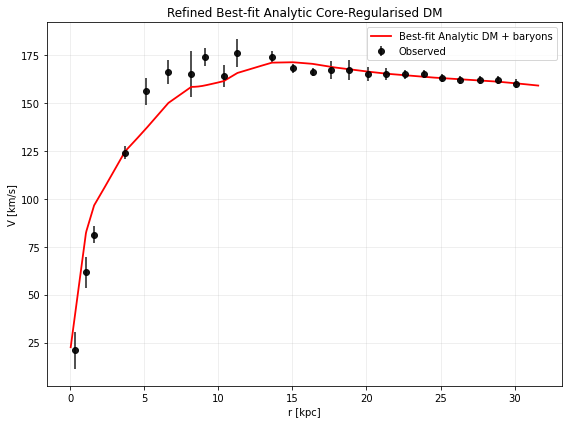}
\caption{The predicted rotation curves after using an optimization
for the SIDM model (\ref{ScaledependentEoSDM}), and the extended
SPARC data for the galaxy NGC1090. We included the rotation curves
of the gas, the disk velocities, the bulge (where present) along
with the SIDM model.} \label{extendedNGC1090}
\end{figure}

Now we shall include contributions to the rotation velocity from
the other components of the galaxy, namely the disk, the gas, and
the bulge if present. In Fig. \ref{extendedNGC1090} we present the
combined rotation curves including all the components of the
galaxy along with the SIDM. As it can be seen, the extended
collisional DM model is marginally viable.
Also in Table \ref{evaluationextendedNGC1090} we present the
optimized values of the free parameters of the SIDM model for
which  we achieve the maximum compatibility with the SPARC data,
for the galaxy NGC1090, and also the resulting reduced
$\chi^2_{red}$ value.
\begin{table}[h!]
\centering \caption{Optimized Parameter Values of the Extended
SIDM model for the Galaxy NGC1090.}
\begin{tabular}{lc}
\hline
Parameter & Value  \\
\hline
$\rho_0 $ ($M_{\odot}/\mathrm{Kpc}^{3}$) & $4.58371\times 10^6$   \\
$K_0$ ($M_{\odot} \,
\mathrm{Kpc}^{-3} \, (\mathrm{km/s})^{2}$) & 6534.69   \\
$ml_{\text{disk}}$ & 0.8333 \\
$ml_{\text{bulge}}$ & 0.5974 \\
$\alpha$ (Kpc) & 21.7871\\
$\chi^2_{red}$ & 2.83686 \\
\hline
\end{tabular}
\label{evaluationextendedNGC1090}
\end{table}

\subsection{The Galaxy NGC2403, Non-viable}

For this galaxy, the optimization method we used, ensures maximum
compatibility of the analytic SIDM model of Eq.
(\ref{ScaledependentEoSDM}) with the SPARC data, if we choose
$\rho_0=9.93149\times 10^7$$M_{\odot}/\mathrm{Kpc}^{3}$ and
$K_0=8809.92
$$M_{\odot} \, \mathrm{Kpc}^{-3} \, (\mathrm{km/s})^{2}$, in which
case the reduced $\chi^2_{red}$ value is $\chi^2_{red}=345.534$.
Also the parameter $\alpha$ in this case is $\alpha=5.43538 $Kpc.

\begin{figure}[h!]
\centering
\includegraphics[width=20pc]{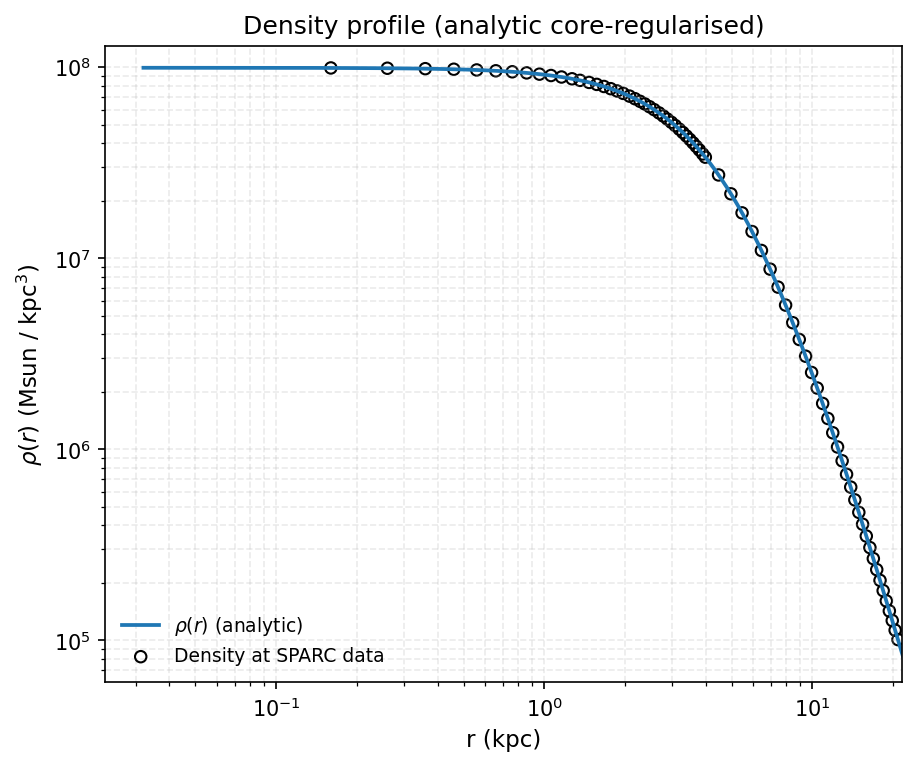}
\caption{The density of the SIDM model of Eq.
(\ref{ScaledependentEoSDM}) for the galaxy NGC2403, versus the
radius.} \label{NGC2403dens}
\end{figure}

\begin{figure}[h!]
\centering
\includegraphics[width=35pc]{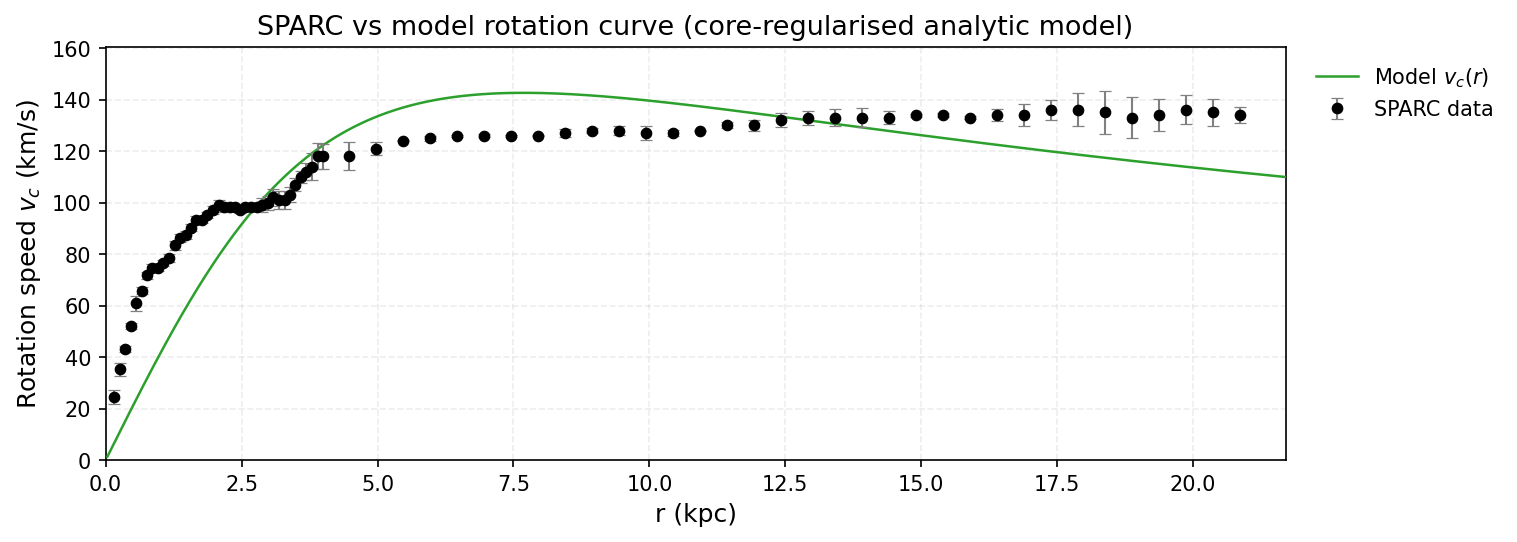}
\caption{The predicted rotation curves for the optimized SIDM
model of Eq. (\ref{ScaledependentEoSDM}), versus the SPARC
observational data for the galaxy NGC2403.} \label{NGC2403}
\end{figure}

In Table \ref{collNGC2403} we present the optimized values of
$K_0$ and $\rho_0$ for the analytic SIDM model of Eq.
(\ref{ScaledependentEoSDM}) for which the maximum compatibility
with the SPARC data is achieved.
\begin{table}[h!]
  \begin{center}
    \caption{SIDM Optimization Values for the galaxy NGC2403}
    \label{collNGC2403}
     \begin{tabular}{|r|r|}
     \hline
      \textbf{Parameter}   & \textbf{Optimization Values}
      \\  \hline
     $\rho_0 $  ($M_{\odot}/\mathrm{Kpc}^{3}$) & $9.93149\times 10^7$
\\  \hline $K_0$ ($M_{\odot} \,
\mathrm{Kpc}^{-3} \, (\mathrm{km/s})^{2}$)& 8809.92
\\  \hline
    \end{tabular}
  \end{center}
\end{table}
In Figs. \ref{NGC2403dens}, \ref{NGC2403} we present the density
of the analytic SIDM model, the predicted rotation curves for the
SIDM model (\ref{ScaledependentEoSDM}), versus the SPARC
observational data and the sound speed, as a function of the
radius respectively. As it can be seen, for this galaxy, the SIDM
model produces non-viable rotation curves which are incompatible
with the SPARC data.

\begin{figure}[h!]
\centering
\includegraphics[width=20pc]{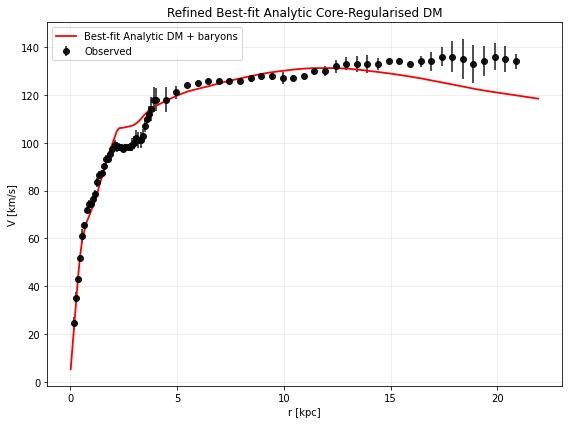}
\caption{The predicted rotation curves after using an optimization
for the SIDM model (\ref{ScaledependentEoSDM}), and the extended
SPARC data for the galaxy NGC2403. We included the rotation curves
of the gas, the disk velocities, the bulge (where present) along
with the SIDM model.} \label{extendedNGC2403}
\end{figure}

Now we shall include contributions to the rotation velocity from
the other components of the galaxy, namely the disk, the gas, and
the bulge if present. In Fig. \ref{extendedNGC2403} we present the
combined rotation curves including all the components of the
galaxy along with the SIDM. As it can be seen, the extended
collisional DM model is non-viable.
Also in Table \ref{evaluationextendedNGC2403} we present the
optimized values of the free parameters of the SIDM model for
which  we achieve the maximum compatibility with the SPARC data,
for the galaxy NGC2403, and also the resulting reduced
$\chi^2_{red}$ value.
\begin{table}[h!]
\centering \caption{Optimized Parameter Values of the Extended
SIDM model for the Galaxy NGC2403.}
\begin{tabular}{lc}
\hline
Parameter & Value  \\
\hline
$\rho_0 $ ($M_{\odot}/\mathrm{Kpc}^{3}$) & $1.71982\times 10^7$   \\
$K_0$ ($M_{\odot} \,
\mathrm{Kpc}^{-3} \, (\mathrm{km/s})^{2}$) & 5194.29   \\
$ml_{\text{disk}}$ & 0.9229 \\
$ml_{\text{bulge}}$ & 0.3255 \\
$\alpha$ (Kpc) & 10.0281\\
$\chi^2_{red}$ & 15.7327 \\
\hline
\end{tabular}
\label{evaluationextendedNGC2403}
\end{table}

\subsection{The Galaxy NGC2683, Non-viable, Extended Viable}

For this galaxy, the optimization method we used, ensures maximum
compatibility of the analytic SIDM model of Eq.
(\ref{ScaledependentEoSDM}) with the SPARC data, if we choose
$\rho_0=1.2279\times 10^8$$M_{\odot}/\mathrm{Kpc}^{3}$ and
$K_0=17675.6
$$M_{\odot} \, \mathrm{Kpc}^{-3} \, (\mathrm{km/s})^{2}$, in which
case the reduced $\chi^2_{red}$ value is $\chi^2_{red}=10.0865$.
Also the parameter $\alpha$ in this case is $\alpha=6.924 $Kpc.

\begin{figure}[h!]
\centering
\includegraphics[width=20pc]{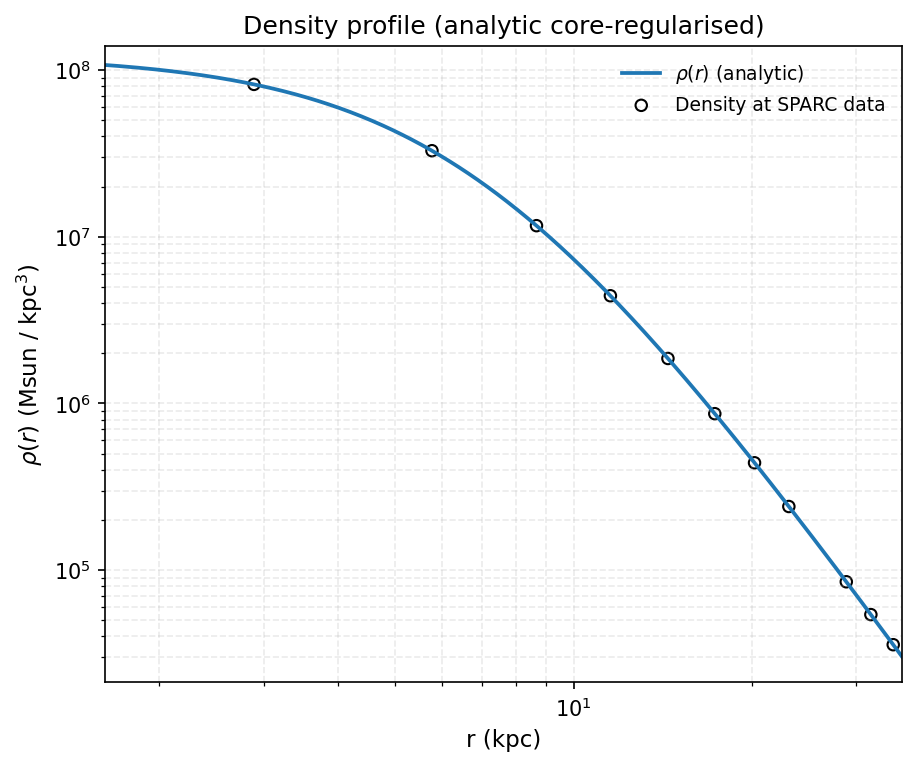}
\caption{The density of the SIDM model of Eq.
(\ref{ScaledependentEoSDM}) for the galaxy NGC2683, versus the
radius.} \label{NGC2683dens}
\end{figure}

\begin{figure}[h!]
\centering
\includegraphics[width=35pc]{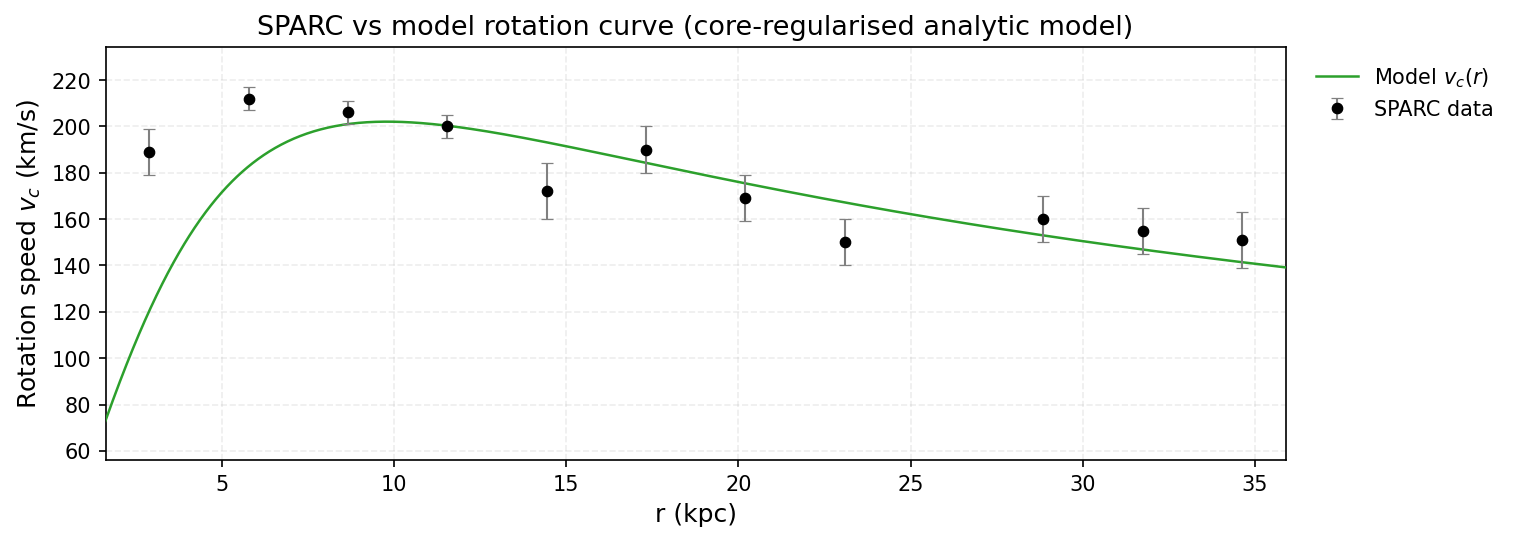}
\caption{The predicted rotation curves for the optimized SIDM
model of Eq. (\ref{ScaledependentEoSDM}), versus the SPARC
observational data for the galaxy NGC2683.} \label{NGC2683}
\end{figure}

In Table \ref{collNGC2683} we present the optimized values of
$K_0$ and $\rho_0$ for the analytic SIDM model of Eq.
(\ref{ScaledependentEoSDM}) for which the maximum compatibility
with the SPARC data is achieved.
\begin{table}[h!]
  \begin{center}
    \caption{SIDM Optimization Values for the galaxy NGC2683}
    \label{collNGC2683}
     \begin{tabular}{|r|r|}
     \hline
      \textbf{Parameter}   & \textbf{Optimization Values}
      \\  \hline
     $\rho_0 $  ($M_{\odot}/\mathrm{Kpc}^{3}$) & $1.2279\times 10^8$
\\  \hline $K_0$ ($M_{\odot} \,
\mathrm{Kpc}^{-3} \, (\mathrm{km/s})^{2}$)& 17675.6
\\  \hline
    \end{tabular}
  \end{center}
\end{table}
In Figs. \ref{NGC2683dens}, \ref{NGC2683} we present the density
of the analytic SIDM model, the predicted rotation curves for the
SIDM model (\ref{ScaledependentEoSDM}), versus the SPARC
observational data and the sound speed, as a function of the
radius respectively. As it can be seen, for this galaxy, the SIDM
model produces non-viable rotation curves which are incompatible
with the SPARC data.

\begin{figure}[h!]
\centering
\includegraphics[width=20pc]{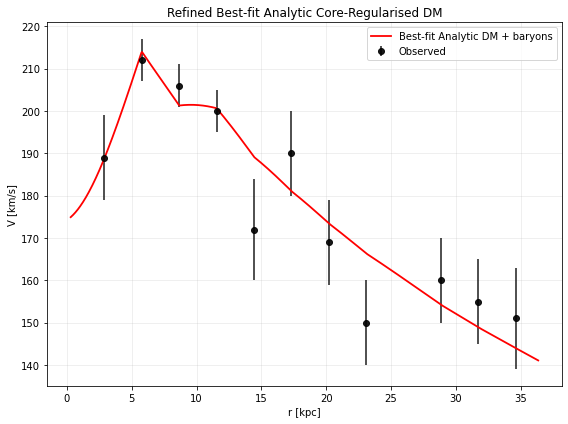}
\caption{The predicted rotation curves after using an optimization
for the SIDM model (\ref{ScaledependentEoSDM}), and the extended
SPARC data for the galaxy NGC2683. We included the rotation curves
of the gas, the disk velocities, the bulge (where present) along
with the SIDM model.} \label{extendedNGC2683}
\end{figure}

Now we shall include contributions to the rotation velocity from
the other components of the galaxy, namely the disk, the gas, and
the bulge if present. In Fig. \ref{extendedNGC2683} we present the
combined rotation curves including all the components of the
galaxy along with the SIDM. As it can be seen, the extended
collisional DM model is non-viable.
Also in Table \ref{evaluationextendedNGC2683} we present the
optimized values of the free parameters of the SIDM model for
which  we achieve the maximum compatibility with the SPARC data,
for the galaxy NGC2683, and also the resulting reduced
$\chi^2_{red}$ value.
\begin{table}[h!]
\centering \caption{Optimized Parameter Values of the Extended
SIDM model for the Galaxy NGC2683.}
\begin{tabular}{lc}
\hline
Parameter & Value  \\
\hline
$\rho_0 $ ($M_{\odot}/\mathrm{Kpc}^{3}$) & $1.5699\times 10^7$   \\
$K_0$ ($M_{\odot} \,
\mathrm{Kpc}^{-3} \, (\mathrm{km/s})^{2}$) & 7502.51  \\
$ml_{\text{disk}}$ & 0.8237 \\
$ml_{\text{bulge}}$ & 0.8055 \\
$\alpha$ (Kpc) & 12.6143 \\
$\chi^2_{red}$ & 1.11827 \\
\hline
\end{tabular}
\label{evaluationextendedNGC2683}
\end{table}

\subsection{The Galaxy NGC2841, Non-viable}

For this galaxy, the optimization method we used, ensures maximum
compatibility of the analytic SIDM model of Eq.
(\ref{ScaledependentEoSDM}) with the SPARC data, if we choose
$\rho_0=1.15657\times 10^8$$M_{\odot}/\mathrm{Kpc}^{3}$ and
$K_0=56258.9
$$M_{\odot} \, \mathrm{Kpc}^{-3} \, (\mathrm{km/s})^{2}$, in which
case the reduced $\chi^2_{red}$ value is $\chi^2_{red}=72.7703$.
Also the parameter $\alpha$ in this case is $\alpha=12.728 $Kpc.

\begin{figure}[h!]
\centering
\includegraphics[width=20pc]{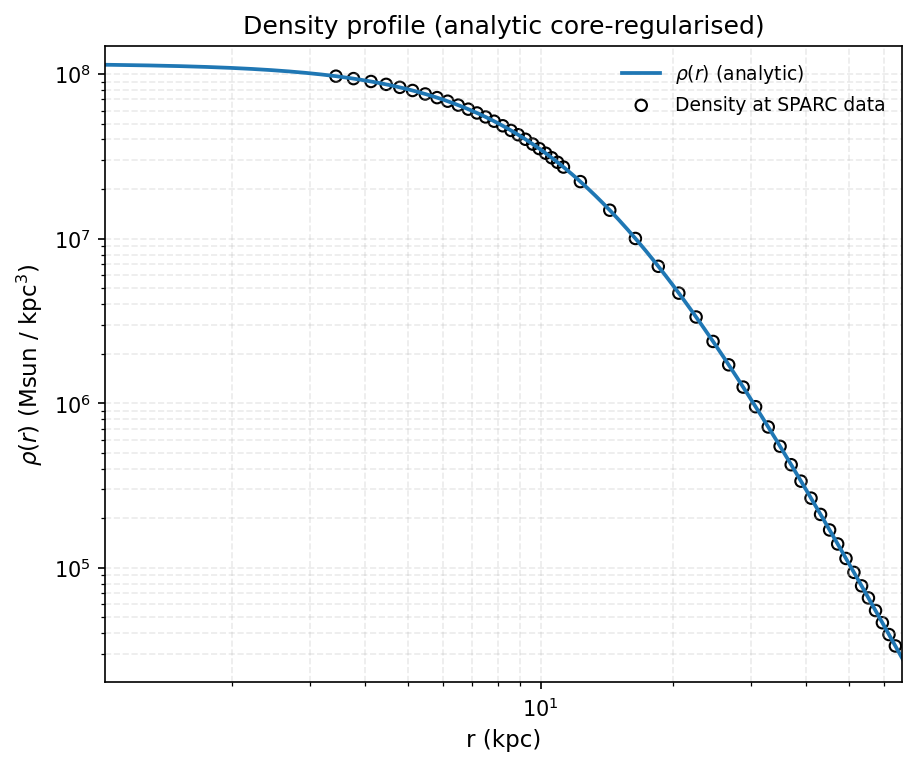}
\caption{The density of the SIDM model of Eq.
(\ref{ScaledependentEoSDM}) for the galaxy NGC2841, versus the
radius.} \label{NGC2841dens}
\end{figure}

\begin{figure}[h!]
\centering
\includegraphics[width=35pc]{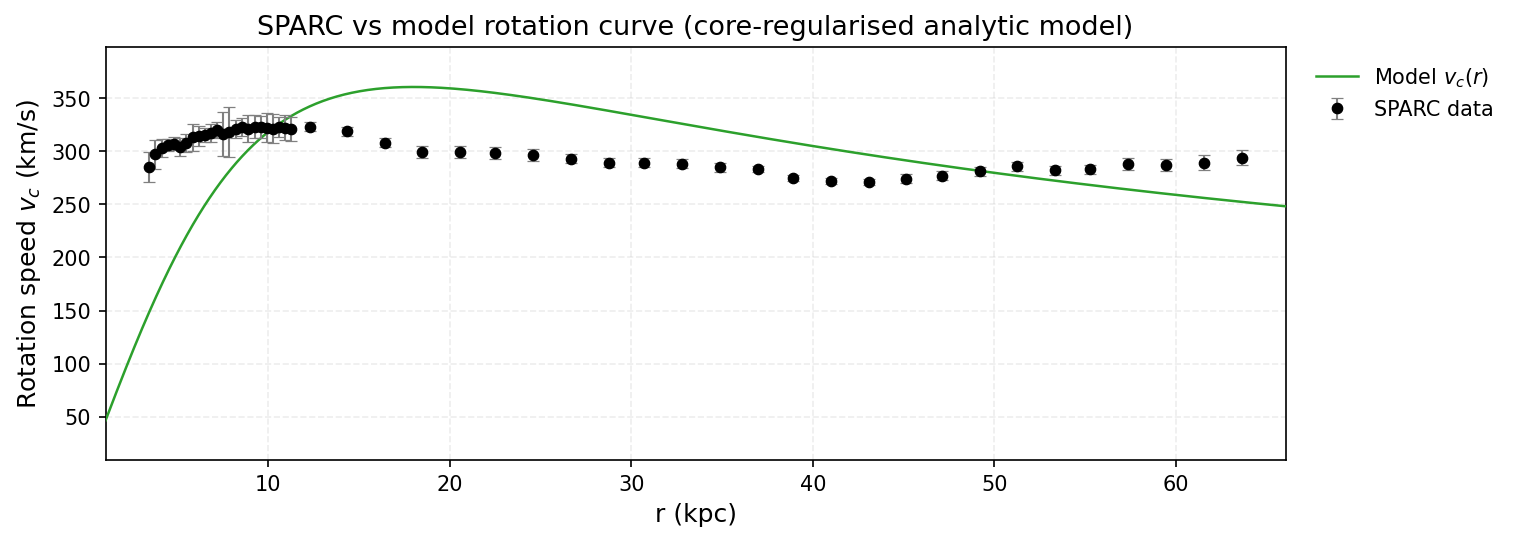}
\caption{The predicted rotation curves for the optimized SIDM
model of Eq. (\ref{ScaledependentEoSDM}), versus the SPARC
observational data for the galaxy NGC2841.} \label{NGC2841}
\end{figure}

In Table \ref{collNGC2841} we present the optimized values of
$K_0$ and $\rho_0$ for the analytic SIDM model of Eq.
(\ref{ScaledependentEoSDM}) for which the maximum compatibility
with the SPARC data is achieved.
\begin{table}[h!]
  \begin{center}
    \caption{SIDM Optimization Values for the galaxy NGC2841}
    \label{collNGC2841}
     \begin{tabular}{|r|r|}
     \hline
      \textbf{Parameter}   & \textbf{Optimization Values}
      \\  \hline
     $\rho_0 $  ($M_{\odot}/\mathrm{Kpc}^{3}$) & $1.15657\times 10^8$
\\  \hline $K_0$ ($M_{\odot} \,
\mathrm{Kpc}^{-3} \, (\mathrm{km/s})^{2}$)& 56258.9
\\  \hline
    \end{tabular}
  \end{center}
\end{table}
In Figs. \ref{NGC2841dens}, \ref{NGC2841}  we present the density
of the analytic SIDM model, the predicted rotation curves for the
SIDM model (\ref{ScaledependentEoSDM}), versus the SPARC
observational data and the sound speed, as a function of the
radius respectively. As it can be seen, for this galaxy, the SIDM
model produces non-viable rotation curves which are incompatible
with the SPARC data.

\begin{figure}[h!]
\centering
\includegraphics[width=20pc]{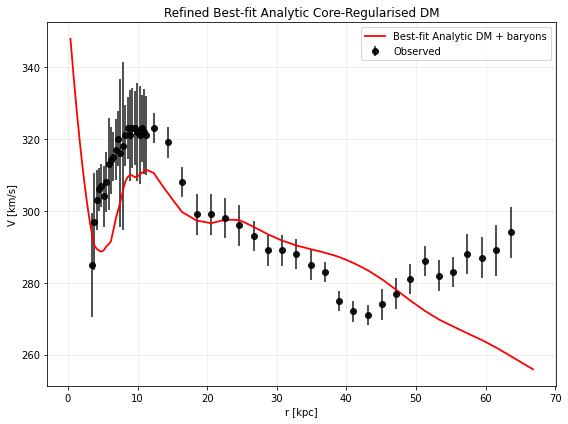}
\caption{The predicted rotation curves after using an optimization
for the SIDM model (\ref{ScaledependentEoSDM}), and the extended
SPARC data for the galaxy NGC2841. We included the rotation curves
of the gas, the disk velocities, the bulge (where present) along
with the SIDM model.} \label{extendedNGC2841}
\end{figure}

Now we shall include contributions to the rotation velocity from
the other components of the galaxy, namely the disk, the gas, and
the bulge if present. In Fig. \ref{extendedNGC2841} we present the
combined rotation curves including all the components of the
galaxy along with the SIDM. As it can be seen, the extended
collisional DM model is non-viable.
Also in Table \ref{evaluationextendedNGC2841} we present the
optimized values of the free parameters of the SIDM model for
which  we achieve the maximum compatibility with the SPARC data,
for the galaxy NGC2841, and also the resulting reduced
$\chi^2_{red}$ value.
\begin{table}[h!]
\centering \caption{Optimized Parameter Values of the Extended
SIDM model for the Galaxy NGC2841.}
\begin{tabular}{lc}
\hline
Parameter & Value  \\
\hline
$\rho_0 $ ($M_{\odot}/\mathrm{Kpc}^{3}$) & $1.0076\times 10^7$   \\
$K_0$ ($M_{\odot} \,
\mathrm{Kpc}^{-3} \, (\mathrm{km/s})^{2}$) & 24413.8   \\
$ml_{\text{disk}}$ & 1 \\
$ml_{\text{bulge}}$ & 0.9 \\
$\alpha$ (Kpc) & 28.4033\\
$\chi^2_{red}$ & 5.87791 \\
\hline
\end{tabular}
\label{evaluationextendedNGC2841}
\end{table}

\subsection{The Galaxy NGC2903, Non-viable, Extended Marginally}

For this galaxy, the optimization method we used, ensures maximum
compatibility of the analytic SIDM model of Eq.
(\ref{ScaledependentEoSDM}) with the SPARC data, if we choose
$\rho_0=3.06984\times 10^8$$M_{\odot}/\mathrm{Kpc}^{3}$ and
$K_0=22970.1
$$M_{\odot} \, \mathrm{Kpc}^{-3} \, (\mathrm{km/s})^{2}$, in which
case the reduced $\chi^2_{red}$ value is $\chi^2_{red}=111.987$.
Also the parameter $\alpha$ in this case is $\alpha=4.992 $Kpc.

\begin{figure}[h!]
\centering
\includegraphics[width=20pc]{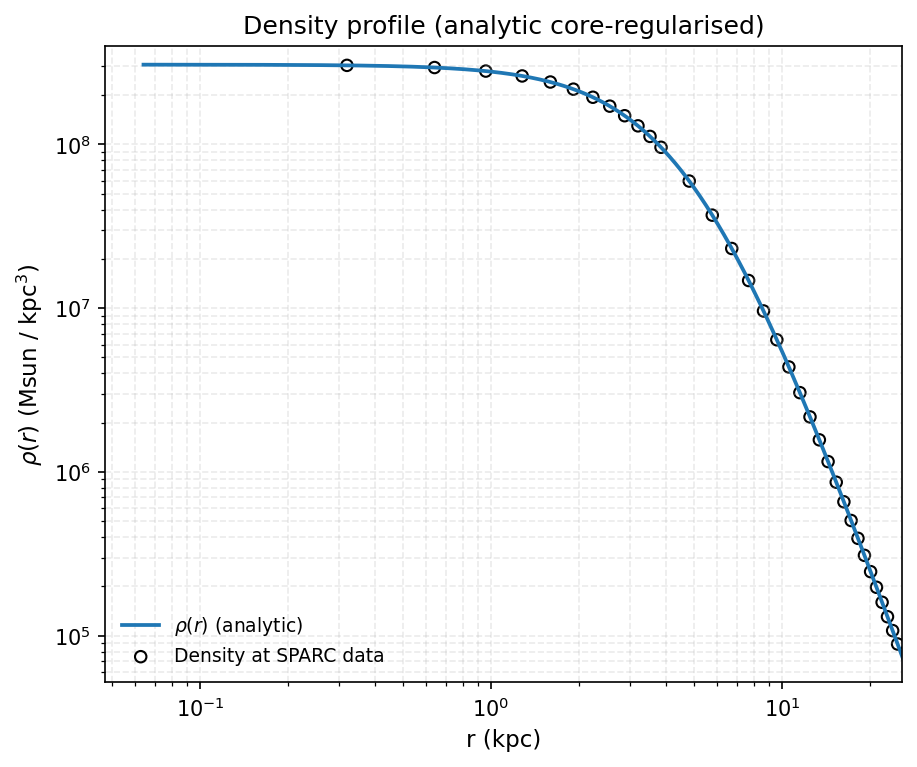}
\caption{The density of the SIDM model of Eq.
(\ref{ScaledependentEoSDM}) for the galaxy NGC2903, versus the
radius.} \label{NGC2903dens}
\end{figure}

\begin{figure}[h!]
\centering
\includegraphics[width=35pc]{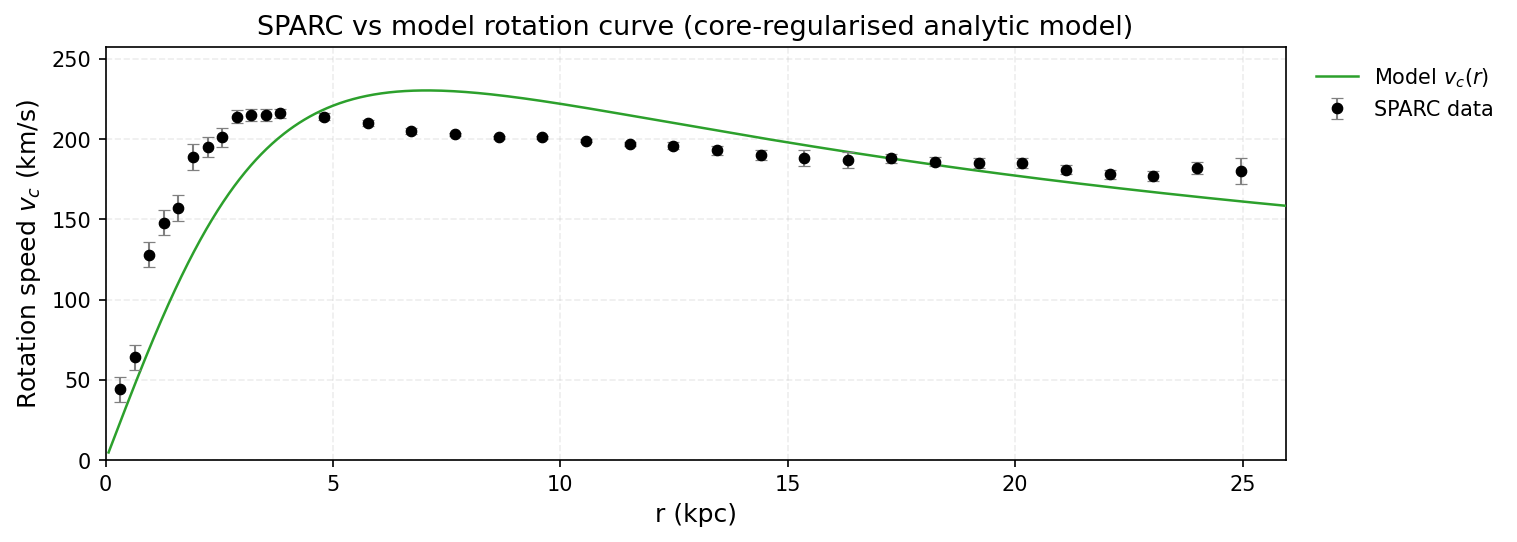}
\caption{The predicted rotation curves for the optimized SIDM
model of Eq. (\ref{ScaledependentEoSDM}), versus the SPARC
observational data for the galaxy NGC2903.} \label{NGC2903}
\end{figure}

In Table \ref{collNGC2903} we present the optimized values of
$K_0$ and $\rho_0$ for the analytic SIDM model of Eq.
(\ref{ScaledependentEoSDM}) for which the maximum compatibility
with the SPARC data is achieved.
\begin{table}[h!]
  \begin{center}
    \caption{SIDM Optimization Values for the galaxy NGC2903}
    \label{collNGC2903}
     \begin{tabular}{|r|r|}
     \hline
      \textbf{Parameter}   & \textbf{Optimization Values}
      \\  \hline
     $\rho_0 $  ($M_{\odot}/\mathrm{Kpc}^{3}$) & $5\times 10^7$
\\  \hline $K_0$ ($M_{\odot} \,
\mathrm{Kpc}^{-3} \, (\mathrm{km/s})^{2}$)& 1250
\\  \hline
    \end{tabular}
  \end{center}
\end{table}
In Figs. \ref{NGC2903dens}, \ref{NGC2903} we present the density
of the analytic SIDM model, the predicted rotation curves for the
SIDM model (\ref{ScaledependentEoSDM}), versus the SPARC
observational data and the sound speed, as a function of the
radius respectively. As it can be seen, for this galaxy, the SIDM
model produces non-viable rotation curves which are incompatible
with the SPARC data.

\begin{figure}[h!]
\centering
\includegraphics[width=20pc]{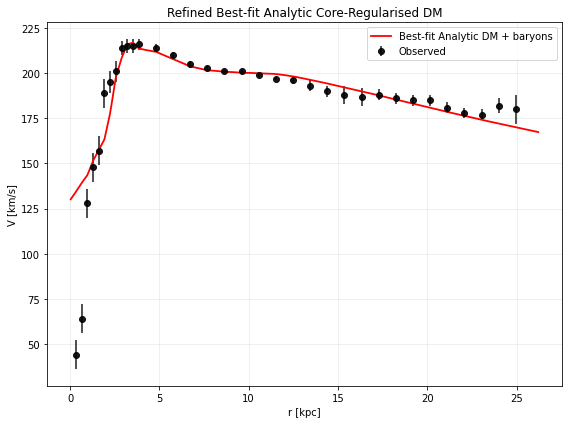}
\caption{The predicted rotation curves after using an optimization
for the SIDM model (\ref{ScaledependentEoSDM}), and the extended
SPARC data for the galaxy NGC2903. We included the rotation curves
of the gas, the disk velocities, the bulge (where present) along
with the SIDM model.} \label{extendedNGC2903}
\end{figure}

Now we shall include contributions to the rotation velocity from
the other components of the galaxy, namely the disk, the gas, and
the bulge if present. In Fig. \ref{extendedNGC2903} we present the
combined rotation curves including all the components of the
galaxy along with the SIDM. As it can be seen, the extended
collisional DM model is non-viable.
Also in Table \ref{evaluationextendedNGC2903} we present the
optimized values of the free parameters of the SIDM model for
which  we achieve the maximum compatibility with the SPARC data,
for the galaxy NGC2903, and also the resulting reduced
$\chi^2_{red}$ value.
\begin{table}[h!]
\centering \caption{Optimized Parameter Values of the Extended
SIDM model for the Galaxy NGC2903.}
\begin{tabular}{lc}
\hline
Parameter & Value  \\
\hline
$\rho_0 $ ($M_{\odot}/\mathrm{Kpc}^{3}$) & $3.04394\times 10^7$   \\
$K_0$ ($M_{\odot} \,
\mathrm{Kpc}^{-3} \, (\mathrm{km/s})^{2}$) & 10665.8   \\
$ml_{\text{disk}}$ & 0.6777 \\
$ml_{\text{bulge}}$ & 0.3 \\
$\alpha$ (Kpc) & 10.8012\\
$\chi^2_{red}$ & 9.03616 \\
\hline
\end{tabular}
\label{evaluationextendedNGC2903}
\end{table}

\subsection{The Galaxy NGC2915}

For this galaxy, the optimization method we used, ensures maximum
compatibility of the analytic SIDM model of Eq.
(\ref{ScaledependentEoSDM}) with the SPARC data, if we choose
$\rho_0=7.76155\times 10^7$$M_{\odot}/\mathrm{Kpc}^{3}$ and
$K_0=3353.32
$$M_{\odot} \, \mathrm{Kpc}^{-3} \, (\mathrm{km/s})^{2}$, in which
case the reduced $\chi^2_{red}$ value is $\chi^2_{red}=0.624307$.
Also the parameter $\alpha$ in this case is $\alpha=3.79327 $Kpc.

\begin{figure}[h!]
\centering
\includegraphics[width=20pc]{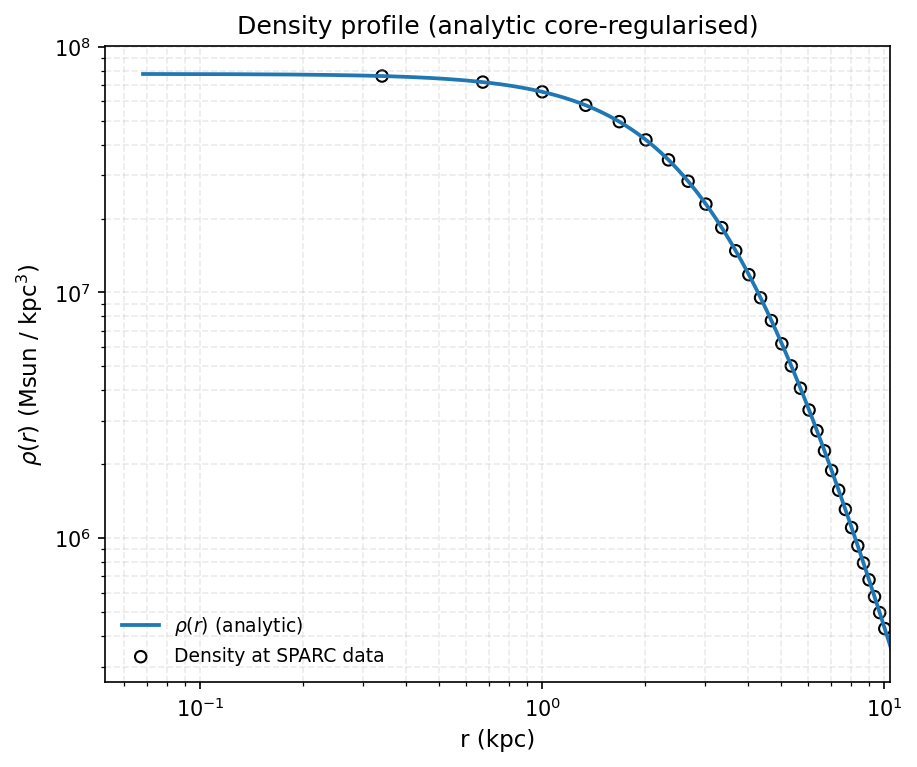}
\caption{The density of the SIDM model of Eq.
(\ref{ScaledependentEoSDM}) for the galaxy NGC2915, versus the
radius.} \label{NGC2915dens}
\end{figure}

\begin{figure}[h!]
\centering
\includegraphics[width=35pc]{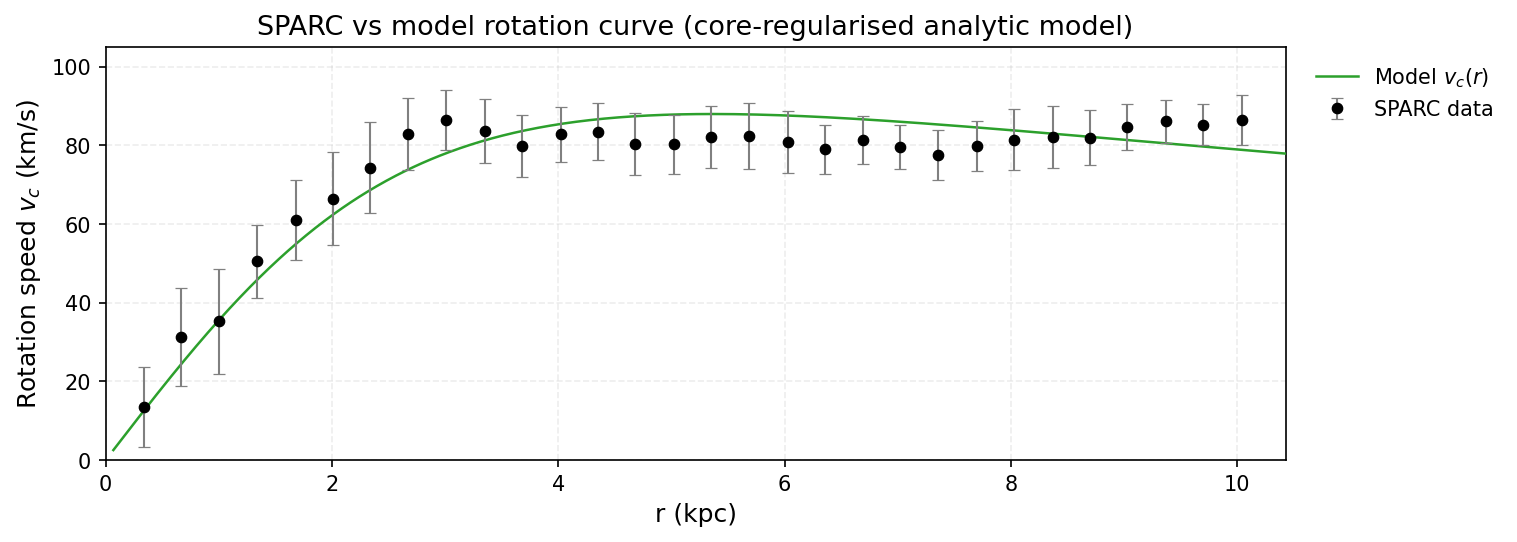}
\caption{The predicted rotation curves for the optimized SIDM
model of Eq. (\ref{ScaledependentEoSDM}), versus the SPARC
observational data for the galaxy NGC2915.} \label{NGC2915}
\end{figure}

In Table \ref{collNGC2915} we present the optimized values of
$K_0$ and $\rho_0$ for the analytic SIDM model of Eq.
(\ref{ScaledependentEoSDM}) for which the maximum compatibility
with the SPARC data is achieved.
\begin{table}[h!]
  \begin{center}
    \caption{SIDM Optimization Values for the galaxy NGC2915}
    \label{collNGC2915}
     \begin{tabular}{|r|r|}
     \hline
      \textbf{Parameter}   & \textbf{Optimization Values}
      \\  \hline
     $\rho_0 $  ($M_{\odot}/\mathrm{Kpc}^{3}$) & $7.76155\times 10^7$
\\  \hline $K_0$ ($M_{\odot} \,
\mathrm{Kpc}^{-3} \, (\mathrm{km/s})^{2}$)& 3353.32
\\  \hline
    \end{tabular}
  \end{center}
\end{table}
In Figs. \ref{NGC2915dens}, \ref{NGC2915} we present the density
of the analytic SIDM model, the predicted rotation curves for the
SIDM model (\ref{ScaledependentEoSDM}), versus the SPARC
observational data and the sound speed, as a function of the
radius respectively. As it can be seen, for this galaxy, the SIDM
model produces viable rotation curves which are compatible with
the SPARC data.

\subsection{The Galaxy NGC2955, Non-viable}

For this galaxy, the optimization method we used, ensures maximum
compatibility of the analytic SIDM model of Eq.
(\ref{ScaledependentEoSDM}) with the SPARC data, if we choose
$\rho_0=2.51138\times 10^8$$M_{\odot}/\mathrm{Kpc}^{3}$ and
$K_0=37862.9
$$M_{\odot} \, \mathrm{Kpc}^{-3} \, (\mathrm{km/s})^{2}$, in which
case the reduced $\chi^2_{red}$ value is $\chi^2_{red}=92.9359$.
Also the parameter $\alpha$ in this case is $\alpha=7.086 $Kpc.

\begin{figure}[h!]
\centering
\includegraphics[width=20pc]{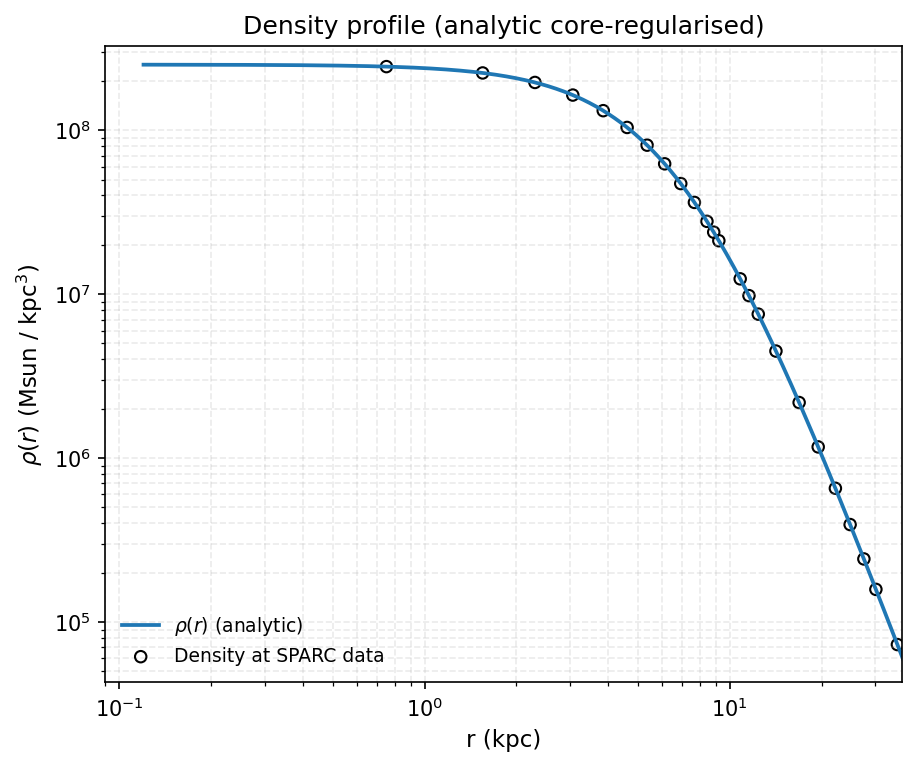}
\caption{The density of the SIDM model of Eq.
(\ref{ScaledependentEoSDM}) for the galaxy NGC2955, versus the
radius.} \label{NGC2955dens}
\end{figure}

\begin{figure}[h!]
\centering
\includegraphics[width=35pc]{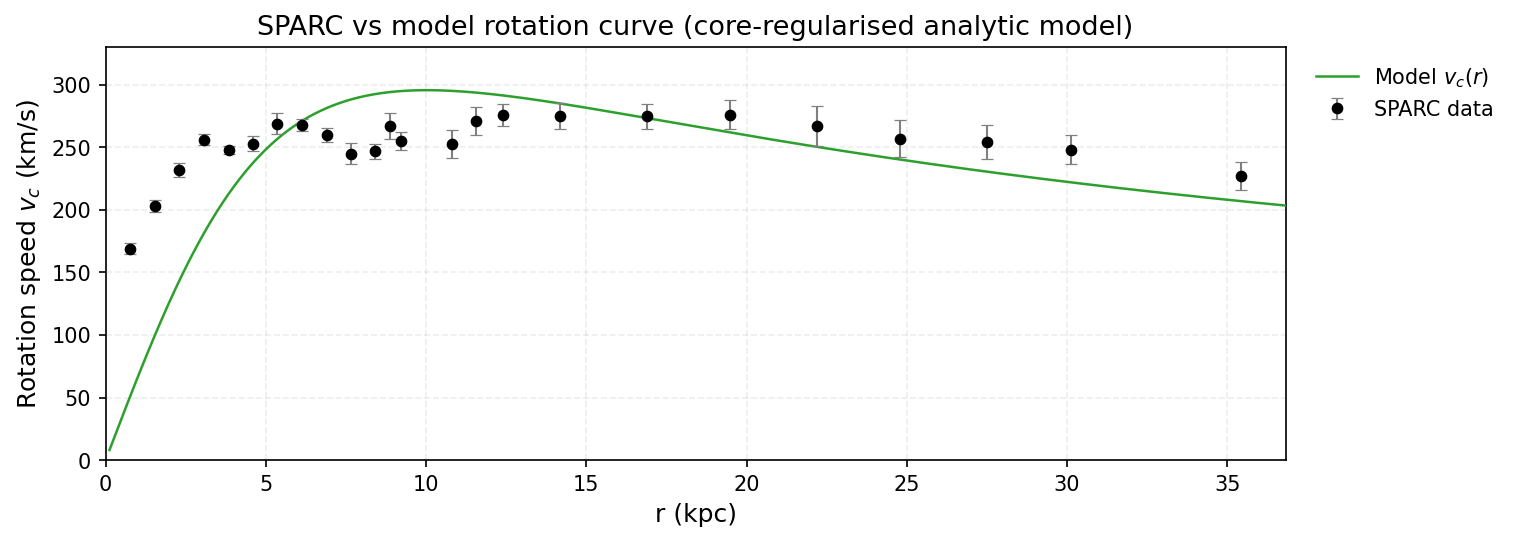}
\caption{The predicted rotation curves for the optimized SIDM
model of Eq. (\ref{ScaledependentEoSDM}), versus the SPARC
observational data for the galaxy NGC2955.} \label{NGC2955}
\end{figure}

In Table \ref{collNGC2955} we present the optimized values of
$K_0$ and $\rho_0$ for the analytic SIDM model of Eq.
(\ref{ScaledependentEoSDM}) for which the maximum compatibility
with the SPARC data is achieved.
\begin{table}[h!]
  \begin{center}
    \caption{SIDM Optimization Values for the galaxy NGC2955}
    \label{collNGC2955}
     \begin{tabular}{|r|r|}
     \hline
      \textbf{Parameter}   & \textbf{Optimization Values}
      \\  \hline
     $\rho_0 $  ($M_{\odot}/\mathrm{Kpc}^{3}$) & $2.51138\times 10^8$
\\  \hline $K_0$ ($M_{\odot} \,
\mathrm{Kpc}^{-3} \, (\mathrm{km/s})^{2}$)& 37862.9
\\  \hline
    \end{tabular}
  \end{center}
\end{table}
In Figs. \ref{NGC2955dens}, \ref{NGC2955} we present the density
of the analytic SIDM model, the predicted rotation curves for the
SIDM model (\ref{ScaledependentEoSDM}), versus the SPARC
observational data and the sound speed, as a function of the
radius respectively. As it can be seen, for this galaxy, the SIDM
model produces non-viable rotation curves which are incompatible
with the SPARC data.

\begin{figure}[h!]
\centering
\includegraphics[width=20pc]{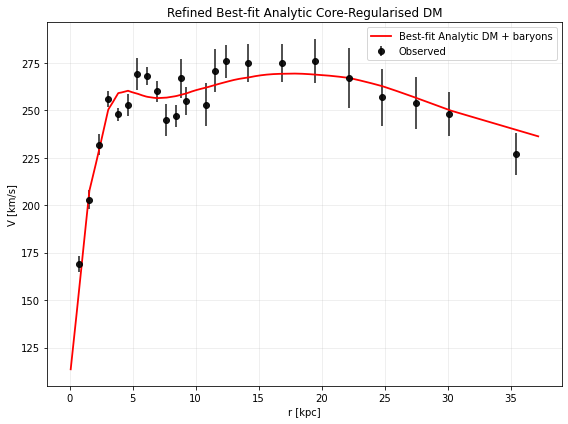}
\caption{The predicted rotation curves after using an optimization
for the SIDM model (\ref{ScaledependentEoSDM}), and the extended
SPARC data for the galaxy NGC2955. We included the rotation curves
of the gas, the disk velocities, the bulge (where present) along
with the SIDM model.} \label{extendedNGC2955}
\end{figure}

Now we shall include contributions to the rotation velocity from
the other components of the galaxy, namely the disk, the gas, and
the bulge if present. In Fig. \ref{extendedNGC2955} we present the
combined rotation curves including all the components of the
galaxy along with the SIDM. As it can be seen, the extended
collisional DM model is non-viable.
Also in Table \ref{evaluationextendedNGC2955} we present the
optimized values of the free parameters of the SIDM model for
which  we achieve the maximum compatibility with the SPARC data,
for the galaxy NGC2955, and also the resulting reduced
$\chi^2_{red}$ value.
\begin{table}[h!]
\centering \caption{Optimized Parameter Values of the Extended
SIDM model for the Galaxy NGC2955.}
\begin{tabular}{lc}
\hline
Parameter & Value  \\
\hline
$\rho_0 $ ($M_{\odot}/\mathrm{Kpc}^{3}$) & $4.13996\times 10^7$   \\
$K_0$ ($M_{\odot} \,
\mathrm{Kpc}^{-3} \, (\mathrm{km/s})^{2}$) & 23227.9   \\
$ml_{\text{disk}}$ & 0.8 \\
$ml_{\text{bulge}}$ & 0.9478 \\
$\alpha$ (Kpc) & 13.6679\\
$\chi^2_{red}$ & 2.14925 \\
\hline
\end{tabular}
\label{evaluationextendedNGC2955}
\end{table}

\subsection{The Galaxy NGC2976 Marginally, Extended Viable}

For this galaxy, the optimization method we used, ensures maximum
compatibility of the analytic SIDM model of Eq.
(\ref{ScaledependentEoSDM}) with the SPARC data, if we choose
$\rho_0=1.87478\times 10^8$$M_{\odot}/\mathrm{Kpc}^{3}$ and
$K_0=4469.65
$$M_{\odot} \, \mathrm{Kpc}^{-3} \, (\mathrm{km/s})^{2}$, in which
case the reduced $\chi^2_{red}$ value is $\chi^2_{red}=0.59403$.
Also the parameter $\alpha$ in this case is $\alpha=2.81781 $Kpc.

\begin{figure}[h!]
\centering
\includegraphics[width=20pc]{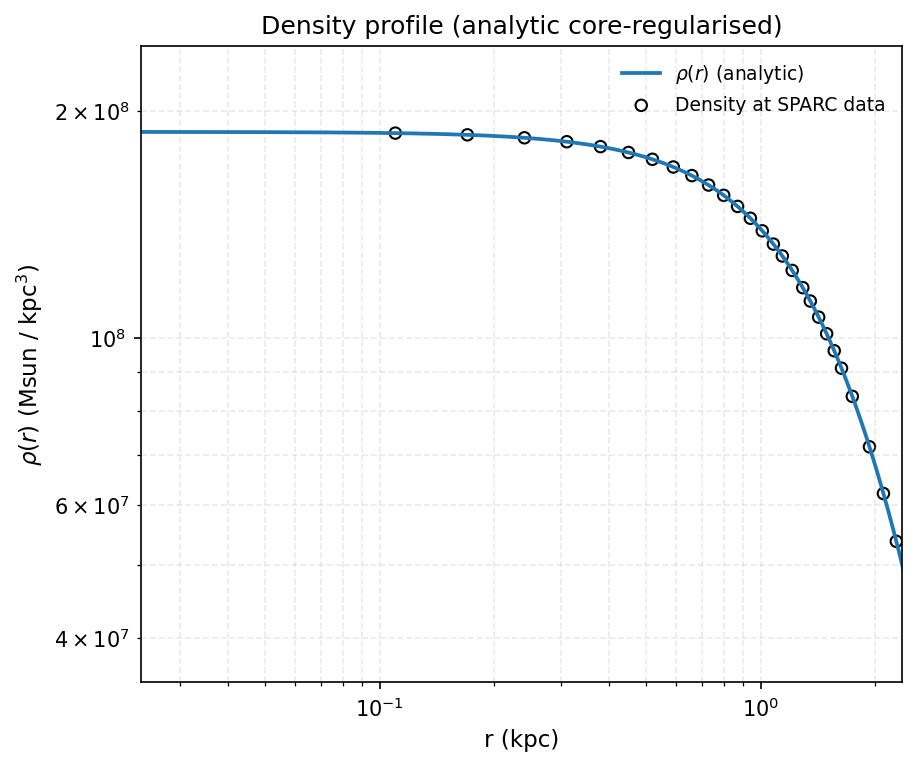}
\caption{The density of the SIDM model of Eq.
(\ref{ScaledependentEoSDM}) for the galaxy NGC2976, versus the
radius.} \label{NGC2976dens}
\end{figure}

\begin{figure}[h!]
\centering
\includegraphics[width=35pc]{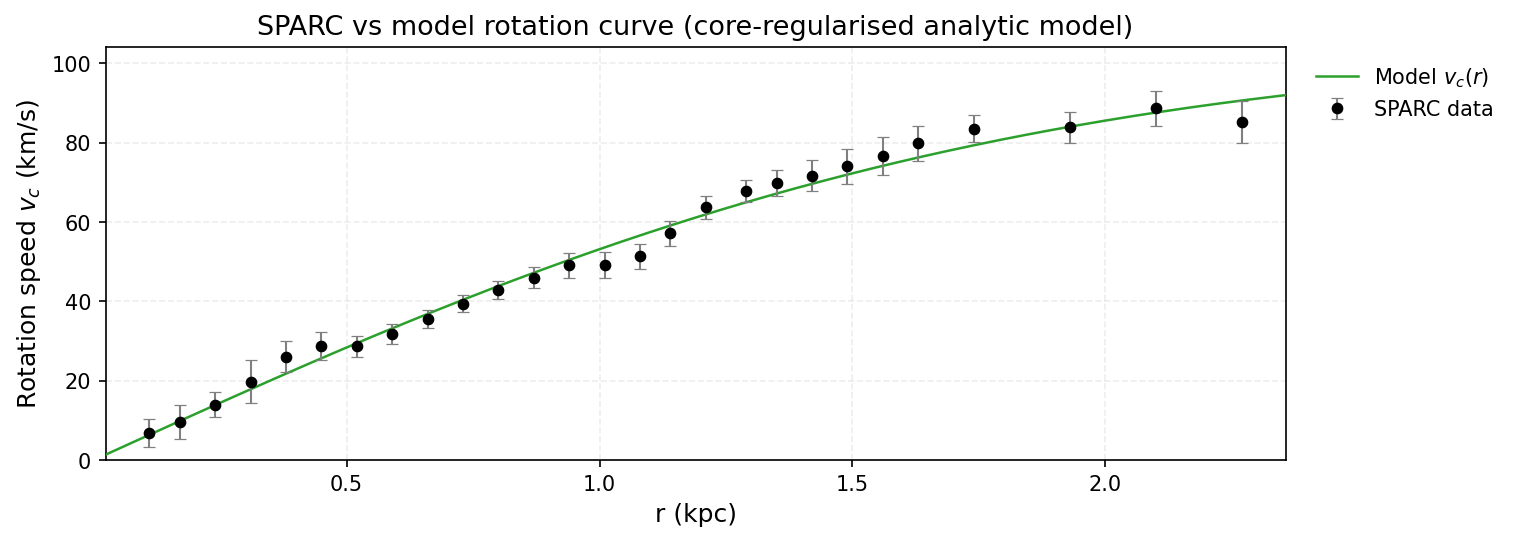}
\caption{The predicted rotation curves for the optimized SIDM
model of Eq. (\ref{ScaledependentEoSDM}), versus the SPARC
observational data for the galaxy NGC2976.} \label{NGC2976}
\end{figure}

In Table \ref{collNGC2976} we present the optimized values of
$K_0$ and $\rho_0$ for the analytic SIDM model of Eq.
(\ref{ScaledependentEoSDM}) for which the maximum compatibility
with the SPARC data is achieved.
\begin{table}[h!]
  \begin{center}
    \caption{SIDM Optimization Values for the galaxy NGC2976}
    \label{collNGC2976}
     \begin{tabular}{|r|r|}
     \hline
      \textbf{Parameter}   & \textbf{Optimization Values}
      \\  \hline
     $\rho_0 $  ($M_{\odot}/\mathrm{Kpc}^{3}$) & $1.87478\times 10^8$
\\  \hline $K_0$ ($M_{\odot} \,
\mathrm{Kpc}^{-3} \, (\mathrm{km/s})^{2}$)& 4469.65
\\  \hline
    \end{tabular}
  \end{center}
\end{table}
In Figs. \ref{NGC2976dens}, \ref{NGC2976} we present the density
of the analytic SIDM model, the predicted rotation curves for the
SIDM model (\ref{ScaledependentEoSDM}), versus the SPARC
observational data and the sound speed, as a function of the
radius respectively. As it can be seen, for this galaxy, the SIDM
model produces viable rotation curves which are marginally
compatible with the SPARC data (off by two data points).

\begin{figure}[h!]
\centering
\includegraphics[width=20pc]{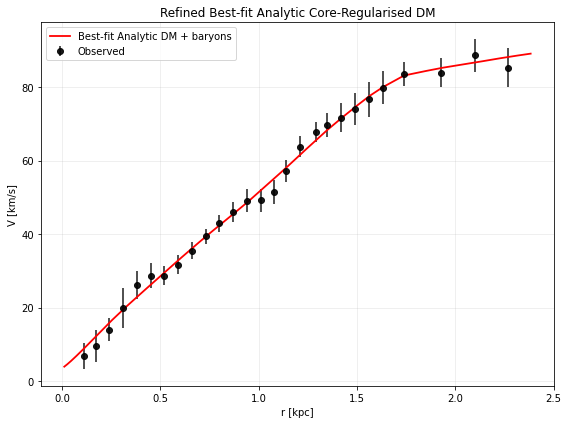}
\caption{The predicted rotation curves after using an optimization
for the SIDM model (\ref{ScaledependentEoSDM}), and the extended
SPARC data for the galaxy NGC2976. We included the rotation curves
of the gas, the disk velocities, the bulge (where present) along
with the SIDM model.} \label{extendedNGC2976}
\end{figure}

Now we shall include contributions to the rotation velocity from
the other components of the galaxy, namely the disk, the gas, and
the bulge if present. In Fig. \ref{extendedNGC2976} we present the
combined rotation curves including all the components of the
galaxy along with the SIDM. As it can be seen, the extended
collisional DM model is viable.
Also in Table \ref{evaluationextendedNGC2976} we present the
optimized values of the free parameters of the SIDM model for
which  we achieve the maximum compatibility with the SPARC data,
for the galaxy NGC2976, and also the resulting reduced
$\chi^2_{red}$ value.
\begin{table}[h!]
\centering \caption{Optimized Parameter Values of the Extended
SIDM model for the Galaxy NGC2976.}
\begin{tabular}{lc}
\hline
Parameter & Value  \\
\hline
$\rho_0 $ ($M_{\odot}/\mathrm{Kpc}^{3}$) & $9.75843\times 10^7$   \\
$K_0$ ($M_{\odot} \,
\mathrm{Kpc}^{-3} \, (\mathrm{km/s})^{2}$) & 2899.61   \\
$ml_{\text{disk}}$ & 0.6165 \\
$ml_{\text{bulge}}$ & 0.0917 \\
$\alpha$ (Kpc) & 3.1454\\
$\chi^2_{red}$ & 0.304637 \\
\hline
\end{tabular}
\label{evaluationextendedNGC2976}
\end{table}

\subsection{The Galaxy NGC2998, Non-viable}

For this galaxy, the optimization method we used, ensures maximum
compatibility of the analytic SIDM model of Eq.
(\ref{ScaledependentEoSDM}) with the SPARC data, if we choose
$\rho_0=1.34663\times 10^8$$M_{\odot}/\mathrm{Kpc}^{3}$ and
$K_0=28911.8
$$M_{\odot} \, \mathrm{Kpc}^{-3} \, (\mathrm{km/s})^{2}$, in which
case the reduced $\chi^2_{red}$ value is $\chi^2_{red}=39.3903$.
Also the parameter $\alpha$ in this case is $\alpha=8.456 $Kpc.

\begin{figure}[h!]
\centering
\includegraphics[width=20pc]{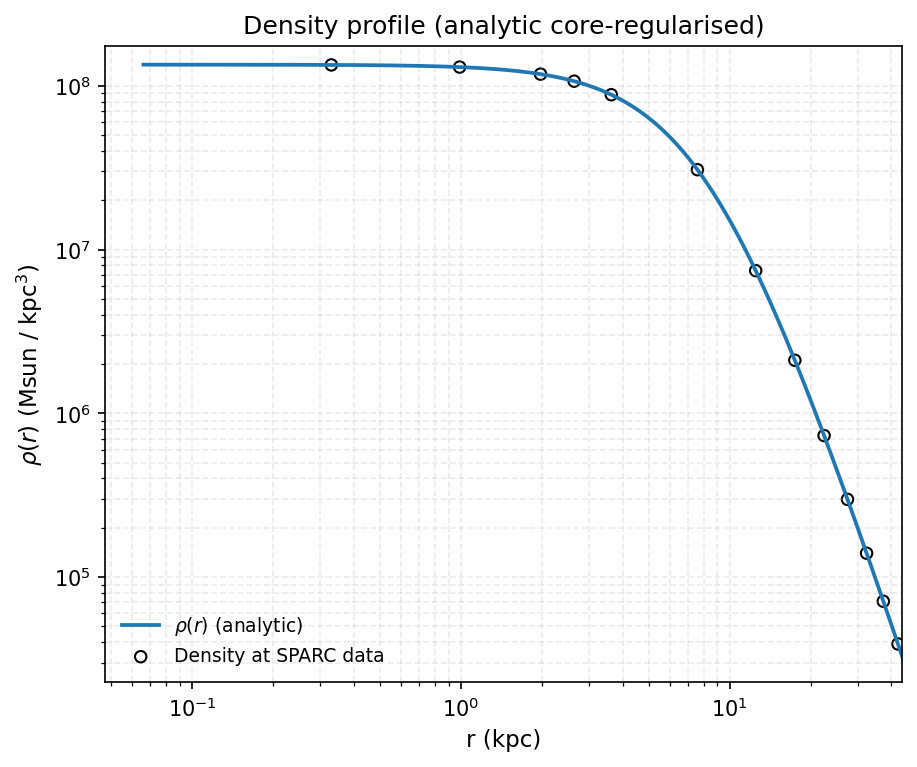}
\caption{The density of the SIDM model of Eq.
(\ref{ScaledependentEoSDM}) for the galaxy NGC2998, versus the
radius.} \label{NGC2998dens}
\end{figure}

\begin{figure}[h!]
\centering
\includegraphics[width=35pc]{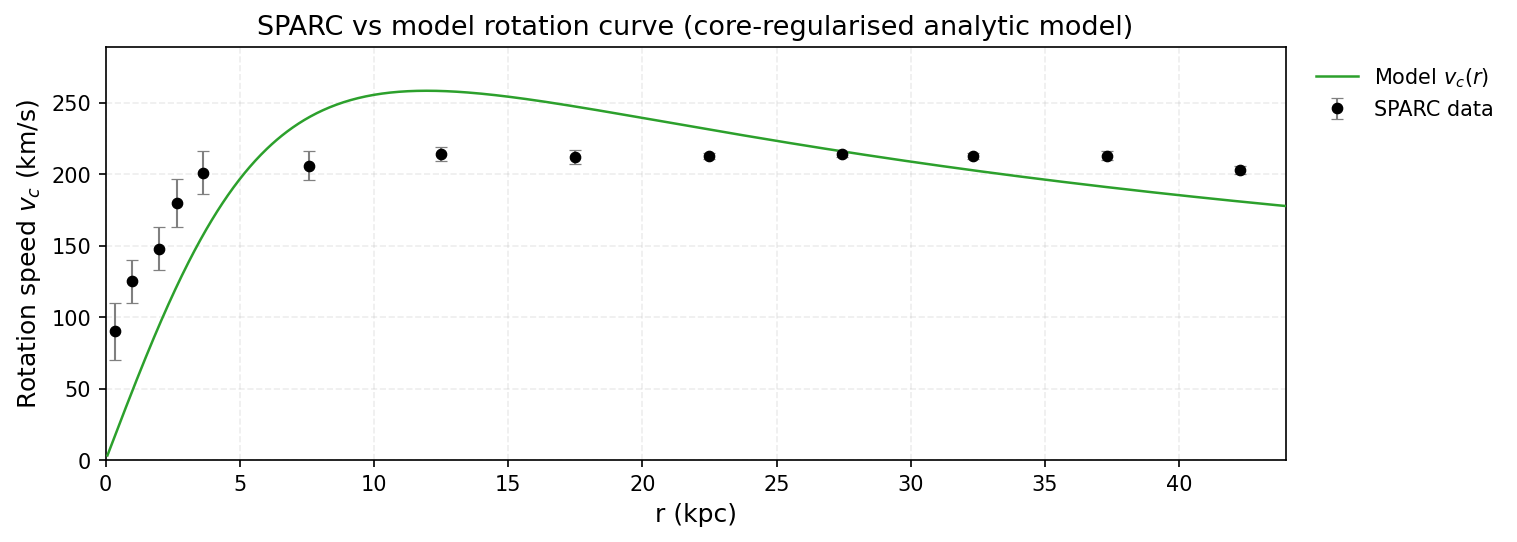}
\caption{The predicted rotation curves for the optimized SIDM
model of Eq. (\ref{ScaledependentEoSDM}), versus the SPARC
observational data for the galaxy NGC2998.} \label{NGC2998}
\end{figure}

In Table \ref{collNGC2998} we present the optimized values of
$K_0$ and $\rho_0$ for the analytic SIDM model of Eq.
(\ref{ScaledependentEoSDM}) for which the maximum compatibility
with the SPARC data is achieved.
\begin{table}[h!]
  \begin{center}
    \caption{SIDM Optimization Values for the galaxy NGC2998}
    \label{collNGC2998}
     \begin{tabular}{|r|r|}
     \hline
      \textbf{Parameter}   & \textbf{Optimization Values}
      \\  \hline
     $\rho_0 $  ($M_{\odot}/\mathrm{Kpc}^{3}$) & $1.34663\times 10^8$
\\  \hline $K_0$ ($M_{\odot} \,
\mathrm{Kpc}^{-3} \, (\mathrm{km/s})^{2}$)& 28911.8
\\  \hline
    \end{tabular}
  \end{center}
\end{table}
In Figs. \ref{NGC2998dens}, \ref{NGC2998} we present the density
of the analytic SIDM model, the predicted rotation curves for the
SIDM model (\ref{ScaledependentEoSDM}), versus the SPARC
observational data and the sound speed, as a function of the
radius respectively. As it can be seen, for this galaxy, the SIDM
model produces non-viable rotation curves which are incompatible
with the SPARC data.

\begin{figure}[h!]
\centering
\includegraphics[width=20pc]{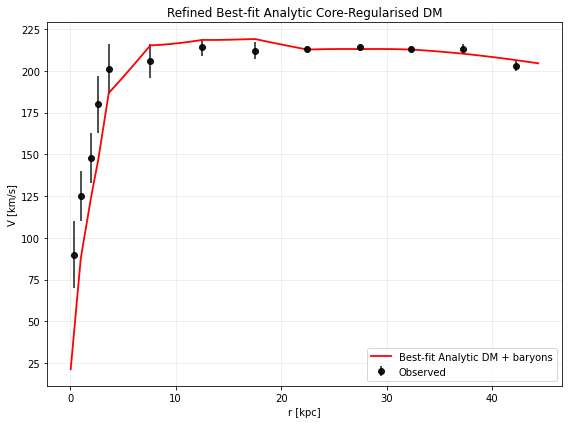}
\caption{The predicted rotation curves after using an optimization
for the SIDM model (\ref{ScaledependentEoSDM}), and the extended
SPARC data for the galaxy NGC2998. We included the rotation curves
of the gas, the disk velocities, the bulge (where present) along
with the SIDM model.} \label{extendedNGC2998}
\end{figure}

Now we shall include contributions to the rotation velocity from
the other components of the galaxy, namely the disk, the gas, and
the bulge if present. In Fig. \ref{extendedNGC2998} we present the
combined rotation curves including all the components of the
galaxy along with the SIDM. As it can be seen, the extended
collisional DM model is non-viable.
Also in Table \ref{evaluationextendedNGC2998} we present the
optimized values of the free parameters of the SIDM model for
which  we achieve the maximum compatibility with the SPARC data,
for the galaxy NGC2998, and also the resulting reduced
$\chi^2_{red}$ value.
\begin{table}[h!]
\centering \caption{Optimized Parameter Values of the Extended
SIDM model for the Galaxy NGC2998.}
\begin{tabular}{lc}
\hline
Parameter & Value  \\
\hline
$\rho_0 $ ($M_{\odot}/\mathrm{Kpc}^{3}$) & $2.81529\times 10^6$   \\
$K_0$ ($M_{\odot} \,
\mathrm{Kpc}^{-3} \, (\mathrm{km/s})^{2}$) & 10365.5   \\
$ml_{\text{disk}}$ & 0.9393 \\
$ml_{\text{bulge}}$ & 0.7347 \\
$\alpha$ (Kpc) & 35.0131\\
$\chi^2_{red}$ & 2.7979 \\
\hline
\end{tabular}
\label{evaluationextendedNGC2998}
\end{table}

\subsection{The Galaxy NGC3109}

For this galaxy, the optimization method we used, ensures maximum
compatibility of the analytic SIDM model of Eq.
(\ref{ScaledependentEoSDM}) with the SPARC data, if we choose
$\rho_0=2.31771\times 10^7$$M_{\odot}/\mathrm{Kpc}^{3}$ and
$K_0=1836.37
$$M_{\odot} \, \mathrm{Kpc}^{-3} \, (\mathrm{km/s})^{2}$, in which
case the reduced $\chi^2_{red}$ value is $\chi^2_{red}=0.360257$.
Also the parameter $\alpha$ in this case is $\alpha=5.13691 $Kpc.

\begin{figure}[h!]
\centering
\includegraphics[width=20pc]{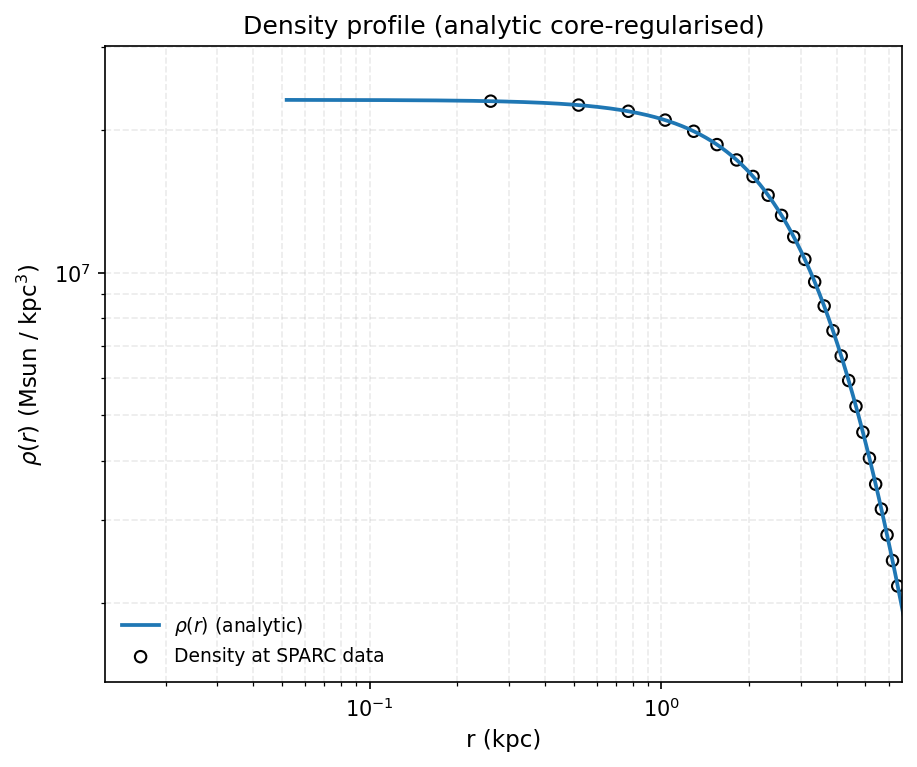}
\caption{The density of the SIDM model of Eq.
(\ref{ScaledependentEoSDM}) for the galaxy NGC3109, versus the
radius.} \label{NGC3109dens}
\end{figure}

\begin{figure}[h!]
\centering
\includegraphics[width=35pc]{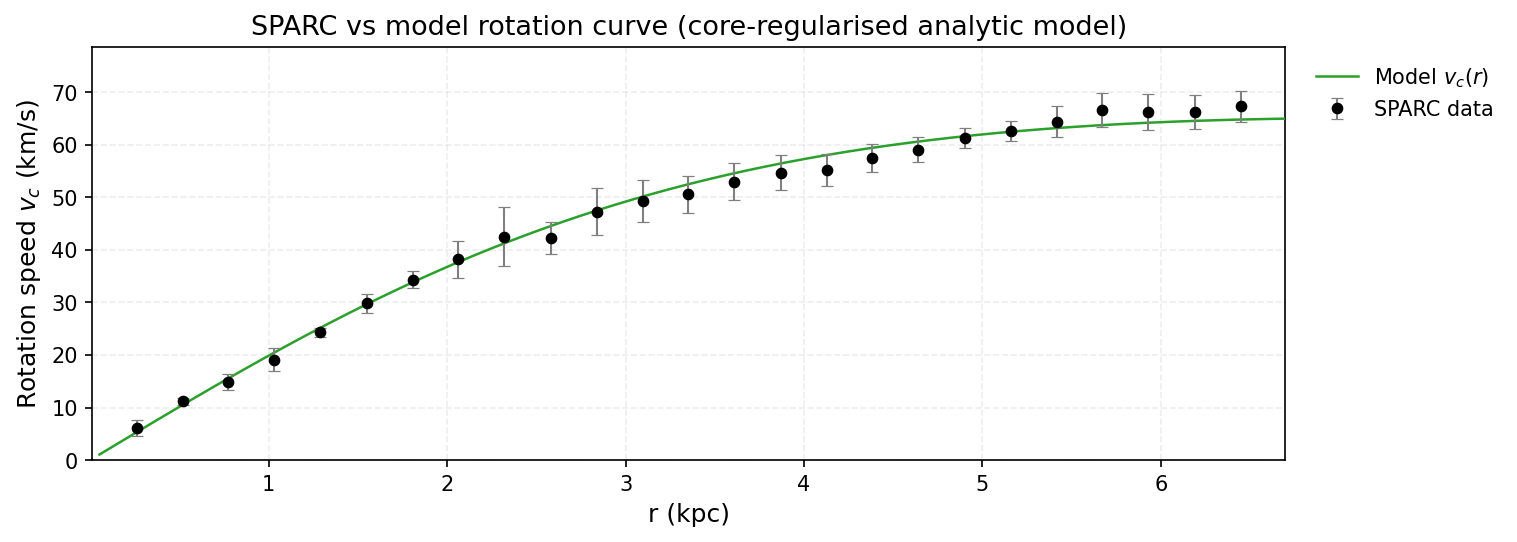}
\caption{The predicted rotation curves for the optimized SIDM
model of Eq. (\ref{ScaledependentEoSDM}), versus the SPARC
observational data for the galaxy NGC3109.} \label{NGC3109}
\end{figure}

In Table \ref{collNGC3109} we present the optimized values of
$K_0$ and $\rho_0$ for the analytic SIDM model of Eq.
(\ref{ScaledependentEoSDM}) for which the maximum compatibility
with the SPARC data is achieved.
\begin{table}[h!]
  \begin{center}
    \caption{SIDM Optimization Values for the galaxy NGC3109}
    \label{collNGC3109}
     \begin{tabular}{|r|r|}
     \hline
      \textbf{Parameter}   & \textbf{Optimization Values}
      \\  \hline
     $\rho_0 $  ($M_{\odot}/\mathrm{Kpc}^{3}$) & $2.31771\times 10^7$
\\  \hline $K_0$ ($M_{\odot} \,
\mathrm{Kpc}^{-3} \, (\mathrm{km/s})^{2}$)& 1836.37
\\  \hline
    \end{tabular}
  \end{center}
\end{table}
In Figs. \ref{NGC3109dens}, \ref{NGC3109} we present the density
of the analytic SIDM model, the predicted rotation curves for the
SIDM model (\ref{ScaledependentEoSDM}), versus the SPARC
observational data and the sound speed, as a function of the
radius respectively. As it can be seen, for this galaxy, the SIDM
model produces viable rotation curves which are compatible with
the SPARC data.

\subsection{The Galaxy NGC3198, Non-viable}

For this galaxy, the optimization method we used, ensures maximum
compatibility of the analytic SIDM model of Eq.
(\ref{ScaledependentEoSDM}) with the SPARC data, if we choose
$\rho_0=5.29788\times 10^7$$M_{\odot}/\mathrm{Kpc}^{3}$ and
$K_0=12494.5
$$M_{\odot} \, \mathrm{Kpc}^{-3} \, (\mathrm{km/s})^{2}$, in which
case the reduced $\chi^2_{red}$ value is $\chi^2_{red}=25.8211$.
Also the parameter $\alpha$ in this case is $\alpha=8.86258 $Kpc.

\begin{figure}[h!]
\centering
\includegraphics[width=20pc]{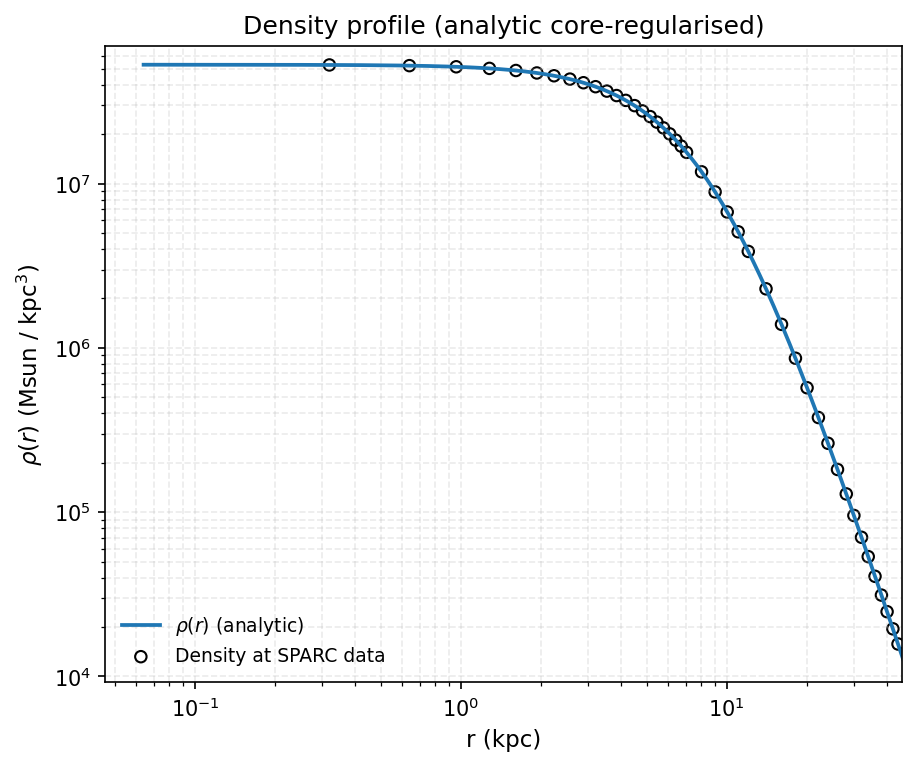}
\caption{The density of the SIDM model of Eq.
(\ref{ScaledependentEoSDM}) for the galaxy NGC3198, versus the
radius.} \label{NGC3198dens}
\end{figure}

\begin{figure}[h!]
\centering
\includegraphics[width=35pc]{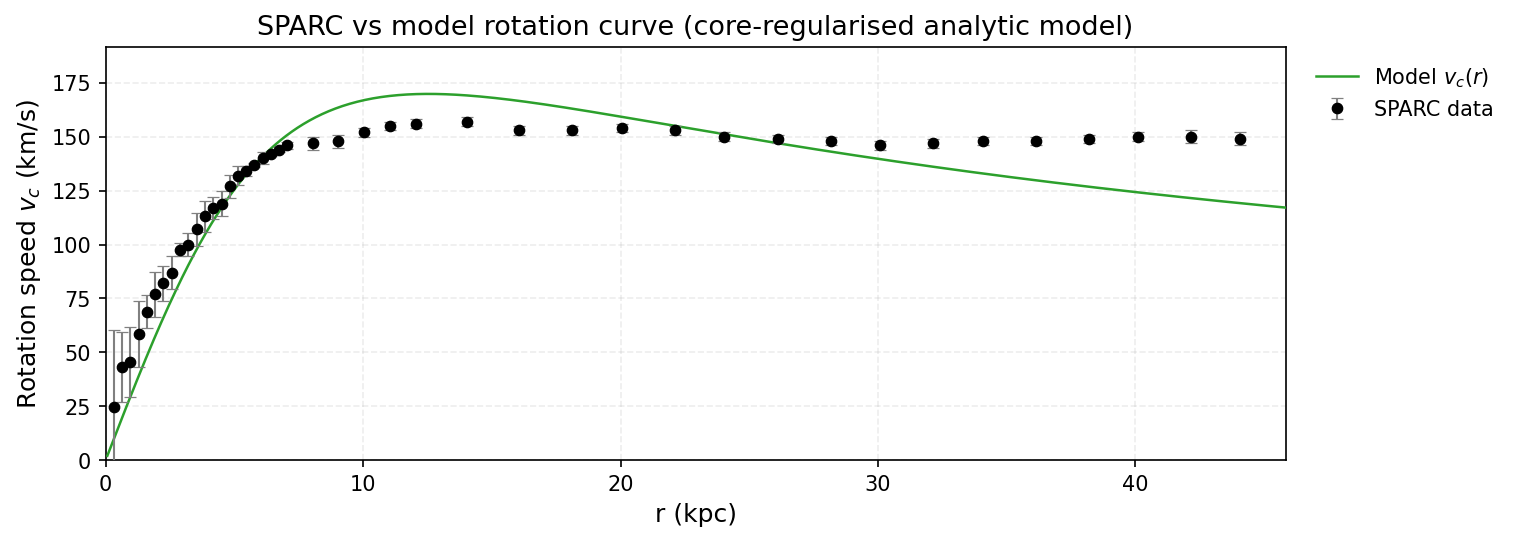}
\caption{The predicted rotation curves for the optimized SIDM
model of Eq. (\ref{ScaledependentEoSDM}), versus the SPARC
observational data for the galaxy NGC3198.} \label{NGC3198}
\end{figure}

In Table \ref{collNGC3198} we present the optimized values of
$K_0$ and $\rho_0$ for the analytic SIDM model of Eq.
(\ref{ScaledependentEoSDM}) for which the maximum compatibility
with the SPARC data is achieved.
\begin{table}[h!]
  \begin{center}
    \caption{SIDM Optimization Values for the galaxy NGC3198}
    \label{collNGC3198}
     \begin{tabular}{|r|r|}
     \hline
      \textbf{Parameter}   & \textbf{Optimization Values}
      \\  \hline
     $\rho_0 $  ($M_{\odot}/\mathrm{Kpc}^{3}$) & $5.29788\times 10^7$
\\  \hline $K_0$ ($M_{\odot} \,
\mathrm{Kpc}^{-3} \, (\mathrm{km/s})^{2}$)& 12494.5
\\  \hline
    \end{tabular}
  \end{center}
\end{table}
In Figs. \ref{NGC3198dens}, \ref{NGC3198} we present the density
of the analytic SIDM model, the predicted rotation curves for the
SIDM model (\ref{ScaledependentEoSDM}), versus the SPARC
observational data and the sound speed, as a function of the
radius respectively. As it can be seen, for this galaxy, the SIDM
model produces non-viable rotation curves which are incompatible
with the SPARC data.

\begin{figure}[h!]
\centering
\includegraphics[width=20pc]{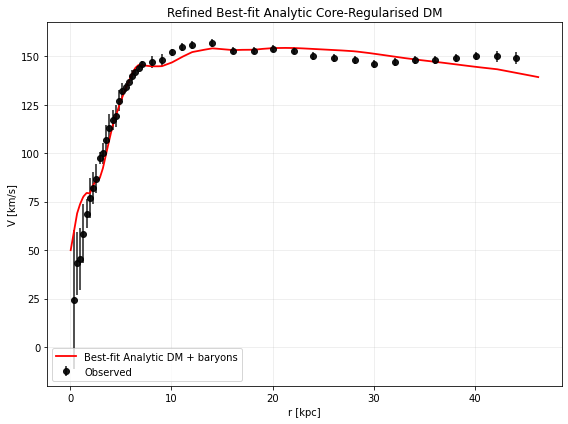}
\caption{The predicted rotation curves after using an optimization
for the SIDM model (\ref{ScaledependentEoSDM}), and the extended
SPARC data for the galaxy NGC3198. We included the rotation curves
of the gas, the disk velocities, the bulge (where present) along
with the SIDM model.} \label{extendedNGC3198}
\end{figure}

Now we shall include contributions to the rotation velocity from
the other components of the galaxy, namely the disk, the gas, and
the bulge if present. In Fig. \ref{extendedNGC3198} we present the
combined rotation curves including all the components of the
galaxy along with the SIDM. As it can be seen, the extended
collisional DM model is non-viable.
Also in Table \ref{evaluationextendedNGC3198} we present the
optimized values of the free parameters of the SIDM model for
which  we achieve the maximum compatibility with the SPARC data,
for the galaxy NGC3198, and also the resulting reduced
$\chi^2_{red}$ value.
\begin{table}[h!]
\centering \caption{Optimized Parameter Values of the Extended
SIDM model for the Galaxy NGC3198.}
\begin{tabular}{lc}
\hline
Parameter & Value  \\
\hline
$\rho_0 $ ($M_{\odot}/\mathrm{Kpc}^{3}$) & $5.18167\times 10^6$   \\
$K_0$ ($M_{\odot} \,
\mathrm{Kpc}^{-3} \, (\mathrm{km/s})^{2}$) & 6879.64   \\
$ml_{\text{disk}}$ & 0.9264 \\
$ml_{\text{bulge}}$ & 0.1394 \\
$\alpha$ (Kpc) & 21.0254\\
$\chi^2_{red}$ & 2.55634 \\
\hline
\end{tabular}
\label{evaluationextendedNGC3198}
\end{table}

\subsection{The Galaxy NGC3521, Non-viable, Extended Viable}

For this galaxy, the optimization method we used, ensures maximum
compatibility of the analytic SIDM model of Eq.
(\ref{ScaledependentEoSDM}) with the SPARC data, if we choose
$\rho_0=6.40035\times 10^8$$M_{\odot}/\mathrm{Kpc}^{3}$ and
$K_0=24191.9
$$M_{\odot} \, \mathrm{Kpc}^{-3} \, (\mathrm{km/s})^{2}$, in which
case the reduced $\chi^2_{red}$ value is $\chi^2_{red}=17.4628$.
Also the parameter $\alpha$ in this case is $\alpha=3.548 $Kpc.

\begin{figure}[h!]
\centering
\includegraphics[width=20pc]{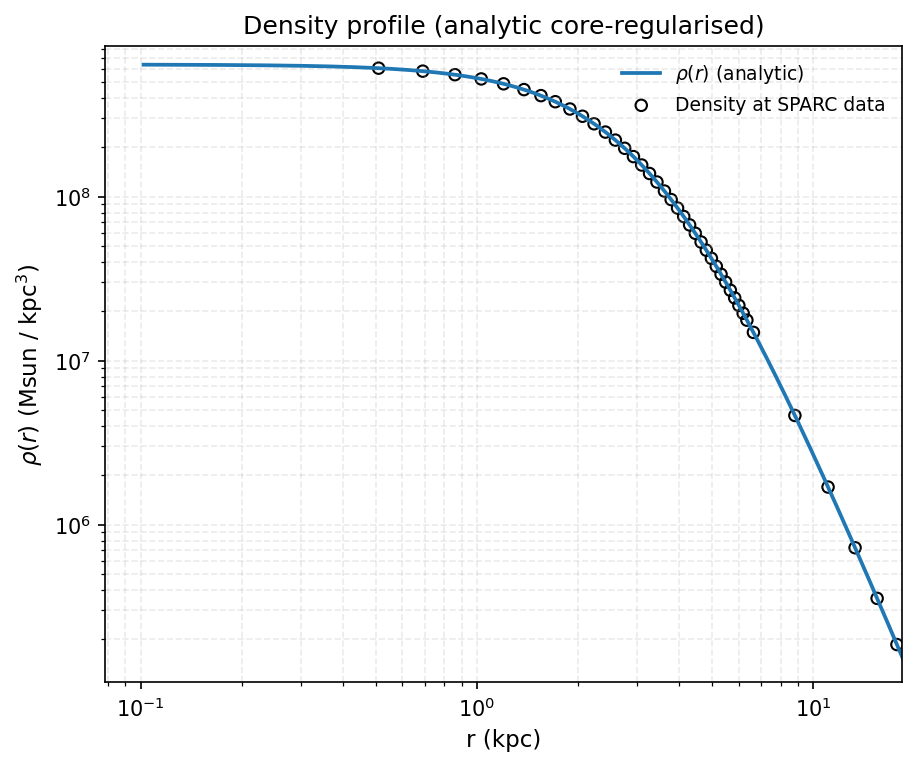}
\caption{The density of the SIDM model of Eq.
(\ref{ScaledependentEoSDM}) for the galaxy NGC3521, versus the
radius.} \label{NGC3521dens}
\end{figure}

\begin{figure}[h!]
\centering
\includegraphics[width=35pc]{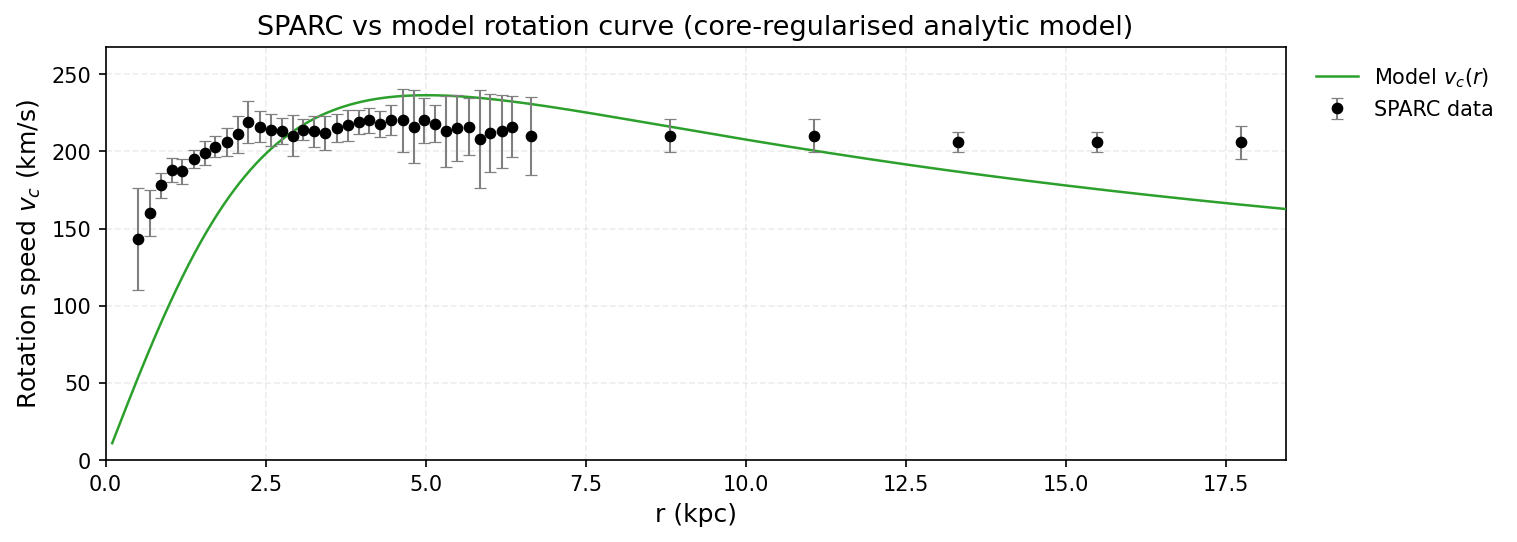}
\caption{The predicted rotation curves for the optimized SIDM
model of Eq. (\ref{ScaledependentEoSDM}), versus the SPARC
observational data for the galaxy NGC3521.} \label{NGC3521}
\end{figure}

In Table \ref{collNGC3521} we present the optimized values of
$K_0$ and $\rho_0$ for the analytic SIDM model of Eq.
(\ref{ScaledependentEoSDM}) for which the maximum compatibility
with the SPARC data is achieved.
\begin{table}[h!]
  \begin{center}
    \caption{SIDM Optimization Values for the galaxy NGC3521}
    \label{collNGC3521}
     \begin{tabular}{|r|r|}
     \hline
      \textbf{Parameter}   & \textbf{Optimization Values}
      \\  \hline
     $\rho_0 $  ($M_{\odot}/\mathrm{Kpc}^{3}$) & $6.40035\times 10^8$
\\  \hline $K_0$ ($M_{\odot} \,
\mathrm{Kpc}^{-3} \, (\mathrm{km/s})^{2}$)& 24191.9
\\  \hline
    \end{tabular}
  \end{center}
\end{table}
In Figs. \ref{NGC3521dens}, \ref{NGC3521} we present the density
of the analytic SIDM model, the predicted rotation curves for the
SIDM model (\ref{ScaledependentEoSDM}), versus the SPARC
observational data and the sound speed, as a function of the
radius respectively. As it can be seen, for this galaxy, the SIDM
model produces non-viable rotation curves which are incompatible
with the SPARC data.

\begin{figure}[h!]
\centering
\includegraphics[width=20pc]{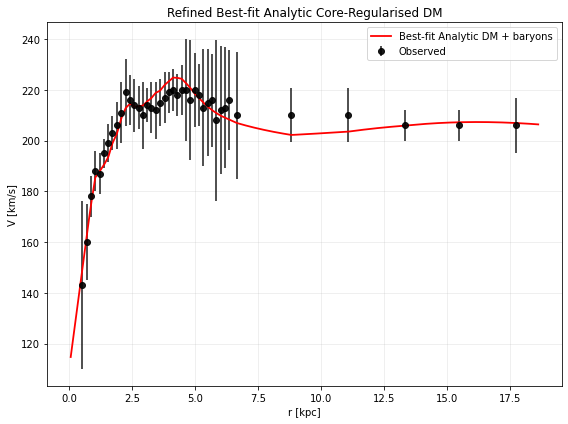}
\caption{The predicted rotation curves after using an optimization
for the SIDM model (\ref{ScaledependentEoSDM}), and the extended
SPARC data for the galaxy NGC3521. We included the rotation curves
of the gas, the disk velocities, the bulge (where present) along
with the SIDM model.} \label{extendedNGC3521}
\end{figure}

Now we shall include contributions to the rotation velocity from
the other components of the galaxy, namely the disk, the gas, and
the bulge if present. In Fig. \ref{extendedNGC3521} we present the
combined rotation curves including all the components of the
galaxy along with the SIDM. As it can be seen, the extended
collisional DM model is viable.
Also in Table \ref{evaluationextendedNGC3521} we present the
optimized values of the free parameters of the SIDM model for
which  we achieve the maximum compatibility with the SPARC data,
for the galaxy NGC3521, and also the resulting reduced
$\chi^2_{red}$ value.
\begin{table}[h!]
\centering \caption{Optimized Parameter Values of the Extended
SIDM model for the Galaxy NGC3521.}
\begin{tabular}{lc}
\hline
Parameter & Value  \\
\hline
$\rho_0 $ ($M_{\odot}/\mathrm{Kpc}^{3}$) & $1.3977\times 10^7$   \\
$K_0$ ($M_{\odot} \,
\mathrm{Kpc}^{-3} \, (\mathrm{km/s})^{2}$) & 13466.1   \\
$ml_{\text{disk}}$ & 0.7622 \\
$ml_{\text{bulge}}$ & 0.3710 \\
$\alpha$ (Kpc) & 17.9106\\
$\chi^2_{red}$ & 0.166192 \\
\hline
\end{tabular}
\label{evaluationextendedNGC3521}
\end{table}

\subsection{The Galaxy NGC3726, Non-viable}

For this galaxy, the optimization method we used, ensures maximum
compatibility of the analytic SIDM model of Eq.
(\ref{ScaledependentEoSDM}) with the SPARC data, if we choose
$\rho_0=4.00421\times 10^7$$M_{\odot}/\mathrm{Kpc}^{3}$ and
$K_0=13094.3
$$M_{\odot} \, \mathrm{Kpc}^{-3} \, (\mathrm{km/s})^{2}$, in which
case the reduced $\chi^2_{red}$ value is $\chi^2_{red}=6.94506$.
Also the parameter $\alpha$ in this case is $\alpha=10.436 $Kpc.

\begin{figure}[h!]
\centering
\includegraphics[width=20pc]{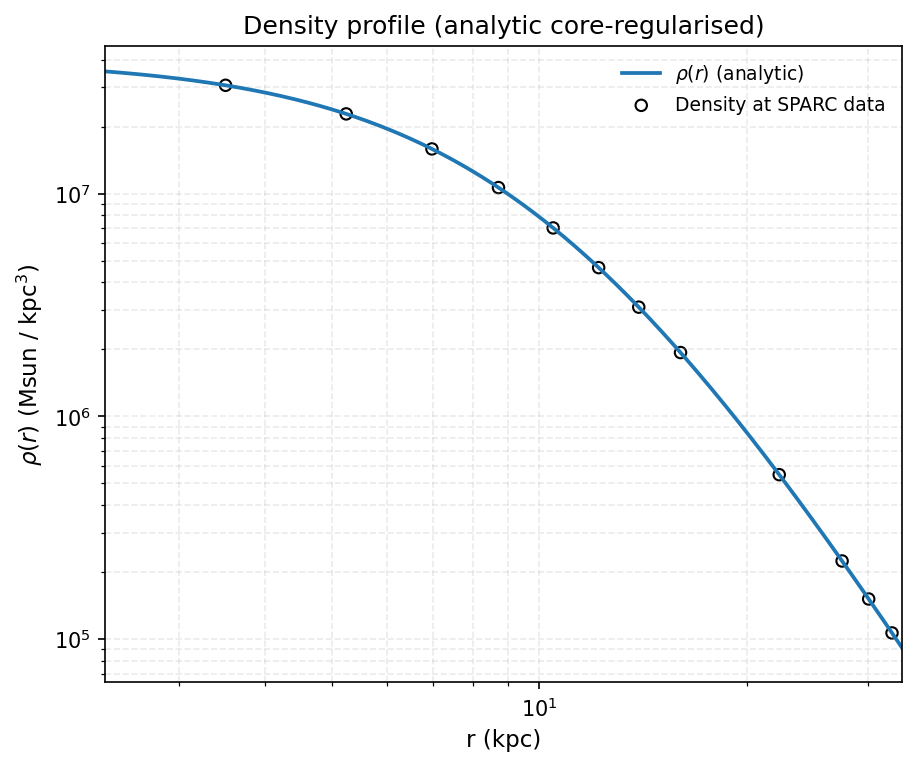}
\caption{The density of the SIDM model of Eq.
(\ref{ScaledependentEoSDM}) for the galaxy NGC3726, versus the
radius.} \label{NGC3726dens}
\end{figure}

\begin{figure}[h!]
\centering
\includegraphics[width=35pc]{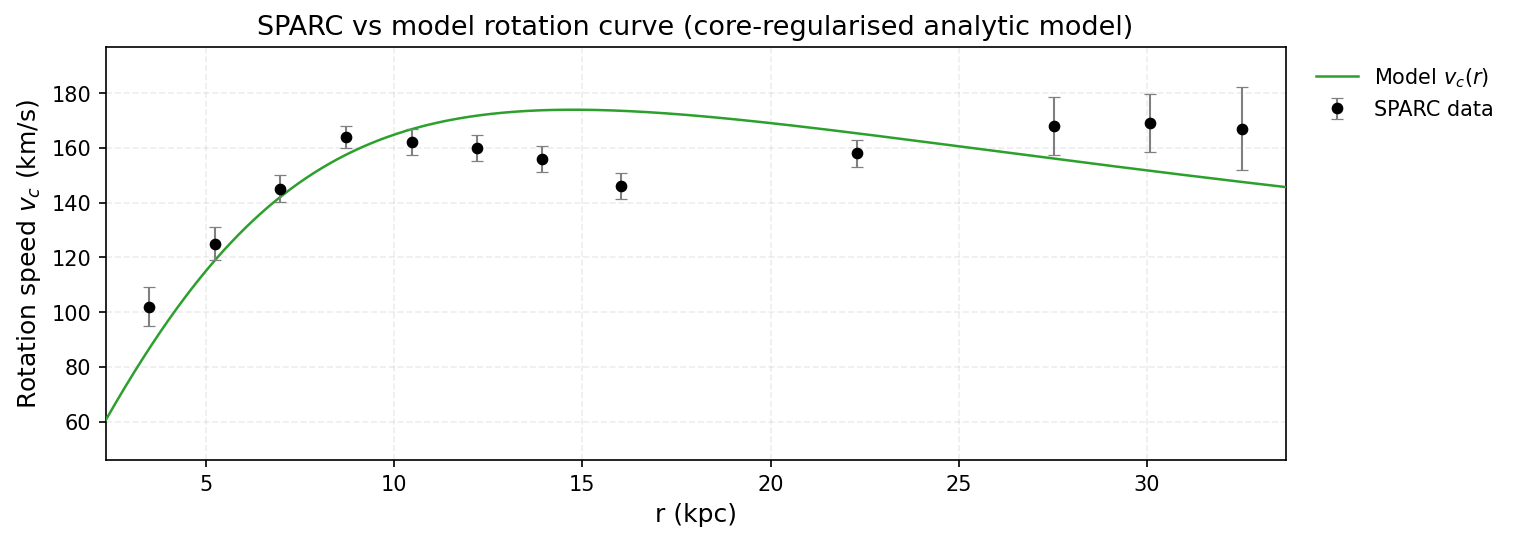}
\caption{The predicted rotation curves for the optimized SIDM
model of Eq. (\ref{ScaledependentEoSDM}), versus the SPARC
observational data for the galaxy NGC3726.} \label{NGC3726}
\end{figure}

In Table \ref{collNGC3726} we present the optimized values of
$K_0$ and $\rho_0$ for the analytic SIDM model of Eq.
(\ref{ScaledependentEoSDM}) for which the maximum compatibility
with the SPARC data is achieved.
\begin{table}[h!]
  \begin{center}
    \caption{SIDM Optimization Values for the galaxy NGC3726}
    \label{collNGC3726}
     \begin{tabular}{|r|r|}
     \hline
      \textbf{Parameter}   & \textbf{Optimization Values}
      \\  \hline
     $\rho_0 $  ($M_{\odot}/\mathrm{Kpc}^{3}$) & $4.00421\times 10^7$
\\  \hline $K_0$ ($M_{\odot} \,
\mathrm{Kpc}^{-3} \, (\mathrm{km/s})^{2}$)& 13094.3
\\  \hline
    \end{tabular}
  \end{center}
\end{table}
In Figs. \ref{NGC3726dens}, \ref{NGC3726} we present the density
of the analytic SIDM model, the predicted rotation curves for the
SIDM model (\ref{ScaledependentEoSDM}), versus the SPARC
observational data and the sound speed, as a function of the
radius respectively. As it can be seen, for this galaxy, the SIDM
model produces non-viable rotation curves which are incompatible
with the SPARC data.

\begin{figure}[h!]
\centering
\includegraphics[width=20pc]{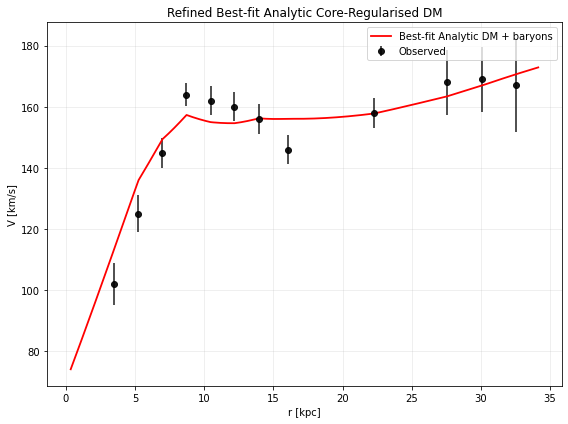}
\caption{The predicted rotation curves after using an optimization
for the SIDM model (\ref{ScaledependentEoSDM}), and the extended
SPARC data for the galaxy NGC3726. We included the rotation curves
of the gas, the disk velocities, the bulge (where present) along
with the SIDM model.} \label{extendedNGC3726}
\end{figure}

Now we shall include contributions to the rotation velocity from
the other components of the galaxy, namely the disk, the gas, and
the bulge if present. In Fig. \ref{extendedNGC3726} we present the
combined rotation curves including all the components of the
galaxy along with the SIDM. As it can be seen, the extended
collisional DM model is non-viable.
Also in Table \ref{evaluationextendedNGC3726} we present the
optimized values of the free parameters of the SIDM model for
which  we achieve the maximum compatibility with the SPARC data,
for the galaxy NGC3726, and also the resulting reduced
$\chi^2_{red}$ value.
\begin{table}[h!]
\centering \caption{Optimized Parameter Values of the Extended
SIDM model for the Galaxy NGC3726.}
\begin{tabular}{lc}
\hline
Parameter & Value  \\
\hline
$\rho_0 $ ($M_{\odot}/\mathrm{Kpc}^{3}$) & $1.7684\times 10^6$   \\
$K_0$ ($M_{\odot} \,
\mathrm{Kpc}^{-3} \, (\mathrm{km/s})^{2}$) & 15334.2   \\
$ml_{\text{disk}}$ & 0.8091 \\
$ml_{\text{bulge}}$ & 0.6023 \\
$\alpha$ (Kpc) & 53.7324\\
$\chi^2_{red}$ & 2.20829 \\
\hline
\end{tabular}
\label{evaluationextendedNGC3726}
\end{table}

\subsection{The Galaxy NGC3741, Non-viable, Extended Viable}

For this galaxy, the optimization method we used, ensures maximum
compatibility of the analytic SIDM model of Eq.
(\ref{ScaledependentEoSDM}) with the SPARC data, if we choose
$\rho_0=2.14808\times 10^7$$M_{\odot}/\mathrm{Kpc}^{3}$ and
$K_0=995.075
$$M_{\odot} \, \mathrm{Kpc}^{-3} \, (\mathrm{km/s})^{2}$, in which
case the reduced $\chi^2_{red}$ value is $\chi^2_{red}=4.39585$.
Also the parameter $\alpha$ in this case is $\alpha=3.92784 $Kpc.

\begin{figure}[h!]
\centering
\includegraphics[width=20pc]{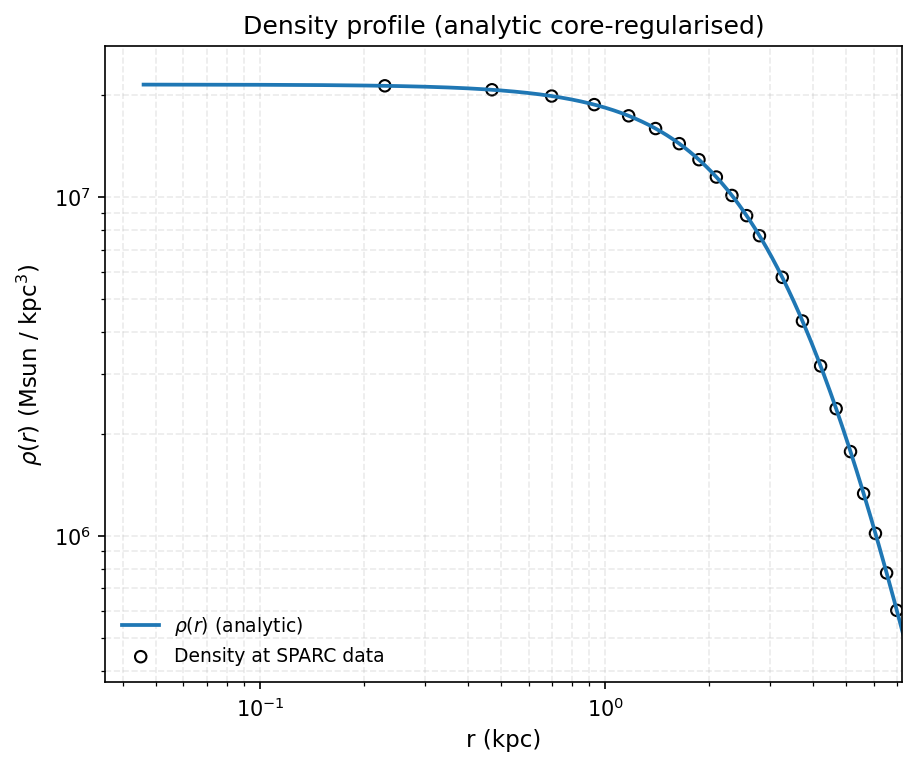}
\caption{The density of the SIDM model of Eq.
(\ref{ScaledependentEoSDM}) for the galaxy NGC3741, versus the
radius.} \label{NGC3741dens}
\end{figure}

\begin{figure}[h!]
\centering
\includegraphics[width=35pc]{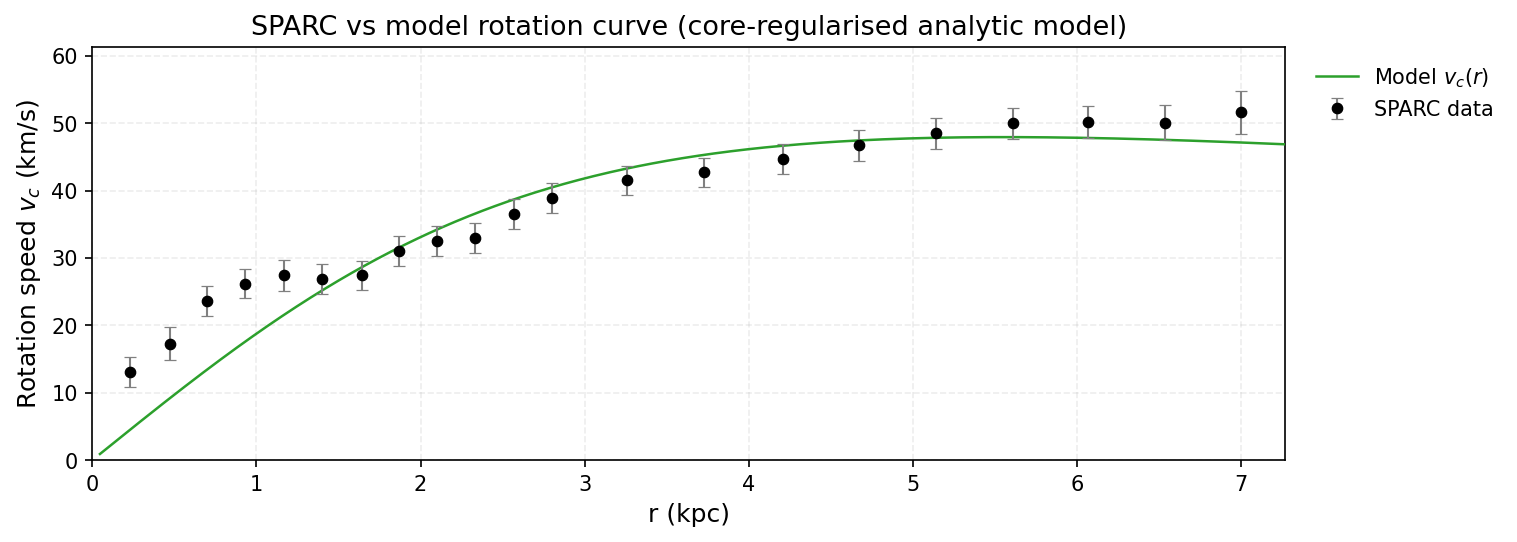}
\caption{The predicted rotation curves for the optimized SIDM
model of Eq. (\ref{ScaledependentEoSDM}), versus the SPARC
observational data for the galaxy NGC3741.} \label{NGC3741}
\end{figure}

In Table \ref{collNGC3741} we present the optimized values of
$K_0$ and $\rho_0$ for the analytic SIDM model of Eq.
(\ref{ScaledependentEoSDM}) for which the maximum compatibility
with the SPARC data is achieved.
\begin{table}[h!]
  \begin{center}
    \caption{SIDM Optimization Values for the galaxy NGC3741}
    \label{collNGC3741}
     \begin{tabular}{|r|r|}
     \hline
      \textbf{Parameter}   & \textbf{Optimization Values}
      \\  \hline
     $\rho_0 $  ($M_{\odot}/\mathrm{Kpc}^{3}$) & $2.14808\times 10^7$
\\  \hline $K_0$ ($M_{\odot} \,
\mathrm{Kpc}^{-3} \, (\mathrm{km/s})^{2}$)& 995.075
\\  \hline
    \end{tabular}
  \end{center}
\end{table}
In Figs. \ref{NGC3741dens}, \ref{NGC3741} we present the density
of the analytic SIDM model, the predicted rotation curves for the
SIDM model (\ref{ScaledependentEoSDM}), versus the SPARC
observational data and the sound speed, as a function of the
radius respectively. As it can be seen, for this galaxy, the SIDM
model produces non-viable rotation curves which are incompatible
with the SPARC data.

\begin{figure}[h!]
\centering
\includegraphics[width=20pc]{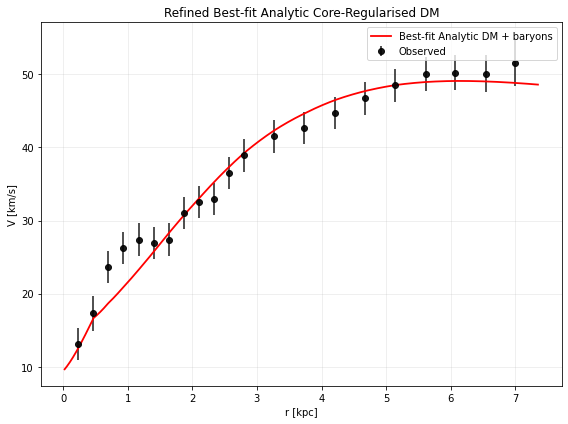}
\caption{The predicted rotation curves after using an optimization
for the SIDM model (\ref{ScaledependentEoSDM}), and the extended
SPARC data for the galaxy NGC3741. We included the rotation curves
of the gas, the disk velocities, the bulge (where present) along
with the SIDM model.} \label{extendedNGC3741}
\end{figure}

Now we shall include contributions to the rotation velocity from
the other components of the galaxy, namely the disk, the gas, and
the bulge if present. In Fig. \ref{extendedNGC3741} we present the
combined rotation curves including all the components of the
galaxy along with the SIDM. As it can be seen, the extended
collisional DM model is non-viable.
Also in Table \ref{evaluationextendedNGC3741} we present the
optimized values of the free parameters of the SIDM model for
which  we achieve the maximum compatibility with the SPARC data,
for the galaxy NGC3741, and also the resulting reduced
$\chi^2_{red}$ value.
\begin{table}[h!]
\centering \caption{Optimized Parameter Values of the Extended
SIDM model for the Galaxy NGC3741.}
\begin{tabular}{lc}
\hline
Parameter & Value  \\
\hline
$\rho_0 $ ($M_{\odot}/\mathrm{Kpc}^{3}$) & $1.58989\times 10^7$   \\
$K_0$ ($M_{\odot} \,
\mathrm{Kpc}^{-3} \, (\mathrm{km/s})^{2}$) & 953.988   \\
$ml_{\text{disk}}$ & 1.0000 \\
$ml_{\text{bulge}}$ & 0.46 \\
$\alpha$ (Kpc) & 4.46975\\
$\chi^2_{red}$ & 1.11831 \\
\hline
\end{tabular}
\label{evaluationextendedNGC3741}
\end{table}

\subsection{The Galaxy NGC3769, Non-viable, Extended Viable}

For this galaxy, the optimization method we used, ensures maximum
compatibility of the analytic SIDM model of Eq.
(\ref{ScaledependentEoSDM}) with the SPARC data, if we choose
$\rho_0=4.54798\times 10^7$$M_{\odot}/\mathrm{Kpc}^{3}$ and
$K_0=7542.73
$$M_{\odot} \, \mathrm{Kpc}^{-3} \, (\mathrm{km/s})^{2}$, in which
case the reduced $\chi^2_{red}$ value is $\chi^2_{red}=5.69727$.
Also the parameter $\alpha$ in this case is $\alpha=7.432 $Kpc.

\begin{figure}[h!]
\centering
\includegraphics[width=20pc]{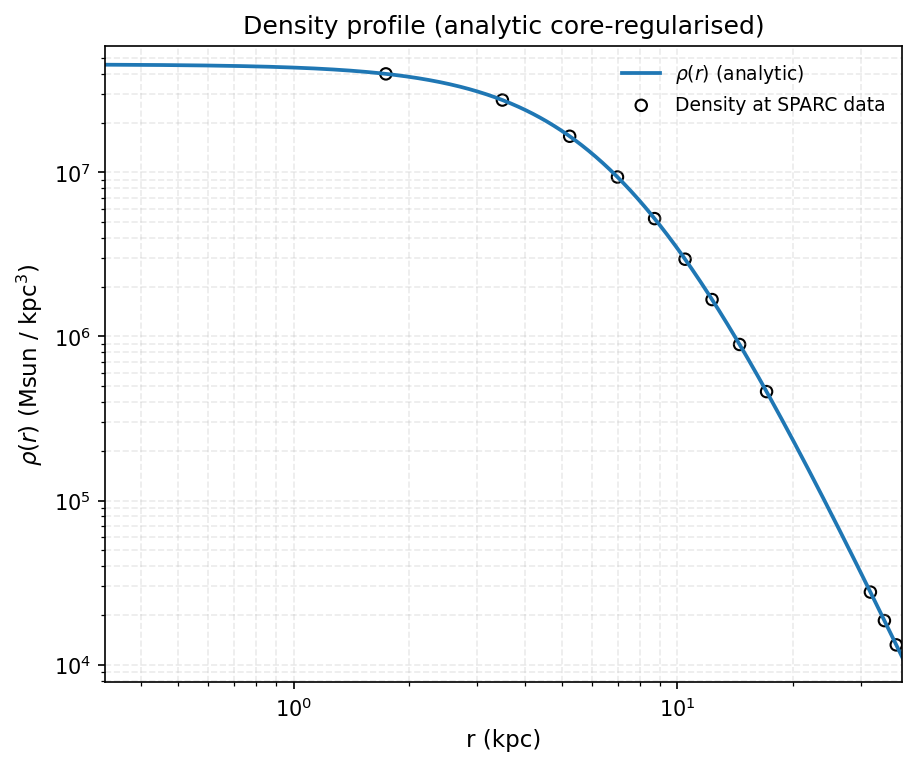}
\caption{The density of the SIDM model of Eq.
(\ref{ScaledependentEoSDM}) for the galaxy NGC3769, versus the
radius.} \label{NGC3769dens}
\end{figure}

\begin{figure}[h!]
\centering
\includegraphics[width=35pc]{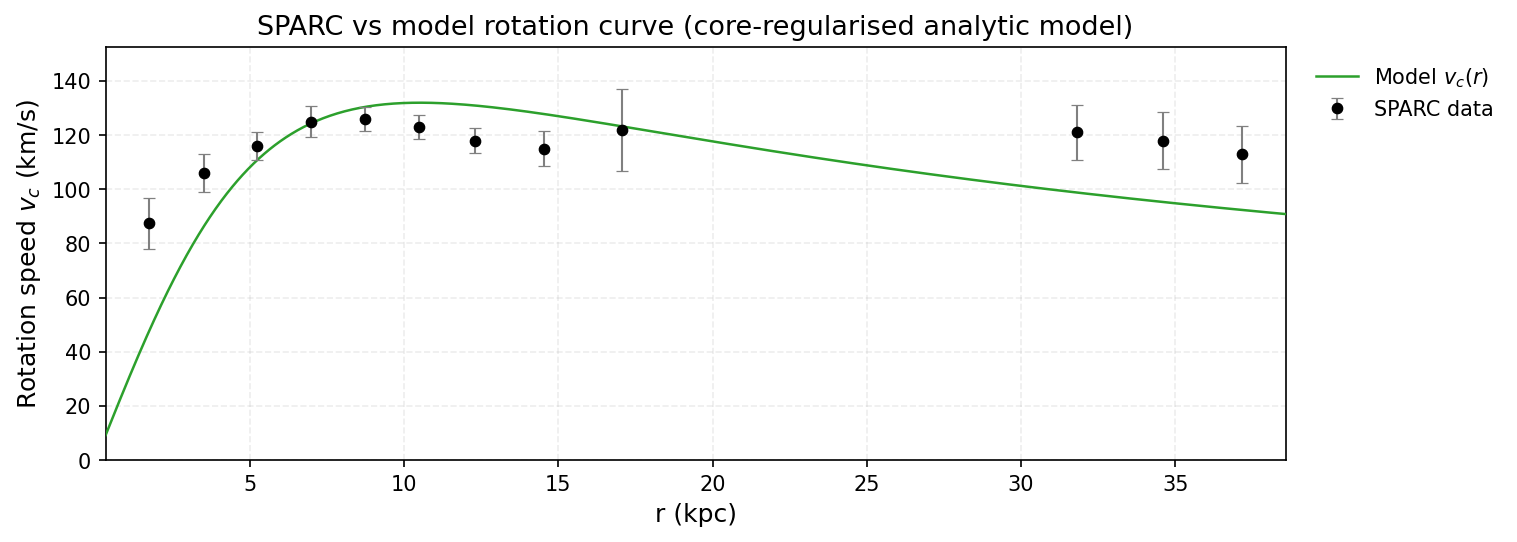}
\caption{The predicted rotation curves for the optimized SIDM
model of Eq. (\ref{ScaledependentEoSDM}), versus the SPARC
observational data for the galaxy NGC3769.} \label{NGC3769}
\end{figure}

In Table \ref{collNGC3769} we present the optimized values of
$K_0$ and $\rho_0$ for the analytic SIDM model of Eq.
(\ref{ScaledependentEoSDM}) for which the maximum compatibility
with the SPARC data is achieved.
\begin{table}[h!]
  \begin{center}
    \caption{SIDM Optimization Values for the galaxy NGC3769}
    \label{collNGC3769}
     \begin{tabular}{|r|r|}
     \hline
      \textbf{Parameter}   & \textbf{Optimization Values}
      \\  \hline
     $\rho_0 $  ($M_{\odot}/\mathrm{Kpc}^{3}$) & $4.54798\times 10^7$
\\  \hline $K_0$ ($M_{\odot} \,
\mathrm{Kpc}^{-3} \, (\mathrm{km/s})^{2}$)& 7542.73
\\  \hline
    \end{tabular}
  \end{center}
\end{table}
In Figs. \ref{NGC3769dens}, \ref{NGC3769} we present the density
of the analytic SIDM model, the predicted rotation curves for the
SIDM model (\ref{ScaledependentEoSDM}), versus the SPARC
observational data and the sound speed, as a function of the
radius respectively. As it can be seen, for this galaxy, the SIDM
model produces non-viable rotation curves which are incompatible
with the SPARC data.

\begin{figure}[h!]
\centering
\includegraphics[width=20pc]{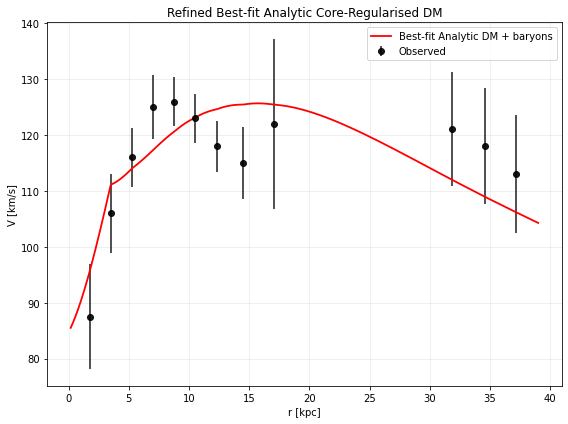}
\caption{The predicted rotation curves after using an optimization
for the SIDM model (\ref{ScaledependentEoSDM}), and the extended
SPARC data for the galaxy NGC3769. We included the rotation curves
of the gas, the disk velocities, the bulge (where present) along
with the SIDM model.} \label{extendedNGC3769}
\end{figure}

Now we shall include contributions to the rotation velocity from
the other components of the galaxy, namely the disk, the gas, and
the bulge if present. In Fig. \ref{extendedNGC3769} we present the
combined rotation curves including all the components of the
galaxy along with the SIDM. As it can be seen, the extended
collisional DM model is viable.
Also in Table \ref{evaluationextendedNGC3769} we present the
optimized values of the free parameters of the SIDM model for
which  we achieve the maximum compatibility with the SPARC data,
for the galaxy NGC3769, and also the resulting reduced
$\chi^2_{red}$ value.
\begin{table}[h!]
\centering \caption{Optimized Parameter Values of the Extended
SIDM model for the Galaxy NGC3769.}
\begin{tabular}{lc}
\hline
Parameter & Value  \\
\hline
$\rho_0 $ ($M_{\odot}/\mathrm{Kpc}^{3}$) & $1.02105\times 10^7$   \\
$K_0$ ($M_{\odot} \,
\mathrm{Kpc}^{-3} \, (\mathrm{km/s})^{2}$) & 5173.24   \\
$ml_{\text{disk}}$ & 0.7280 \\
$ml_{\text{bulge}}$ & 0.42 \\
$\alpha$ (Kpc) & 12.9884\\
$\chi^2_{red}$ & 1.44246 \\
\hline
\end{tabular}
\label{evaluationextendedNGC3769}
\end{table}

\subsection{The Galaxy NGC3877}

For this galaxy, the optimization method we used, ensures maximum
compatibility of the analytic SIDM model of Eq.
(\ref{ScaledependentEoSDM}) with the SPARC data, if we choose
$\rho_0=1.42995\times 10^8$$M_{\odot}/\mathrm{Kpc}^{3}$ and
$K_0=12511.1
$$M_{\odot} \, \mathrm{Kpc}^{-3} \, (\mathrm{km/s})^{2}$, in which
case the reduced $\chi^2_{red}$ value is $\chi^2_{red}=0.523333$.
Also the parameter $\alpha$ in this case is $\alpha=5.39806 $Kpc.

\begin{figure}[h!]
\centering
\includegraphics[width=20pc]{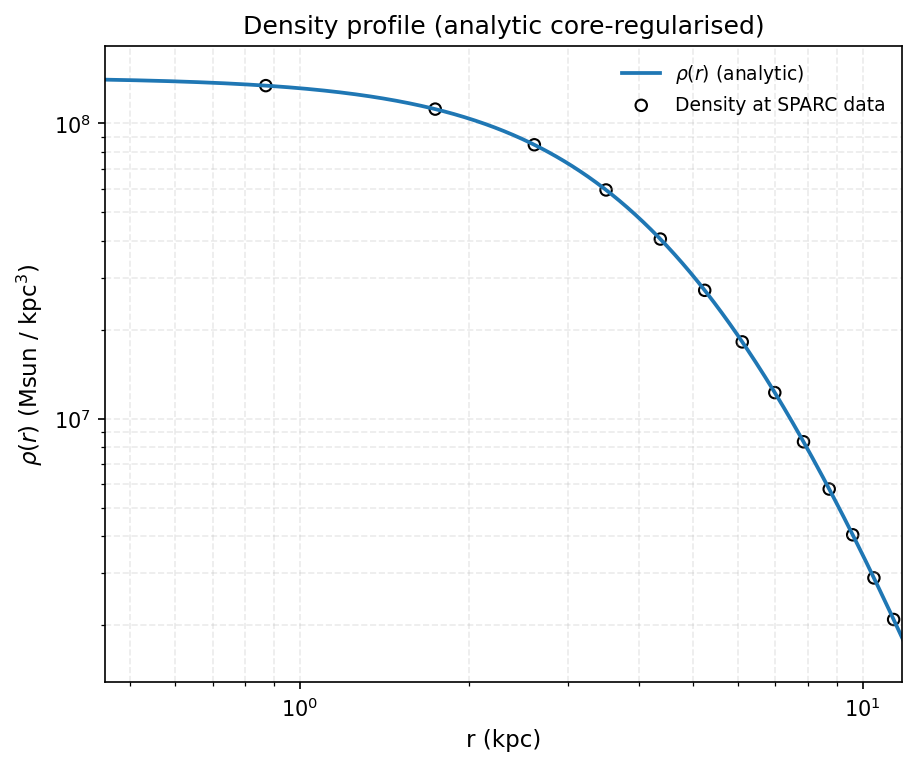}
\caption{The density of the SIDM model of Eq.
(\ref{ScaledependentEoSDM}) for the galaxy NGC3877, versus the
radius.} \label{NGC3877dens}
\end{figure}

\begin{figure}[h!]
\centering
\includegraphics[width=35pc]{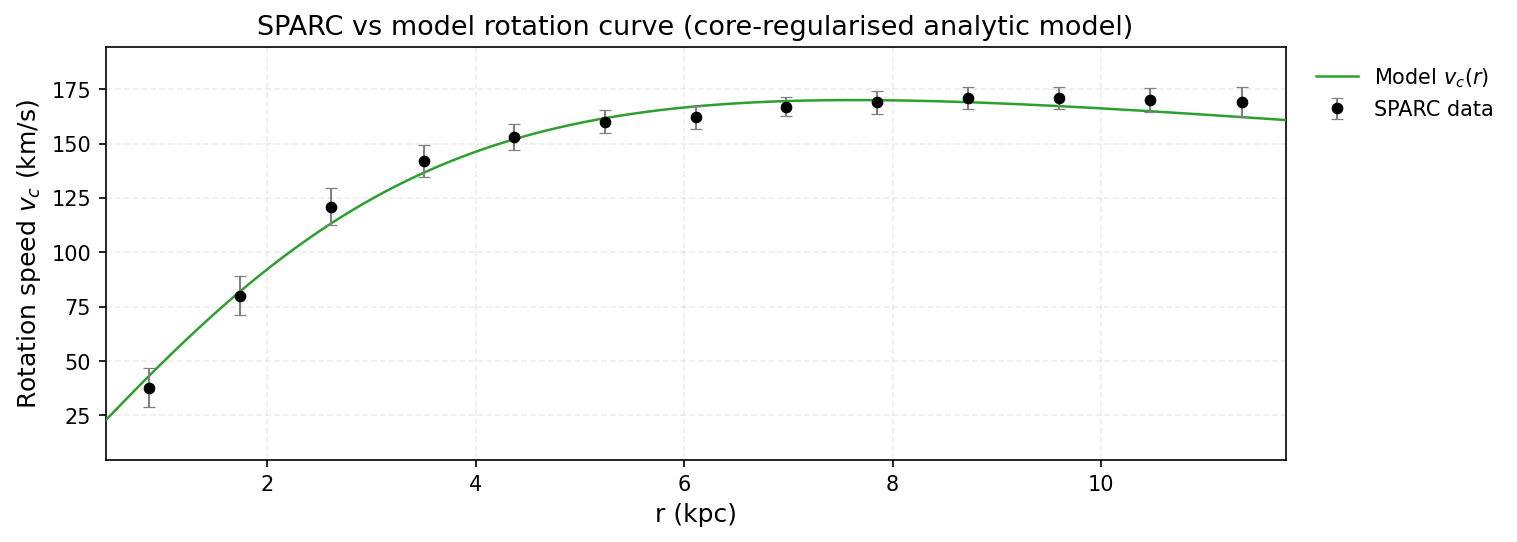}
\caption{The predicted rotation curves for the optimized SIDM
model of Eq. (\ref{ScaledependentEoSDM}), versus the SPARC
observational data for the galaxy NGC3877.} \label{NGC3877}
\end{figure}

In Table \ref{collNGC3877} we present the optimized values of
$K_0$ and $\rho_0$ for the analytic SIDM model of Eq.
(\ref{ScaledependentEoSDM}) for which the maximum compatibility
with the SPARC data is achieved.
\begin{table}[h!]
  \begin{center}
    \caption{SIDM Optimization Values for the galaxy NGC3877}
    \label{collNGC3877}
     \begin{tabular}{|r|r|}
     \hline
      \textbf{Parameter}   & \textbf{Optimization Values}
      \\  \hline
     $\rho_0 $  ($M_{\odot}/\mathrm{Kpc}^{3}$) & $1.42995\times 10^7$
\\  \hline $K_0$ ($M_{\odot} \,
\mathrm{Kpc}^{-3} \, (\mathrm{km/s})^{2}$)& 12511.1
\\  \hline
    \end{tabular}
  \end{center}
\end{table}
In Figs. \ref{NGC3877dens}, \ref{NGC3877} we present the density
of the analytic SIDM model, the predicted rotation curves for the
SIDM model (\ref{ScaledependentEoSDM}), versus the SPARC
observational data and the sound speed, as a function of the
radius respectively. As it can be seen, for this galaxy, the SIDM
model produces viable rotation curves which are compatible with
the SPARC data.

\subsection{The Galaxy NGC3893, Non-viable, Extended Viable}

For this galaxy, the optimization method we used, ensures maximum
compatibility of the analytic SIDM model of Eq.
(\ref{ScaledependentEoSDM}) with the SPARC data, if we choose
$\rho_0=2.88976\times 10^8$$M_{\odot}/\mathrm{Kpc}^{3}$ and
$K_0=17896.5
$$M_{\odot} \, \mathrm{Kpc}^{-3} \, (\mathrm{km/s})^{2}$, in which
case the reduced $\chi^2_{red}$ value is $\chi^2_{red}=4.67278$.
Also the parameter $\alpha$ in this case is $\alpha=4.54155 $Kpc.

\begin{figure}[h!]
\centering
\includegraphics[width=20pc]{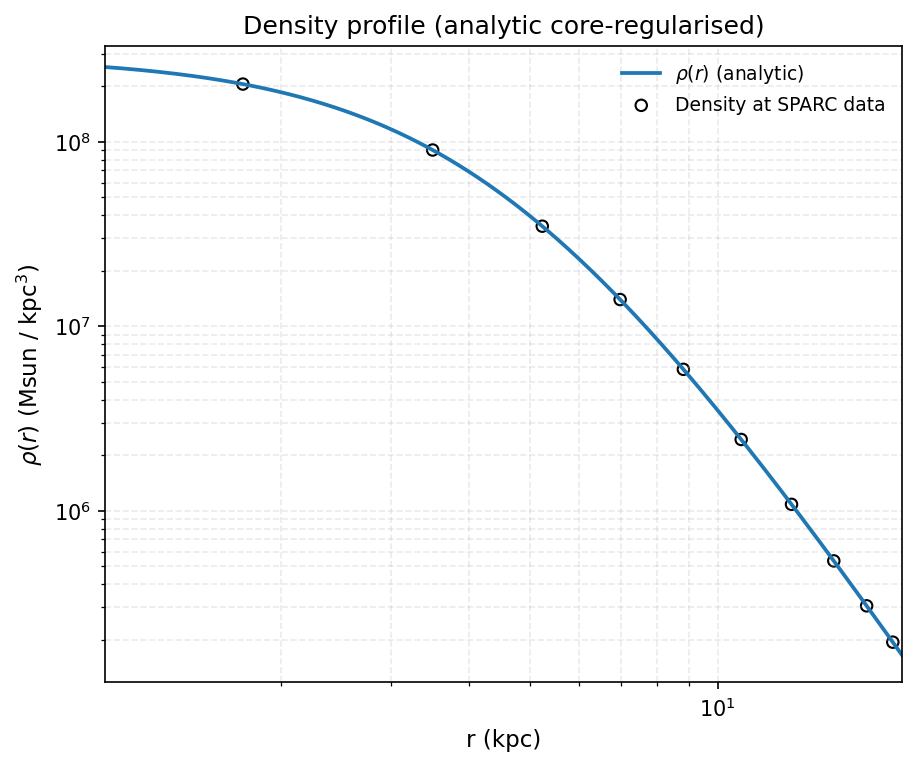}
\caption{The density of the SIDM model of Eq.
(\ref{ScaledependentEoSDM}) for the galaxy NGC3893, versus the
radius.} \label{NGC3893dens}
\end{figure}

\begin{figure}[h!]
\centering
\includegraphics[width=35pc]{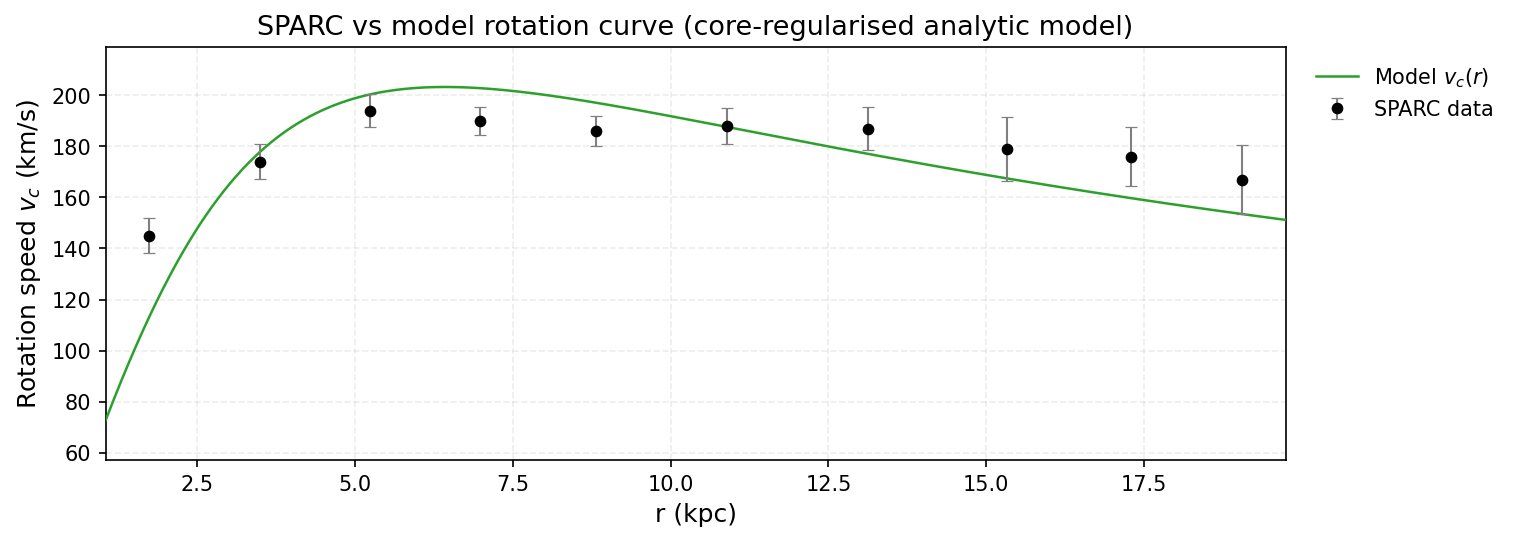}
\caption{The predicted rotation curves for the optimized SIDM
model of Eq. (\ref{ScaledependentEoSDM}), versus the SPARC
observational data for the galaxy NGC3893.} \label{NGC3893}
\end{figure}

In Table \ref{collNGC3893} we present the optimized values of
$K_0$ and $\rho_0$ for the analytic SIDM model of Eq.
(\ref{ScaledependentEoSDM}) for which the maximum compatibility
with the SPARC data is achieved.
\begin{table}[h!]
  \begin{center}
    \caption{SIDM Optimization Values for the galaxy NGC3893}
    \label{collNGC3893}
     \begin{tabular}{|r|r|}
     \hline
      \textbf{Parameter}   & \textbf{Optimization Values}
      \\  \hline
     $\rho_0 $  ($M_{\odot}/\mathrm{Kpc}^{3}$) & $2.88976\times 10^8$
\\  \hline $K_0$ ($M_{\odot} \,
\mathrm{Kpc}^{-3} \, (\mathrm{km/s})^{2}$)& 17896.5
\\  \hline
    \end{tabular}
  \end{center}
\end{table}
In Figs. \ref{NGC3893dens}, \ref{NGC3893} we present the density
of the analytic SIDM model, the predicted rotation curves for the
SIDM model (\ref{ScaledependentEoSDM}), versus the SPARC
observational data and the sound speed, as a function of the
radius respectively. As it can be seen, for this galaxy, the SIDM
model produces non-viable rotation curves which are incompatible
with the SPARC data.

\begin{figure}[h!]
\centering
\includegraphics[width=20pc]{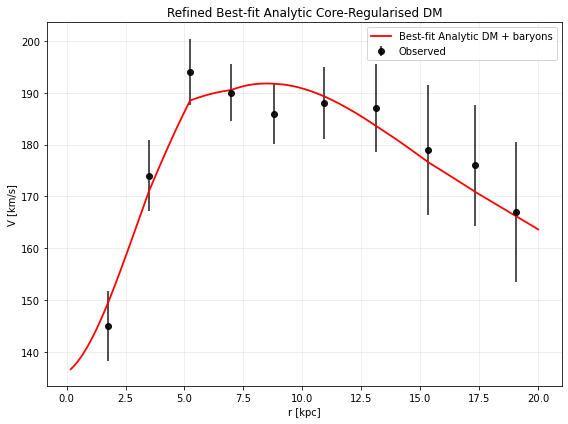}
\caption{The predicted rotation curves after using an optimization
for the SIDM model (\ref{ScaledependentEoSDM}), and the extended
SPARC data for the galaxy NGC3893. We included the rotation curves
of the gas, the disk velocities, the bulge (where present) along
with the SIDM model.} \label{extendedNGC3893}
\end{figure}

Now we shall include contributions to the rotation velocity from
the other components of the galaxy, namely the disk, the gas, and
the bulge if present. In Fig. \ref{extendedNGC3893} we present the
combined rotation curves including all the components of the
galaxy along with the SIDM. As it can be seen, the extended
collisional DM model is non-viable.
Also in Table \ref{evaluationextendedNGC3893} we present the
optimized values of the free parameters of the SIDM model for
which  we achieve the maximum compatibility with the SPARC data,
for the galaxy NGC3893, and also the resulting reduced
$\chi^2_{red}$ value.
\begin{table}[h!]
\centering \caption{Optimized Parameter Values of the Extended
SIDM model for the Galaxy NGC3893.}
\begin{tabular}{lc}
\hline
Parameter & Value  \\
\hline
$\rho_0 $ ($M_{\odot}/\mathrm{Kpc}^{3}$) & $4.35752\times 10^7$   \\
$K_0$ ($M_{\odot} \,
\mathrm{Kpc}^{-3} \, (\mathrm{km/s})^{2}$) & 9258.6   \\
$ml_{\text{disk}}$ & 0.6949 \\
$ml_{\text{bulge}}$ & 0.4635 \\
$\alpha$ (Kpc) & 8.41103\\
$\chi^2_{red}$ & 0.454429 \\
\hline
\end{tabular}
\label{evaluationextendedNGC3893}
\end{table}

\subsection{The Galaxy NGC3917}

For this galaxy, the optimization method we used, ensures maximum
compatibility of the analytic SIDM model of Eq.
(\ref{ScaledependentEoSDM}) with the SPARC data, if we choose
$\rho_0=5.60942\times 10^7$$M_{\odot}/\mathrm{Kpc}^{3}$ and
$K_0=8230.72
$$M_{\odot} \, \mathrm{Kpc}^{-3} \, (\mathrm{km/s})^{2}$, in which
case the reduced $\chi^2_{red}$ value is $\chi^2_{red}=0.617376$.
Also the parameter $\alpha$ in this case is $\alpha=6.99054 $Kpc.

\begin{figure}[h!]
\centering
\includegraphics[width=20pc]{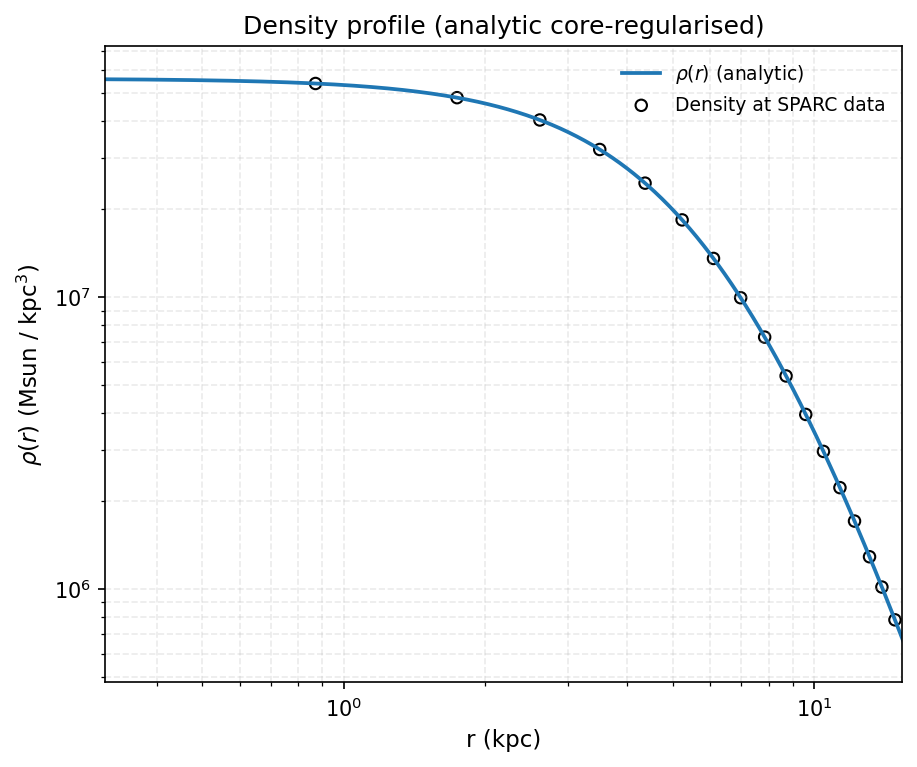}
\caption{The density of the SIDM model of Eq.
(\ref{ScaledependentEoSDM}) for the galaxy NGC3917, versus the
radius.} \label{NGC3917dens}
\end{figure}

\begin{figure}[h!]
\centering
\includegraphics[width=35pc]{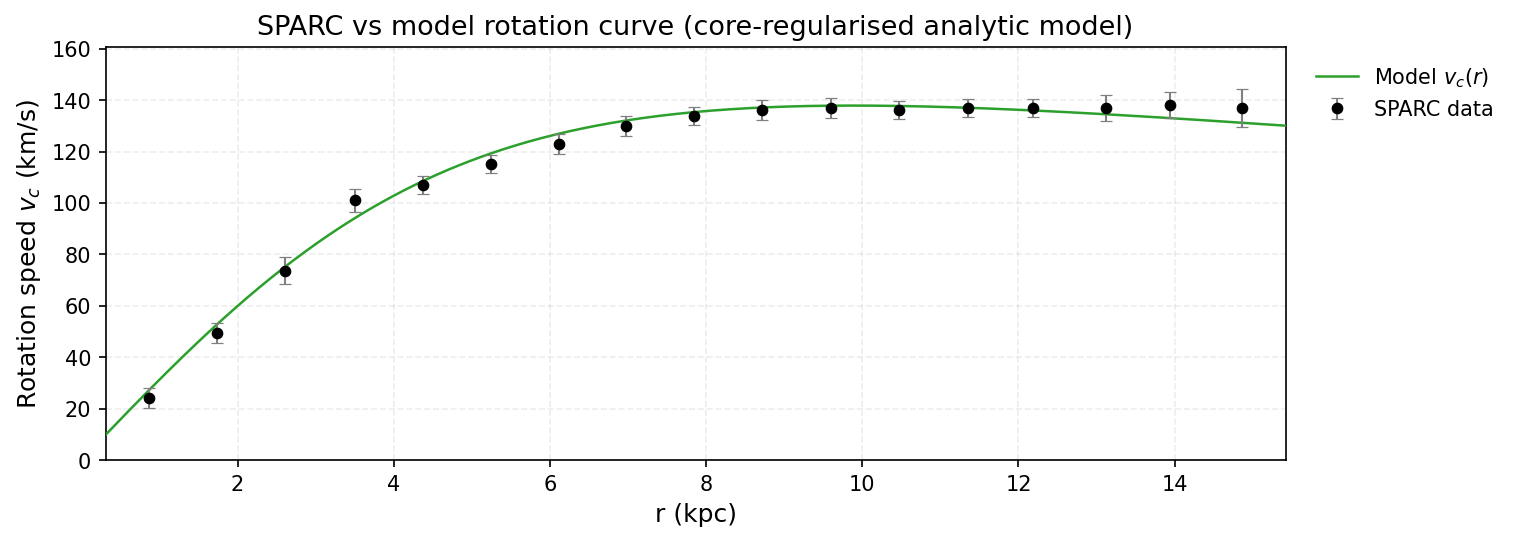}
\caption{The predicted rotation curves for the optimized SIDM
model of Eq. (\ref{ScaledependentEoSDM}), versus the SPARC
observational data for the galaxy NGC3917.} \label{NGC3917}
\end{figure}

In Table \ref{collNGC3917} we present the optimized values of
$K_0$ and $\rho_0$ for the analytic SIDM model of Eq.
(\ref{ScaledependentEoSDM}) for which the maximum compatibility
with the SPARC data is achieved.
\begin{table}[h!]
  \begin{center}
    \caption{SIDM Optimization Values for the galaxy NGC3917}
    \label{collNGC3917}
     \begin{tabular}{|r|r|}
     \hline
      \textbf{Parameter}   & \textbf{Optimization Values}
      \\  \hline
     $\rho_0 $  ($M_{\odot}/\mathrm{Kpc}^{3}$) & $5.60942\times 10^7$
\\  \hline $K_0$ ($M_{\odot} \,
\mathrm{Kpc}^{-3} \, (\mathrm{km/s})^{2}$)& 8230.72
\\  \hline
    \end{tabular}
  \end{center}
\end{table}
In Figs. \ref{NGC3917dens}, \ref{NGC3917} we present the density
of the analytic SIDM model, the predicted rotation curves for the
SIDM model (\ref{ScaledependentEoSDM}), versus the SPARC
observational data and the sound speed, as a function of the
radius respectively. As it can be seen, for this galaxy, the SIDM
model produces viable rotation curves which are compatible with
the SPARC data.

\subsection{The Galaxy NGC3949}

For this galaxy, the optimization method we used, ensures maximum
compatibility of the analytic SIDM model of Eq.
(\ref{ScaledependentEoSDM}) with the SPARC data, if we choose
$\rho_0=3.85388\times 10^8$$M_{\odot}/\mathrm{Kpc}^{3}$ and
$K_0=10995.9
$$M_{\odot} \, \mathrm{Kpc}^{-3} \, (\mathrm{km/s})^{2}$, in which
case the reduced $\chi^2_{red}$ value is $\chi^2_{red}=0.667849$.
Also the parameter $\alpha$ in this case is $\alpha=3.0826 $Kpc.

\begin{figure}[h!]
\centering
\includegraphics[width=20pc]{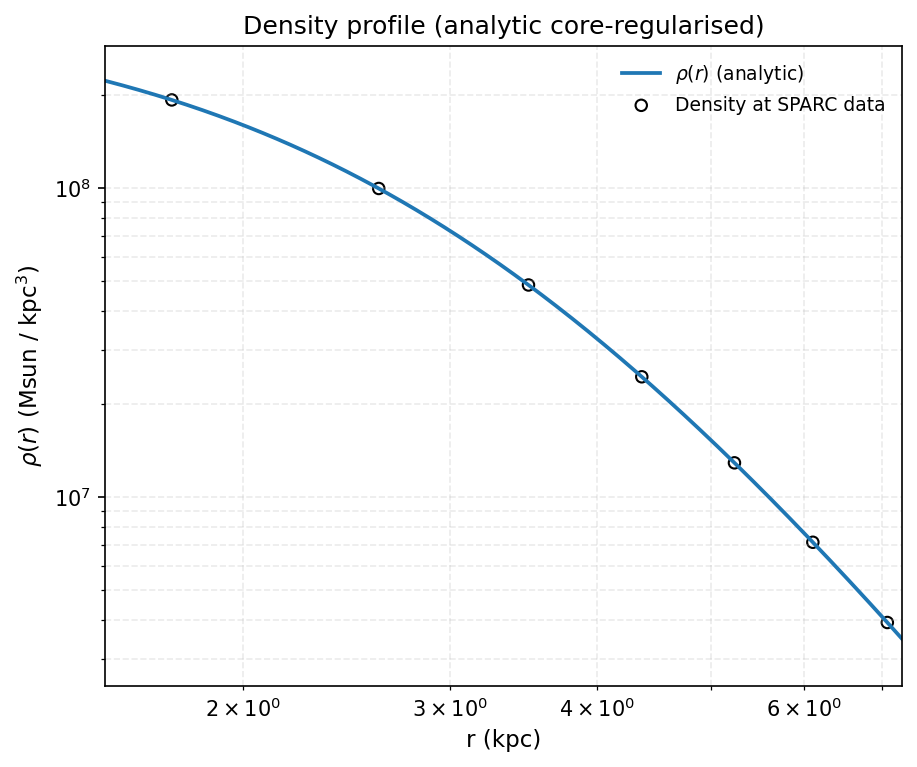}
\caption{The density of the SIDM model of Eq.
(\ref{ScaledependentEoSDM}) for the galaxy NGC3949, versus the
radius.} \label{NGC3949dens}
\end{figure}

\begin{figure}[h!]
\centering
\includegraphics[width=35pc]{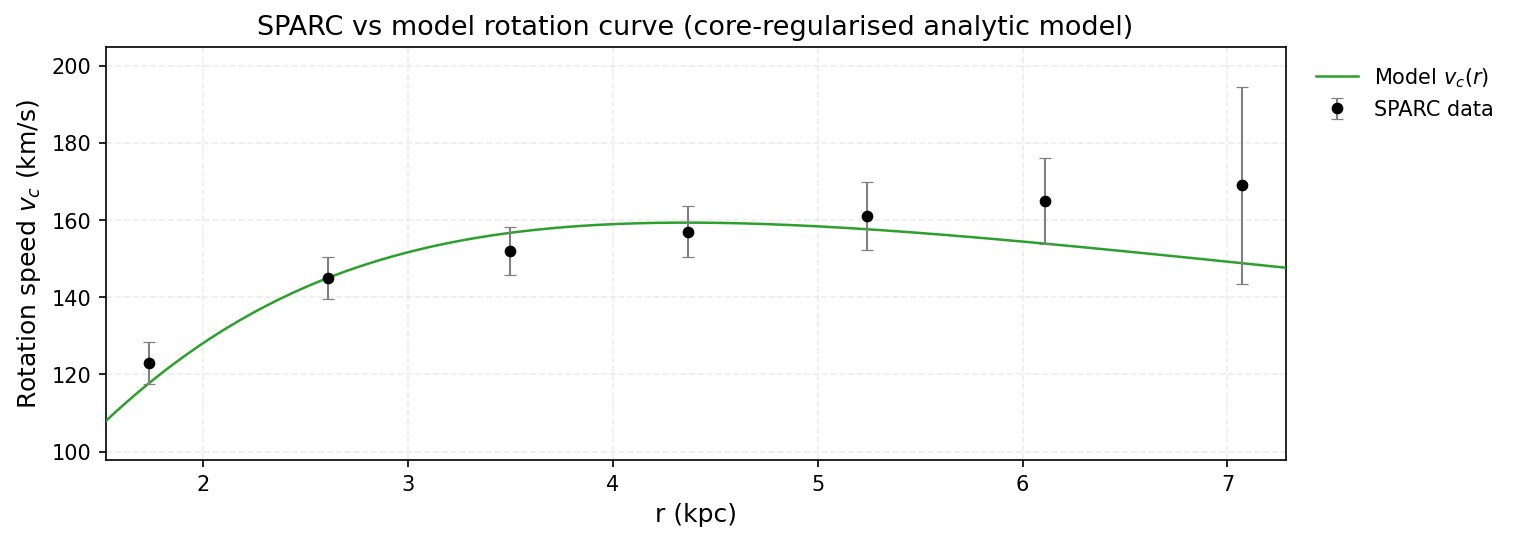}
\caption{The predicted rotation curves for the optimized SIDM
model of Eq. (\ref{ScaledependentEoSDM}), versus the SPARC
observational data for the galaxy NGC3949.} \label{NGC3949}
\end{figure}

In Table \ref{collNGC3949} we present the optimized values of
$K_0$ and $\rho_0$ for the analytic SIDM model of Eq.
(\ref{ScaledependentEoSDM}) for which the maximum compatibility
with the SPARC data is achieved.
\begin{table}[h!]
  \begin{center}
    \caption{SIDM Optimization Values for the galaxy NGC3949}
    \label{collNGC3949}
     \begin{tabular}{|r|r|}
     \hline
      \textbf{Parameter}   & \textbf{Optimization Values}
      \\  \hline
     $\rho_0 $  ($M_{\odot}/\mathrm{Kpc}^{3}$) & $3.85388\times 10^8$
\\  \hline $K_0$ ($M_{\odot} \,
\mathrm{Kpc}^{-3} \, (\mathrm{km/s})^{2}$)& 10995.9
\\  \hline
    \end{tabular}
  \end{center}
\end{table}
In Figs. \ref{NGC3949dens}, \ref{NGC3949} we present the density
of the analytic SIDM model, the predicted rotation curves for the
SIDM model (\ref{ScaledependentEoSDM}), versus the SPARC
observational data and the sound speed, as a function of the
radius respectively. As it can be seen, for this galaxy, the SIDM
model produces viable rotation curves which are compatible with
the SPARC data.

\subsection{The Galaxy NGC3953, Non-viable}

For this galaxy, the optimization method we used, ensures maximum
compatibility of the analytic SIDM model of Eq.
(\ref{ScaledependentEoSDM}) with the SPARC data, if we choose
$\rho_0=2.03016\times 10^8$$M_{\odot}/\mathrm{Kpc}^{3}$ and
$K_0=22728.9
$$M_{\odot} \, \mathrm{Kpc}^{-3} \, (\mathrm{km/s})^{2}$, in which
case the reduced $\chi^2_{red}$ value is $\chi^2_{red}=1.68893$.
Also the parameter $\alpha$ in this case is $\alpha=6.10626 $Kpc.

\begin{figure}[h!]
\centering
\includegraphics[width=20pc]{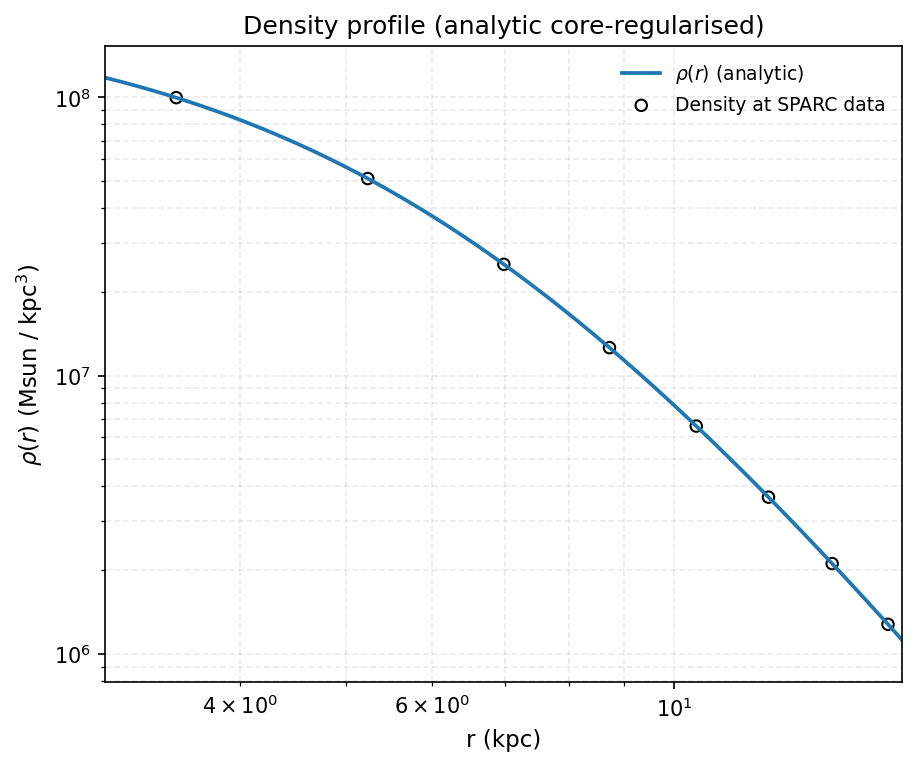}
\caption{The density of the SIDM model of Eq.
(\ref{ScaledependentEoSDM}) for the galaxy NGC3953, versus the
radius.} \label{NGC3953dens}
\end{figure}

\begin{figure}[h!]
\centering
\includegraphics[width=35pc]{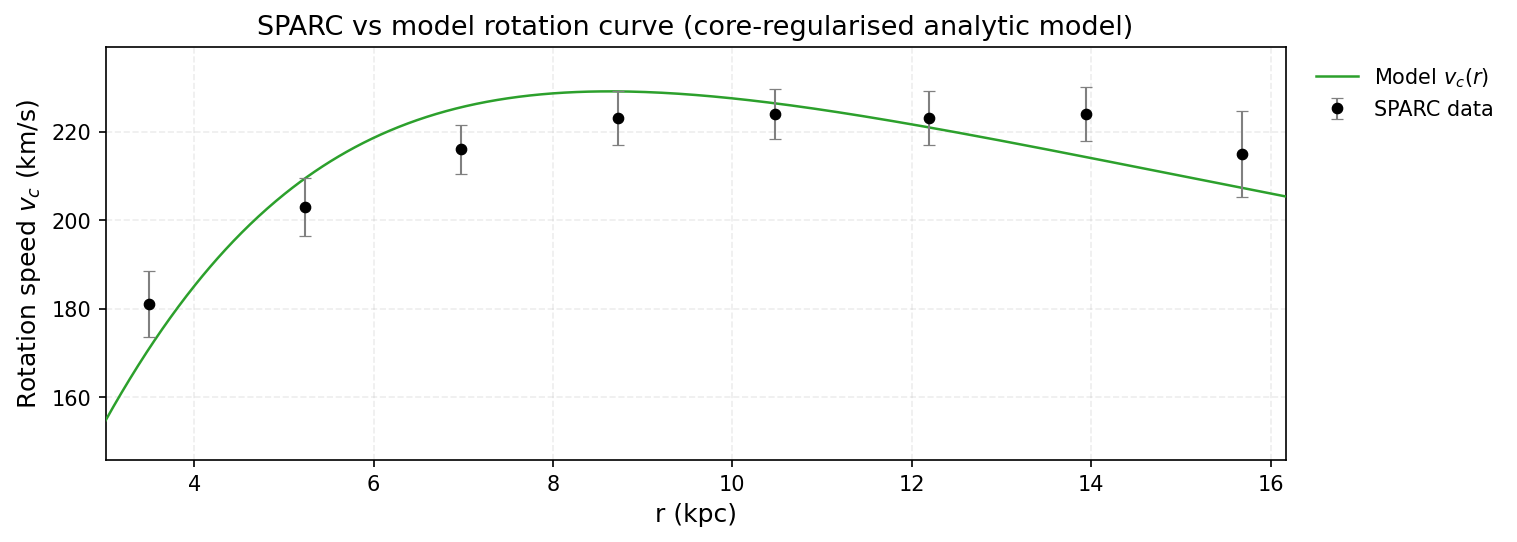}
\caption{The predicted rotation curves for the optimized SIDM
model of Eq. (\ref{ScaledependentEoSDM}), versus the SPARC
observational data for the galaxy NGC3953.} \label{NGC3953}
\end{figure}

In Table \ref{collNGC3953} we present the optimized values of
$K_0$ and $\rho_0$ for the analytic SIDM model of Eq.
(\ref{ScaledependentEoSDM}) for which the maximum compatibility
with the SPARC data is achieved.
\begin{table}[h!]
  \begin{center}
    \caption{SIDM Optimization Values for the galaxy NGC3953}
    \label{collNGC3953}
     \begin{tabular}{|r|r|}
     \hline
      \textbf{Parameter}   & \textbf{Optimization Values}
      \\  \hline
     $\rho_0 $  ($M_{\odot}/\mathrm{Kpc}^{3}$) & $2.03016\times 10^8$
\\  \hline $K_0$ ($M_{\odot} \,
\mathrm{Kpc}^{-3} \, (\mathrm{km/s})^{2}$)& 22728.9
\\  \hline
    \end{tabular}
  \end{center}
\end{table}
In Figs. \ref{NGC3953dens}, \ref{NGC3953} we present the density
of the analytic SIDM model, the predicted rotation curves for the
SIDM model (\ref{ScaledependentEoSDM}), versus the SPARC
observational data and the sound speed, as a function of the
radius respectively. As it can be seen, for this galaxy, the SIDM
model produces non-viable rotation curves which are incompatible
with the SPARC data.


\subsection{The Galaxy NGC3972}

For this galaxy, the optimization method we used, ensures maximum
compatibility of the analytic SIDM model of Eq.
(\ref{ScaledependentEoSDM}) with the SPARC data, if we choose
$\rho_0=7.82466\times 10^7$$M_{\odot}/\mathrm{Kpc}^{3}$ and
$K_0=7267.09
$$M_{\odot} \, \mathrm{Kpc}^{-3} \, (\mathrm{km/s})^{2}$, in which
case the reduced $\chi^2_{red}$ value is $\chi^2_{red}=1.70186$.
Also the parameter $\alpha$ in this case is $\alpha=5.56158 $Kpc.

\begin{figure}[h!]
\centering
\includegraphics[width=20pc]{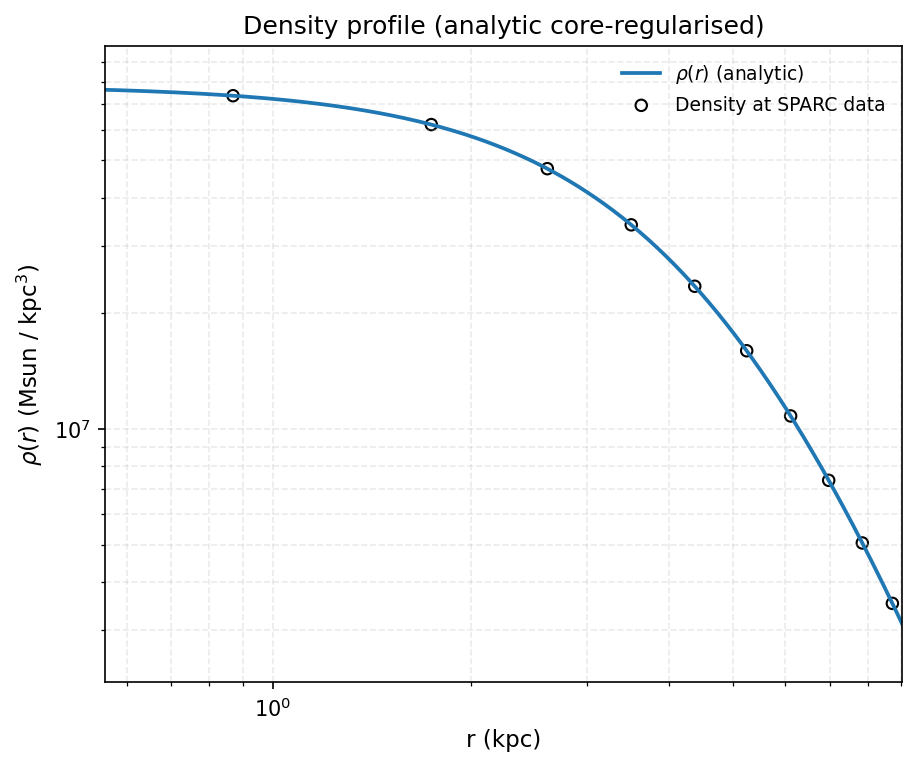}
\caption{The density of the SIDM model of Eq.
(\ref{ScaledependentEoSDM}) for the galaxy NGC3972, versus the
radius.} \label{NGC3972dens}
\end{figure}

\begin{figure}[h!]
\centering
\includegraphics[width=35pc]{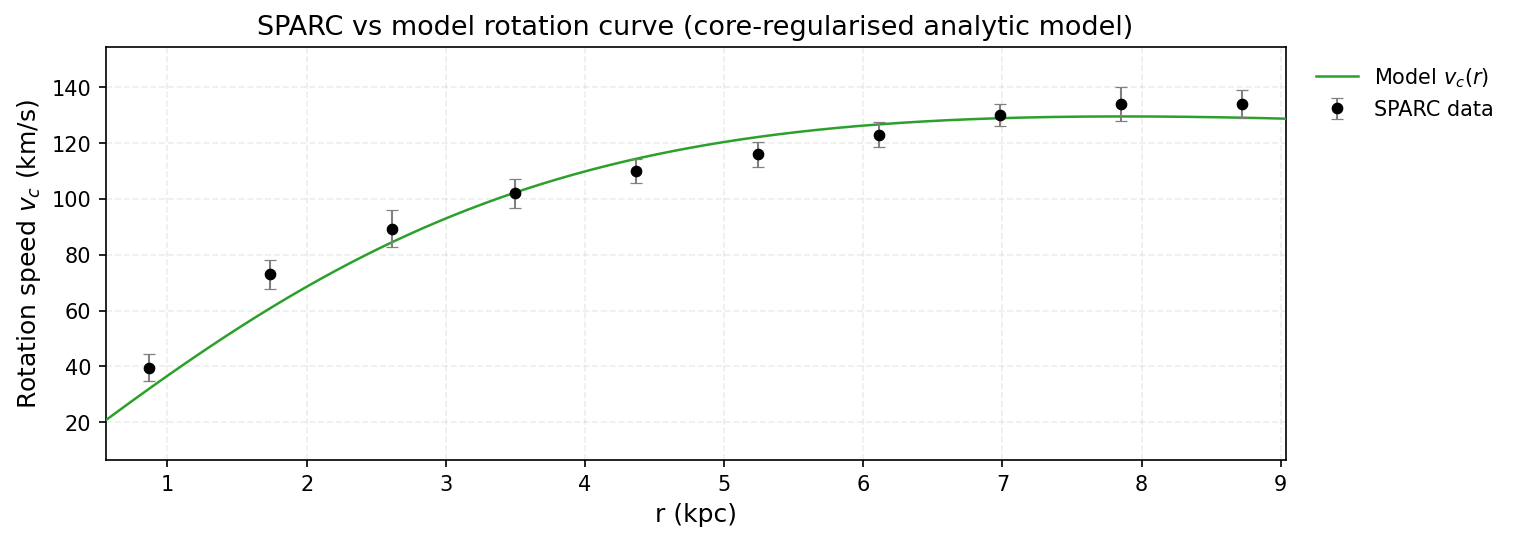}
\caption{The predicted rotation curves for the optimized SIDM
model of Eq. (\ref{ScaledependentEoSDM}), versus the SPARC
observational data for the galaxy NGC3972.} \label{NGC3972}
\end{figure}

In Table \ref{collNGC3972} we present the optimized values of
$K_0$ and $\rho_0$ for the analytic SIDM model of Eq.
(\ref{ScaledependentEoSDM}) for which the maximum compatibility
with the SPARC data is achieved.
\begin{table}[h!]
  \begin{center}
    \caption{SIDM Optimization Values for the galaxy NGC3972}
    \label{collNGC3972}
     \begin{tabular}{|r|r|}
     \hline
      \textbf{Parameter}   & \textbf{Optimization Values}
      \\  \hline
     $\rho_0 $  ($M_{\odot}/\mathrm{Kpc}^{3}$) & $7.82466\times 10^7$
\\  \hline $K_0$ ($M_{\odot} \,
\mathrm{Kpc}^{-3} \, (\mathrm{km/s})^{2}$)& 7267.09
\\  \hline
    \end{tabular}
  \end{center}
\end{table}
In Figs. \ref{NGC3972dens}, \ref{NGC3972} we present the density
of the analytic SIDM model, the predicted rotation curves for the
SIDM model (\ref{ScaledependentEoSDM}), versus the SPARC
observational data and the sound speed, as a function of the
radius respectively. As it can be seen, for this galaxy, the SIDM
model produces non-viable rotation curves which are incompatible
with the SPARC data.

\begin{figure}[h!]
\centering
\includegraphics[width=20pc]{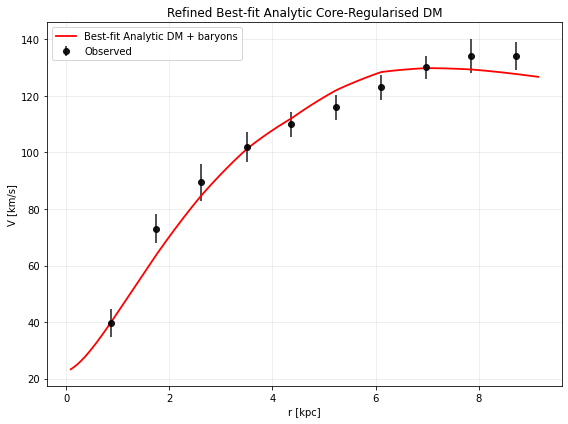}
\caption{The predicted rotation curves after using an optimization
for the SIDM model (\ref{ScaledependentEoSDM}), and the extended
SPARC data for the galaxy NGC3972. We included the rotation curves
of the gas, the disk velocities, the bulge (where present) along
with the SIDM model.} \label{extendedNGC3972}
\end{figure}

Now we shall include contributions to the rotation velocity from
the other components of the galaxy, namely the disk, the gas, and
the bulge if present. In Fig. \ref{extendedNGC3972} we present the
combined rotation curves including all the components of the
galaxy along with the SIDM. As it can be seen, the extended
collisional DM model is non-viable.
Also in Table \ref{evaluationextendedNGC3972} we present the
optimized values of the free parameters of the SIDM model for
which  we achieve the maximum compatibility with the SPARC data,
for the galaxy NGC3972, and also the resulting reduced
$\chi^2_{red}$ value.
\begin{table}[h!]
\centering \caption{Optimized Parameter Values of the Extended
SIDM model for the Galaxy NGC3972.}
\begin{tabular}{lc}
\hline
Parameter & Value  \\
\hline
$\rho_0 $ ($M_{\odot}/\mathrm{Kpc}^{3}$) & $5.40336\times 10^7$   \\
$K_0$ ($M_{\odot} \,
\mathrm{Kpc}^{-3} \, (\mathrm{km/s})^{2}$) & 5279.85   \\
$ml_{\text{disk}}$ & 0.6178 \\
$ml_{\text{bulge}}$ & 0.0361 \\
$\alpha$ (Kpc) & 5.70394\\
$\chi^2_{red}$ & 1.60993 \\
\hline
\end{tabular}
\label{evaluationextendedNGC3972}
\end{table}

\subsection{The Galaxy NGC3992, Marginally}

For this galaxy, the optimization method we used, ensures maximum
compatibility of the analytic SIDM model of Eq.
(\ref{ScaledependentEoSDM}) with the SPARC data, if we choose
$\rho_0=6.59039\times 10^7$$M_{\odot}/\mathrm{Kpc}^{3}$ and
$K_0=33644.4
$$M_{\odot} \, \mathrm{Kpc}^{-3} \, (\mathrm{km/s})^{2}$, in which
case the reduced $\chi^2_{red}$ value is $\chi^2_{red}=1.84261$.
Also the parameter $\alpha$ in this case is $\alpha=13.0392 $Kpc.

\begin{figure}[h!]
\centering
\includegraphics[width=20pc]{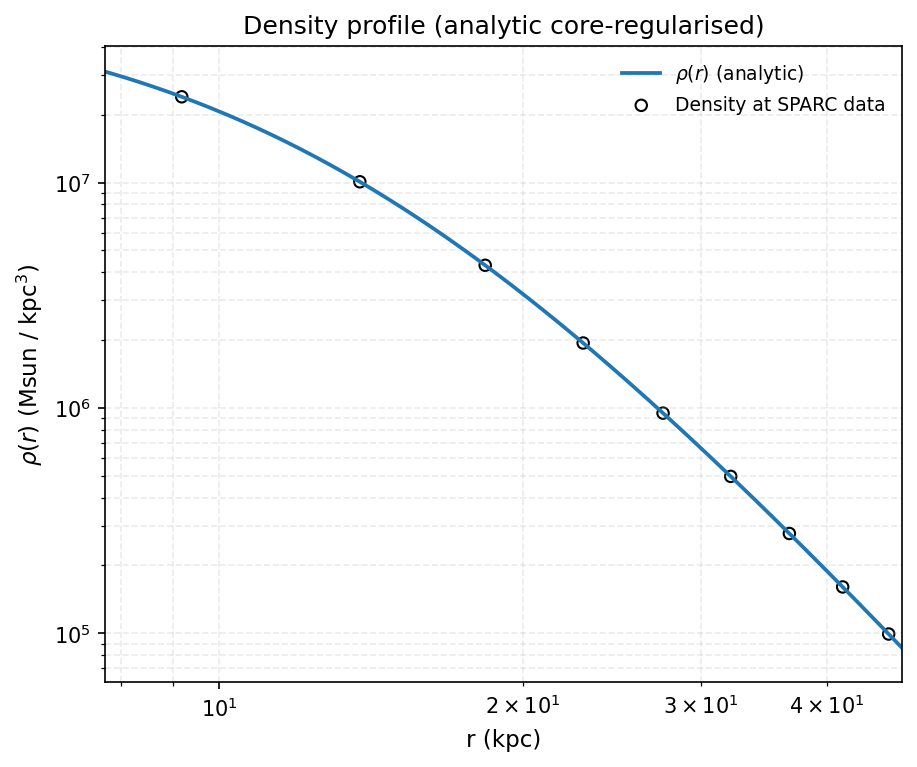}
\caption{The density of the SIDM model of Eq.
(\ref{ScaledependentEoSDM}) for the galaxy NGC3992, versus the
radius.} \label{NGC3992dens}
\end{figure}

\begin{figure}[h!]
\centering
\includegraphics[width=35pc]{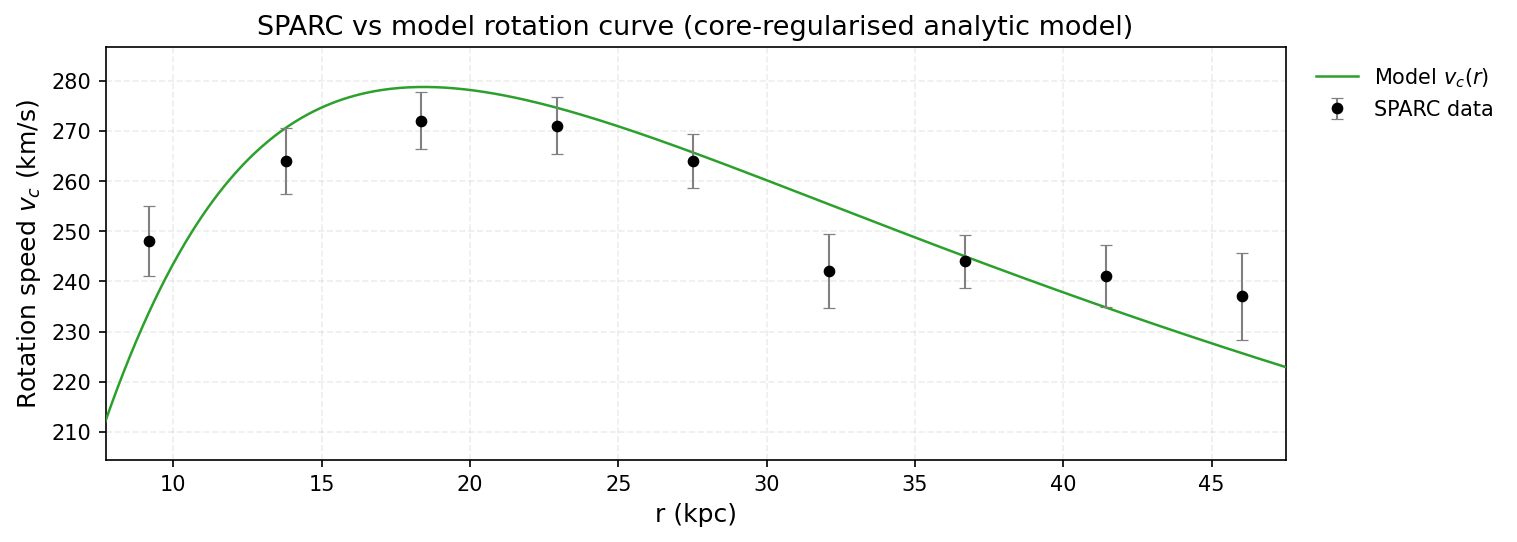}
\caption{The predicted rotation curves for the optimized SIDM
model of Eq. (\ref{ScaledependentEoSDM}), versus the SPARC
observational data for the galaxy NGC3992.} \label{NGC3992}
\end{figure}

In Table \ref{collNGC3992} we present the optimized values of
$K_0$ and $\rho_0$ for the analytic SIDM model of Eq.
(\ref{ScaledependentEoSDM}) for which the maximum compatibility
with the SPARC data is achieved.
\begin{table}[h!]
  \begin{center}
    \caption{SIDM Optimization Values for the galaxy NGC3992}
    \label{collNGC3992}
     \begin{tabular}{|r|r|}
     \hline
      \textbf{Parameter}   & \textbf{Optimization Values}
      \\  \hline
     $\rho_0 $  ($M_{\odot}/\mathrm{Kpc}^{3}$) & $6.59039\times 10^7$
\\  \hline $K_0$ ($M_{\odot} \,
\mathrm{Kpc}^{-3} \, (\mathrm{km/s})^{2}$)& 33644.4
\\  \hline
    \end{tabular}
  \end{center}
\end{table}
In Figs. \ref{NGC3992dens}, \ref{NGC3992} we present the density
of the analytic SIDM model, the predicted rotation curves for the
SIDM model (\ref{ScaledependentEoSDM}), versus the SPARC
observational data and the sound speed, as a function of the
radius respectively. As it can be seen, for this galaxy, the SIDM
model produces marginally viable rotation curves which are
incompatible with the SPARC data.

\begin{figure}[h!]
\centering
\includegraphics[width=20pc]{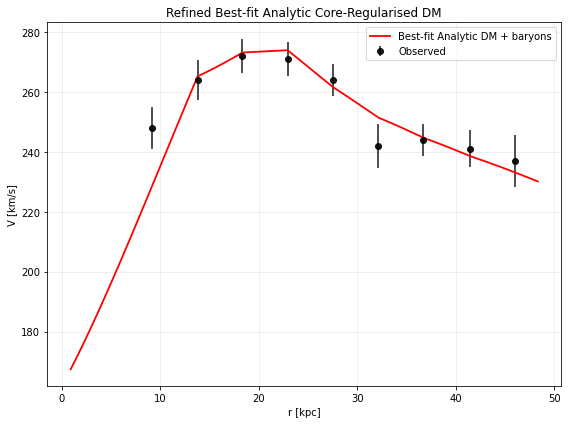}
\caption{The predicted rotation curves after using an optimization
for the SIDM model (\ref{ScaledependentEoSDM}), and the extended
SPARC data for the galaxy NGC3992. We included the rotation curves
of the gas, the disk velocities, the bulge (where present) along
with the SIDM model.} \label{extendedNGC3992}
\end{figure}

Now we shall include contributions to the rotation velocity from
the other components of the galaxy, namely the disk, the gas, and
the bulge if present. In Fig. \ref{extendedNGC3992} we present the
combined rotation curves including all the components of the
galaxy along with the SIDM. As it can be seen, the extended
collisional DM model is marginally viable.
Also in Table \ref{evaluationextendedNGC3992} we present the
optimized values of the free parameters of the SIDM model for
which  we achieve the maximum compatibility with the SPARC data,
for the galaxy NGC3992, and also the resulting reduced
$\chi^2_{red}$ value.
\begin{table}[h!]
\centering \caption{Optimized Parameter Values of the Extended
SIDM model for the Galaxy NGC3992.}
\begin{tabular}{lc}
\hline
Parameter & Value  \\
\hline
$\rho_0 $ ($M_{\odot}/\mathrm{Kpc}^{3}$) & $3.41415\times 10^6$   \\
$K_0$ ($M_{\odot} \,
\mathrm{Kpc}^{-3} \, (\mathrm{km/s})^{2}$) & 12823.2   \\
$ml_{\text{disk}}$ & 1 \\
$ml_{\text{bulge}}$ & 0.8274 \\
$\alpha$ (Kpc) & 35.3632\\
$\chi^2_{red}$ & 2.03714  \\
\hline
\end{tabular}
\label{evaluationextendedNGC3992}
\end{table}

\subsection{The Galaxy NGC4010}

For this galaxy, the optimization method we used, ensures maximum
compatibility of the analytic SIDM model of Eq.
(\ref{ScaledependentEoSDM}) with the SPARC data, if we choose
$\rho_0=4.92231\times 10^7$$M_{\odot}/\mathrm{Kpc}^{3}$ and
$K_0=6892.94
$$M_{\odot} \, \mathrm{Kpc}^{-3} \, (\mathrm{km/s})^{2}$, in which
case the reduced $\chi^2_{red}$ value is $\chi^2_{red}=1.17299$.
Also the parameter $\alpha$ in this case is $\alpha=6.82918 $Kpc.

\begin{figure}[h!]
\centering
\includegraphics[width=20pc]{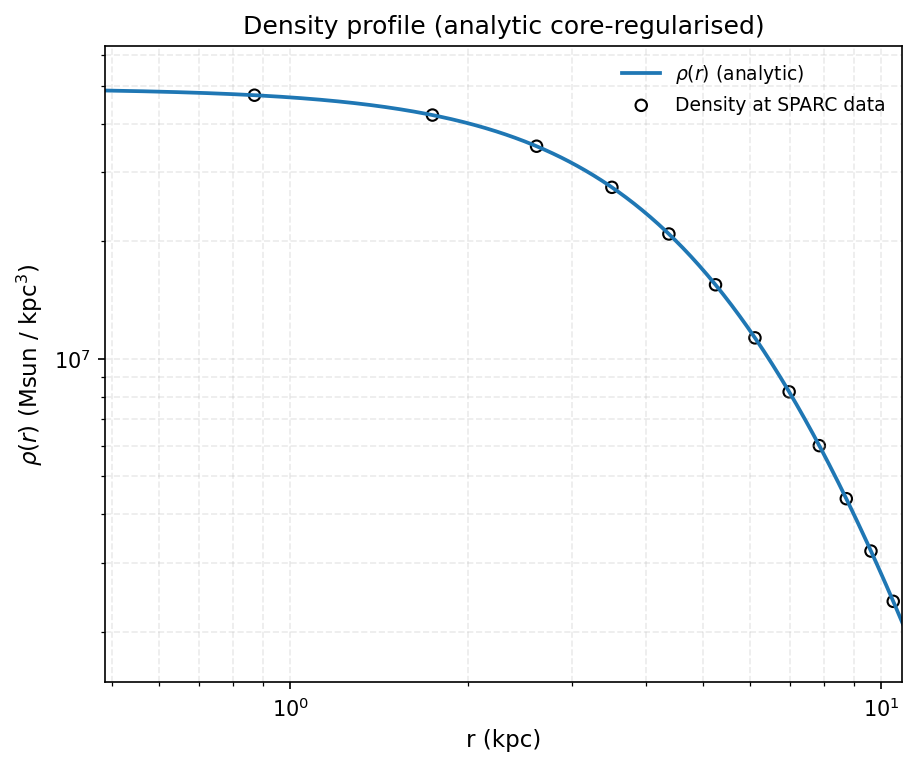}
\caption{The density of the SIDM model of Eq.
(\ref{ScaledependentEoSDM}) for the galaxy NGC4010, versus the
radius.} \label{NGC4010dens}
\end{figure}

\begin{figure}[h!]
\centering
\includegraphics[width=35pc]{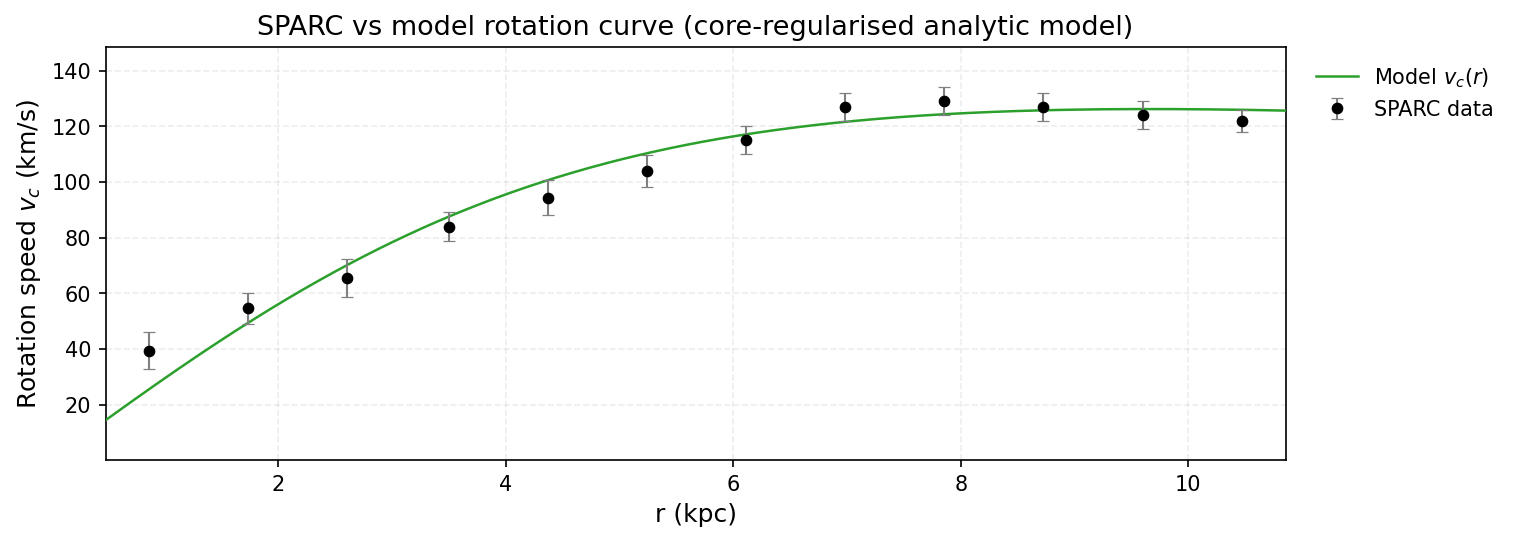}
\caption{The predicted rotation curves for the optimized SIDM
model of Eq. (\ref{ScaledependentEoSDM}), versus the SPARC
observational data for the galaxy NGC4010.} \label{NGC4010}
\end{figure}

In Table \ref{collNGC4010} we present the optimized values of
$K_0$ and $\rho_0$ for the analytic SIDM model of Eq.
(\ref{ScaledependentEoSDM}) for which the maximum compatibility
with the SPARC data is achieved.
\begin{table}[h!]
  \begin{center}
    \caption{SIDM Optimization Values for the galaxy NGC4010}
    \label{collNGC4010}
     \begin{tabular}{|r|r|}
     \hline
      \textbf{Parameter}   & \textbf{Optimization Values}
      \\  \hline
     $\rho_0 $  ($M_{\odot}/\mathrm{Kpc}^{3}$) & $4.92231\times 10^7$
\\  \hline $K_0$ ($M_{\odot} \,
\mathrm{Kpc}^{-3} \, (\mathrm{km/s})^{2}$)& 6892.94
\\  \hline
    \end{tabular}
  \end{center}
\end{table}
In Figs. \ref{NGC4010dens}, \ref{NGC4010} we present the density
of the analytic SIDM model, the predicted rotation curves for the
SIDM model (\ref{ScaledependentEoSDM}), versus the SPARC
observational data and the sound speed, as a function of the
radius respectively. As it can be seen, for this galaxy, the SIDM
model produces viable rotation curves which are compatible with
the SPARC data.

\subsection{The Galaxy NGC4013, Non-viable, Extended Viable}

For this galaxy, the optimization method we used, ensures maximum
compatibility of the analytic SIDM model of Eq.
(\ref{ScaledependentEoSDM}) with the SPARC data, if we choose
$\rho_0=1.10219\times 10^8$$M_{\odot}/\mathrm{Kpc}^{3}$ and
$K_0=16800.3
$$M_{\odot} \, \mathrm{Kpc}^{-3} \, (\mathrm{km/s})^{2}$, in which
case the reduced $\chi^2_{red}$ value is $\chi^2_{red}=8.57117$.
Also the parameter $\alpha$ in this case is $\alpha=7.12495 $Kpc.

\begin{figure}[h!]
\centering
\includegraphics[width=20pc]{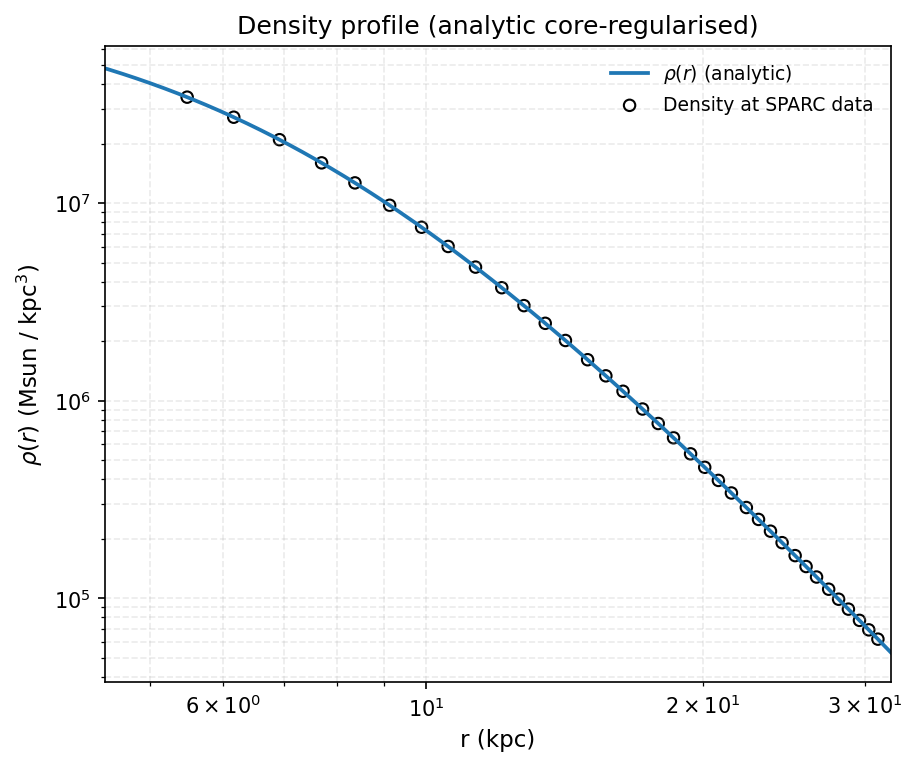}
\caption{The density of the SIDM model of Eq.
(\ref{ScaledependentEoSDM}) for the galaxy NGC4013, versus the
radius.} \label{NGC4013dens}
\end{figure}

\begin{figure}[h!]
\centering
\includegraphics[width=35pc]{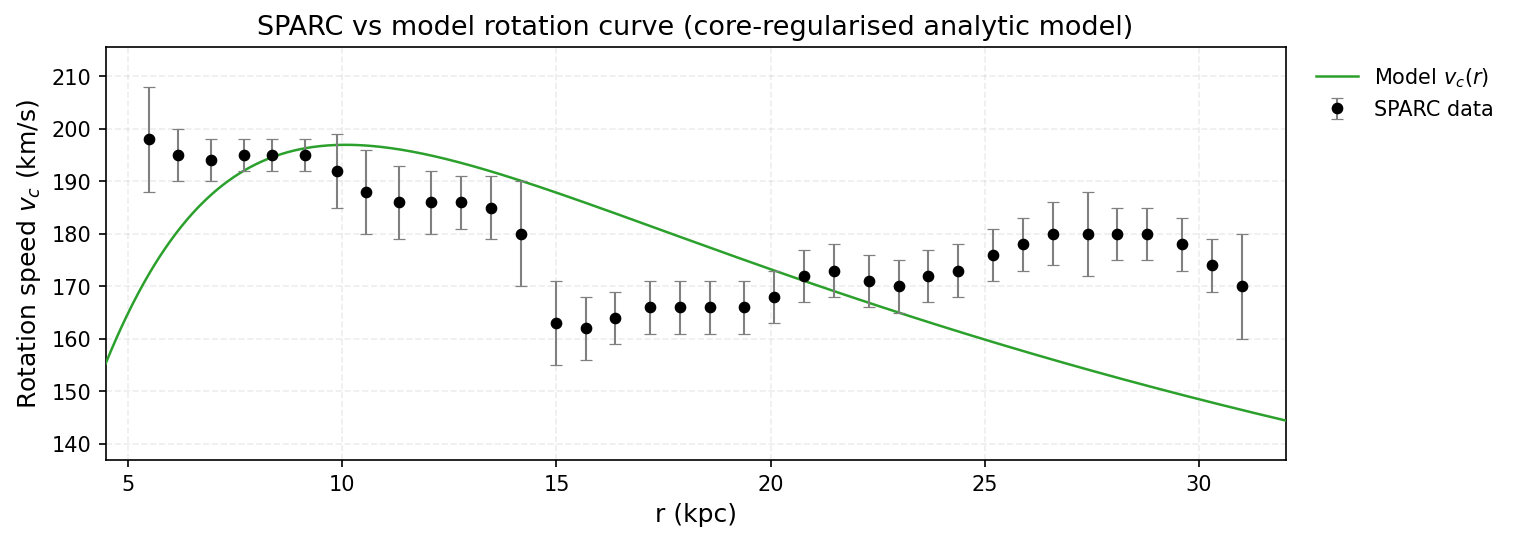}
\caption{The predicted rotation curves for the optimized SIDM
model of Eq. (\ref{ScaledependentEoSDM}), versus the SPARC
observational data for the galaxy NGC4013.} \label{NGC4013}
\end{figure}

In Table \ref{collNGC4013} we present the optimized values of
$K_0$ and $\rho_0$ for the analytic SIDM model of Eq.
(\ref{ScaledependentEoSDM}) for which the maximum compatibility
with the SPARC data is achieved.
\begin{table}[h!]
  \begin{center}
    \caption{SIDM Optimization Values for the galaxy NGC4013}
    \label{collNGC4013}
     \begin{tabular}{|r|r|}
     \hline
      \textbf{Parameter}   & \textbf{Optimization Values}
      \\  \hline
     $\rho_0 $  ($M_{\odot}/\mathrm{Kpc}^{3}$) & $1.10219\times 10^8$
\\  \hline $K_0$ ($M_{\odot} \,
\mathrm{Kpc}^{-3} \, (\mathrm{km/s})^{2}$)& 16800.3
\\  \hline
    \end{tabular}
  \end{center}
\end{table}
In Figs. \ref{NGC4013dens}, \ref{NGC4013} we present the density
of the analytic SIDM model, the predicted rotation curves for the
SIDM model (\ref{ScaledependentEoSDM}), versus the SPARC
observational data and the sound speed, as a function of the
radius respectively. As it can be seen, for this galaxy, the SIDM
model produces non-viable rotation curves which are incompatible
with the SPARC data.

\begin{figure}[h!]
\centering
\includegraphics[width=20pc]{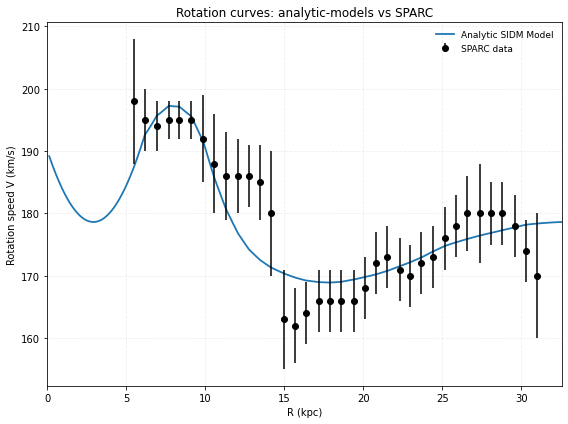}
\caption{The predicted rotation curves after using an optimization
for the SIDM model (\ref{ScaledependentEoSDM}), and the extended
SPARC data for the galaxy NGC4013. We included the rotation curves
of the gas, the disk velocities, the bulge (where present) along
with the SIDM model.} \label{extendedNGC4013}
\end{figure}

Now we shall include contributions to the rotation velocity from
the other components of the galaxy, namely the disk, the gas, and
the bulge if present. In Fig. \ref{extendedNGC4013} we present the
combined rotation curves including all the components of the
galaxy along with the SIDM. As it can be seen, the extended
collisional DM model is non-viable.
Also in Table \ref{evaluationextendedNGC4013} we present the
optimized values of the free parameters of the SIDM model for
which  we achieve the maximum compatibility with the SPARC data,
for the galaxy NGC4013, and also the resulting reduced
$\chi^2_{red}$ value.
\begin{table}[h!]
\centering \caption{Optimized Parameter Values of the Extended
SIDM model for the Galaxy NGC4013.}
\begin{tabular}{lc}
\hline
Parameter & Value  \\
\hline
$\rho_0 $ ($M_{\odot}/\mathrm{Kpc}^{3}$) & $3.47372\times 10^6$   \\
$K_0$ ($M_{\odot} \,
\mathrm{Kpc}^{-3} \, (\mathrm{km/s})^{2}$) & 11416.9   \\
$ml_{\text{disk}}$ & 0.7877  \\
$ml_{\text{bulge}}$ & 1.0000 \\
$\alpha$ (Kpc) & 33.0805\\
$\chi^2_{red}$ & 0.787128 \\
\hline
\end{tabular}
\label{evaluationextendedNGC4013}
\end{table}

\subsection{The Galaxy NGC4051, Non-viable, Extended Viable}

For this galaxy, the optimization method we used, ensures maximum
compatibility of the analytic SIDM model of Eq.
(\ref{ScaledependentEoSDM}) with the SPARC data, if we choose
$\rho_0=1.14949\times 10^8$$M_{\odot}/\mathrm{Kpc}^{3}$ and
$K_0=10589.6
$$M_{\odot} \, \mathrm{Kpc}^{-3} \, (\mathrm{km/s})^{2}$, in which
case the reduced $\chi^2_{red}$ value is $\chi^2_{red}=2.83023$.
Also the parameter $\alpha$ in this case is $\alpha=5.5391 $Kpc.

\begin{figure}[h!]
\centering
\includegraphics[width=20pc]{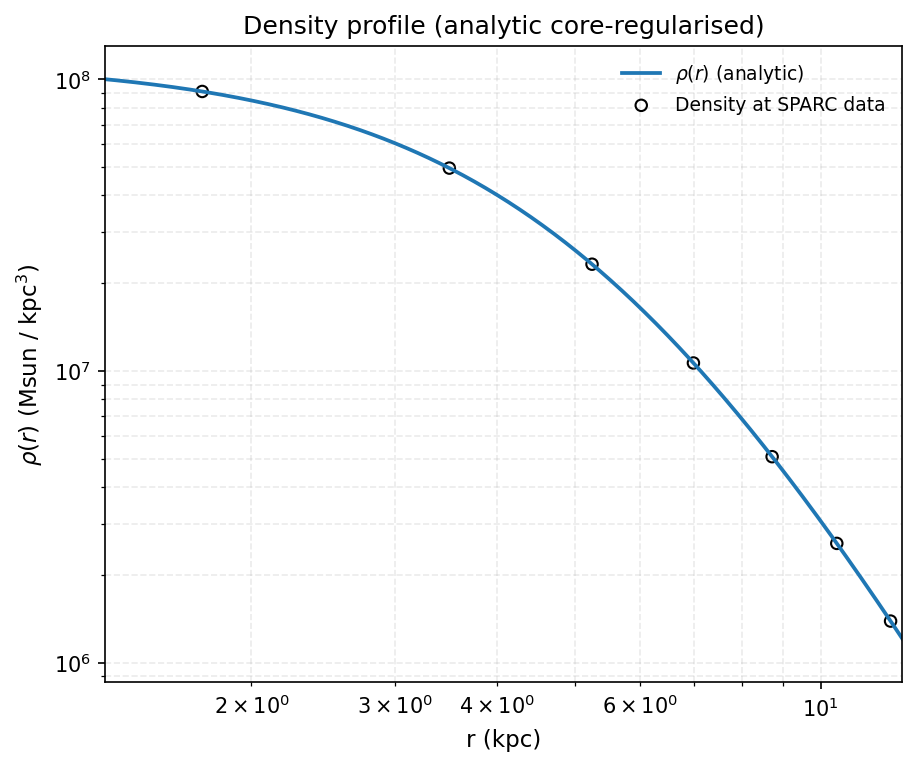}
\caption{The density of the SIDM model of Eq.
(\ref{ScaledependentEoSDM}) for the galaxy NGC4051, versus the
radius.} \label{NGC4051dens}
\end{figure}

\begin{figure}[h!]
\centering
\includegraphics[width=35pc]{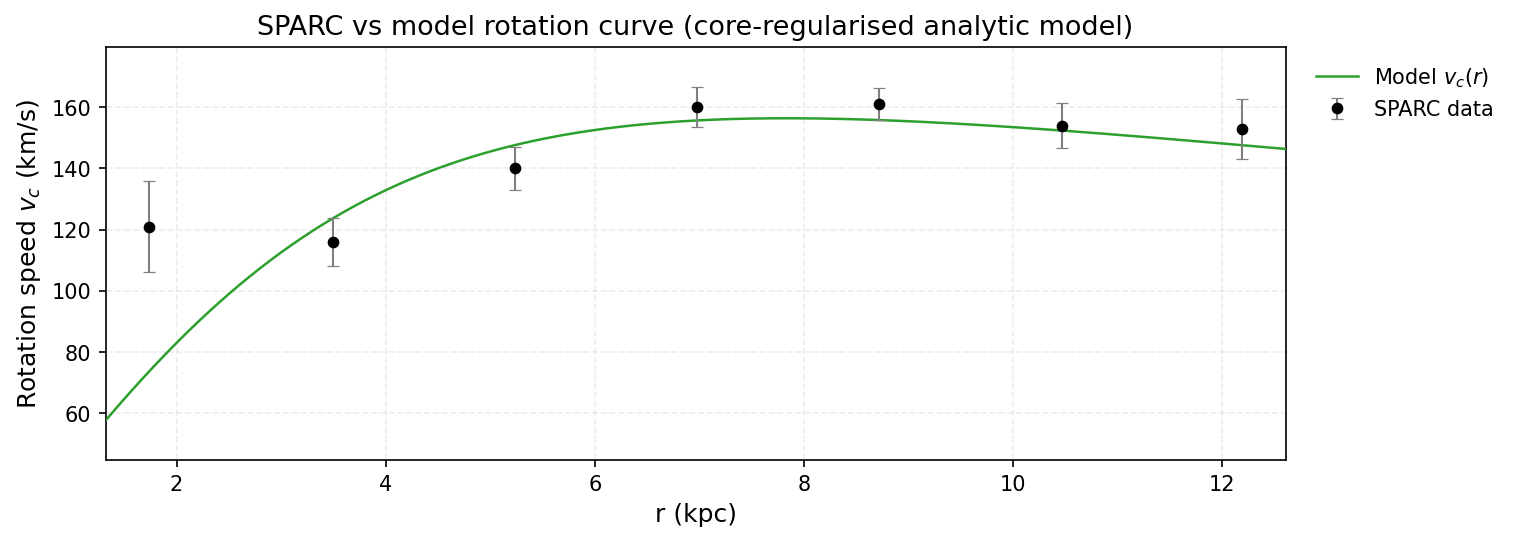}
\caption{The predicted rotation curves for the optimized SIDM
model of Eq. (\ref{ScaledependentEoSDM}), versus the SPARC
observational data for the galaxy NGC4051.} \label{NGC4051}
\end{figure}

In Table \ref{collNGC4051} we present the optimized values of
$K_0$ and $\rho_0$ for the analytic SIDM model of Eq.
(\ref{ScaledependentEoSDM}) for which the maximum compatibility
with the SPARC data is achieved.
\begin{table}[h!]
  \begin{center}
    \caption{SIDM Optimization Values for the galaxy NGC4051}
    \label{collNGC4051}
     \begin{tabular}{|r|r|}
     \hline
      \textbf{Parameter}   & \textbf{Optimization Values}
      \\  \hline
     $\rho_0 $  ($M_{\odot}/\mathrm{Kpc}^{3}$) & $1.14949\times 10^8$
\\  \hline $K_0$ ($M_{\odot} \,
\mathrm{Kpc}^{-3} \, (\mathrm{km/s})^{2}$)& 10589.6
\\  \hline
    \end{tabular}
  \end{center}
\end{table}
In Figs. \ref{NGC4051dens}, \ref{NGC4051} we present the density
of the analytic SIDM model, the predicted rotation curves for the
SIDM model (\ref{ScaledependentEoSDM}), versus the SPARC
observational data and the sound speed, as a function of the
radius respectively. As it can be seen, for this galaxy, the SIDM
model produces non-viable rotation curves which are incompatible
with the SPARC data.

\begin{figure}[h!]
\centering
\includegraphics[width=20pc]{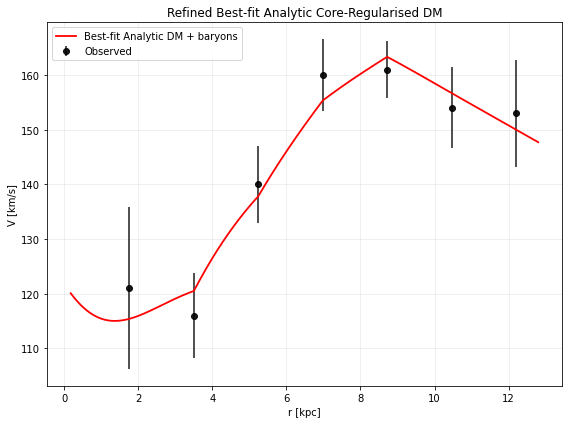}
\caption{The predicted rotation curves after using an optimization
for the SIDM model (\ref{ScaledependentEoSDM}), and the extended
SPARC data for the galaxy NGC4051. We included the rotation curves
of the gas, the disk velocities, the bulge (where present) along
with the SIDM model.} \label{extendedNGC4051}
\end{figure}

Now we shall include contributions to the rotation velocity from
the other components of the galaxy, namely the disk, the gas, and
the bulge if present. In Fig. \ref{extendedNGC4051} we present the
combined rotation curves including all the components of the
galaxy along with the SIDM. As it can be seen, the extended
collisional DM model is non-viable.
Also in Table \ref{evaluationextendedNGC4051} we present the
optimized values of the free parameters of the SIDM model for
which  we achieve the maximum compatibility with the SPARC data,
for the galaxy NGC4051, and also the resulting reduced
$\chi^2_{red}$ value.
\begin{table}[h!]
\centering \caption{Optimized Parameter Values of the Extended
SIDM model for the Galaxy NGC4051.}
\begin{tabular}{lc}
\hline
Parameter & Value  \\
\hline
$\rho_0 $ ($M_{\odot}/\mathrm{Kpc}^{3}$) & $5.50877\times 10^7$   \\
$K_0$ ($M_{\odot} \,
\mathrm{Kpc}^{-3} \, (\mathrm{km/s})^{2}$) & 5011.09   \\
$ml_{\text{disk}}$ & 0.6124 \\
$ml_{\text{bulge}}$ & 0.2543 \\
$\alpha$ (Kpc) & 5.50345\\
$\chi^2_{red}$ & 0.497737 \\
\hline
\end{tabular}
\label{evaluationextendedNGC4051}
\end{table}

\subsection{The Galaxy NGC4068}

For this galaxy, the optimization method we used, ensures maximum
compatibility of the analytic SIDM model of Eq.
(\ref{ScaledependentEoSDM}) with the SPARC data, if we choose
$\rho_0=3.13381\times 10^7$$M_{\odot}/\mathrm{Kpc}^{3}$ and
$K_0=915.044
$$M_{\odot} \, \mathrm{Kpc}^{-3} \, (\mathrm{km/s})^{2}$, in which
case the reduced $\chi^2_{red}$ value is $\chi^2_{red}=0.637966$.
Also the parameter $\alpha$ in this case is $\alpha=3.11843 $Kpc.

\begin{figure}[h!]
\centering
\includegraphics[width=20pc]{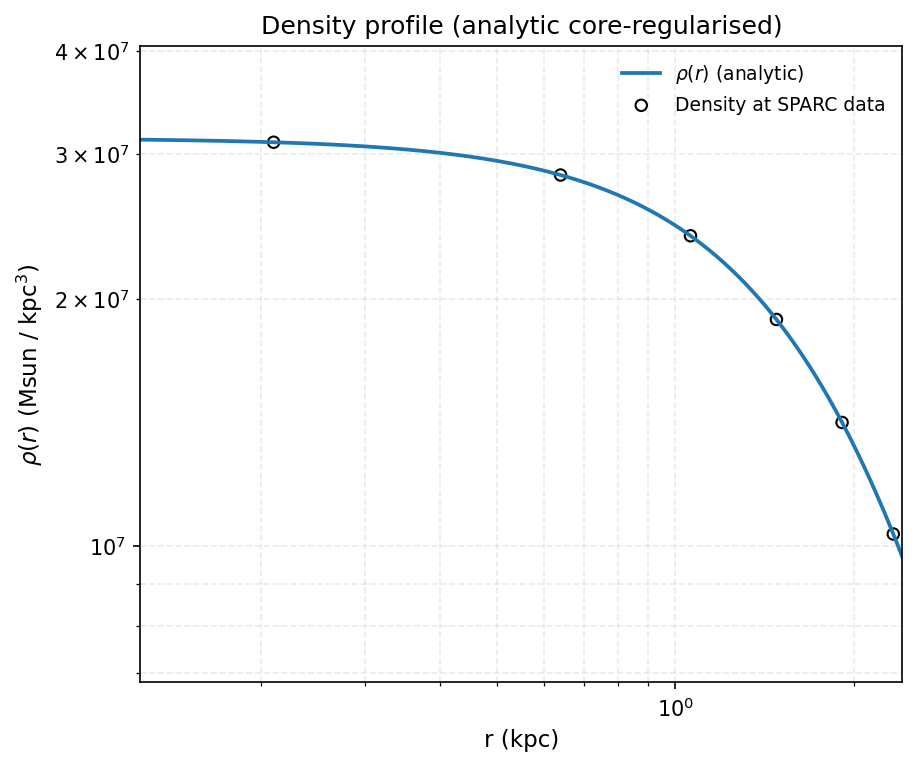}
\caption{The density of the SIDM model of Eq.
(\ref{ScaledependentEoSDM}) for the galaxy NGC4068, versus the
radius.} \label{NGC4068dens}
\end{figure}

\begin{figure}[h!]
\centering
\includegraphics[width=35pc]{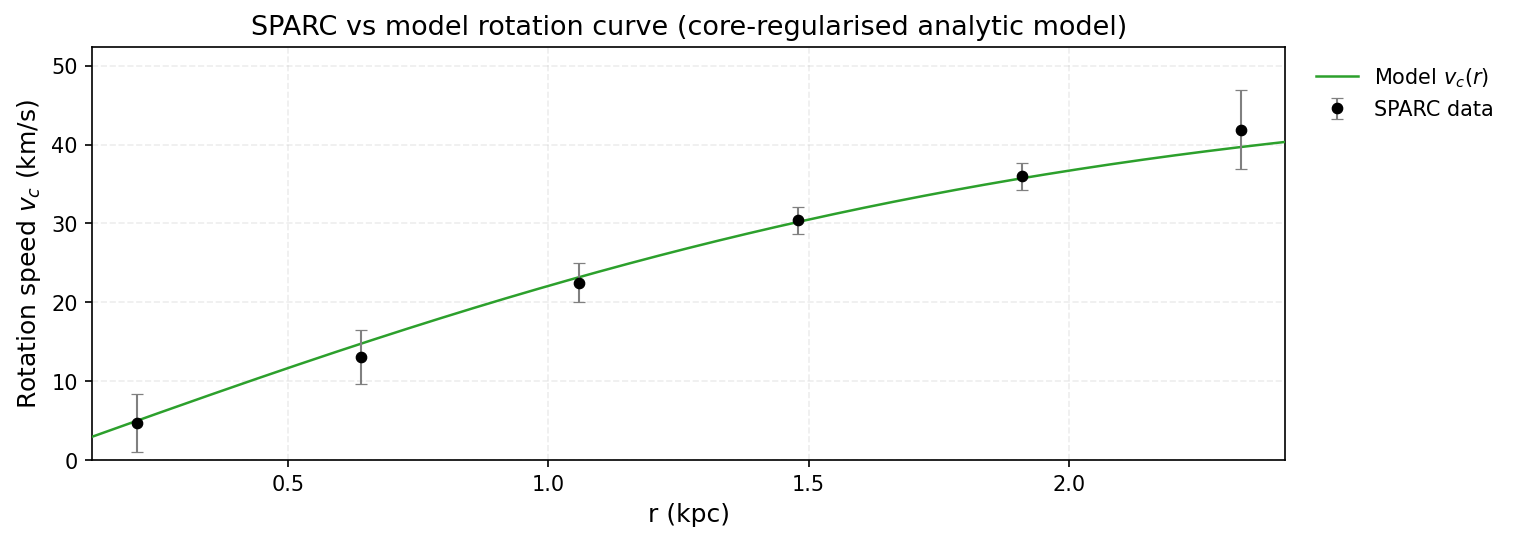}
\caption{The predicted rotation curves for the optimized SIDM
model of Eq. (\ref{ScaledependentEoSDM}), versus the SPARC
observational data for the galaxy NGC4068.} \label{NGC4068}
\end{figure}

In Table \ref{collNGC4068} we present the optimized values of
$K_0$ and $\rho_0$ for the analytic SIDM model of Eq.
(\ref{ScaledependentEoSDM}) for which the maximum compatibility
with the SPARC data is achieved.
\begin{table}[h!]
  \begin{center}
    \caption{SIDM Optimization Values for the galaxy NGC4068}
    \label{collNGC4068}
     \begin{tabular}{|r|r|}
     \hline
      \textbf{Parameter}   & \textbf{Optimization Values}
      \\  \hline
     $\rho_0 $  ($M_{\odot}/\mathrm{Kpc}^{3}$) & $3.13381\times 10^7$
\\  \hline $K_0$ ($M_{\odot} \,
\mathrm{Kpc}^{-3} \, (\mathrm{km/s})^{2}$)& 915.044
\\  \hline
    \end{tabular}
  \end{center}
\end{table}
In Figs. \ref{NGC4068dens}, \ref{NGC4068} we present the density
of the analytic SIDM model, the predicted rotation curves for the
SIDM model (\ref{ScaledependentEoSDM}), versus the SPARC
observational data and the sound speed, as a function of the
radius respectively. As it can be seen, for this galaxy, the SIDM
model produces viable rotation curves which are compatible with
the SPARC data.

\subsection{The Galaxy NGC4085}

For this galaxy, the optimization method we used, ensures maximum
compatibility of the analytic SIDM model of Eq.
(\ref{ScaledependentEoSDM}) with the SPARC data, if we choose
$\rho_0=1.49325\times 10^8$$M_{\odot}/\mathrm{Kpc}^{3}$ and
$K_0=7851.63
$$M_{\odot} \, \mathrm{Kpc}^{-3} \, (\mathrm{km/s})^{2}$, in which
case the reduced $\chi^2_{red}$ value is $\chi^2_{red}=0.24862$.
Also the parameter $\alpha$ in this case is $\alpha=4.1847 $Kpc.

\begin{figure}[h!]
\centering
\includegraphics[width=20pc]{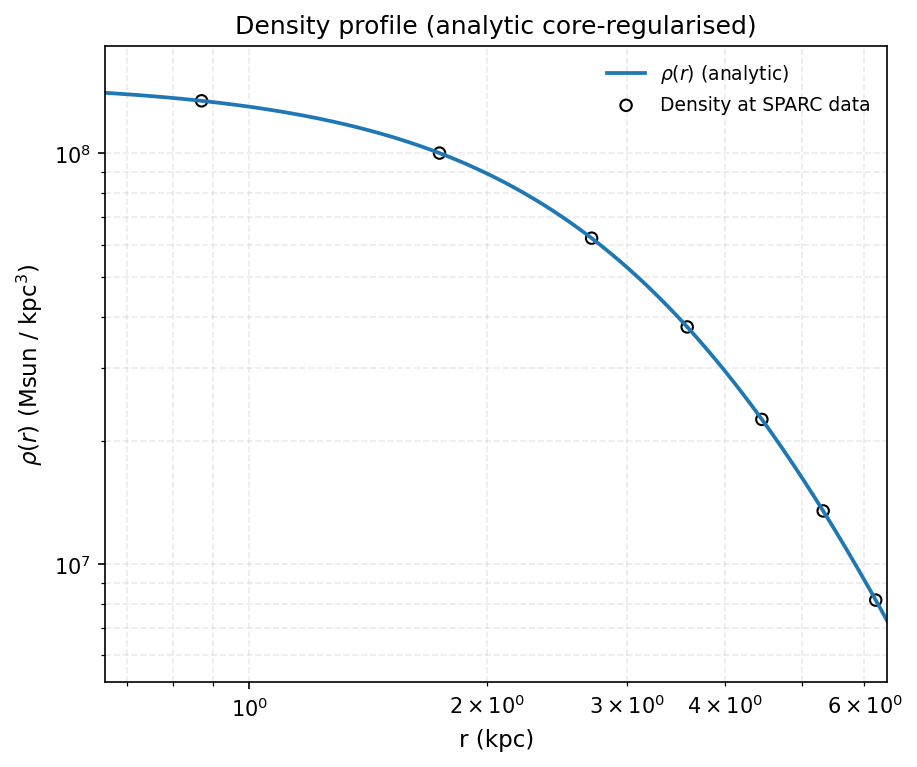}
\caption{The density of the SIDM model of Eq.
(\ref{ScaledependentEoSDM}) for the galaxy NGC4085, versus the
radius.} \label{NGC4085dens}
\end{figure}

\begin{figure}[h!]
\centering
\includegraphics[width=35pc]{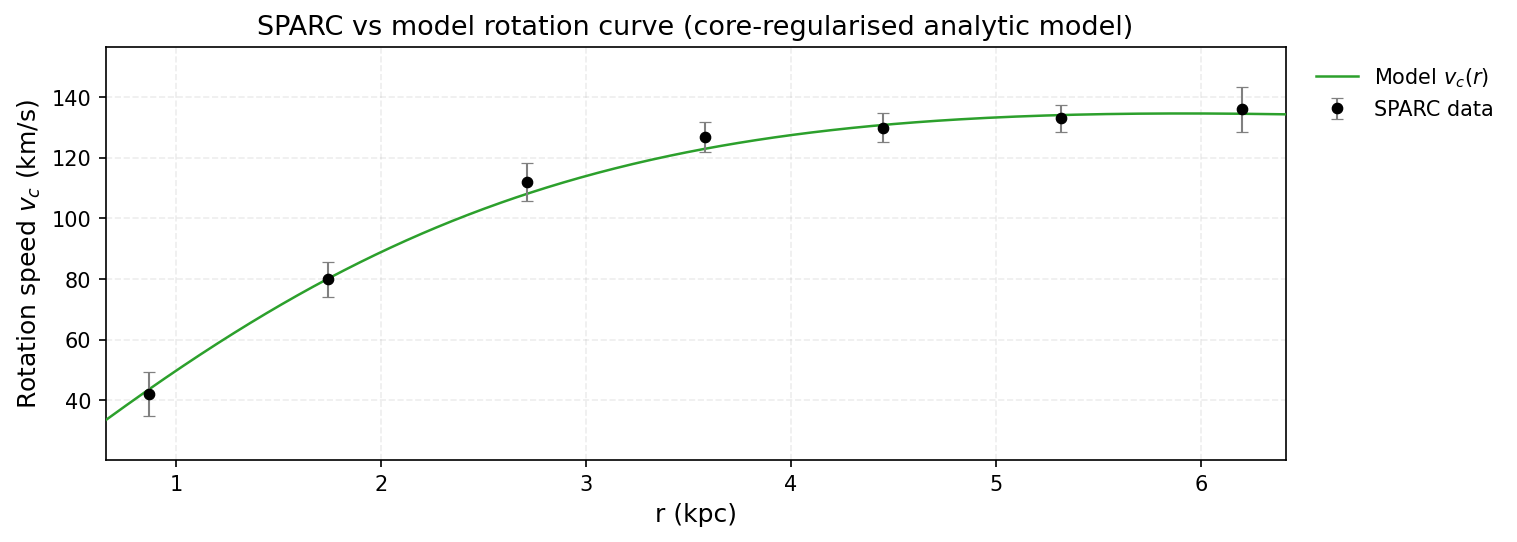}
\caption{The predicted rotation curves for the optimized SIDM
model of Eq. (\ref{ScaledependentEoSDM}), versus the SPARC
observational data for the galaxy NGC4085.} \label{NGC4085}
\end{figure}

In Table \ref{collNGC4085} we present the optimized values of
$K_0$ and $\rho_0$ for the analytic SIDM model of Eq.
(\ref{ScaledependentEoSDM}) for which the maximum compatibility
with the SPARC data is achieved.
\begin{table}[h!]
  \begin{center}
    \caption{SIDM Optimization Values for the galaxy NGC4085}
    \label{collNGC4085}
     \begin{tabular}{|r|r|}
     \hline
      \textbf{Parameter}   & \textbf{Optimization Values}
      \\  \hline
     $\rho_0 $  ($M_{\odot}/\mathrm{Kpc}^{3}$) & $1.49325\times 10^8$
\\  \hline $K_0$ ($M_{\odot} \,
\mathrm{Kpc}^{-3} \, (\mathrm{km/s})^{2}$)& 7851.63
\\  \hline
    \end{tabular}
  \end{center}
\end{table}
In Figs. \ref{NGC4085dens}, \ref{NGC4085} we present the density
of the analytic SIDM model, the predicted rotation curves for the
SIDM model (\ref{ScaledependentEoSDM}), versus the SPARC
observational data and the sound speed, as a function of the
radius respectively. As it can be seen, for this galaxy, the SIDM
model produces viable rotation curves which are compatible with
the SPARC data.

\subsection{The Galaxy NGC4088, Viable}

For this galaxy, the optimization method we used, ensures maximum
compatibility of the analytic SIDM model of Eq.
(\ref{ScaledependentEoSDM}) with the SPARC data, if we choose
$\rho_0=1.09358\times 10^8$$M_{\odot}/\mathrm{Kpc}^{3}$ and
$K_0=14986
$$M_{\odot} \, \mathrm{Kpc}^{-3} \, (\mathrm{km/s})^{2}$, in which
case the reduced $\chi^2_{red}$ value is $\chi^2_{red}=1.26962$.
Also the parameter $\alpha$ in this case is $\alpha=6.75568 $Kpc.

\begin{figure}[h!]
\centering
\includegraphics[width=20pc]{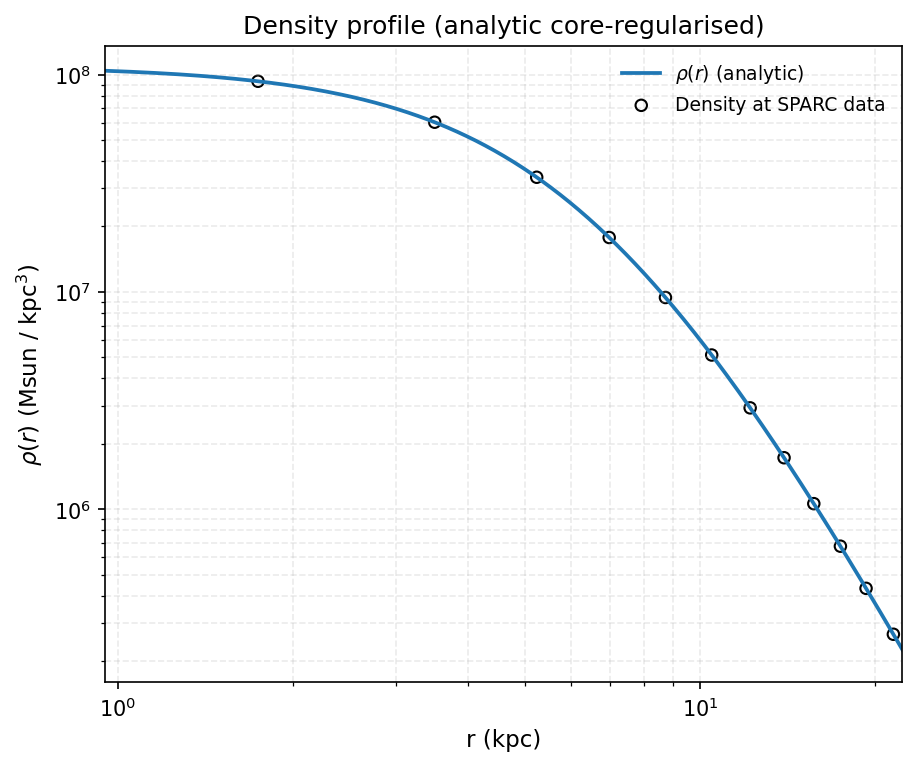}
\caption{The density of the SIDM model of Eq.
(\ref{ScaledependentEoSDM}) for the galaxy NGC4088, versus the
radius.} \label{NGC4088dens}
\end{figure}

\begin{figure}[h!]
\centering
\includegraphics[width=35pc]{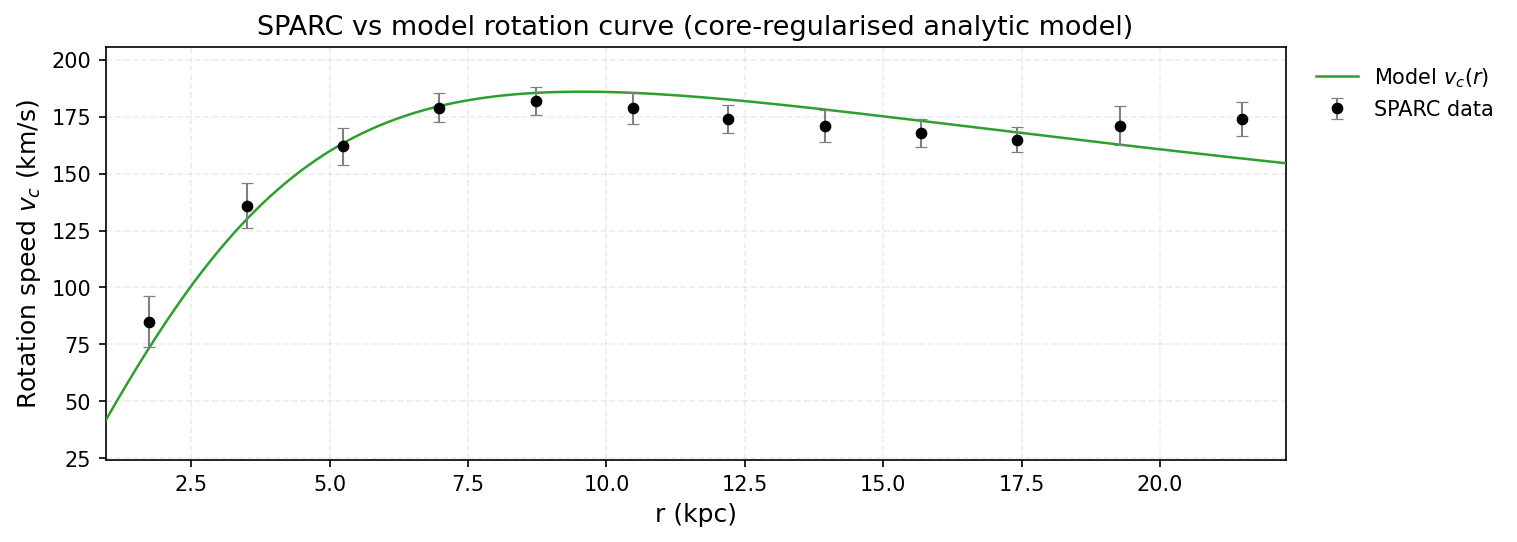}
\caption{The predicted rotation curves for the optimized SIDM
model of Eq. (\ref{ScaledependentEoSDM}), versus the SPARC
observational data for the galaxy NGC4088.} \label{NGC4088}
\end{figure}

In Table \ref{collNGC4088} we present the optimized values of
$K_0$ and $\rho_0$ for the analytic SIDM model of Eq.
(\ref{ScaledependentEoSDM}) for which the maximum compatibility
with the SPARC data is achieved.
\begin{table}[h!]
  \begin{center}
    \caption{SIDM Optimization Values for the galaxy NGC4088}
    \label{collNGC4088}
     \begin{tabular}{|r|r|}
     \hline
      \textbf{Parameter}   & \textbf{Optimization Values}
      \\  \hline
     $\rho_0 $  ($M_{\odot}/\mathrm{Kpc}^{3}$) & $1.09358\times 10^8$
\\  \hline $K_0$ ($M_{\odot} \,
\mathrm{Kpc}^{-3} \, (\mathrm{km/s})^{2}$)& 14986
\\  \hline
    \end{tabular}
  \end{center}
\end{table}
In Figs. \ref{NGC4088dens}, \ref{NGC4088} we present the density
of the analytic SIDM model, the predicted rotation curves for the
SIDM model (\ref{ScaledependentEoSDM}), versus the SPARC
observational data and the sound speed, as a function of the
radius respectively. As it can be seen, for this galaxy, the SIDM
model produces marginally viable rotation curves which are
marginally compatible with the SPARC data.


\subsection{The Galaxy NGC4100}

For this galaxy, the optimization method we used, ensures maximum
compatibility of the analytic SIDM model of Eq.
(\ref{ScaledependentEoSDM}) with the SPARC data, if we choose
$\rho_0=1.4466\times 10^8$$M_{\odot}/\mathrm{Kpc}^{3}$ and
$K_0=16706.4
$$M_{\odot} \, \mathrm{Kpc}^{-3} \, (\mathrm{km/s})^{2}$, in which
case the reduced $\chi^2_{red}$ value is $\chi^2_{red}=0.322466$.
Also the parameter $\alpha$ in this case is $\alpha=6.2018 $Kpc.

\begin{figure}[h!]
\centering
\includegraphics[width=20pc]{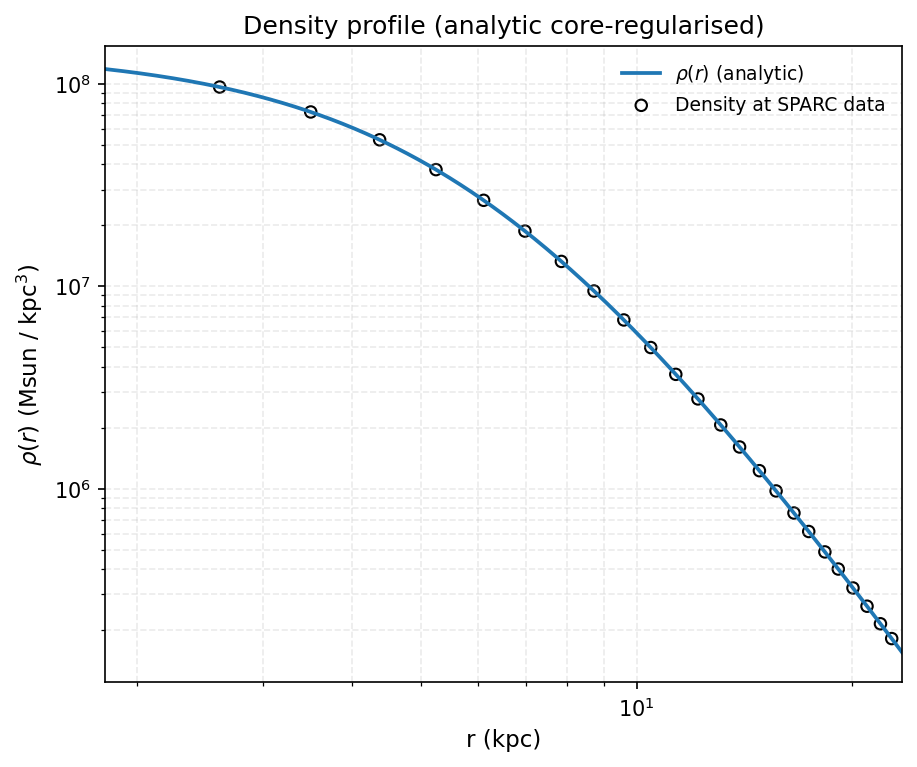}
\caption{The density of the SIDM model of Eq.
(\ref{ScaledependentEoSDM}) for the galaxy NGC4100, versus the
radius.} \label{NGC4100dens}
\end{figure}

\begin{figure}[h!]
\centering
\includegraphics[width=35pc]{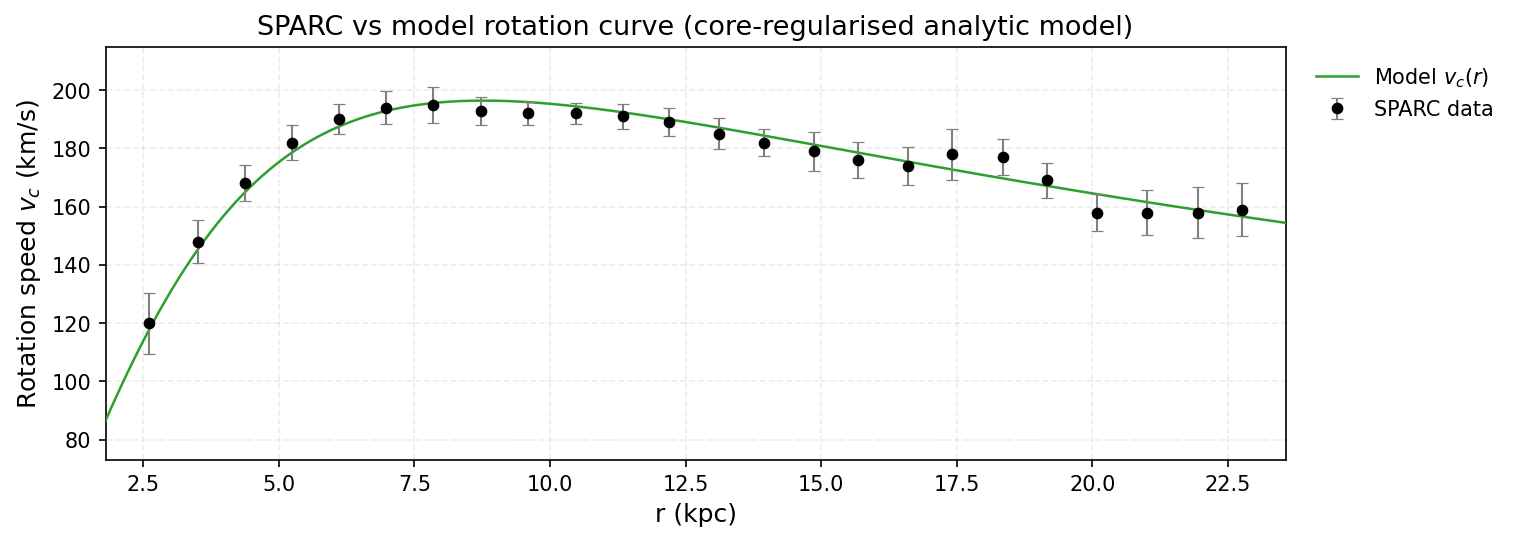}
\caption{The predicted rotation curves for the optimized SIDM
model of Eq. (\ref{ScaledependentEoSDM}), versus the SPARC
observational data for the galaxy NGC4100.} \label{NGC4100}
\end{figure}

In Table \ref{collNGC4100} we present the optimized values of
$K_0$ and $\rho_0$ for the analytic SIDM model of Eq.
(\ref{ScaledependentEoSDM}) for which the maximum compatibility
with the SPARC data is achieved.
\begin{table}[h!]
  \begin{center}
    \caption{SIDM Optimization Values for the galaxy NGC4100}
    \label{collNGC4100}
     \begin{tabular}{|r|r|}
     \hline
      \textbf{Parameter}   & \textbf{Optimization Values}
      \\  \hline
     $\rho_0 $  ($M_{\odot}/\mathrm{Kpc}^{3}$) & $1.4466\times 10^8$
\\  \hline $K_0$ ($M_{\odot} \,
\mathrm{Kpc}^{-3} \, (\mathrm{km/s})^{2}$)& 16706.4
\\  \hline
    \end{tabular}
  \end{center}
\end{table}
In Figs. \ref{NGC4100dens}, \ref{NGC4100} we present the density
of the analytic SIDM model, the predicted rotation curves for the
SIDM model (\ref{ScaledependentEoSDM}), versus the SPARC
observational data and the sound speed, as a function of the
radius respectively. As it can be seen, for this galaxy, the SIDM
model produces viable rotation curves which are compatible with
the SPARC data.

\subsection{The Galaxy NGC4138}

For this galaxy, the optimization method we used, ensures maximum
compatibility of the analytic SIDM model of Eq.
(\ref{ScaledependentEoSDM}) with the SPARC data, if we choose
$\rho_0=5.01961\times 10^8$$M_{\odot}/\mathrm{Kpc}^{3}$ and
$K_0=17200.2
$$M_{\odot} \, \mathrm{Kpc}^{-3} \, (\mathrm{km/s})^{2}$, in which
case the reduced $\chi^2_{red}$ value is $\chi^2_{red}=0.43578$.
Also the parameter $\alpha$ in this case is $\alpha=3.32603 $Kpc.

\begin{figure}[h!]
\centering
\includegraphics[width=20pc]{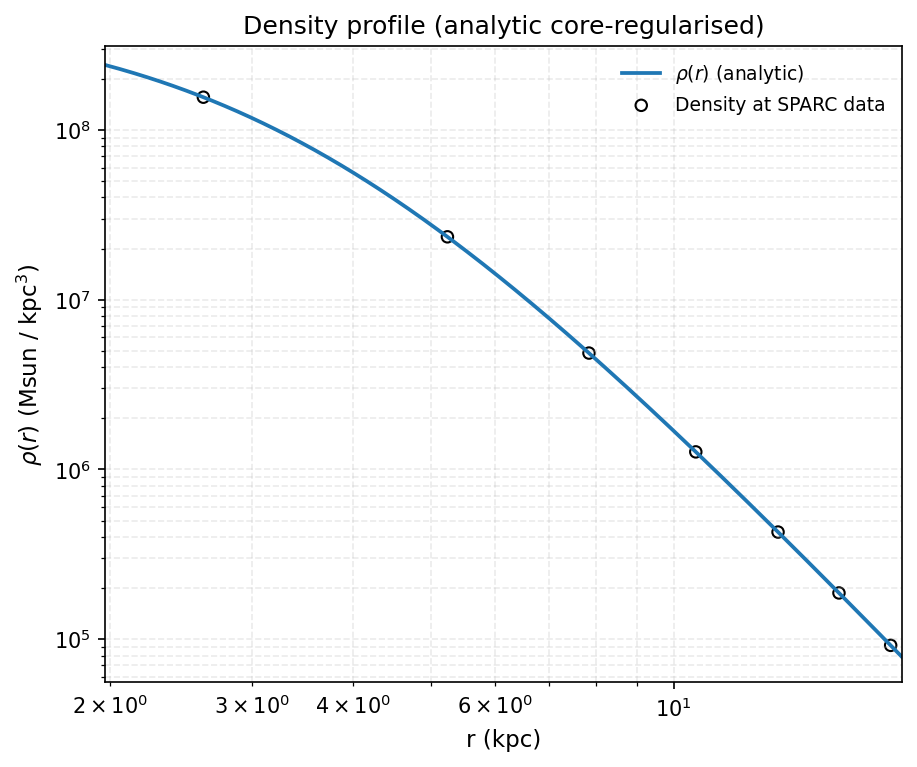}
\caption{The density of the SIDM model of Eq.
(\ref{ScaledependentEoSDM}) for the galaxy NGC4138, versus the
radius.} \label{NGC4138dens}
\end{figure}

\begin{figure}[h!]
\centering
\includegraphics[width=35pc]{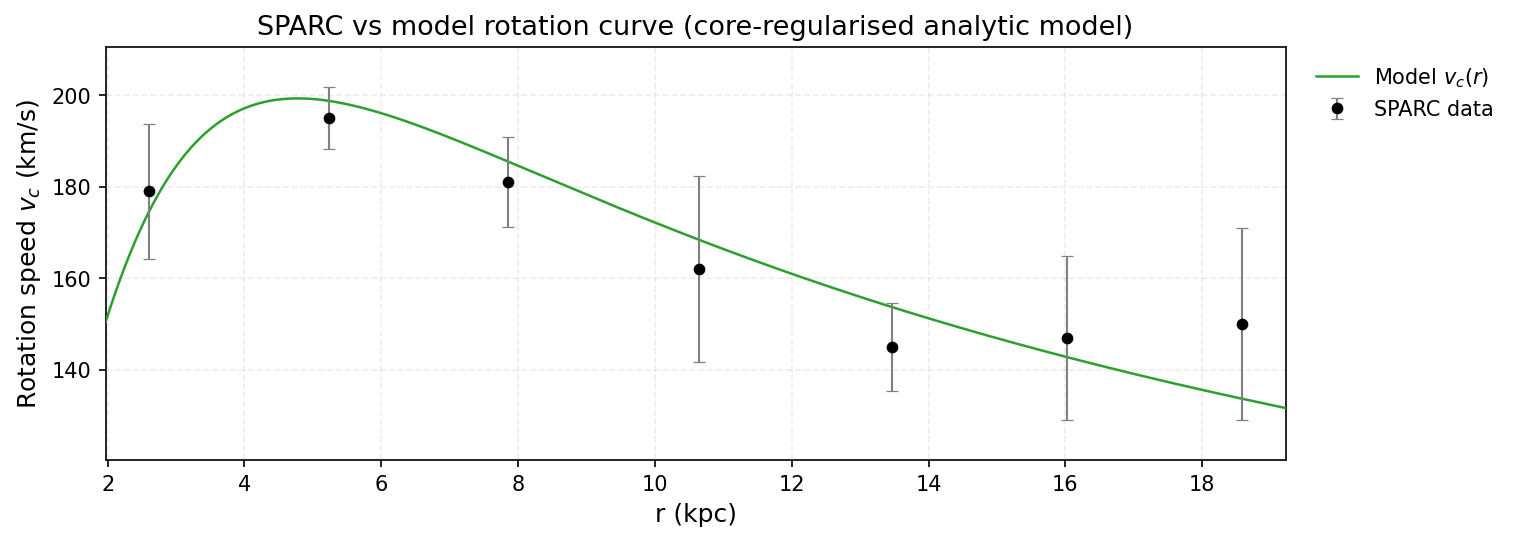}
\caption{The predicted rotation curves for the optimized SIDM
model of Eq. (\ref{ScaledependentEoSDM}), versus the SPARC
observational data for the galaxy NGC4138.} \label{NGC4138}
\end{figure}

In Table \ref{collNGC4138} we present the optimized values of
$K_0$ and $\rho_0$ for the analytic SIDM model of Eq.
(\ref{ScaledependentEoSDM}) for which the maximum compatibility
with the SPARC data is achieved.
\begin{table}[h!]
  \begin{center}
    \caption{SIDM Optimization Values for the galaxy NGC4138}
    \label{collNGC4138}
     \begin{tabular}{|r|r|}
     \hline
      \textbf{Parameter}   & \textbf{Optimization Values}
      \\  \hline
     $\rho_0 $  ($M_{\odot}/\mathrm{Kpc}^{3}$) & $5.01961\times 10^8$
\\  \hline $K_0$ ($M_{\odot} \,
\mathrm{Kpc}^{-3} \, (\mathrm{km/s})^{2}$)& 17200.2
\\  \hline
    \end{tabular}
  \end{center}
\end{table}
In Figs. \ref{NGC4138dens}, \ref{NGC4138} we present the density
of the analytic SIDM model, the predicted rotation curves for the
SIDM model (\ref{ScaledependentEoSDM}), versus the SPARC
observational data and the sound speed, as a function of the
radius respectively. As it can be seen, for this galaxy, the SIDM
model produces viable rotation curves which are compatible with
the SPARC data.

\subsection{The Galaxy NGC4157, Non-viable, Extended Viable}

For this galaxy, the optimization method we used, ensures maximum
compatibility of the analytic SIDM model of Eq.
(\ref{ScaledependentEoSDM}) with the SPARC data, if we choose
$\rho_0=1.37718\times 10^8$$M_{\odot}/\mathrm{Kpc}^{3}$ and
$K_0=18971.8
$$M_{\odot} \, \mathrm{Kpc}^{-3} \, (\mathrm{km/s})^{2}$, in which
case the reduced $\chi^2_{red}$ value is $\chi^2_{red}=3.04141$.
Also the parameter $\alpha$ in this case is $\alpha=6.77345 $Kpc.

\begin{figure}[h!]
\centering
\includegraphics[width=20pc]{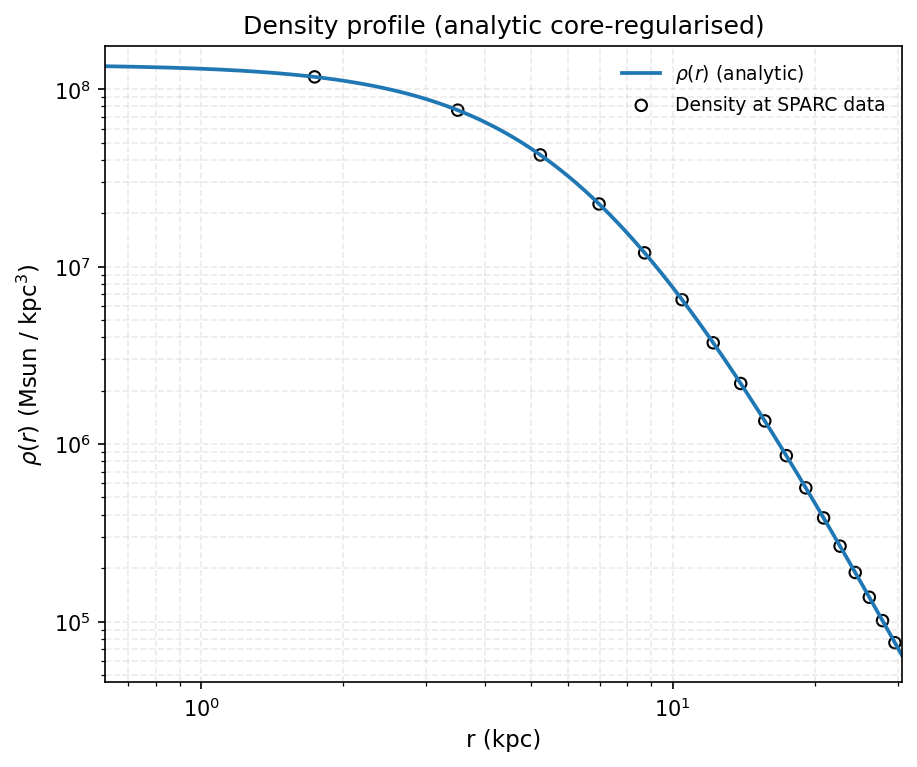}
\caption{The density of the SIDM model of Eq.
(\ref{ScaledependentEoSDM}) for the galaxy NGC4157, versus the
radius.} \label{NGC4157dens}
\end{figure}

\begin{figure}[h!]
\centering
\includegraphics[width=35pc]{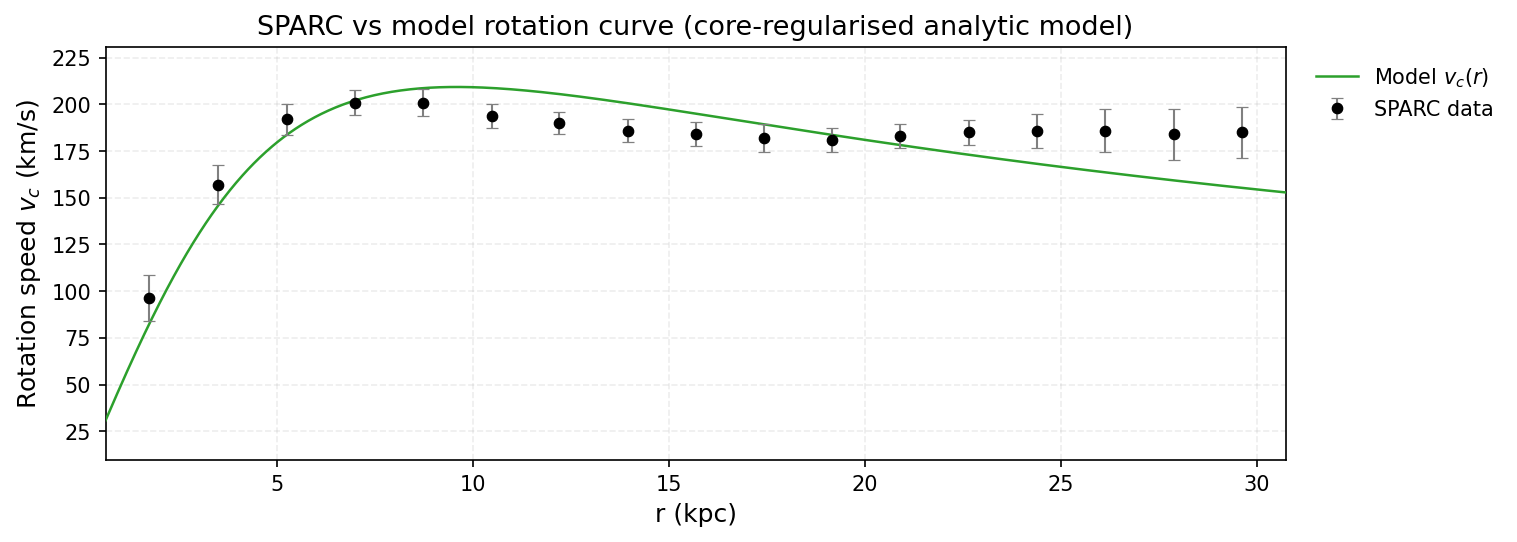}
\caption{The predicted rotation curves for the optimized SIDM
model of Eq. (\ref{ScaledependentEoSDM}), versus the SPARC
observational data for the galaxy NGC4157.} \label{NGC4157}
\end{figure}

In Table \ref{collNGC4157} we present the optimized values of
$K_0$ and $\rho_0$ for the analytic SIDM model of Eq.
(\ref{ScaledependentEoSDM}) for which the maximum compatibility
with the SPARC data is achieved.
\begin{table}[h!]
  \begin{center}
    \caption{SIDM Optimization Values for the galaxy NGC4157}
    \label{collNGC4157}
     \begin{tabular}{|r|r|}
     \hline
      \textbf{Parameter}   & \textbf{Optimization Values}
      \\  \hline
     $\rho_0 $  ($M_{\odot}/\mathrm{Kpc}^{3}$) & $1.37718\times 10^8$
\\  \hline $K_0$ ($M_{\odot} \,
\mathrm{Kpc}^{-3} \, (\mathrm{km/s})^{2}$)& 18971.8
\\  \hline
    \end{tabular}
  \end{center}
\end{table}
In Figs. \ref{NGC4157dens}, \ref{NGC4157} we present the density
of the analytic SIDM model, the predicted rotation curves for the
SIDM model (\ref{ScaledependentEoSDM}), versus the SPARC
observational data and the sound speed, as a function of the
radius respectively. As it can be seen, for this galaxy, the SIDM
model produces non-viable rotation curves which are incompatible
with the SPARC data.

\begin{figure}[h!]
\centering
\includegraphics[width=20pc]{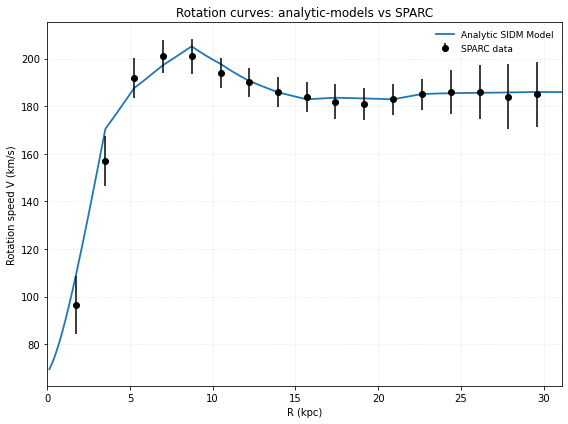}
\caption{The predicted rotation curves after using an optimization
for the SIDM model (\ref{ScaledependentEoSDM}), and the extended
SPARC data for the galaxy NGC4157. We included the rotation curves
of the gas, the disk velocities, the bulge (where present) along
with the SIDM model.} \label{extendedNGC4157}
\end{figure}

Now we shall include contributions to the rotation velocity from
the other components of the galaxy, namely the disk, the gas, and
the bulge if present. In Fig. \ref{extendedNGC4157} we present the
combined rotation curves including all the components of the
galaxy along with the SIDM. As it can be seen, the extended
collisional DM model is non-viable.
Also in Table \ref{evaluationextendedNGC4157} we present the
optimized values of the free parameters of the SIDM model for
which  we achieve the maximum compatibility with the SPARC data,
for the galaxy NGC4157, and also the resulting reduced
$\chi^2_{red}$ value.
\begin{table}[h!]
\centering \caption{Optimized Parameter Values of the Extended
SIDM model for the Galaxy NGC4157.}
\begin{tabular}{lc}
\hline
Parameter & Value  \\
\hline
$\rho_0 $ ($M_{\odot}/\mathrm{Kpc}^{3}$) & $4.02447\times 10^6$   \\
$K_0$ ($M_{\odot} \,
\mathrm{Kpc}^{-3} \, (\mathrm{km/s})^{2}$) & 11427.5   \\
$ml_{\text{disk}}$ & 0.7542 \\
$ml_{\text{bulge}}$ & 0.653 \\
$\alpha$ (Kpc) & 30.748\\
$\chi^2_{red}$ & 0.329047 \\
\hline
\end{tabular}
\label{evaluationextendedNGC4157}
\end{table}

\subsection{The Galaxy NGC4183, Non-viable}

For this galaxy, the optimization method we used, ensures maximum
compatibility of the analytic SIDM model of Eq.
(\ref{ScaledependentEoSDM}) with the SPARC data, if we choose
$\rho_0=4.15754\times 10^7$$M_{\odot}/\mathrm{Kpc}^{3}$ and
$K_0=5923.15
$$M_{\odot} \, \mathrm{Kpc}^{-3} \, (\mathrm{km/s})^{2}$, in which
case the reduced $\chi^2_{red}$ value is $\chi^2_{red}=1.72865$.
Also the parameter $\alpha$ in this case is $\alpha=6.88825 $Kpc.

\begin{figure}[h!]
\centering
\includegraphics[width=20pc]{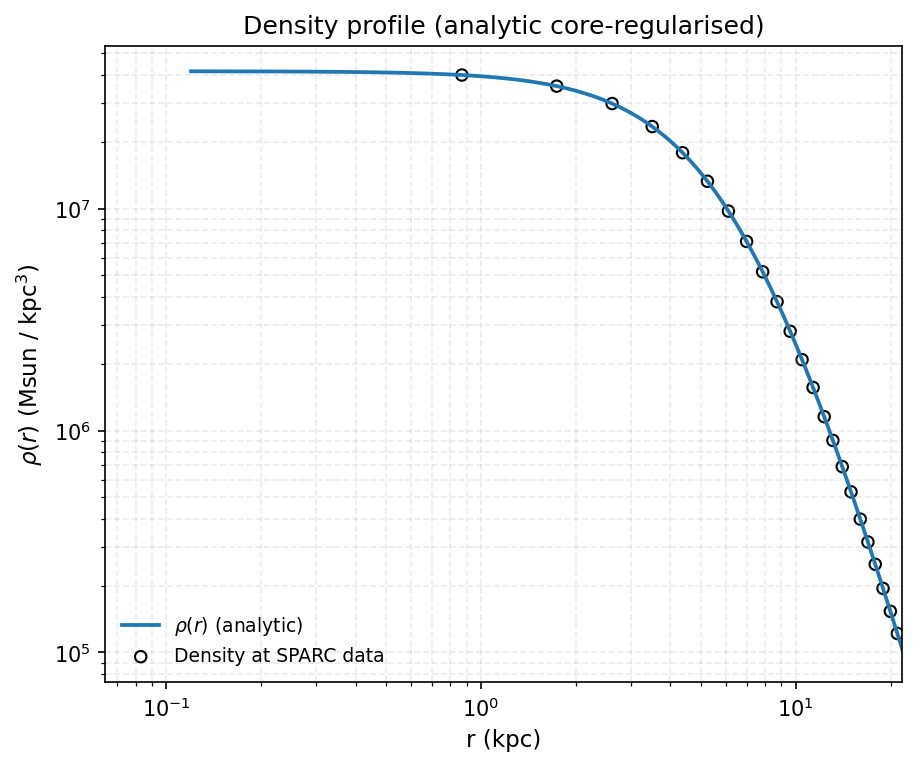}
\caption{The density of the SIDM model of Eq.
(\ref{ScaledependentEoSDM}) for the galaxy NGC4183, versus the
radius.} \label{NGC4183dens}
\end{figure}

\begin{figure}[h!]
\centering
\includegraphics[width=35pc]{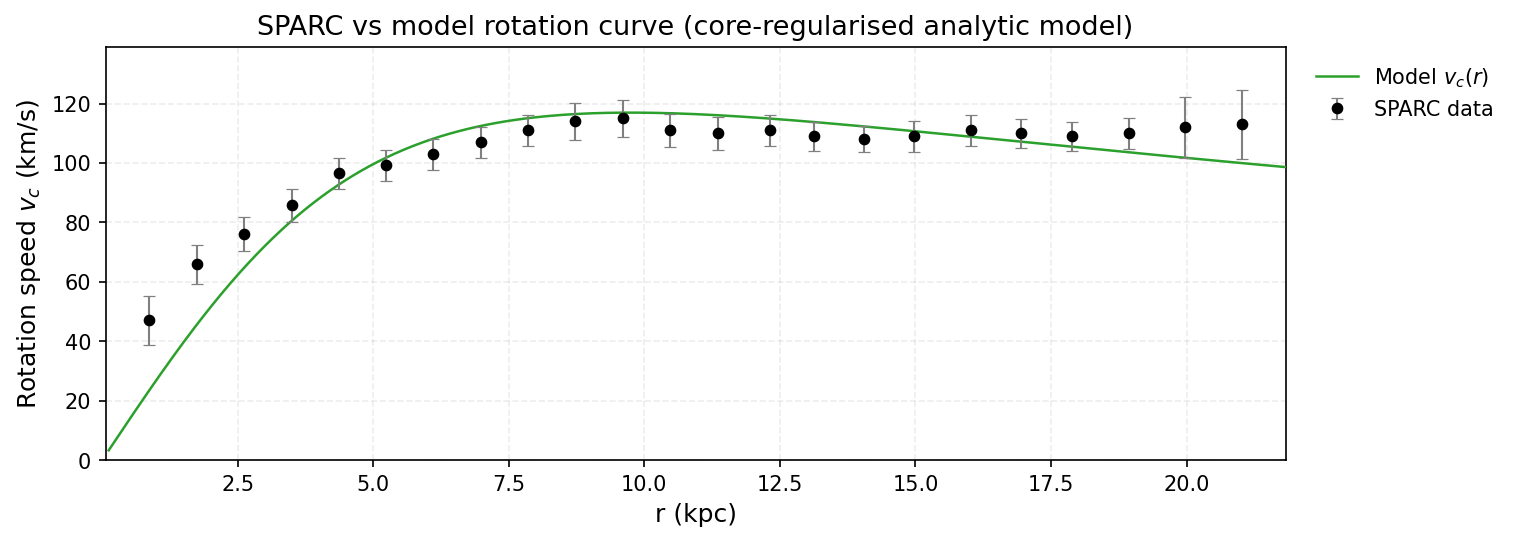}
\caption{The predicted rotation curves for the optimized SIDM
model of Eq. (\ref{ScaledependentEoSDM}), versus the SPARC
observational data for the galaxy NGC4183.} \label{NGC4183}
\end{figure}

In Table \ref{collNGC4183} we present the optimized values of
$K_0$ and $\rho_0$ for the analytic SIDM model of Eq.
(\ref{ScaledependentEoSDM}) for which the maximum compatibility
with the SPARC data is achieved.
\begin{table}[h!]
  \begin{center}
    \caption{SIDM Optimization Values for the galaxy NGC4183}
    \label{collNGC4183}
     \begin{tabular}{|r|r|}
     \hline
      \textbf{Parameter}   & \textbf{Optimization Values}
      \\  \hline
     $\rho_0 $  ($M_{\odot}/\mathrm{Kpc}^{3}$) & $4.15754\times 10^7$
\\  \hline $K_0$ ($M_{\odot} \,
\mathrm{Kpc}^{-3} \, (\mathrm{km/s})^{2}$)& 5923.15
\\  \hline
    \end{tabular}
  \end{center}
\end{table}
In Figs. \ref{NGC4183dens}, \ref{NGC4183} we present the density
of the analytic SIDM model, the predicted rotation curves for the
SIDM model (\ref{ScaledependentEoSDM}), versus the SPARC
observational data and the sound speed, as a function of the
radius respectively. As it can be seen, for this galaxy, the SIDM
model produces non-viable rotation curves which are incompatible
with the SPARC data.

\begin{figure}[h!]
\centering
\includegraphics[width=20pc]{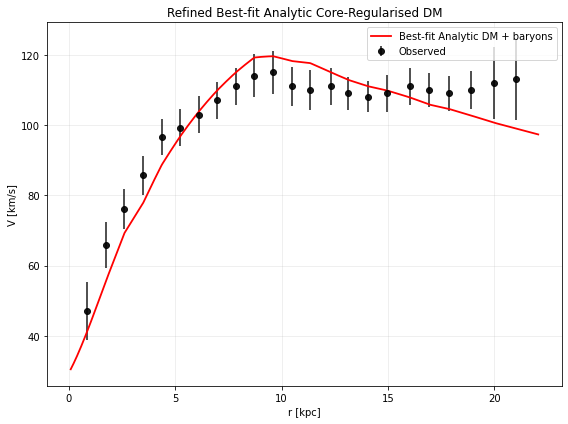}
\caption{The predicted rotation curves after using an optimization
for the SIDM model (\ref{ScaledependentEoSDM}), and the extended
SPARC data for the galaxy NGC4183. We included the rotation curves
of the gas, the disk velocities, the bulge (where present) along
with the SIDM model.} \label{extendedNGC4183}
\end{figure}

Now we shall include contributions to the rotation velocity from
the other components of the galaxy, namely the disk, the gas, and
the bulge if present. In Fig. \ref{extendedNGC4183} we present the
combined rotation curves including all the components of the
galaxy along with the SIDM. As it can be seen, the extended
collisional DM model is non-viable.
Also in Table \ref{evaluationextendedNGC4183} we present the
optimized values of the free parameters of the SIDM model for
which  we achieve the maximum compatibility with the SPARC data,
for the galaxy NGC4183, and also the resulting reduced
$\chi^2_{red}$ value.
\begin{table}[h!]
\centering \caption{Optimized Parameter Values of the Extended
SIDM model for the Galaxy NGC4183.}
\begin{tabular}{lc}
\hline
Parameter & Value  \\
\hline
$\rho_0 $ ($M_{\odot}/\mathrm{Kpc}^{3}$) & $1.32016\times 10^7$   \\
$K_0$ ($M_{\odot} \,
\mathrm{Kpc}^{-3} \, (\mathrm{km/s})^{2}$) & 3076.59   \\
$ml_{\text{disk}}$ & 1 \\
$ml_{\text{bulge}}$ & 0.0720 \\
$\alpha$ (Kpc) & 8.80883\\
$\chi^2_{red}$ & 1.20285 \\
\hline
\end{tabular}
\label{evaluationextendedNGC4183}
\end{table}

\subsection{The Galaxy NGC4217}

For this galaxy, the optimization method we used, ensures maximum
compatibility of the analytic SIDM model of Eq.
(\ref{ScaledependentEoSDM}) with the SPARC data, if we choose
$\rho_0=1.84193\times 10^8$$M_{\odot}/\mathrm{Kpc}^{3}$ and
$K_0=16273.8
$$M_{\odot} \, \mathrm{Kpc}^{-3} \, (\mathrm{km/s})^{2}$, in which
case the reduced $\chi^2_{red}$ value is $\chi^2_{red}=1.04953$.
Also the parameter $\alpha$ in this case is $\alpha=5.42449 $Kpc.

\begin{figure}[h!]
\centering
\includegraphics[width=20pc]{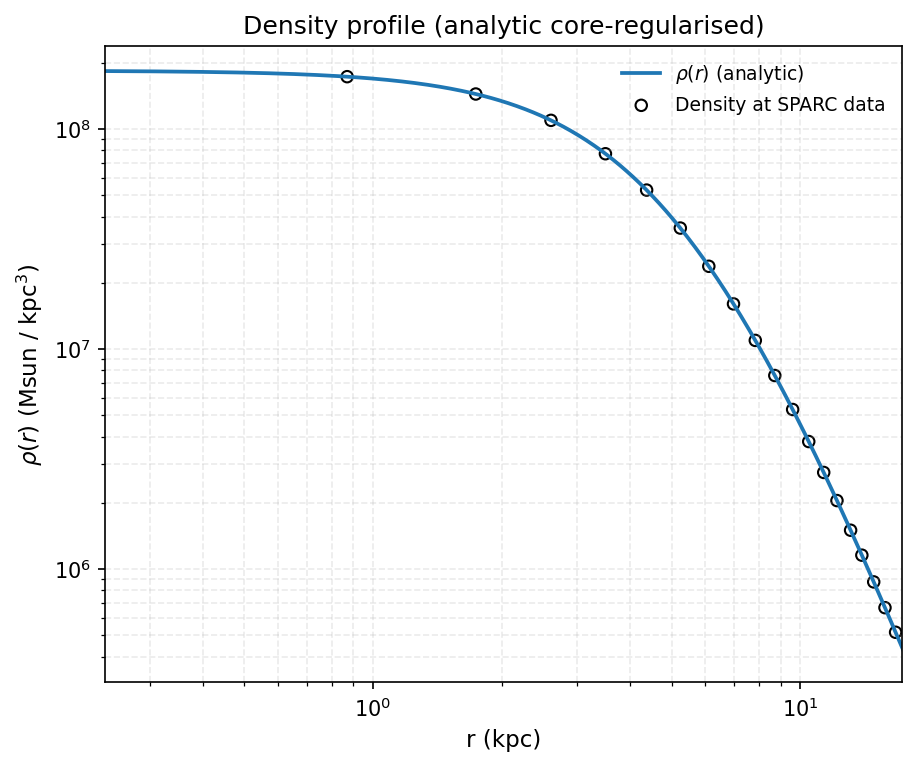}
\caption{The density of the SIDM model of Eq.
(\ref{ScaledependentEoSDM}) for the galaxy NGC4217, versus the
radius.} \label{NGC4217dens}
\end{figure}

\begin{figure}[h!]
\centering
\includegraphics[width=35pc]{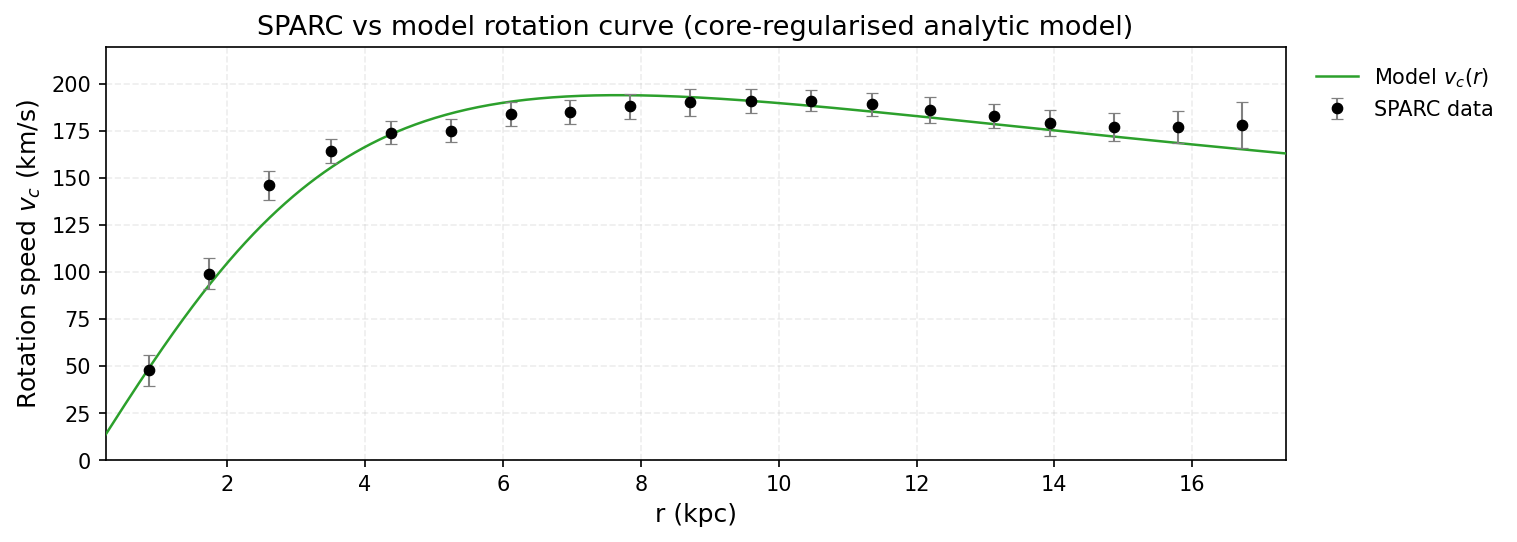}
\caption{The predicted rotation curves for the optimized SIDM
model of Eq. (\ref{ScaledependentEoSDM}), versus the SPARC
observational data for the galaxy NGC4217.} \label{NGC4217}
\end{figure}

In Table \ref{collNGC4217} we present the optimized values of
$K_0$ and $\rho_0$ for the analytic SIDM model of Eq.
(\ref{ScaledependentEoSDM}) for which the maximum compatibility
with the SPARC data is achieved.
\begin{table}[h!]
  \begin{center}
    \caption{SIDM Optimization Values for the galaxy NGC4217}
    \label{collNGC4217}
     \begin{tabular}{|r|r|}
     \hline
      \textbf{Parameter}   & \textbf{Optimization Values}
      \\  \hline
     $\rho_0 $  ($M_{\odot}/\mathrm{Kpc}^{3}$) & $5\times 10^7$
\\  \hline $K_0$ ($M_{\odot} \,
\mathrm{Kpc}^{-3} \, (\mathrm{km/s})^{2}$)& 1250
\\  \hline
    \end{tabular}
  \end{center}
\end{table}
In Figs. \ref{NGC4217dens}, \ref{NGC4217} we present the density
of the analytic SIDM model, the predicted rotation curves for the
SIDM model (\ref{ScaledependentEoSDM}), versus the SPARC
observational data and the sound speed, as a function of the
radius respectively. As it can be seen, for this galaxy, the SIDM
model produces marginally viable rotation curves which are
marginally compatible with the SPARC data.


\subsection{The Galaxy NGC4559, Non-viable, Extended Viable}

For this galaxy, the optimization method we used, ensures maximum
compatibility of the analytic SIDM model of Eq.
(\ref{ScaledependentEoSDM}) with the SPARC data, if we choose
$\rho_0=4.10065\times 10^7$$M_{\odot}/\mathrm{Kpc}^{3}$ and
$K_0=6819.38
$$M_{\odot} \, \mathrm{Kpc}^{-3} \, (\mathrm{km/s})^{2}$, in which
case the reduced $\chi^2_{red}$ value is $\chi^2_{red}=2.81172$.
Also the parameter $\alpha$ in this case is $\alpha=7.44212 $Kpc.

\begin{figure}[h!]
\centering
\includegraphics[width=20pc]{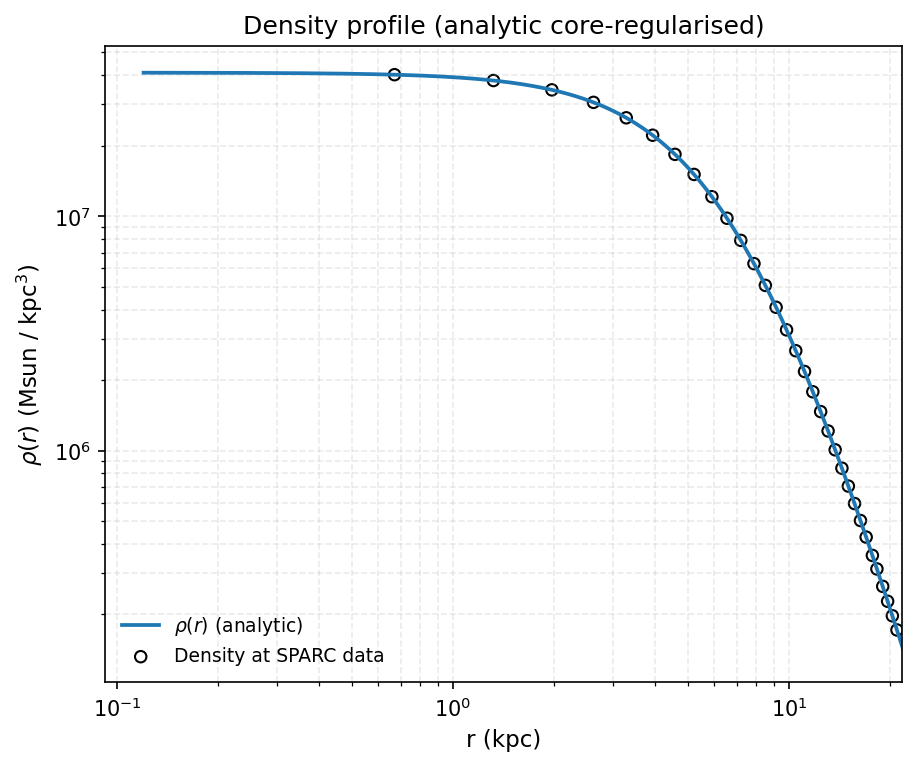}
\caption{The density of the SIDM model of Eq.
(\ref{ScaledependentEoSDM}) for the galaxy NGC4559, versus the
radius.} \label{NGC4559dens}
\end{figure}

\begin{figure}[h!]
\centering
\includegraphics[width=35pc]{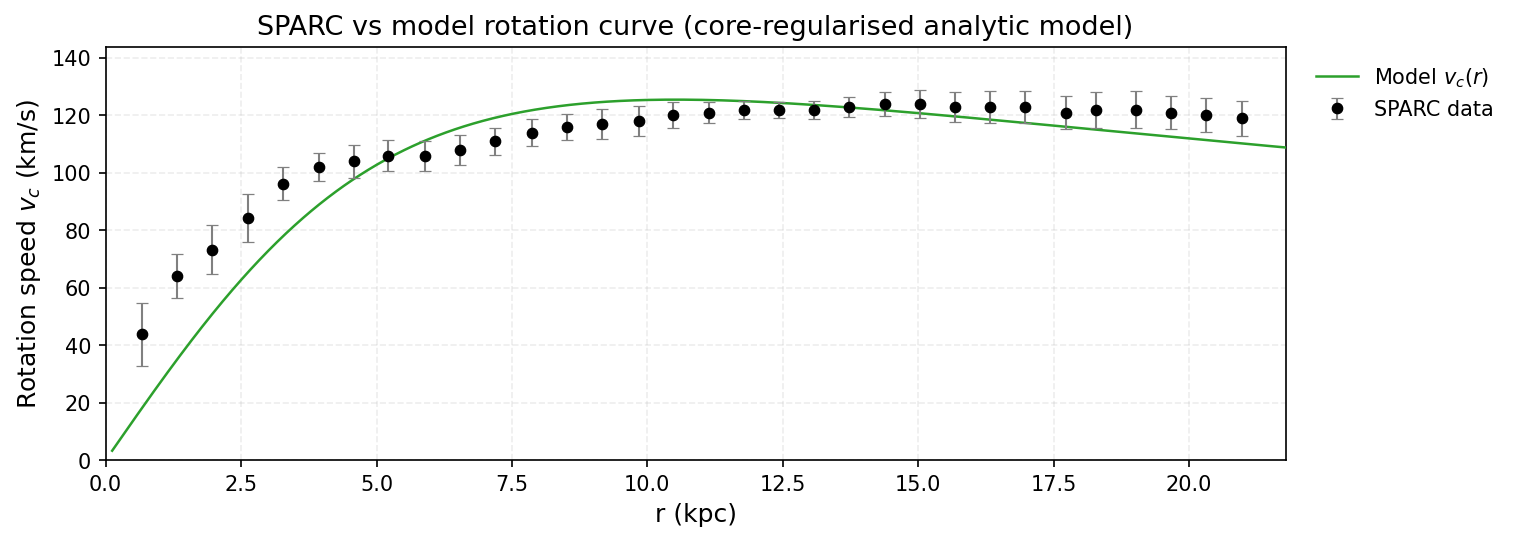}
\caption{The predicted rotation curves for the optimized SIDM
model of Eq. (\ref{ScaledependentEoSDM}), versus the SPARC
observational data for the galaxy NGC4559.} \label{NGC4559}
\end{figure}

In Table \ref{collNGC4559} we present the optimized values of
$K_0$ and $\rho_0$ for the analytic SIDM model of Eq.
(\ref{ScaledependentEoSDM}) for which the maximum compatibility
with the SPARC data is achieved.
\begin{table}[h!]
  \begin{center}
    \caption{SIDM Optimization Values for the galaxy NGC4559}
    \label{collNGC4559}
     \begin{tabular}{|r|r|}
     \hline
      \textbf{Parameter}   & \textbf{Optimization Values}
      \\  \hline
     $\rho_0 $  ($M_{\odot}/\mathrm{Kpc}^{3}$) & $4.10065\times 10^7$
\\  \hline $K_0$ ($M_{\odot} \,
\mathrm{Kpc}^{-3} \, (\mathrm{km/s})^{2}$)& 6819.38
\\  \hline
    \end{tabular}
  \end{center}
\end{table}
In Figs. \ref{NGC4559dens}, \ref{NGC4559} we present the density
of the analytic SIDM model, the predicted rotation curves for the
SIDM model (\ref{ScaledependentEoSDM}), versus the SPARC
observational data and the sound speed, as a function of the
radius respectively. As it can be seen, for this galaxy, the SIDM
model produces non-viable rotation curves which are incompatible
with the SPARC data.

\begin{figure}[h!]
\centering
\includegraphics[width=20pc]{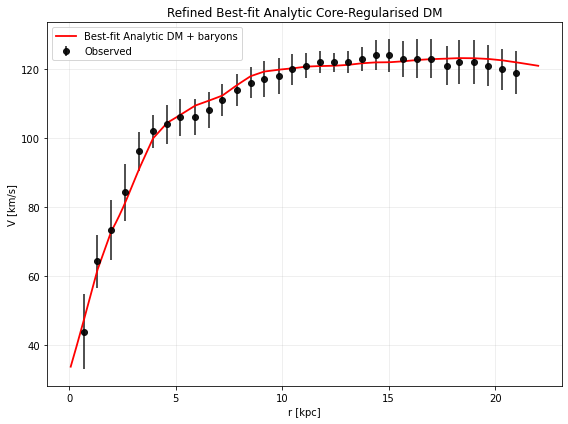}
\caption{The predicted rotation curves after using an optimization
for the SIDM model (\ref{ScaledependentEoSDM}), and the extended
SPARC data for the galaxy NGC4559. We included the rotation curves
of the gas, the disk velocities, the bulge (where present) along
with the SIDM model.} \label{extendedNGC4559}
\end{figure}

Now we shall include contributions to the rotation velocity from
the other components of the galaxy, namely the disk, the gas, and
the bulge if present. In Fig. \ref{extendedNGC4559} we present the
combined rotation curves including all the components of the
galaxy along with the SIDM. As it can be seen, the extended
collisional DM model is non-viable.
Also in Table \ref{evaluationextendedNGC4559} we present the
optimized values of the free parameters of the SIDM model for
which  we achieve the maximum compatibility with the SPARC data,
for the galaxy NGC4559, and also the resulting reduced
$\chi^2_{red}$ value.
\begin{table}[h!]
\centering \caption{Optimized Parameter Values of the Extended
SIDM model for the Galaxy NGC4559.}
\begin{tabular}{lc}
\hline
Parameter & Value  \\
\hline
$\rho_0 $ ($M_{\odot}/\mathrm{Kpc}^{3}$) & $7.1517\times 10^6$   \\
$K_0$ ($M_{\odot} \,
\mathrm{Kpc}^{-3} \, (\mathrm{km/s})^{2}$) & 4169.72   \\
$ml_{\text{disk}}$ & 0.7826 \\
$ml_{\text{bulge}}$ & 0.1341 \\
$\alpha$ (Kpc) & 13.933\\
$\chi^2_{red}$ & 0.15606 \\
\hline
\end{tabular}
\label{evaluationextendedNGC4559}
\end{table}

\subsection{The Galaxy NGC5005, Non-viable, Extended Viable}

For this galaxy, the optimization method we used, ensures maximum
compatibility of the analytic SIDM model of Eq.
(\ref{ScaledependentEoSDM}) with the SPARC data, if we choose
$\rho_0=1.1626\times 10^9$$M_{\odot}/\mathrm{Kpc}^{3}$ and
$K_0=35101.4
$$M_{\odot} \, \mathrm{Kpc}^{-3} \, (\mathrm{km/s})^{2}$, in which
case the reduced $\chi^2_{red}$ value is $\chi^2_{red}=4.28754$.
Also the parameter $\alpha$ in this case is $\alpha=3.17101 $Kpc.

\begin{figure}[h!]
\centering
\includegraphics[width=20pc]{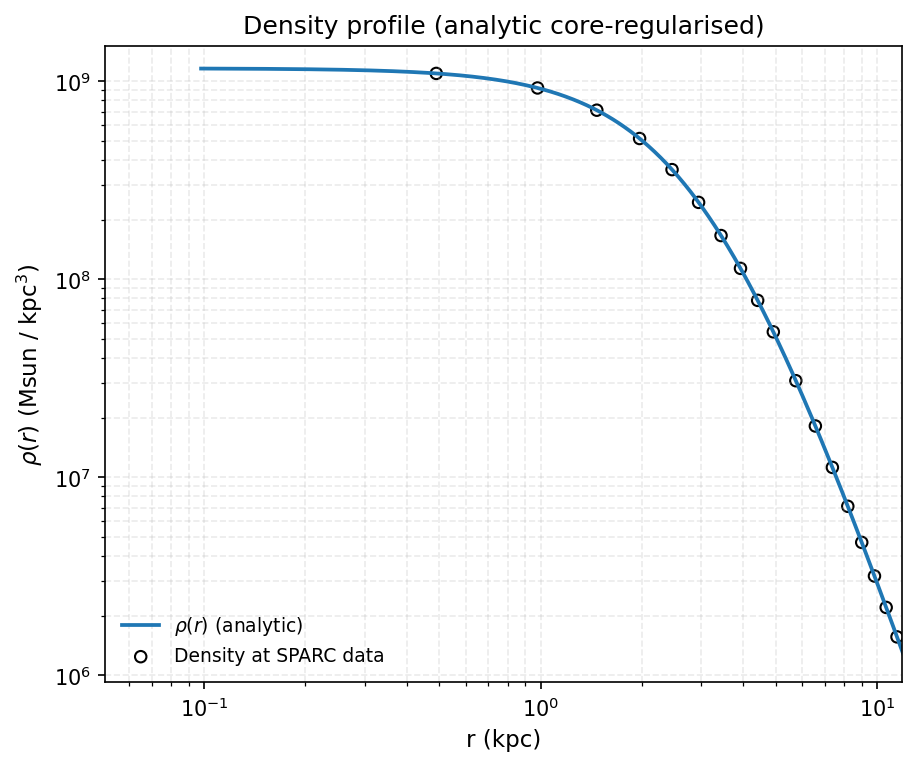}
\caption{The density of the SIDM model of Eq.
(\ref{ScaledependentEoSDM}) for the galaxy NGC5005, versus the
radius.} \label{NGC5005dens}
\end{figure}

\begin{figure}[h!]
\centering
\includegraphics[width=35pc]{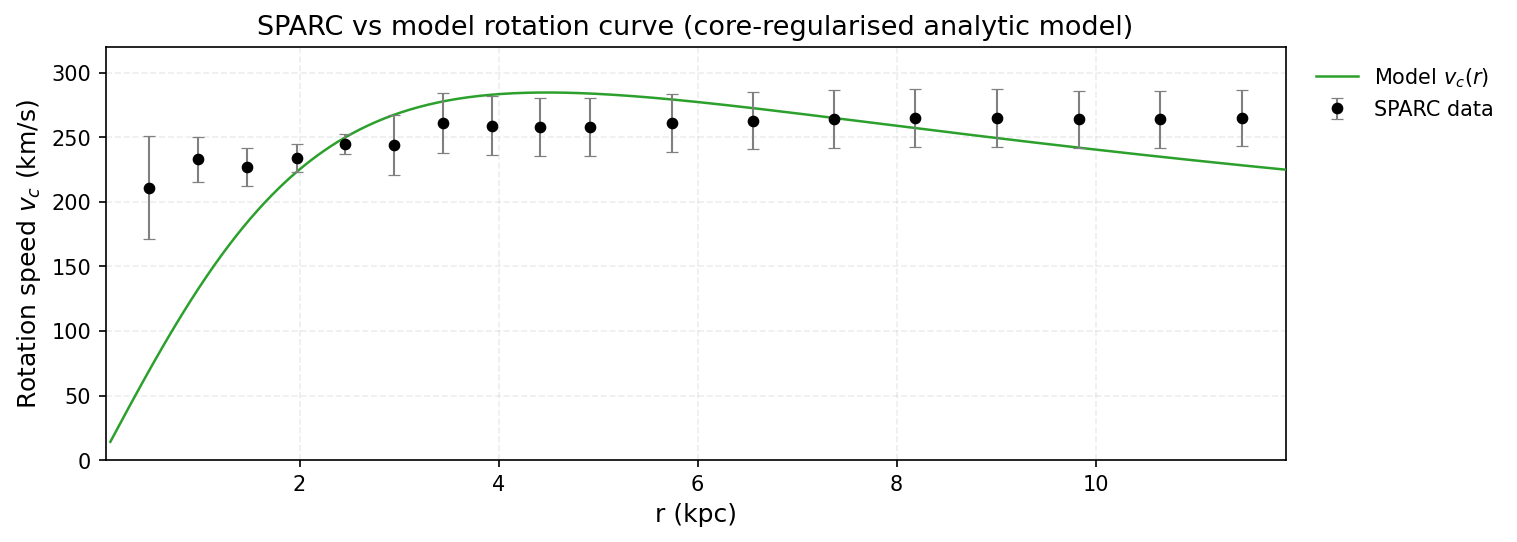}
\caption{The predicted rotation curves for the optimized SIDM
model of Eq. (\ref{ScaledependentEoSDM}), versus the SPARC
observational data for the galaxy NGC5005.} \label{NGC5005}
\end{figure}

In Table \ref{collNGC5005} we present the optimized values of
$K_0$ and $\rho_0$ for the analytic SIDM model of Eq.
(\ref{ScaledependentEoSDM}) for which the maximum compatibility
with the SPARC data is achieved.
\begin{table}[h!]
  \begin{center}
    \caption{SIDM Optimization Values for the galaxy NGC5005}
    \label{collNGC5005}
     \begin{tabular}{|r|r|}
     \hline
      \textbf{Parameter}   & \textbf{Optimization Values}
      \\  \hline
     $\rho_0 $  ($M_{\odot}/\mathrm{Kpc}^{3}$) & $1.1626\times 10^9$
\\  \hline $K_0$ ($M_{\odot} \,
\mathrm{Kpc}^{-3} \, (\mathrm{km/s})^{2}$)& 35101.4
\\  \hline
    \end{tabular}
  \end{center}
\end{table}
In Figs. \ref{NGC5005dens}, \ref{NGC5005} we present the density
of the analytic SIDM model, the predicted rotation curves for the
SIDM model (\ref{ScaledependentEoSDM}), versus the SPARC
observational data and the sound speed, as a function of the
radius respectively. As it can be seen, for this galaxy, the SIDM
model produces non-viable rotation curves which are incompatible
with the SPARC data.

\begin{figure}[h!]
\centering
\includegraphics[width=20pc]{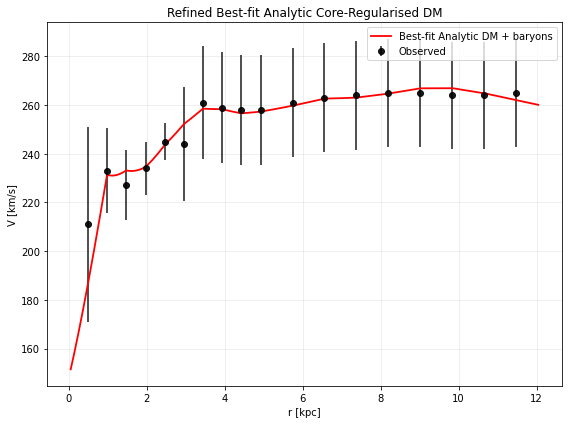}
\caption{The predicted rotation curves after using an optimization
for the SIDM model (\ref{ScaledependentEoSDM}), and the extended
SPARC data for the galaxy NGC5005. We included the rotation curves
of the gas, the disk velocities, the bulge (where present) along
with the SIDM model.} \label{extendedNGC5005}
\end{figure}

Now we shall include contributions to the rotation velocity from
the other components of the galaxy, namely the disk, the gas, and
the bulge if present. In Fig. \ref{extendedNGC5005} we present the
combined rotation curves including all the components of the
galaxy along with the SIDM. As it can be seen, the extended
collisional DM model is viable.
Also in Table \ref{evaluationextendedNGC5005} we present the
optimized values of the free parameters of the SIDM model for
which  we achieve the maximum compatibility with the SPARC data,
for the galaxy NGC5005, and also the resulting reduced
$\chi^2_{red}$ value.
\begin{table}[h!]
\centering \caption{Optimized Parameter Values of the Extended
SIDM model for the Galaxy NGC5005.}
\begin{tabular}{lc}
\hline
Parameter & Value  \\
\hline
$\rho_0 $ ($M_{\odot}/\mathrm{Kpc}^{3}$) & $1.51217\times 10^7$   \\
$K_0$ ($M_{\odot} \,
\mathrm{Kpc}^{-3} \, (\mathrm{km/s})^{2}$) & 17393.3   \\
$ml_{\text{disk}}$ & 0.7760 \\
$ml_{\text{bulge}}$ & 0.348 \\
$\alpha$ (Kpc) & 19.5698\\
$\chi^2_{red}$ & 0.0565593 \\
\hline
\end{tabular}
\label{evaluationextendedNGC5005}
\end{table}

\subsection{The Galaxy NGC5033, Non-viable}

For this galaxy, the optimization method we used, ensures maximum
compatibility of the analytic SIDM model of Eq.
(\ref{ScaledependentEoSDM}) with the SPARC data, if we choose
$\rho_0=1.1056\times 10^8$$M_{\odot}/\mathrm{Kpc}^{3}$ and
$K_0=26887.6
$$M_{\odot} \, \mathrm{Kpc}^{-3} \, (\mathrm{km/s})^{2}$, in which
case the reduced $\chi^2_{red}$ value is $\chi^2_{red}=59.3191$.
Also the parameter $\alpha$ in this case is $\alpha=8.99971 $Kpc.

\begin{figure}[h!]
\centering
\includegraphics[width=20pc]{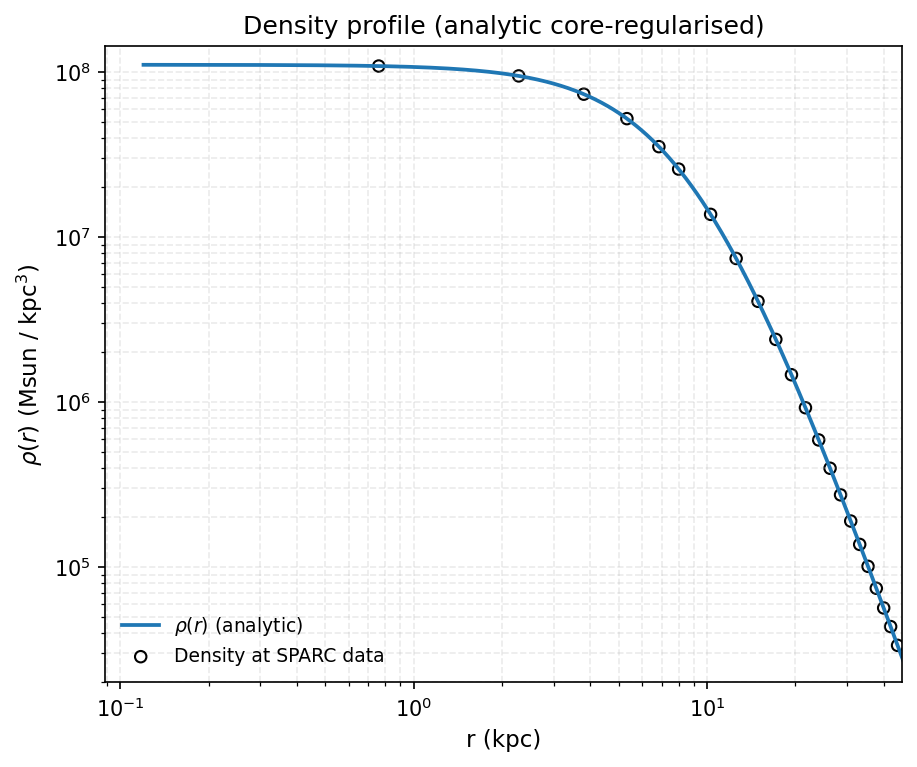}
\caption{The density of the SIDM model of Eq.
(\ref{ScaledependentEoSDM}) for the galaxy NGC5033, versus the
radius.} \label{NGC5033dens}
\end{figure}

\begin{figure}[h!]
\centering
\includegraphics[width=35pc]{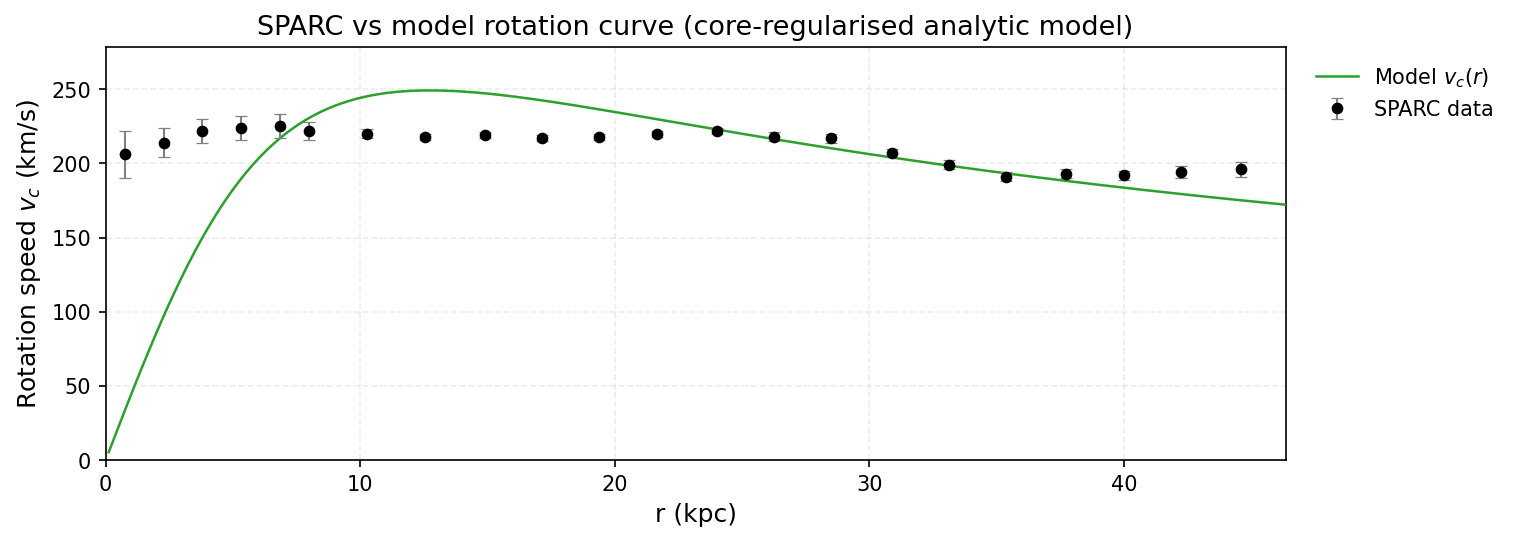}
\caption{The predicted rotation curves for the optimized SIDM
model of Eq. (\ref{ScaledependentEoSDM}), versus the SPARC
observational data for the galaxy NGC5033.} \label{NGC5033}
\end{figure}

In Table \ref{collNGC5033} we present the optimized values of
$K_0$ and $\rho_0$ for the analytic SIDM model of Eq.
(\ref{ScaledependentEoSDM}) for which the maximum compatibility
with the SPARC data is achieved.
\begin{table}[h!]
  \begin{center}
    \caption{SIDM Optimization Values for the galaxy NGC5033}
    \label{collNGC5033}
     \begin{tabular}{|r|r|}
     \hline
      \textbf{Parameter}   & \textbf{Optimization Values}
      \\  \hline
     $\rho_0 $  ($M_{\odot}/\mathrm{Kpc}^{3}$) & $1.1056\times 10^8$
\\  \hline $K_0$ ($M_{\odot} \,
\mathrm{Kpc}^{-3} \, (\mathrm{km/s})^{2}$)& 26887.6
\\  \hline
    \end{tabular}
  \end{center}
\end{table}
In Figs. \ref{NGC5033dens}, \ref{NGC5033} we present the density
of the analytic SIDM model, the predicted rotation curves for the
SIDM model (\ref{ScaledependentEoSDM}), versus the SPARC
observational data and the sound speed, as a function of the
radius respectively. As it can be seen, for this galaxy, the SIDM
model produces non-viable rotation curves which are incompatible
with the SPARC data.

\begin{figure}[h!]
\centering
\includegraphics[width=20pc]{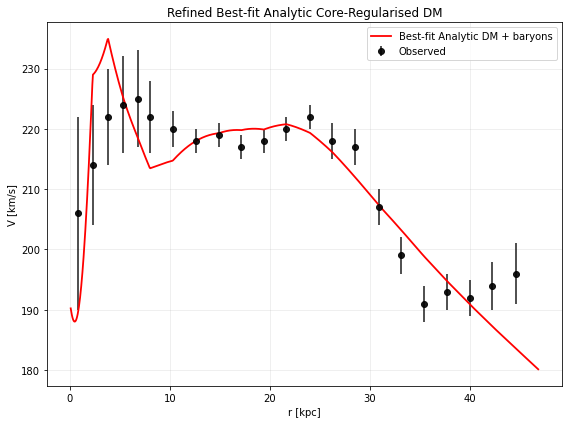}
\caption{The predicted rotation curves after using an optimization
for the SIDM model (\ref{ScaledependentEoSDM}), and the extended
SPARC data for the galaxy NGC5033. We included the rotation curves
of the gas, the disk velocities, the bulge (where present) along
with the SIDM model.} \label{extendedNGC5033}
\end{figure}

Now we shall include contributions to the rotation velocity from
the other components of the galaxy, namely the disk, the gas, and
the bulge if present. In Fig. \ref{extendedNGC5033} we present the
combined rotation curves including all the components of the
galaxy along with the SIDM. As it can be seen, the extended
collisional DM model is non-viable.
Also in Table \ref{evaluationextendedNGC5033} we present the
optimized values of the free parameters of the SIDM model for
which  we achieve the maximum compatibility with the SPARC data,
for the galaxy NGC5033, and also the resulting reduced
$\chi^2_{red}$ value.
\begin{table}[h!]
\centering \caption{Optimized Parameter Values of the Extended
SIDM model for the Galaxy NGC5033.}
\begin{tabular}{lc}
\hline
Parameter & Value  \\
\hline
$\rho_0 $ ($M_{\odot}/\mathrm{Kpc}^{3}$) & $1.57462\times 10^7$   \\
$K_0$ ($M_{\odot} \,
\mathrm{Kpc}^{-3} \, (\mathrm{km/s})^{2}$) & 13356.2   \\
$ml_{\text{disk}}$ & 0.8484 \\
$ml_{\text{bulge}}$ & 0.6147 \\
$\alpha$ (Kpc) & 16.8055\\
$\chi^2_{red}$ & 2.12808 \\
\hline
\end{tabular}
\label{evaluationextendedNGC5033}
\end{table}

\subsection{The Galaxy NGC5055, Non-viable}

For this galaxy, the optimization method we used, ensures maximum
compatibility of the analytic SIDM model of Eq.
(\ref{ScaledependentEoSDM}) with the SPARC data, if we choose
$\rho_0=6.4709\times 10^7$$M_{\odot}/\mathrm{Kpc}^{3}$ and
$K_0=23160.6
$$M_{\odot} \, \mathrm{Kpc}^{-3} \, (\mathrm{km/s})^{2}$, in which
case the reduced $\chi^2_{red}$ value is $\chi^2_{red}=900.8$.
Also the parameter $\alpha$ in this case is $\alpha=10.918 $Kpc.

\begin{figure}[h!]
\centering
\includegraphics[width=20pc]{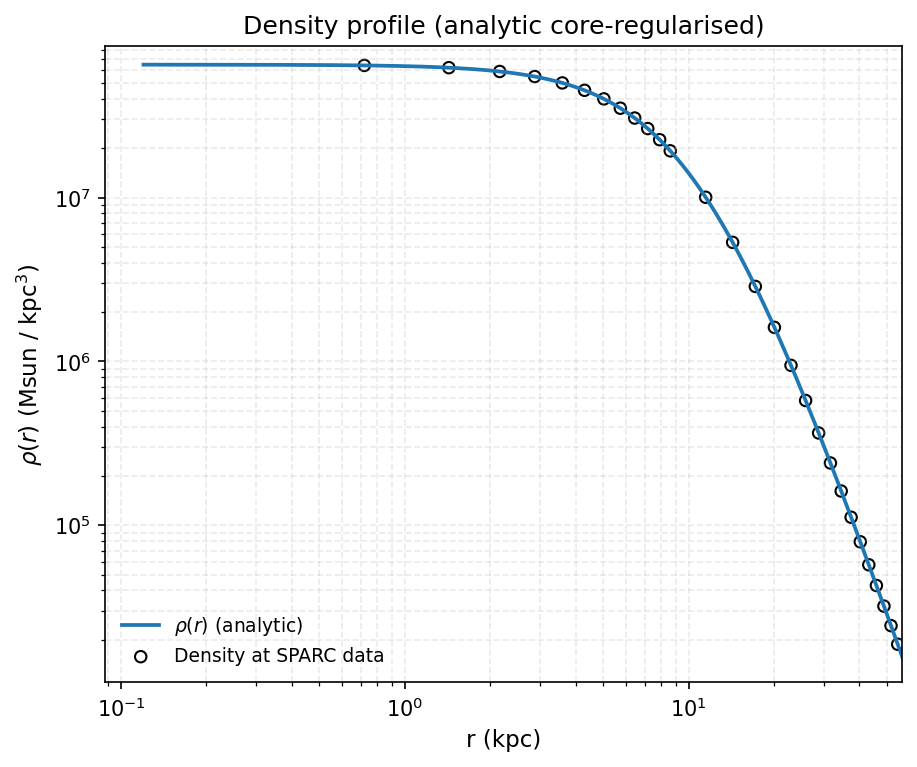}
\caption{The density of the SIDM model of Eq.
(\ref{ScaledependentEoSDM}) for the galaxy NGC5055, versus the
radius.} \label{NGC5055dens}
\end{figure}

\begin{figure}[h!]
\centering
\includegraphics[width=35pc]{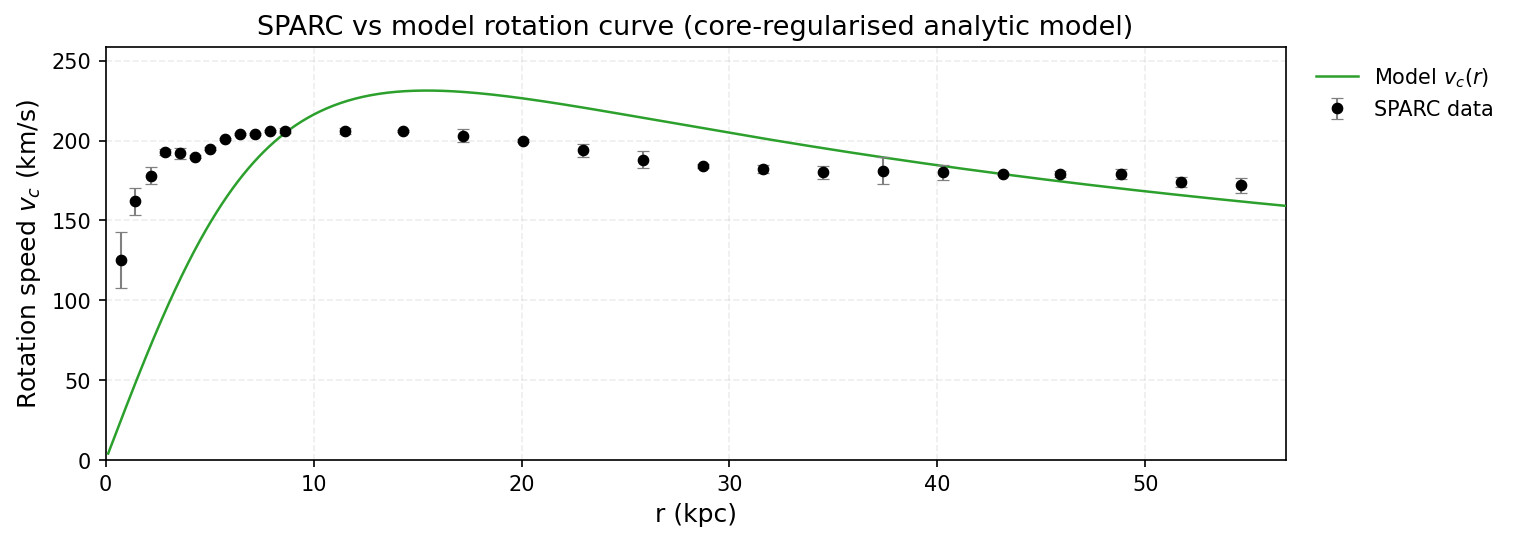}
\caption{The predicted rotation curves for the optimized SIDM
model of Eq. (\ref{ScaledependentEoSDM}), versus the SPARC
observational data for the galaxy NGC5055.} \label{NGC5055}
\end{figure}

In Table \ref{collNGC5055} we present the optimized values of
$K_0$ and $\rho_0$ for the analytic SIDM model of Eq.
(\ref{ScaledependentEoSDM}) for which the maximum compatibility
with the SPARC data is achieved.
\begin{table}[h!]
  \begin{center}
    \caption{SIDM Optimization Values for the galaxy NGC5055}
    \label{collNGC5055}
     \begin{tabular}{|r|r|}
     \hline
      \textbf{Parameter}   & \textbf{Optimization Values}
      \\  \hline
     $\rho_0 $  ($M_{\odot}/\mathrm{Kpc}^{3}$) & $6.4709\times 10^7$
\\  \hline $K_0$ ($M_{\odot} \,
\mathrm{Kpc}^{-3} \, (\mathrm{km/s})^{2}$)& 23160.6
\\  \hline
    \end{tabular}
  \end{center}
\end{table}
In Figs. \ref{NGC5055dens}, \ref{NGC5055} we present the density
of the analytic SIDM model, the predicted rotation curves for the
SIDM model (\ref{ScaledependentEoSDM}), versus the SPARC
observational data and the sound speed, as a function of the
radius respectively. As it can be seen, for this galaxy, the SIDM
model produces non-viable rotation curves which are incompatible
with the SPARC data.

\begin{figure}[h!]
\centering
\includegraphics[width=20pc]{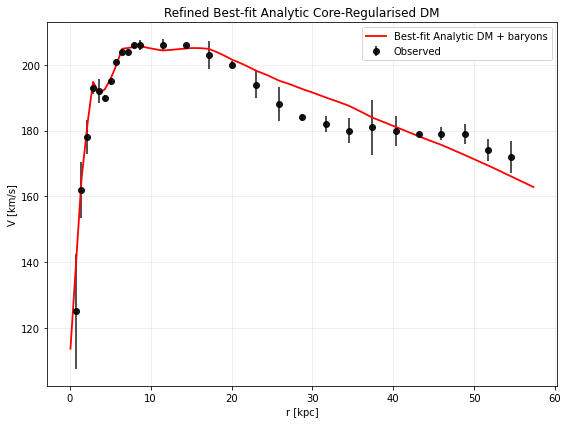}
\caption{The predicted rotation curves after using an optimization
for the SIDM model (\ref{ScaledependentEoSDM}), and the extended
SPARC data for the galaxy NGC5055. We included the rotation curves
of the gas, the disk velocities, the bulge (where present) along
with the SIDM model.} \label{extendedNGC5055}
\end{figure}

Now we shall include contributions to the rotation velocity from
the other components of the galaxy, namely the disk, the gas, and
the bulge if present. In Fig. \ref{extendedNGC5055} we present the
combined rotation curves including all the components of the
galaxy along with the SIDM. As it can be seen, the extended
collisional DM model is non-viable.
Also in Table \ref{evaluationextendedNGC5055} we present the
optimized values of the free parameters of the SIDM model for
which  we achieve the maximum compatibility with the SPARC data,
for the galaxy NGC5055, and also the resulting reduced
$\chi^2_{red}$ value.
\begin{table}[h!]
\centering \caption{Optimized Parameter Values of the Extended
SIDM model for the Galaxy NGC5055.}
\begin{tabular}{lc}
\hline
Parameter & Value  \\
\hline
$\rho_0 $ ($M_{\odot}/\mathrm{Kpc}^{3}$) & $7.63898\times 10^6$   \\
$K_0$ ($M_{\odot} \,
\mathrm{Kpc}^{-3} \, (\mathrm{km/s})^{2}$) & 10626.9   \\
$ml_{\text{disk}}$ & 0.6838 \\
$ml_{\text{bulge}}$ & 0.1214 \\
$\alpha$ (Kpc) & 21.5219\\
$\chi^2_{red}$ & 4.79553 \\
\hline
\end{tabular}
\label{evaluationextendedNGC5055}
\end{table}

\subsection{The Galaxy NGC5371, Non-viable}

For this galaxy, the optimization method we used, ensures maximum
compatibility of the analytic SIDM model of Eq.
(\ref{ScaledependentEoSDM}) with the SPARC data, if we choose
$\rho_0=9.64957\times 10^7$$M_{\odot}/\mathrm{Kpc}^{3}$ and
$K_0=28585.2
$$M_{\odot} \, \mathrm{Kpc}^{-3} \, (\mathrm{km/s})^{2}$, in which
case the reduced $\chi^2_{red}$ value is $\chi^2_{red}=47.002$.
Also the parameter $\alpha$ in this case is $\alpha=9.9327 $Kpc.

\begin{figure}[h!]
\centering
\includegraphics[width=20pc]{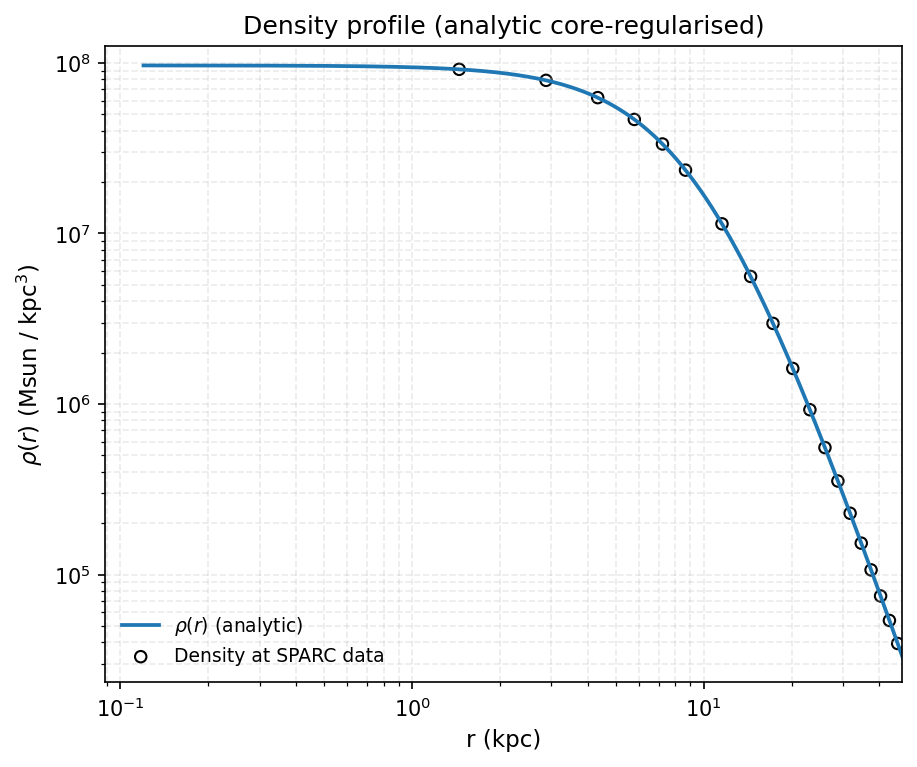}
\caption{The density of the SIDM model of Eq.
(\ref{ScaledependentEoSDM}) for the galaxy NGC5371, versus the
radius.} \label{NGC5371dens}
\end{figure}

\begin{figure}[h!]
\centering
\includegraphics[width=35pc]{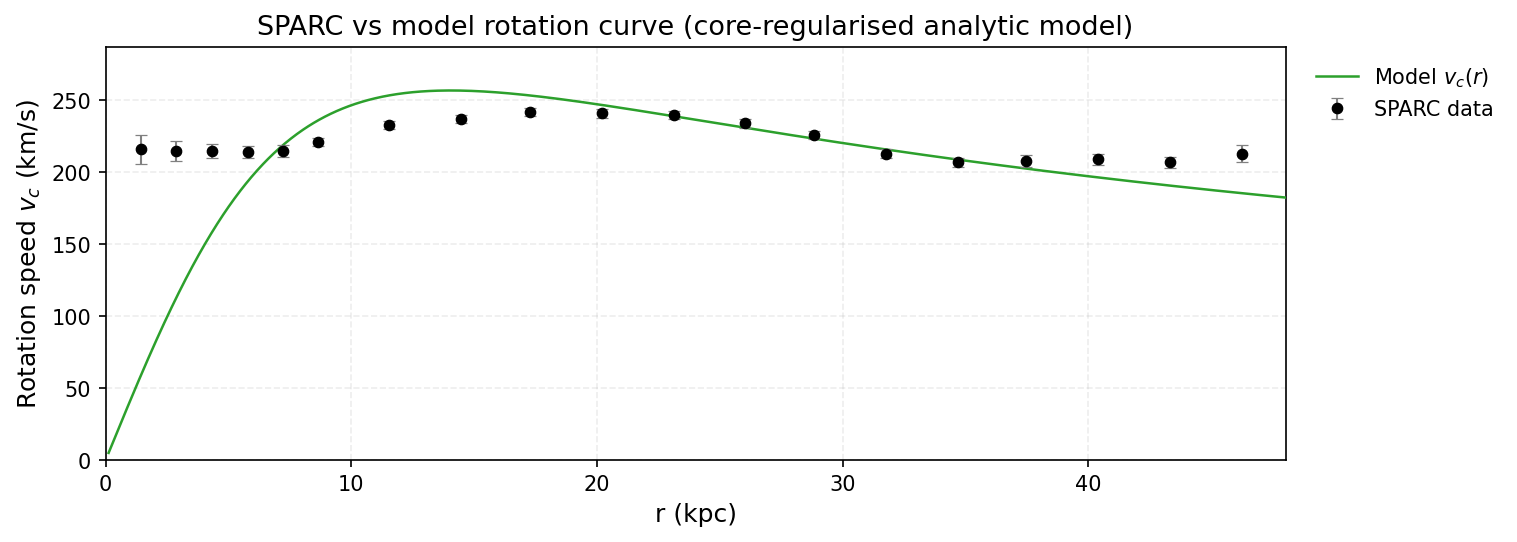}
\caption{The predicted rotation curves for the optimized SIDM
model of Eq. (\ref{ScaledependentEoSDM}), versus the SPARC
observational data for the galaxy NGC5371.} \label{NGC5371}
\end{figure}

In Table \ref{collNGC5371} we present the optimized values of
$K_0$ and $\rho_0$ for the analytic SIDM model of Eq.
(\ref{ScaledependentEoSDM}) for which the maximum compatibility
with the SPARC data is achieved.
\begin{table}[h!]
  \begin{center}
    \caption{SIDM Optimization Values for the galaxy NGC5371}
    \label{collNGC5371}
     \begin{tabular}{|r|r|}
     \hline
      \textbf{Parameter}   & \textbf{Optimization Values}
      \\  \hline
     $\rho_0 $  ($M_{\odot}/\mathrm{Kpc}^{3}$) & $9.64957\times 10^7$
\\  \hline $K_0$ ($M_{\odot} \,
\mathrm{Kpc}^{-3} \, (\mathrm{km/s})^{2}$)& 28585.2
\\  \hline
    \end{tabular}
  \end{center}
\end{table}
In Figs. \ref{NGC5371dens}, \ref{NGC5371} we present the density
of the analytic SIDM model, the predicted rotation curves for the
SIDM model (\ref{ScaledependentEoSDM}), versus the SPARC
observational data and the sound speed, as a function of the
radius respectively. As it can be seen, for this galaxy, the SIDM
model produces non-viable rotation curves which are incompatible
with the SPARC data.

\begin{figure}[h!]
\centering
\includegraphics[width=20pc]{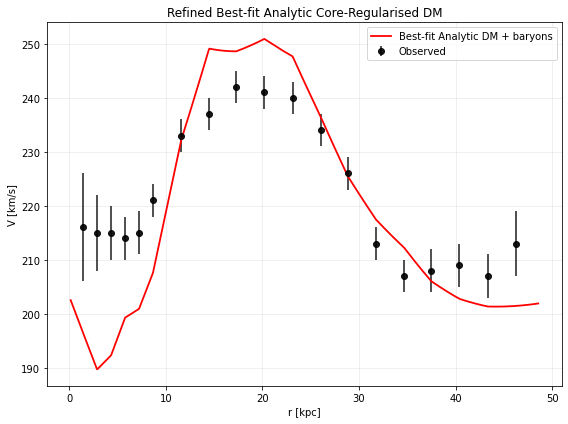}
\caption{The predicted rotation curves after using an optimization
for the SIDM model (\ref{ScaledependentEoSDM}), and the extended
SPARC data for the galaxy NGC5371. We included the rotation curves
of the gas, the disk velocities, the bulge (where present) along
with the SIDM model.} \label{extendedNGC5371}
\end{figure}

Now we shall include contributions to the rotation velocity from
the other components of the galaxy, namely the disk, the gas, and
the bulge if present. In Fig. \ref{extendedNGC5371} we present the
combined rotation curves including all the components of the
galaxy along with the SIDM. As it can be seen, the extended
collisional DM model is non-viable.
Also in Table \ref{evaluationextendedNGC5371} we present the
optimized values of the free parameters of the SIDM model for
which  we achieve the maximum compatibility with the SPARC data,
for the galaxy NGC5371, and also the resulting reduced
$\chi^2_{red}$ value.
\begin{table}[h!]
\centering \caption{Optimized Parameter Values of the Extended
SIDM model for the Galaxy NGC5371.}
\begin{tabular}{lc}
\hline
Parameter & Value  \\
\hline
$\rho_0 $ ($M_{\odot}/\mathrm{Kpc}^{3}$) & $4.3558\times 10^7$   \\
$K_0$ ($M_{\odot} \,
\mathrm{Kpc}^{-3} \, (\mathrm{km/s})^{2}$) & 40678   \\
$ml_{\text{disk}}$ & 0.8149 \\
$ml_{\text{bulge}}$ & 0.567 \\
$\alpha$ (Kpc) & 58.3994\\
$\chi^2_{red}$ & 9.09415 \\
\hline
\end{tabular}
\label{evaluationextendedNGC5371}
\end{table}

\subsection{The Galaxy NGC5585, Non-viable}

For this galaxy, the optimization method we used, ensures maximum
compatibility of the analytic SIDM model of Eq.
(\ref{ScaledependentEoSDM}) with the SPARC data, if we choose
$\rho_0=4.78164\times 10^7$$M_{\odot}/\mathrm{Kpc}^{3}$ and
$K_0=3635.48
$$M_{\odot} \, \mathrm{Kpc}^{-3} \, (\mathrm{km/s})^{2}$, in which
case the reduced $\chi^2_{red}$ value is $\chi^2_{red}=51.6287$.
Also the parameter $\alpha$ in this case is $\alpha=5.03203 $Kpc.

\begin{figure}[h!]
\centering
\includegraphics[width=20pc]{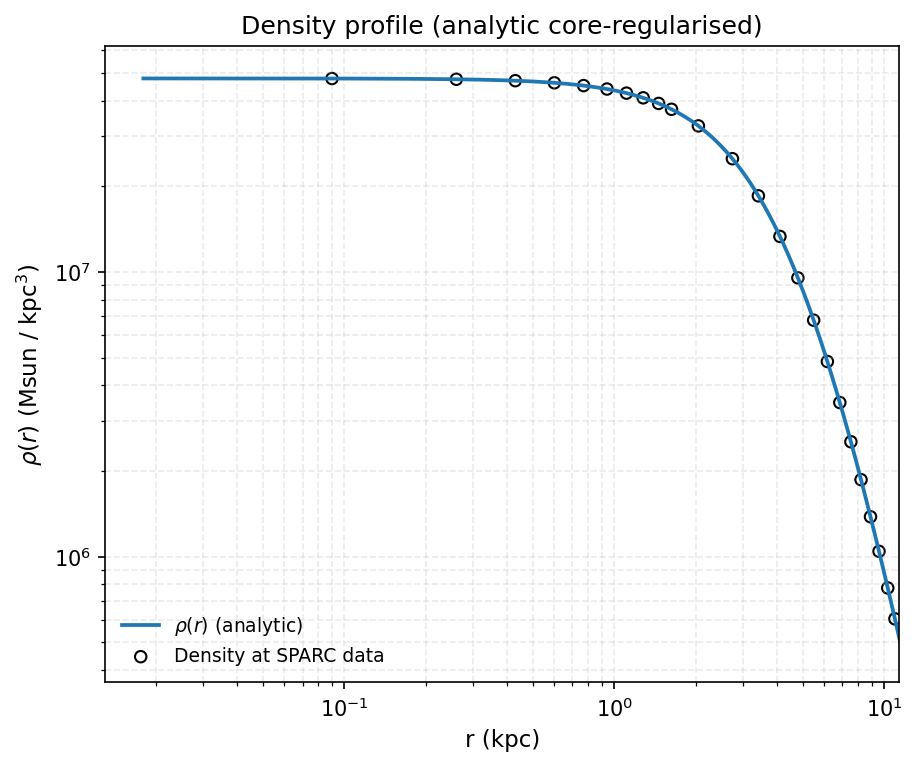}
\caption{The density of the SIDM model of Eq.
(\ref{ScaledependentEoSDM}) for the galaxy NGC5585, versus the
radius.} \label{NGC5585dens}
\end{figure}

\begin{figure}[h!]
\centering
\includegraphics[width=35pc]{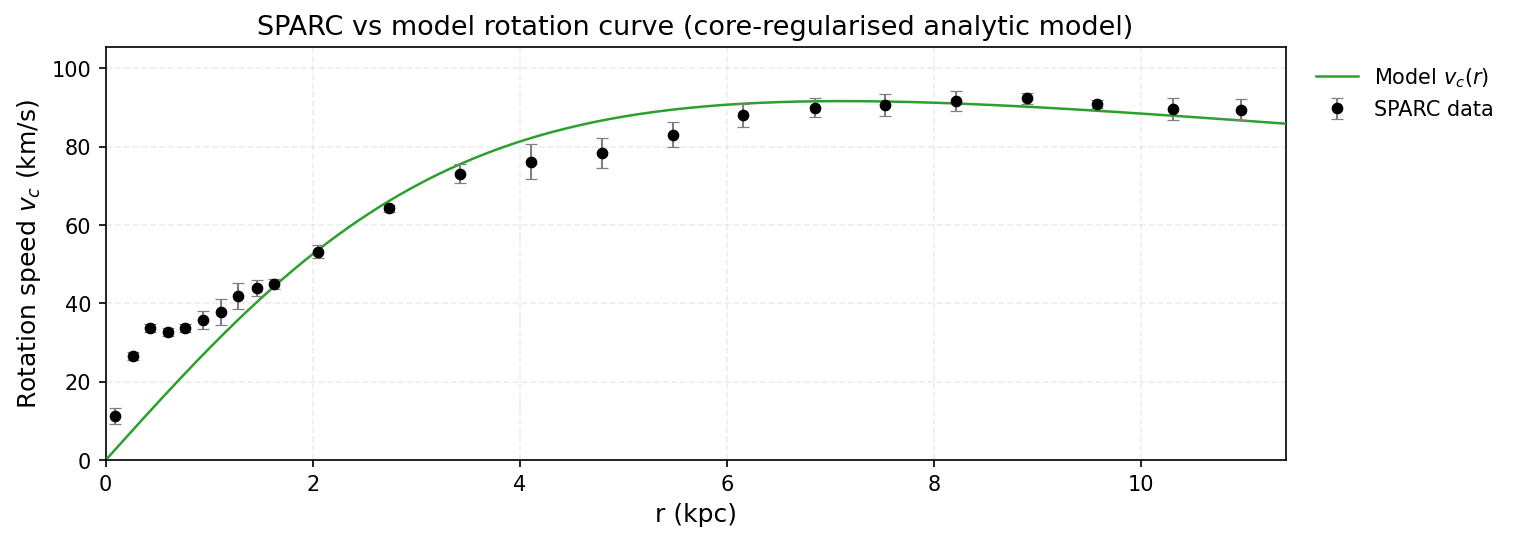}
\caption{The predicted rotation curves for the optimized SIDM
model of Eq. (\ref{ScaledependentEoSDM}), versus the SPARC
observational data for the galaxy NGC5585.} \label{NGC5585}
\end{figure}

In Table \ref{collNGC5585} we present the optimized values of
$K_0$ and $\rho_0$ for the analytic SIDM model of Eq.
(\ref{ScaledependentEoSDM}) for which the maximum compatibility
with the SPARC data is achieved.
\begin{table}[h!]
  \begin{center}
    \caption{SIDM Optimization Values for the galaxy NGC5585}
    \label{collNGC5585}
     \begin{tabular}{|r|r|}
     \hline
      \textbf{Parameter}   & \textbf{Optimization Values}
      \\  \hline
     $\rho_0 $  ($M_{\odot}/\mathrm{Kpc}^{3}$) & $4.78164\times 10^7$
\\  \hline $K_0$ ($M_{\odot} \,
\mathrm{Kpc}^{-3} \, (\mathrm{km/s})^{2}$)& 3635.48
\\  \hline
    \end{tabular}
  \end{center}
\end{table}
In Figs. \ref{NGC5585dens}, \ref{NGC5585} we present the density
of the analytic SIDM model, the predicted rotation curves for the
SIDM model (\ref{ScaledependentEoSDM}), versus the SPARC
observational data and the sound speed, as a function of the
radius respectively. As it can be seen, for this galaxy, the SIDM
model produces non-viable rotation curves which are incompatible
with the SPARC data.

\begin{figure}[h!]
\centering
\includegraphics[width=20pc]{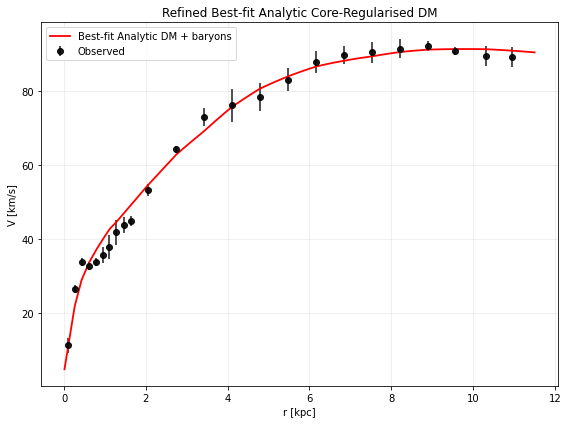}
\caption{The predicted rotation curves after using an optimization
for the SIDM model (\ref{ScaledependentEoSDM}), and the extended
SPARC data for the galaxy NGC5585. We included the rotation curves
of the gas, the disk velocities, the bulge (where present) along
with the SIDM model.} \label{extendedNGC5585}
\end{figure}

Now we shall include contributions to the rotation velocity from
the other components of the galaxy, namely the disk, the gas, and
the bulge if present. In Fig. \ref{extendedNGC5585} we present the
combined rotation curves including all the components of the
galaxy along with the SIDM. As it can be seen, the extended
collisional DM model is non-viable.
Also in Table \ref{evaluationextendedNGC5585} we present the
optimized values of the free parameters of the SIDM model for
which  we achieve the maximum compatibility with the SPARC data,
for the galaxy NGC5585, and also the resulting reduced
$\chi^2_{red}$ value.
\begin{table}[h!]
\centering \caption{Optimized Parameter Values of the Extended
SIDM model for the Galaxy NGC5585.}
\begin{tabular}{lc}
\hline
Parameter & Value  \\
\hline
$\rho_0 $ ($M_{\odot}/\mathrm{Kpc}^{3}$) & $1.65374\times 10^7$   \\
$K_0$ ($M_{\odot} \,
\mathrm{Kpc}^{-3} \, (\mathrm{km/s})^{2}$) & 2801.69   \\
$ml_{\text{disk}}$ & 0.7915 \\
$ml_{\text{bulge}}$ & 0.2957 \\
$\alpha$ (Kpc) & 7.51057\\
$\chi^2_{red}$ & 4.01573 \\
\hline
\end{tabular}
\label{evaluationextendedNGC5585}
\end{table}

\subsection{The Galaxy NGC5907, Non-viable}

For this galaxy, the optimization method we used, ensures maximum
compatibility of the analytic SIDM model of Eq.
(\ref{ScaledependentEoSDM}) with the SPARC data, if we choose
$\rho_0=7.04918\times 10^7$$M_{\odot}/\mathrm{Kpc}^{3}$ and
$K_0=26855.6
$$M_{\odot} \, \mathrm{Kpc}^{-3} \, (\mathrm{km/s})^{2}$, in which
case the reduced $\chi^2_{red}$ value is $\chi^2_{red}=34.0134$.
Also the parameter $\alpha$ in this case is $\alpha=11.2642 $Kpc.

\begin{figure}[h!]
\centering
\includegraphics[width=20pc]{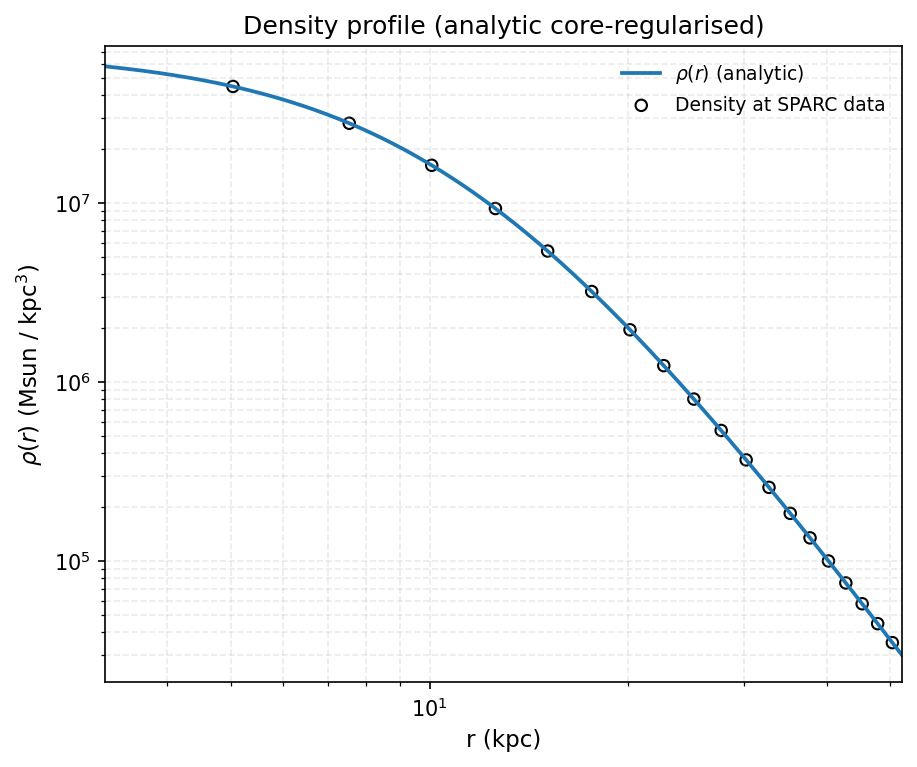}
\caption{The density of the SIDM model of Eq.
(\ref{ScaledependentEoSDM}) for the galaxy NGC5907, versus the
radius.} \label{NGC5907dens}
\end{figure}

\begin{figure}[h!]
\centering
\includegraphics[width=35pc]{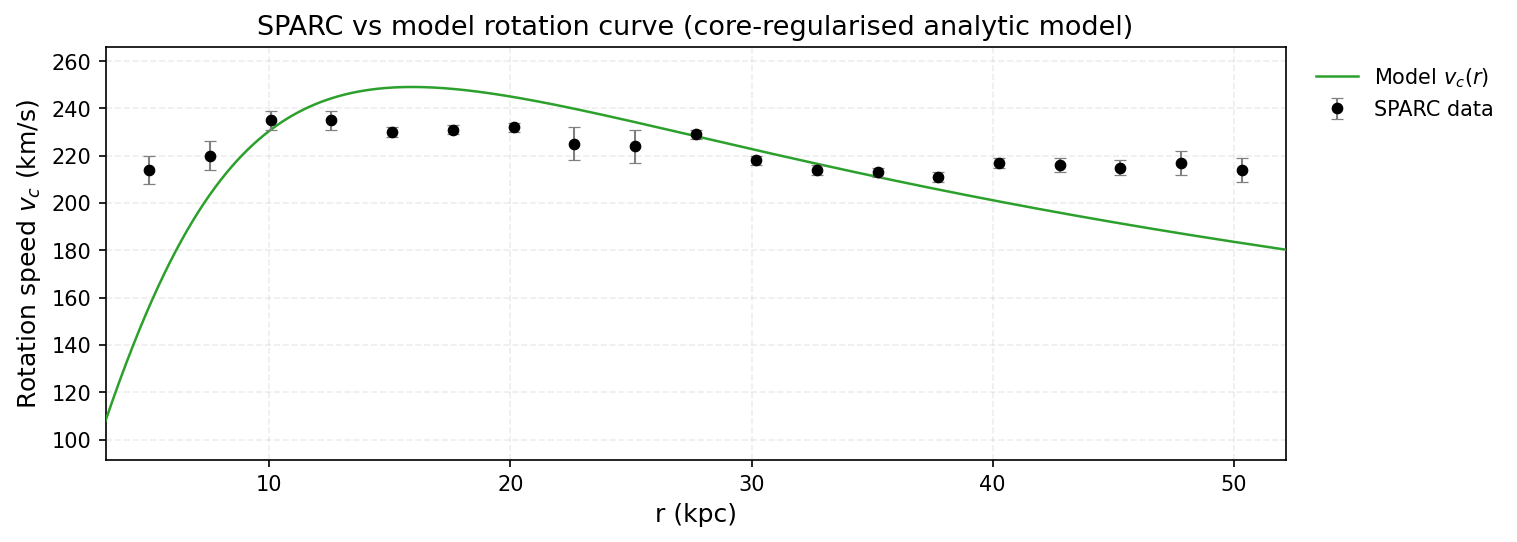}
\caption{The predicted rotation curves for the optimized SIDM
model of Eq. (\ref{ScaledependentEoSDM}), versus the SPARC
observational data for the galaxy NGC5907.} \label{NGC5907}
\end{figure}

In Table \ref{collNGC5907} we present the optimized values of
$K_0$ and $\rho_0$ for the analytic SIDM model of Eq.
(\ref{ScaledependentEoSDM}) for which the maximum compatibility
with the SPARC data is achieved.
\begin{table}[h!]
  \begin{center}
    \caption{SIDM Optimization Values for the galaxy NGC5907}
    \label{collNGC5907}
     \begin{tabular}{|r|r|}
     \hline
      \textbf{Parameter}   & \textbf{Optimization Values}
      \\  \hline
     $\rho_0 $  ($M_{\odot}/\mathrm{Kpc}^{3}$) & $7.04918\times 10^7$
\\  \hline $K_0$ ($M_{\odot} \,
\mathrm{Kpc}^{-3} \, (\mathrm{km/s})^{2}$)& 26855.6
\\  \hline
    \end{tabular}
  \end{center}
\end{table}
In Figs. \ref{NGC5907dens}, \ref{NGC5907} we present the density
of the analytic SIDM model, the predicted rotation curves for the
SIDM model (\ref{ScaledependentEoSDM}), versus the SPARC
observational data and the sound speed, as a function of the
radius respectively. As it can be seen, for this galaxy, the SIDM
model produces non-viable rotation curves which are incompatible
with the SPARC data.

\begin{figure}[h!]
\centering
\includegraphics[width=20pc]{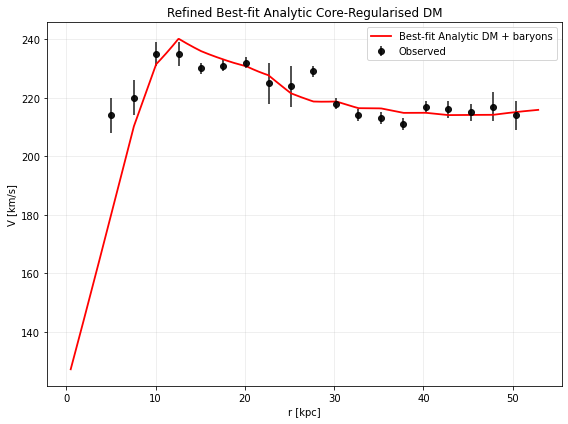}
\caption{The predicted rotation curves after using an optimization
for the SIDM model (\ref{ScaledependentEoSDM}), and the extended
SPARC data for the galaxy NGC5907. We included the rotation curves
of the gas, the disk velocities, the bulge (where present) along
with the SIDM model.} \label{extendedNGC5907}
\end{figure}

Now we shall include contributions to the rotation velocity from
the other components of the galaxy, namely the disk, the gas, and
the bulge if present. In Fig. \ref{extendedNGC5907} we present the
combined rotation curves including all the components of the
galaxy along with the SIDM. As it can be seen, the extended
collisional DM model is non-viable.
Also in Table \ref{evaluationextendedNGC5907} we present the
optimized values of the free parameters of the SIDM model for
which  we achieve the maximum compatibility with the SPARC data,
for the galaxy NGC5907, and also the resulting reduced
$\chi^2_{red}$ value.
\begin{table}[h!]
\centering \caption{Optimized Parameter Values of the Extended
SIDM model for the Galaxy NGC5907.}
\begin{tabular}{lc}
\hline
Parameter & Value  \\
\hline
$\rho_0 $ ($M_{\odot}/\mathrm{Kpc}^{3}$) & $1.29681\times 10^6$   \\
$K_0$ ($M_{\odot} \,
\mathrm{Kpc}^{-3} \, (\mathrm{km/s})^{2}$) & 13880.2   \\
$ml_{\text{disk}}$ & 0.9975 \\
$ml_{\text{bulge}}$ & 0.6240 \\
$\alpha$ (Kpc) & 59.6974\\
$\chi^2_{red}$ & 5.62445 \\
\hline
\end{tabular}
\label{evaluationextendedNGC5907}
\end{table}

\subsection{The Galaxy NGC5985, Non-viable}

For this galaxy, the optimization method we used, ensures maximum
compatibility of the analytic SIDM model of Eq.
(\ref{ScaledependentEoSDM}) with the SPARC data, if we choose
$\rho_0=1.73176\times 10^8$$M_{\odot}/\mathrm{Kpc}^{3}$ and
$K_0=42340
$$M_{\odot} \, \mathrm{Kpc}^{-3} \, (\mathrm{km/s})^{2}$, in which
case the reduced $\chi^2_{red}$ value is $\chi^2_{red}=14.2938$.
Also the parameter $\alpha$ in this case is $\alpha=9.02365 $Kpc.

\begin{figure}[h!]
\centering
\includegraphics[width=20pc]{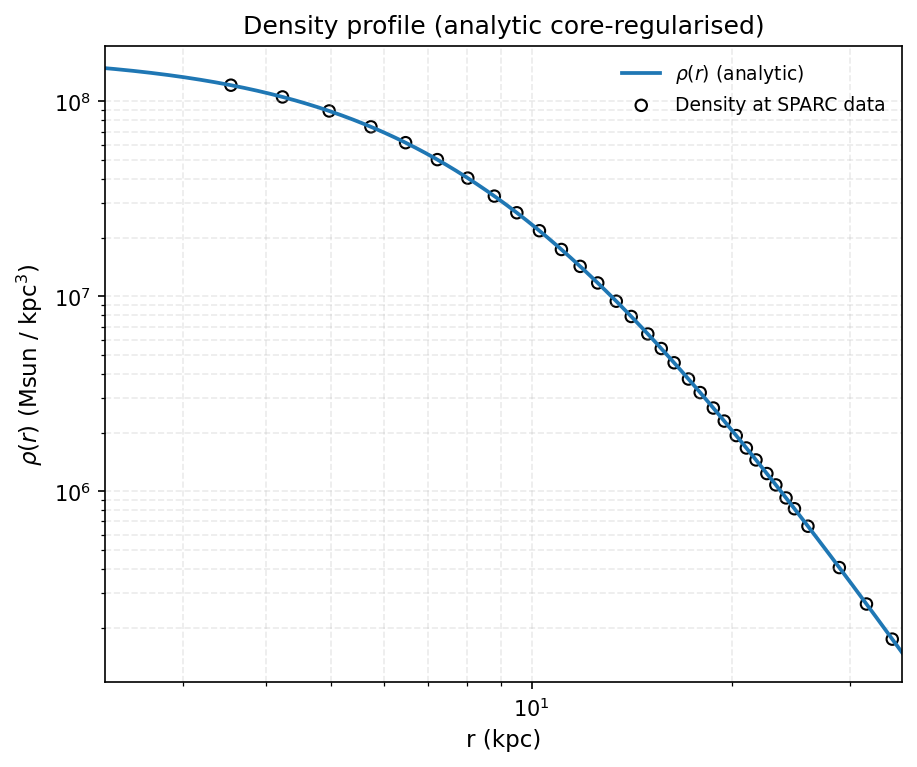}
\caption{The density of the SIDM model of Eq.
(\ref{ScaledependentEoSDM}) for the galaxy NGC5985, versus the
radius.} \label{NGC5985dens}
\end{figure}

\begin{figure}[h!]
\centering
\includegraphics[width=35pc]{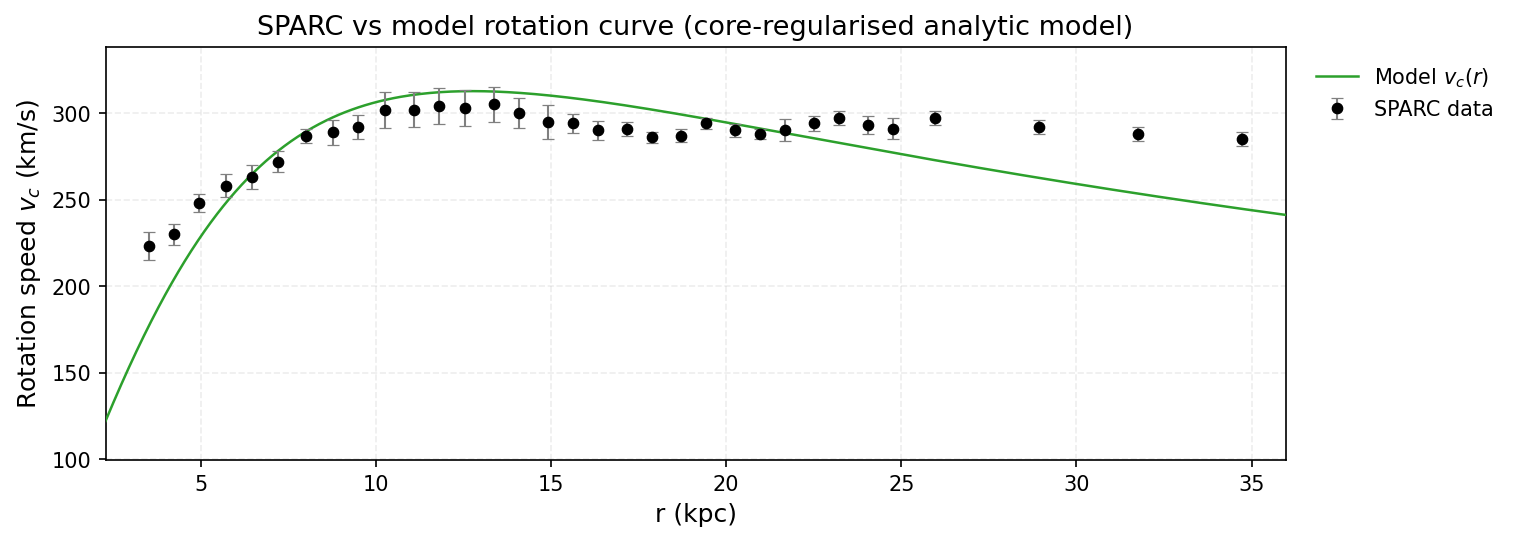}
\caption{The predicted rotation curves for the optimized SIDM
model of Eq. (\ref{ScaledependentEoSDM}), versus the SPARC
observational data for the galaxy NGC5985.} \label{NGC5985}
\end{figure}

In Table \ref{collNGC5985} we present the optimized values of
$K_0$ and $\rho_0$ for the analytic SIDM model of Eq.
(\ref{ScaledependentEoSDM}) for which the maximum compatibility
with the SPARC data is achieved.
\begin{table}[h!]
  \begin{center}
    \caption{SIDM Optimization Values for the galaxy NGC5985}
    \label{collNGC5985}
     \begin{tabular}{|r|r|}
     \hline
      \textbf{Parameter}   & \textbf{Optimization Values}
      \\  \hline
     $\rho_0 $  ($M_{\odot}/\mathrm{Kpc}^{3}$) & $1.73176\times 10^8$
\\  \hline $K_0$ ($M_{\odot} \,
\mathrm{Kpc}^{-3} \, (\mathrm{km/s})^{2}$)& 42340
\\  \hline
    \end{tabular}
  \end{center}
\end{table}
In Figs. \ref{NGC5985dens}, \ref{NGC5985} we present the density
of the analytic SIDM model, the predicted rotation curves for the
SIDM model (\ref{ScaledependentEoSDM}), versus the SPARC
observational data and the sound speed, as a function of the
radius respectively. As it can be seen, for this galaxy, the SIDM
model produces non-viable rotation curves which are incompatible
with the SPARC data.

\begin{figure}[h!]
\centering
\includegraphics[width=20pc]{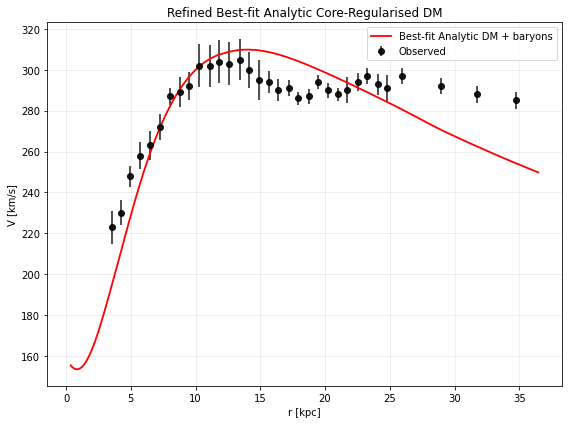}
\caption{The predicted rotation curves after using an optimization
for the SIDM model (\ref{ScaledependentEoSDM}), and the extended
SPARC data for the galaxy NGC5985. We included the rotation curves
of the gas, the disk velocities, the bulge (where present) along
with the SIDM model.} \label{extendedNGC5985}
\end{figure}

Now we shall include contributions to the rotation velocity from
the other components of the galaxy, namely the disk, the gas, and
the bulge if present. In Fig. \ref{extendedNGC5985} we present the
combined rotation curves including all the components of the
galaxy along with the SIDM. As it can be seen, the extended
collisional DM model is non-viable.
Also in Table \ref{evaluationextendedNGC5985} we present the
optimized values of the free parameters of the SIDM model for
which  we achieve the maximum compatibility with the SPARC data,
for the galaxy NGC5985, and also the resulting reduced
$\chi^2_{red}$ value.
\begin{table}[h!]
\centering \caption{Optimized Parameter Values of the Extended
SIDM model for the Galaxy NGC5985.}
\begin{tabular}{lc}
\hline
Parameter & Value  \\
\hline
$\rho_0 $ ($M_{\odot}/\mathrm{Kpc}^{3}$) & $1.35395\times 10^8$   \\
$K_0$ ($M_{\odot} \,
\mathrm{Kpc}^{-3} \, (\mathrm{km/s})^{2}$) & 40097.1   \\
$ml_{\text{disk}}$ & 1 \\
$ml_{\text{bulge}}$ & 1 \\
$\alpha$ (Kpc) & 9.9300\\
$\chi^2_{red}$ & 10.2045 \\
\hline
\end{tabular}
\label{evaluationextendedNGC5985}
\end{table}

\subsection{The Galaxy NGC6015, Non-viable}

For this galaxy, the optimization method we used, ensures maximum
compatibility of the analytic SIDM model of Eq.
(\ref{ScaledependentEoSDM}) with the SPARC data, if we choose
$\rho_0=1.38858\times 10^8$$M_{\odot}/\mathrm{Kpc}^{3}$ and
$K_0=15142
$$M_{\odot} \, \mathrm{Kpc}^{-3} \, (\mathrm{km/s})^{2}$, in which
case the reduced $\chi^2_{red}$ value is $\chi^2_{red}=59.7361$.
Also the parameter $\alpha$ in this case is $\alpha=6.0264 $Kpc.

\begin{figure}[h!]
\centering
\includegraphics[width=20pc]{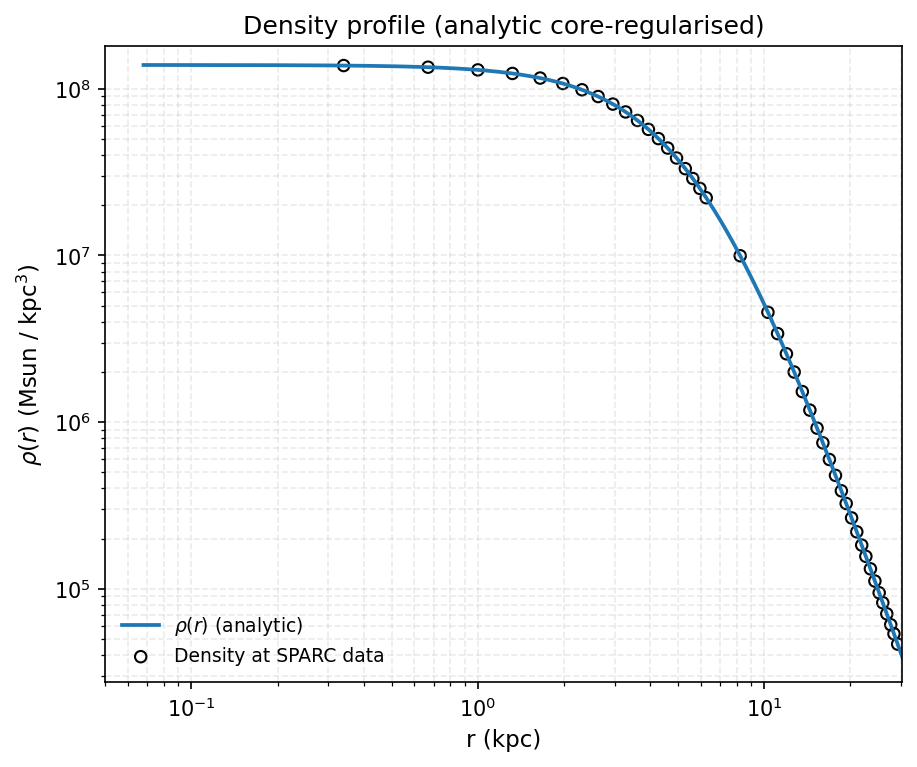}
\caption{The density of the SIDM model of Eq.
(\ref{ScaledependentEoSDM}) for the galaxy NGC6015, versus the
radius.} \label{NGC6015dens}
\end{figure}

\begin{figure}[h!]
\centering
\includegraphics[width=35pc]{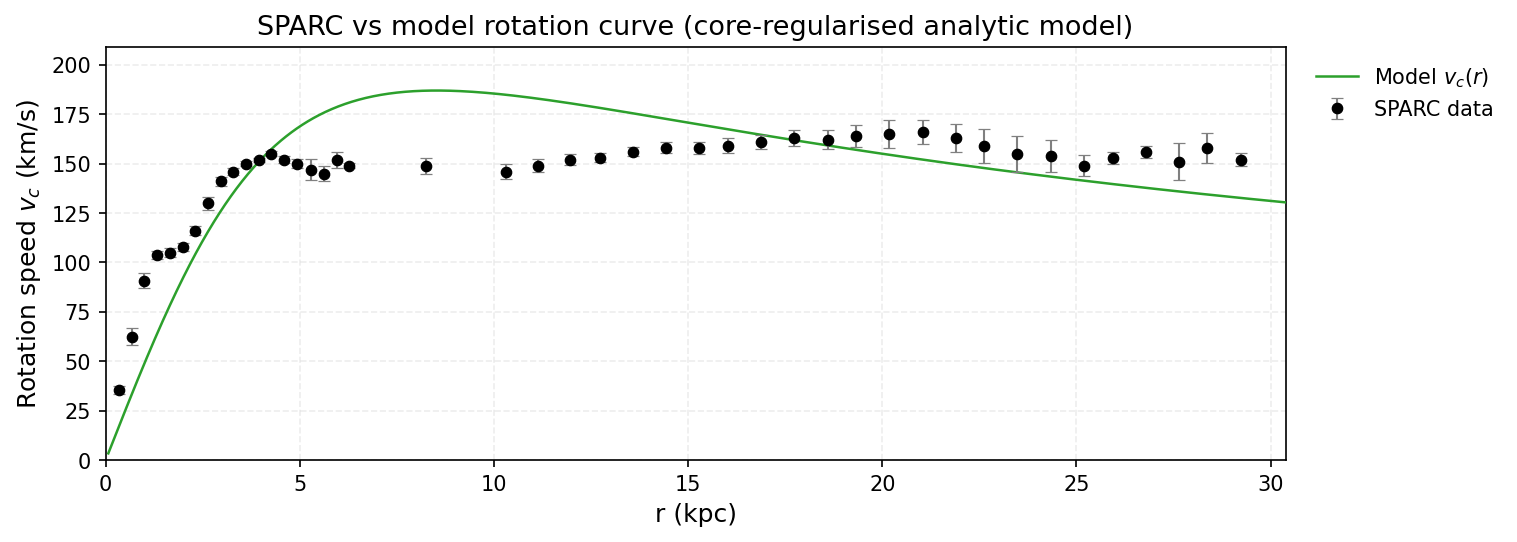}
\caption{The predicted rotation curves for the optimized SIDM
model of Eq. (\ref{ScaledependentEoSDM}), versus the SPARC
observational data for the galaxy NGC6015.} \label{NGC6015}
\end{figure}

In Table \ref{collNGC6015} we present the optimized values of
$K_0$ and $\rho_0$ for the analytic SIDM model of Eq.
(\ref{ScaledependentEoSDM}) for which the maximum compatibility
with the SPARC data is achieved.
\begin{table}[h!]
  \begin{center}
    \caption{SIDM Optimization Values for the galaxy NGC6015}
    \label{collNGC6015}
     \begin{tabular}{|r|r|}
     \hline
      \textbf{Parameter}   & \textbf{Optimization Values}
      \\  \hline
     $\rho_0 $  ($M_{\odot}/\mathrm{Kpc}^{3}$) & $1.38858\times 10^8$
\\  \hline $K_0$ ($M_{\odot} \,
\mathrm{Kpc}^{-3} \, (\mathrm{km/s})^{2}$)& 15142
\\  \hline
    \end{tabular}
  \end{center}
\end{table}
In Figs. \ref{NGC6015dens}, \ref{NGC6015} we present the density
of the analytic SIDM model, the predicted rotation curves for the
SIDM model (\ref{ScaledependentEoSDM}), versus the SPARC
observational data and the sound speed, as a function of the
radius respectively. As it can be seen, for this galaxy, the SIDM
model produces non-viable rotation curves which are incompatible
with the SPARC data.

\begin{figure}[h!]
\centering
\includegraphics[width=20pc]{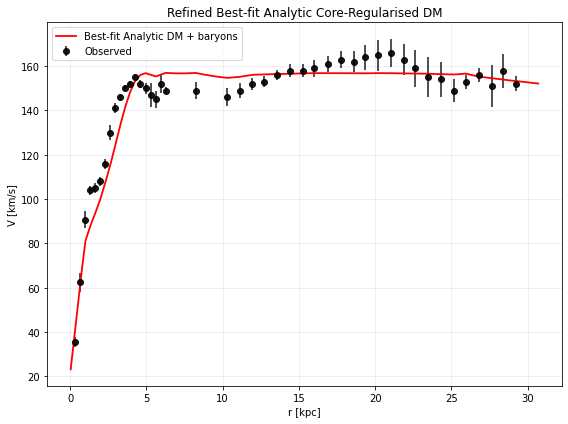}
\caption{The predicted rotation curves after using an optimization
for the SIDM model (\ref{ScaledependentEoSDM}), and the extended
SPARC data for the galaxy NGC6015. We included the rotation curves
of the gas, the disk velocities, the bulge (where present) along
with the SIDM model.} \label{extendedNGC6015}
\end{figure}

Now we shall include contributions to the rotation velocity from
the other components of the galaxy, namely the disk, the gas, and
the bulge if present. In Fig. \ref{extendedNGC6015} we present the
combined rotation curves including all the components of the
galaxy along with the SIDM. As it can be seen, the extended
collisional DM model is non-viable.
Also in Table \ref{evaluationextendedNGC6015} we present the
optimized values of the free parameters of the SIDM model for
which  we achieve the maximum compatibility with the SPARC data,
for the galaxy NGC6015, and also the resulting reduced
$\chi^2_{red}$ value.
\begin{table}[h!]
\centering \caption{Optimized Parameter Values of the Extended
SIDM model for the Galaxy NGC6015.}
\begin{tabular}{lc}
\hline
Parameter & Value  \\
\hline
$\rho_0 $ ($M_{\odot}/\mathrm{Kpc}^{3}$) & $5.11985\times 10^8$   \\
$K_0$ ($M_{\odot} \,
\mathrm{Kpc}^{-3} \, (\mathrm{km/s})^{2}$) & 7447.45   \\
$ml_{\text{disk}}$ & 1 \\
$ml_{\text{bulge}}$ & 0.6368 \\
$\alpha$ (Kpc) & 22.0075\\
$\chi^2_{red}$ & 10.2067 \\
\hline
\end{tabular}
\label{evaluationextendedNGC6015}
\end{table}

\subsection{The Galaxy NGC6195, Non-viable, Extended Viable}

For this galaxy, the optimization method we used, ensures maximum
compatibility of the analytic SIDM model of Eq.
(\ref{ScaledependentEoSDM}) with the SPARC data, if we choose
$\rho_0=1.66739\times 10^8$$M_{\odot}/\mathrm{Kpc}^{3}$ and
$K_0=30314
$$M_{\odot} \, \mathrm{Kpc}^{-3} \, (\mathrm{km/s})^{2}$, in which
case the reduced $\chi^2_{red}$ value is $\chi^2_{red}=102.22$.
Also the parameter $\alpha$ in this case is $\alpha=7.78133 $Kpc.

\begin{figure}[h!]
\centering
\includegraphics[width=20pc]{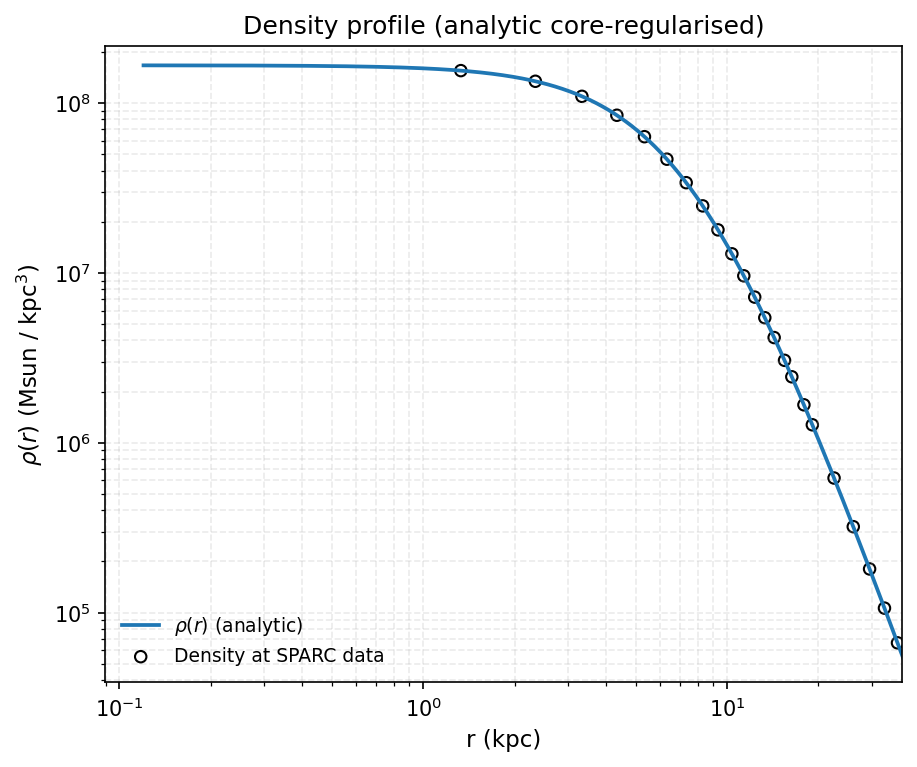}
\caption{The density of the SIDM model of Eq.
(\ref{ScaledependentEoSDM}) for the galaxy NGC6195, versus the
radius.} \label{NGC6195dens}
\end{figure}

\begin{figure}[h!]
\centering
\includegraphics[width=35pc]{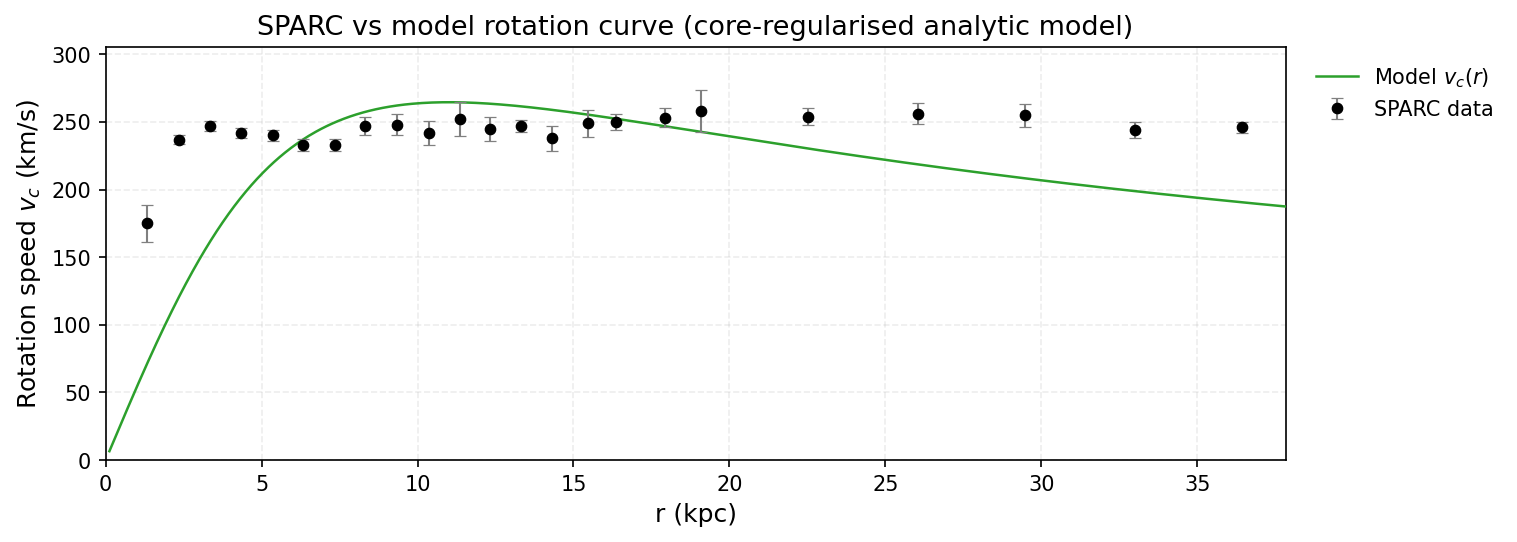}
\caption{The predicted rotation curves for the optimized SIDM
model of Eq. (\ref{ScaledependentEoSDM}), versus the SPARC
observational data for the galaxy NGC6195.} \label{NGC6195}
\end{figure}

In Table \ref{collNGC6195} we present the optimized values of
$K_0$ and $\rho_0$ for the analytic SIDM model of Eq.
(\ref{ScaledependentEoSDM}) for which the maximum compatibility
with the SPARC data is achieved.
\begin{table}[h!]
  \begin{center}
    \caption{SIDM Optimization Values for the galaxy NGC6195}
    \label{collNGC6195}
     \begin{tabular}{|r|r|}
     \hline
      \textbf{Parameter}   & \textbf{Optimization Values}
      \\  \hline
     $\rho_0 $  ($M_{\odot}/\mathrm{Kpc}^{3}$) & $1.66739\times 10^8$
\\  \hline $K_0$ ($M_{\odot} \,
\mathrm{Kpc}^{-3} \, (\mathrm{km/s})^{2}$)& 30314
\\  \hline
    \end{tabular}
  \end{center}
\end{table}
In Figs. \ref{NGC6195dens}, \ref{NGC6195} we present the density
of the analytic SIDM model, the predicted rotation curves for the
SIDM model (\ref{ScaledependentEoSDM}), versus the SPARC
observational data and the sound speed, as a function of the
radius respectively. As it can be seen, for this galaxy, the SIDM
model produces non-viable rotation curves which are incompatible
with the SPARC data.

\begin{figure}[h!]
\centering
\includegraphics[width=20pc]{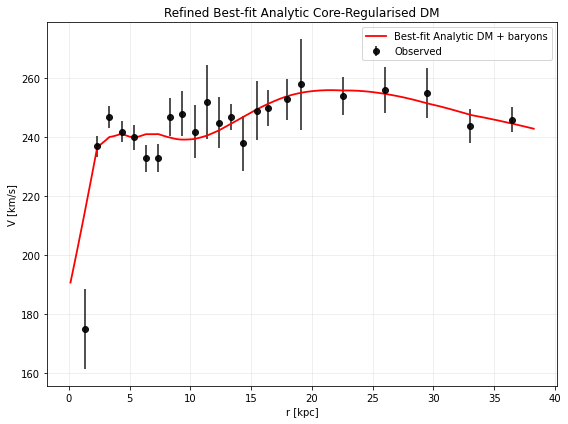}
\caption{The predicted rotation curves after using an optimization
for the SIDM model (\ref{ScaledependentEoSDM}), and the extended
SPARC data for the galaxy NGC6195. We included the rotation curves
of the gas, the disk velocities, the bulge (where present) along
with the SIDM model.} \label{extendedNGC6195}
\end{figure}

Now we shall include contributions to the rotation velocity from
the other components of the galaxy, namely the disk, the gas, and
the bulge if present. In Fig. \ref{extendedNGC6195} we present the
combined rotation curves including all the components of the
galaxy along with the SIDM. As it can be seen, the extended
collisional DM model is non-viable.
Also in Table \ref{evaluationextendedNGC6195} we present the
optimized values of the free parameters of the SIDM model for
which  we achieve the maximum compatibility with the SPARC data,
for the galaxy NGC6195, and also the resulting reduced
$\chi^2_{red}$ value.
\begin{table}[h!]
\centering \caption{Optimized Parameter Values of the Extended
SIDM model for the Galaxy NGC6195.}
\begin{tabular}{lc}
\hline
Parameter & Value  \\
\hline
$\rho_0 $ ($M_{\odot}/\mathrm{Kpc}^{3}$) & $1.56515\times 10^7$   \\
$K_0$ ($M_{\odot} \,
\mathrm{Kpc}^{-3} \, (\mathrm{km/s})^{2}$) & 20264.7   \\
$ml_{\text{disk}}$ & 0.8 \\
$ml_{\text{bulge}}$ & 0.9037 \\
$\alpha$ (Kpc) & 20.7629\\
$\chi^2_{red}$ & 1.25343 \\
\hline
\end{tabular}
\label{evaluationextendedNGC6195}
\end{table}

\subsection{The Galaxy NGC6503, Non-viable}

For this galaxy, the optimization method we used, ensures maximum
compatibility of the analytic SIDM model of Eq.
(\ref{ScaledependentEoSDM}) with the SPARC data, if we choose
$\rho_0=6.71565\times 10^7$$M_{\odot}/\mathrm{Kpc}^{3}$ and
$K_0=7133.48
$$M_{\odot} \, \mathrm{Kpc}^{-3} \, (\mathrm{km/s})^{2}$, in which
case the reduced $\chi^2_{red}$ value is $\chi^2_{red}=59.6713$.
Also the parameter $\alpha$ in this case is $\alpha=5.94781 $Kpc.

\begin{figure}[h!]
\centering
\includegraphics[width=20pc]{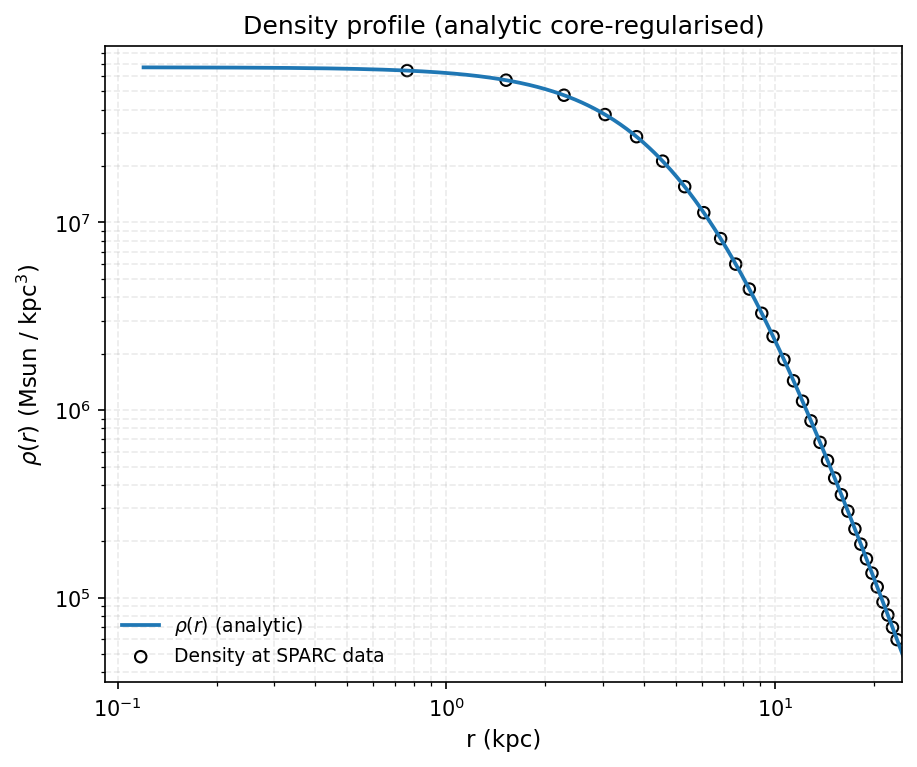}
\caption{The density of the SIDM model of Eq.
(\ref{ScaledependentEoSDM}) for the galaxy NGC6503, versus the
radius.} \label{NGC6503dens}
\end{figure}

\begin{figure}[h!]
\centering
\includegraphics[width=35pc]{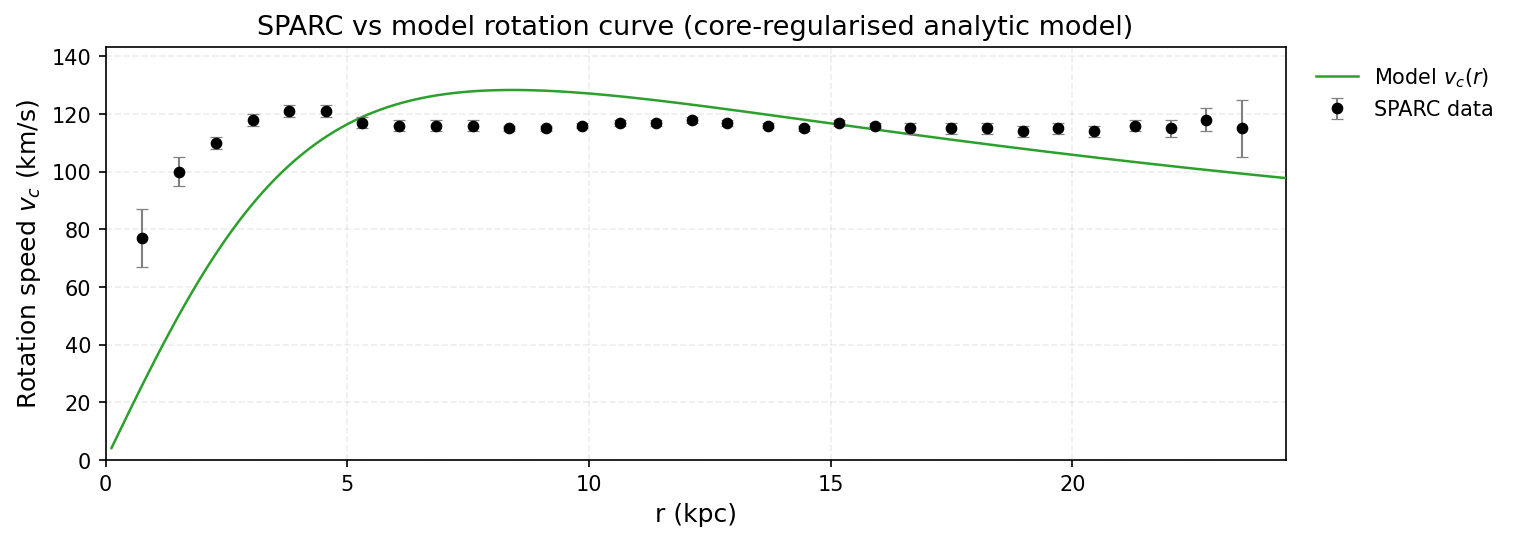}
\caption{The predicted rotation curves for the optimized SIDM
model of Eq. (\ref{ScaledependentEoSDM}), versus the SPARC
observational data for the galaxy NGC6503.} \label{NGC6503}
\end{figure}

In Table \ref{collNGC6503} we present the optimized values of
$K_0$ and $\rho_0$ for the analytic SIDM model of Eq.
(\ref{ScaledependentEoSDM}) for which the maximum compatibility
with the SPARC data is achieved.
\begin{table}[h!]
  \begin{center}
    \caption{SIDM Optimization Values for the galaxy NGC6503}
    \label{collNGC6503}
     \begin{tabular}{|r|r|}
     \hline
      \textbf{Parameter}   & \textbf{Optimization Values}
      \\  \hline
     $\rho_0 $  ($M_{\odot}/\mathrm{Kpc}^{3}$) & $6.71565\times 10^7$
\\  \hline $K_0$ ($M_{\odot} \,
\mathrm{Kpc}^{-3} \, (\mathrm{km/s})^{2}$)& 7133.48
\\  \hline
    \end{tabular}
  \end{center}
\end{table}
In Figs. \ref{NGC6503dens}, \ref{NGC6503} we present the density
of the analytic SIDM model, the predicted rotation curves for the
SIDM model (\ref{ScaledependentEoSDM}), versus the SPARC
observational data and the sound speed, as a function of the
radius respectively. As it can be seen, for this galaxy, the SIDM
model produces non-viable rotation curves which are incompatible
with the SPARC data.

\begin{figure}[h!]
\centering
\includegraphics[width=20pc]{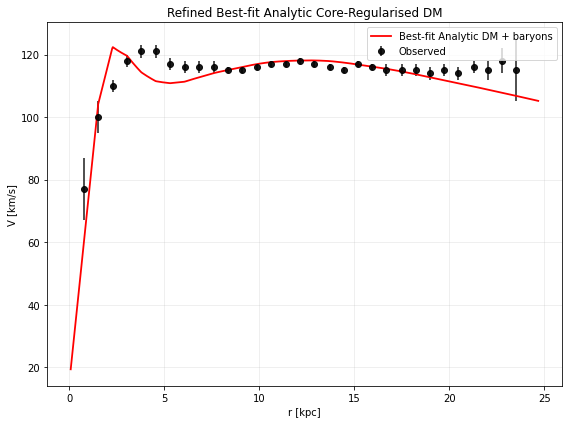}
\caption{The predicted rotation curves after using an optimization
for the SIDM model (\ref{ScaledependentEoSDM}), and the extended
SPARC data for the galaxy NGC6503. We included the rotation curves
of the gas, the disk velocities, the bulge (where present) along
with the SIDM model.} \label{extendedNGC6503}
\end{figure}

Now we shall include contributions to the rotation velocity from
the other components of the galaxy, namely the disk, the gas, and
the bulge if present. In Fig. \ref{extendedNGC6503} we present the
combined rotation curves including all the components of the
galaxy along with the SIDM. As it can be seen, the extended
collisional DM model is non-viable.
Also in Table \ref{evaluationextendedNGC6503} we present the
optimized values of the free parameters of the SIDM model for
which  we achieve the maximum compatibility with the SPARC data,
for the galaxy NGC6503, and also the resulting reduced
$\chi^2_{red}$ value.
\begin{table}[h!]
\centering \caption{Optimized Parameter Values of the Extended
SIDM model for the Galaxy NGC6503.}
\begin{tabular}{lc}
\hline
Parameter & Value  \\
\hline
$\rho_0 $ ($M_{\odot}/\mathrm{Kpc}^{3}$) & $1.31402\times 10^7$   \\
$K_0$ ($M_{\odot} \,
\mathrm{Kpc}^{-3} \, (\mathrm{km/s})^{2}$) & 4458.35  \\
$ml_{\text{disk}}$ & 0.8219 \\
$ml_{\text{bulge}}$ & 0.5576 \\
$\alpha$ (Kpc) & 10.6288 \\
$\chi^2_{red}$ & 5.05801 \\
\hline
\end{tabular}
\label{evaluationextendedNGC6503}
\end{table}

\subsection{The Galaxy NGC6674}

For this galaxy, the optimization method we used, ensures maximum
compatibility of the analytic SIDM model of Eq.
(\ref{ScaledependentEoSDM}) with the SPARC data, if we choose
$\rho_0=5.4478\times 10^7$$M_{\odot}/\mathrm{Kpc}^{3}$ and
$K_0=34306.5
$$M_{\odot} \, \mathrm{Kpc}^{-3} \, (\mathrm{km/s})^{2}$, in which
case the reduced $\chi^2_{red}$ value is $\chi^2_{red}=79.118$.
Also the parameter $\alpha$ in this case is $\alpha=14.482 $Kpc.

\begin{figure}[h!]
\centering
\includegraphics[width=20pc]{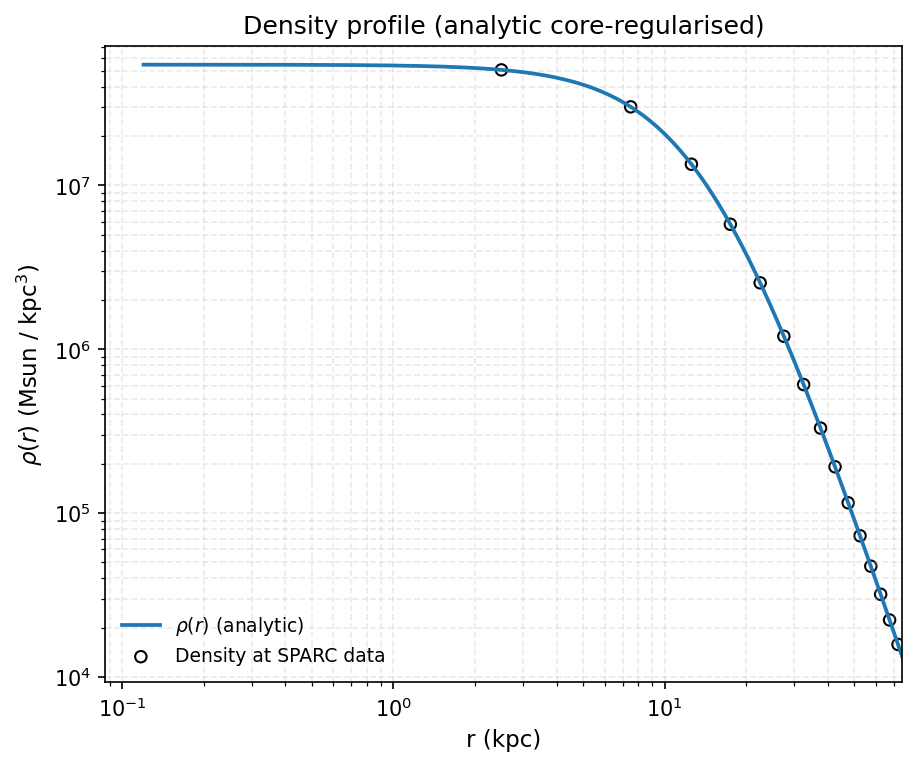}
\caption{The density of the SIDM model of Eq.
(\ref{ScaledependentEoSDM}) for the galaxy NGC6674, versus the
radius.} \label{NGC6674dens}
\end{figure}

\begin{figure}[h!]
\centering
\includegraphics[width=35pc]{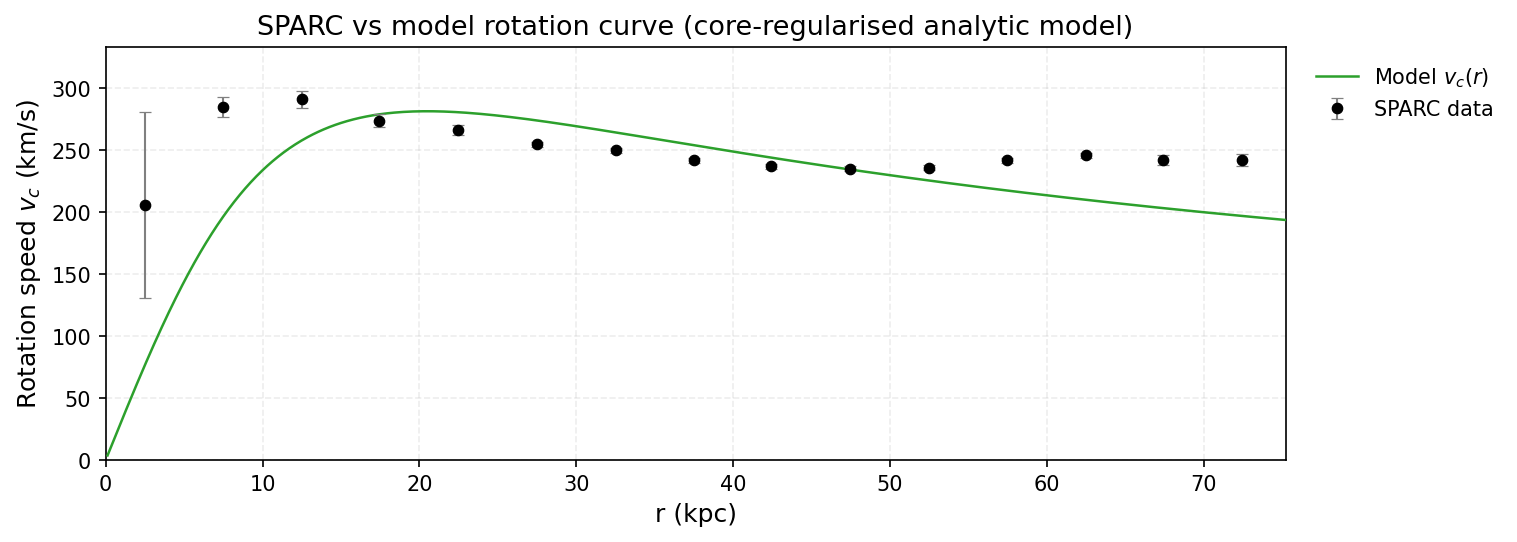}
\caption{The predicted rotation curves for the optimized SIDM
model of Eq. (\ref{ScaledependentEoSDM}), versus the SPARC
observational data for the galaxy NGC6674.} \label{NGC6674}
\end{figure}

In Table \ref{collNGC6674} we present the optimized values of
$K_0$ and $\rho_0$ for the analytic SIDM model of Eq.
(\ref{ScaledependentEoSDM}) for which the maximum compatibility
with the SPARC data is achieved.
\begin{table}[h!]
  \begin{center}
    \caption{SIDM Optimization Values for the galaxy NGC6674}
    \label{collNGC6674}
     \begin{tabular}{|r|r|}
     \hline
      \textbf{Parameter}   & \textbf{Optimization Values}
      \\  \hline
     $\rho_0 $  ($M_{\odot}/\mathrm{Kpc}^{3}$) & $5.4478\times 10^7$
\\  \hline $K_0$ ($M_{\odot} \,
\mathrm{Kpc}^{-3} \, (\mathrm{km/s})^{2}$)& 34306.5
\\  \hline
    \end{tabular}
  \end{center}
\end{table}
In Figs. \ref{NGC6674dens}, \ref{NGC6674} we present the density
of the analytic SIDM model, the predicted rotation curves for the
SIDM model (\ref{ScaledependentEoSDM}), versus the SPARC
observational data and the sound speed, as a function of the
radius respectively. As it can be seen, for this galaxy, the SIDM
model produces non-viable rotation curves which are incompatible
with the SPARC data.

\begin{figure}[h!]
\centering
\includegraphics[width=20pc]{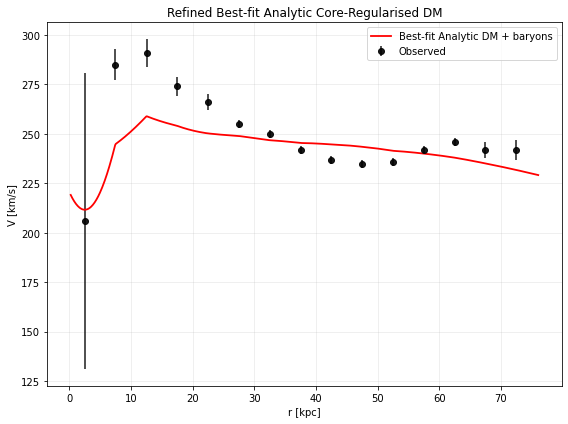}
\caption{The predicted rotation curves after using an optimization
for the SIDM model (\ref{ScaledependentEoSDM}), and the extended
SPARC data for the galaxy NGC6674. We included the rotation curves
of the gas, the disk velocities, the bulge (where present) along
with the SIDM model.} \label{extendedNGC6674}
\end{figure}

Now we shall include contributions to the rotation velocity from
the other components of the galaxy, namely the disk, the gas, and
the bulge if present. In Fig. \ref{extendedNGC6674} we present the
combined rotation curves including all the components of the
galaxy along with the SIDM. As it can be seen, the extended
collisional DM model is non-viable.
Also in Table \ref{evaluationextendedNGC6674} we present the
optimized values of the free parameters of the SIDM model for
which  we achieve the maximum compatibility with the SPARC data,
for the galaxy NGC6674, and also the resulting reduced
$\chi^2_{red}$ value.
\begin{table}[h!]
\centering \caption{Optimized Parameter Values of the Extended
SIDM model for the Galaxy NGC6674.}
\begin{tabular}{lc}
\hline
Parameter & Value  \\
\hline
$\rho_0 $ ($M_{\odot}/\mathrm{Kpc}^{3}$) & $2.47914\times 10^6$   \\
$K_0$ ($M_{\odot} \,
\mathrm{Kpc}^{-3} \, (\mathrm{km/s})^{2}$) & 16756   \\
$ml_{\text{disk}}$ & 1 \\
$ml_{\text{bulge}}$ & 1\\
$\alpha$ (Kpc) & 47.4385\\
$\chi^2_{red}$ & 14.223 \\
\hline
\end{tabular}
\label{evaluationextendedNGC6674}
\end{table}

\subsection{The Galaxy NGC6789}

For this galaxy, the optimization method we used, ensures maximum
compatibility of the analytic SIDM model of Eq.
(\ref{ScaledependentEoSDM}) with the SPARC data, if we choose
$\rho_0=6.72725\times 10^8$$M_{\odot}/\mathrm{Kpc}^{3}$ and
$K_0=1573.45
$$M_{\odot} \, \mathrm{Kpc}^{-3} \, (\mathrm{km/s})^{2}$, in which
case the reduced $\chi^2_{red}$ value is $\chi^2_{red}=0.626789$.
Also the parameter $\alpha$ in this case is $\alpha=0.882589 $Kpc.

\begin{figure}[h!]
\centering
\includegraphics[width=20pc]{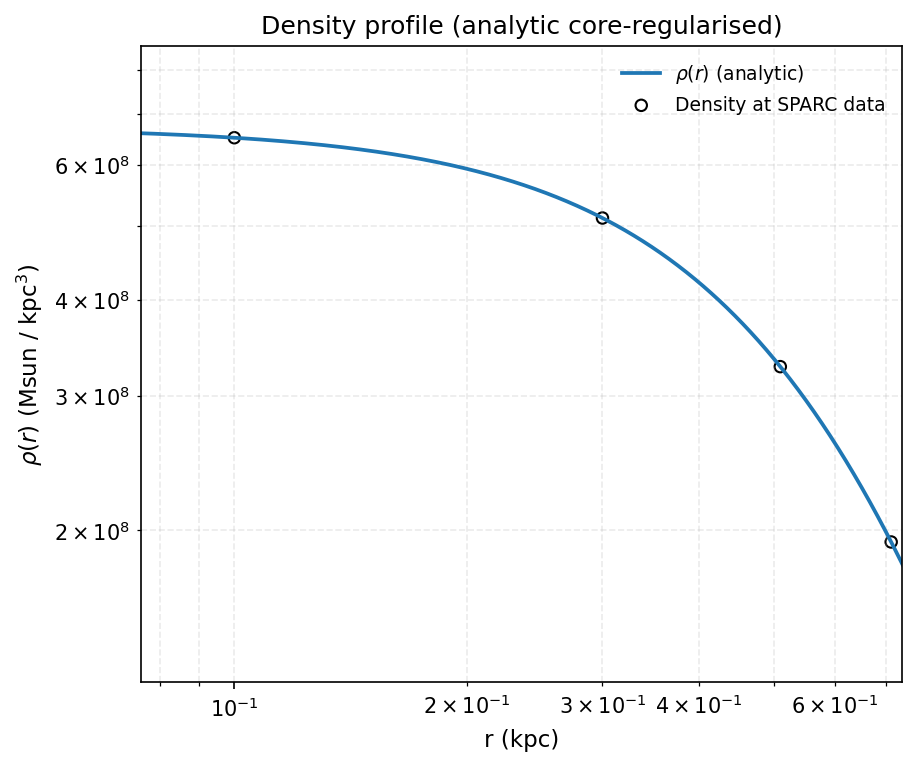}
\caption{The density of the SIDM model of Eq.
(\ref{ScaledependentEoSDM}) for the galaxy NGC6789, versus the
radius.} \label{NGC6789dens}
\end{figure}

\begin{figure}[h!]
\centering
\includegraphics[width=35pc]{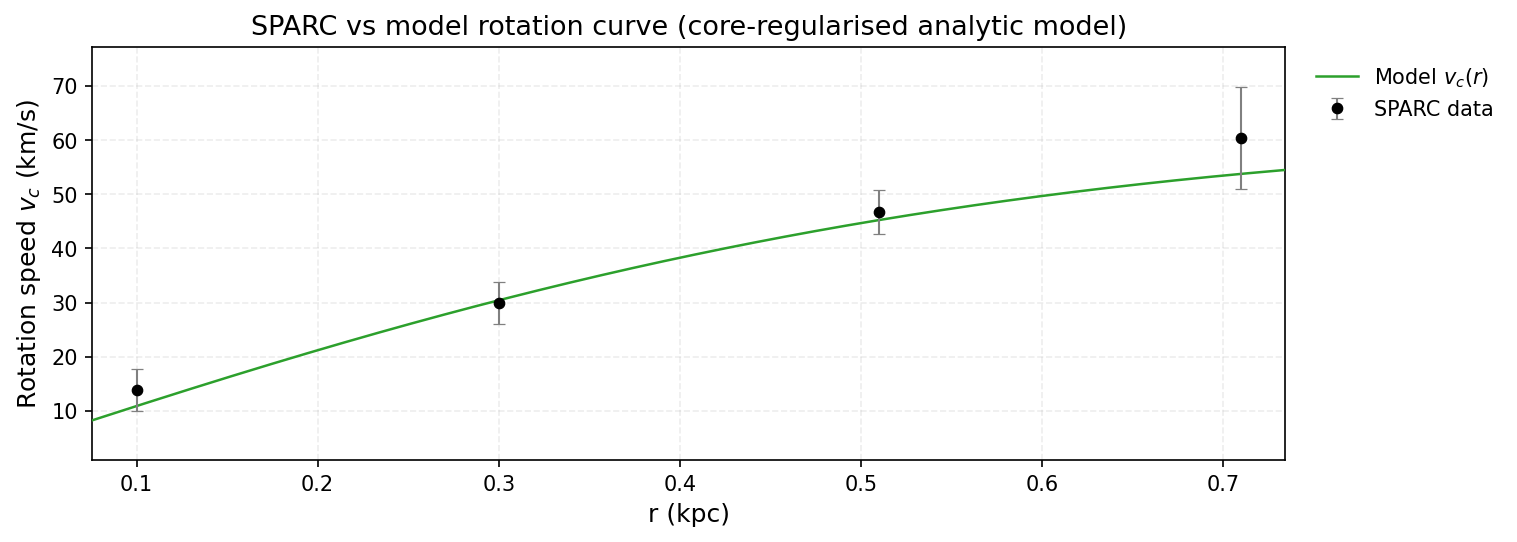}
\caption{The predicted rotation curves for the optimized SIDM
model of Eq. (\ref{ScaledependentEoSDM}), versus the SPARC
observational data for the galaxy NGC6789.} \label{NGC6789}
\end{figure}

In Table \ref{collNGC6789} we present the optimized values of
$K_0$ and $\rho_0$ for the analytic SIDM model of Eq.
(\ref{ScaledependentEoSDM}) for which the maximum compatibility
with the SPARC data is achieved.
\begin{table}[h!]
  \begin{center}
    \caption{SIDM Optimization Values for the galaxy NGC6789}
    \label{collNGC6789}
     \begin{tabular}{|r|r|}
     \hline
      \textbf{Parameter}   & \textbf{Optimization Values}
      \\  \hline
     $\rho_0 $  ($M_{\odot}/\mathrm{Kpc}^{3}$) & $6.72725\times 10^8$
\\  \hline $K_0$ ($M_{\odot} \,
\mathrm{Kpc}^{-3} \, (\mathrm{km/s})^{2}$)& 1573.45
\\  \hline
    \end{tabular}
  \end{center}
\end{table}
In Figs. \ref{NGC6789dens}, \ref{NGC6789} we present the density
of the analytic SIDM model, the predicted rotation curves for the
SIDM model (\ref{ScaledependentEoSDM}), versus the SPARC
observational data and the sound speed, as a function of the
radius respectively. As it can be seen, for this galaxy, the SIDM
model produces viable rotation curves which are compatible with
the SPARC data.

\subsection{The Galaxy NGC7331, Non-viable}

For this galaxy, the optimization method we used, ensures maximum
compatibility of the analytic SIDM model of Eq.
(\ref{ScaledependentEoSDM}) with the SPARC data, if we choose
$\rho_0=2.20061\times 10^8$$M_{\odot}/\mathrm{Kpc}^{3}$ and
$K_0=34846.1
$$M_{\odot} \, \mathrm{Kpc}^{-3} \, (\mathrm{km/s})^{2}$, in which
case the reduced $\chi^2_{red}$ value is $\chi^2_{red}=35.4787$.
Also the parameter $\alpha$ in this case is $\alpha=7.262 $Kpc.

\begin{figure}[h!]
\centering
\includegraphics[width=20pc]{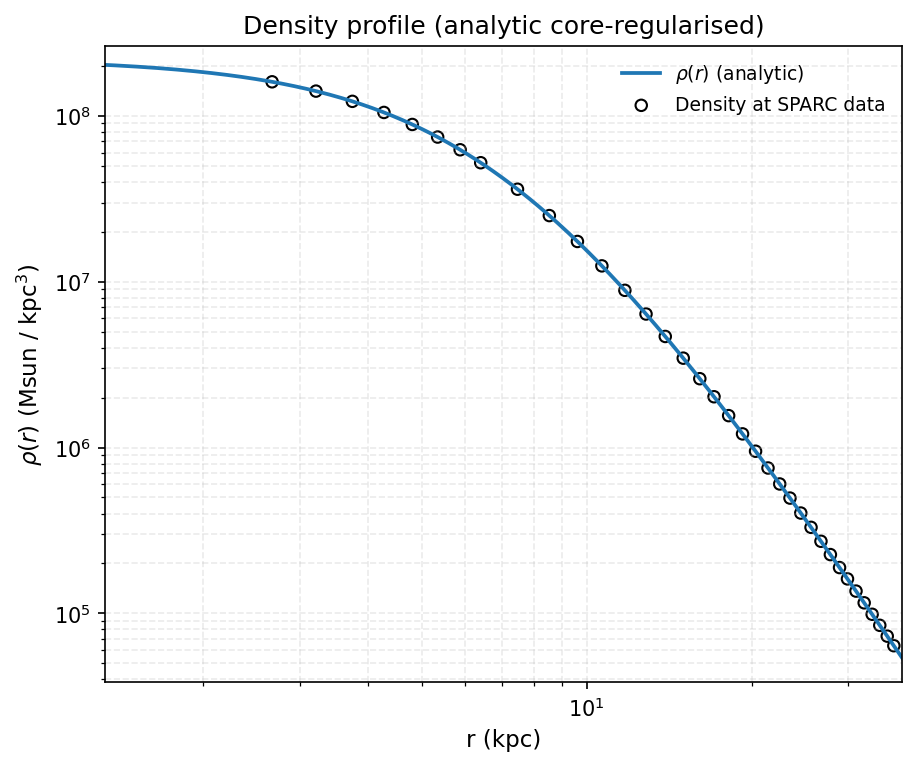}
\caption{The density of the SIDM model of Eq.
(\ref{ScaledependentEoSDM}) for the galaxy NGC7331, versus the
radius.} \label{NGC7331dens}
\end{figure}

\begin{figure}[h!]
\centering
\includegraphics[width=35pc]{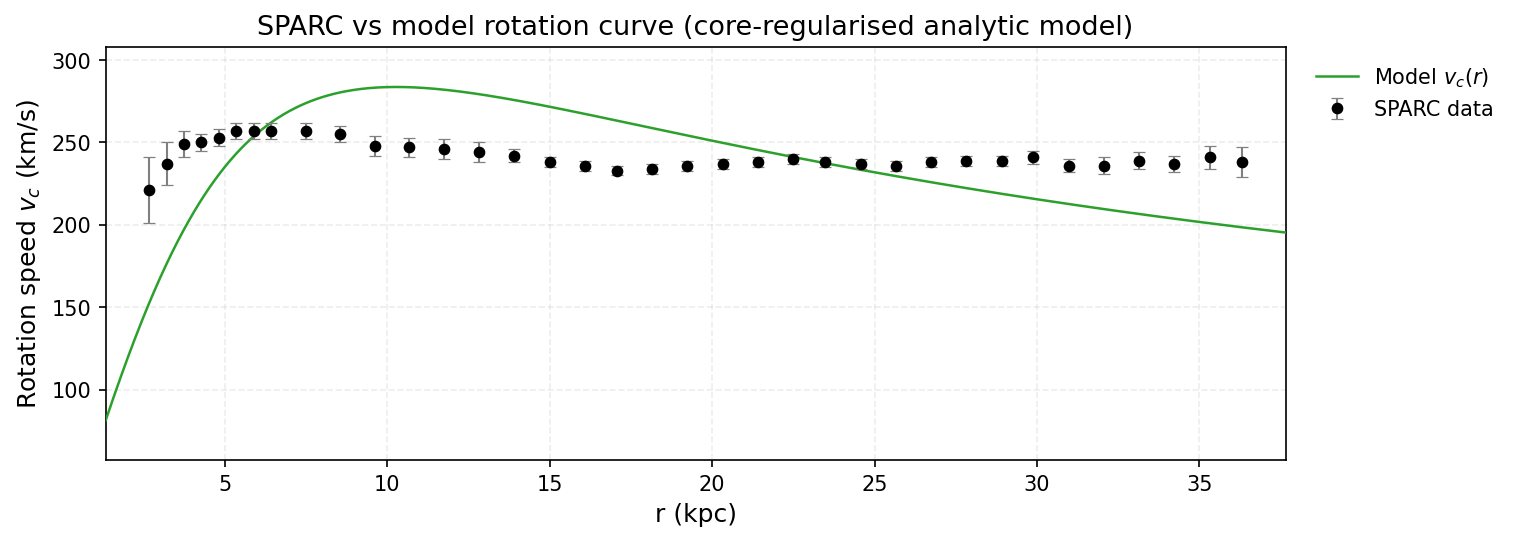}
\caption{The predicted rotation curves for the optimized SIDM
model of Eq. (\ref{ScaledependentEoSDM}), versus the SPARC
observational data for the galaxy NGC7331.} \label{NGC7331}
\end{figure}

In Table \ref{collNGC7331} we present the optimized values of
$K_0$ and $\rho_0$ for the analytic SIDM model of Eq.
(\ref{ScaledependentEoSDM}) for which the maximum compatibility
with the SPARC data is achieved.
\begin{table}[h!]
  \begin{center}
    \caption{SIDM Optimization Values for the galaxy NGC7331}
    \label{collNGC7331}
     \begin{tabular}{|r|r|}
     \hline
      \textbf{Parameter}   & \textbf{Optimization Values}
      \\  \hline
     $\rho_0 $  ($M_{\odot}/\mathrm{Kpc}^{3}$) & $2.20061\times 10^8$
\\  \hline $K_0$ ($M_{\odot} \,
\mathrm{Kpc}^{-3} \, (\mathrm{km/s})^{2}$)& 34846.1
\\  \hline
    \end{tabular}
  \end{center}
\end{table}
In Figs. \ref{NGC7331dens}, \ref{NGC7331} we present the density
of the analytic SIDM model, the predicted rotation curves for the
SIDM model (\ref{ScaledependentEoSDM}), versus the SPARC
observational data and the sound speed, as a function of the
radius respectively. As it can be seen, for this galaxy, the SIDM
model produces non-viable rotation curves which are incompatible
with the SPARC data.

\begin{figure}[h!]
\centering
\includegraphics[width=20pc]{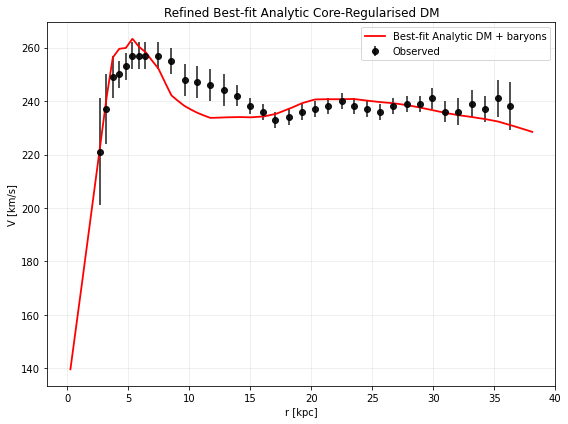}
\caption{The predicted rotation curves after using an optimization
for the SIDM model (\ref{ScaledependentEoSDM}), and the extended
SPARC data for the galaxy NGC7331. We included the rotation curves
of the gas, the disk velocities, the bulge (where present) along
with the SIDM model.} \label{extendedNGC7331}
\end{figure}

Now we shall include contributions to the rotation velocity from
the other components of the galaxy, namely the disk, the gas, and
the bulge if present. In Fig. \ref{extendedNGC7331} we present the
combined rotation curves including all the components of the
galaxy along with the SIDM. As it can be seen, the extended
collisional DM model is non-viable.
Also in Table \ref{evaluationextendedNGC7331} we present the
optimized values of the free parameters of the SIDM model for
which  we achieve the maximum compatibility with the SPARC data,
for the galaxy NGC7331, and also the resulting reduced
$\chi^2_{red}$ value.
\begin{table}[h!]
\centering \caption{Optimized Parameter Values of the Extended
SIDM model for the Galaxy NGC7331.}
\begin{tabular}{lc}
\hline
Parameter & Value  \\
\hline
$\rho_0 $ ($M_{\odot}/\mathrm{Kpc}^{3}$) & $7.12149\times 10^6$   \\
$K_0$ ($M_{\odot} \,
\mathrm{Kpc}^{-3} \, (\mathrm{km/s})^{2}$) & 15912.5  \\
$ml_{\text{disk}}$ & 0.6909 \\
$ml_{\text{bulge}}$ & 0.5 \\
$\alpha$ (Kpc) & 27.2759\\
$\chi^2_{red}$ & 1.5433 \\
\hline
\end{tabular}
\label{evaluationextendedNGC7331}
\end{table}

\subsection{The Galaxy NGC7793, Non-viable, Extended Viable}

For this galaxy, the optimization method we used, ensures maximum
compatibility of the analytic SIDM model of Eq.
(\ref{ScaledependentEoSDM}) with the SPARC data, if we choose
$\rho_0=3.09584\times 10^8$$M_{\odot}/\mathrm{Kpc}^{3}$ and
$K_0=4953.05
$$M_{\odot} \, \mathrm{Kpc}^{-3} \, (\mathrm{km/s})^{2}$, in which
case the reduced $\chi^2_{red}$ value is $\chi^2_{red}=3.75772$.
Also the parameter $\alpha$ in this case is $\alpha=2.30833 $Kpc.

\begin{figure}[h!]
\centering
\includegraphics[width=20pc]{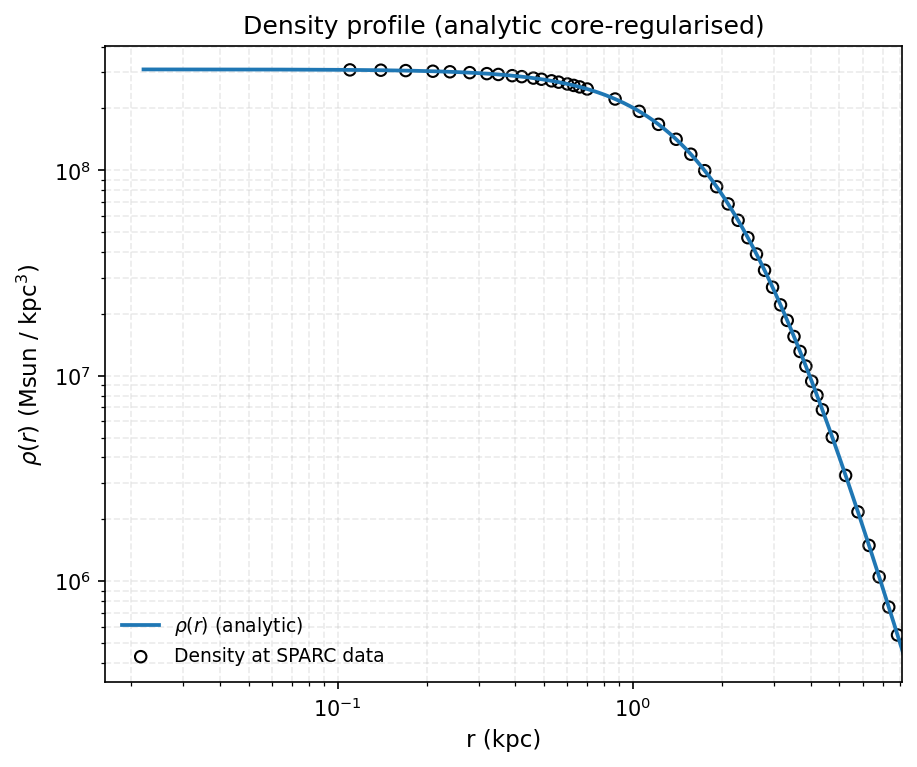}
\caption{The density of the SIDM model of Eq.
(\ref{ScaledependentEoSDM}) for the galaxy NGC7793, versus the
radius.} \label{NGC7793dens}
\end{figure}

\begin{figure}[h!]
\centering
\includegraphics[width=35pc]{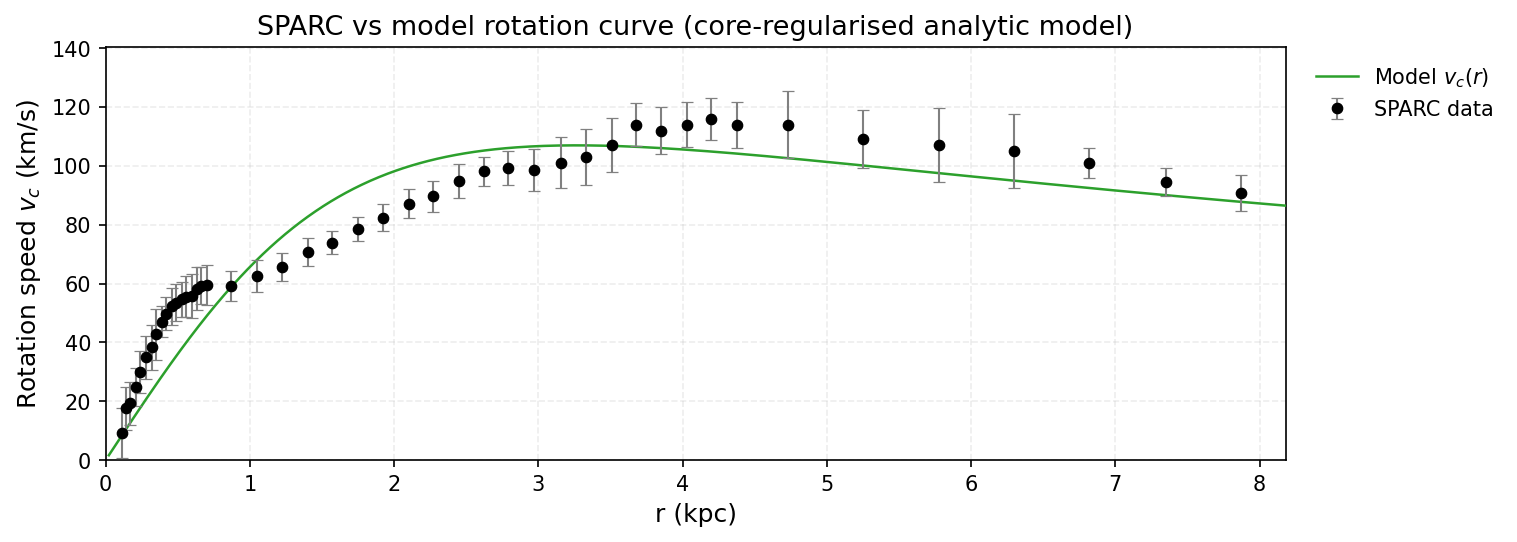}
\caption{The predicted rotation curves for the optimized SIDM
model of Eq. (\ref{ScaledependentEoSDM}), versus the SPARC
observational data for the galaxy NGC7793.} \label{NGC7793}
\end{figure}

In Table \ref{collNGC7793} we present the optimized values of
$K_0$ and $\rho_0$ for the analytic SIDM model of Eq.
(\ref{ScaledependentEoSDM}) for which the maximum compatibility
with the SPARC data is achieved.
\begin{table}[h!]
  \begin{center}
    \caption{SIDM Optimization Values for the galaxy NGC7793}
    \label{collNGC7793}
     \begin{tabular}{|r|r|}
     \hline
      \textbf{Parameter}   & \textbf{Optimization Values}
      \\  \hline
     $\rho_0 $  ($M_{\odot}/\mathrm{Kpc}^{3}$) & $3.09584\times 10^8$
\\  \hline $K_0$ ($M_{\odot} \,
\mathrm{Kpc}^{-3} \, (\mathrm{km/s})^{2}$)& 4953.05
\\  \hline
    \end{tabular}
  \end{center}
\end{table}
In Figs. \ref{NGC7793dens}, \ref{NGC7793} we present the density
of the analytic SIDM model, the predicted rotation curves for the
SIDM model (\ref{ScaledependentEoSDM}), versus the SPARC
observational data and the sound speed, as a function of the
radius respectively. As it can be seen, for this galaxy, the SIDM
model produces non-viable rotation curves which are incompatible
with the SPARC data.

\begin{figure}[h!]
\centering
\includegraphics[width=20pc]{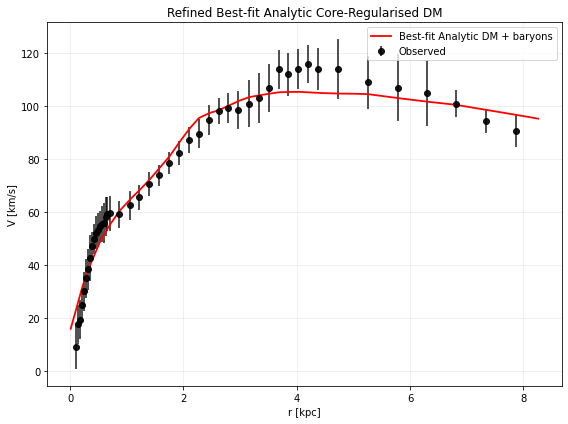}
\caption{The predicted rotation curves after using an optimization
for the SIDM model (\ref{ScaledependentEoSDM}), and the extended
SPARC data for the galaxy NGC7793. We included the rotation curves
of the gas, the disk velocities, the bulge (where present) along
with the SIDM model.} \label{extendedNGC7793}
\end{figure}

Now we shall include contributions to the rotation velocity from
the other components of the galaxy, namely the disk, the gas, and
the bulge if present. In Fig. \ref{extendedNGC7793} we present the
combined rotation curves including all the components of the
galaxy along with the SIDM. As it can be seen, the extended
collisional DM model is deemed viable.
Also in Table \ref{evaluationextendedNGC7793} we present the
optimized values of the free parameters of the SIDM model for
which  we achieve the maximum compatibility with the SPARC data,
for the galaxy NGC7793, and also the resulting reduced
$\chi^2_{red}$ value.
\begin{table}[h!]
\centering \caption{Optimized Parameter Values of the Extended
SIDM model for the Galaxy NGC7793.}
\begin{tabular}{lc}
\hline
Parameter & Value  \\
\hline
$\rho_0 $ ($M_{\odot}/\mathrm{Kpc}^{3}$) & $2.86207\times 10^7$   \\
$K_0$ ($M_{\odot} \,
\mathrm{Kpc}^{-3} \, (\mathrm{km/s})^{2}$) & 2200.74   \\
$ml_{\text{disk}}$ & 0.8641 \\
$ml_{\text{bulge}}$ & 0.2 \\
$\alpha$ (Kpc) & 5.05989\\
$\chi^2_{red}$ & 0.699734 \\
\hline
\end{tabular}
\label{evaluationextendedNGC7793}
\end{table}

\subsection{The Galaxy NGC7814, Non-viable, Extended Viable}

For this galaxy, the optimization method we used, ensures maximum
compatibility of the analytic SIDM model of Eq.
(\ref{ScaledependentEoSDM}) with the SPARC data, if we choose
$\rho_0=5.34343\times 10^8$$M_{\odot}/\mathrm{Kpc}^{3}$ and
$K_0=27059.5
$$M_{\odot} \, \mathrm{Kpc}^{-3} \, (\mathrm{km/s})^{2}$, in which
case the reduced $\chi^2_{red}$ value is $\chi^2_{red}=30.4685$.
Also the parameter $\alpha$ in this case is $\alpha=4.10677 $Kpc.

\begin{figure}[h!]
\centering
\includegraphics[width=20pc]{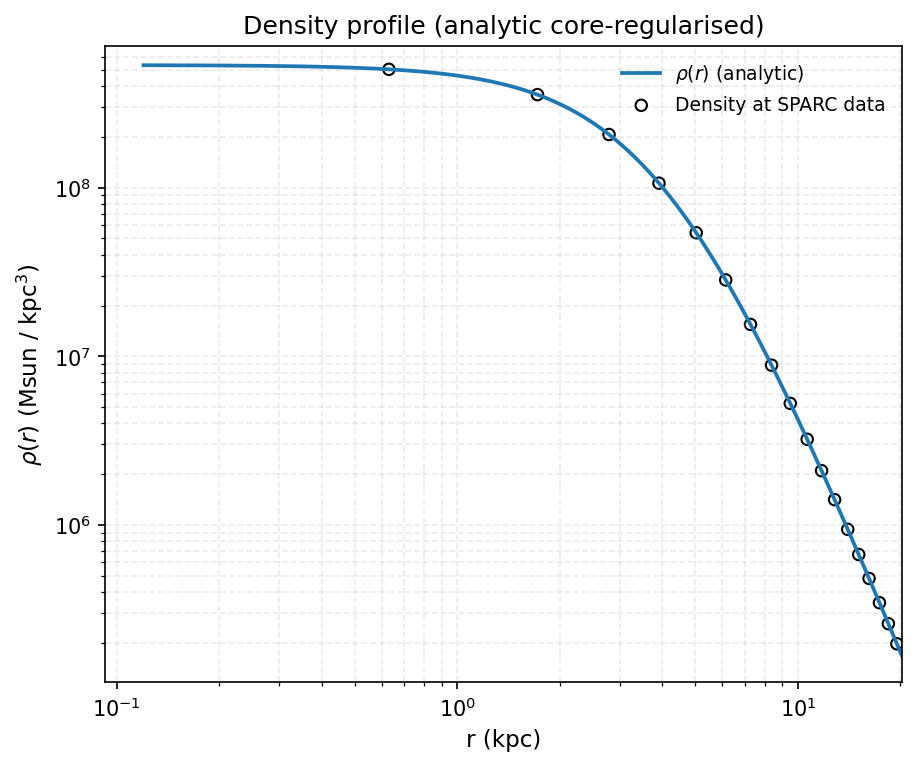}
\caption{The density of the SIDM model of Eq.
(\ref{ScaledependentEoSDM}) for the galaxy NGC7814, versus the
radius.} \label{NGC7814dens}
\end{figure}

\begin{figure}[h!]
\centering
\includegraphics[width=35pc]{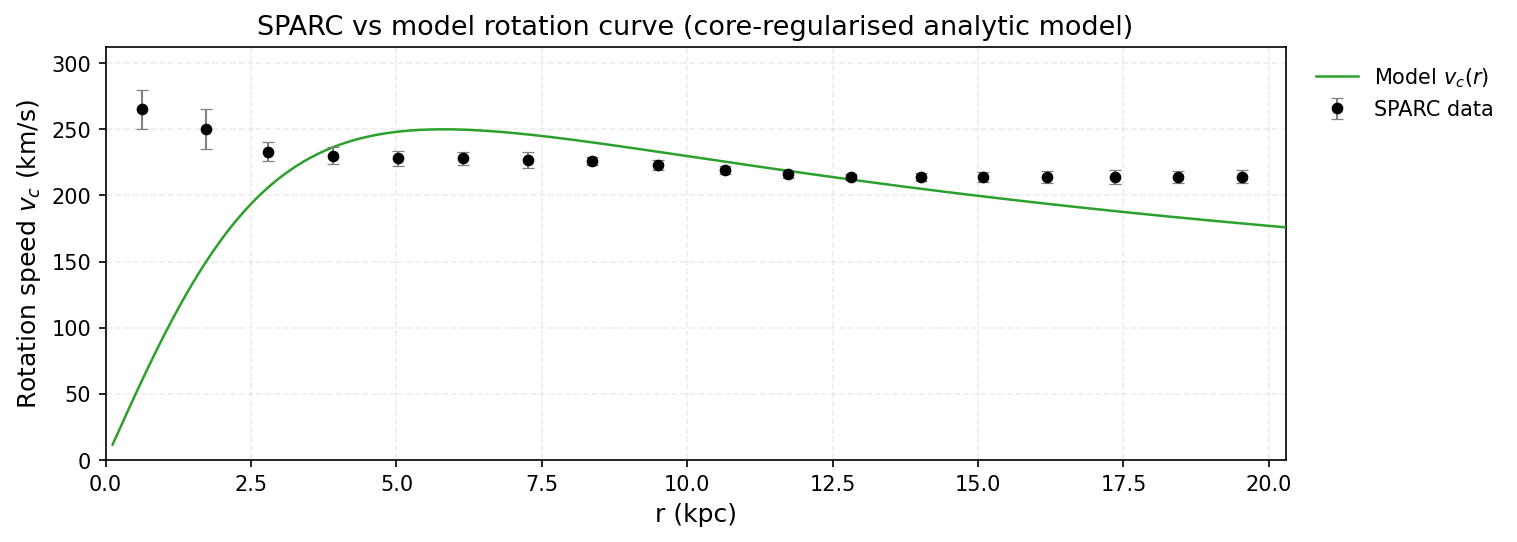}
\caption{The predicted rotation curves for the optimized SIDM
model of Eq. (\ref{ScaledependentEoSDM}), versus the SPARC
observational data for the galaxy NGC7814.} \label{NGC7814}
\end{figure}

In Table \ref{collNGC7814} we present the optimized values of
$K_0$ and $\rho_0$ for the analytic SIDM model of Eq.
(\ref{ScaledependentEoSDM}) for which the maximum compatibility
with the SPARC data is achieved.
\begin{table}[h!]
  \begin{center}
    \caption{SIDM Optimization Values for the galaxy NGC7814}
    \label{collNGC7814}
     \begin{tabular}{|r|r|}
     \hline
      \textbf{Parameter}   & \textbf{Optimization Values}
      \\  \hline
     $\rho_0 $  ($M_{\odot}/\mathrm{Kpc}^{3}$) & $5.34343\times 10^8$
\\  \hline $K_0$ ($M_{\odot} \,
\mathrm{Kpc}^{-3} \, (\mathrm{km/s})^{2}$)& 27059.5
\\  \hline
    \end{tabular}
  \end{center}
\end{table}
In Figs. \ref{NGC7814dens}, \ref{NGC7814} we present the density
of the analytic SIDM model, the predicted rotation curves for the
SIDM model (\ref{ScaledependentEoSDM}), versus the SPARC
observational data and the sound speed, as a function of the
radius respectively. As it can be seen, for this galaxy, the SIDM
model produces non-viable rotation curves which are incompatible
with the SPARC data.

\begin{figure}[h!]
\centering
\includegraphics[width=20pc]{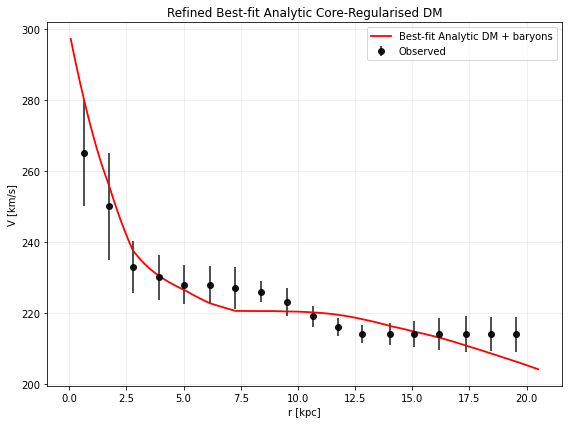}
\caption{The predicted rotation curves after using an optimization
for the SIDM model (\ref{ScaledependentEoSDM}), and the extended
SPARC data for the galaxy NGC7814. We included the rotation curves
of the gas, the disk velocities, the bulge (where present) along
with the SIDM model.} \label{extendedNGC7814}
\end{figure}

Now we shall include contributions to the rotation velocity from
the other components of the galaxy, namely the disk, the gas, and
the bulge if present. In Fig. \ref{extendedNGC7814} we present the
combined rotation curves including all the components of the
galaxy along with the SIDM. As it can be seen, the extended
collisional DM model is non-viable.
Also in Table \ref{evaluationextendedNGC7814} we present the
optimized values of the free parameters of the SIDM model for
which  we achieve the maximum compatibility with the SPARC data,
for the galaxy NGC7814, and also the resulting reduced
$\chi^2_{red}$ value.
\begin{table}[h!]
\centering \caption{Optimized Parameter Values of the Extended
SIDM model for the Galaxy NGC7814.}
\begin{tabular}{lc}
\hline
Parameter & Value  \\
\hline
$\rho_0 $ ($M_{\odot}/\mathrm{Kpc}^{3}$) & $2.58268\times 10^7$   \\
$K_0$ ($M_{\odot} \,
\mathrm{Kpc}^{-3} \, (\mathrm{km/s})^{2}$) & 12568.5   \\
$ml_{\text{disk}}$ & 1 \\
$ml_{\text{bulge}}$ & 0.8033 \\
$\alpha$ (Kpc) & 12.7293\\
$\chi^2_{red}$ & 1.23637 \\
\hline
\end{tabular}
\label{evaluationextendedNGC7814}
\end{table}

\subsection{The Galaxy UGC00891}

For this galaxy, the optimization method we used, ensures maximum
compatibility of the analytic SIDM model of Eq.
(\ref{ScaledependentEoSDM}) with the SPARC data, if we choose
$\rho_0=1.53512\times 10^7$$M_{\odot}/\mathrm{Kpc}^{3}$ and
$K_0=1741.03
$$M_{\odot} \, \mathrm{Kpc}^{-3} \, (\mathrm{km/s})^{2}$, in which
case the reduced $\chi^2_{red}$ value is $\chi^2_{red}=0.681321$.
Also the parameter $\alpha$ in this case is $\alpha=6.14586 $Kpc.

\begin{figure}[h!]
\centering
\includegraphics[width=20pc]{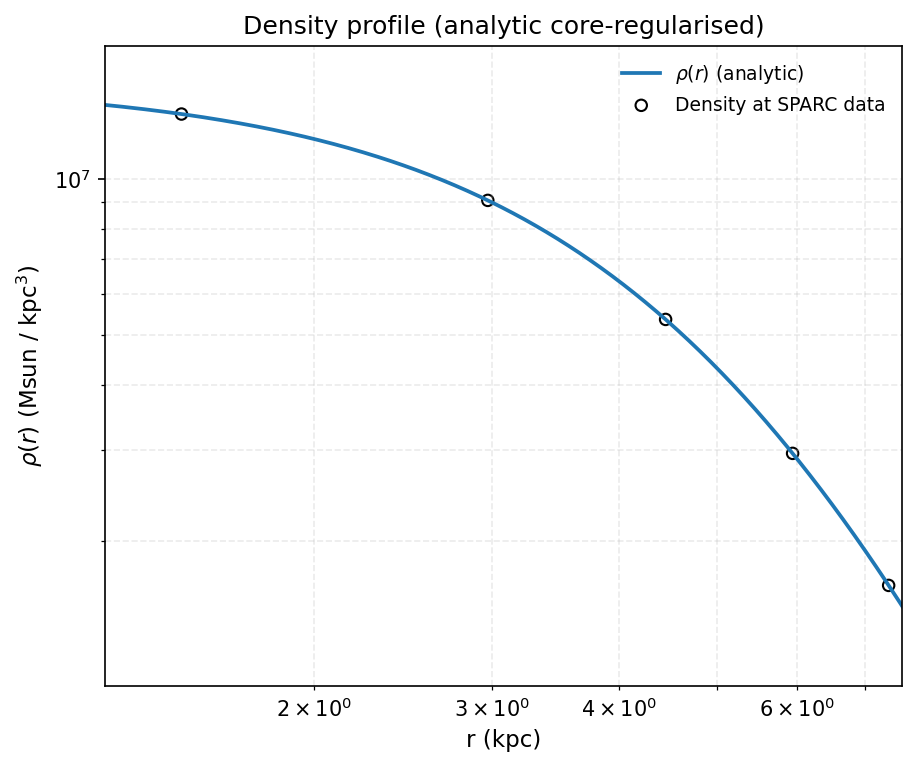}
\caption{The density of the SIDM model of Eq.
(\ref{ScaledependentEoSDM}) for the galaxy UGC00891, versus the
radius.} \label{UGC00891dens}
\end{figure}

\begin{figure}[h!]
\centering
\includegraphics[width=35pc]{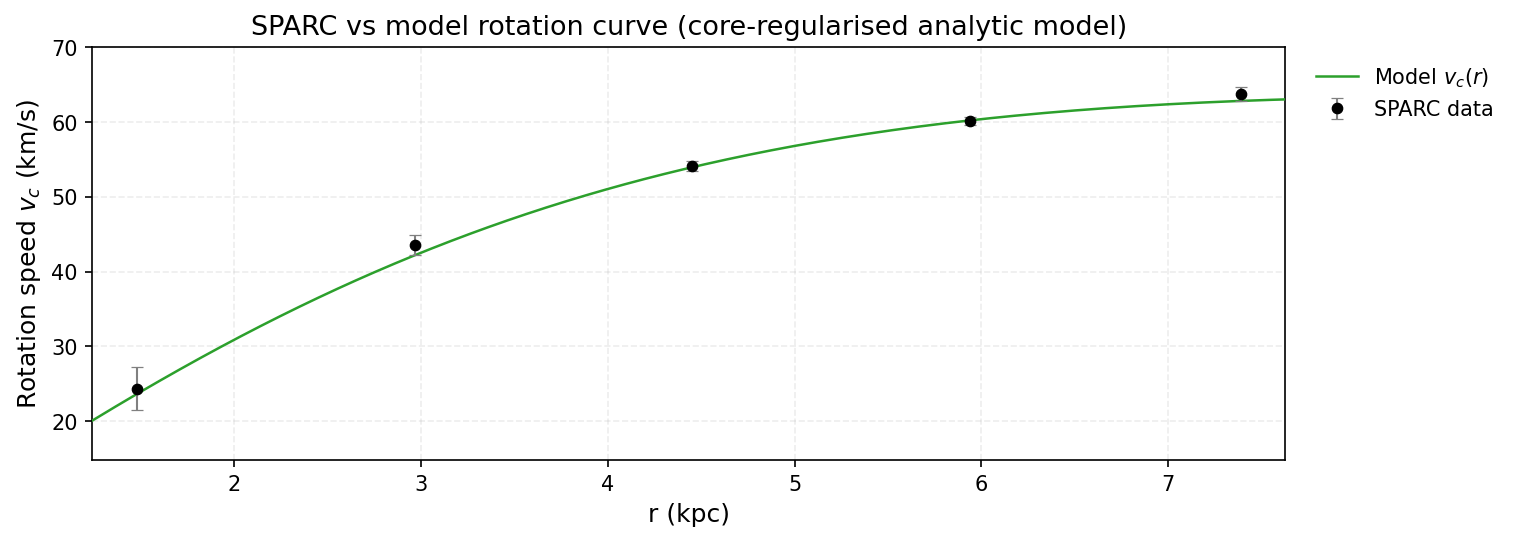}
\caption{The predicted rotation curves for the optimized SIDM
model of Eq. (\ref{ScaledependentEoSDM}), versus the SPARC
observational data for the galaxy UGC00891.} \label{UGC00891}
\end{figure}

In Table \ref{collUGC00891} we present the optimized values of
$K_0$ and $\rho_0$ for the analytic SIDM model of Eq.
(\ref{ScaledependentEoSDM}) for which the maximum compatibility
with the SPARC data is achieved.
\begin{table}[h!]
  \begin{center}
    \caption{SIDM Optimization Values for the galaxy UGC00891}
    \label{collUGC00891}
     \begin{tabular}{|r|r|}
     \hline
      \textbf{Parameter}   & \textbf{Optimization Values}
      \\  \hline
     $\rho_0 $  ($M_{\odot}/\mathrm{Kpc}^{3}$) & $1.53512\times 10^7$
\\  \hline $K_0$ ($M_{\odot} \,
\mathrm{Kpc}^{-3} \, (\mathrm{km/s})^{2}$)& 1741.03
\\  \hline
    \end{tabular}
  \end{center}
\end{table}
In Figs. \ref{UGC00891dens}, \ref{UGC00891}  we present the
density of the analytic SIDM model, the predicted rotation curves
for the SIDM model (\ref{ScaledependentEoSDM}), versus the SPARC
observational data and the sound speed, as a function of the
radius respectively. As it can be seen, for this galaxy, the SIDM
model produces viable rotation curves which are compatible with
the SPARC data.

\subsection{The Galaxy UGC01281}

For this galaxy, the optimization method we used, ensures maximum
compatibility of the analytic SIDM model of Eq.
(\ref{ScaledependentEoSDM}) with the SPARC data, if we choose
$\rho_0=2.85574\times 10^7$$M_{\odot}/\mathrm{Kpc}^{3}$ and
$K_0=1350.75
$$M_{\odot} \, \mathrm{Kpc}^{-3} \, (\mathrm{km/s})^{2}$, in which
case the reduced $\chi^2_{red}$ value is $\chi^2_{red}=0.138773$.
Also the parameter $\alpha$ in this case is $\alpha=4.27865 $Kpc.

\begin{figure}[h!]
\centering
\includegraphics[width=20pc]{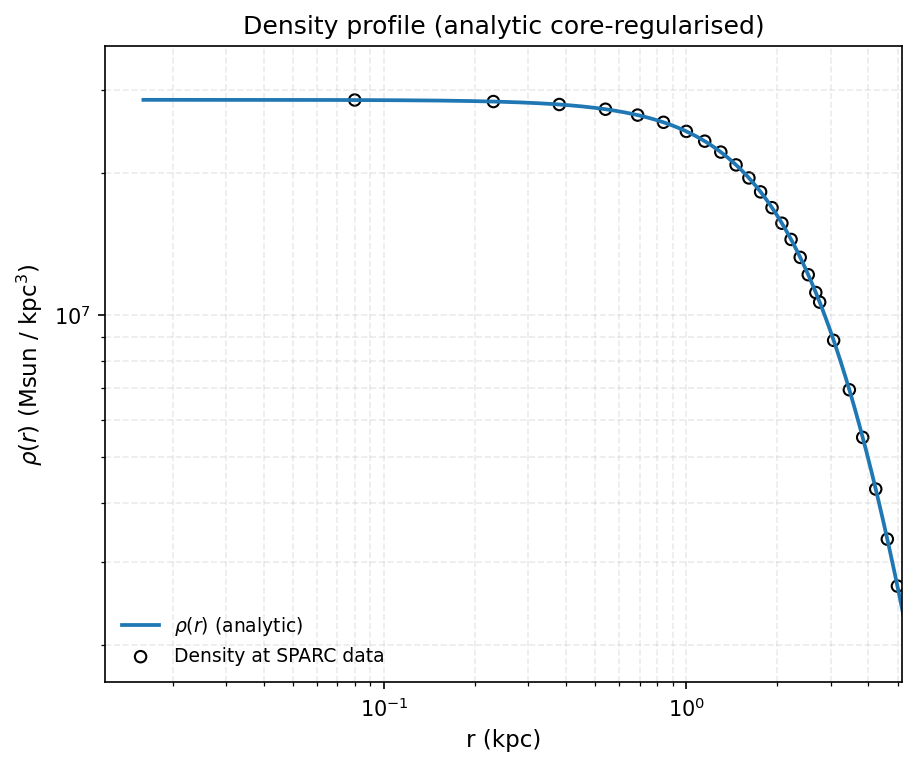}
\caption{The density of the SIDM model of Eq.
(\ref{ScaledependentEoSDM}) for the galaxy UGC01281, versus the
radius.} \label{UGC01281dens}
\end{figure}

\begin{figure}[h!]
\centering
\includegraphics[width=35pc]{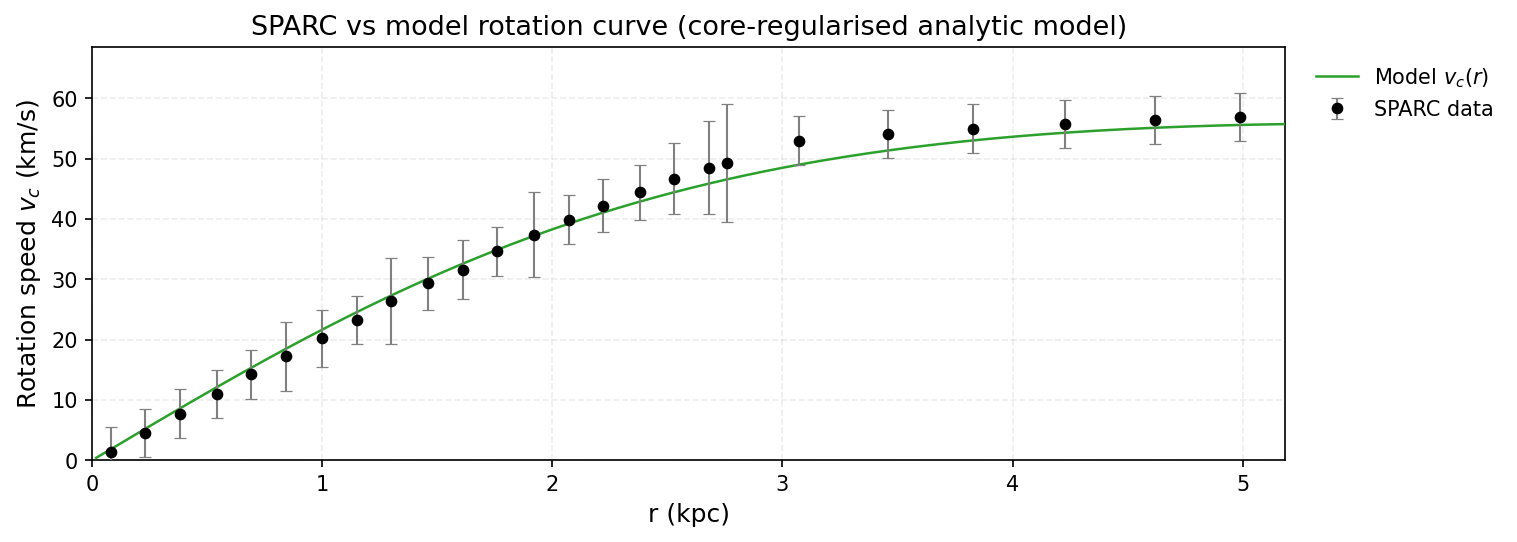}
\caption{The predicted rotation curves for the optimized SIDM
model of Eq. (\ref{ScaledependentEoSDM}), versus the SPARC
observational data for the galaxy UGC01281.} \label{UGC01281}
\end{figure}

In Table \ref{collUGC01281} we present the optimized values of
$K_0$ and $\rho_0$ for the analytic SIDM model of Eq.
(\ref{ScaledependentEoSDM}) for which the maximum compatibility
with the SPARC data is achieved.
\begin{table}[h!]
  \begin{center}
    \caption{SIDM Optimization Values for the galaxy UGC01281}
    \label{collUGC01281}
     \begin{tabular}{|r|r|}
     \hline
      \textbf{Parameter}   & \textbf{Optimization Values}
      \\  \hline
     $\rho_0 $  ($M_{\odot}/\mathrm{Kpc}^{3}$) & $2.85574\times 10^7$
\\  \hline $K_0$ ($M_{\odot} \,
\mathrm{Kpc}^{-3} \, (\mathrm{km/s})^{2}$)& 1350.75
\\  \hline
    \end{tabular}
  \end{center}
\end{table}
In Figs. \ref{UGC01281dens}, \ref{UGC01281}  we present the
density of the analytic SIDM model, the predicted rotation curves
for the SIDM model (\ref{ScaledependentEoSDM}), versus the SPARC
observational data and the sound speed, as a function of the
radius respectively. As it can be seen, for this galaxy, the SIDM
model produces viable rotation curves which are compatible with
the SPARC data.

\subsection{The Galaxy UGC02259, Non-viable}

For this galaxy, the optimization method we used, ensures maximum
compatibility of the analytic SIDM model of Eq.
(\ref{ScaledependentEoSDM}) with the SPARC data, if we choose
$\rho_0=9.76276\times 10^7$$M_{\odot}/\mathrm{Kpc}^{3}$ and
$K_0=3430.77
$$M_{\odot} \, \mathrm{Kpc}^{-3} \, (\mathrm{km/s})^{2}$, in which
case the reduced $\chi^2_{red}$ value is $\chi^2_{red}=9.77349$.
Also the parameter $\alpha$ in this case is $\alpha=3.42105 $Kpc.

\begin{figure}[h!]
\centering
\includegraphics[width=20pc]{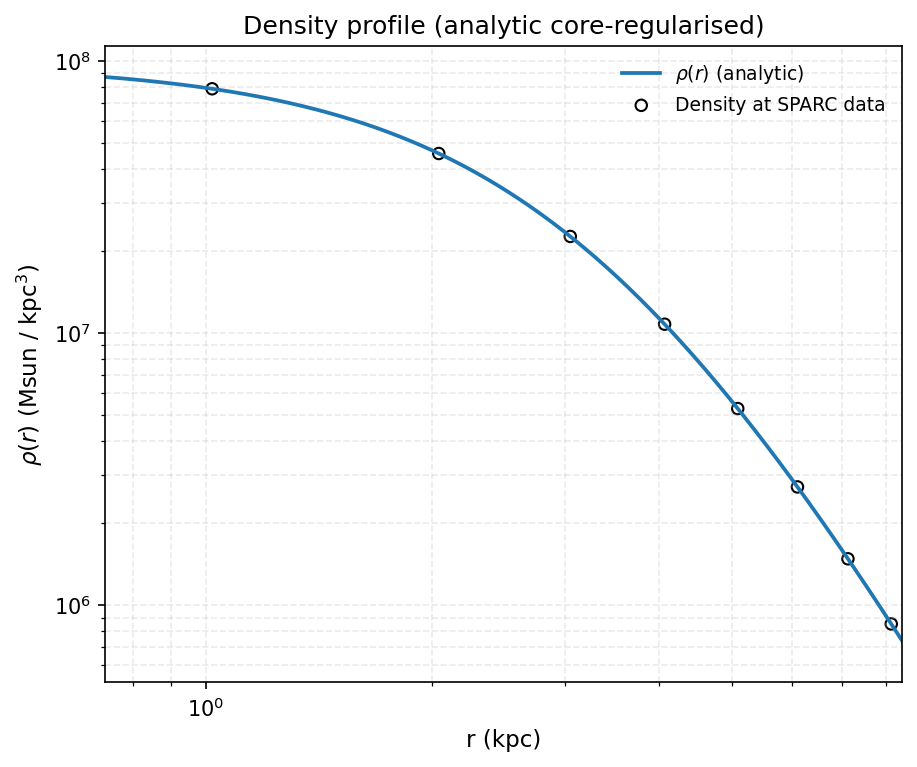}
\caption{The density of the SIDM model of Eq.
(\ref{ScaledependentEoSDM}) for the galaxy UGC02259, versus the
radius.} \label{UGC02259dens}
\end{figure}

\begin{figure}[h!]
\centering
\includegraphics[width=35pc]{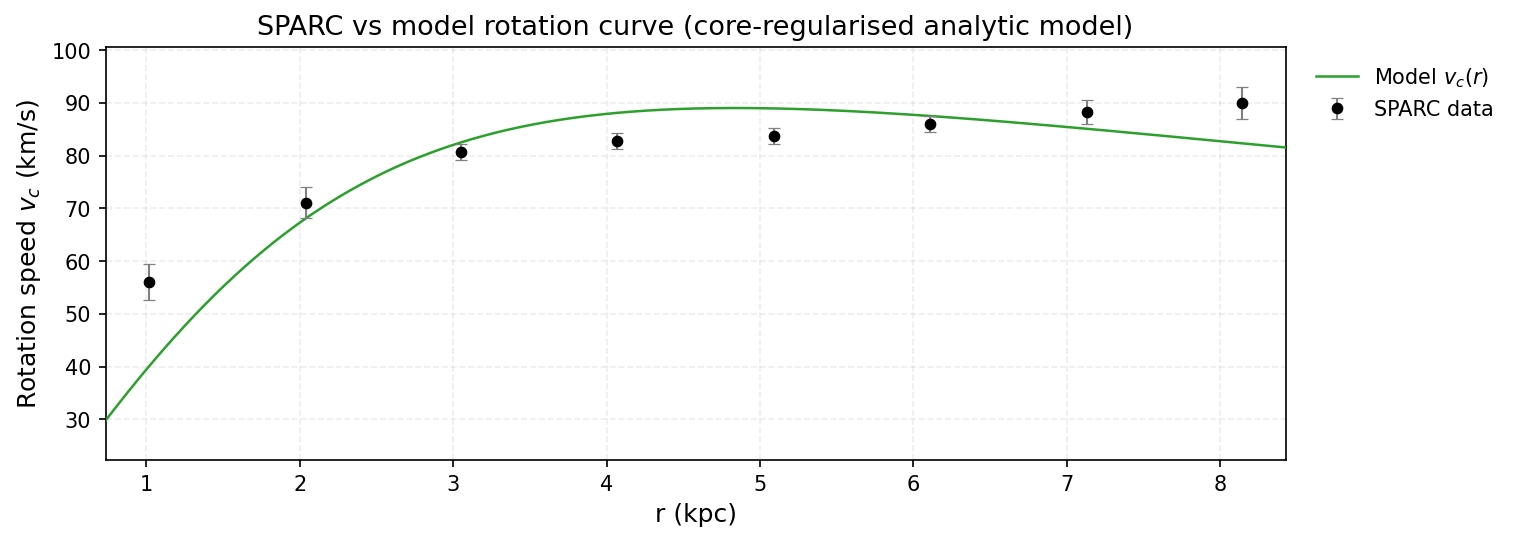}
\caption{The predicted rotation curves for the optimized SIDM
model of Eq. (\ref{ScaledependentEoSDM}), versus the SPARC
observational data for the galaxy UGC02259.} \label{UGC02259}
\end{figure}

In Table \ref{collUGC02259} we present the optimized values of
$K_0$ and $\rho_0$ for the analytic SIDM model of Eq.
(\ref{ScaledependentEoSDM}) for which the maximum compatibility
with the SPARC data is achieved.
\begin{table}[h!]
  \begin{center}
    \caption{SIDM Optimization Values for the galaxy UGC02259}
    \label{collUGC02259}
     \begin{tabular}{|r|r|}
     \hline
      \textbf{Parameter}   & \textbf{Optimization Values}
      \\  \hline
     $\rho_0 $  ($M_{\odot}/\mathrm{Kpc}^{3}$) & $9.76276\times 10^7$
\\  \hline $K_0$ ($M_{\odot} \,
\mathrm{Kpc}^{-3} \, (\mathrm{km/s})^{2}$)& 3430.771250
\\  \hline
    \end{tabular}
  \end{center}
\end{table}
In Figs. \ref{UGC02259dens}, \ref{UGC02259}  we present the
density of the analytic SIDM model, the predicted rotation curves
for the SIDM model (\ref{ScaledependentEoSDM}), versus the SPARC
observational data and the sound speed, as a function of the
radius respectively. As it can be seen, for this galaxy, the SIDM
model produces non-viable rotation curves which are incompatible
with the SPARC data.

\begin{figure}[h!]
\centering
\includegraphics[width=20pc]{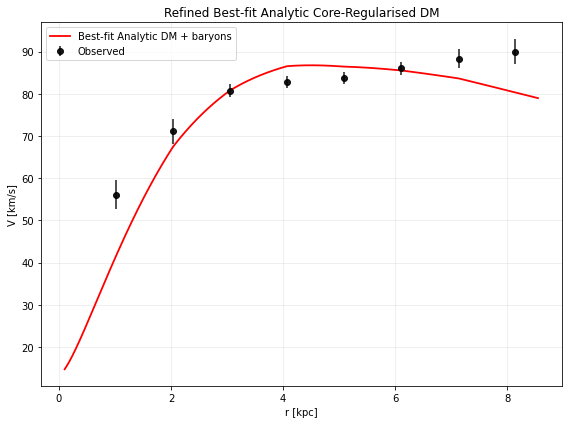}
\caption{The predicted rotation curves after using an optimization
for the SIDM model (\ref{ScaledependentEoSDM}), and the extended
SPARC data for the galaxy UGC02259. We included the rotation
curves of the gas, the disk velocities, the bulge (where present)
along with the SIDM model.} \label{extendedUGC02259}
\end{figure}

Now we shall include contributions to the rotation velocity from
the other components of the galaxy, namely the disk, the gas, and
the bulge if present. In Fig. \ref{extendedUGC02259} we present
the combined rotation curves including all the components of the
galaxy along with the SIDM. As it can be seen, the extended
collisional DM model is non-viable.
Also in Table \ref{evaluationextendedUGC02259} we present the
optimized values of the free parameters of the SIDM model for
which  we achieve the maximum compatibility with the SPARC data,
for the galaxy UGC02259, and also the resulting reduced
$\chi^2_{red}$ value.
\begin{table}[h!]
\centering \caption{Optimized Parameter Values of the Extended
SIDM model for the Galaxy UGC02259.}
\begin{tabular}{lc}
\hline
Parameter & Value  \\
\hline
$\rho_0 $ ($M_{\odot}/\mathrm{Kpc}^{3}$) & $6.71904\times 10^7$   \\
$K_0$ ($M_{\odot} \,
\mathrm{Kpc}^{-3} \, (\mathrm{km/s})^{2}$) & 2361.61   \\
$ml_{\text{disk}}$ & 1 \\
$ml_{\text{bulge}}$ & 0.4079 \\
$\alpha$ (Kpc) & 3.42095\\
$\chi^2_{red}$ & 10.9448 \\
\hline
\end{tabular}
\label{evaluationextendedUGC02259}
\end{table}

\subsection{The Galaxy UGC02487, Non-viable}

For this galaxy, the optimization method we used, ensures maximum
compatibility of the analytic SIDM model of Eq.
(\ref{ScaledependentEoSDM}) with the SPARC data, if we choose
$\rho_0=9.07148\times 10^7$$M_{\odot}/\mathrm{Kpc}^{3}$ and
$K_0=70393.6
$$M_{\odot} \, \mathrm{Kpc}^{-3} \, (\mathrm{km/s})^{2}$, in which
case the reduced $\chi^2_{red}$ value is $\chi^2_{red}=24.7828$.
Also the parameter $\alpha$ in this case is $\alpha=16.076 $Kpc.

\begin{figure}[h!]
\centering
\includegraphics[width=20pc]{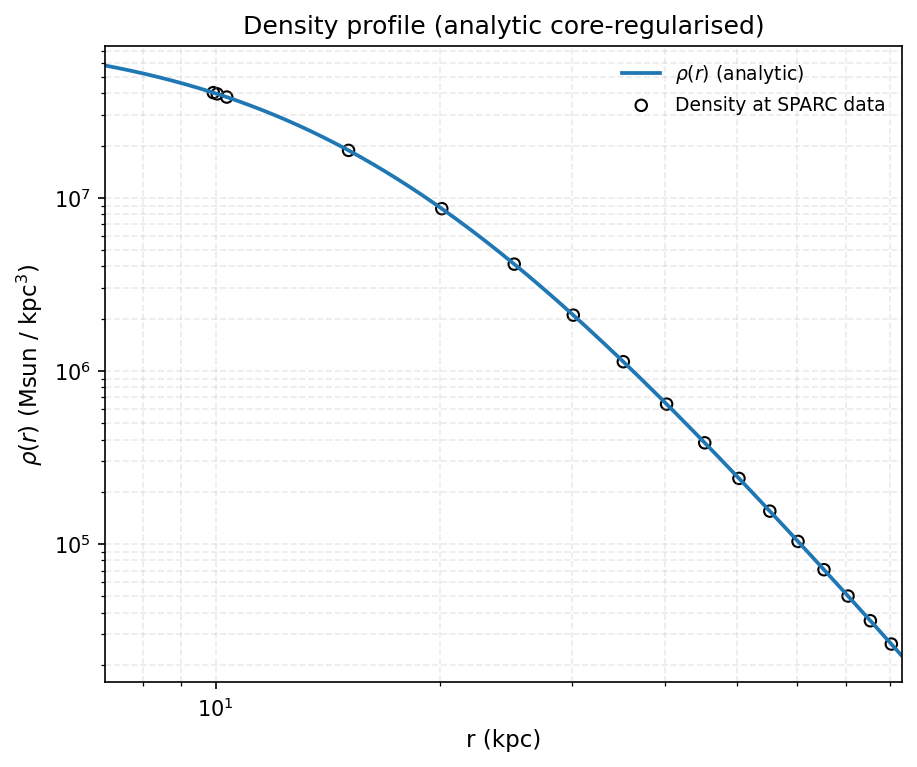}
\caption{The density of the SIDM model of Eq.
(\ref{ScaledependentEoSDM}) for the galaxy UGC02487, versus the
radius.} \label{UGC02487dens}
\end{figure}

\begin{figure}[h!]
\centering
\includegraphics[width=35pc]{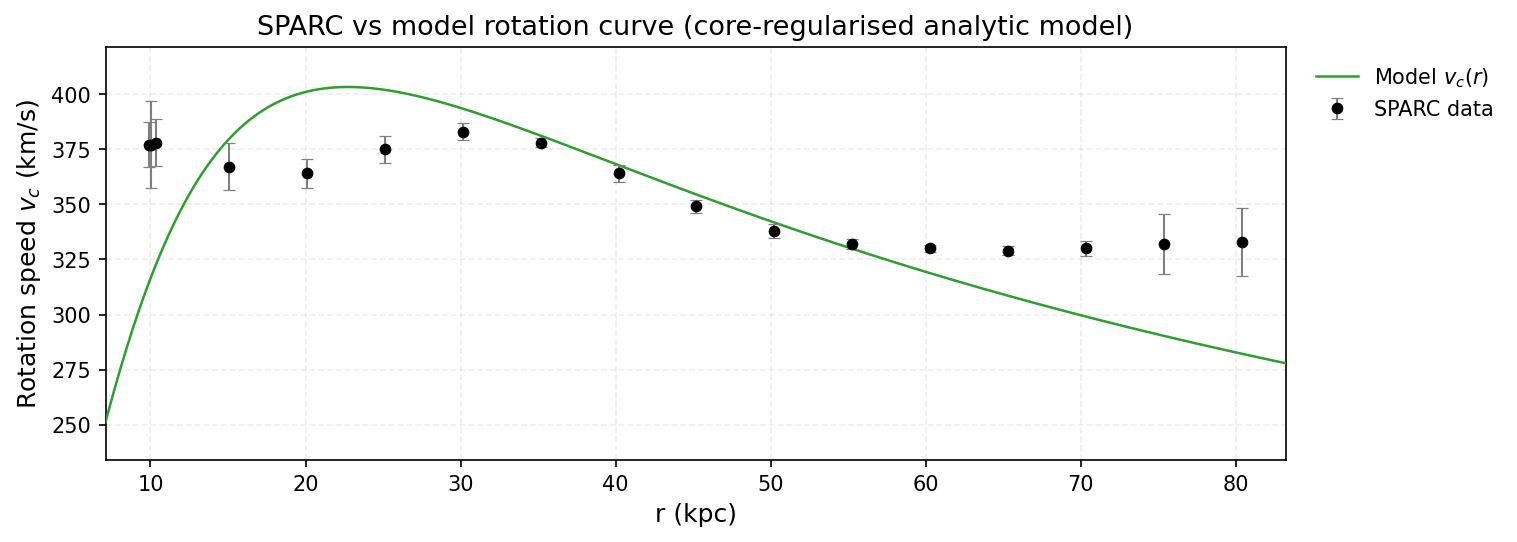}
\caption{The predicted rotation curves for the optimized SIDM
model of Eq. (\ref{ScaledependentEoSDM}), versus the SPARC
observational data for the galaxy UGC02487.} \label{UGC02487}
\end{figure}

In Table \ref{collUGC02487} we present the optimized values of
$K_0$ and $\rho_0$ for the analytic SIDM model of Eq.
(\ref{ScaledependentEoSDM}) for which the maximum compatibility
with the SPARC data is achieved.
\begin{table}[h!]
  \begin{center}
    \caption{SIDM Optimization Values for the galaxy UGC02487}
    \label{collUGC02487}
     \begin{tabular}{|r|r|}
     \hline
      \textbf{Parameter}   & \textbf{Optimization Values}
      \\  \hline
     $\rho_0 $  ($M_{\odot}/\mathrm{Kpc}^{3}$) & $9.07148\times 10^7$
\\  \hline $K_0$ ($M_{\odot} \,
\mathrm{Kpc}^{-3} \, (\mathrm{km/s})^{2}$)& 70393.6
\\  \hline
    \end{tabular}
  \end{center}
\end{table}
In Figs. \ref{UGC02487dens}, \ref{UGC02487}  we present the
density of the analytic SIDM model, the predicted rotation curves
for the SIDM model (\ref{ScaledependentEoSDM}), versus the SPARC
observational data and the sound speed, as a function of the
radius respectively. As it can be seen, for this galaxy, the SIDM
model produces non-viable rotation curves which are incompatible
with the SPARC data.

\begin{figure}[h!]
\centering
\includegraphics[width=20pc]{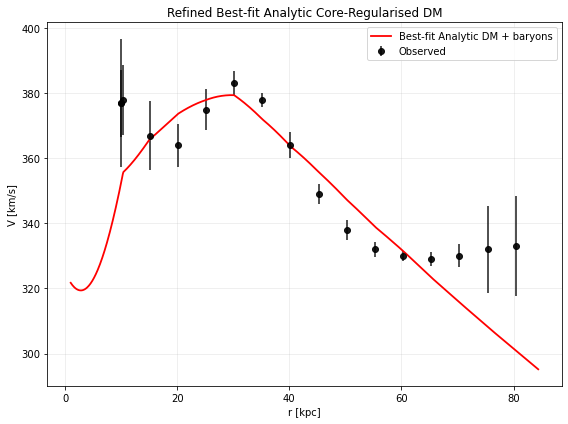}
\caption{The predicted rotation curves after using an optimization
for the SIDM model (\ref{ScaledependentEoSDM}), and the extended
SPARC data for the galaxy UGC02487. We included the rotation
curves of the gas, the disk velocities, the bulge (where present)
along with the SIDM model.} \label{extendedUGC02487}
\end{figure}

Now we shall include contributions to the rotation velocity from
the other components of the galaxy, namely the disk, the gas, and
the bulge if present. In Fig. \ref{extendedUGC02487} we present
the combined rotation curves including all the components of the
galaxy along with the SIDM. As it can be seen, the extended
collisional DM model is non-viable.
Also in Table \ref{evaluationextendedUGC02487} we present the
optimized values of the free parameters of the SIDM model for
which  we achieve the maximum compatibility with the SPARC data,
for the galaxy UGC02487, and also the resulting reduced
$\chi^2_{red}$ value.
\begin{table}[h!]
\centering \caption{Optimized Parameter Values of the Extended
SIDM model for the Galaxy UGC02487.}
\begin{tabular}{lc}
\hline
Parameter & Value  \\
\hline
$\rho_0 $ ($M_{\odot}/\mathrm{Kpc}^{3}$) & $9.62207\times 10^6$   \\
$K_0$ ($M_{\odot} \,
\mathrm{Kpc}^{-3} \, (\mathrm{km/s})^{2}$) & 31456.4   \\
$ml_{\text{disk}}$ & 1 \\
$ml_{\text{bulge}}$ & 1 \\
$\alpha$ (Kpc) & 32.9926\\
$\chi^2_{red}$ & 5.98761 \\
\hline
\end{tabular}
\label{evaluationextendedUGC02487}
\end{table}

\subsection{The Galaxy UGC02885, Non-viable, Extended Viable}

For this galaxy, the optimization method we used, ensures maximum
compatibility of the analytic SIDM model of Eq.
(\ref{ScaledependentEoSDM}) with the SPARC data, if we choose
$\rho_0=2.19424\times 10^7$$M_{\odot}/\mathrm{Kpc}^{3}$ and
$K_0=41500.7
$$M_{\odot} \, \mathrm{Kpc}^{-3} \, (\mathrm{km/s})^{2}$, in which
case the reduced $\chi^2_{red}$ value is $\chi^2_{red}=79.0084$.
Also the parameter $\alpha$ in this case is $\alpha=25.0979 $Kpc.

\begin{figure}[h!]
\centering
\includegraphics[width=20pc]{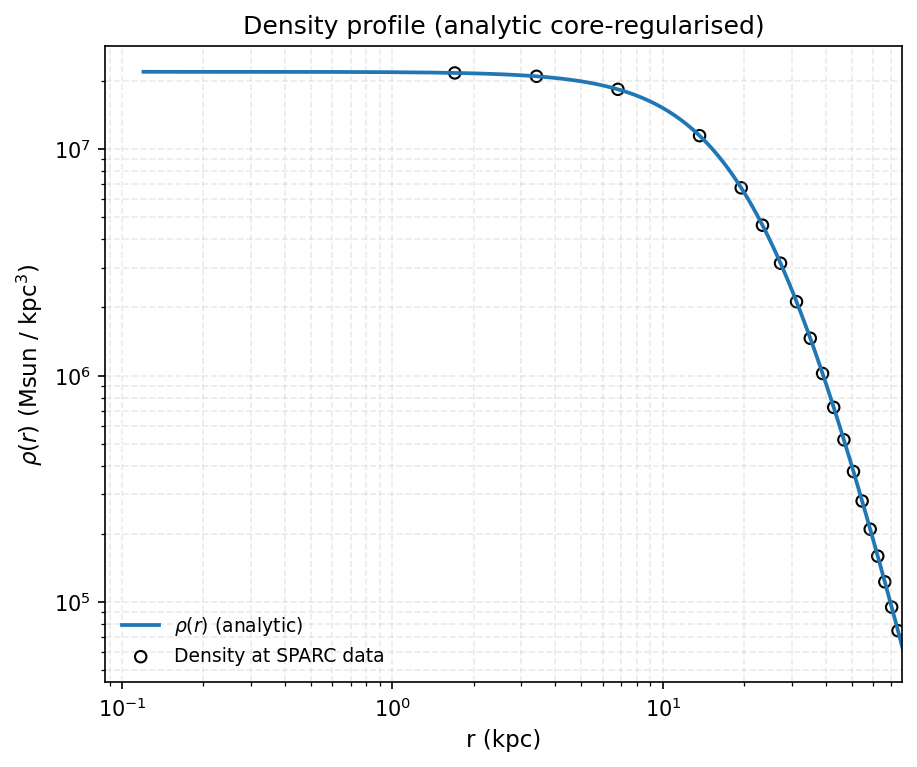}
\caption{The density of the SIDM model of Eq.
(\ref{ScaledependentEoSDM}) for the galaxy UGC02885, versus the
radius.} \label{UGC02885dens}
\end{figure}

\begin{figure}[h!]
\centering
\includegraphics[width=35pc]{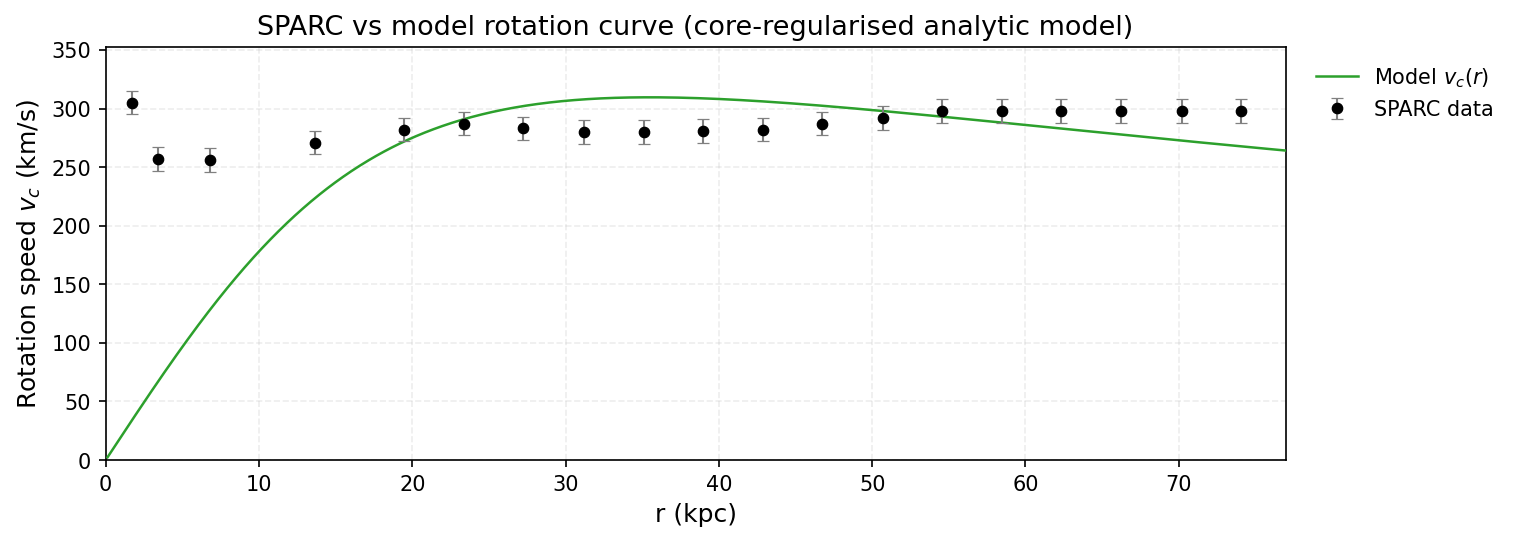}
\caption{The predicted rotation curves for the optimized SIDM
model of Eq. (\ref{ScaledependentEoSDM}), versus the SPARC
observational data for the galaxy UGC02885.} \label{UGC02885}
\end{figure}

In Table \ref{collUGC02885} we present the optimized values of
$K_0$ and $\rho_0$ for the analytic SIDM model of Eq.
(\ref{ScaledependentEoSDM}) for which the maximum compatibility
with the SPARC data is achieved.
\begin{table}[h!]
  \begin{center}
    \caption{SIDM Optimization Values for the galaxy UGC02885}
    \label{collUGC02885}
     \begin{tabular}{|r|r|}
     \hline
      \textbf{Parameter}   & \textbf{Optimization Values}
      \\  \hline
     $\rho_0 $  ($M_{\odot}/\mathrm{Kpc}^{3}$) & $2.19424\times 10^7$
\\  \hline $K_0$ ($M_{\odot} \,
\mathrm{Kpc}^{-3} \, (\mathrm{km/s})^{2}$)& 41500.7
\\  \hline
    \end{tabular}
  \end{center}
\end{table}
In Figs. \ref{UGC02885dens}, \ref{UGC02885}  we present the
density of the analytic SIDM model, the predicted rotation curves
for the SIDM model (\ref{ScaledependentEoSDM}), versus the SPARC
observational data and the sound speed, as a function of the
radius respectively. As it can be seen, for this galaxy, the SIDM
model produces non-viable rotation curves which are incompatible
with the SPARC data.

\begin{figure}[h!]
\centering
\includegraphics[width=20pc]{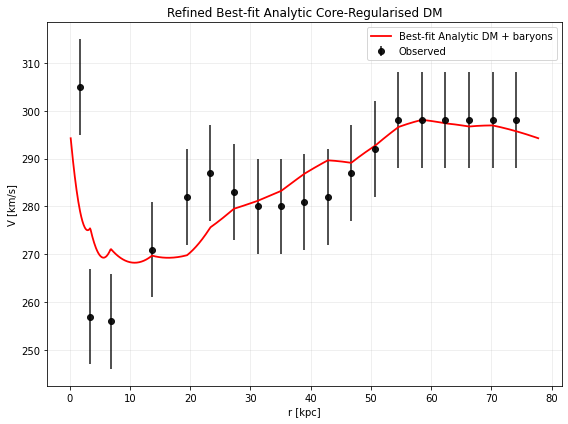}
\caption{The predicted rotation curves after using an optimization
for the SIDM model (\ref{ScaledependentEoSDM}), and the extended
SPARC data for the galaxy UGC02885. We included the rotation
curves of the gas, the disk velocities, the bulge (where present)
along with the SIDM model.} \label{extendedUGC02885}
\end{figure}

Now we shall include contributions to the rotation velocity from
the other components of the galaxy, namely the disk, the gas, and
the bulge if present. In Fig. \ref{extendedUGC02885} we present
the combined rotation curves including all the components of the
galaxy along with the SIDM. As it can be seen, the extended
collisional DM model is viable.
Also in Table \ref{evaluationextendedUGC02885} we present the
optimized values of the free parameters of the SIDM model for
which  we achieve the maximum compatibility with the SPARC data,
for the galaxy UGC02885, and also the resulting reduced
$\chi^2_{red}$ value.
\begin{table}[h!]
\centering \caption{Optimized Parameter Values of the Extended
SIDM model for the Galaxy UGC02885.}
\begin{tabular}{lc}
\hline
Parameter & Value  \\
\hline
$\rho_0 $ ($M_{\odot}/\mathrm{Kpc}^{3}$) & $2.30924\times 10^6$   \\
$K_0$ ($M_{\odot} \,
\mathrm{Kpc}^{-3} \, (\mathrm{km/s})^{2}$) & 25290.3   \\
$ml_{\text{disk}}$ & 0.9148 \\
$ml_{\text{bulge}}$ & 1 \\
$\alpha$ (Kpc) & 60.3863\\
$\chi^2_{red}$ & 1.10113 \\
\hline
\end{tabular}
\label{evaluationextendedUGC02885}
\end{table}

\subsection{The Galaxy UGC02916, Non-viable}

For this galaxy, the optimization method we used, ensures maximum
compatibility of the analytic SIDM model of Eq.
(\ref{ScaledependentEoSDM}) with the SPARC data, if we choose
$\rho_0=1.32108\times 10^8$$M_{\odot}/\mathrm{Kpc}^{3}$ and
$K_0=22911.6
$$M_{\odot} \, \mathrm{Kpc}^{-3} \, (\mathrm{km/s})^{2}$, in which
case the reduced $\chi^2_{red}$ value is $\chi^2_{red}=434.539$.
Also the parameter $\alpha$ in this case is $\alpha=7.60001  $Kpc.

\begin{figure}[h!]
\centering
\includegraphics[width=20pc]{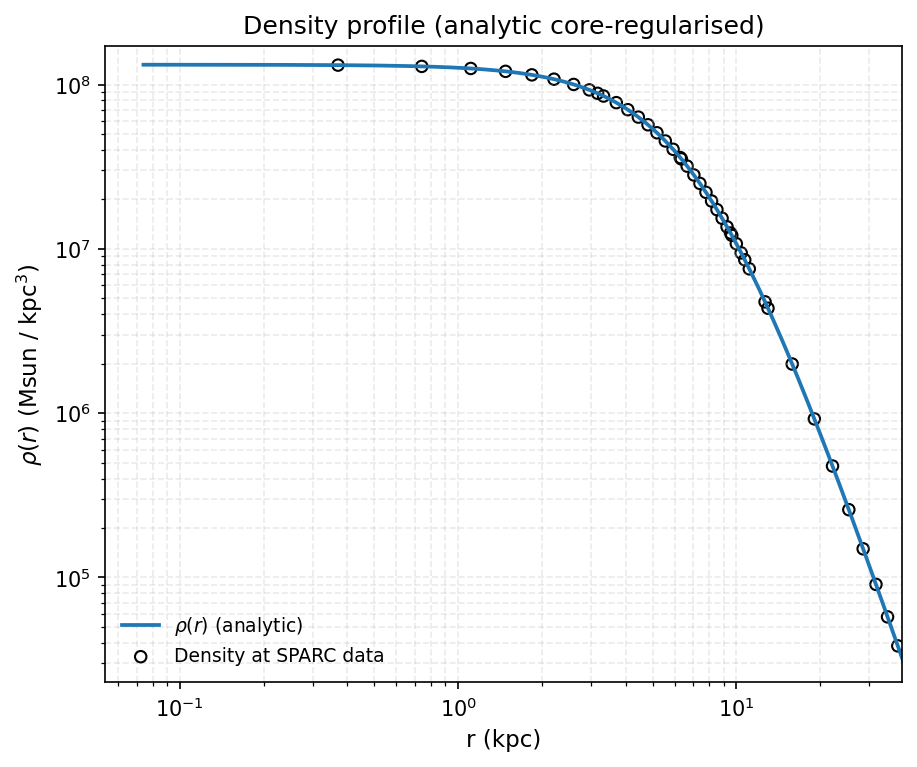}
\caption{The density of the SIDM model of Eq.
(\ref{ScaledependentEoSDM}) for the galaxy UGC02916, versus the
radius.} \label{UGC02916dens}
\end{figure}

\begin{figure}[h!]
\centering
\includegraphics[width=35pc]{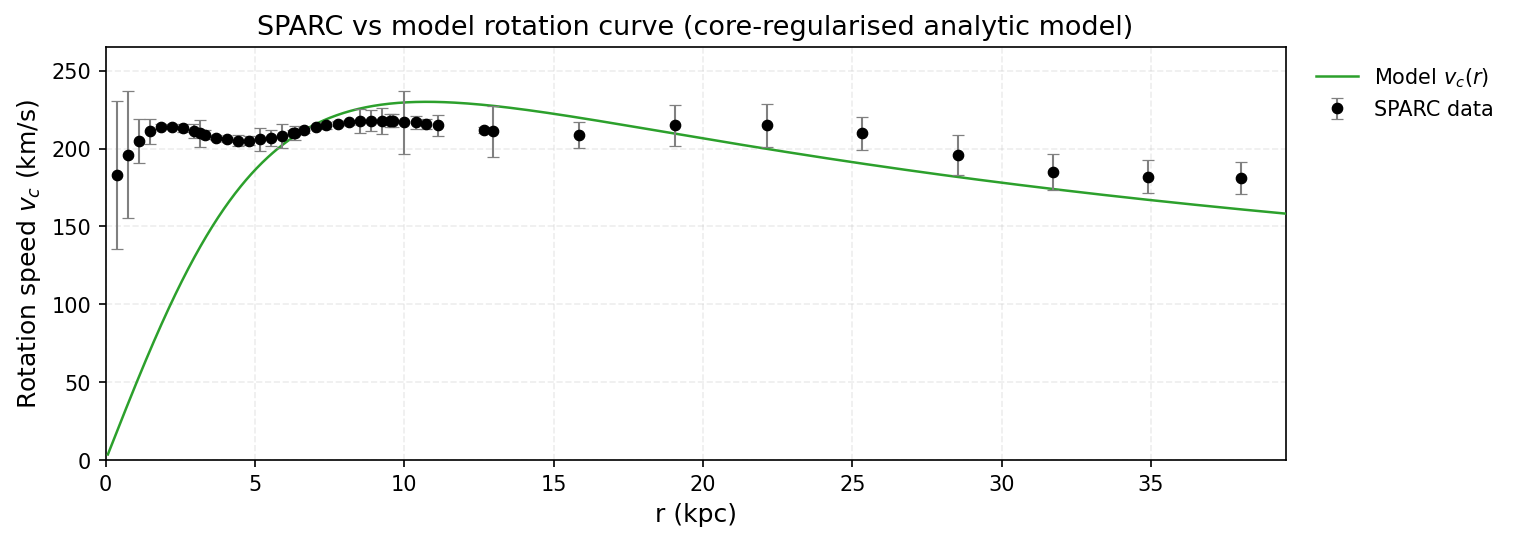}
\caption{The predicted rotation curves for the optimized SIDM
model of Eq. (\ref{ScaledependentEoSDM}), versus the SPARC
observational data for the galaxy UGC02916.} \label{UGC02916}
\end{figure}

In Table \ref{collUGC02916} we present the optimized values of
$K_0$ and $\rho_0$ for the analytic SIDM model of Eq.
(\ref{ScaledependentEoSDM}) for which the maximum compatibility
with the SPARC data is achieved.
\begin{table}[h!]
  \begin{center}
    \caption{SIDM Optimization Values for the galaxy UGC02916}
    \label{collUGC02916}
     \begin{tabular}{|r|r|}
     \hline
      \textbf{Parameter}   & \textbf{Optimization Values}
      \\  \hline
     $\rho_0 $  ($M_{\odot}/\mathrm{Kpc}^{3}$) & $1.32108\times 10^8$
\\  \hline $K_0$ ($M_{\odot} \,
\mathrm{Kpc}^{-3} \, (\mathrm{km/s})^{2}$)& 22911.6
\\  \hline
    \end{tabular}
  \end{center}
\end{table}
In Figs. \ref{UGC02916dens}, \ref{UGC02916}  we present the
density of the analytic SIDM model, the predicted rotation curves
for the SIDM model (\ref{ScaledependentEoSDM}), versus the SPARC
observational data and the sound speed, as a function of the
radius respectively. As it can be seen, for this galaxy, the SIDM
model produces non-viable rotation curves which are incompatible
with the SPARC data.

\begin{figure}[h!]
\centering
\includegraphics[width=20pc]{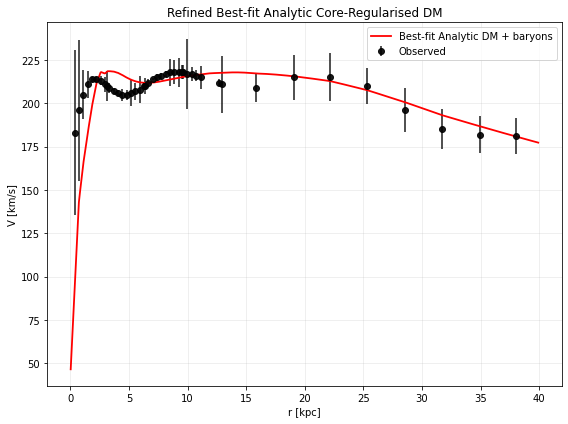}
\caption{The predicted rotation curves after using an optimization
for the SIDM model (\ref{ScaledependentEoSDM}), and the extended
SPARC data for the galaxy UGC02916. We included the rotation
curves of the gas, the disk velocities, the bulge (where present)
along with the SIDM model.} \label{extendedUGC02916}
\end{figure}

Now we shall include contributions to the rotation velocity from
the other components of the galaxy, namely the disk, the gas, and
the bulge if present. In Fig. \ref{extendedUGC02916} we present
the combined rotation curves including all the components of the
galaxy along with the SIDM. As it can be seen, the extended
collisional DM model is non-viable.
Also in Table \ref{evaluationextendedUGC02916} we present the
optimized values of the free parameters of the SIDM model for
which  we achieve the maximum compatibility with the SPARC data,
for the galaxy UGC02916, and also the resulting reduced
$\chi^2_{red}$ value.
\begin{table}[h!]
\centering \caption{Optimized Parameter Values of the Extended
SIDM model for the Galaxy UGC02916.}
\begin{tabular}{lc}
\hline
Parameter & Value  \\
\hline
$\rho_0 $ ($M_{\odot}/\mathrm{Kpc}^{3}$) & $3.66857\times 10^7$   \\
$K_0$ ($M_{\odot} \,
\mathrm{Kpc}^{-3} \, (\mathrm{km/s})^{2}$) & 14588.2   \\
$ml_{\text{disk}}$ & 1 \\
$ml_{\text{bulge}}$ & 0.7 \\
$\alpha$ (Kpc) & 11.5067\\
$\chi^2_{red}$ & 8.16682 \\
\hline
\end{tabular}
\label{evaluationextendedUGC02916}
\end{table}

\subsection{The Galaxy UGC02953, Non-viable}

For this galaxy, the optimization method we used, ensures maximum
compatibility of the analytic SIDM model of Eq.
(\ref{ScaledependentEoSDM}) with the SPARC data, if we choose
$\rho_0=1.21853\times 10^8$$M_{\odot}/\mathrm{Kpc}^{3}$ and
$K_0=56967.3
$$M_{\odot} \, \mathrm{Kpc}^{-3} \, (\mathrm{km/s})^{2}$, in which
case the reduced $\chi^2_{red}$ value is $\chi^2_{red}=934.878$.
Also the parameter $\alpha$ in this case is $\alpha=12.478 $Kpc.

\begin{figure}[h!]
\centering
\includegraphics[width=20pc]{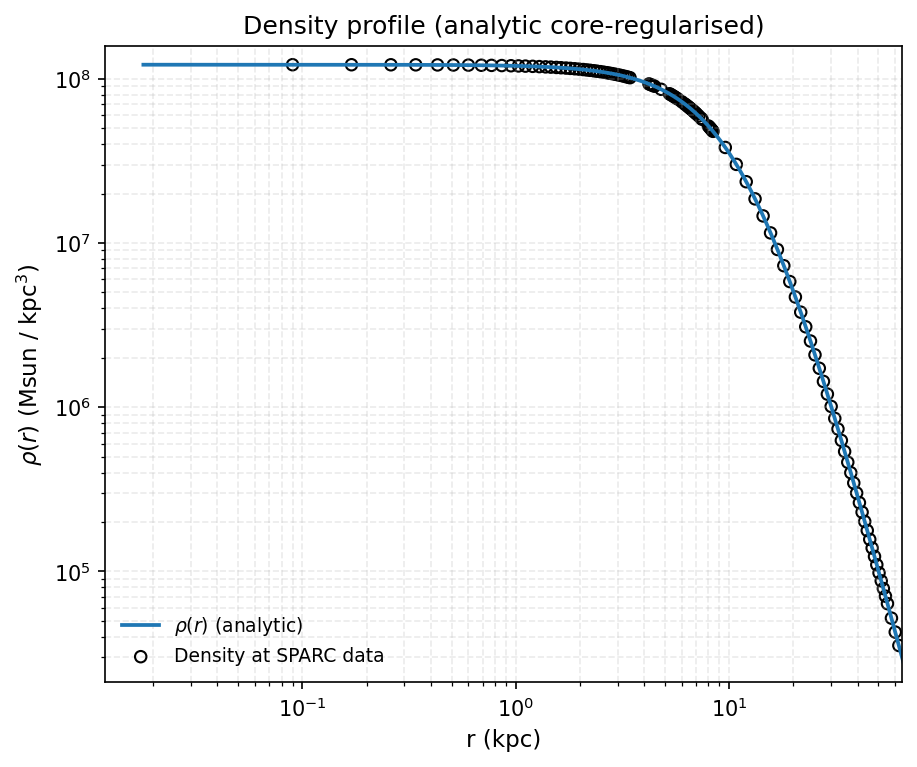}
\caption{The density of the SIDM model of Eq.
(\ref{ScaledependentEoSDM}) for the galaxy UGC02953, versus the
radius.} \label{UGC02953dens}
\end{figure}

\begin{figure}[h!]
\centering
\includegraphics[width=35pc]{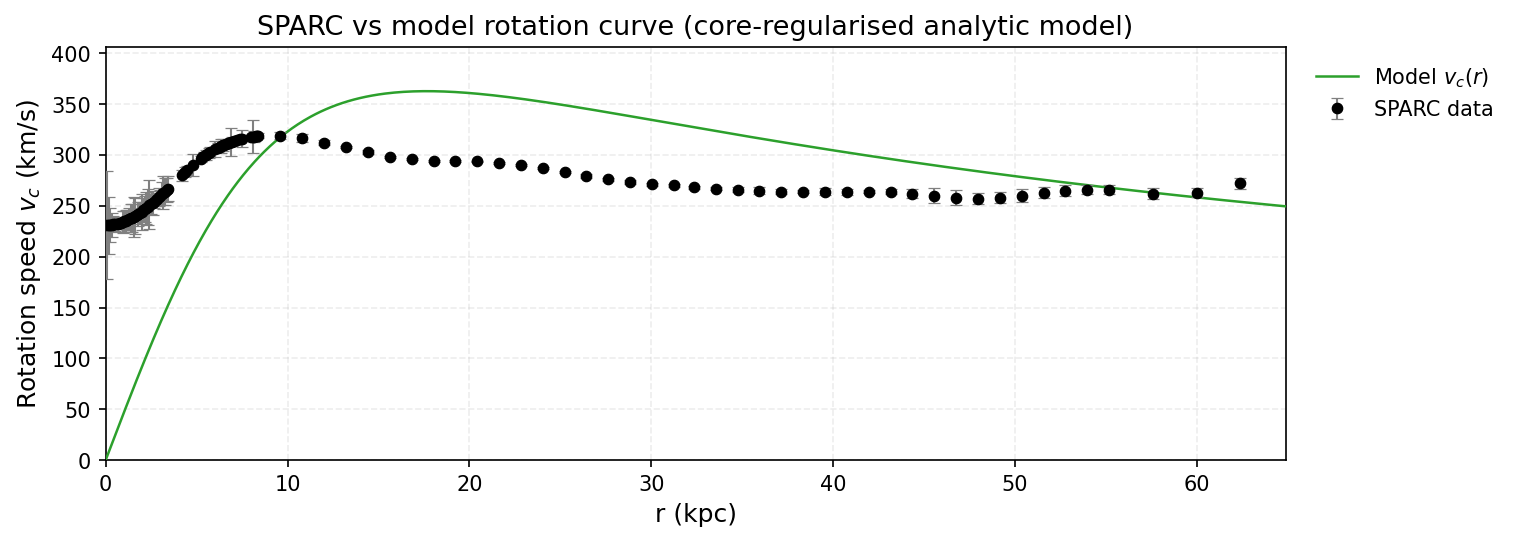}
\caption{The predicted rotation curves for the optimized SIDM
model of Eq. (\ref{ScaledependentEoSDM}), versus the SPARC
observational data for the galaxy UGC02953.} \label{UGC02953}
\end{figure}

In Table \ref{collUGC02953} we present the optimized values of
$K_0$ and $\rho_0$ for the analytic SIDM model of Eq.
(\ref{ScaledependentEoSDM}) for which the maximum compatibility
with the SPARC data is achieved.
\begin{table}[h!]
  \begin{center}
    \caption{SIDM Optimization Values for the galaxy UGC02953}
    \label{collUGC02953}
     \begin{tabular}{|r|r|}
     \hline
      \textbf{Parameter}   & \textbf{Optimization Values}
      \\  \hline
     $\rho_0 $  ($M_{\odot}/\mathrm{Kpc}^{3}$) & $1.21853\times 10^8$
\\  \hline $K_0$ ($M_{\odot} \,
\mathrm{Kpc}^{-3} \, (\mathrm{km/s})^{2}$)& 56967.3
\\  \hline
    \end{tabular}
  \end{center}
\end{table}
In Figs. \ref{UGC02953dens}, \ref{UGC02953}  we present the
density of the analytic SIDM model, the predicted rotation curves
for the SIDM model (\ref{ScaledependentEoSDM}), versus the SPARC
observational data and the sound speed, as a function of the
radius respectively. As it can be seen, for this galaxy, the SIDM
model produces non-viable rotation curves which are incompatible
with the SPARC data.

\begin{figure}[h!]
\centering
\includegraphics[width=20pc]{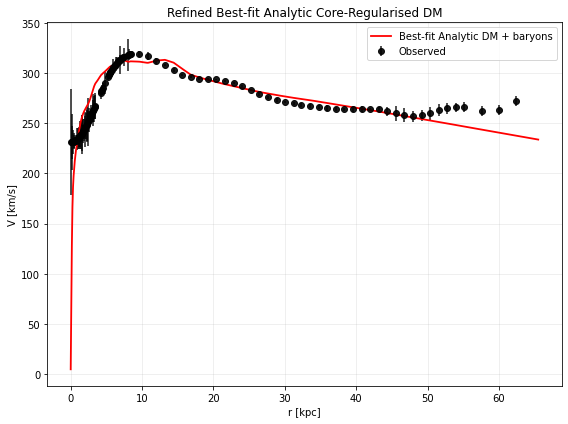}
\caption{The predicted rotation curves after using an optimization
for the SIDM model (\ref{ScaledependentEoSDM}), and the extended
SPARC data for the galaxy UGC02953. We included the rotation
curves of the gas, the disk velocities, the bulge (where present)
along with the SIDM model.} \label{extendedUGC02953}
\end{figure}

Now we shall include contributions to the rotation velocity from
the other components of the galaxy, namely the disk, the gas, and
the bulge if present. In Fig. \ref{extendedUGC02953} we present
the combined rotation curves including all the components of the
galaxy along with the SIDM. As it can be seen, the extended
collisional DM model is non-viable.
Also in Table \ref{evaluationextendedUGC02953} we present the
optimized values of the free parameters of the SIDM model for
which  we achieve the maximum compatibility with the SPARC data,
for the galaxy UGC02953, and also the resulting reduced
$\chi^2_{red}$ value.
\begin{table}[h!]
\centering \caption{Optimized Parameter Values of the Extended
SIDM model for the Galaxy UGC02953.}
\begin{tabular}{lc}
\hline
Parameter & Value  \\
\hline
$\rho_0 $ ($M_{\odot}/\mathrm{Kpc}^{3}$) & $8.2663\times 10^6$   \\
$K_0$ ($M_{\odot} \,
\mathrm{Kpc}^{-3} \, (\mathrm{km/s})^{2}$) & 19898.9   \\
$ml_{\text{disk}}$ & 0.9064 \\
$ml_{\text{bulge}}$ & 0.8139 \\
$\alpha$ (Kpc) & 28.311\\
$\chi^2_{red}$ & 7.09491 \\
\hline
\end{tabular}
\label{evaluationextendedUGC02953}
\end{table}

\subsection{The Galaxy UGC03205, Non-viable}

For this galaxy, the optimization method we used, ensures maximum
compatibility of the analytic SIDM model of Eq.
(\ref{ScaledependentEoSDM}) with the SPARC data, if we choose
$\rho_0=1.3911\times 10^8$$M_{\odot}/\mathrm{Kpc}^{3}$ and
$K_0=26786
$$M_{\odot} \, \mathrm{Kpc}^{-3} \, (\mathrm{km/s})^{2}$, in which
case the reduced $\chi^2_{red}$ value is $\chi^2_{red}=34.3376$.
Also the parameter $\alpha$ in this case is $\alpha=8.008 $Kpc.

\begin{figure}[h!]
\centering
\includegraphics[width=20pc]{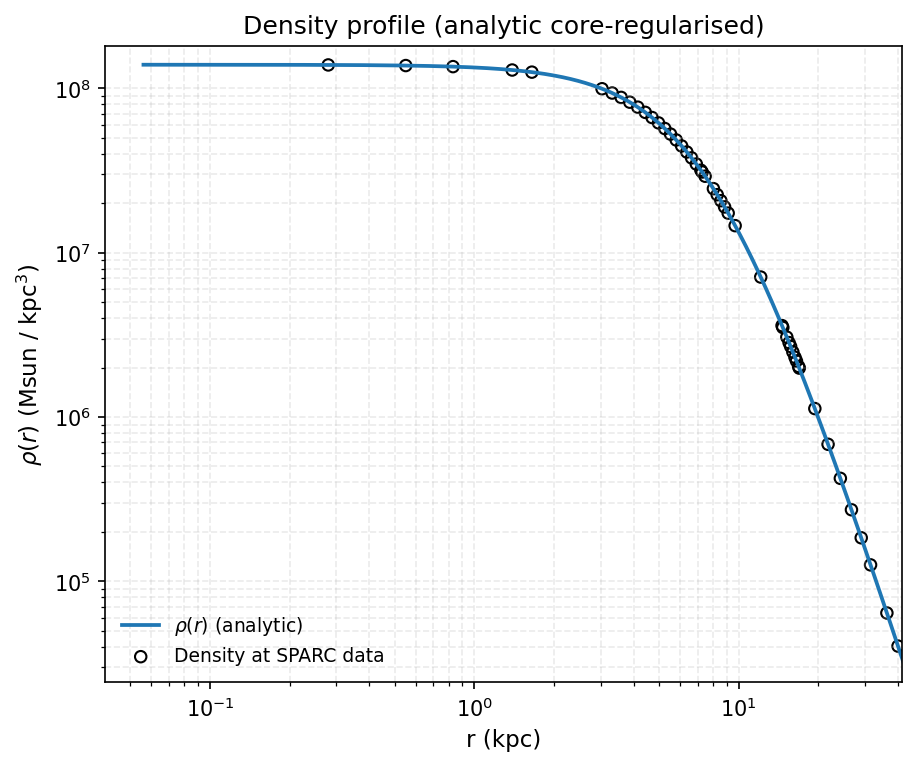}
\caption{The density of the SIDM model of Eq.
(\ref{ScaledependentEoSDM}) for the galaxy UGC03205, versus the
radius.} \label{UGC03205dens}
\end{figure}

\begin{figure}[h!]
\centering
\includegraphics[width=35pc]{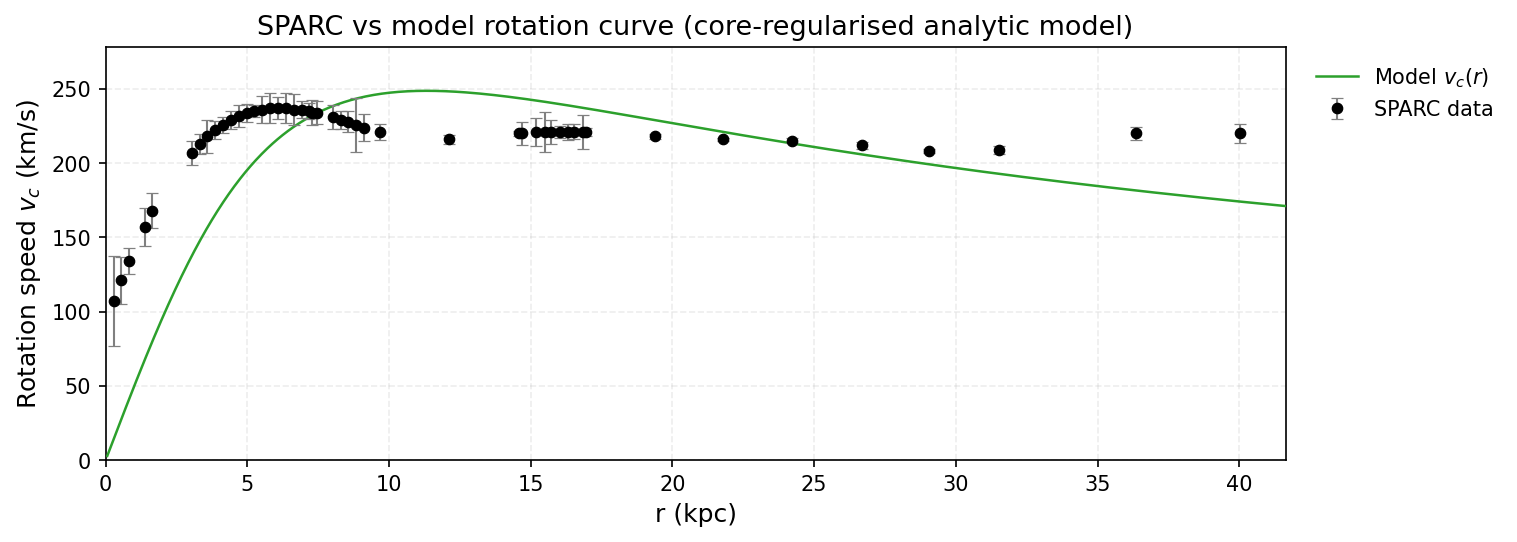}
\caption{The predicted rotation curves for the optimized SIDM
model of Eq. (\ref{ScaledependentEoSDM}), versus the SPARC
observational data for the galaxy UGC03205.} \label{UGC03205}
\end{figure}

In Table \ref{collUGC03205} we present the optimized values of
$K_0$ and $\rho_0$ for the analytic SIDM model of Eq.
(\ref{ScaledependentEoSDM}) for which the maximum compatibility
with the SPARC data is achieved.
\begin{table}[h!]
  \begin{center}
    \caption{SIDM Optimization Values for the galaxy UGC03205}
    \label{collUGC03205}
     \begin{tabular}{|r|r|}
     \hline
      \textbf{Parameter}   & \textbf{Optimization Values}
      \\  \hline
     $\rho_0 $  ($M_{\odot}/\mathrm{Kpc}^{3}$) & $1.3911\times 10^8$
\\  \hline $K_0$ ($M_{\odot} \,
\mathrm{Kpc}^{-3} \, (\mathrm{km/s})^{2}$)& 26786
\\  \hline
    \end{tabular}
  \end{center}
\end{table}
In Figs. \ref{UGC03205dens}, \ref{UGC03205} we present the density
of the analytic SIDM model, the predicted rotation curves for the
SIDM model (\ref{ScaledependentEoSDM}), versus the SPARC
observational data and the sound speed, as a function of the
radius respectively. As it can be seen, for this galaxy, the SIDM
model produces non-viable rotation curves which are incompatible
with the SPARC data.

\begin{figure}[h!]
\centering
\includegraphics[width=20pc]{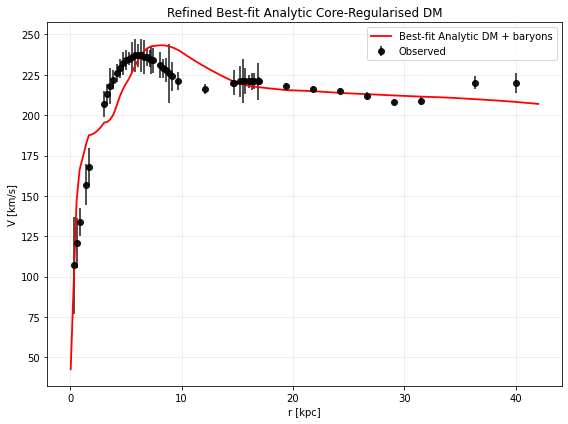}
\caption{The predicted rotation curves after using an optimization
for the SIDM model (\ref{ScaledependentEoSDM}), and the extended
SPARC data for the galaxy UGC03205. We included the rotation
curves of the gas, the disk velocities, the bulge (where present)
along with the SIDM model.} \label{extendedUGC03205}
\end{figure}

Now we shall include contributions to the rotation velocity from
the other components of the galaxy, namely the disk, the gas, and
the bulge if present. In Fig. \ref{extendedUGC03205} we present
the combined rotation curves including all the components of the
galaxy along with the SIDM. As it can be seen, the extended
collisional DM model is non-viable.
Also in Table \ref{evaluationextendedUGC03205} we present the
optimized values of the free parameters of the SIDM model for
which  we achieve the maximum compatibility with the SPARC data,
for the galaxy UGC03205, and also the resulting reduced
$\chi^2_{red}$ value.
\begin{table}[h!]
\centering \caption{Optimized Parameter Values of the Extended
SIDM model for the Galaxy UGC03205.}
\begin{tabular}{lc}
\hline
Parameter & Value  \\
\hline
$\rho_0 $ ($M_{\odot}/\mathrm{Kpc}^{3}$) & $3.45378\times 10^6$   \\
$K_0$ ($M_{\odot} \,
\mathrm{Kpc}^{-3} \, (\mathrm{km/s})^{2}$) & 13033.5   \\
$ml_{\text{disk}}$ & 0.9924 \\
$ml_{\text{bulge}}$ & 1 \\
$\alpha$ (Kpc) & 35.4469\\
$\chi^2_{red}$ & 3.95292 \\
\hline
\end{tabular}
\label{evaluationextendedUGC03205}
\end{table}

\subsection{The Galaxy UGC03546, Non-viable, Extended Viable}

For this galaxy, the optimization method we used, ensures maximum
compatibility of the analytic SIDM model of Eq.
(\ref{ScaledependentEoSDM}) with the SPARC data, if we choose
$\rho_0=2.26423\times 10^8$$M_{\odot}/\mathrm{Kpc}^{3}$ and
$K_0=23234.7
$$M_{\odot} \, \mathrm{Kpc}^{-3} \, (\mathrm{km/s})^{2}$, in which
case the reduced $\chi^2_{red}$ value is $\chi^2_{red}=98.132$.
Also the parameter $\alpha$ in this case is $\alpha=5.846 $Kpc.

\begin{figure}[h!]
\centering
\includegraphics[width=20pc]{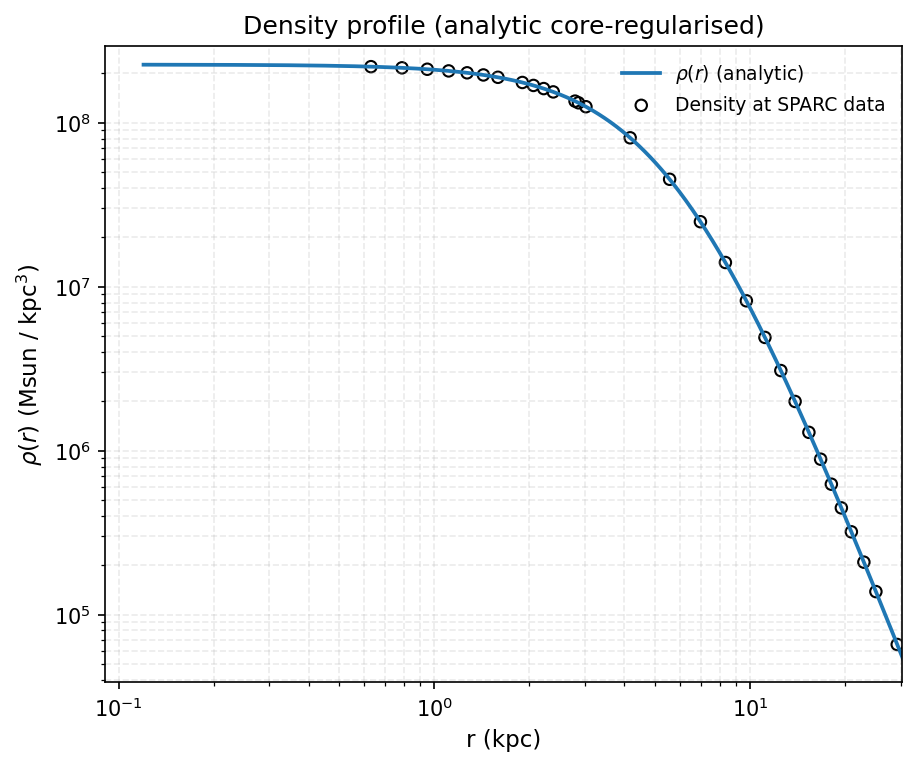}
\caption{The density of the SIDM model of Eq.
(\ref{ScaledependentEoSDM}) for the galaxy UGC03546, versus the
radius.} \label{UGC03546dens}
\end{figure}

\begin{figure}[h!]
\centering
\includegraphics[width=35pc]{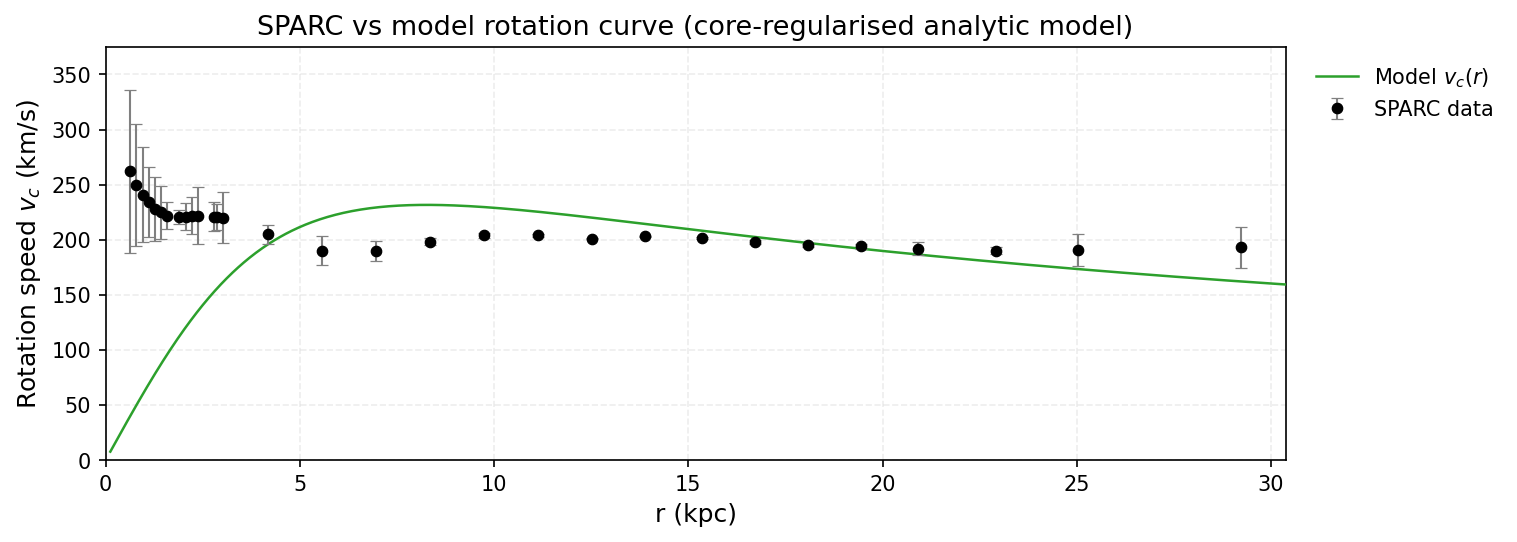}
\caption{The predicted rotation curves for the optimized SIDM
model of Eq. (\ref{ScaledependentEoSDM}), versus the SPARC
observational data for the galaxy UGC03546.} \label{UGC03546}
\end{figure}

In Table \ref{collUGC03546} we present the optimized values of
$K_0$ and $\rho_0$ for the analytic SIDM model of Eq.
(\ref{ScaledependentEoSDM}) for which the maximum compatibility
with the SPARC data is achieved.
\begin{table}[h!]
  \begin{center}
    \caption{SIDM Optimization Values for the galaxy UGC03546}
    \label{collUGC03546}
     \begin{tabular}{|r|r|}
     \hline
      \textbf{Parameter}   & \textbf{Optimization Values}
      \\  \hline
     $\rho_0 $  ($M_{\odot}/\mathrm{Kpc}^{3}$) & $2.26423\times 10^8$
\\  \hline $K_0$ ($M_{\odot} \,
\mathrm{Kpc}^{-3} \, (\mathrm{km/s})^{2}$)& 23234.7
\\  \hline
    \end{tabular}
  \end{center}
\end{table}
In Figs. \ref{UGC03546dens}, \ref{UGC03546}  we present the
density of the analytic SIDM model, the predicted rotation curves
for the SIDM model (\ref{ScaledependentEoSDM}), versus the SPARC
observational data and the sound speed, as a function of the
radius respectively. As it can be seen, for this galaxy, the SIDM
model produces non-viable rotation curves which are incompatible
with the SPARC data.

\begin{figure}[h!]
\centering
\includegraphics[width=20pc]{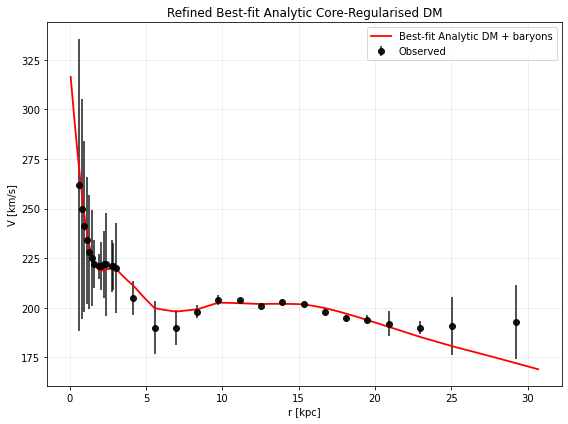}
\caption{The predicted rotation curves after using an optimization
for the SIDM model (\ref{ScaledependentEoSDM}), and the extended
SPARC data for the galaxy UGC03546. We included the rotation
curves of the gas, the disk velocities, the bulge (where present)
along with the SIDM model.} \label{extendedUGC03546}
\end{figure}

Now we shall include contributions to the rotation velocity from
the other components of the galaxy, namely the disk, the gas, and
the bulge if present. In Fig. \ref{extendedUGC03546} we present
the combined rotation curves including all the components of the
galaxy along with the SIDM. As it can be seen, the extended
collisional DM model is viable.
Also in Table \ref{evaluationextendedUGC03546} we present the
optimized values of the free parameters of the SIDM model for
which  we achieve the maximum compatibility with the SPARC data,
for the galaxy UGC03546, and also the resulting reduced
$\chi^2_{red}$ value.
\begin{table}[h!]
\centering \caption{Optimized Parameter Values of the Extended
SIDM model for the Galaxy UGC03546.}
\begin{tabular}{lc}
\hline
Parameter & Value  \\
\hline
$\rho_0 $ ($M_{\odot}/\mathrm{Kpc}^{3}$) & $1.62481\times 10^7$   \\
$K_0$ ($M_{\odot} \,
\mathrm{Kpc}^{-3} \, (\mathrm{km/s})^{2}$) & 9543.66   \\
$ml_{\text{disk}}$ & 0.7826 \\
$ml_{\text{bulge}}$ & 0.6702 \\
$\alpha$ (Kpc) & 13.9847\\
$\chi^2_{red}$ & 0.638214 \\
\hline
\end{tabular}
\label{evaluationextendedUGC03546}
\end{table}

\subsection{The Galaxy UGC03580, Non-viable}

For this galaxy, the optimization method we used, ensures maximum
compatibility of the analytic SIDM model of Eq.
(\ref{ScaledependentEoSDM}) with the SPARC data, if we choose
$\rho_0=8.61957\times 10^7$$M_{\odot}/\mathrm{Kpc}^{3}$ and
$K_0=7580.54
$$M_{\odot} \, \mathrm{Kpc}^{-3} \, (\mathrm{km/s})^{2}$, in which
case the reduced $\chi^2_{red}$ value is $\chi^2_{red}=75.2683$.
Also the parameter $\alpha$ in this case is $\alpha=5.412 $Kpc.

\begin{figure}[h!]
\centering
\includegraphics[width=20pc]{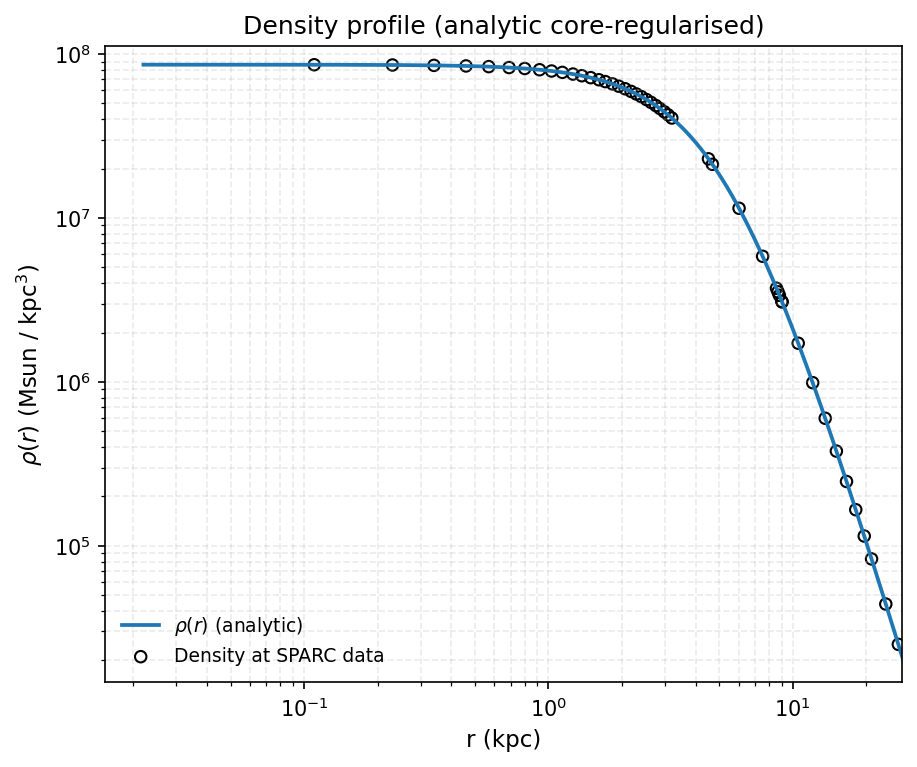}
\caption{The density of the SIDM model of Eq.
(\ref{ScaledependentEoSDM}) for the galaxy UGC03580, versus the
radius.} \label{UGC03580dens}
\end{figure}

\begin{figure}[h!]
\centering
\includegraphics[width=35pc]{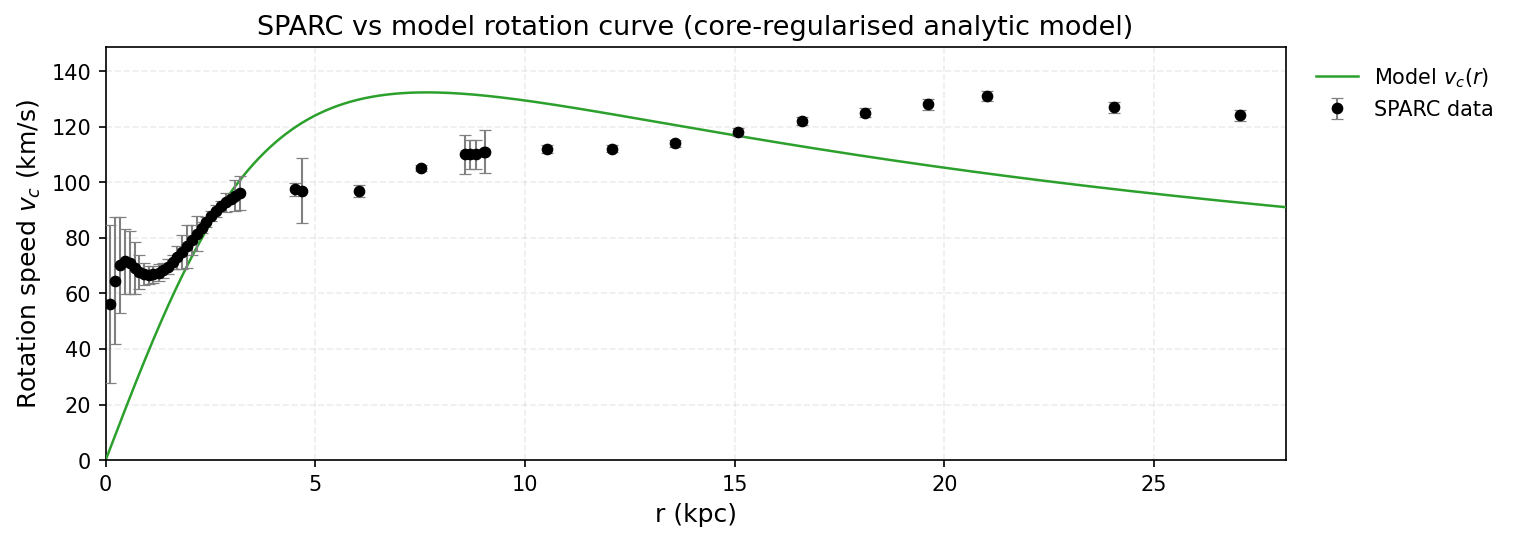}
\caption{The predicted rotation curves for the optimized SIDM
model of Eq. (\ref{ScaledependentEoSDM}), versus the SPARC
observational data for the galaxy UGC03580.} \label{UGC03580}
\end{figure}

In Table \ref{collUGC03580} we present the optimized values of
$K_0$ and $\rho_0$ for the analytic SIDM model of Eq.
(\ref{ScaledependentEoSDM}) for which the maximum compatibility
with the SPARC data is achieved.
\begin{table}[h!]
  \begin{center}
    \caption{SIDM Optimization Values for the galaxy UGC03580}
    \label{collUGC03580}
     \begin{tabular}{|r|r|}
     \hline
      \textbf{Parameter}   & \textbf{Optimization Values}
      \\  \hline
     $\rho_0 $  ($M_{\odot}/\mathrm{Kpc}^{3}$) & $8.61957\times 10^7$
\\  \hline $K_0$ ($M_{\odot} \,
\mathrm{Kpc}^{-3} \, (\mathrm{km/s})^{2}$)& 7580.54
\\  \hline
    \end{tabular}
  \end{center}
\end{table}
In Figs. \ref{UGC03580dens}, \ref{UGC03580} we present the density
of the analytic SIDM model, the predicted rotation curves for the
SIDM model (\ref{ScaledependentEoSDM}), versus the SPARC
observational data and the sound speed, as a function of the
radius respectively. As it can be seen, for this galaxy, the SIDM
model produces non-viable rotation curves which are incompatible
with the SPARC data.

\begin{figure}[h!]
\centering
\includegraphics[width=20pc]{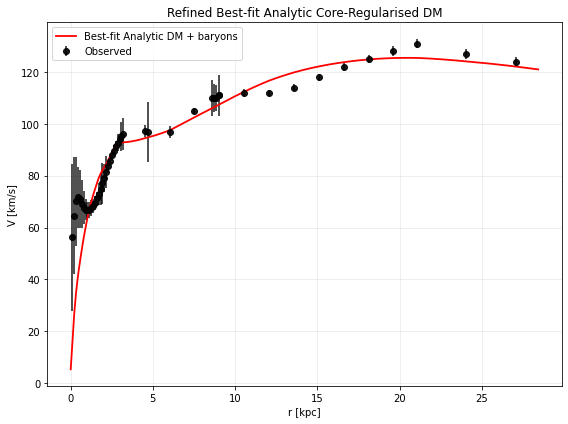}
\caption{The predicted rotation curves after using an optimization
for the SIDM model (\ref{ScaledependentEoSDM}), and the extended
SPARC data for the galaxy UGC03580. We included the rotation
curves of the gas, the disk velocities, the bulge (where present)
along with the SIDM model.} \label{extendedUGC03580}
\end{figure}

Now we shall include contributions to the rotation velocity from
the other components of the galaxy, namely the disk, the gas, and
the bulge if present. In Fig. \ref{extendedUGC03580} we present
the combined rotation curves including all the components of the
galaxy along with the SIDM. As it can be seen, the extended
collisional DM model is non-viable.
Also in Table \ref{evaluationextendedUGC03580} we present the
optimized values of the free parameters of the SIDM model for
which  we achieve the maximum compatibility with the SPARC data,
for the galaxy UGC03580, and also the resulting reduced
$\chi^2_{red}$ value.
\begin{table}[h!]
\centering \caption{Optimized Parameter Values of the Extended
SIDM model for the Galaxy UGC03580.}
\begin{tabular}{lc}
\hline
Parameter & Value  \\
\hline
$\rho_0 $ ($M_{\odot}/\mathrm{Kpc}^{3}$) & $6.43606\times 10^6$   \\
$K_0$ ($M_{\odot} \,
\mathrm{Kpc}^{-3} \, (\mathrm{km/s})^{2}$) & 5409   \\
$ml_{\text{disk}}$ & 0.8941 \\
$ml_{\text{bulge}}$ & 0.2673 \\
$\alpha$ (Kpc) & 16.728\\
$\chi^2_{red}$ & 3.22713 \\
\hline
\end{tabular}
\label{evaluationextendedUGC03580}
\end{table}

\subsection{The Galaxy UGC04278, Non-viable, Extended Viable}

For this galaxy, the optimization method we used, ensures maximum
compatibility of the analytic SIDM model of Eq.
(\ref{ScaledependentEoSDM}) with the SPARC data, if we choose
$\rho_0=2.39591\times 10^7$$M_{\odot}/\mathrm{Kpc}^{3}$ and
$K_0=3846.21
$$M_{\odot} \, \mathrm{Kpc}^{-3} \, (\mathrm{km/s})^{2}$, in which
case the reduced $\chi^2_{red}$ value is $\chi^2_{red}=1.79109$.
Also the parameter $\alpha$ in this case is $\alpha=7.31193 $Kpc.

\begin{figure}[h!]
\centering
\includegraphics[width=20pc]{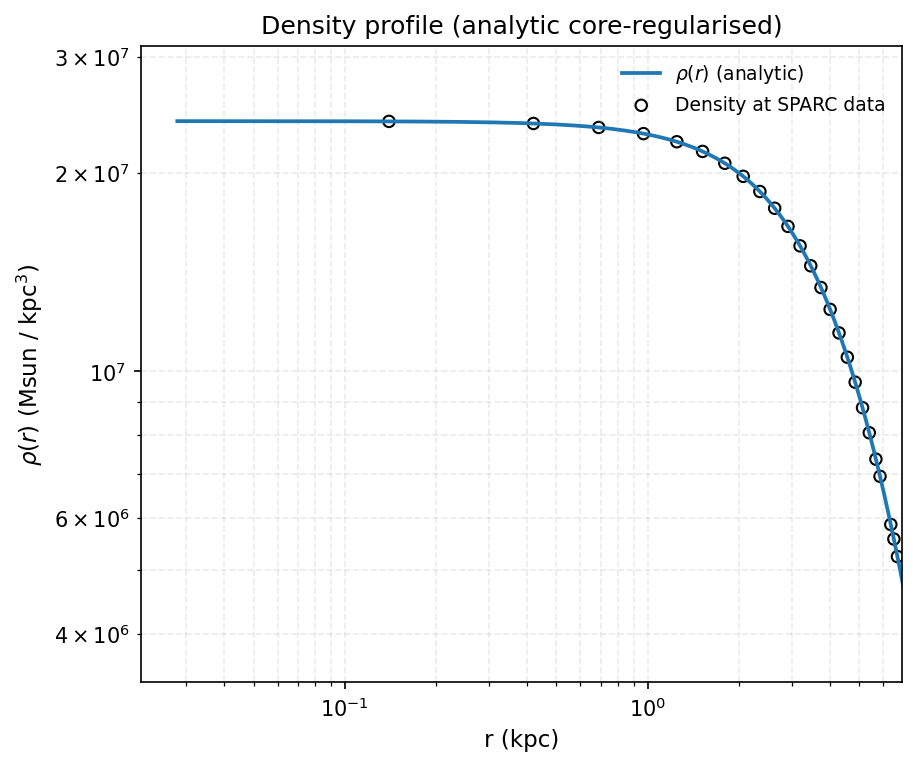}
\caption{The density of the SIDM model of Eq.
(\ref{ScaledependentEoSDM}) for the galaxy UGC04278, versus the
radius.} \label{UGC04278dens}
\end{figure}

\begin{figure}[h!]
\centering
\includegraphics[width=35pc]{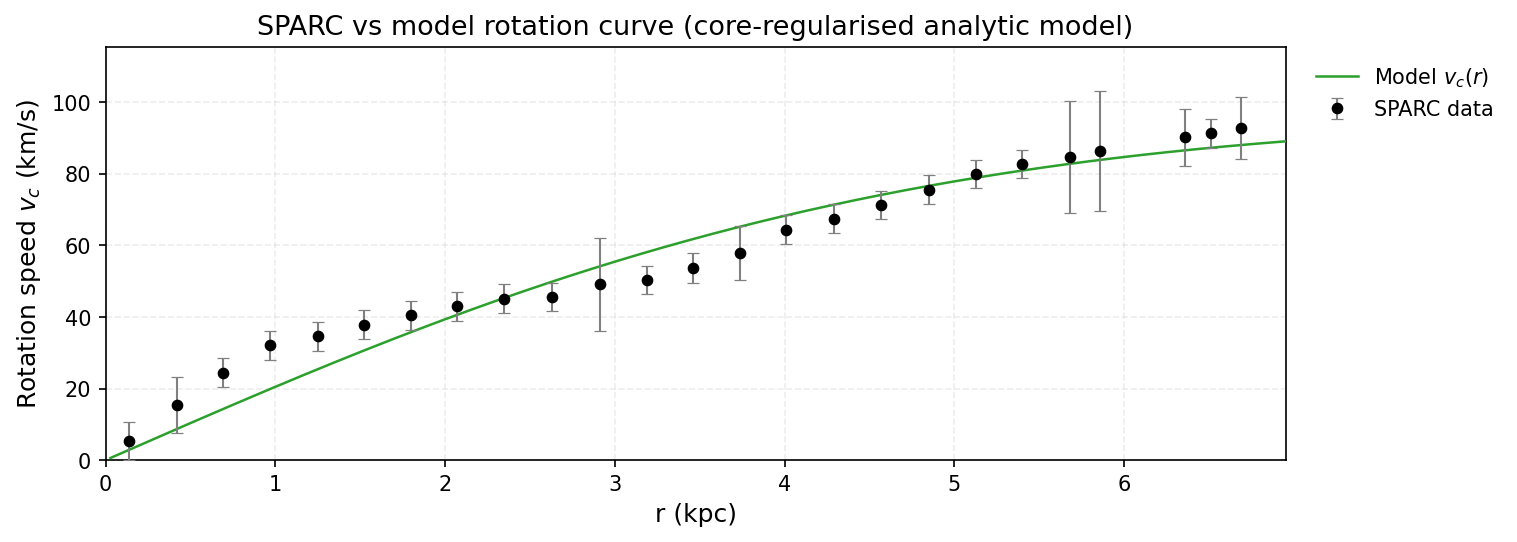}
\caption{The predicted rotation curves for the optimized SIDM
model of Eq. (\ref{ScaledependentEoSDM}), versus the SPARC
observational data for the galaxy UGC04278.} \label{UGC04278}
\end{figure}

In Table \ref{collUGC04278} we present the optimized values of
$K_0$ and $\rho_0$ for the analytic SIDM model of Eq.
(\ref{ScaledependentEoSDM}) for which the maximum compatibility
with the SPARC data is achieved.
\begin{table}[h!]
  \begin{center}
    \caption{SIDM Optimization Values for the galaxy UGC04278}
    \label{collUGC04278}
     \begin{tabular}{|r|r|}
     \hline
      \textbf{Parameter}   & \textbf{Optimization Values}
      \\  \hline
     $\rho_0 $  ($M_{\odot}/\mathrm{Kpc}^{3}$) & $2.39591\times 10^7$
\\  \hline $K_0$ ($M_{\odot} \,
\mathrm{Kpc}^{-3} \, (\mathrm{km/s})^{2}$)& 3846.21
\\  \hline
    \end{tabular}
  \end{center}
\end{table}
In Figs. \ref{UGC04278dens}, \ref{UGC04278}  we present the
density of the analytic SIDM model, the predicted rotation curves
for the SIDM model (\ref{ScaledependentEoSDM}), versus the SPARC
observational data and the sound speed, as a function of the
radius respectively. As it can be seen, for this galaxy, the SIDM
model produces non-viable rotation curves which are incompatible
with the SPARC data.

\begin{figure}[h!]
\centering
\includegraphics[width=20pc]{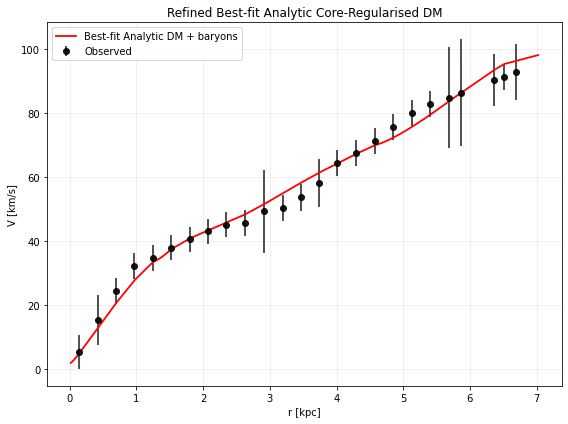}
\caption{The predicted rotation curves after using an optimization
for the SIDM model (\ref{ScaledependentEoSDM}), and the extended
SPARC data for the galaxy UGC04278. We included the rotation
curves of the gas, the disk velocities, the bulge (where present)
along with the SIDM model.} \label{extendedUGC04278}
\end{figure}

Now we shall include contributions to the rotation velocity from
the other components of the galaxy, namely the disk, the gas, and
the bulge if present. In Fig. \ref{extendedUGC04278} we present
the combined rotation curves including all the components of the
galaxy along with the SIDM. As it can be seen, the extended
collisional DM model is non-viable.
Also in Table \ref{evaluationextendedUGC04278} we present the
optimized values of the free parameters of the SIDM model for
which  we achieve the maximum compatibility with the SPARC data,
for the galaxy UGC04278, and also the resulting reduced
$\chi^2_{red}$ value.
\begin{table}[h!]
\centering \caption{Optimized Parameter Values of the Extended
SIDM model for the Galaxy UGC04278.}
\begin{tabular}{lc}
\hline
Parameter & Value  \\
\hline
$\rho_0 $ ($M_{\odot}/\mathrm{Kpc}^{3}$) & $1.11917\times 10^7$   \\
$K_0$ ($M_{\odot} \,
\mathrm{Kpc}^{-3} \, (\mathrm{km/s})^{2}$) & 6514.27   \\
$ml_{\text{disk}}$ & 1 \\
$ml_{\text{bulge}}$ & 0.5354 \\
$\alpha$ (Kpc) & 13.9214\\
$\chi^2_{red}$ & 0.45359 \\
\hline
\end{tabular}
\label{evaluationextendedUGC04278}
\end{table}

\subsection{The Galaxy UGC04325}

For this galaxy, the optimization method we used, ensures maximum
compatibility of the analytic SIDM model of Eq.
(\ref{ScaledependentEoSDM}) with the SPARC data, if we choose
$\rho_0=1.8532\times 10^8$$M_{\odot}/\mathrm{Kpc}^{3}$ and
$K_0=3763.46
$$M_{\odot} \, \mathrm{Kpc}^{-3} \, (\mathrm{km/s})^{2}$, in which
case the reduced $\chi^2_{red}$ value is $\chi^2_{red}=0.706892$.
Also the parameter $\alpha$ in this case is $\alpha=2.60065 $Kpc.

\begin{figure}[h!]
\centering
\includegraphics[width=20pc]{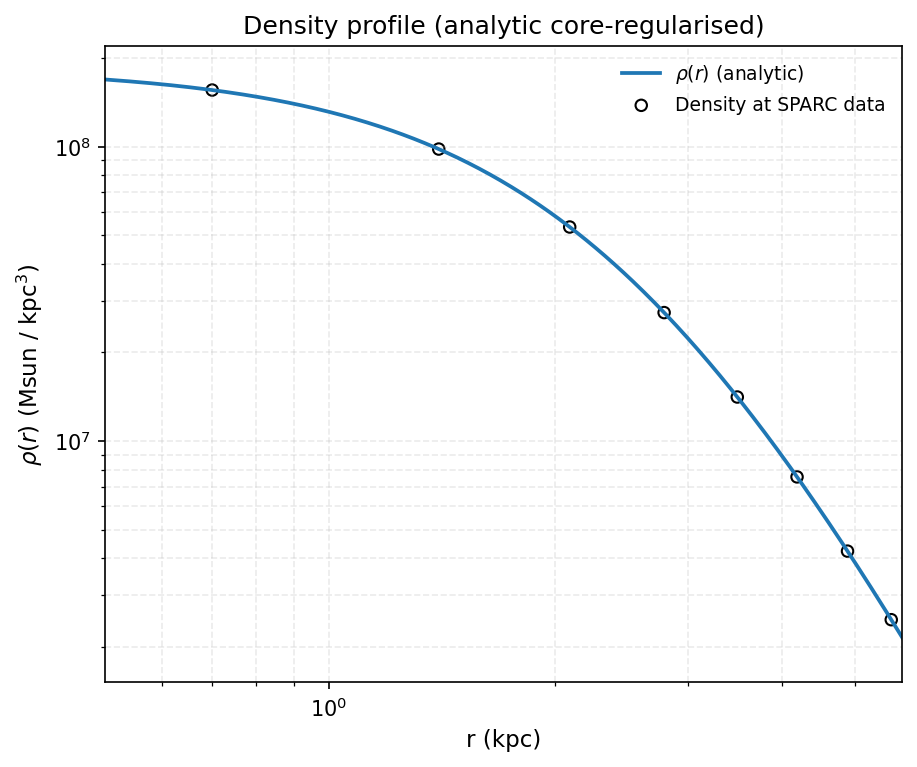}
\caption{The density of the SIDM model of Eq.
(\ref{ScaledependentEoSDM}) for the galaxy UGC04325, versus the
radius.} \label{UGC04325dens}
\end{figure}

\begin{figure}[h!]
\centering
\includegraphics[width=35pc]{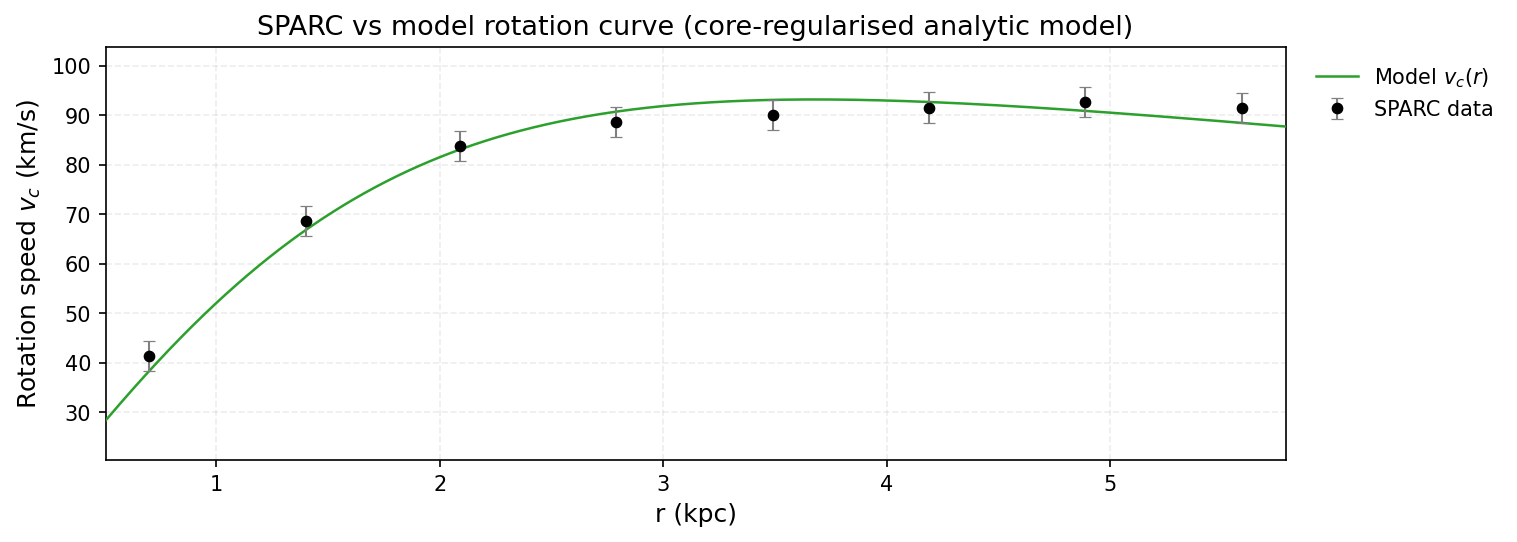}
\caption{The predicted rotation curves for the optimized SIDM
model of Eq. (\ref{ScaledependentEoSDM}), versus the SPARC
observational data for the galaxy UGC04325.} \label{UGC04325}
\end{figure}

In Table \ref{collUGC04325} we present the optimized values of
$K_0$ and $\rho_0$ for the analytic SIDM model of Eq.
(\ref{ScaledependentEoSDM}) for which the maximum compatibility
with the SPARC data is achieved.
\begin{table}[h!]
  \begin{center}
    \caption{SIDM Optimization Values for the galaxy UGC04325}
    \label{collUGC04325}
     \begin{tabular}{|r|r|}
     \hline
      \textbf{Parameter}   & \textbf{Optimization Values}
      \\  \hline
     $\rho_0 $  ($M_{\odot}/\mathrm{Kpc}^{3}$) & $1.8532\times 10^8$
\\  \hline $K_0$ ($M_{\odot} \,
\mathrm{Kpc}^{-3} \, (\mathrm{km/s})^{2}$)& 3763.46
\\  \hline
    \end{tabular}
  \end{center}
\end{table}
In Figs. \ref{UGC04325dens}, \ref{UGC04325}  we present the
density of the analytic SIDM model, the predicted rotation curves
for the SIDM model (\ref{ScaledependentEoSDM}), versus the SPARC
observational data and the sound speed, as a function of the
radius respectively. As it can be seen, for this galaxy, the SIDM
model produces viable rotation curves which are compatible with
the SPARC data.

\subsection{The Galaxy UGC04483}

For this galaxy, the optimization method we used, ensures maximum
compatibility of the analytic SIDM model of Eq.
(\ref{ScaledependentEoSDM}) with the SPARC data, if we choose
$\rho_0=1.20775\times 10^8$$M_{\odot}/\mathrm{Kpc}^{3}$ and
$K_0=253.812
$$M_{\odot} \, \mathrm{Kpc}^{-3} \, (\mathrm{km/s})^{2}$, in which
case the reduced $\chi^2_{red}$ value is $\chi^2_{red}=0.395683$.
Also the parameter $\alpha$ in this case is $\alpha=0.836599 $Kpc.

\begin{figure}[h!]
\centering
\includegraphics[width=20pc]{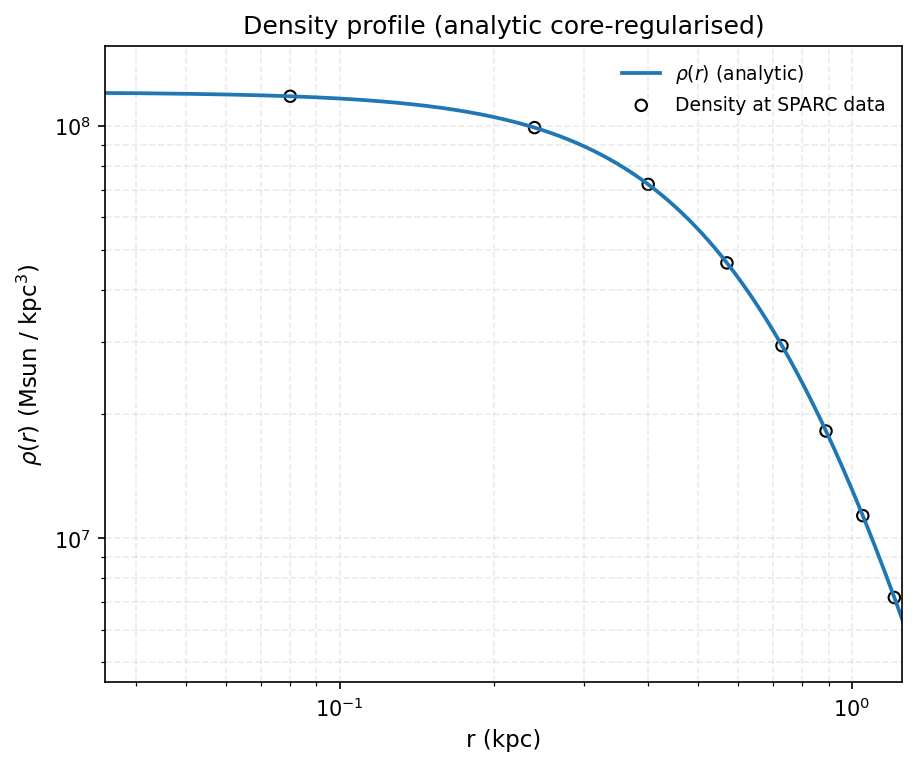}
\caption{The density of the SIDM model of Eq.
(\ref{ScaledependentEoSDM}) for the galaxy UGC04483, versus the
radius.} \label{UGC04483dens}
\end{figure}

\begin{figure}[h!]
\centering
\includegraphics[width=35pc]{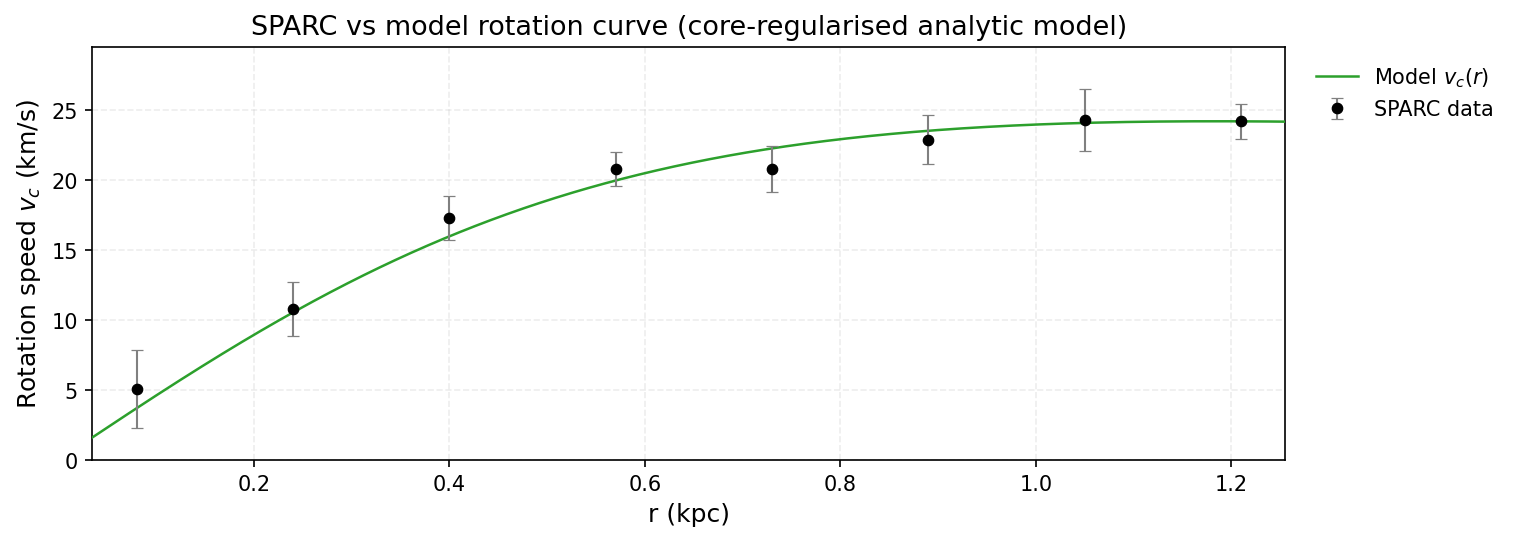}
\caption{The predicted rotation curves for the optimized SIDM
model of Eq. (\ref{ScaledependentEoSDM}), versus the SPARC
observational data for the galaxy UGC04483.} \label{UGC04483}
\end{figure}

In Table \ref{collUGC04483} we present the optimized values of
$K_0$ and $\rho_0$ for the analytic SIDM model of Eq.
(\ref{ScaledependentEoSDM}) for which the maximum compatibility
with the SPARC data is achieved.
\begin{table}[h!]
  \begin{center}
    \caption{SIDM Optimization Values for the galaxy UGC04483}
    \label{collUGC04483}
     \begin{tabular}{|r|r|}
     \hline
      \textbf{Parameter}   & \textbf{Optimization Values}
      \\  \hline
     $\rho_0 $  ($M_{\odot}/\mathrm{Kpc}^{3}$) & $1.20775\times 10^8$
\\  \hline $K_0$ ($M_{\odot} \,
\mathrm{Kpc}^{-3} \, (\mathrm{km/s})^{2}$)& 253.812
\\  \hline
    \end{tabular}
  \end{center}
\end{table}
In Figs. \ref{UGC04483dens}, \ref{UGC04483} we present the density
of the analytic SIDM model, the predicted rotation curves for the
SIDM model (\ref{ScaledependentEoSDM}), versus the SPARC
observational data and the sound speed, as a function of the
radius respectively. As it can be seen, for this galaxy, the SIDM
model produces viable rotation curves which are compatible with
the SPARC data.

\subsection{The Galaxy UGC04499}

For this galaxy, the optimization method we used, ensures maximum
compatibility of the analytic SIDM model of Eq.
(\ref{ScaledependentEoSDM}) with the SPARC data, if we choose
$\rho_0=3.70754\times 10^7$$M_{\odot}/\mathrm{Kpc}^{3}$ and
$K_0=2303.64
$$M_{\odot} \, \mathrm{Kpc}^{-3} \, (\mathrm{km/s})^{2}$, in which
case the reduced $\chi^2_{red}$ value is $\chi^2_{red}=0.756822$.
Also the parameter $\alpha$ in this case is $\alpha=4.549 $Kpc.

\begin{figure}[h!]
\centering
\includegraphics[width=20pc]{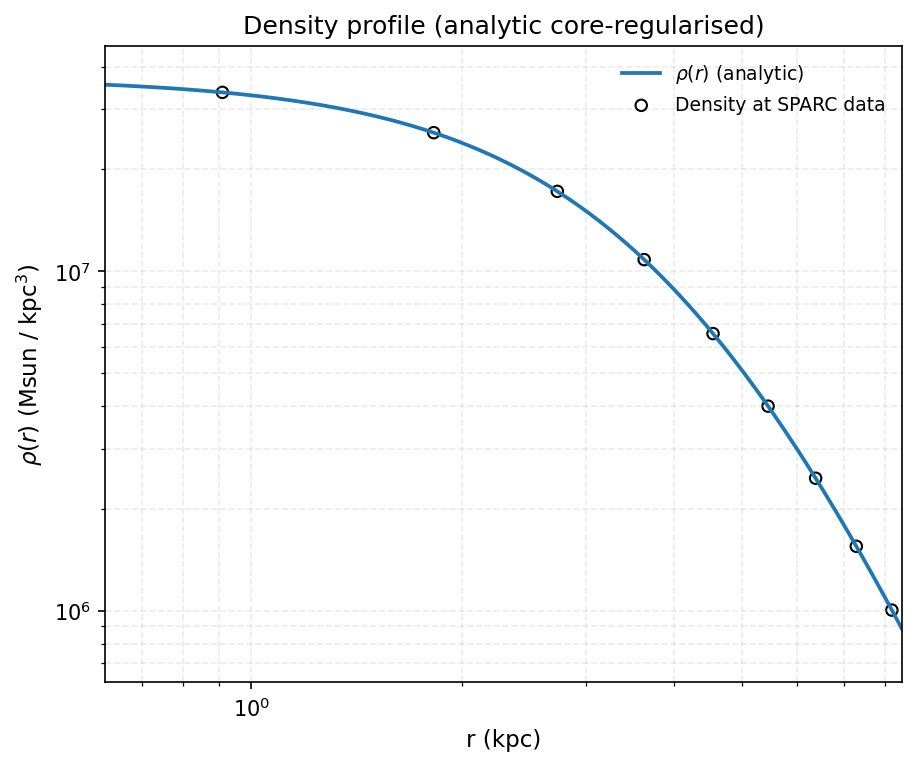}
\caption{The density of the SIDM model of Eq.
(\ref{ScaledependentEoSDM}) for the galaxy UGC04499, versus the
radius.} \label{UGC04499dens}
\end{figure}

\begin{figure}[h!]
\centering
\includegraphics[width=35pc]{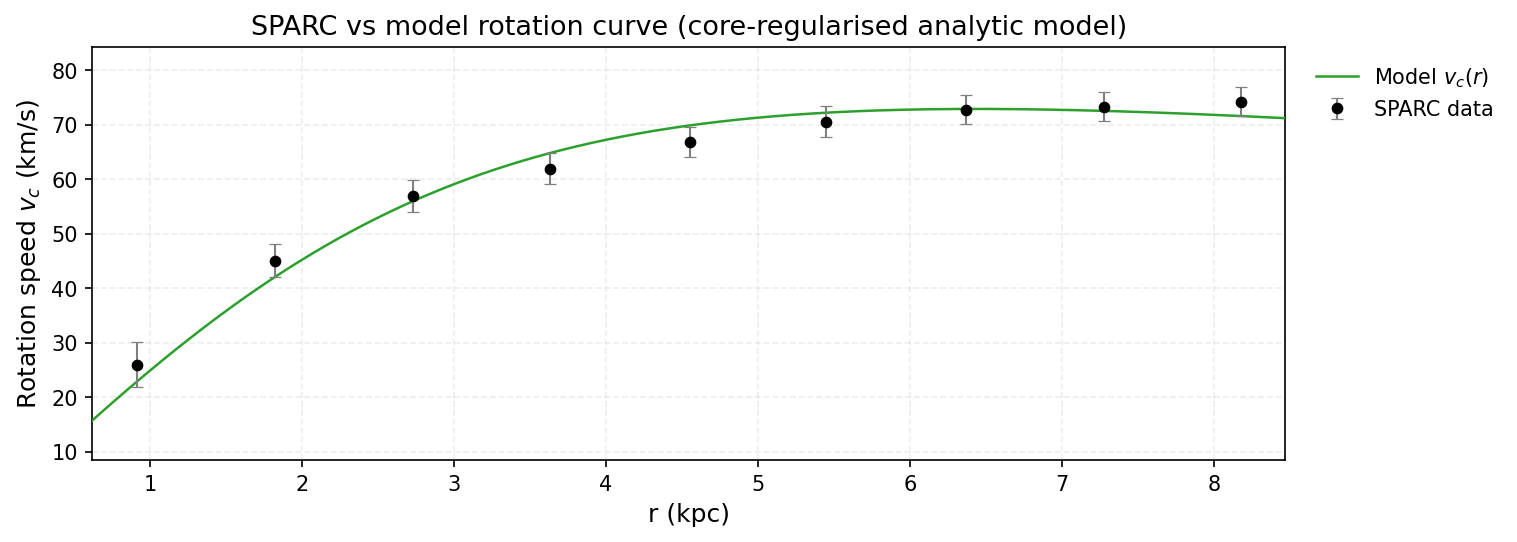}
\caption{The predicted rotation curves for the optimized SIDM
model of Eq. (\ref{ScaledependentEoSDM}), versus the SPARC
observational data for the galaxy UGC04499.} \label{UGC04499}
\end{figure}

In Table \ref{collUGC04499} we present the optimized values of
$K_0$ and $\rho_0$ for the analytic SIDM model of Eq.
(\ref{ScaledependentEoSDM}) for which the maximum compatibility
with the SPARC data is achieved.
\begin{table}[h!]
  \begin{center}
    \caption{SIDM Optimization Values for the galaxy UGC04499}
    \label{collUGC04499}
     \begin{tabular}{|r|r|}
     \hline
      \textbf{Parameter}   & \textbf{Optimization Values}
      \\  \hline
     $\rho_0 $  ($M_{\odot}/\mathrm{Kpc}^{3}$) & $3.70754\times 10^7$
\\  \hline $K_0$ ($M_{\odot} \,
\mathrm{Kpc}^{-3} \, (\mathrm{km/s})^{2}$)& 2303.64
\\  \hline
    \end{tabular}
  \end{center}
\end{table}
In Figs. \ref{UGC04499dens}, \ref{UGC04499} we present the density
of the analytic SIDM model, the predicted rotation curves for the
SIDM model (\ref{ScaledependentEoSDM}), versus the SPARC
observational data and the sound speed, as a function of the
radius respectively. As it can be seen, for this galaxy, the SIDM
model produces viable rotation curves which are compatible with
the SPARC data.

\subsection{The Galaxy UGC05005}

For this galaxy, the optimization method we used, ensures maximum
compatibility of the analytic SIDM model of Eq.
(\ref{ScaledependentEoSDM}) with the SPARC data, if we choose
$\rho_0=6.43428\times 10^6$$M_{\odot}/\mathrm{Kpc}^{3}$ and
$K_0=4016.81
$$M_{\odot} \, \mathrm{Kpc}^{-3} \, (\mathrm{km/s})^{2}$, in which
case the reduced $\chi^2_{red}$ value is $\chi^2_{red}=0.285774$.
Also the parameter $\alpha$ in this case is $\alpha=14.4192 $Kpc.

\begin{figure}[h!]
\centering
\includegraphics[width=20pc]{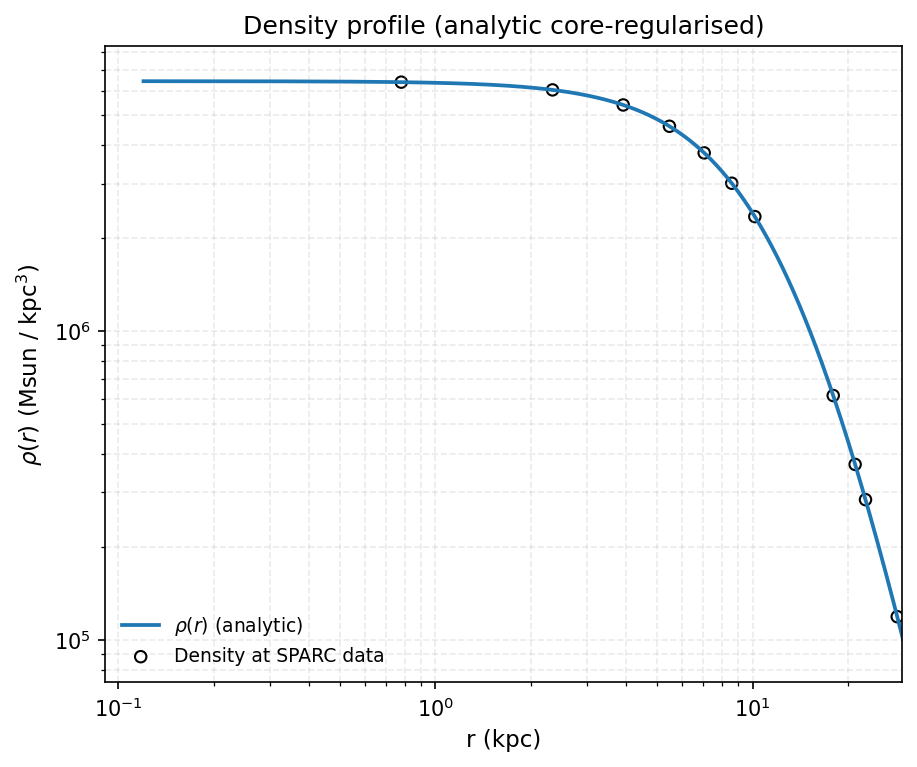}
\caption{The density of the SIDM model of Eq.
(\ref{ScaledependentEoSDM}) for the galaxy UGC05005, versus the
radius.} \label{UGC05005dens}
\end{figure}

\begin{figure}[h!]
\centering
\includegraphics[width=35pc]{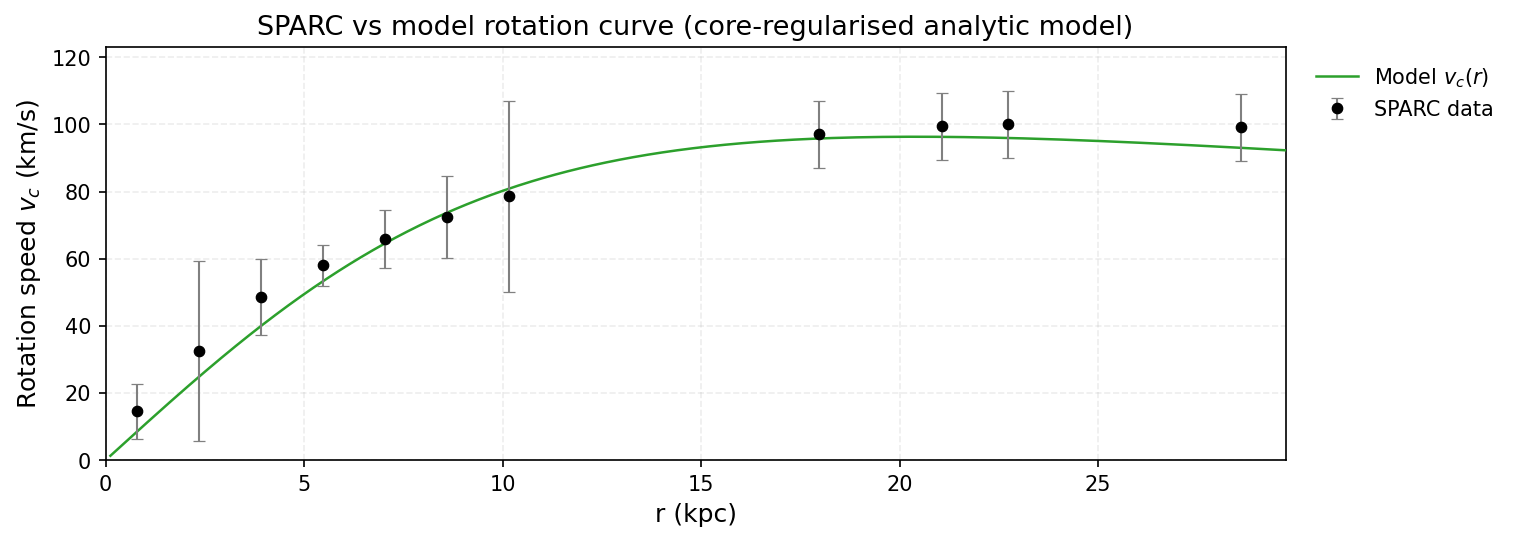}
\caption{The predicted rotation curves for the optimized SIDM
model of Eq. (\ref{ScaledependentEoSDM}), versus the SPARC
observational data for the galaxy UGC05005.} \label{UGC05005}
\end{figure}

In Table \ref{collUGC05005} we present the optimized values of
$K_0$ and $\rho_0$ for the analytic SIDM model of Eq.
(\ref{ScaledependentEoSDM}) for which the maximum compatibility
with the SPARC data is achieved.
\begin{table}[h!]
  \begin{center}
    \caption{SIDM Optimization Values for the galaxy UGC05005}
    \label{collUGC05005}
     \begin{tabular}{|r|r|}
     \hline
      \textbf{Parameter}   & \textbf{Optimization Values}
      \\  \hline
     $\rho_0 $  ($M_{\odot}/\mathrm{Kpc}^{3}$) & $6.43428\times 10^6$
\\  \hline $K_0$ ($M_{\odot} \,
\mathrm{Kpc}^{-3} \, (\mathrm{km/s})^{2}$)& 4016.81
\\  \hline
    \end{tabular}
  \end{center}
\end{table}
In Figs. \ref{UGC05005dens}, \ref{UGC05005}  we present the
density of the analytic SIDM model, the predicted rotation curves
for the SIDM model (\ref{ScaledependentEoSDM}), versus the SPARC
observational data and the sound speed, as a function of the
radius respectively. As it can be seen, for this galaxy, the SIDM
model produces viable rotation curves which are compatible with
the SPARC data.

\subsection{The Galaxy UGC05253, Non-viable, Extended Viable}

For this galaxy, the optimization method we used, ensures maximum
compatibility of the analytic SIDM model of Eq.
(\ref{ScaledependentEoSDM}) with the SPARC data, if we choose
$\rho_0=1.23585\times 10^8$$M_{\odot}/\mathrm{Kpc}^{3}$ and
$K_0=42151.6
$$M_{\odot} \, \mathrm{Kpc}^{-3} \, (\mathrm{km/s})^{2}$, in which
case the reduced $\chi^2_{red}$ value is $\chi^2_{red}=1141.15$.
Also the parameter $\alpha$ in this case is $\alpha=10.658 $Kpc.

\begin{figure}[h!]
\centering
\includegraphics[width=20pc]{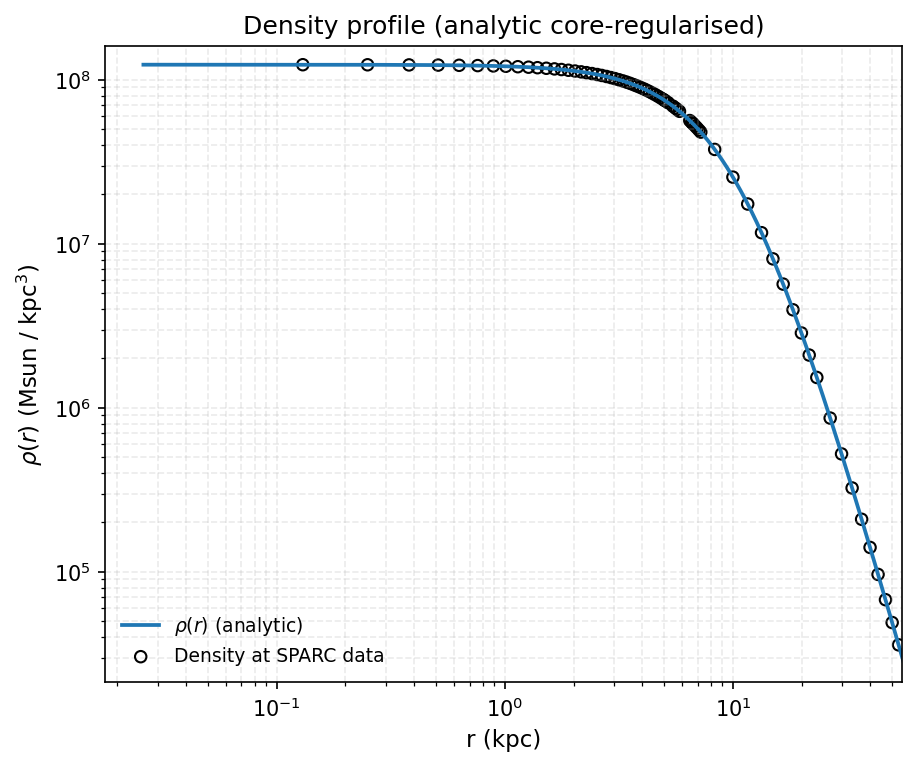}
\caption{The density of the SIDM model of Eq.
(\ref{ScaledependentEoSDM}) for the galaxy UGC05253, versus the
radius.} \label{UGC05253dens}
\end{figure}

\begin{figure}[h!]
\centering
\includegraphics[width=35pc]{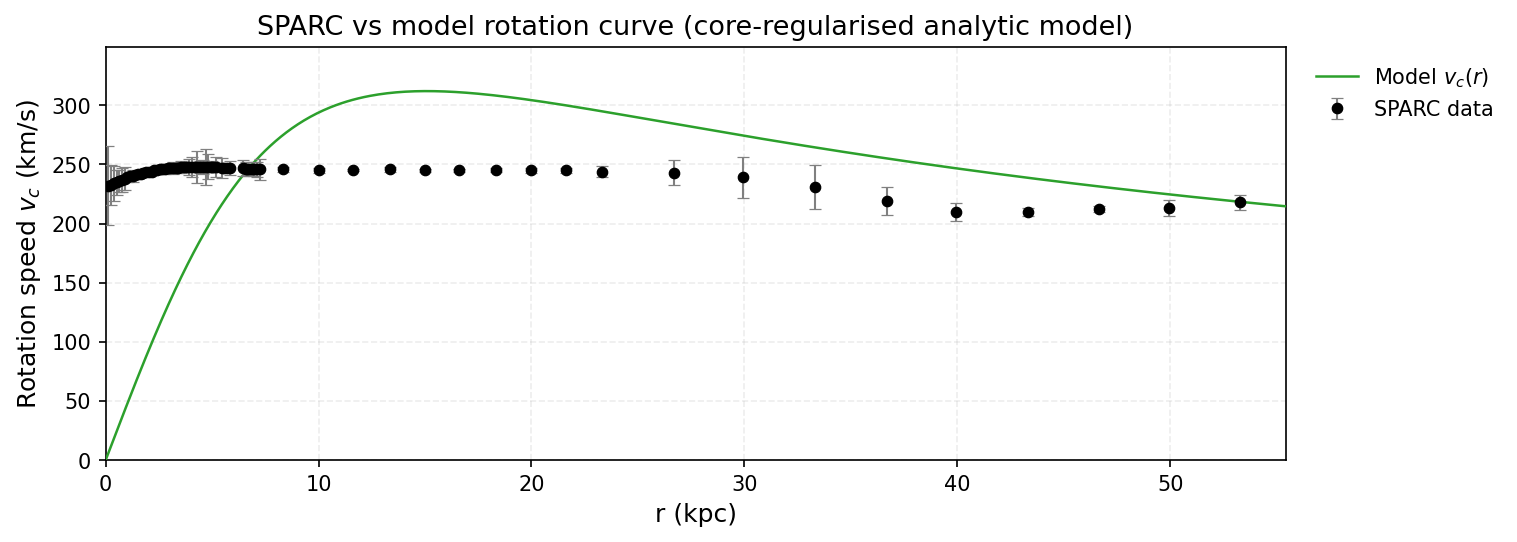}
\caption{The predicted rotation curves for the optimized SIDM
model of Eq. (\ref{ScaledependentEoSDM}), versus the SPARC
observational data for the galaxy UGC05253.} \label{UGC05253}
\end{figure}

In Table \ref{collUGC05253} we present the optimized values of
$K_0$ and $\rho_0$ for the analytic SIDM model of Eq.
(\ref{ScaledependentEoSDM}) for which the maximum compatibility
with the SPARC data is achieved.
\begin{table}[h!]
  \begin{center}
    \caption{SIDM Optimization Values for the galaxy UGC05253}
    \label{collUGC05253}
     \begin{tabular}{|r|r|}
     \hline
      \textbf{Parameter}   & \textbf{Optimization Values}
      \\  \hline
     $\rho_0 $  ($M_{\odot}/\mathrm{Kpc}^{3}$) & $1.23585\times 10^8$
\\  \hline $K_0$ ($M_{\odot} \,
\mathrm{Kpc}^{-3} \, (\mathrm{km/s})^{2}$)& 42151.6
\\  \hline
    \end{tabular}
  \end{center}
\end{table}
In Figs. \ref{UGC05253dens}, \ref{UGC05253}  we present the
density of the analytic SIDM model, the predicted rotation curves
for the SIDM model (\ref{ScaledependentEoSDM}), versus the SPARC
observational data and the sound speed, as a function of the
radius respectively. As it can be seen, for this galaxy, the SIDM
model produces non-viable rotation curves which are incompatible
with the SPARC data.

\begin{figure}[h!]
\centering
\includegraphics[width=20pc]{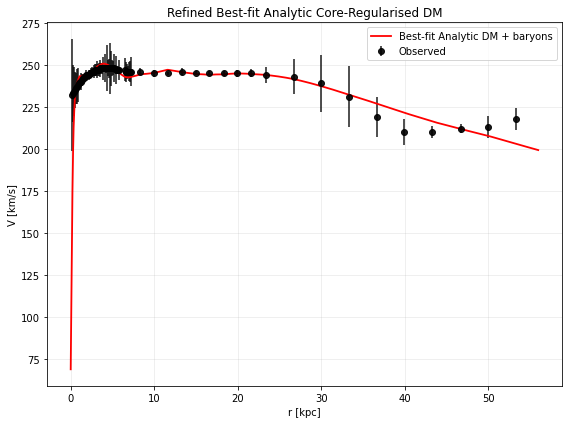}
\caption{The predicted rotation curves after using an optimization
for the SIDM model (\ref{ScaledependentEoSDM}), and the extended
SPARC data for the galaxy UGC05253. We included the rotation
curves of the gas, the disk velocities, the bulge (where present)
along with the SIDM model.} \label{extendedUGC05253}
\end{figure}

Now we shall include contributions to the rotation velocity from
the other components of the galaxy, namely the disk, the gas, and
the bulge if present. In Fig. \ref{extendedUGC05253} we present
the combined rotation curves including all the components of the
galaxy along with the SIDM. As it can be seen, the extended
collisional DM model is viable.
Also in Table \ref{evaluationextendedUGC05253} we present the
optimized values of the free parameters of the SIDM model for
which  we achieve the maximum compatibility with the SPARC data,
for the galaxy UGC05253, and also the resulting reduced
$\chi^2_{red}$ value.
\begin{table}[h!]
\centering \caption{Optimized Parameter Values of the Extended
SIDM model for the Galaxy UGC05253.}
\begin{tabular}{lc}
\hline
Parameter & Value  \\
\hline
$\rho_0 $ ($M_{\odot}/\mathrm{Kpc}^{3}$) & $1.35094\times 10^7$   \\
$K_0$ ($M_{\odot} \,
\mathrm{Kpc}^{-3} \, (\mathrm{km/s})^{2}$) & 15849.8   \\
$ml_{\text{disk}}$ & 0.8062 \\
$ml_{\text{bulge}}$ & 0.8488 \\
$\alpha$ (Kpc) & 19.7647\\
$\chi^2_{red}$ & 0.799952 \\
\hline
\end{tabular}
\label{evaluationextendedUGC05253}
\end{table}

\subsection{The Galaxy UGC05414}

For this galaxy, the optimization method we used, ensures maximum
compatibility of the analytic SIDM model of Eq.
(\ref{ScaledependentEoSDM}) with the SPARC data, if we choose
$\rho_0=4.14699\times 10^7$$M_{\odot}/\mathrm{Kpc}^{3}$ and
$K_0=1615.04
$$M_{\odot} \, \mathrm{Kpc}^{-3} \, (\mathrm{km/s})^{2}$, in which
case the reduced $\chi^2_{red}$ value is $\chi^2_{red}=0.458671$.
Also the parameter $\alpha$ in this case is $\alpha=3.60143 $Kpc.

\begin{figure}[h!]
\centering
\includegraphics[width=20pc]{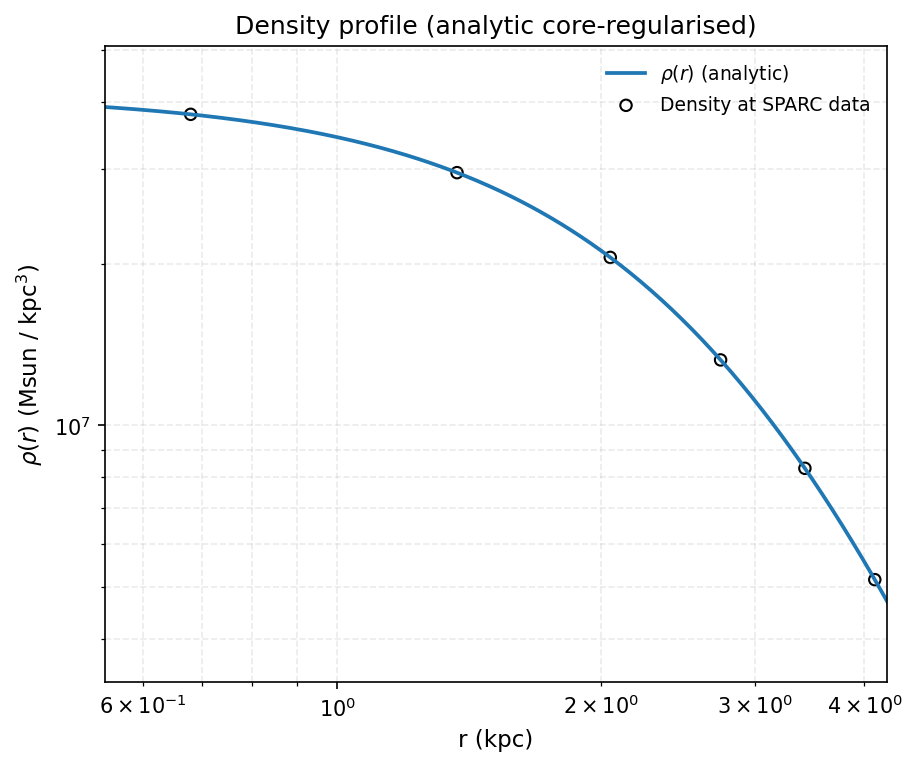}
\caption{The density of the SIDM model of Eq.
(\ref{ScaledependentEoSDM}) for the galaxy UGC05414, versus the
radius.} \label{UGC05414dens}
\end{figure}

\begin{figure}[h!]
\centering
\includegraphics[width=35pc]{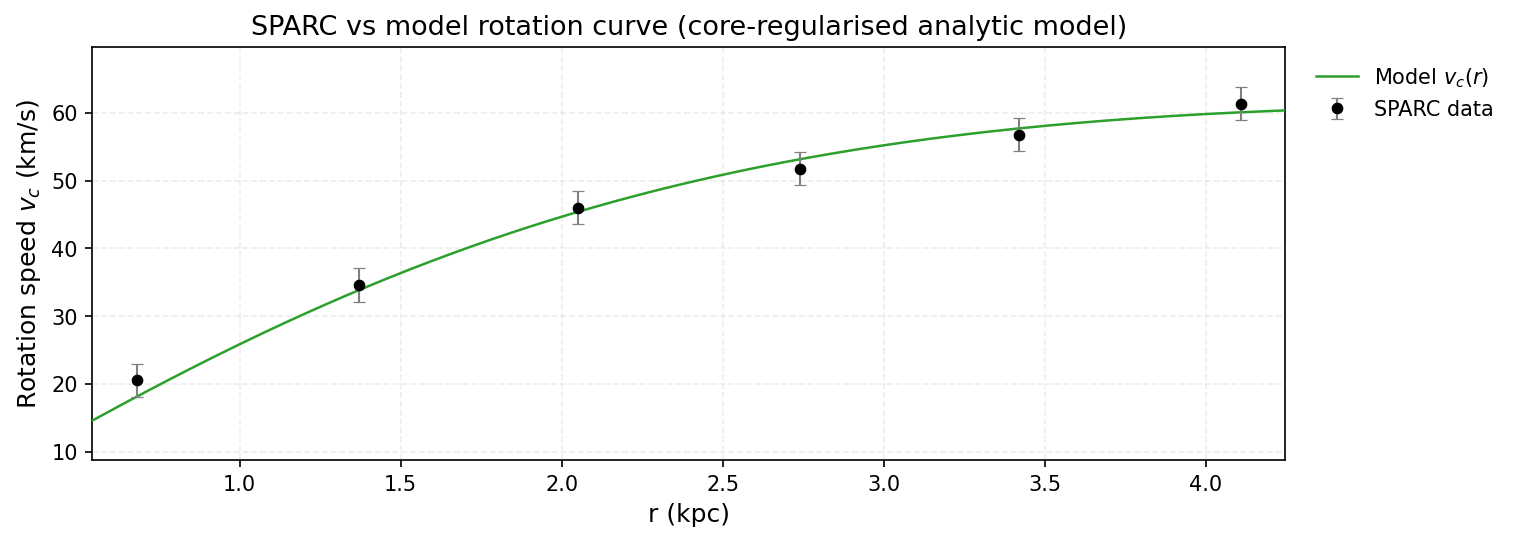}
\caption{The predicted rotation curves for the optimized SIDM
model of Eq. (\ref{ScaledependentEoSDM}), versus the SPARC
observational data for the galaxy UGC05414.} \label{UGC05414}
\end{figure}

In Table \ref{collUGC05414} we present the optimized values of
$K_0$ and $\rho_0$ for the analytic SIDM model of Eq.
(\ref{ScaledependentEoSDM}) for which the maximum compatibility
with the SPARC data is achieved.
\begin{table}[h!]
  \begin{center}
    \caption{SIDM Optimization Values for the galaxy UGC05414}
    \label{collUGC05414}
     \begin{tabular}{|r|r|}
     \hline
      \textbf{Parameter}   & \textbf{Optimization Values}
      \\  \hline
     $\rho_0 $  ($M_{\odot}/\mathrm{Kpc}^{3}$) & $4.14699\times 10^7$
\\  \hline $K_0$ ($M_{\odot} \,
\mathrm{Kpc}^{-3} \, (\mathrm{km/s})^{2}$)& 1615.04
\\  \hline
    \end{tabular}
  \end{center}
\end{table}
In Figs. \ref{UGC05414dens}, \ref{UGC05414}  we present the
density of the analytic SIDM model, the predicted rotation curves
for the SIDM model (\ref{ScaledependentEoSDM}), versus the SPARC
observational data and the sound speed, as a function of the
radius respectively. As it can be seen, for this galaxy, the SIDM
model produces viable rotation curves which are compatible with
the SPARC data.

\subsection{The Galaxy UGC05716, Non-viable}

For this galaxy, the optimization method we used, ensures maximum
compatibility of the analytic SIDM model of Eq.
(\ref{ScaledependentEoSDM}) with the SPARC data, if we choose
$\rho_0=2.00932\times 10^7$$M_{\odot}/\mathrm{Kpc}^{3}$ and
$K_0=2336.07
$$M_{\odot} \, \mathrm{Kpc}^{-3} \, (\mathrm{km/s})^{2}$, in which
case the reduced $\chi^2_{red}$ value is $\chi^2_{red}=20.0144$.
Also the parameter $\alpha$ in this case is $\alpha=6.22257 $Kpc.

\begin{figure}[h!]
\centering
\includegraphics[width=20pc]{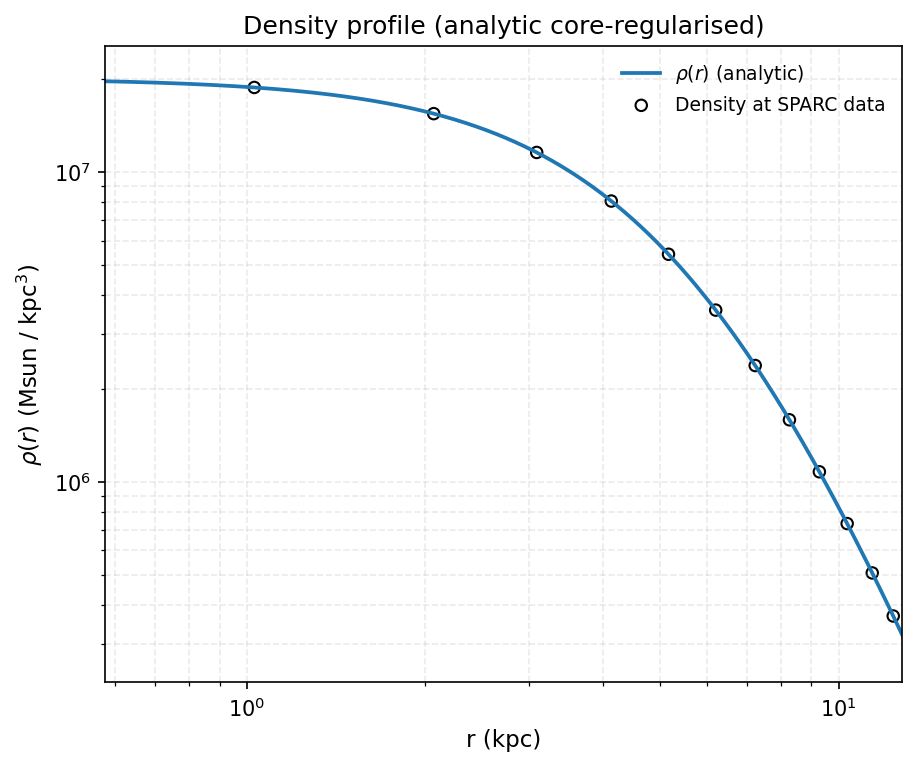}
\caption{The density of the SIDM model of Eq.
(\ref{ScaledependentEoSDM}) for the galaxy UGC05716, versus the
radius.} \label{UGC05716dens}
\end{figure}

\begin{figure}[h!]
\centering
\includegraphics[width=20pc]{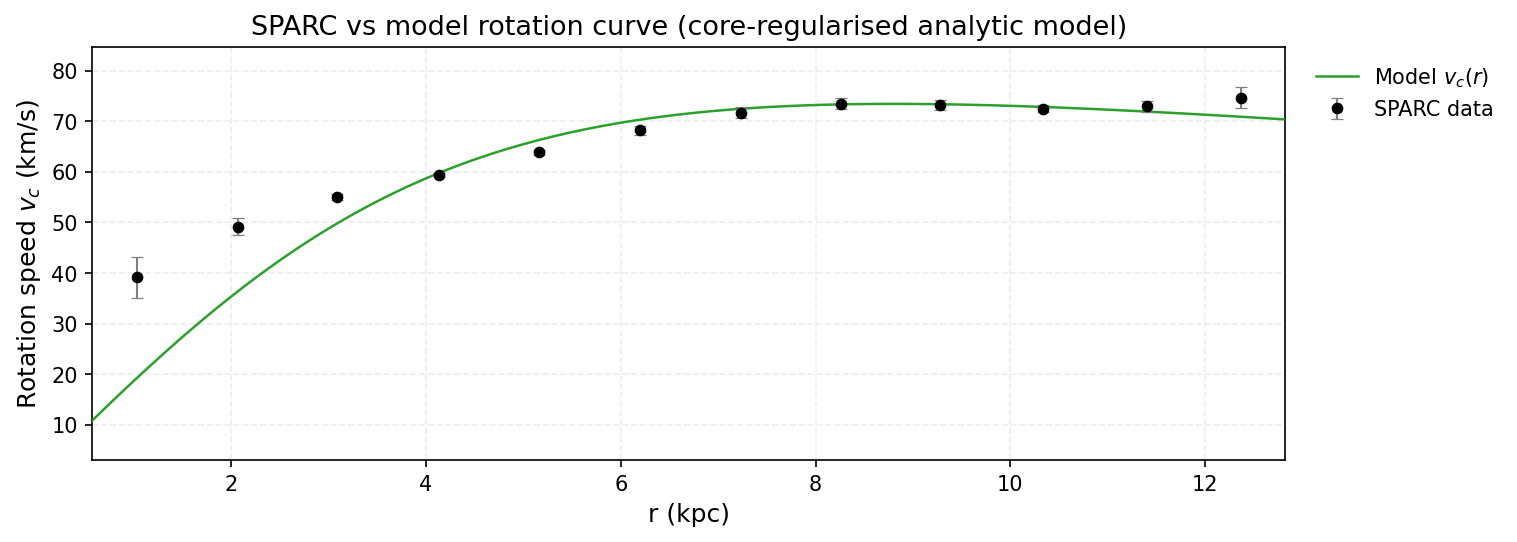}
\caption{The predicted rotation curves for the optimized SIDM
model of Eq. (\ref{ScaledependentEoSDM}), versus the SPARC
observational data for the galaxy UGC05716.} \label{UGC05716}
\end{figure}

In Table \ref{collUGC05716} we present the optimized values of
$K_0$ and $\rho_0$ for the analytic SIDM model of Eq.
(\ref{ScaledependentEoSDM}) for which the maximum compatibility
with the SPARC data is achieved.
\begin{table}[h!]
  \begin{center}
    \caption{SIDM Optimization Values for the galaxy UGC05716}
    \label{collUGC05716}
     \begin{tabular}{|r|r|}
     \hline
      \textbf{Parameter}   & \textbf{Optimization Values}
      \\  \hline
     $\rho_0 $  ($M_{\odot}/\mathrm{Kpc}^{3}$) & $2.00932\times 10^7$
\\  \hline $K_0$ ($M_{\odot} \,
\mathrm{Kpc}^{-3} \, (\mathrm{km/s})^{2}$)& 2336.07
\\  \hline
    \end{tabular}
  \end{center}
\end{table}
In Figs. \ref{UGC05716dens}, \ref{UGC05716}  we present the
density of the analytic SIDM model, the predicted rotation curves
for the SIDM model (\ref{ScaledependentEoSDM}), versus the SPARC
observational data and the sound speed, as a function of the
radius respectively. As it can be seen, for this galaxy, the SIDM
model produces non-viable rotation curves which are incompatible
with the SPARC data.

\begin{figure}[h!]
\centering
\includegraphics[width=20pc]{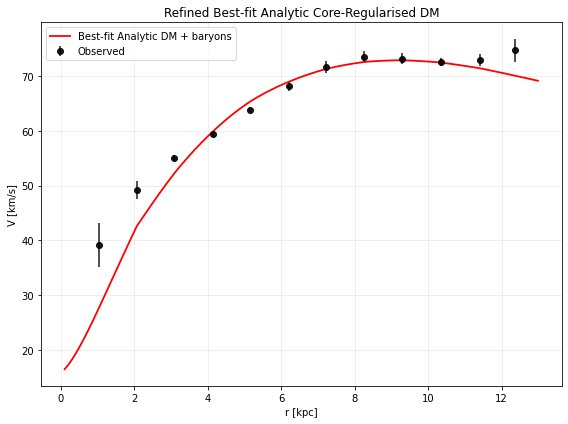}
\caption{The predicted rotation curves after using an optimization
for the SIDM model (\ref{ScaledependentEoSDM}), and the extended
SPARC data for the galaxy UGC05716. We included the rotation
curves of the gas, the disk velocities, the bulge (where present)
along with the SIDM model.} \label{extendedUGC05716}
\end{figure}

Now we shall include contributions to the rotation velocity from
the other components of the galaxy, namely the disk, the gas, and
the bulge if present. In Fig. \ref{extendedUGC05716} we present
the combined rotation curves including all the components of the
galaxy along with the SIDM. As it can be seen, the extended
collisional DM model is non-viable.
Also in Table \ref{evaluationextendedUGC05716} we present the
optimized values of the free parameters of the SIDM model for
which  we achieve the maximum compatibility with the SPARC data,
for the galaxy UGC05716, and also the resulting reduced
$\chi^2_{red}$ value.
\begin{table}[h!]
\centering \caption{Optimized Parameter Values of the Extended
SIDM model for the Galaxy UGC05716.}
\begin{tabular}{lc}
\hline
Parameter & Value  \\
\hline
$\rho_0 $ ($M_{\odot}/\mathrm{Kpc}^{3}$) & $1.50901\times 10^7$   \\
$K_0$ ($M_{\odot} \,
\mathrm{Kpc}^{-3} \, (\mathrm{km/s})^{2}$) & 1811.55   \\
$ml_{\text{disk}}$ & 1 \\
$ml_{\text{bulge}}$ & 0.3007 \\
$\alpha$ (Kpc) & 6.32229\\
$\chi^2_{red}$ & 8.33583\\
\hline
\end{tabular}
\label{evaluationextendedUGC05716}
\end{table}

\subsection{The Galaxy UGC05721, Non-viable, Extended Viable}

For this galaxy, the optimization method we used, ensures maximum
compatibility of the analytic SIDM model of Eq.
(\ref{ScaledependentEoSDM}) with the SPARC data, if we choose
$\rho_0=3.62066\times 10^8$$M_{\odot}/\mathrm{Kpc}^{3}$ and
$K_0=3269.08
$$M_{\odot} \, \mathrm{Kpc}^{-3} \, (\mathrm{km/s})^{2}$, in which
case the reduced $\chi^2_{red}$ value is $\chi^2_{red}=2.11638$.
Also the parameter $\alpha$ in this case is $\alpha=1.73408 $Kpc.

\begin{figure}[h!]
\centering
\includegraphics[width=20pc]{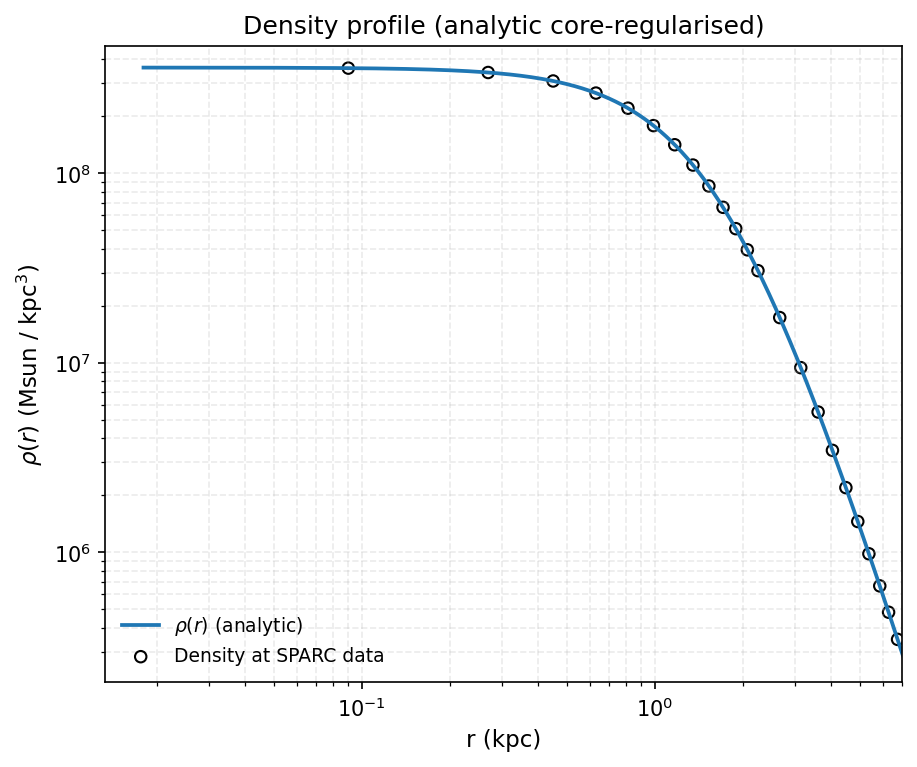}
\caption{The density of the SIDM model of Eq.
(\ref{ScaledependentEoSDM}) for the galaxy UGC05721, versus the
radius.} \label{UGC05721dens}
\end{figure}

\begin{figure}[h!]
\centering
\includegraphics[width=35pc]{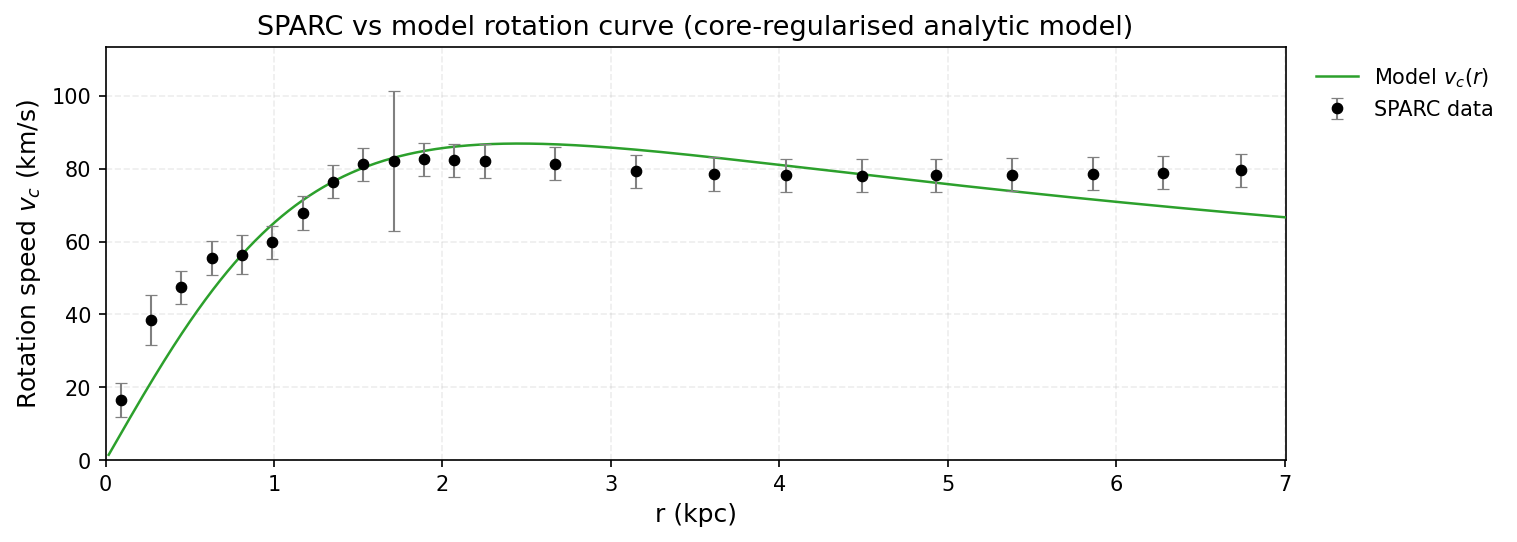}
\caption{The predicted rotation curves for the optimized SIDM
model of Eq. (\ref{ScaledependentEoSDM}), versus the SPARC
observational data for the galaxy UGC05721.} \label{UGC05721}
\end{figure}

In Table \ref{collUGC05721} we present the optimized values of
$K_0$ and $\rho_0$ for the analytic SIDM model of Eq.
(\ref{ScaledependentEoSDM}) for which the maximum compatibility
with the SPARC data is achieved.
\begin{table}[h!]
  \begin{center}
    \caption{SIDM Optimization Values for the galaxy UGC05721}
    \label{collUGC05721}
     \begin{tabular}{|r|r|}
     \hline
      \textbf{Parameter}   & \textbf{Optimization Values}
      \\  \hline
     $\rho_0 $  ($M_{\odot}/\mathrm{Kpc}^{3}$) & $3.62066\times 10^8$
\\  \hline $K_0$ ($M_{\odot} \,
\mathrm{Kpc}^{-3} \, (\mathrm{km/s})^{2}$)& 3269.08
\\  \hline
    \end{tabular}
  \end{center}
\end{table}
In Figs. \ref{UGC05721dens}, \ref{UGC05721}  we present the
density of the analytic SIDM model, the predicted rotation curves
for the SIDM model (\ref{ScaledependentEoSDM}), versus the SPARC
observational data and the sound speed, as a function of the
radius respectively. As it can be seen, for this galaxy, the SIDM
model produces non-viable rotation curves which are incompatible
with the SPARC data.

\begin{figure}[h!]
\centering
\includegraphics[width=20pc]{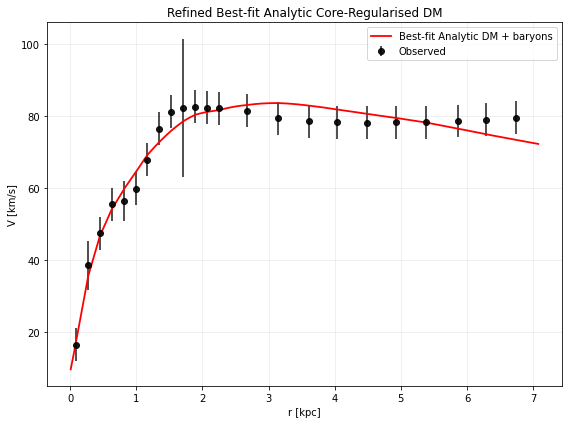}
\caption{The predicted rotation curves after using an optimization
for the SIDM model (\ref{ScaledependentEoSDM}), and the extended
SPARC data for the galaxy UGC05721. We included the rotation
curves of the gas, the disk velocities, the bulge (where present)
along with the SIDM model.} \label{extendedUGC05721}
\end{figure}

Now we shall include contributions to the rotation velocity from
the other components of the galaxy, namely the disk, the gas, and
the bulge if present. In Fig. \ref{extendedUGC05721} we present
the combined rotation curves including all the components of the
galaxy along with the SIDM. As it can be seen, the extended
collisional DM model is non-viable.
Also in Table \ref{evaluationextendedUGC05721} we present the
optimized values of the free parameters of the SIDM model for
which  we achieve the maximum compatibility with the SPARC data,
for the galaxy UGC05721, and also the resulting reduced
$\chi^2_{red}$ value.
\begin{table}[h!]
\centering \caption{Optimized Parameter Values of the Extended
SIDM model for the Galaxy UGC05721.}
\begin{tabular}{lc}
\hline
Parameter & Value  \\
\hline
$\rho_0 $ ($M_{\odot}/\mathrm{Kpc}^{3}$) & $1.50375\times 10^8$   \\
$K_0$ ($M_{\odot} \,
\mathrm{Kpc}^{-3} \, (\mathrm{km/s})^{2}$) & 2511.14   \\
$ml_{\text{disk}}$ & 1 \\
$ml_{\text{bulge}}$ & 0.4923 \\
$\alpha$ (Kpc) & 2.358\\
$\chi^2_{red}$ & 0.512261 \\
\hline
\end{tabular}
\label{evaluationextendedUGC05721}
\end{table}

\subsection{The Galaxy UGC05750}

For this galaxy, the optimization method we used, ensures maximum
compatibility of the analytic SIDM model of Eq.
(\ref{ScaledependentEoSDM}) with the SPARC data, if we choose
$\rho_0=1\times 10^7$$M_{\odot}/\mathrm{Kpc}^{3}$ and $K_0=2398.38
$$M_{\odot} \, \mathrm{Kpc}^{-3} \, (\mathrm{km/s})^{2}$, in which
case the reduced $\chi^2_{red}$ value is $\chi^2_{red}=0.38913$.
Also the parameter $\alpha$ in this case is $\alpha=11.3881 $Kpc.

\begin{figure}[h!]
\centering
\includegraphics[width=20pc]{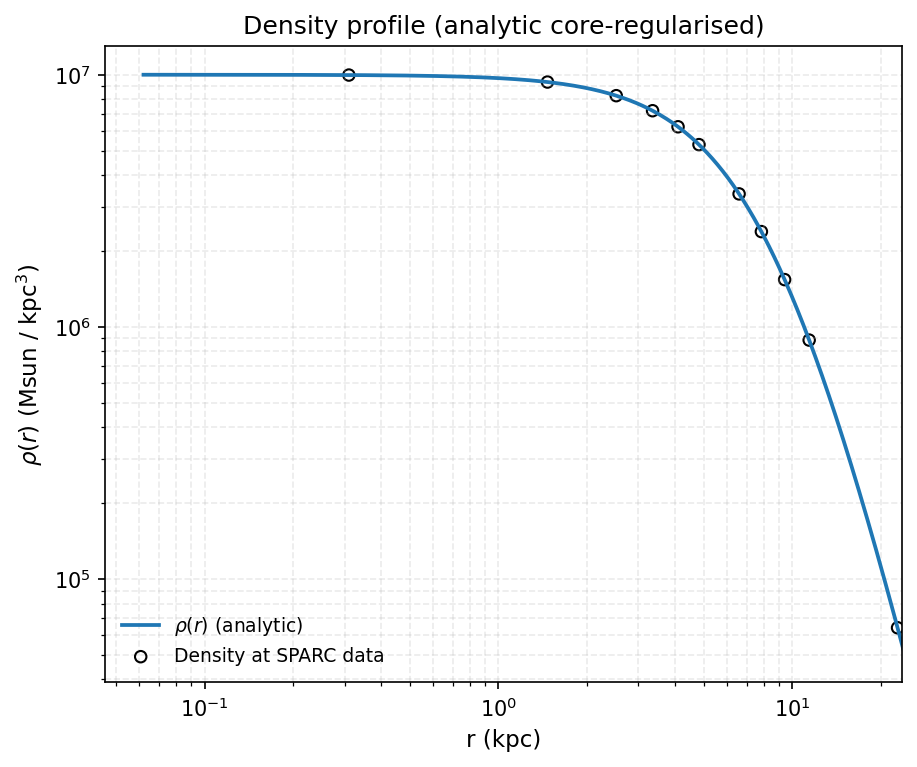}
\caption{The density of the SIDM model of Eq.
(\ref{ScaledependentEoSDM}) for the galaxy UGC05750, versus the
radius.} \label{UGC05750dens}
\end{figure}

\begin{figure}[h!]
\centering
\includegraphics[width=35pc]{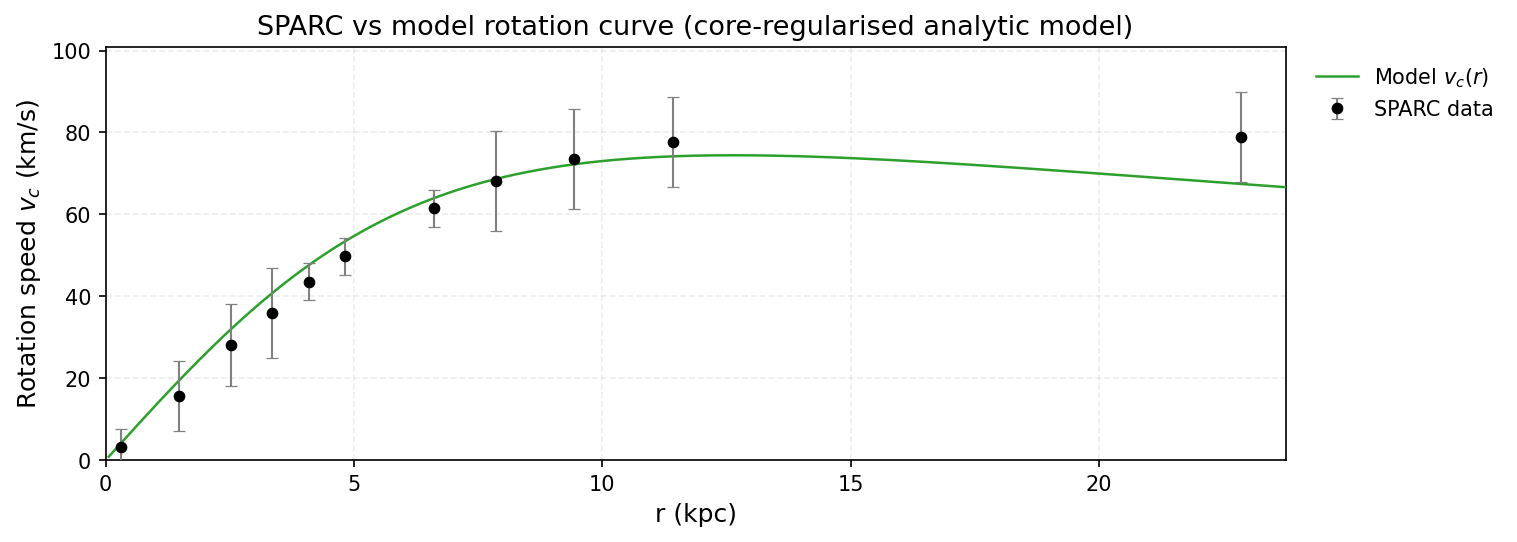}
\caption{The predicted rotation curves for the optimized SIDM
model of Eq. (\ref{ScaledependentEoSDM}), versus the SPARC
observational data for the galaxy UGC05750.} \label{UGC05750}
\end{figure}

In Table \ref{collUGC05750} we present the optimized values of
$K_0$ and $\rho_0$ for the analytic SIDM model of Eq.
(\ref{ScaledependentEoSDM}) for which the maximum compatibility
with the SPARC data is achieved.
\begin{table}[h!]
  \begin{center}
    \caption{SIDM Optimization Values for the galaxy UGC05750}
    \label{collUGC05750}
     \begin{tabular}{|r|r|}
     \hline
      \textbf{Parameter}   & \textbf{Optimization Values}
      \\  \hline
     $\rho_0 $  ($M_{\odot}/\mathrm{Kpc}^{3}$) & $1\times 10^7$
\\  \hline $K_0$ ($M_{\odot} \,
\mathrm{Kpc}^{-3} \, (\mathrm{km/s})^{2}$)& 2398.38
\\  \hline
    \end{tabular}
  \end{center}
\end{table}
In Figs. \ref{UGC05750dens}, \ref{UGC05750}  we present the
density of the analytic SIDM model, the predicted rotation curves
for the SIDM model (\ref{ScaledependentEoSDM}), versus the SPARC
observational data and the sound speed, as a function of the
radius respectively. As it can be seen, for this galaxy, the SIDM
model produces viable rotation curves which are compatible with
the SPARC data.

\subsection{The Galaxy UGC05764, Marginally  Viable}

For this galaxy, the optimization method we used, ensures maximum
compatibility of the analytic SIDM model of Eq.
(\ref{ScaledependentEoSDM}) with the SPARC data, if we choose
$\rho_0=1.36328\times 10^8$$M_{\odot}/\mathrm{Kpc}^{3}$ and
$K_0=1273.63
$$M_{\odot} \, \mathrm{Kpc}^{-3} \, (\mathrm{km/s})^{2}$, in which
case the reduced $\chi^2_{red}$ value is $\chi^2_{red}=4.29775$.
Also the parameter $\alpha$ in this case is $\alpha=1.76392 $Kpc.

\begin{figure}[h!]
\centering
\includegraphics[width=20pc]{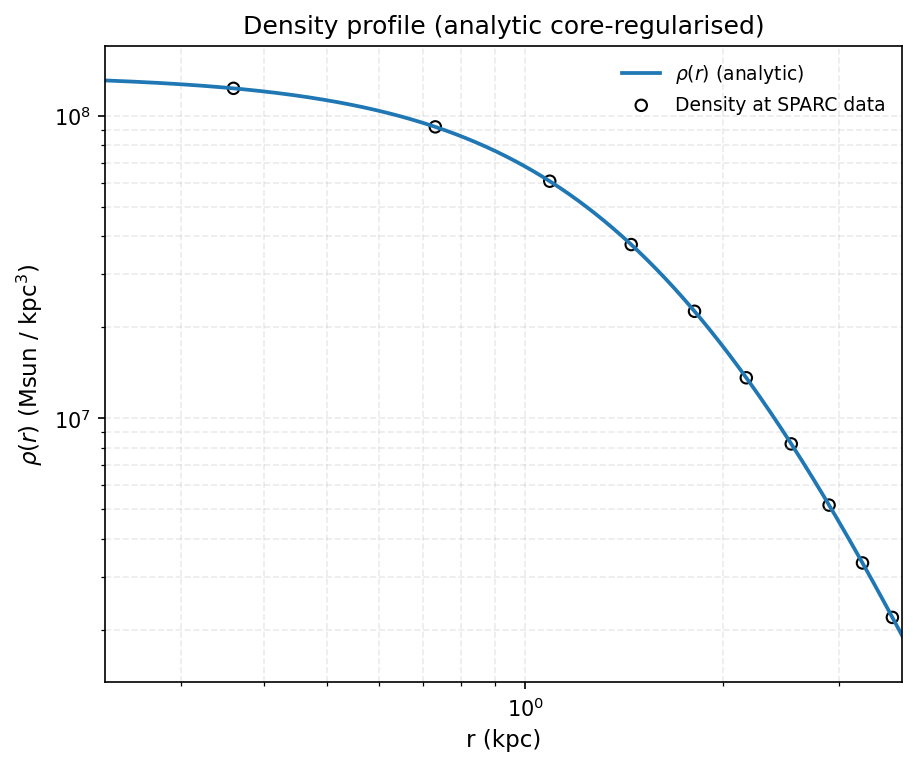}
\caption{The density of the SIDM model of Eq.
(\ref{ScaledependentEoSDM}) for the galaxy UGC05764, versus the
radius.} \label{UGC05764dens}
\end{figure}

\begin{figure}[h!]
\centering
\includegraphics[width=35pc]{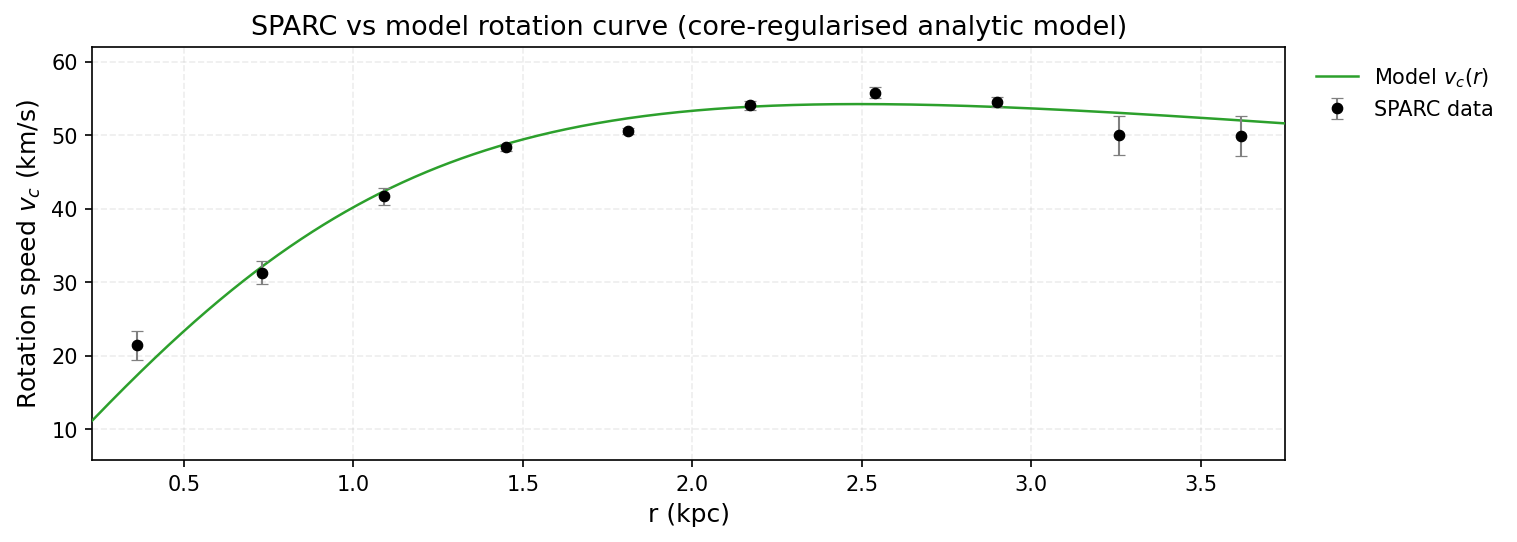}
\caption{The predicted rotation curves for the optimized SIDM
model of Eq. (\ref{ScaledependentEoSDM}), versus the SPARC
observational data for the galaxy UGC05764.} \label{UGC05764}
\end{figure}

In Table \ref{collUGC05764} we present the optimized values of
$K_0$ and $\rho_0$ for the analytic SIDM model of Eq.
(\ref{ScaledependentEoSDM}) for which the maximum compatibility
with the SPARC data is achieved.
\begin{table}[h!]
  \begin{center}
    \caption{SIDM Optimization Values for the galaxy UGC05764}
    \label{collUGC05764}
     \begin{tabular}{|r|r|}
     \hline
      \textbf{Parameter}   & \textbf{Optimization Values}
      \\  \hline
     $\rho_0 $  ($M_{\odot}/\mathrm{Kpc}^{3}$) & $1.36328\times 10^8$
\\  \hline $K_0$ ($M_{\odot} \,
\mathrm{Kpc}^{-3} \, (\mathrm{km/s})^{2}$)& 1273.63
\\  \hline
    \end{tabular}
  \end{center}
\end{table}
In Figs. \ref{UGC05764dens}, \ref{UGC05764}  we present the
density of the analytic SIDM model, the predicted rotation curves
for the SIDM model (\ref{ScaledependentEoSDM}), versus the SPARC
observational data and the sound speed, as a function of the
radius respectively. As it can be seen, for this galaxy, the SIDM
model produces marginally viable rotation curves which are
marginally compatible with the SPARC data.

\begin{figure}[h!]
\centering
\includegraphics[width=20pc]{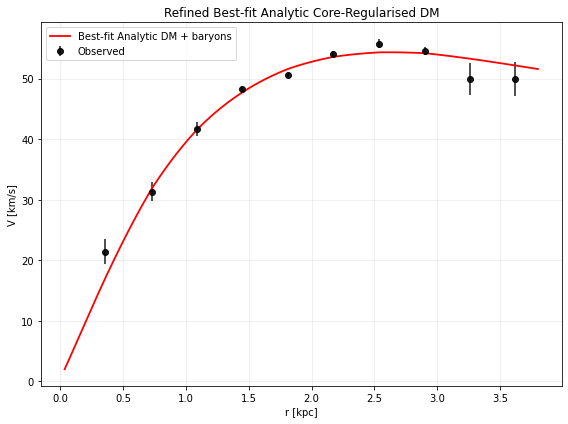}
\caption{The predicted rotation curves after using an optimization
for the SIDM model (\ref{ScaledependentEoSDM}), and the extended
SPARC data for the galaxy UGC05764. We included the rotation
curves of the gas, the disk velocities, the bulge (where present)
along with the SIDM model.} \label{extendedUGC05764}
\end{figure}

Now we shall include contributions to the rotation velocity from
the other components of the galaxy, namely the disk, the gas, and
the bulge if present. In Fig. \ref{extendedUGC05764} we present
the combined rotation curves including all the components of the
galaxy along with the SIDM. As it can be seen, the extended
collisional DM model is marginally viable.
Also in Table \ref{evaluationextendedUGC05764} we present the
optimized values of the free parameters of the SIDM model for
which  we achieve the maximum compatibility with the SPARC data,
for the galaxy UGC05764, and also the resulting reduced
$\chi^2_{red}$ value.
\begin{table}[h!]
\centering \caption{Optimized Parameter Values of the Extended
SIDM model for the Galaxy UGC05764.}
\begin{tabular}{lc}
\hline
Parameter & Value  \\
\hline
$\rho_0 $ ($M_{\odot}/\mathrm{Kpc}^{3}$) & $1.16291\times 10^8$   \\
$K_0$ ($M_{\odot} \,
\mathrm{Kpc}^{-3} \, (\mathrm{km/s})^{2}$) & 1049.6  \\
$ml_{\text{disk}}$ & 1 \\
$ml_{\text{bulge}}$ & 0.3790 \\
$\alpha$ (Kpc) & 1.73355\\
$\chi^2_{red}$ & 3.44125 \\
\hline
\end{tabular}
\label{evaluationextendedUGC05764}
\end{table}

\subsection{The Galaxy UGC05829}

For this galaxy, the optimization method we used, ensures maximum
compatibility of the analytic SIDM model of Eq.
(\ref{ScaledependentEoSDM}) with the SPARC data, if we choose
$\rho_0=1.89055\times 10^7$$M_{\odot}/\mathrm{Kpc}^{3}$ and
$K_0=1787.68
$$M_{\odot} \, \mathrm{Kpc}^{-3} \, (\mathrm{km/s})^{2}$, in which
case the reduced $\chi^2_{red}$ value is $\chi^2_{red}=0.608563$.
Also the parameter $\alpha$ in this case is $\alpha=5.6118 $Kpc.

\begin{figure}[h!]
\centering
\includegraphics[width=20pc]{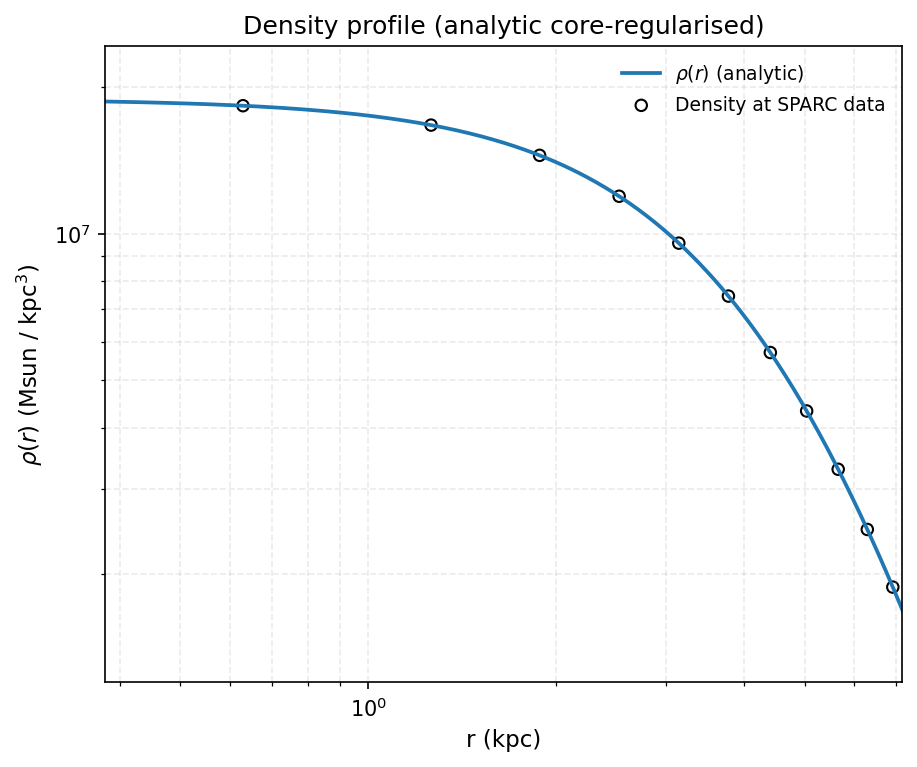}
\caption{The density of the SIDM model of Eq.
(\ref{ScaledependentEoSDM}) for the galaxy UGC05829, versus the
radius.} \label{UGC05829dens}
\end{figure}

\begin{figure}[h!]
\centering
\includegraphics[width=35pc]{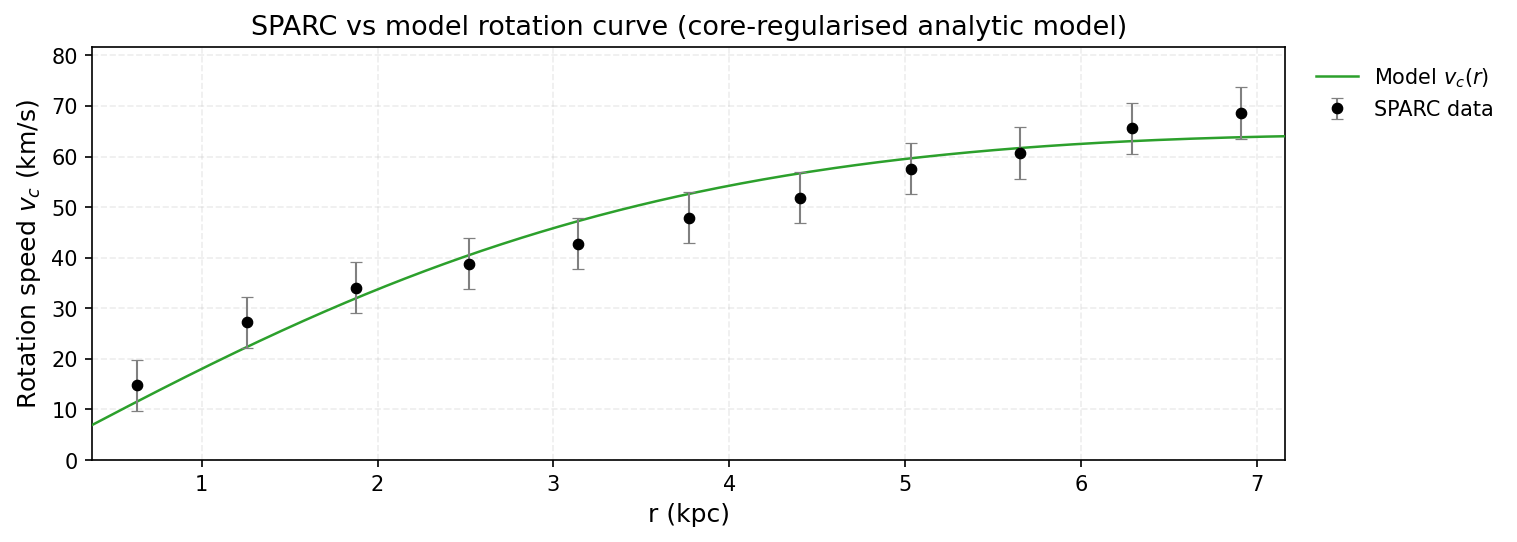}
\caption{The predicted rotation curves for the optimized SIDM
model of Eq. (\ref{ScaledependentEoSDM}), versus the SPARC
observational data for the galaxy UGC05829.} \label{UGC05829}
\end{figure}

In Table \ref{collUGC05829} we present the optimized values of
$K_0$ and $\rho_0$ for the analytic SIDM model of Eq.
(\ref{ScaledependentEoSDM}) for which the maximum compatibility
with the SPARC data is achieved.
\begin{table}[h!]
  \begin{center}
    \caption{SIDM Optimization Values for the galaxy UGC05829}
    \label{collUGC05829}
     \begin{tabular}{|r|r|}
     \hline
      \textbf{Parameter}   & \textbf{Optimization Values}
      \\  \hline
     $\rho_0 $  ($M_{\odot}/\mathrm{Kpc}^{3}$) & $1.89055\times 10^7$
\\  \hline $K_0$ ($M_{\odot} \,
\mathrm{Kpc}^{-3} \, (\mathrm{km/s})^{2}$)& 1787.68
\\  \hline
    \end{tabular}
  \end{center}
\end{table}
In Figs. \ref{UGC05829dens}, \ref{UGC05829}  we present the
density of the analytic SIDM model, the predicted rotation curves
for the SIDM model (\ref{ScaledependentEoSDM}), versus the SPARC
observational data and the sound speed, as a function of the
radius respectively. As it can be seen, for this galaxy, the SIDM
model produces viable rotation curves which are compatible with
the SPARC data.

\subsection{The Galaxy UGC05918}

For this galaxy, the optimization method we used, ensures maximum
compatibility of the analytic SIDM model of Eq.
(\ref{ScaledependentEoSDM}) with the SPARC data, if we choose
$\rho_0=3.4607\times 10^7$$M_{\odot}/\mathrm{Kpc}^{3}$ and
$K_0=869.36
$$M_{\odot} \, \mathrm{Kpc}^{-3} \, (\mathrm{km/s})^{2}$, in which
case the reduced $\chi^2_{red}$ value is $\chi^2_{red}=0.494426$.
Also the parameter $\alpha$ in this case is $\alpha=2.89247 $Kpc.

\begin{figure}[h!]
\centering
\includegraphics[width=20pc]{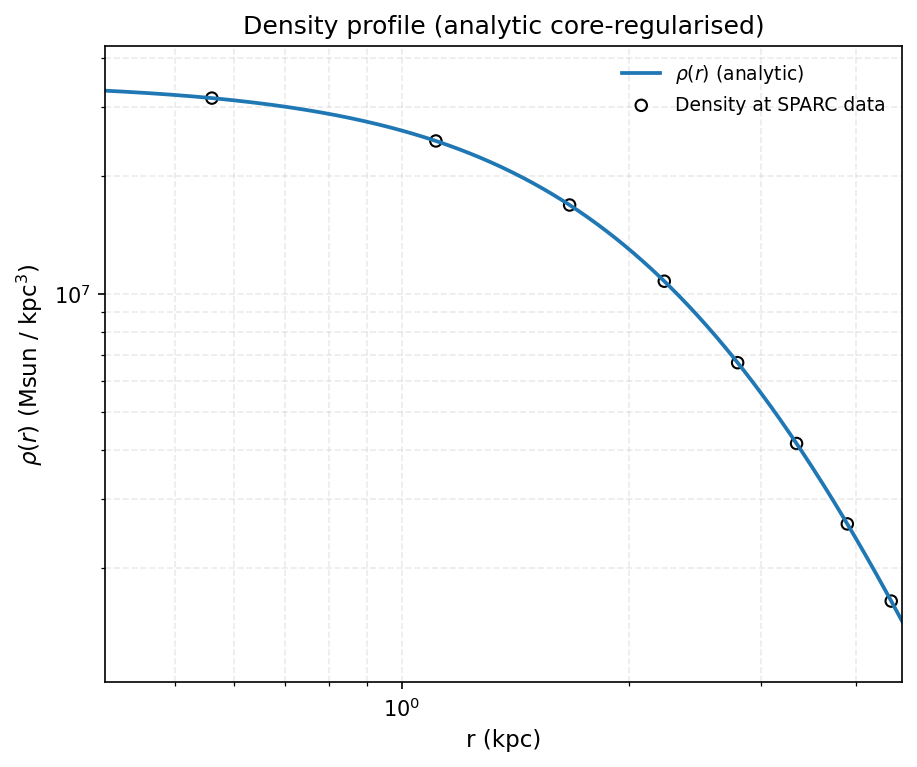}
\caption{The density of the SIDM model of Eq.
(\ref{ScaledependentEoSDM}) for the galaxy UGC05918, versus the
radius.} \label{UGC05918dens}
\end{figure}

\begin{figure}[h!]
\centering
\includegraphics[width=35pc]{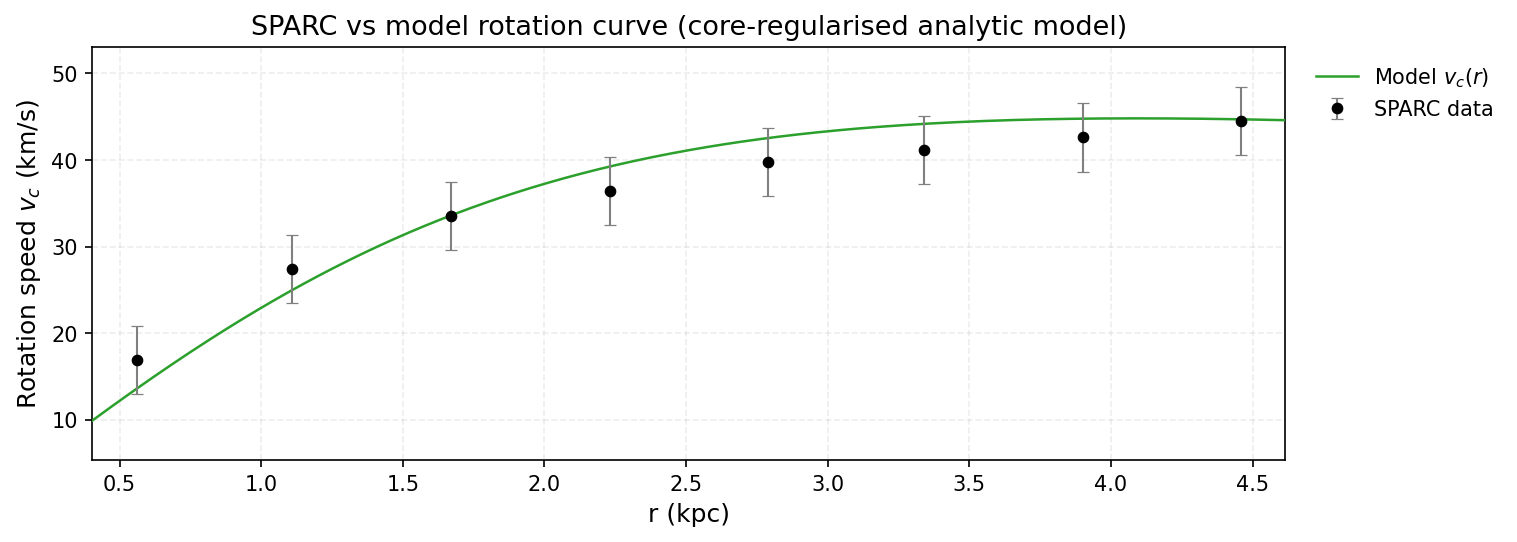}
\caption{The predicted rotation curves for the optimized SIDM
model of Eq. (\ref{ScaledependentEoSDM}), versus the SPARC
observational data for the galaxy UGC05918.} \label{UGC05918}
\end{figure}

In Table \ref{collUGC05918} we present the optimized values of
$K_0$ and $\rho_0$ for the analytic SIDM model of Eq.
(\ref{ScaledependentEoSDM}) for which the maximum compatibility
with the SPARC data is achieved.
\begin{table}[h!]
  \begin{center}
    \caption{SIDM Optimization Values for the galaxy UGC05918}
    \label{collUGC05918}
     \begin{tabular}{|r|r|}
     \hline
      \textbf{Parameter}   & \textbf{Optimization Values}
      \\  \hline
     $\rho_0 $  ($M_{\odot}/\mathrm{Kpc}^{3}$) & $3.4607\times 10^7$
\\  \hline $K_0$ ($M_{\odot} \,
\mathrm{Kpc}^{-3} \, (\mathrm{km/s})^{2}$)& 869.36
\\  \hline
    \end{tabular}
  \end{center}
\end{table}
In Figs. \ref{UGC05918dens}, \ref{UGC05918}  we present the
density of the analytic SIDM model, the predicted rotation curves
for the SIDM model (\ref{ScaledependentEoSDM}), versus the SPARC
observational data and the sound speed, as a function of the
radius respectively. As it can be seen, for this galaxy, the SIDM
model produces viable rotation curves which are compatible with
the SPARC data.

\subsection{The Galaxy UGC05986}

For this galaxy, the optimization method we used, ensures maximum
compatibility of the analytic SIDM model of Eq.
(\ref{ScaledependentEoSDM}) with the SPARC data, if we choose
$\rho_0=1.01159\times 10^8$$M_{\odot}/\mathrm{Kpc}^{3}$ and
$K_0=5654.85
$$M_{\odot} \, \mathrm{Kpc}^{-3} \, (\mathrm{km/s})^{2}$, in which
case the reduced $\chi^2_{red}$ value is $\chi^2_{red}=0.564247$.
Also the parameter $\alpha$ in this case is $\alpha=4.31479 $Kpc.

\begin{figure}[h!]
\centering
\includegraphics[width=20pc]{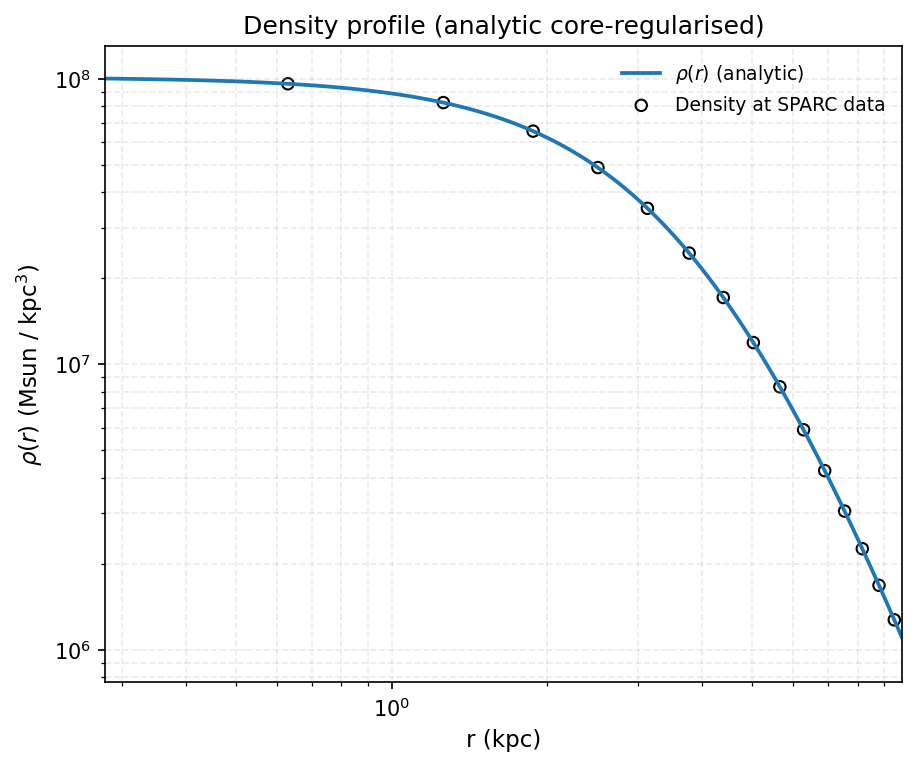}
\caption{The density of the SIDM model of Eq.
(\ref{ScaledependentEoSDM}) for the galaxy UGC05986, versus the
radius.} \label{UGC05986dens}
\end{figure}

\begin{figure}[h!]
\centering
\includegraphics[width=35pc]{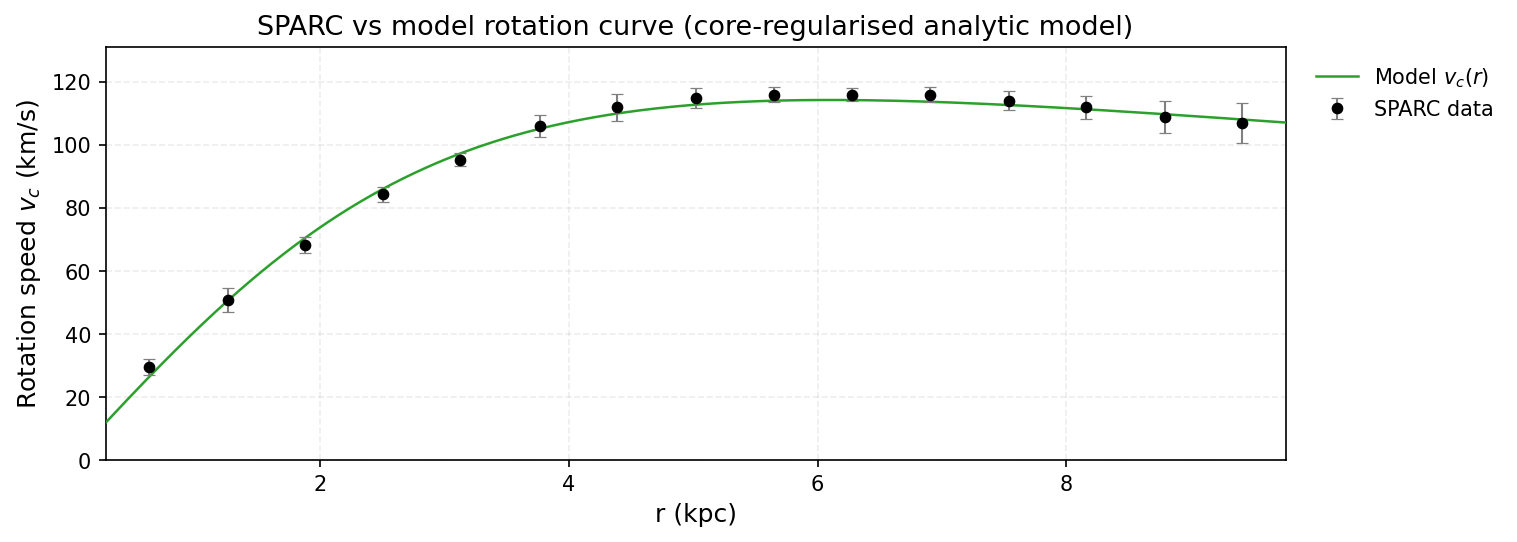}
\caption{The predicted rotation curves for the optimized SIDM
model of Eq. (\ref{ScaledependentEoSDM}), versus the SPARC
observational data for the galaxy UGC05986.} \label{UGC05986}
\end{figure}

In Table \ref{collUGC05986} we present the optimized values of
$K_0$ and $\rho_0$ for the analytic SIDM model of Eq.
(\ref{ScaledependentEoSDM}) for which the maximum compatibility
with the SPARC data is achieved.
\begin{table}[h!]
  \begin{center}
    \caption{SIDM Optimization Values for the galaxy UGC05986}
    \label{collUGC05986}
     \begin{tabular}{|r|r|}
     \hline
      \textbf{Parameter}   & \textbf{Optimization Values}
      \\  \hline
     $\rho_0 $  ($M_{\odot}/\mathrm{Kpc}^{3}$) & $1.01159\times 10^8$
\\  \hline $K_0$ ($M_{\odot} \,
\mathrm{Kpc}^{-3} \, (\mathrm{km/s})^{2}$)& 5654.85
\\  \hline
    \end{tabular}
  \end{center}
\end{table}
In Figs. \ref{UGC05986dens}, \ref{UGC05986}  we present the
density of the analytic SIDM model, the predicted rotation curves
for the SIDM model (\ref{ScaledependentEoSDM}), versus the SPARC
observational data and the sound speed, as a function of the
radius respectively. As it can be seen, for this galaxy, the SIDM
model produces viable rotation curves which are compatible with
the SPARC data.

\subsection{The Galaxy UGC06399}

For this galaxy, the optimization method we used, ensures maximum
compatibility of the analytic SIDM model of Eq.
(\ref{ScaledependentEoSDM}) with the SPARC data, if we choose
$\rho_0=3.97388\times 10^7$$M_{\odot}/\mathrm{Kpc}^{3}$ and
$K_0=3557.78
$$M_{\odot} \, \mathrm{Kpc}^{-3} \, (\mathrm{km/s})^{2}$, in which
case the reduced $\chi^2_{red}$ value is $\chi^2_{red}=0.619603$.
Also the parameter $\alpha$ in this case is $\alpha=5.46051 $Kpc.

\begin{figure}[h!]
\centering
\includegraphics[width=20pc]{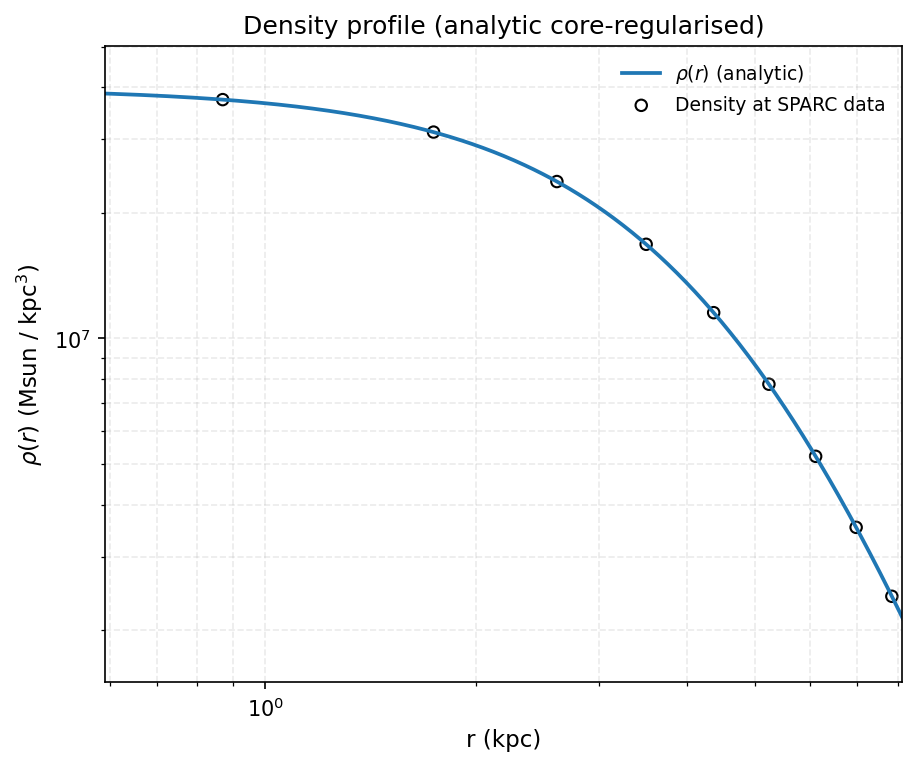}
\caption{The density of the SIDM model of Eq.
(\ref{ScaledependentEoSDM}) for the galaxy UGC06399, versus the
radius.} \label{UGC06399dens}
\end{figure}

\begin{figure}[h!]
\centering
\includegraphics[width=35pc]{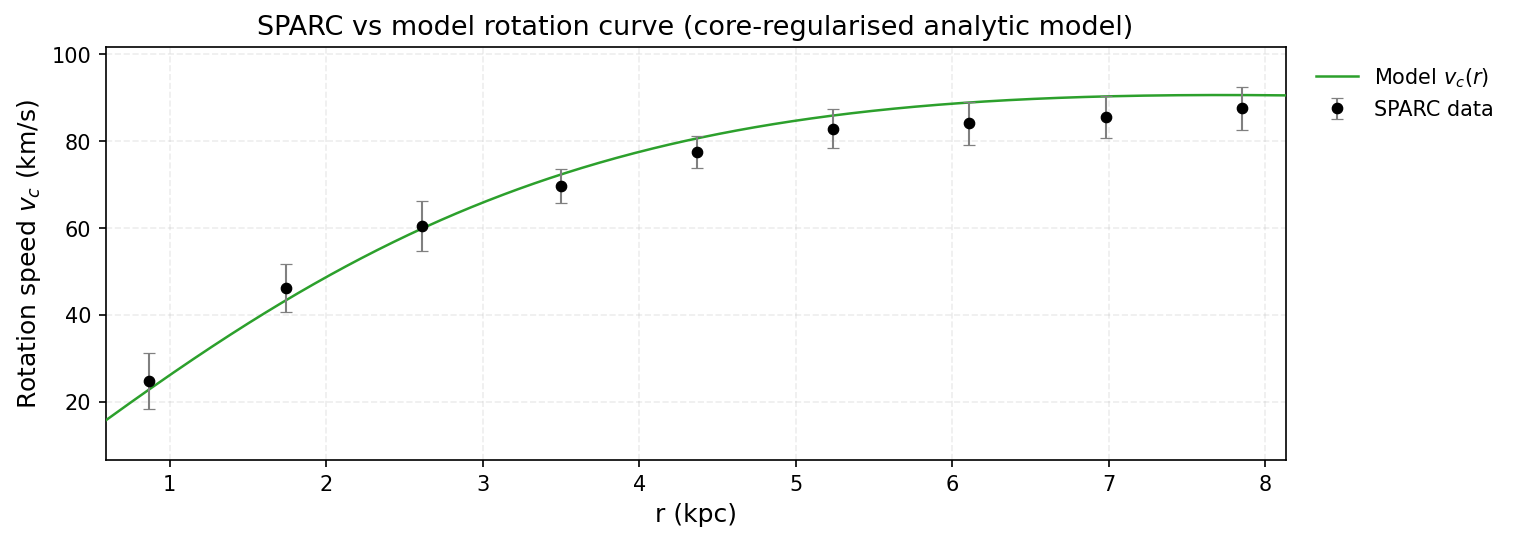}
\caption{The predicted rotation curves for the optimized SIDM
model of Eq. (\ref{ScaledependentEoSDM}), versus the SPARC
observational data for the galaxy UGC06399.} \label{UGC06399}
\end{figure}

In Table \ref{collUGC06399} we present the optimized values of
$K_0$ and $\rho_0$ for the analytic SIDM model of Eq.
(\ref{ScaledependentEoSDM}) for which the maximum compatibility
with the SPARC data is achieved.
\begin{table}[h!]
  \begin{center}
    \caption{SIDM Optimization Values for the galaxy UGC06399}
    \label{collUGC06399}
     \begin{tabular}{|r|r|}
     \hline
      \textbf{Parameter}   & \textbf{Optimization Values}
      \\  \hline
     $\rho_0 $  ($M_{\odot}/\mathrm{Kpc}^{3}$) & $3.97388\times 10^7$
\\  \hline $K_0$ ($M_{\odot} \,
\mathrm{Kpc}^{-3} \, (\mathrm{km/s})^{2}$)& 3557.78
\\  \hline
    \end{tabular}
  \end{center}
\end{table}
In Figs. \ref{UGC06399dens}, \ref{UGC06399}  we present the
density of the analytic SIDM model, the predicted rotation curves
for the SIDM model (\ref{ScaledependentEoSDM}), versus the SPARC
observational data and the sound speed, as a function of the
radius respectively. As it can be seen, for this galaxy, the SIDM
model produces viable rotation curves which are compatible with
the SPARC data.

\subsection{The Galaxy UGC06446, Non-viable, Extended Viable}

For this galaxy, the optimization method we used, ensures maximum
compatibility of the analytic SIDM model of Eq.
(\ref{ScaledependentEoSDM}) with the SPARC data, if we choose
$\rho_0=6.37837\times 10^7$$M_{\odot}/\mathrm{Kpc}^{3}$ and
$K_0=2994.32
$$M_{\odot} \, \mathrm{Kpc}^{-3} \, (\mathrm{km/s})^{2}$, in which
case the reduced $\chi^2_{red}$ value is $\chi^2_{red}=2.09777$.
Also the parameter $\alpha$ in this case is $\alpha=3.95408 $Kpc.

\begin{figure}[h!]
\centering
\includegraphics[width=20pc]{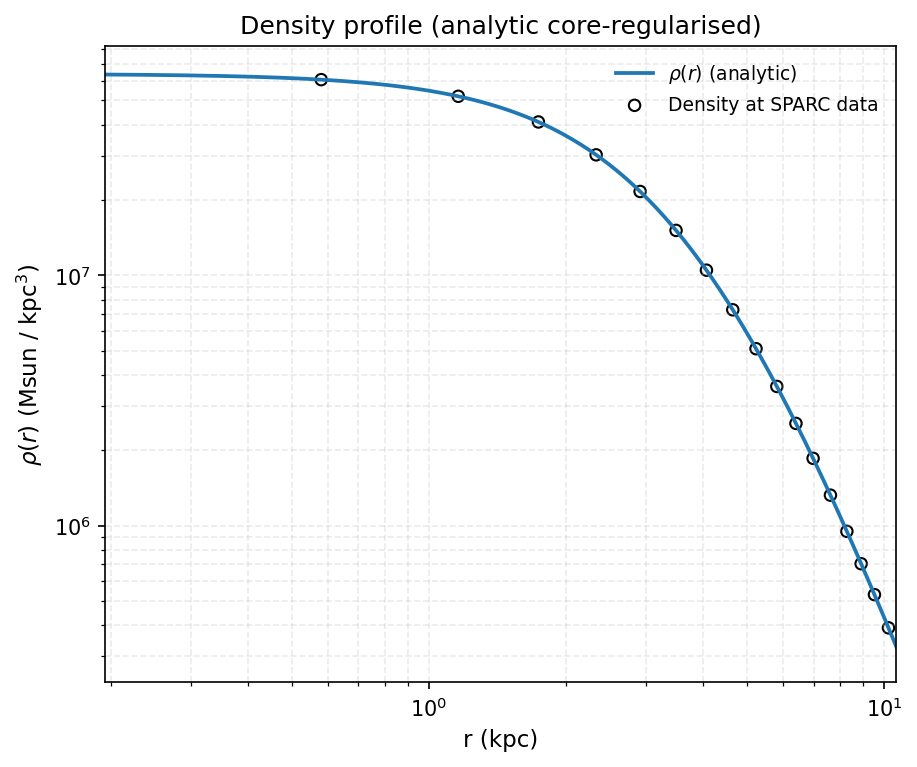}
\caption{The density of the SIDM model of Eq.
(\ref{ScaledependentEoSDM}) for the galaxy UGC06446, versus the
radius.} \label{UGC06446dens}
\end{figure}

\begin{figure}[h!]
\centering
\includegraphics[width=35pc]{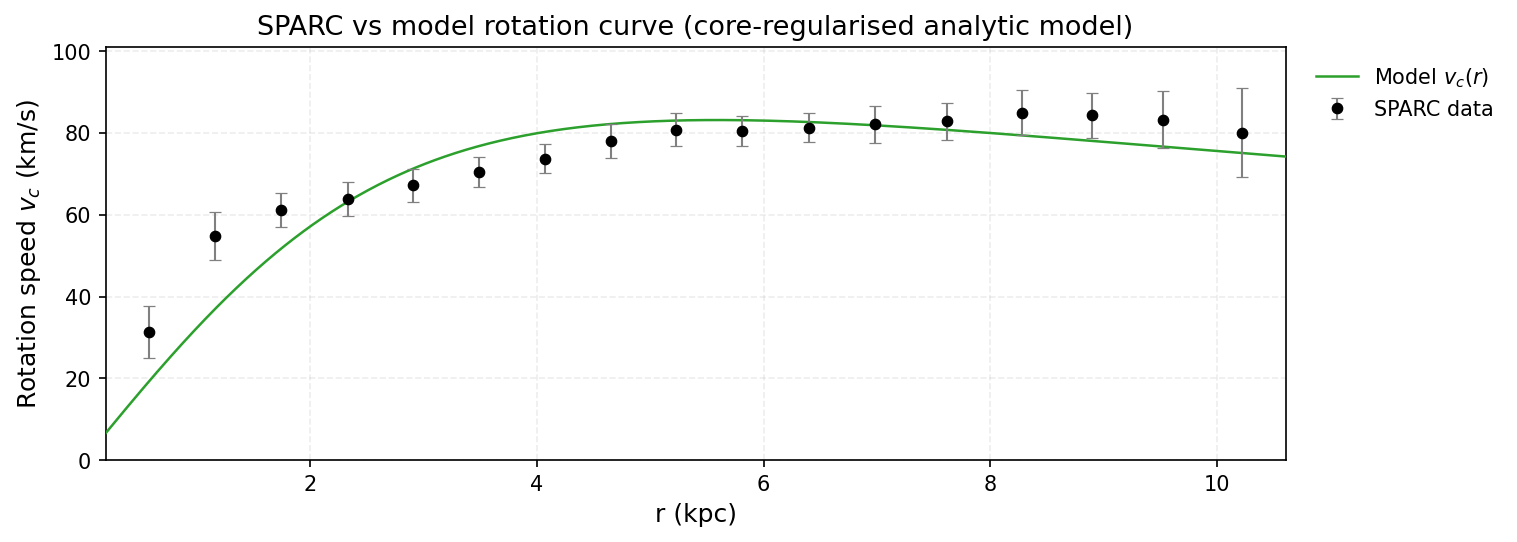}
\caption{The predicted rotation curves for the optimized SIDM
model of Eq. (\ref{ScaledependentEoSDM}), versus the SPARC
observational data for the galaxy UGC06446.} \label{UGC06446}
\end{figure}

In Table \ref{collUGC06446} we present the optimized values of
$K_0$ and $\rho_0$ for the analytic SIDM model of Eq.
(\ref{ScaledependentEoSDM}) for which the maximum compatibility
with the SPARC data is achieved.
\begin{table}[h!]
  \begin{center}
    \caption{SIDM Optimization Values for the galaxy UGC06446}
    \label{collUGC06446}
     \begin{tabular}{|r|r|}
     \hline
      \textbf{Parameter}   & \textbf{Optimization Values}
      \\  \hline
     $\rho_0 $  ($M_{\odot}/\mathrm{Kpc}^{3}$) & $6.37837\times 10^7$
\\  \hline $K_0$ ($M_{\odot} \,
\mathrm{Kpc}^{-3} \, (\mathrm{km/s})^{2}$)& 2994.32
\\  \hline
    \end{tabular}
  \end{center}
\end{table}
In Figs. \ref{UGC06446dens}, \ref{UGC06446}  we present the
density of the analytic SIDM model, the predicted rotation curves
for the SIDM model (\ref{ScaledependentEoSDM}), versus the SPARC
observational data and the sound speed, as a function of the
radius respectively. As it can be seen, for this galaxy, the SIDM
model produces non-viable rotation curves which are incompatible
with the SPARC data.

\begin{figure}[h!]
\centering
\includegraphics[width=20pc]{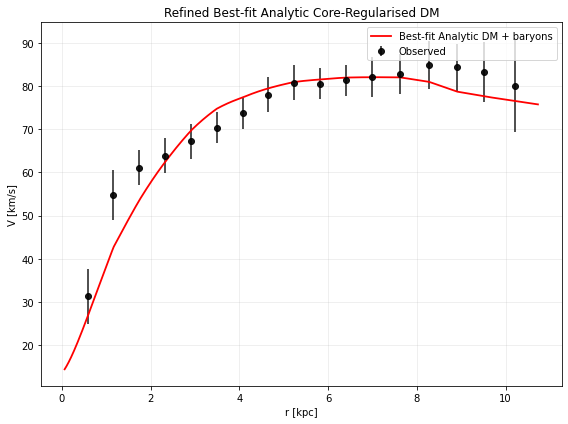}
\caption{The predicted rotation curves after using an optimization
for the SIDM model (\ref{ScaledependentEoSDM}), and the extended
SPARC data for the galaxy UGC06446. We included the rotation
curves of the gas, the disk velocities, the bulge (where present)
along with the SIDM model.} \label{extendedUGC06446}
\end{figure}

Now we shall include contributions to the rotation velocity from
the other components of the galaxy, namely the disk, the gas, and
the bulge if present. In Fig. \ref{extendedUGC06446} we present
the combined rotation curves including all the components of the
galaxy along with the SIDM. As it can be seen, the extended
collisional DM model is viable.
Also in Table \ref{evaluationextendedUGC06446} we present the
optimized values of the free parameters of the SIDM model for
which  we achieve the maximum compatibility with the SPARC data,
for the galaxy UGC06446, and also the resulting reduced
$\chi^2_{red}$ value.
\begin{table}[h!]
\centering \caption{Optimized Parameter Values of the Extended
SIDM model for the Galaxy UGC06446.}
\begin{tabular}{lc}
\hline
Parameter & Value  \\
\hline
$\rho_0 $ ($M_{\odot}/\mathrm{Kpc}^{3}$) & $4.41183\times 10^7$   \\
$K_0$ ($M_{\odot} \,
\mathrm{Kpc}^{-3} \, (\mathrm{km/s})^{2}$) & 2230.52   \\
$ml_{\text{disk}}$ & 1 \\
$ml_{\text{bulge}}$ & 0.2324 \\
$\alpha$ (Kpc) & 4.10289\\
$\chi^2_{red}$ & 1.0631 \\
\hline
\end{tabular}
\label{evaluationextendedUGC06446}
\end{table}

\subsection{The Galaxy UGC06614, Non-viable, Extended Viable}

For this galaxy, the optimization method we used, ensures maximum
compatibility of the analytic SIDM model of Eq.
(\ref{ScaledependentEoSDM}) with the SPARC data, if we choose
$\rho_0=1.9692\times 10^7$$M_{\odot}/\mathrm{Kpc}^{3}$ and
$K_0=19839.8
$$M_{\odot} \, \mathrm{Kpc}^{-3} \, (\mathrm{km/s})^{2}$, in which
case the reduced $\chi^2_{red}$ value is $\chi^2_{red}=7.7828$.
Also the parameter $\alpha$ in this case is $\alpha=18.3179 $Kpc.

\begin{figure}[h!]
\centering
\includegraphics[width=20pc]{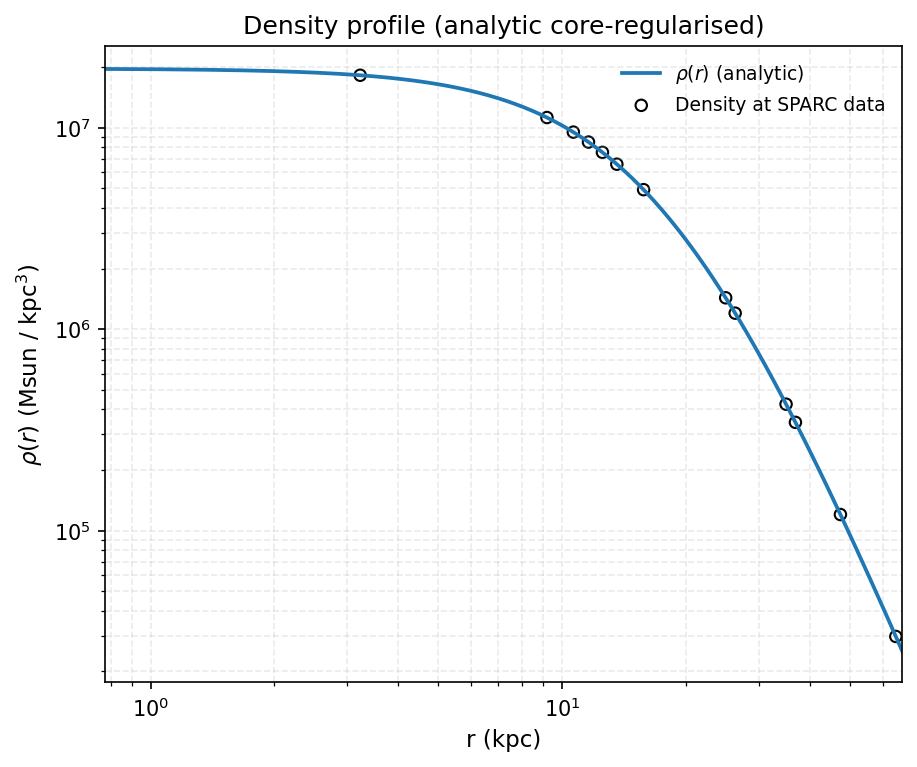}
\caption{The density of the SIDM model of Eq.
(\ref{ScaledependentEoSDM}) for the galaxy UGC06614, versus the
radius.} \label{UGC06614dens}
\end{figure}

\begin{figure}[h!]
\centering
\includegraphics[width=35pc]{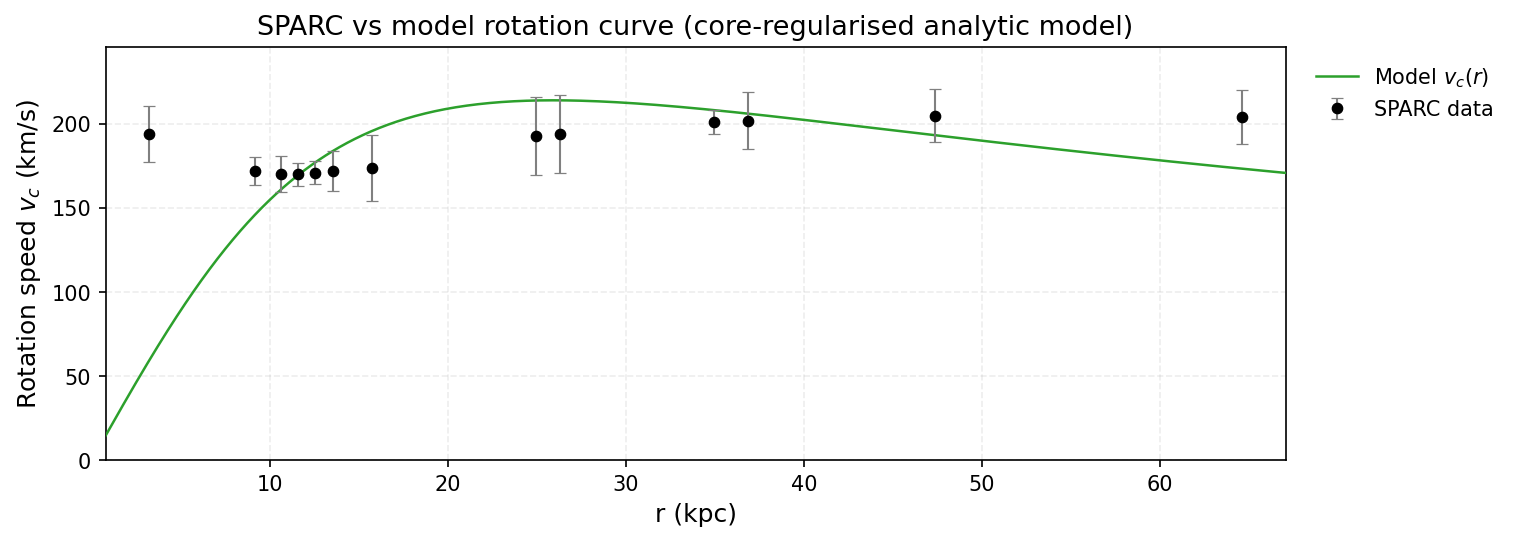}
\caption{The predicted rotation curves for the optimized SIDM
model of Eq. (\ref{ScaledependentEoSDM}), versus the SPARC
observational data for the galaxy UGC06614.} \label{UGC06614}
\end{figure}

In Table \ref{collUGC06614} we present the optimized values of
$K_0$ and $\rho_0$ for the analytic SIDM model of Eq.
(\ref{ScaledependentEoSDM}) for which the maximum compatibility
with the SPARC data is achieved.
\begin{table}[h!]
  \begin{center}
    \caption{SIDM Optimization Values for the galaxy UGC06614}
    \label{collUGC06614}
     \begin{tabular}{|r|r|}
     \hline
      \textbf{Parameter}   & \textbf{Optimization Values}
      \\  \hline
     $\rho_0 $  ($M_{\odot}/\mathrm{Kpc}^{3}$) & $1.9692\times 10^7$
\\  \hline $K_0$ ($M_{\odot} \,
\mathrm{Kpc}^{-3} \, (\mathrm{km/s})^{2}$)& 19839.8
\\  \hline
    \end{tabular}
  \end{center}
\end{table}
In Figs. \ref{UGC06614dens}, \ref{UGC06614}  we present the
density of the analytic SIDM model, the predicted rotation curves
for the SIDM model (\ref{ScaledependentEoSDM}), versus the SPARC
observational data and the sound speed, as a function of the
radius respectively. As it can be seen, for this galaxy, the SIDM
model produces non-viable rotation curves which are incompatible
with the SPARC data.

\begin{figure}[h!]
\centering
\includegraphics[width=20pc]{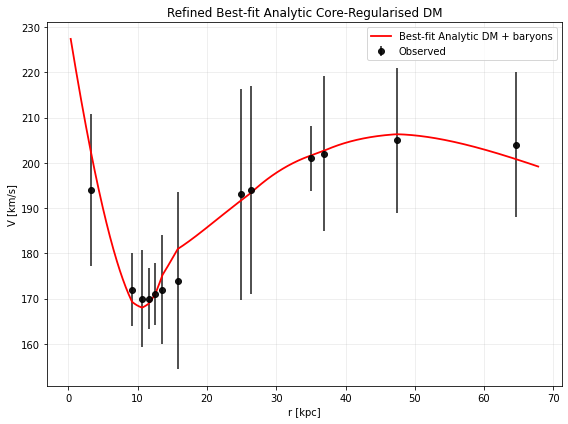}
\caption{The predicted rotation curves after using an optimization
for the SIDM model (\ref{ScaledependentEoSDM}), and the extended
SPARC data for the galaxy UGC06614. We included the rotation
curves of the gas, the disk velocities, the bulge (where present)
along with the SIDM model.} \label{extendedUGC06614}
\end{figure}

Now we shall include contributions to the rotation velocity from
the other components of the galaxy, namely the disk, the gas, and
the bulge if present. In Fig. \ref{extendedUGC06614} we present
the combined rotation curves including all the components of the
galaxy along with the SIDM. As it can be seen, the extended
collisional DM model is viable.
Also in Table \ref{evaluationextendedUGC06614} we present the
optimized values of the free parameters of the SIDM model for
which  we achieve the maximum compatibility with the SPARC data,
for the galaxy UGC06614, and also the resulting reduced
$\chi^2_{red}$ value.
\begin{table}[h!]
\centering \caption{Optimized Parameter Values of the Extended
SIDM model for the Galaxy UGC06614.}
\begin{tabular}{lc}
\hline
Parameter & Value  \\
\hline
$\rho_0 $ ($M_{\odot}/\mathrm{Kpc}^{3}$) & $3.82792\times 10^6$   \\
$K_0$ ($M_{\odot} \,
\mathrm{Kpc}^{-3} \, (\mathrm{km/s})^{2}$) & 14400.6   \\
$ml_{\text{disk}}$ & 0.6687 \\
$ml_{\text{bulge}}$ & 0.8008 \\
$\alpha$ (Kpc) & 35.3919\\
$\chi^2_{red}$ & 0.760491 \\
\hline
\end{tabular}
\label{evaluationextendedUGC06614}
\end{table}

\subsection{The Galaxy UGC06667}

For this galaxy, the optimization method we used, ensures maximum
compatibility of the analytic SIDM model of Eq.
(\ref{ScaledependentEoSDM}) with the SPARC data, if we choose
$\rho_0=4.63803\times 10^7$$M_{\odot}/\mathrm{Kpc}^{3}$ and
$K_0=3075.79
$$M_{\odot} \, \mathrm{Kpc}^{-3} \, (\mathrm{km/s})^{2}$, in which
case the reduced $\chi^2_{red}$ value is $\chi^2_{red}=0.566936$.
Also the parameter $\alpha$ in this case is $\alpha=5.03668 $Kpc.

\begin{figure}[h!]
\centering
\includegraphics[width=20pc]{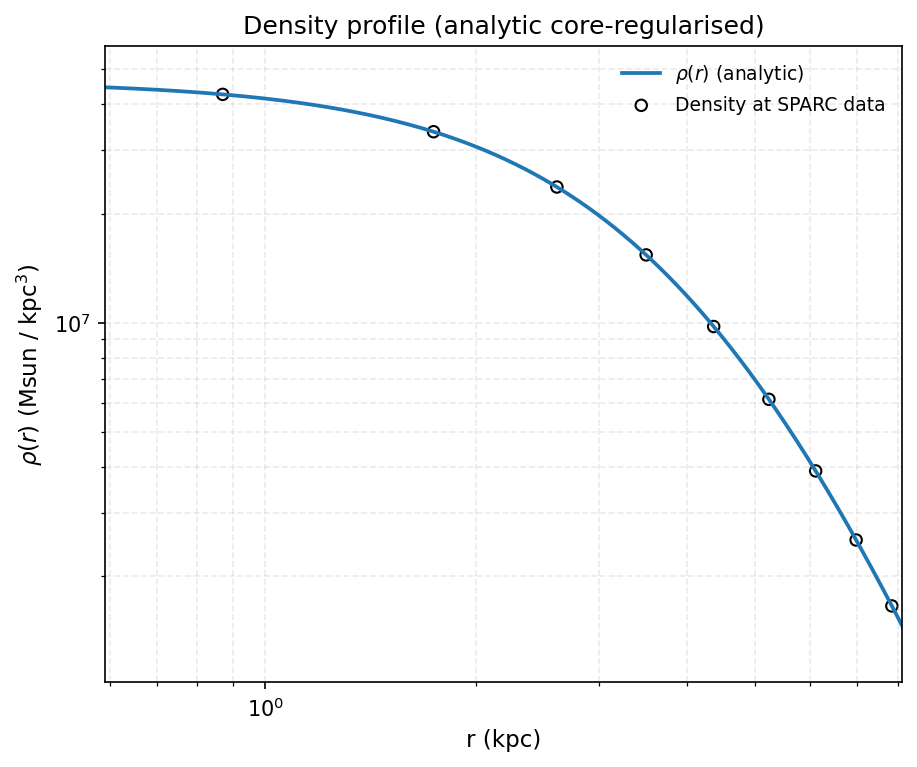}
\caption{The density of the SIDM model of Eq.
(\ref{ScaledependentEoSDM}) for the galaxy UGC06667, versus the
radius.} \label{UGC06667dens}
\end{figure}

\begin{figure}[h!]
\centering
\includegraphics[width=35pc]{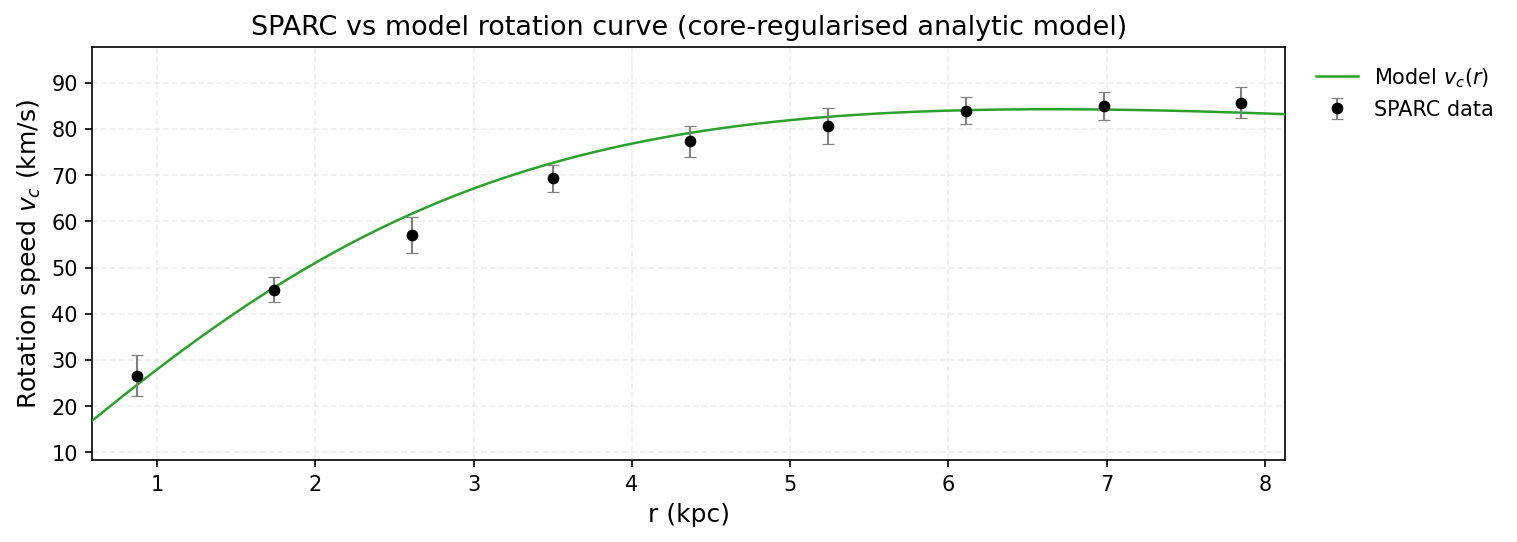}
\caption{The predicted rotation curves for the optimized SIDM
model of Eq. (\ref{ScaledependentEoSDM}), versus the SPARC
observational data for the galaxy UGC06667.} \label{UGC06667}
\end{figure}

In Table \ref{collUGC06667} we present the optimized values of
$K_0$ and $\rho_0$ for the analytic SIDM model of Eq.
(\ref{ScaledependentEoSDM}) for which the maximum compatibility
with the SPARC data is achieved.
\begin{table}[h!]
  \begin{center}
    \caption{SIDM Optimization Values for the galaxy UGC06667}
    \label{collUGC06667}
     \begin{tabular}{|r|r|}
     \hline
      \textbf{Parameter}   & \textbf{Optimization Values}
      \\  \hline
     $\rho_0 $  ($M_{\odot}/\mathrm{Kpc}^{3}$) & $4.63803\times 10^7$
\\  \hline $K_0$ ($M_{\odot} \,
\mathrm{Kpc}^{-3} \, (\mathrm{km/s})^{2}$)& 3075.79
\\  \hline
    \end{tabular}
  \end{center}
\end{table}
In Figs. \ref{UGC06667dens}, \ref{UGC06667}  we present the
density of the analytic SIDM model, the predicted rotation curves
for the SIDM model (\ref{ScaledependentEoSDM}), versus the SPARC
observational data and the sound speed, as a function of the
radius respectively. As it can be seen, for this galaxy, the SIDM
model produces viable rotation curves which are compatible with
the SPARC data.

\subsection{The Galaxy UGC06786, Non-viable}

For this galaxy, the optimization method we used, ensures maximum
compatibility of the analytic SIDM model of Eq.
(\ref{ScaledependentEoSDM}) with the SPARC data, if we choose
$\rho_0=1.90042\times 10^8$$M_{\odot}/\mathrm{Kpc}^{3}$ and
$K_0=26463.3
$$M_{\odot} \, \mathrm{Kpc}^{-3} \, (\mathrm{km/s})^{2}$, in which
case the reduced $\chi^2_{red}$ value is $\chi^2_{red}=52.6091$.
Also the parameter $\alpha$ in this case is $\alpha=6.81 $Kpc.

\begin{figure}[h!]
\centering
\includegraphics[width=20pc]{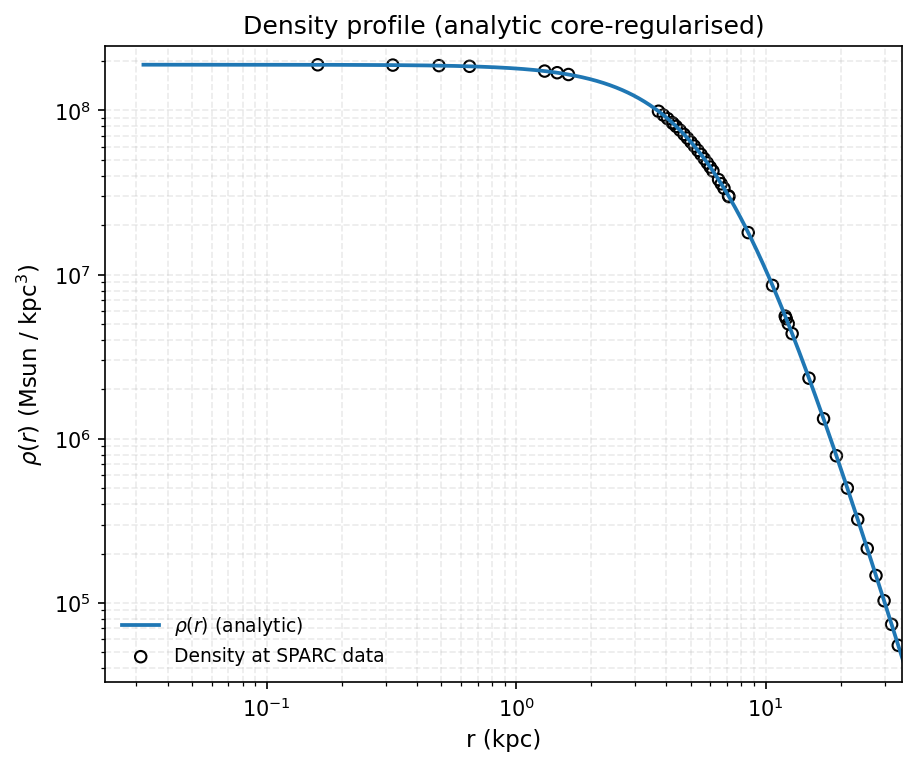}
\caption{The density of the SIDM model of Eq.
(\ref{ScaledependentEoSDM}) for the galaxy UGC06786, versus the
radius.} \label{UGC06786dens}
\end{figure}

\begin{figure}[h!]
\centering
\includegraphics[width=20pc]{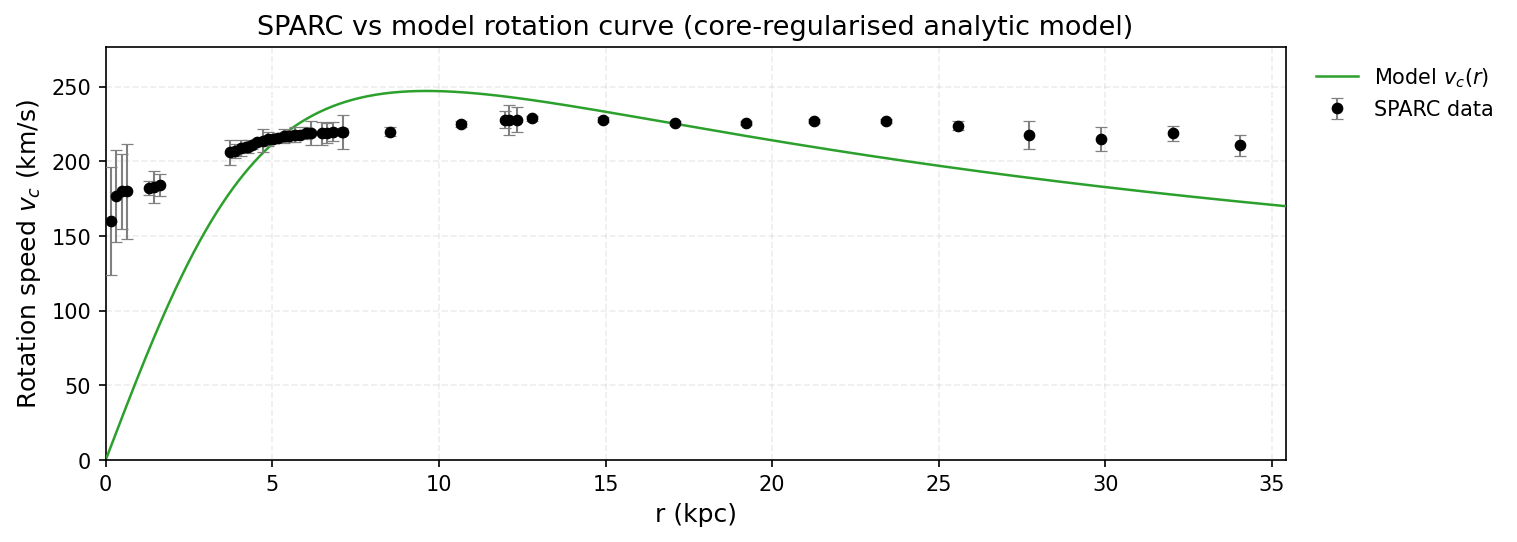}
\caption{The predicted rotation curves for the optimized SIDM
model of Eq. (\ref{ScaledependentEoSDM}), versus the SPARC
observational data for the galaxy UGC06786.} \label{UGC06786}
\end{figure}

In Table \ref{collUGC06786} we present the optimized values of
$K_0$ and $\rho_0$ for the analytic SIDM model of Eq.
(\ref{ScaledependentEoSDM}) for which the maximum compatibility
with the SPARC data is achieved.
\begin{table}[h!]
  \begin{center}
    \caption{SIDM Optimization Values for the galaxy UGC06786}
    \label{collUGC06786}
     \begin{tabular}{|r|r|}
     \hline
      \textbf{Parameter}   & \textbf{Optimization Values}
      \\  \hline
     $\rho_0 $  ($M_{\odot}/\mathrm{Kpc}^{3}$) & $1.90042\times 10^8$
\\  \hline $K_0$ ($M_{\odot} \,
\mathrm{Kpc}^{-3} \, (\mathrm{km/s})^{2}$)& 26463.3
\\  \hline
    \end{tabular}
  \end{center}
\end{table}
In Figs. \ref{UGC06786dens}, \ref{UGC06786} we present the density
of the analytic SIDM model, the predicted rotation curves for the
SIDM model (\ref{ScaledependentEoSDM}), versus the SPARC
observational data and the sound speed, as a function of the
radius respectively. As it can be seen, for this galaxy, the SIDM
model produces non-viable rotation curves which are incompatible
with the SPARC data.

\begin{figure}[h!]
\centering
\includegraphics[width=20pc]{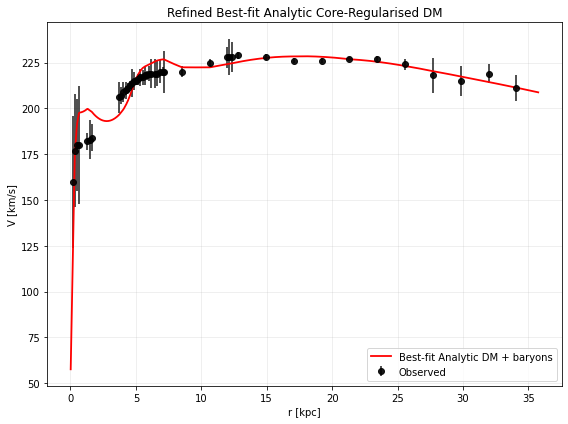}
\caption{The predicted rotation curves after using an optimization
for the SIDM model (\ref{ScaledependentEoSDM}), and the extended
SPARC data for the galaxy UGC06786. We included the rotation
curves of the gas, the disk velocities, the bulge (where present)
along with the SIDM model.} \label{extendedUGC06786}
\end{figure}

Now we shall include contributions to the rotation velocity from
the other components of the galaxy, namely the disk, the gas, and
the bulge if present. In Fig. \ref{extendedUGC06786} we present
the combined rotation curves including all the components of the
galaxy along with the SIDM. As it can be seen, the extended
collisional DM model is non-viable.
Also in Table \ref{evaluationextendedUGC06786} we present the
optimized values of the free parameters of the SIDM model for
which  we achieve the maximum compatibility with the SPARC data,
for the galaxy UGC06786, and also the resulting reduced
$\chi^2_{red}$ value.
\begin{table}[h!]
\centering \caption{Optimized Parameter Values of the Extended
SIDM model for the Galaxy UGC06786.}
\begin{tabular}{lc}
\hline
Parameter & Value  \\
\hline
$\rho_0 $ ($M_{\odot}/\mathrm{Kpc}^{3}$) & $1.45289\times 10^7$   \\
$K_0$ ($M_{\odot} \,
\mathrm{Kpc}^{-3} \, (\mathrm{km/s})^{2}$) & 15487.9   \\
$ml_{\text{disk}}$ & 1 \\
$ml_{\text{bulge}}$ & 0.9395 \\
$\alpha$ (Kpc) & 18.8398\\
$\chi^2_{red}$ & 2.38481 \\
\hline
\end{tabular}
\label{evaluationextendedUGC06786}
\end{table}

\subsection{The Galaxy UGC06787, Non-viable}

For this galaxy, the optimization method we used, ensures maximum
compatibility of the analytic SIDM model of Eq.
(\ref{ScaledependentEoSDM}) with the SPARC data, if we choose
$\rho_0=1.86717\times 10^8$$M_{\odot}/\mathrm{Kpc}^{3}$ and
$K_0=31016.7
$$M_{\odot} \, \mathrm{Kpc}^{-3} \, (\mathrm{km/s})^{2}$, in which
case the reduced $\chi^2_{red}$ value is $\chi^2_{red}=177.21$.
Also the parameter $\alpha$ in this case is $\alpha=7.438 $Kpc.

\begin{figure}[h!]
\centering
\includegraphics[width=20pc]{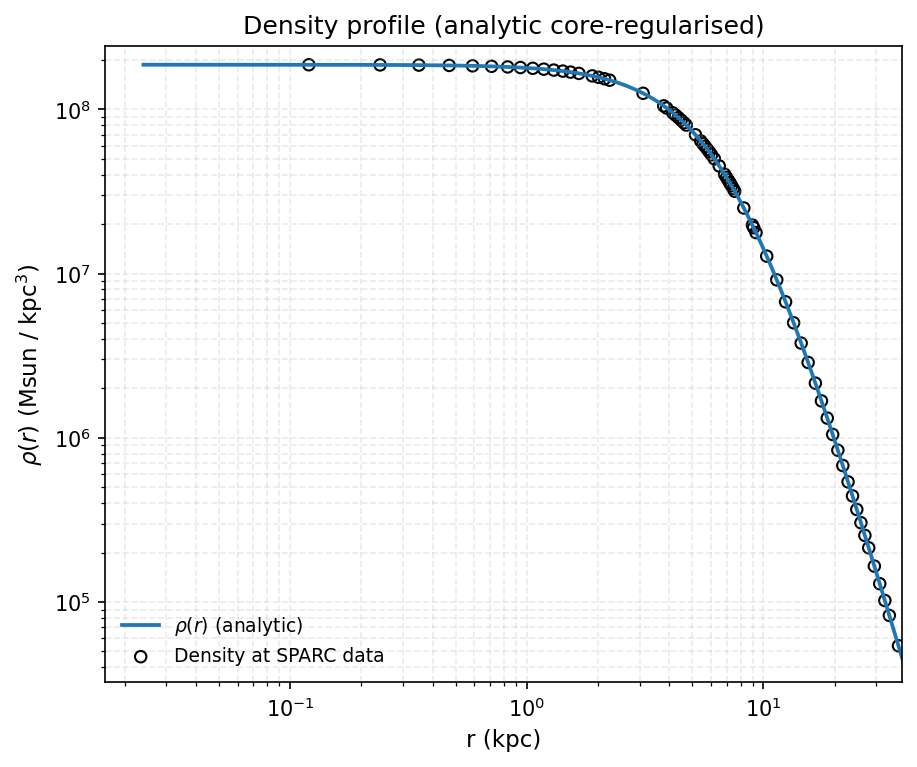}
\caption{The density of the SIDM model of Eq.
(\ref{ScaledependentEoSDM}) for the galaxy UGC06787, versus the
radius.} \label{UGC06787dens}
\end{figure}

\begin{figure}[h!]
\centering
\includegraphics[width=35pc]{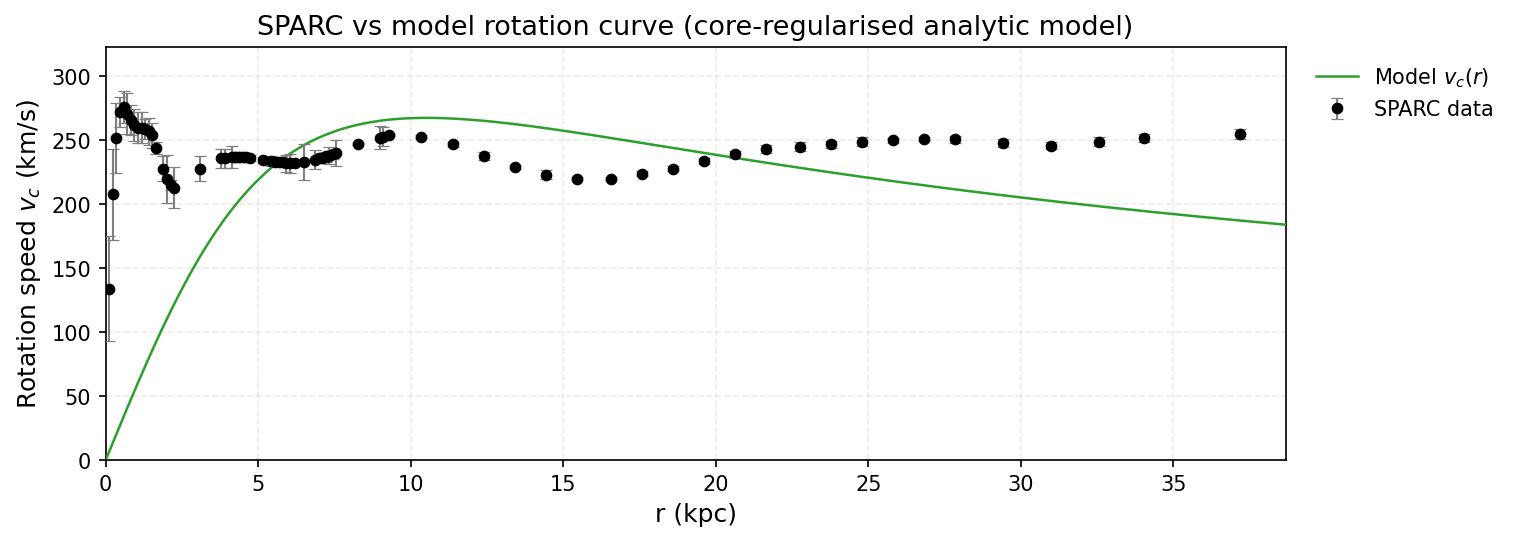}
\caption{The predicted rotation curves for the optimized SIDM
model of Eq. (\ref{ScaledependentEoSDM}), versus the SPARC
observational data for the galaxy UGC06787.} \label{UGC06787}
\end{figure}

In Table \ref{collUGC06787} we present the optimized values of
$K_0$ and $\rho_0$ for the analytic SIDM model of Eq.
(\ref{ScaledependentEoSDM}) for which the maximum compatibility
with the SPARC data is achieved.
\begin{table}[h!]
  \begin{center}
    \caption{SIDM Optimization Values for the galaxy UGC06787}
    \label{collUGC06787}
     \begin{tabular}{|r|r|}
     \hline
      \textbf{Parameter}   & \textbf{Optimization Values}
      \\  \hline
     $\rho_0 $  ($M_{\odot}/\mathrm{Kpc}^{3}$) & $1.86717\times 10^8$
\\  \hline $K_0$ ($M_{\odot} \,
\mathrm{Kpc}^{-3} \, (\mathrm{km/s})^{2}$)& 31016.7
\\  \hline
    \end{tabular}
  \end{center}
\end{table}
In Figs. \ref{UGC06787dens}, \ref{UGC06787}  we present the
density of the analytic SIDM model, the predicted rotation curves
for the SIDM model (\ref{ScaledependentEoSDM}), versus the SPARC
observational data and the sound speed, as a function of the
radius respectively. As it can be seen, for this galaxy, the SIDM
model produces non-viable rotation curves which are incompatible
with the SPARC data.

\begin{figure}[h!]
\centering
\includegraphics[width=20pc]{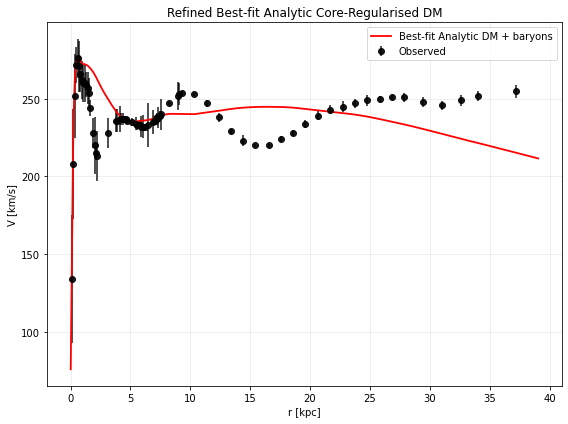}
\caption{The predicted rotation curves after using an optimization
for the SIDM model (\ref{ScaledependentEoSDM}), and the extended
SPARC data for the galaxy UGC06787. We included the rotation
curves of the gas, the disk velocities, the bulge (where present)
along with the SIDM model.} \label{extendedUGC06787}
\end{figure}

Now we shall include contributions to the rotation velocity from
the other components of the galaxy, namely the disk, the gas, and
the bulge if present. In Fig. \ref{extendedUGC06787} we present
the combined rotation curves including all the components of the
galaxy along with the SIDM. As it can be seen, the extended
collisional DM model is non-viable.
Also in Table \ref{evaluationextendedUGC06787} we present the
optimized values of the free parameters of the SIDM model for
which  we achieve the maximum compatibility with the SPARC data,
for the galaxy UGC06787, and also the resulting reduced
$\chi^2_{red}$ value.
\begin{table}[h!]
\centering \caption{Optimized Parameter Values of the Extended
SIDM model for the Galaxy UGC06787.}
\begin{tabular}{lc}
\hline
Parameter & Value  \\
\hline
$\rho_0 $ ($M_{\odot}/\mathrm{Kpc}^{3}$) & $2.1286\times 10^7$   \\
$K_0$ ($M_{\odot} \,
\mathrm{Kpc}^{-3} \, (\mathrm{km/s})^{2}$) & 17225.7   \\
$ml_{\text{disk}}$ & 1 \\
$ml_{\text{bulge}}$ & 0.8451 \\
$\alpha$ (Kpc) & 16.4149\\
$\chi^2_{red}$ & 38.8218 \\
\hline
\end{tabular}
\label{evaluationextendedUGC06787}
\end{table}

\subsection{The Galaxy UGC06818}

For this galaxy, the optimization method we used, ensures maximum
compatibility of the analytic SIDM model of Eq.
(\ref{ScaledependentEoSDM}) with the SPARC data, if we choose
$\rho_0=1.63491\times 10^7$$M_{\odot}/\mathrm{Kpc}^{3}$ and
$K_0=2632.02
$$M_{\odot} \, \mathrm{Kpc}^{-3} \, (\mathrm{km/s})^{2}$, in which
case the reduced $\chi^2_{red}$ value is $\chi^2_{red}=1.00372$.
Also the parameter $\alpha$ in this case is $\alpha=7.32232 $Kpc.

\begin{figure}[h!]
\centering
\includegraphics[width=20pc]{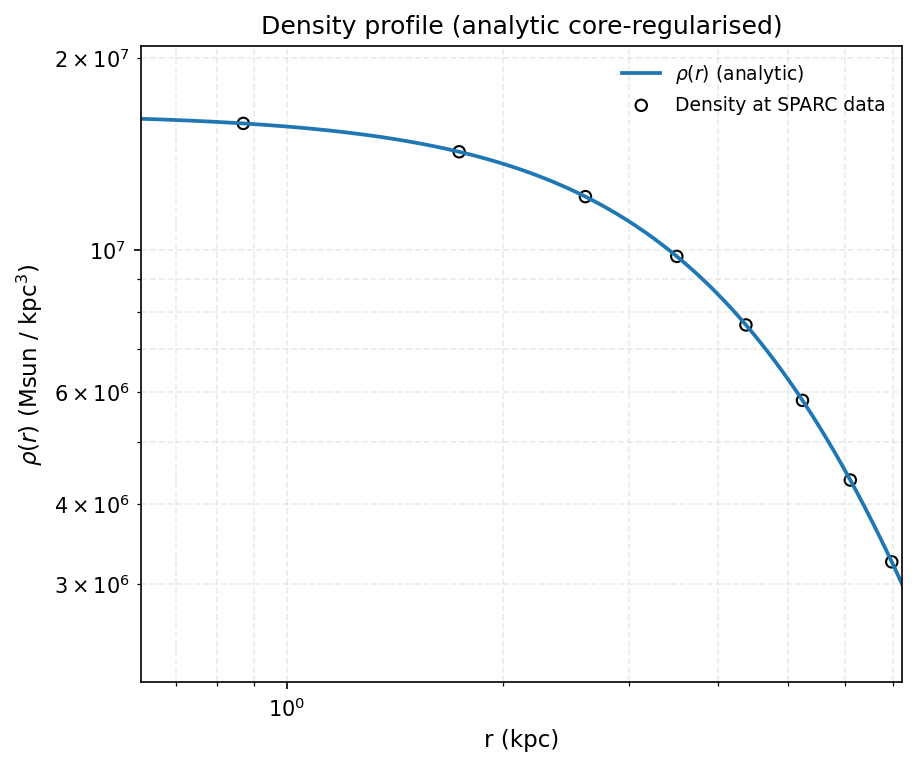}
\caption{The density of the SIDM model of Eq.
(\ref{ScaledependentEoSDM}) for the galaxy UGC06818, versus the
radius.} \label{UGC06818dens}
\end{figure}

\begin{figure}[h!]
\centering
\includegraphics[width=35pc]{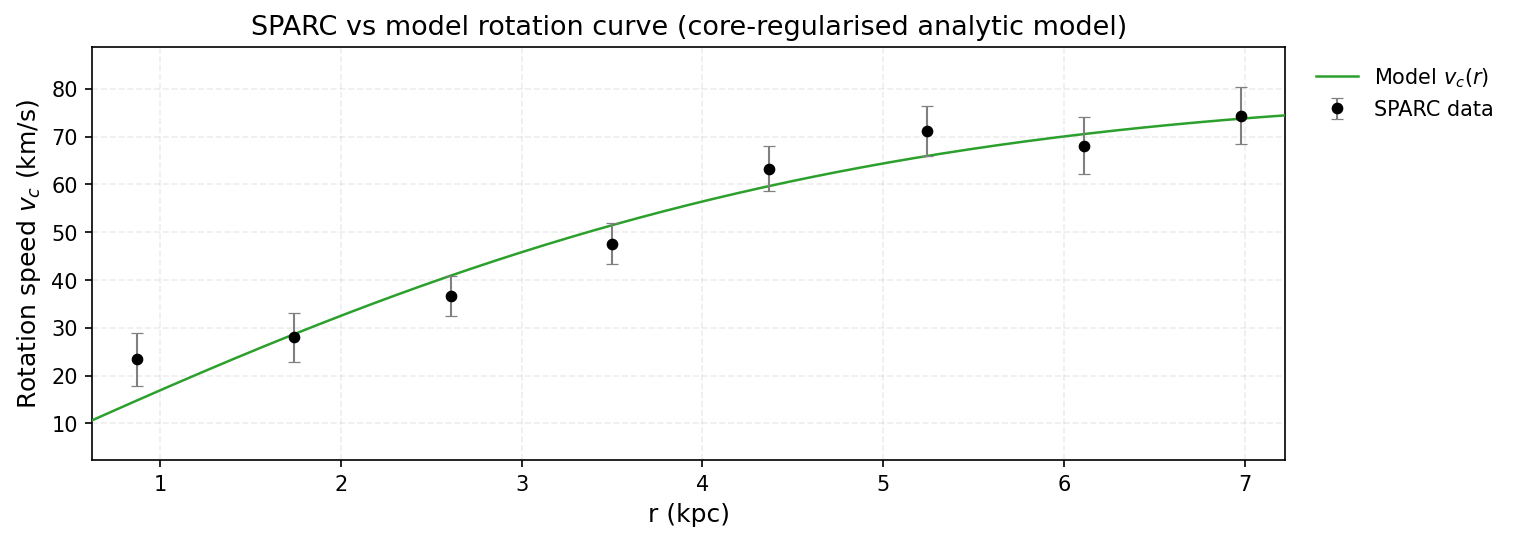}
\caption{The predicted rotation curves for the optimized SIDM
model of Eq. (\ref{ScaledependentEoSDM}), versus the SPARC
observational data for the galaxy UGC06818.} \label{UGC06818}
\end{figure}

In Table \ref{collUGC06818} we present the optimized values of
$K_0$ and $\rho_0$ for the analytic SIDM model of Eq.
(\ref{ScaledependentEoSDM}) for which the maximum compatibility
with the SPARC data is achieved.
\begin{table}[h!]
  \begin{center}
    \caption{SIDM Optimization Values for the galaxy UGC06818}
    \label{collUGC06818}
     \begin{tabular}{|r|r|}
     \hline
      \textbf{Parameter}   & \textbf{Optimization Values}
      \\  \hline
     $\rho_0 $  ($M_{\odot}/\mathrm{Kpc}^{3}$) & $1.63491\times 10^7$
\\  \hline $K_0$ ($M_{\odot} \,
\mathrm{Kpc}^{-3} \, (\mathrm{km/s})^{2}$)& 2632.02
\\  \hline
    \end{tabular}
  \end{center}
\end{table}
In Figs. \ref{UGC06818dens}, \ref{UGC06818}  we present the
density of the analytic SIDM model, the predicted rotation curves
for the SIDM model (\ref{ScaledependentEoSDM}), versus the SPARC
observational data and the sound speed, as a function of the
radius respectively. As it can be seen, for this galaxy, the SIDM
model produces viable rotation curves which are compatible with
the SPARC data.

\begin{figure}[h!]
\centering
\includegraphics[width=20pc]{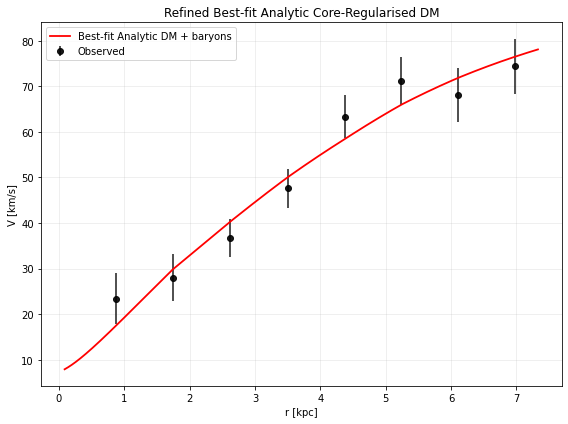}
\caption{The predicted rotation curves after using an optimization
for the SIDM model (\ref{ScaledependentEoSDM}), and the extended
SPARC data for the galaxy UGC06818. We included the rotation
curves of the gas, the disk velocities, the bulge (where present)
along with the SIDM model.} \label{extendedUGC06818}
\end{figure}

Now we shall include contributions to the rotation velocity from
the other components of the galaxy, namely the disk, the gas, and
the bulge if present. In Fig. \ref{extendedUGC06818} we present
the combined rotation curves including all the components of the
galaxy along with the SIDM. As it can be seen, the extended
collisional DM model is viable.
Also in Table \ref{evaluationextendedUGC06818} we present the
optimized values of the free parameters of the SIDM model for
which  we achieve the maximum compatibility with the SPARC data,
for the galaxy UGC06818, and also the resulting reduced
$\chi^2_{red}$ value.
\begin{table}[h!]
\centering \caption{Optimized Parameter Values of the Extended
SIDM model for the Galaxy UGC06818.}
\begin{tabular}{lc}
\hline
Parameter & Value  \\
\hline
$\rho_0 $ ($M_{\odot}/\mathrm{Kpc}^{3}$) & $1.14068\times 10^7$   \\
$K_0$ ($M_{\odot} \,
\mathrm{Kpc}^{-3} \, (\mathrm{km/s})^{2}$) & 3036.28   \\
$ml_{\text{disk}}$ & 0.4313 \\
$ml_{\text{bulge}}$ & 0.3118 \\
$\alpha$ (Kpc) & 9.41424\\
$\chi^2_{red}$ & 1.2134 \\
\hline
\end{tabular}
\label{evaluationextendedUGC06818}
\end{table}

\subsection{The Galaxy UGC06917, Non-viable, Extended Viable}

For this galaxy, the optimization method we used, ensures maximum
compatibility of the analytic SIDM model of Eq.
(\ref{ScaledependentEoSDM}) with the SPARC data, if we choose
$\rho_0=5.06612\times 10^7$$M_{\odot}/\mathrm{Kpc}^{3}$ and
$K_0=4883.69
$$M_{\odot} \, \mathrm{Kpc}^{-3} \, (\mathrm{km/s})^{2}$, in which
case the reduced $\chi^2_{red}$ value is $\chi^2_{red}=1.65275$.
Also the parameter $\alpha$ in this case is $\alpha=5.66614 $Kpc.

\begin{figure}[h!]
\centering
\includegraphics[width=20pc]{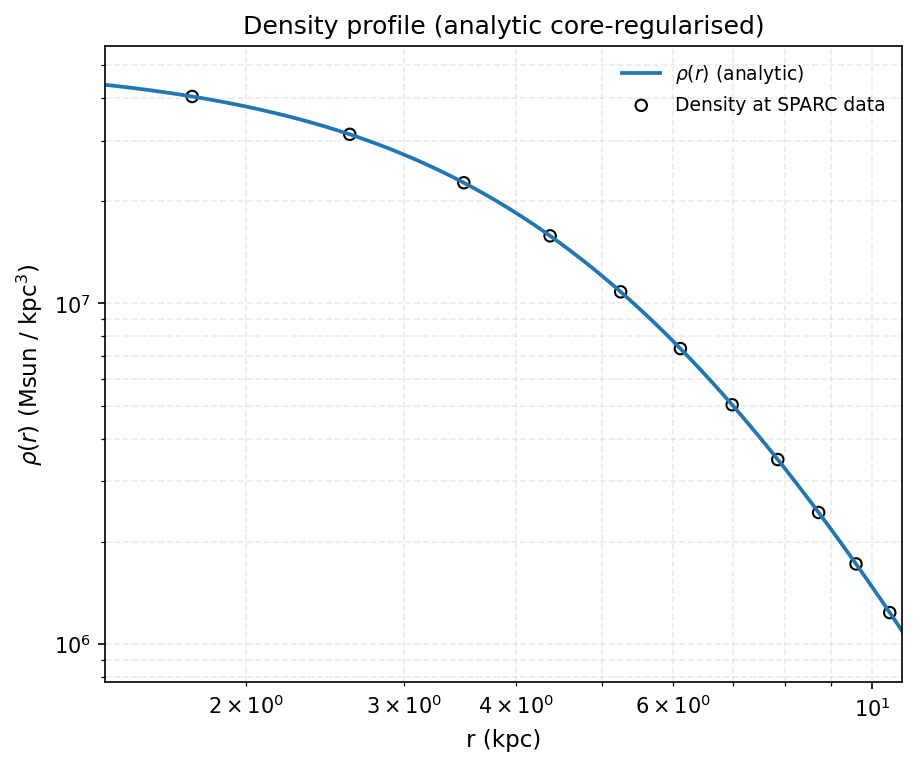}
\caption{The density of the SIDM model of Eq.
(\ref{ScaledependentEoSDM}) for the galaxy UGC06917, versus the
radius.} \label{UGC06917dens}
\end{figure}

\begin{figure}[h!]
\centering
\includegraphics[width=35pc]{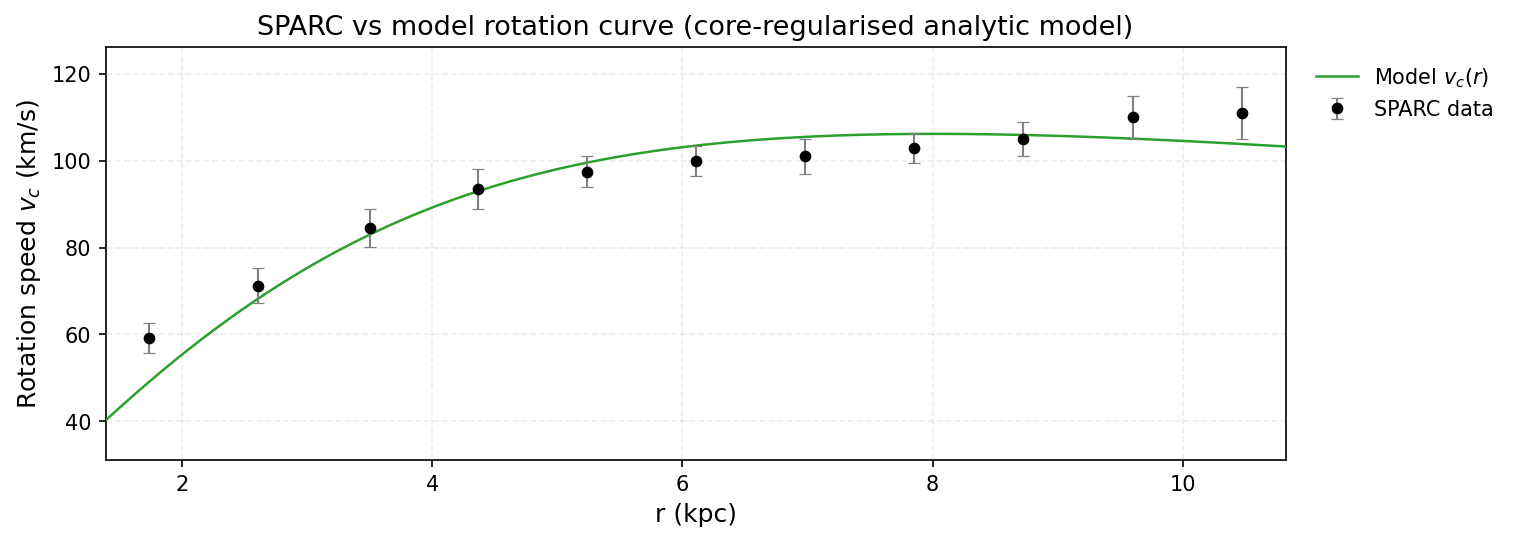}
\caption{The predicted rotation curves for the optimized SIDM
model of Eq. (\ref{ScaledependentEoSDM}), versus the SPARC
observational data for the galaxy UGC06917.} \label{UGC06917}
\end{figure}

In Table \ref{collUGC06917} we present the optimized values of
$K_0$ and $\rho_0$ for the analytic SIDM model of Eq.
(\ref{ScaledependentEoSDM}) for which the maximum compatibility
with the SPARC data is achieved.
\begin{table}[h!]
  \begin{center}
    \caption{SIDM Optimization Values for the galaxy UGC06917}
    \label{collUGC06917}
     \begin{tabular}{|r|r|}
     \hline
      \textbf{Parameter}   & \textbf{Optimization Values}
      \\  \hline
     $\rho_0 $  ($M_{\odot}/\mathrm{Kpc}^{3}$) & $5.06612\times 10^7$
\\  \hline $K_0$ ($M_{\odot} \,
\mathrm{Kpc}^{-3} \, (\mathrm{km/s})^{2}$)& 4883.69
\\  \hline
    \end{tabular}
  \end{center}
\end{table}
In Figs. \ref{UGC06917dens}, \ref{UGC06917}  we present the
density of the analytic SIDM model, the predicted rotation curves
for the SIDM model (\ref{ScaledependentEoSDM}), versus the SPARC
observational data and the sound speed, as a function of the
radius respectively. As it can be seen, for this galaxy, the SIDM
model produces non-viable rotation curves which are incompatible
with the SPARC data.

\begin{figure}[h!]
\centering
\includegraphics[width=20pc]{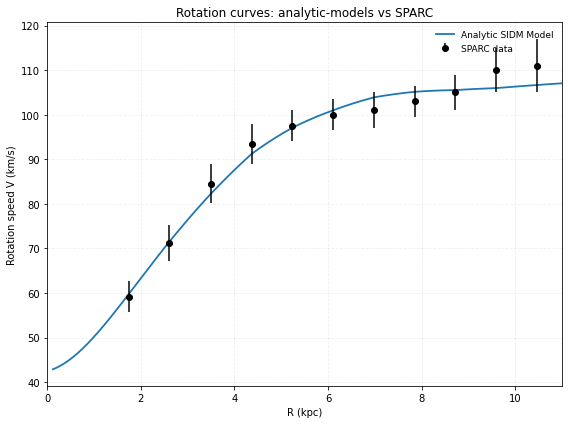}
\caption{The predicted rotation curves after using an optimization
for the SIDM model (\ref{ScaledependentEoSDM}), and the extended
SPARC data for the galaxy UGC06917. We included the rotation
curves of the gas, the disk velocities, the bulge (where present)
along with the SIDM model.} \label{extendedUGC06917}
\end{figure}

Now we shall include contributions to the rotation velocity from
the other components of the galaxy, namely the disk, the gas, and
the bulge if present. In Fig. \ref{extendedUGC06917} we present
the combined rotation curves including all the components of the
galaxy along with the SIDM. As it can be seen, the extended
collisional DM model is non-viable.
Also in Table \ref{evaluationextendedUGC06917} we present the
optimized values of the free parameters of the SIDM model for
which  we achieve the maximum compatibility with the SPARC data,
for the galaxy UGC06917, and also the resulting reduced
$\chi^2_{red}$ value.
\begin{table}[h!]
\centering \caption{Optimized Parameter Values of the Extended
SIDM model for the Galaxy UGC06917.}
\begin{tabular}{lc}
\hline
Parameter & Value  \\
\hline
$\rho_0 $ ($M_{\odot}/\mathrm{Kpc}^{3}$) & $2.73664\times 10^7$   \\
$K_0$ ($M_{\odot} \,
\mathrm{Kpc}^{-3} \, (\mathrm{km/s})^{2}$) & 3283.4   \\
$ml_{\text{disk}}$ & 0.8401 \\
$ml_{\text{bulge}}$ & 0.1058 \\
$\alpha$ (Kpc) & 6.32046\\
$\chi^2_{red}$ & 0.39102 \\
\hline
\end{tabular}
\label{evaluationextendedUGC06917}
\end{table}

\subsection{The Galaxy UGC06923}

For this galaxy, the optimization method we used, ensures maximum
compatibility of the analytic SIDM model of Eq.
(\ref{ScaledependentEoSDM}) with the SPARC data, if we choose
$\rho_0=8.29599\times 10^7$$M_{\odot}/\mathrm{Kpc}^{3}$ and
$K_0=2622.3
$$M_{\odot} \, \mathrm{Kpc}^{-3} \, (\mathrm{km/s})^{2}$, in which
case the reduced $\chi^2_{red}$ value is $\chi^2_{red}=0.649073$.
Also the parameter $\alpha$ in this case is $\alpha=3.24458 $Kpc.

\begin{figure}[h!]
\centering
\includegraphics[width=20pc]{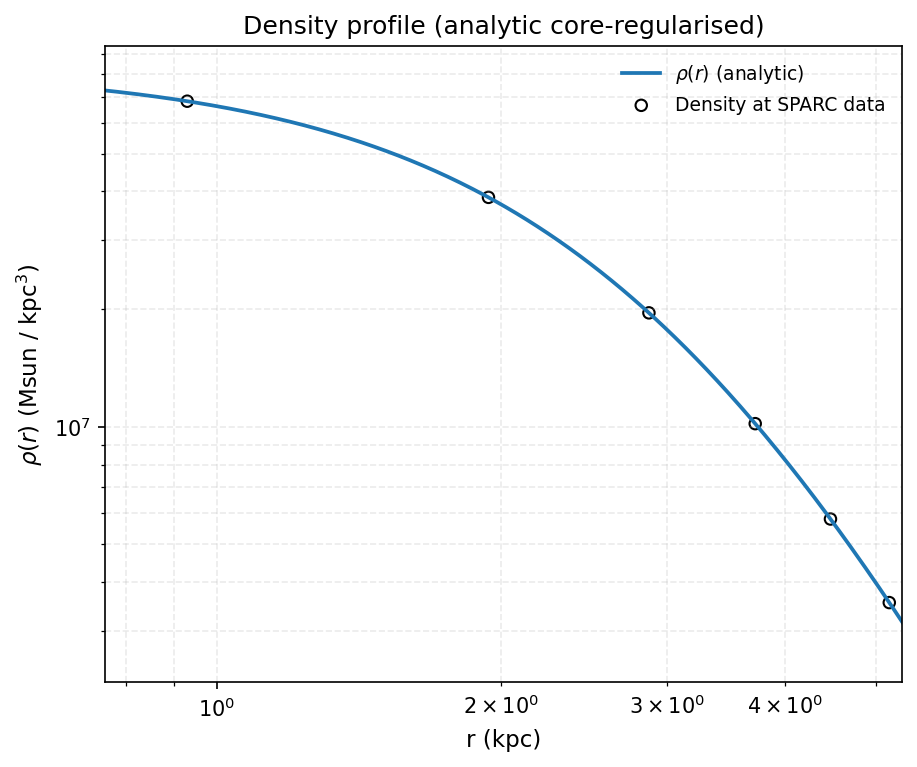}
\caption{The density of the SIDM model of Eq.
(\ref{ScaledependentEoSDM}) for the galaxy UGC06923, versus the
radius.} \label{UGC06923dens}
\end{figure}

\begin{figure}[h!]
\centering
\includegraphics[width=35pc]{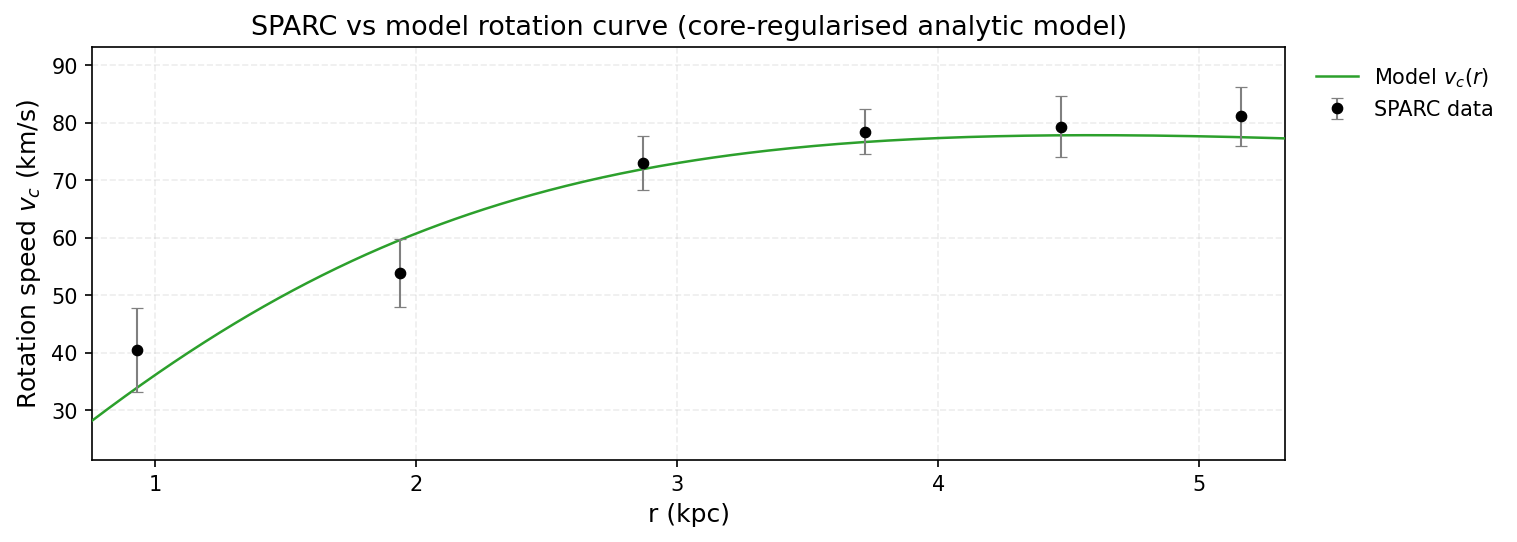}
\caption{The predicted rotation curves for the optimized SIDM
model of Eq. (\ref{ScaledependentEoSDM}), versus the SPARC
observational data for the galaxy UGC06923.} \label{UGC06923}
\end{figure}

In Table \ref{collUGC06923} we present the optimized values of
$K_0$ and $\rho_0$ for the analytic SIDM model of Eq.
(\ref{ScaledependentEoSDM}) for which the maximum compatibility
with the SPARC data is achieved.
\begin{table}[h!]
  \begin{center}
    \caption{SIDM Optimization Values for the galaxy UGC06923}
    \label{collUGC06923}
     \begin{tabular}{|r|r|}
     \hline
      \textbf{Parameter}   & \textbf{Optimization Values}
      \\  \hline
     $\rho_0 $  ($M_{\odot}/\mathrm{Kpc}^{3}$) & $8.29599\times 10^7$
\\  \hline $K_0$ ($M_{\odot} \,
\mathrm{Kpc}^{-3} \, (\mathrm{km/s})^{2}$)& 2622.3
\\  \hline
    \end{tabular}
  \end{center}
\end{table}
In Figs. \ref{UGC06923dens}, \ref{UGC06923}  we present the
density of the analytic SIDM model, the predicted rotation curves
for the SIDM model (\ref{ScaledependentEoSDM}), versus the SPARC
observational data and the sound speed, as a function of the
radius respectively. As it can be seen, for this galaxy, the SIDM
model produces viable rotation curves which are compatible with
the SPARC data.

\subsection{The Galaxy UGC06930, Marginally Viable, Extended Viable}

For this galaxy, the optimization method we used, ensures maximum
compatibility of the analytic SIDM model of Eq.
(\ref{ScaledependentEoSDM}) with the SPARC data, if we choose
$\rho_0=4.08457\times 10^7$$M_{\odot}/\mathrm{Kpc}^{3}$ and
$K_0=5165.97
$$M_{\odot} \, \mathrm{Kpc}^{-3} \, (\mathrm{km/s})^{2}$, in which
case the reduced $\chi^2_{red}$ value is $\chi^2_{red}=1.19127$.
Also the parameter $\alpha$ in this case is $\alpha=6.49013 $Kpc.

\begin{figure}[h!]
\centering
\includegraphics[width=20pc]{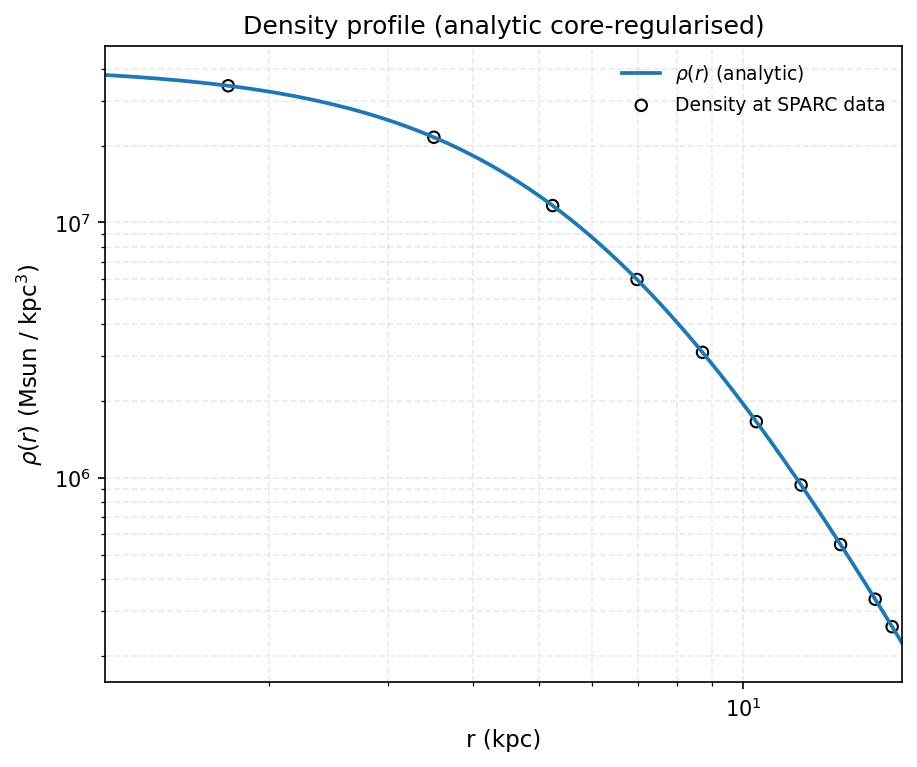}
\caption{The density of the SIDM model of Eq.
(\ref{ScaledependentEoSDM}) for the galaxy UGC06930, versus the
radius.} \label{UGC06930dens}
\end{figure}

\begin{figure}[h!]
\centering
\includegraphics[width=35pc]{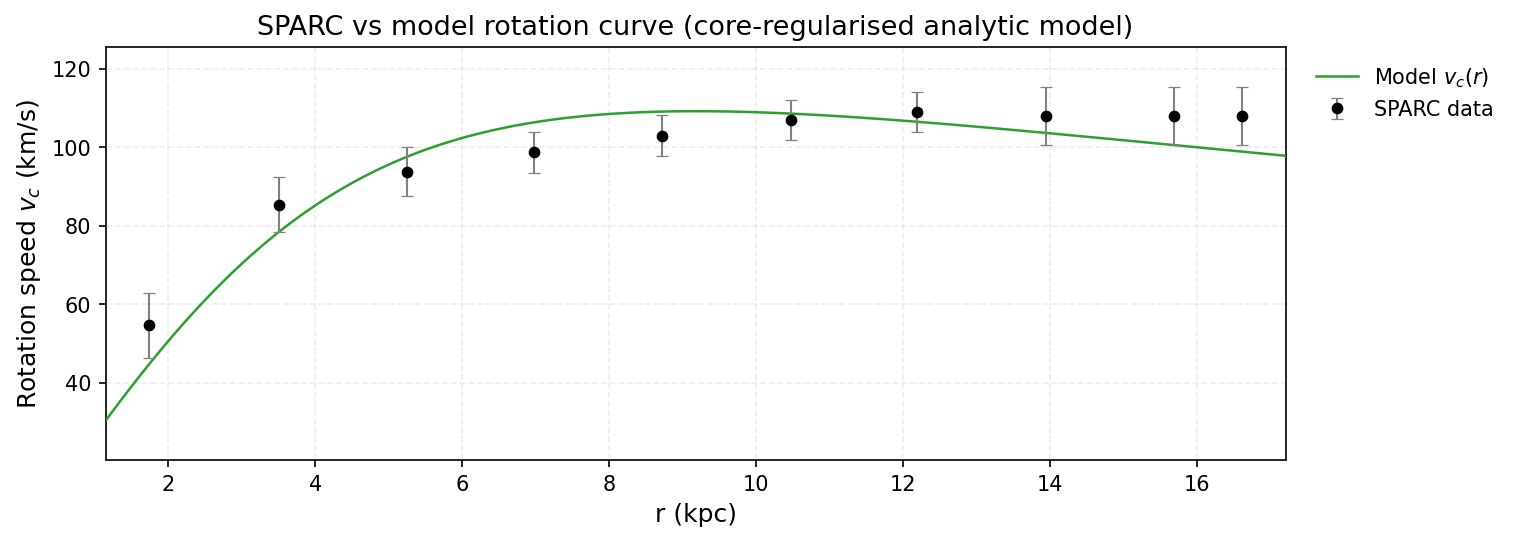}
\caption{The predicted rotation curves for the optimized SIDM
model of Eq. (\ref{ScaledependentEoSDM}), versus the SPARC
observational data for the galaxy UGC06930.} \label{UGC06930}
\end{figure}

In Table \ref{collUGC06930} we present the optimized values of
$K_0$ and $\rho_0$ for the analytic SIDM model of Eq.
(\ref{ScaledependentEoSDM}) for which the maximum compatibility
with the SPARC data is achieved.
\begin{table}[h!]
  \begin{center}
    \caption{SIDM Optimization Values for the galaxy UGC06930}
    \label{collUGC06930}
     \begin{tabular}{|r|r|}
     \hline
      \textbf{Parameter}   & \textbf{Optimization Values}
      \\  \hline
     $\rho_0 $  ($M_{\odot}/\mathrm{Kpc}^{3}$) & $4.08457\times 10^7$
\\  \hline $K_0$ ($M_{\odot} \,
\mathrm{Kpc}^{-3} \, (\mathrm{km/s})^{2}$)& 5165.97
\\  \hline
    \end{tabular}
  \end{center}
\end{table}
In Figs. \ref{UGC06930dens}, \ref{UGC06930}  we present the
density of the analytic SIDM model, the predicted rotation curves
for the SIDM model (\ref{ScaledependentEoSDM}), versus the SPARC
observational data and the sound speed, as a function of the
radius respectively. As it can be seen, for this galaxy, the SIDM
model produces marginally viable rotation curves which are
marginally compatible with the SPARC data.

\begin{figure}[h!]
\centering
\includegraphics[width=20pc]{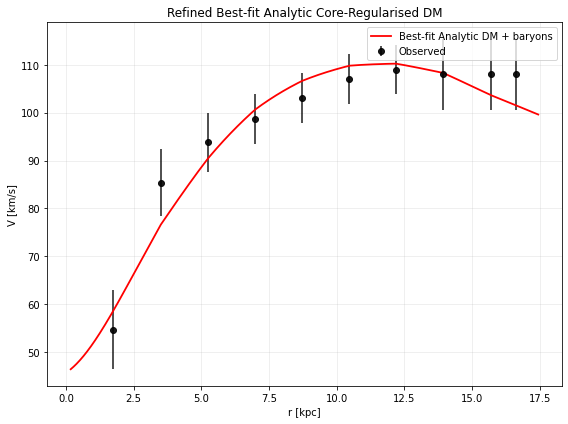}
\caption{The predicted rotation curves after using an optimization
for the SIDM model (\ref{ScaledependentEoSDM}), and the extended
SPARC data for the galaxy UGC06930. We included the rotation
curves of the gas, the disk velocities, the bulge (where present)
along with the SIDM model.} \label{extendedUGC06930}
\end{figure}

Now we shall include contributions to the rotation velocity from
the other components of the galaxy, namely the disk, the gas, and
the bulge if present. In Fig. \ref{extendedUGC06930} we present
the combined rotation curves including all the components of the
galaxy along with the SIDM. As it can be seen, the extended
collisional DM model is viable.
Also in Table \ref{evaluationextendedUGC06930} we present the
optimized values of the free parameters of the SIDM model for
which  we achieve the maximum compatibility with the SPARC data,
for the galaxy UGC06930, and also the resulting reduced
$\chi^2_{red}$ value.
\begin{table}[h!]
\centering \caption{Optimized Parameter Values of the Extended
SIDM model for the Galaxy UGC06930.}
\begin{tabular}{lc}
\hline
Parameter & Value  \\
\hline
$\rho_0 $ ($M_{\odot}/\mathrm{Kpc}^{3}$) & $1.28708\times 10^7$   \\
$K_0$ ($M_{\odot} \,
\mathrm{Kpc}^{-3} \, (\mathrm{km/s})^{2}$) & 2789.07   \\
$ml_{\text{disk}}$ & 1 \\
$ml_{\text{bulge}}$ & 0.1717 \\
$\alpha$ (Kpc) & 8.4942\\
$\chi^2_{red}$ & 0.695298 \\
\hline
\end{tabular}
\label{evaluationextendedUGC06930}
\end{table}

\subsection{The Galaxy UGC06983, Non-viable, Extended Viable}

For this galaxy, the optimization method we used, ensures maximum
compatibility of the analytic SIDM model of Eq.
(\ref{ScaledependentEoSDM}) with the SPARC data, if we choose
$\rho_0=5.54619\times 10^7$$M_{\odot}/\mathrm{Kpc}^{3}$ and
$K_0=5224.15
$$M_{\odot} \, \mathrm{Kpc}^{-3} \, (\mathrm{km/s})^{2}$, in which
case the reduced $\chi^2_{red}$ value is $\chi^2_{red}=1.92876$.
Also the parameter $\alpha$ in this case is $\alpha=5.60094 $Kpc.

\begin{figure}[h!]
\centering
\includegraphics[width=20pc]{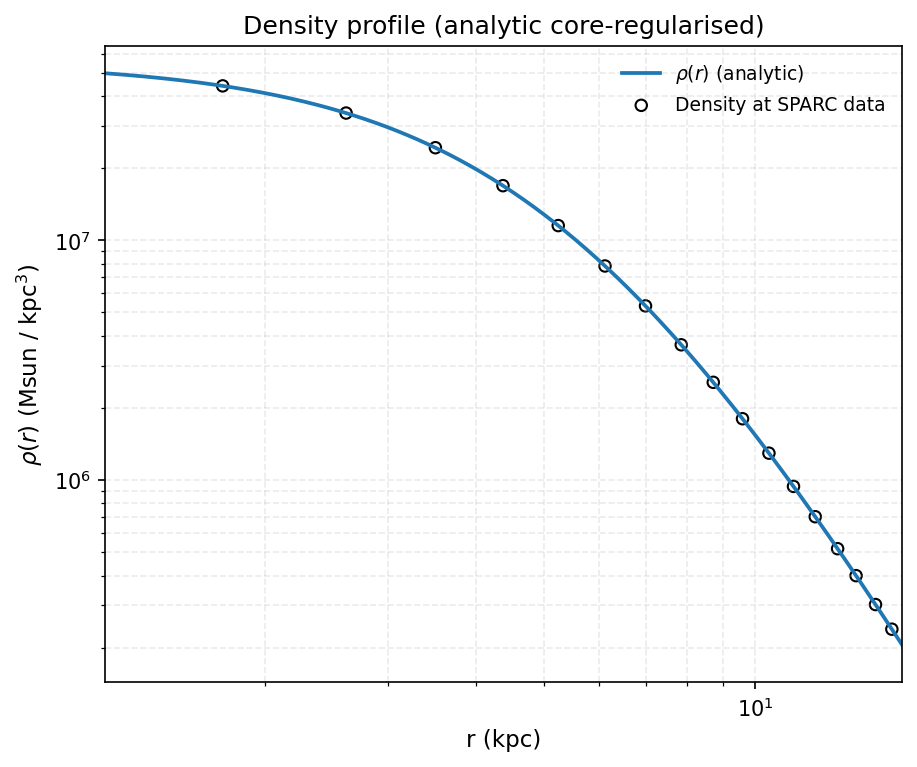}
\caption{The density of the SIDM model of Eq.
(\ref{ScaledependentEoSDM}) for the galaxy UGC06983, versus the
radius.} \label{UGC06983dens}
\end{figure}

\begin{figure}[h!]
\centering
\includegraphics[width=35pc]{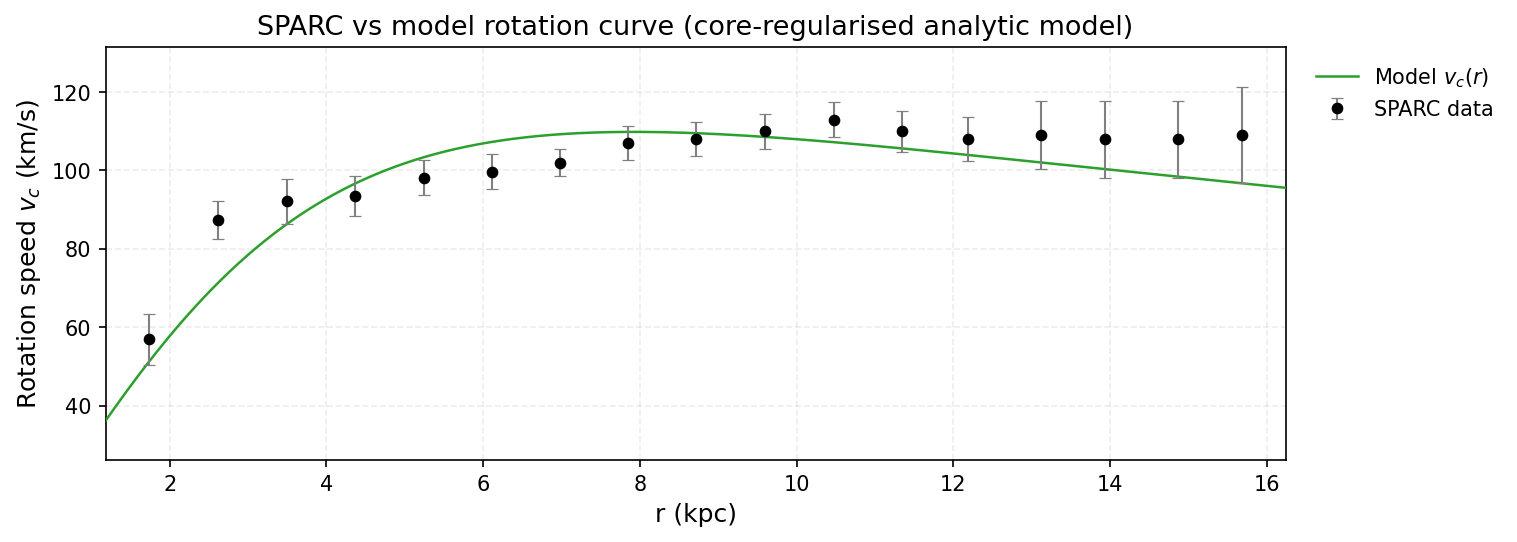}
\caption{The predicted rotation curves for the optimized SIDM
model of Eq. (\ref{ScaledependentEoSDM}), versus the SPARC
observational data for the galaxy UGC06983.} \label{UGC06983}
\end{figure}

In Table \ref{collUGC06983} we present the optimized values of
$K_0$ and $\rho_0$ for the analytic SIDM model of Eq.
(\ref{ScaledependentEoSDM}) for which the maximum compatibility
with the SPARC data is achieved.
\begin{table}[h!]
  \begin{center}
    \caption{SIDM Optimization Values for the galaxy UGC06983}
    \label{collUGC06983}
     \begin{tabular}{|r|r|}
     \hline
      \textbf{Parameter}   & \textbf{Optimization Values}
      \\  \hline
     $\rho_0 $  ($M_{\odot}/\mathrm{Kpc}^{3}$) & $5.54619\times 10^7$
\\  \hline $K_0$ ($M_{\odot} \,
\mathrm{Kpc}^{-3} \, (\mathrm{km/s})^{2}$)& 5224.15
\\  \hline
    \end{tabular}
  \end{center}
\end{table}
In Figs. \ref{UGC06983dens}, \ref{UGC06983}  we present the
density of the analytic SIDM model, the predicted rotation curves
for the SIDM model (\ref{ScaledependentEoSDM}), versus the SPARC
observational data and the sound speed, as a function of the
radius respectively. As it can be seen, for this galaxy, the SIDM
model produces non-viable rotation curves which are incompatible
with the SPARC data.

\begin{figure}[h!]
\centering
\includegraphics[width=20pc]{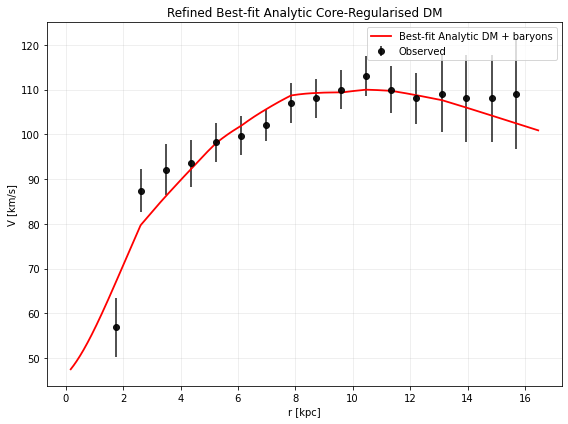}
\caption{The predicted rotation curves after using an optimization
for the SIDM model (\ref{ScaledependentEoSDM}), and the extended
SPARC data for the galaxy UGC06983. We included the rotation
curves of the gas, the disk velocities, the bulge (where present)
along with the SIDM model.} \label{extendedUGC06983}
\end{figure}

Now we shall include contributions to the rotation velocity from
the other components of the galaxy, namely the disk, the gas, and
the bulge if present. In Fig. \ref{extendedUGC06983} we present
the combined rotation curves including all the components of the
galaxy along with the SIDM. As it can be seen, the extended
collisional DM model is viable.
Also in Table \ref{evaluationextendedUGC06983} we present the
optimized values of the free parameters of the SIDM model for
which  we achieve the maximum compatibility with the SPARC data,
for the galaxy UGC06983, and also the resulting reduced
$\chi^2_{red}$ value.
\begin{table}[h!]
\centering \caption{Optimized Parameter Values of the Extended
SIDM model for the Galaxy UGC06983.}
\begin{tabular}{lc}
\hline
Parameter & Value  \\
\hline
$\rho_0 $ ($M_{\odot}/\mathrm{Kpc}^{3}$) & $2.32153\times 10^7$   \\
$K_0$ ($M_{\odot} \,
\mathrm{Kpc}^{-3} \, (\mathrm{km/s})^{2}$) & 3544.87   \\
$ml_{\text{disk}}$ & 1 \\
$ml_{\text{bulge}}$ & 0.0396 \\
$\alpha$ (Kpc) & 7.13032\\
$\chi^2_{red}$ & 0.653901 \\
\hline
\end{tabular}
\label{evaluationextendedUGC06983}
\end{table}

\subsection{The Galaxy UGC07089}

For this galaxy, the optimization method we used, ensures maximum
compatibility of the analytic SIDM model of Eq.
(\ref{ScaledependentEoSDM}) with the SPARC data, if we choose
$\rho_0=2.26195\times 10^7$$M_{\odot}/\mathrm{Kpc}^{3}$ and
$K_0=2262.44
$$M_{\odot} \, \mathrm{Kpc}^{-3} \, (\mathrm{km/s})^{2}$, in which
case the reduced $\chi^2_{red}$ value is $\chi^2_{red}=0.608198$.
Also the parameter $\alpha$ in this case is $\alpha=5.77161 $Kpc.

\begin{figure}[h!]
\centering
\includegraphics[width=20pc]{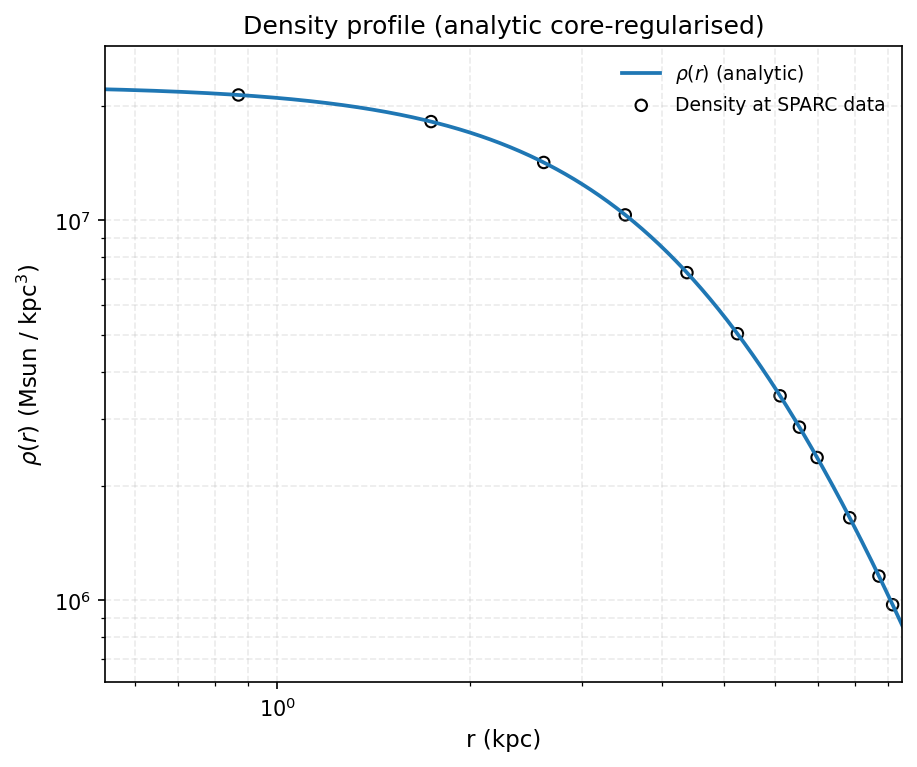}
\caption{The density of the SIDM model of Eq.
(\ref{ScaledependentEoSDM}) for the galaxy UGC07089, versus the
radius.} \label{UGC07089dens}
\end{figure}

\begin{figure}[h!]
\centering
\includegraphics[width=35pc]{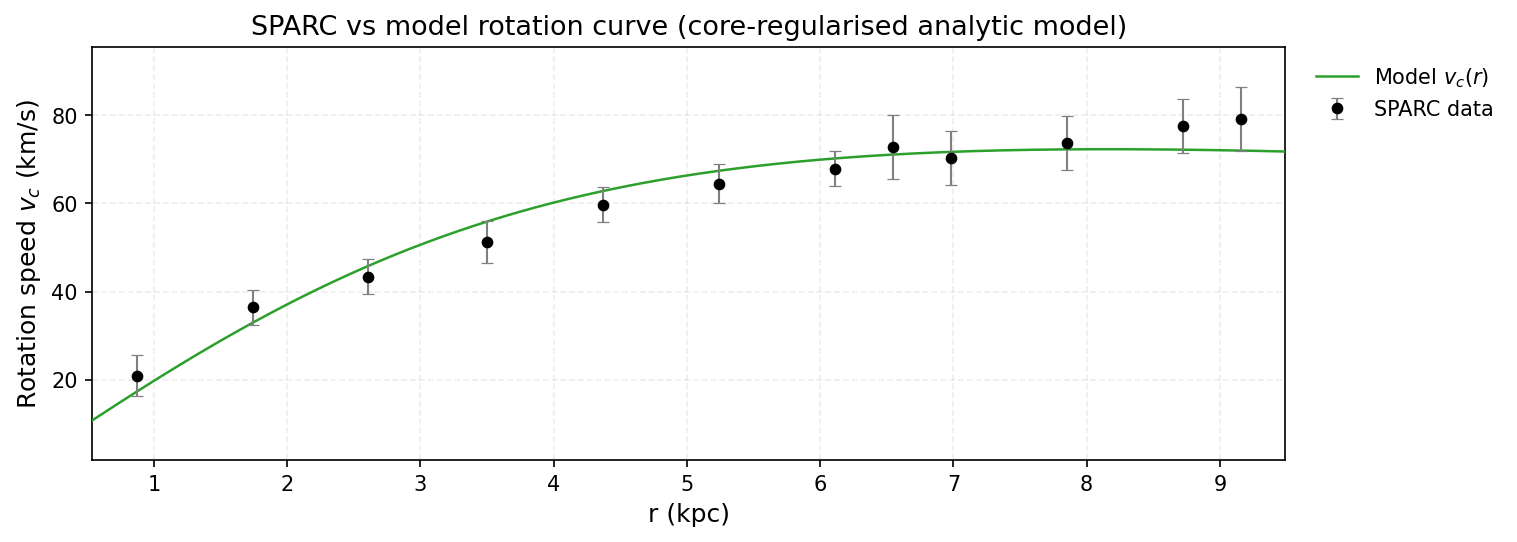}
\caption{The predicted rotation curves for the optimized SIDM
model of Eq. (\ref{ScaledependentEoSDM}), versus the SPARC
observational data for the galaxy UGC07089.} \label{UGC07089}
\end{figure}

In Table \ref{collUGC07089} we present the optimized values of
$K_0$ and $\rho_0$ for the analytic SIDM model of Eq.
(\ref{ScaledependentEoSDM}) for which the maximum compatibility
with the SPARC data is achieved.
\begin{table}[h!]
  \begin{center}
    \caption{SIDM Optimization Values for the galaxy UGC07089}
    \label{collUGC07089}
     \begin{tabular}{|r|r|}
     \hline
      \textbf{Parameter}   & \textbf{Optimization Values}
      \\  \hline
     $\rho_0 $  ($M_{\odot}/\mathrm{Kpc}^{3}$) & $2.26195\times 10^7$
\\  \hline $K_0$ ($M_{\odot} \,
\mathrm{Kpc}^{-3} \, (\mathrm{km/s})^{2}$)& 2262.44
\\  \hline
    \end{tabular}
  \end{center}
\end{table}
In Figs. \ref{UGC07089dens}, \ref{UGC07089}  we present the
density of the analytic SIDM model, the predicted rotation curves
for the SIDM model (\ref{ScaledependentEoSDM}), versus the SPARC
observational data and the sound speed, as a function of the
radius respectively. As it can be seen, for this galaxy, the SIDM
model produces viable rotation curves which are compatible with
the SPARC data.

\subsection{The Galaxy UGC07125, Non-viable, Extended Viable}

For this galaxy, the optimization method we used, ensures maximum
compatibility of the analytic SIDM model of Eq.
(\ref{ScaledependentEoSDM}) with the SPARC data, if we choose
$\rho_0=9.09695\times 10^6$$M_{\odot}/\mathrm{Kpc}^{3}$ and
$K_0=1825.39
$$M_{\odot} \, \mathrm{Kpc}^{-3} \, (\mathrm{km/s})^{2}$, in which
case the reduced $\chi^2_{red}$ value is $\chi^2_{red}=2.54759$.
Also the parameter $\alpha$ in this case is $\alpha=8.17486 $Kpc.

\begin{figure}[h!]
\centering
\includegraphics[width=20pc]{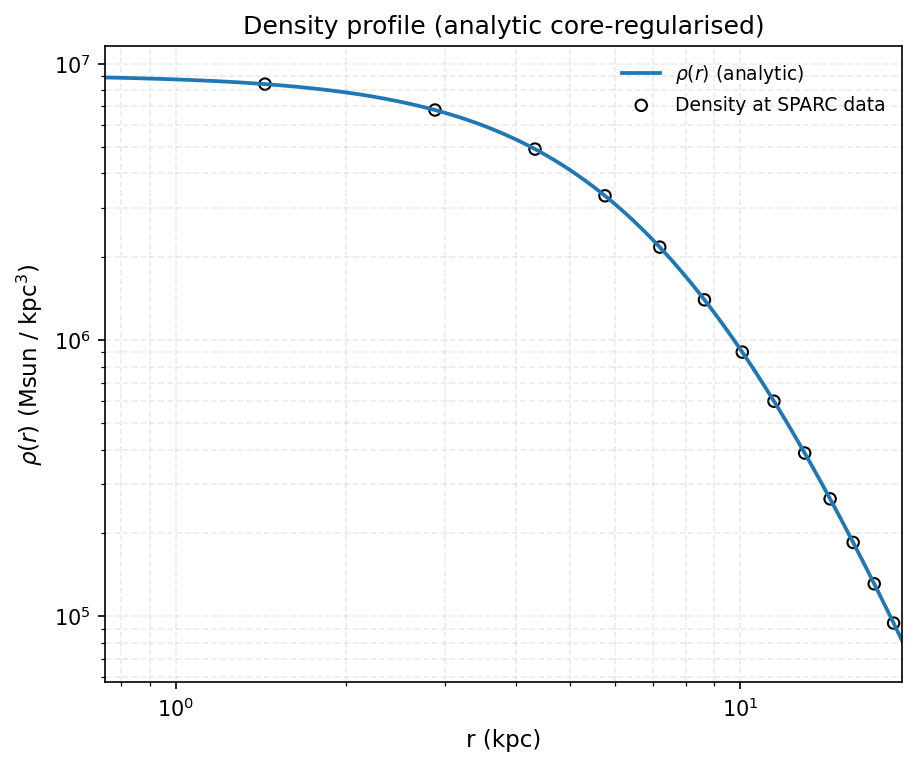}
\caption{The density of the SIDM model of Eq.
(\ref{ScaledependentEoSDM}) for the galaxy UGC07125, versus the
radius.} \label{UGC07125dens}
\end{figure}

\begin{figure}[h!]
\centering
\includegraphics[width=35pc]{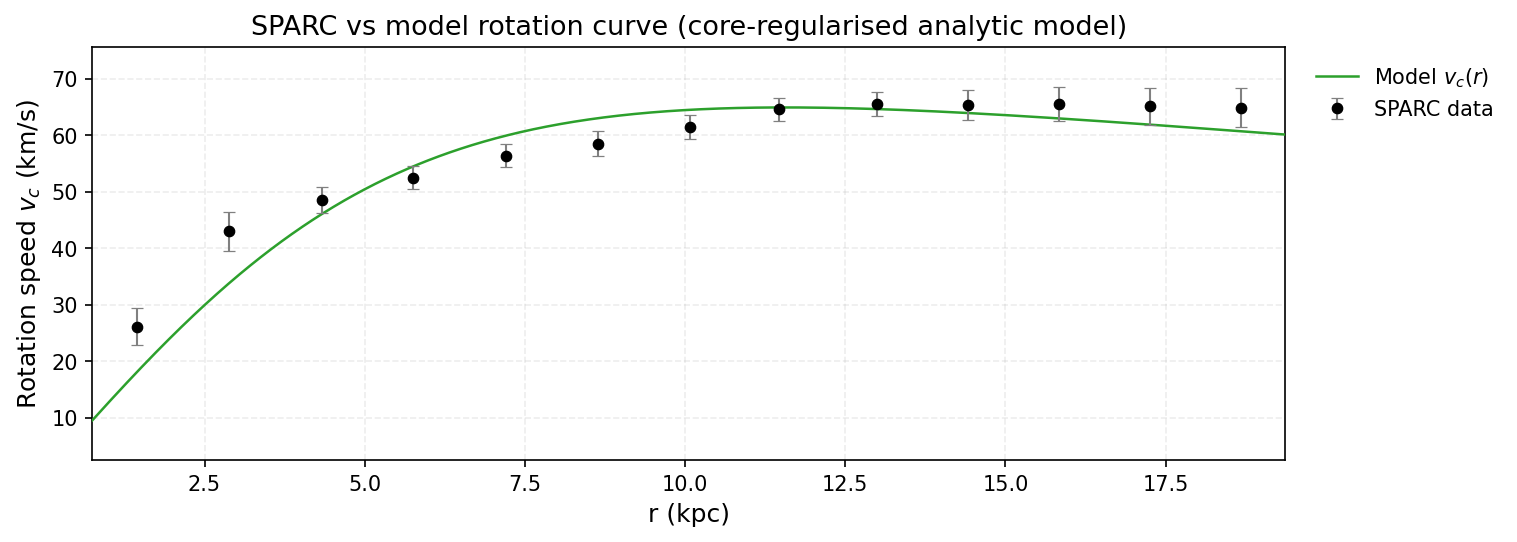}
\caption{The predicted rotation curves for the optimized SIDM
model of Eq. (\ref{ScaledependentEoSDM}), versus the SPARC
observational data for the galaxy UGC07125.} \label{UGC07125}
\end{figure}

In Table \ref{collUGC07125} we present the optimized values of
$K_0$ and $\rho_0$ for the analytic SIDM model of Eq.
(\ref{ScaledependentEoSDM}) for which the maximum compatibility
with the SPARC data is achieved.
\begin{table}[h!]
  \begin{center}
    \caption{SIDM Optimization Values for the galaxy UGC07125}
    \label{collUGC07125}
     \begin{tabular}{|r|r|}
     \hline
      \textbf{Parameter}   & \textbf{Optimization Values}
      \\  \hline
     $\rho_0 $  ($M_{\odot}/\mathrm{Kpc}^{3}$) & $9.09695\times 10^6$
\\  \hline $K_0$ ($M_{\odot} \,
\mathrm{Kpc}^{-3} \, (\mathrm{km/s})^{2}$)& 1825.39
\\  \hline
    \end{tabular}
  \end{center}
\end{table}
In Figs. \ref{UGC07125dens}, \ref{UGC07125}  we present the
density of the analytic SIDM model, the predicted rotation curves
for the SIDM model (\ref{ScaledependentEoSDM}), versus the SPARC
observational data and the sound speed, as a function of the
radius respectively. As it can be seen, for this galaxy, the SIDM
model produces non-viable rotation curves which are incompatible
with the SPARC data.

\begin{figure}[h!]
\centering
\includegraphics[width=20pc]{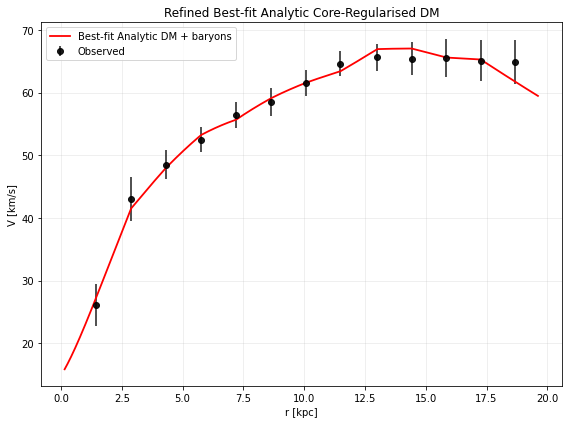}
\caption{The predicted rotation curves after using an optimization
for the SIDM model (\ref{ScaledependentEoSDM}), and the extended
SPARC data for the galaxy UGC07125. We included the rotation
curves of the gas, the disk velocities, the bulge (where present)
along with the SIDM model.} \label{extendedUGC07125}
\end{figure}

Now we shall include contributions to the rotation velocity from
the other components of the galaxy, namely the disk, the gas, and
the bulge if present. In Fig. \ref{extendedUGC07125} we present
the combined rotation curves including all the components of the
galaxy along with the SIDM. As it can be seen, the extended
collisional DM model is viable.
Also in Table \ref{evaluationextendedUGC07125} we present the
optimized values of the free parameters of the SIDM model for
which  we achieve the maximum compatibility with the SPARC data,
for the galaxy UGC07125, and also the resulting reduced
$\chi^2_{red}$ value.
\begin{table}[h!]
\centering \caption{Optimized Parameter Values of the Extended
SIDM model for the Galaxy UGC07125.}
\begin{tabular}{lc}
\hline
Parameter & Value  \\
\hline
$\rho_0 $ ($M_{\odot}/\mathrm{Kpc}^{3}$) & $4.01917\times 10^6$   \\
$K_0$ ($M_{\odot} \,
\mathrm{Kpc}^{-3} \, (\mathrm{km/s})^{2}$) & 1013.34   \\
$ml_{\text{disk}}$ & 0.8239 \\
$ml_{\text{bulge}}$ & 0.129 \\
$\alpha$ (Kpc) & 9.16234\\
$\chi^2_{red}$ & 0.296445 \\
\hline
\end{tabular}
\label{evaluationextendedUGC07125}
\end{table}

\subsection{The Galaxy UGC07151, Non-viable, Extended Viable}

For this galaxy, the optimization method we used, ensures maximum
compatibility of the analytic SIDM model of Eq.
(\ref{ScaledependentEoSDM}) with the SPARC data, if we choose
$\rho_0=1.05704\times 10^8$$M_{\odot}/\mathrm{Kpc}^{3}$ and
$K_0=2266.67
$$M_{\odot} \, \mathrm{Kpc}^{-3} \, (\mathrm{km/s})^{2}$, in which
case the reduced $\chi^2_{red}$ value is $\chi^2_{red}=2.42944$.
Also the parameter $\alpha$ in this case is $\alpha=2.67239 $Kpc.

\begin{figure}[h!]
\centering
\includegraphics[width=20pc]{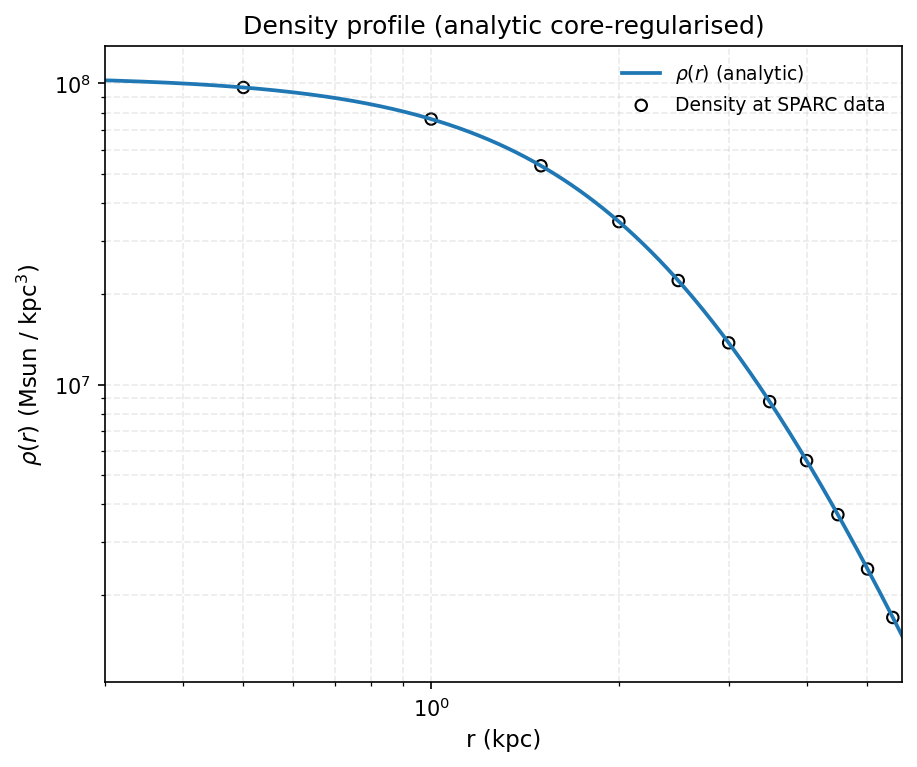}
\caption{The density of the SIDM model of Eq.
(\ref{ScaledependentEoSDM}) for the galaxy UGC07151, versus the
radius.} \label{UGC07151dens}
\end{figure}

\begin{figure}[h!]
\centering
\includegraphics[width=35pc]{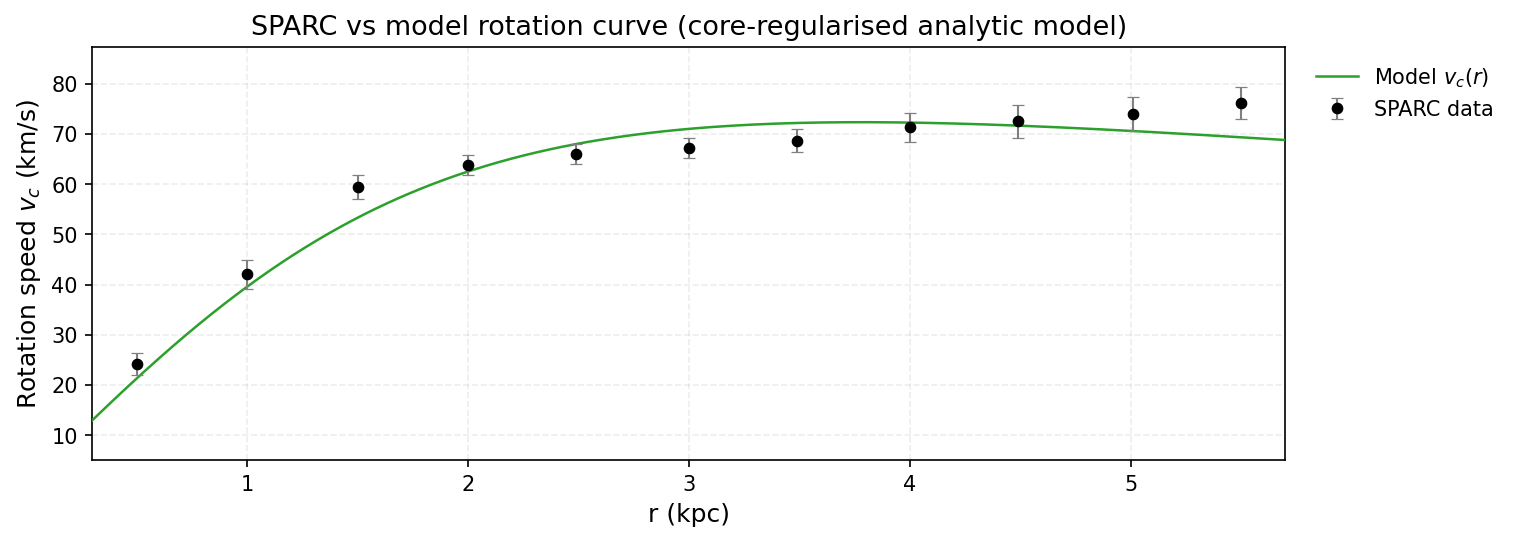}
\caption{The predicted rotation curves for the optimized SIDM
model of Eq. (\ref{ScaledependentEoSDM}), versus the SPARC
observational data for the galaxy UGC07151.} \label{UGC07151}
\end{figure}

In Table \ref{collUGC07151} we present the optimized values of
$K_0$ and $\rho_0$ for the analytic SIDM model of Eq.
(\ref{ScaledependentEoSDM}) for which the maximum compatibility
with the SPARC data is achieved.
\begin{table}[h!]
  \begin{center}
    \caption{SIDM Optimization Values for the galaxy UGC07151}
    \label{collUGC07151}
     \begin{tabular}{|r|r|}
     \hline
      \textbf{Parameter}   & \textbf{Optimization Values}
      \\  \hline
     $\rho_0 $  ($M_{\odot}/\mathrm{Kpc}^{3}$) & $1.05704\times 10^8$
\\  \hline $K_0$ ($M_{\odot} \,
\mathrm{Kpc}^{-3} \, (\mathrm{km/s})^{2}$)& 2266.67
\\  \hline
    \end{tabular}
  \end{center}
\end{table}
In Figs. \ref{UGC07151dens}, \ref{UGC07151}  we present the
density of the analytic SIDM model, the predicted rotation curves
for the SIDM model (\ref{ScaledependentEoSDM}), versus the SPARC
observational data and the sound speed, as a function of the
radius respectively. As it can be seen, for this galaxy, the SIDM
model produces non-viable rotation curves which are incompatible
with the SPARC data.

\begin{figure}[h!]
\centering
\includegraphics[width=20pc]{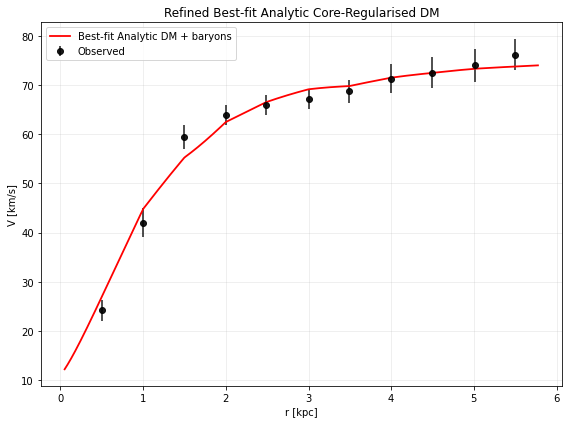}
\caption{The predicted rotation curves after using an optimization
for the SIDM model (\ref{ScaledependentEoSDM}), and the extended
SPARC data for the galaxy UGC07151. We included the rotation
curves of the gas, the disk velocities, the bulge (where present)
along with the SIDM model.} \label{extendedUGC07151}
\end{figure}

Now we shall include contributions to the rotation velocity from
the other components of the galaxy, namely the disk, the gas, and
the bulge if present. In Fig. \ref{extendedUGC07151} we present
the combined rotation curves including all the components of the
galaxy along with the SIDM. As it can be seen, the extended
collisional DM model is non-viable.
Also in Table \ref{evaluationextendedUGC07151} we present the
optimized values of the free parameters of the SIDM model for
which  we achieve the maximum compatibility with the SPARC data,
for the galaxy UGC07151, and also the resulting reduced
$\chi^2_{red}$ value.
\begin{table}[h!]
\centering \caption{Optimized Parameter Values of the Extended
SIDM model for the Galaxy UGC07151.}
\begin{tabular}{lc}
\hline
Parameter & Value  \\
\hline
$\rho_0 $ ($M_{\odot}/\mathrm{Kpc}^{3}$) & $3.46961\times 10^7$   \\
$K_0$ ($M_{\odot} \,
\mathrm{Kpc}^{-3} \, (\mathrm{km/s})^{2}$) & 1104.72  \\
$ml_{\text{disk}}$ & 0.9669 \\
$ml_{\text{bulge}}$ &0.3661\\
$\alpha$ (Kpc) & 3.25598\\
$\chi^2_{red}$ & 1.10662 \\
\hline
\end{tabular}
\label{evaluationextendedUGC07151}
\end{table}

\subsection{The Galaxy UGC07232}

For this galaxy, the optimization method we used, ensures maximum
compatibility of the analytic SIDM model of Eq.
(\ref{ScaledependentEoSDM}) with the SPARC data, if we choose
$\rho_0=3.17341\times 10^8$$M_{\odot}/\mathrm{Kpc}^{3}$ and
$K_0=862.3
$$M_{\odot} \, \mathrm{Kpc}^{-3} \, (\mathrm{km/s})^{2}$, in which
case the reduced $\chi^2_{red}$ value is $\chi^2_{red}=0.582015$.
Also the parameter $\alpha$ in this case is $\alpha=0.951298 $Kpc.

\begin{figure}[h!]
\centering
\includegraphics[width=20pc]{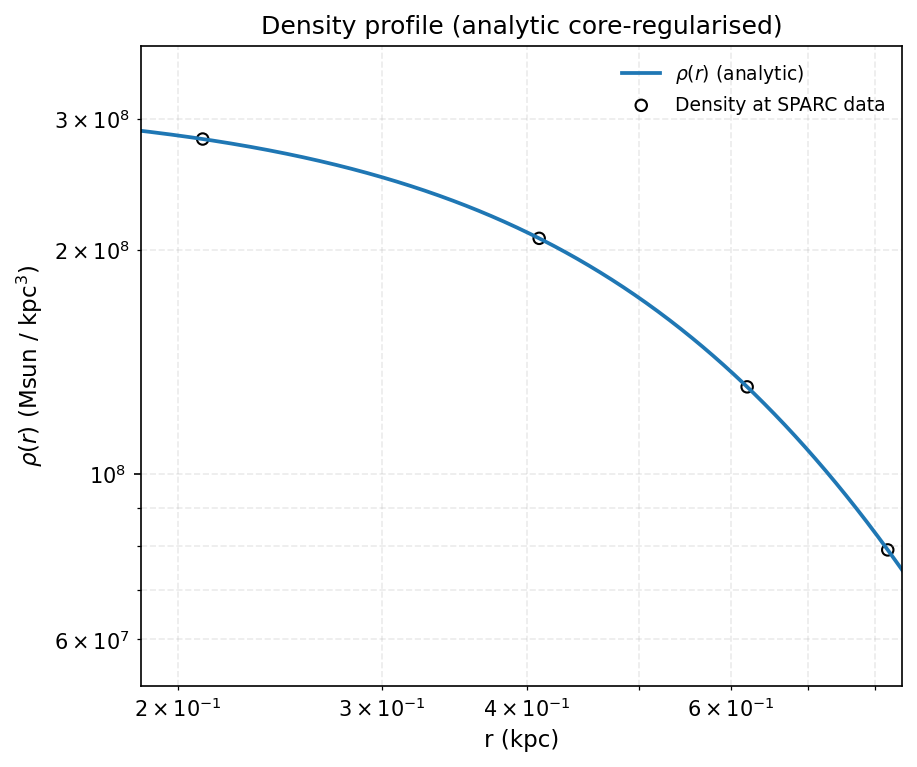}
\caption{The density of the SIDM model of Eq.
(\ref{ScaledependentEoSDM}) for the galaxy UGC07232, versus the
radius.} \label{UGC07232dens}
\end{figure}

\begin{figure}[h!]
\centering
\includegraphics[width=35pc]{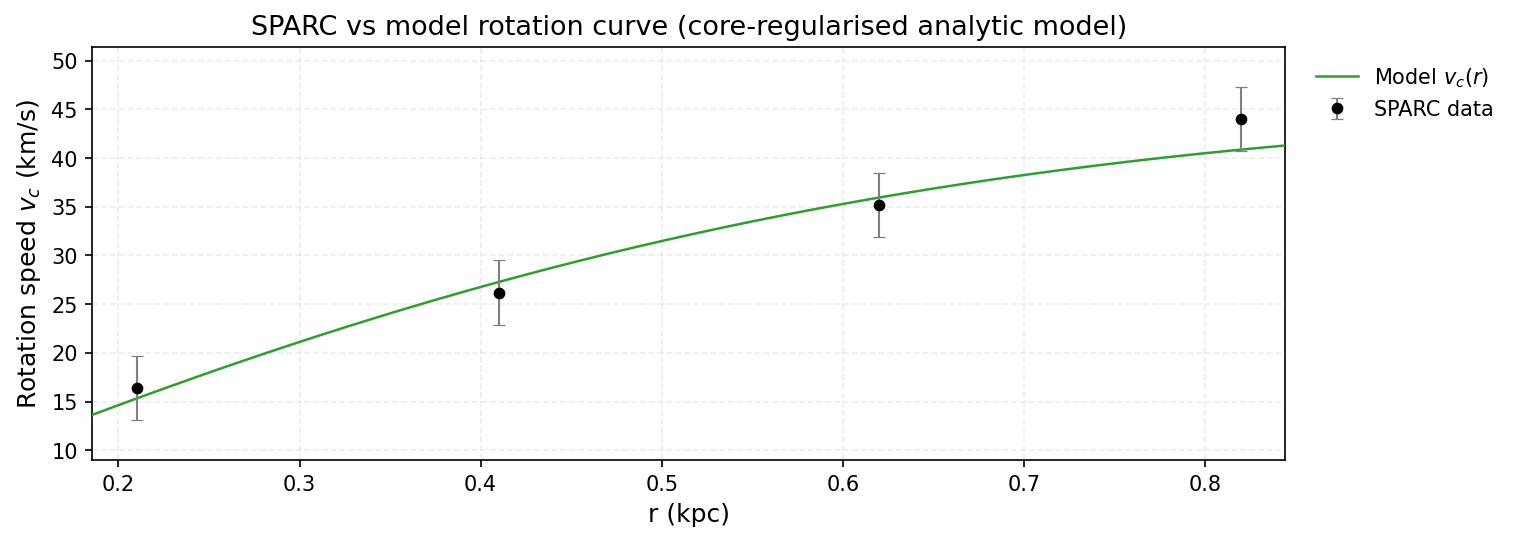}
\caption{The predicted rotation curves for the optimized SIDM
model of Eq. (\ref{ScaledependentEoSDM}), versus the SPARC
observational data for the galaxy UGC07232.} \label{UGC07232}
\end{figure}

In Table \ref{collUGC07232} we present the optimized values of
$K_0$ and $\rho_0$ for the analytic SIDM model of Eq.
(\ref{ScaledependentEoSDM}) for which the maximum compatibility
with the SPARC data is achieved.
\begin{table}[h!]
  \begin{center}
    \caption{SIDM Optimization Values for the galaxy UGC07232}
    \label{collUGC07232}
     \begin{tabular}{|r|r|}
     \hline
      \textbf{Parameter}   & \textbf{Optimization Values}
      \\  \hline
     $\rho_0 $  ($M_{\odot}/\mathrm{Kpc}^{3}$) & $3.17341\times 10^8$
\\  \hline $K_0$ ($M_{\odot} \,
\mathrm{Kpc}^{-3} \, (\mathrm{km/s})^{2}$)& 862.3
\\  \hline
    \end{tabular}
  \end{center}
\end{table}
In Figs. \ref{UGC07232dens}, \ref{UGC07232}  we present the
density of the analytic SIDM model, the predicted rotation curves
for the SIDM model (\ref{ScaledependentEoSDM}), versus the SPARC
observational data and the sound speed, as a function of the
radius respectively. As it can be seen, for this galaxy, the SIDM
model produces viable rotation curves which are compatible with
the SPARC data.

\subsection{The Galaxy UGC07261, Marginally Viable, Extended Viable}

For this galaxy, the optimization method we used, ensures maximum
compatibility of the analytic SIDM model of Eq.
(\ref{ScaledependentEoSDM}) with the SPARC data, if we choose
$\rho_0=1.03253\times 10^8$$M_{\odot}/\mathrm{Kpc}^{3}$ and
$K_0=2455.77
$$M_{\odot} \, \mathrm{Kpc}^{-3} \, (\mathrm{km/s})^{2}$, in which
case the reduced $\chi^2_{red}$ value is $\chi^2_{red}=1.2108$.
Also the parameter $\alpha$ in this case is $\alpha=2.81444 $Kpc.

\begin{figure}[h!]
\centering
\includegraphics[width=20pc]{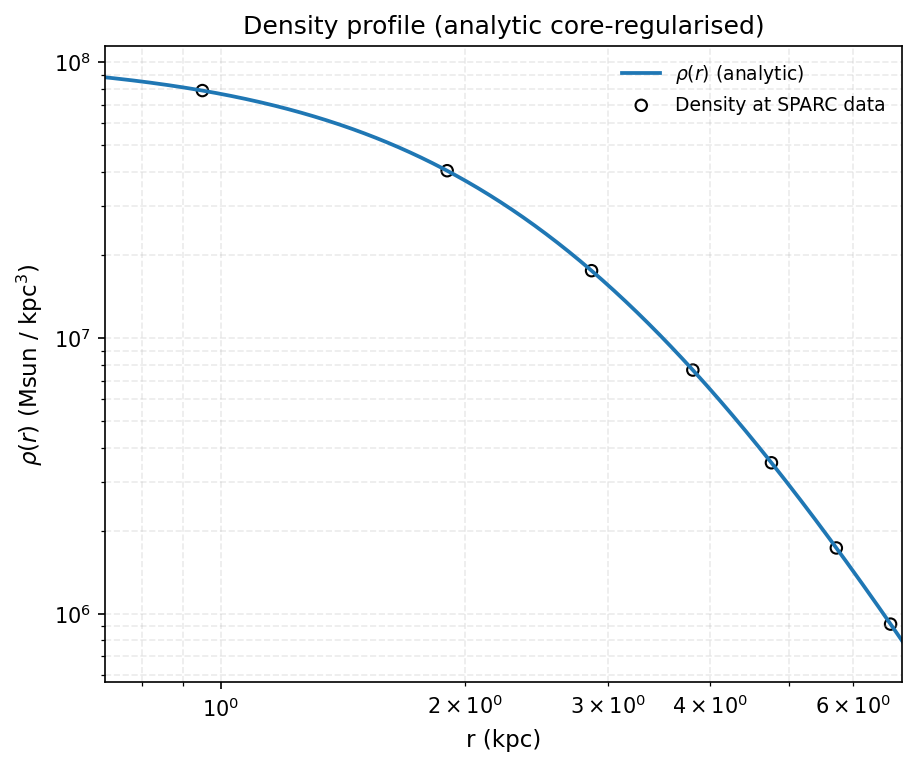}
\caption{The density of the SIDM model of Eq.
(\ref{ScaledependentEoSDM}) for the galaxy UGC07261, versus the
radius.} \label{UGC07261dens}
\end{figure}

\begin{figure}[h!]
\centering
\includegraphics[width=20pc]{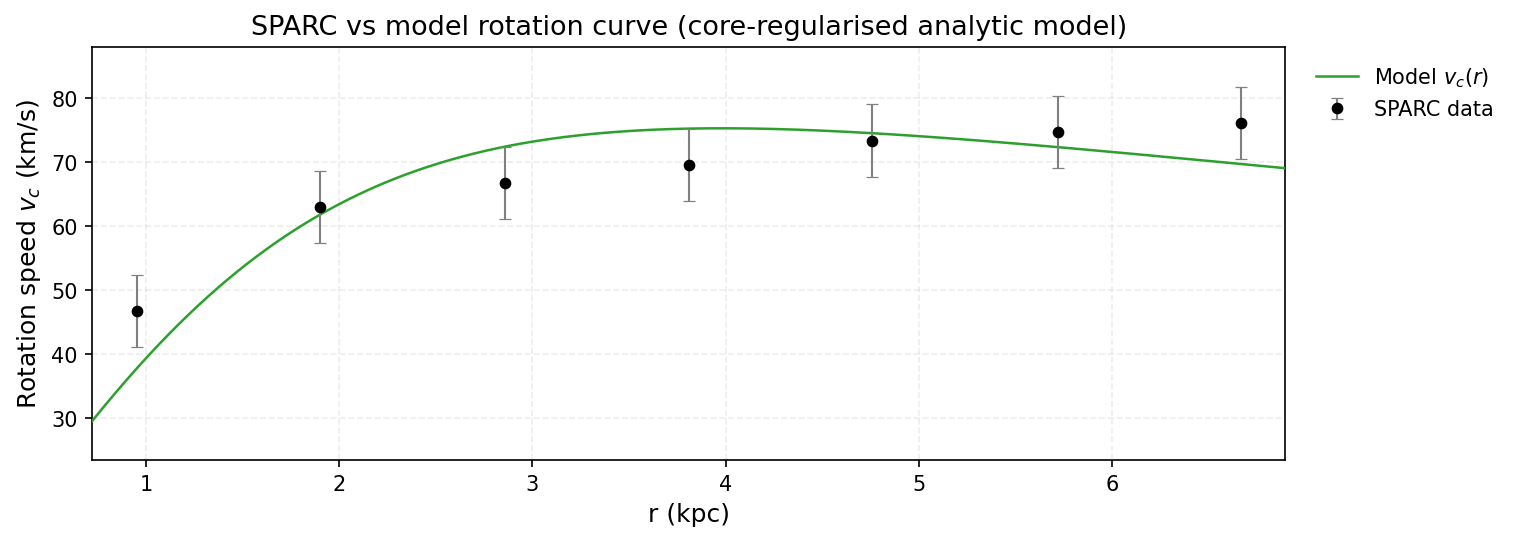}
\caption{The predicted rotation curves for the optimized SIDM
model of Eq. (\ref{ScaledependentEoSDM}), versus the SPARC
observational data for the galaxy UGC07261.} \label{UGC07261}
\end{figure}

In Table \ref{collUGC07261} we present the optimized values of
$K_0$ and $\rho_0$ for the analytic SIDM model of Eq.
(\ref{ScaledependentEoSDM}) for which the maximum compatibility
with the SPARC data is achieved.
\begin{table}[h!]
  \begin{center}
    \caption{SIDM Optimization Values for the galaxy UGC07261}
    \label{collUGC07261}
     \begin{tabular}{|r|r|}
     \hline
      \textbf{Parameter}   & \textbf{Optimization Values}
      \\  \hline
     $\rho_0 $  ($M_{\odot}/\mathrm{Kpc}^{3}$) & $1.03253\times 10^8$
\\  \hline $K_0$ ($M_{\odot} \,
\mathrm{Kpc}^{-3} \, (\mathrm{km/s})^{2}$)& 2455.77
\\  \hline
    \end{tabular}
  \end{center}
\end{table}
In Figs. \ref{UGC07261dens}, \ref{UGC07261}  we present the
density of the analytic SIDM model, the predicted rotation curves
for the SIDM model (\ref{ScaledependentEoSDM}), versus the SPARC
observational data and the sound speed, as a function of the
radius respectively. As it can be seen, for this galaxy, the SIDM
model produces marginally viable rotation curves which are
marginally compatible with the SPARC data.

\begin{figure}[h!]
\centering
\includegraphics[width=20pc]{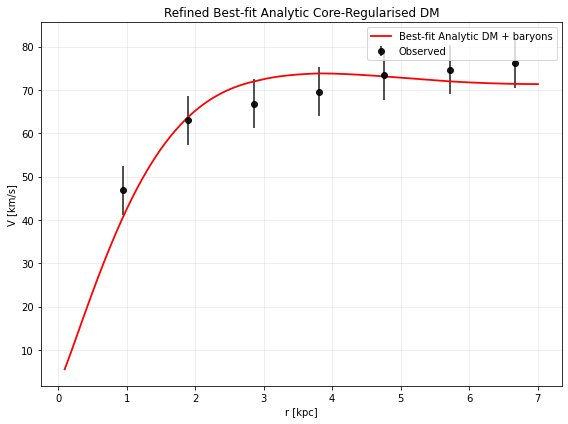}
\caption{The predicted rotation curves after using an optimization
for the SIDM model (\ref{ScaledependentEoSDM}), and the extended
SPARC data for the galaxy UGC07261. We included the rotation
curves of the gas, the disk velocities, the bulge (where present)
along with the SIDM model.} \label{extendedUGC07261}
\end{figure}

Now we shall include contributions to the rotation velocity from
the other components of the galaxy, namely the disk, the gas, and
the bulge if present. In Fig. \ref{extendedUGC07261} we present
the combined rotation curves including all the components of the
galaxy along with the SIDM. As it can be seen, the extended
collisional DM model is marginally viable.
Also in Table \ref{evaluationextendedUGC07261} we present the
optimized values of the free parameters of the SIDM model for
which  we achieve the maximum compatibility with the SPARC data,
for the galaxy UGC07261, and also the resulting reduced
$\chi^2_{red}$ value.
\begin{table}[h!]
\centering \caption{Optimized Parameter Values of the Extended
SIDM model for the Galaxy UGC07261.}
\begin{tabular}{lc}
\hline
Parameter & Value  \\
\hline
$\rho_0 $ ($M_{\odot}/\mathrm{Kpc}^{3}$) & $1.24235\times 10^8$   \\
$K_0$ ($M_{\odot} \,
\mathrm{Kpc}^{-3} \, (\mathrm{km/s})^{2}$) & 2298.32   \\
$ml_{\text{disk}}$ & 1 \\
$ml_{\text{bulge}}$ & 0.2 \\
$\alpha$ (Kpc) & 2.48188\\
$\chi^2_{red}$ & 1.15754 \\
\hline
\end{tabular}
\label{evaluationextendedUGC07261}
\end{table}

\subsection{The Galaxy UGC07323, Marginally Viable, Extended Viable}

For this galaxy, the optimization method we used, ensures maximum
compatibility of the analytic SIDM model of Eq.
(\ref{ScaledependentEoSDM}) with the SPARC data, if we choose
$\rho_0=4.19782\times 10^7$$M_{\odot}/\mathrm{Kpc}^{3}$ and
$K_0=2944.37
$$M_{\odot} \, \mathrm{Kpc}^{-3} \, (\mathrm{km/s})^{2}$, in which
case the reduced $\chi^2_{red}$ value is $\chi^2_{red}=1.31901$.
Also the parameter $\alpha$ in this case is $\alpha=4.8332 $Kpc.

\begin{figure}[h!]
\centering
\includegraphics[width=20pc]{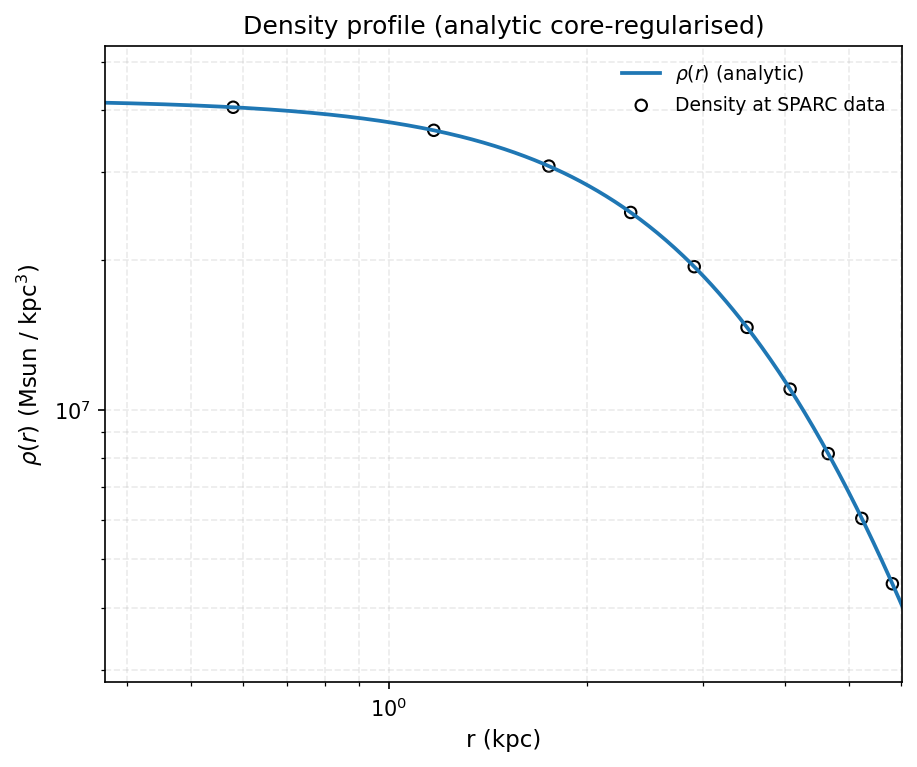}
\caption{The density of the SIDM model of Eq.
(\ref{ScaledependentEoSDM}) for the galaxy UGC07323, versus the
radius.} \label{UGC07323dens}
\end{figure}

\begin{figure}[h!]
\centering
\includegraphics[width=35pc]{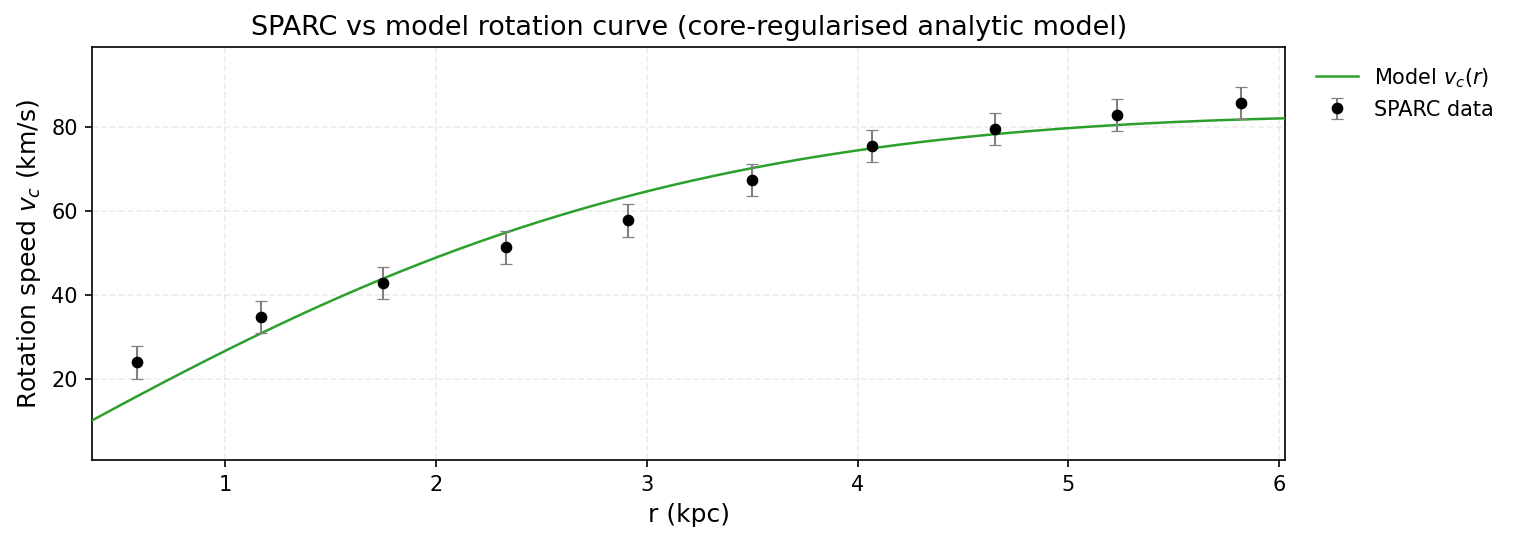}
\caption{The predicted rotation curves for the optimized SIDM
model of Eq. (\ref{ScaledependentEoSDM}), versus the SPARC
observational data for the galaxy UGC07323.} \label{UGC07323}
\end{figure}

In Table \ref{collUGC07323} we present the optimized values of
$K_0$ and $\rho_0$ for the analytic SIDM model of Eq.
(\ref{ScaledependentEoSDM}) for which the maximum compatibility
with the SPARC data is achieved.
\begin{table}[h!]
  \begin{center}
    \caption{SIDM Optimization Values for the galaxy UGC07323}
    \label{collUGC07323}
     \begin{tabular}{|r|r|}
     \hline
      \textbf{Parameter}   & \textbf{Optimization Values}
      \\  \hline
     $\rho_0 $  ($M_{\odot}/\mathrm{Kpc}^{3}$) & $4.19782\times 10^7$
\\  \hline $K_0$ ($M_{\odot} \,
\mathrm{Kpc}^{-3} \, (\mathrm{km/s})^{2}$)& 2944.37
\\  \hline
    \end{tabular}
  \end{center}
\end{table}
In Figs. \ref{UGC07323dens}, \ref{UGC07323}  we present the
density of the analytic SIDM model, the predicted rotation curves
for the SIDM model (\ref{ScaledependentEoSDM}), versus the SPARC
observational data and the sound speed, as a function of the
radius respectively. As it can be seen, for this galaxy, the SIDM
model produces marginally viable rotation curves which are
marginally compatible with the SPARC data.

\begin{figure}[h!]
\centering
\includegraphics[width=20pc]{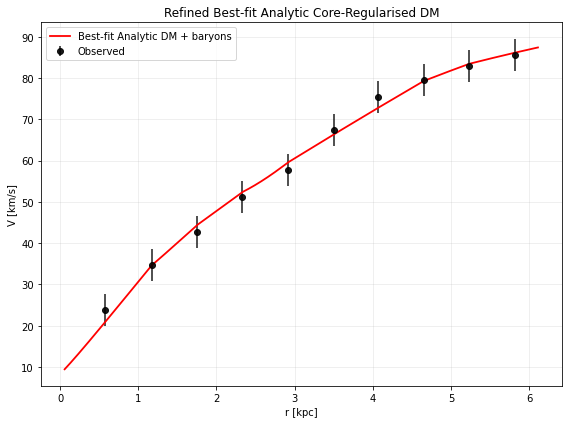}
\caption{The predicted rotation curves after using an optimization
for the SIDM model (\ref{ScaledependentEoSDM}), and the extended
SPARC data for the galaxy UGC07323. We included the rotation
curves of the gas, the disk velocities, the bulge (where present)
along with the SIDM model.} \label{extendedUGC07323}
\end{figure}

Now we shall include contributions to the rotation velocity from
the other components of the galaxy, namely the disk, the gas, and
the bulge if present. In Fig. \ref{extendedUGC07323} we present
the combined rotation curves including all the components of the
galaxy along with the SIDM. As it can be seen, the extended
collisional DM model is marginally viable.
Also in Table \ref{evaluationextendedUGC07323} we present the
optimized values of the free parameters of the SIDM model for
which  we achieve the maximum compatibility with the SPARC data,
for the galaxy UGC07323, and also the resulting reduced
$\chi^2_{red}$ value.
\begin{table}[h!]
\centering \caption{Optimized Parameter Values of the Extended
SIDM model for the Galaxy UGC07323.}
\begin{tabular}{lc}
\hline
Parameter & Value  \\
\hline
$\rho_0 $ ($M_{\odot}/\mathrm{Kpc}^{3}$) & $9.54047\times 10^6$   \\
$K_0$ ($M_{\odot} \,
\mathrm{Kpc}^{-3} \, (\mathrm{km/s})^{2}$) & 2693.68   \\
$ml_{\text{disk}}$ & 0.8887 \\
$ml_{\text{bulge}}$ & 0.4027 \\
$\alpha$ (Kpc) & 9.69581\\
$\chi^2_{red}$ & 0.263756 \\
\hline
\end{tabular}
\label{evaluationextendedUGC07323}
\end{table}

\subsection{The Galaxy UGC07399, Non-viable}

For this galaxy, the optimization method we used, ensures maximum
compatibility of the analytic SIDM model of Eq.
(\ref{ScaledependentEoSDM}) with the SPARC data, if we choose
$\rho_0=1.84111\times 10^8$$M_{\odot}/\mathrm{Kpc}^{3}$ and
$K_0=4391.78
$$M_{\odot} \, \mathrm{Kpc}^{-3} \, (\mathrm{km/s})^{2}$, in which
case the reduced $\chi^2_{red}$ value is $\chi^2_{red}=6.30798$.
Also the parameter $\alpha$ in this case is $\alpha=2.81859 $Kpc.

\begin{figure}[h!]
\centering
\includegraphics[width=20pc]{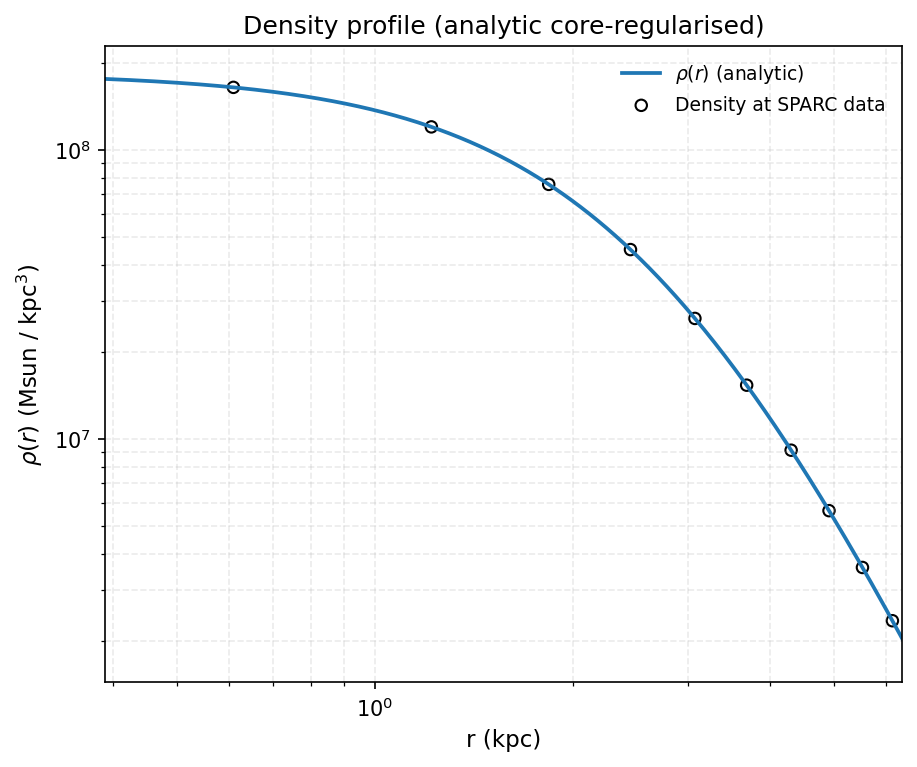}
\caption{The density of the SIDM model of Eq.
(\ref{ScaledependentEoSDM}) for the galaxy UGC07399, versus the
radius.} \label{UGC07399dens}
\end{figure}

\begin{figure}[h!]
\centering
\includegraphics[width=35pc]{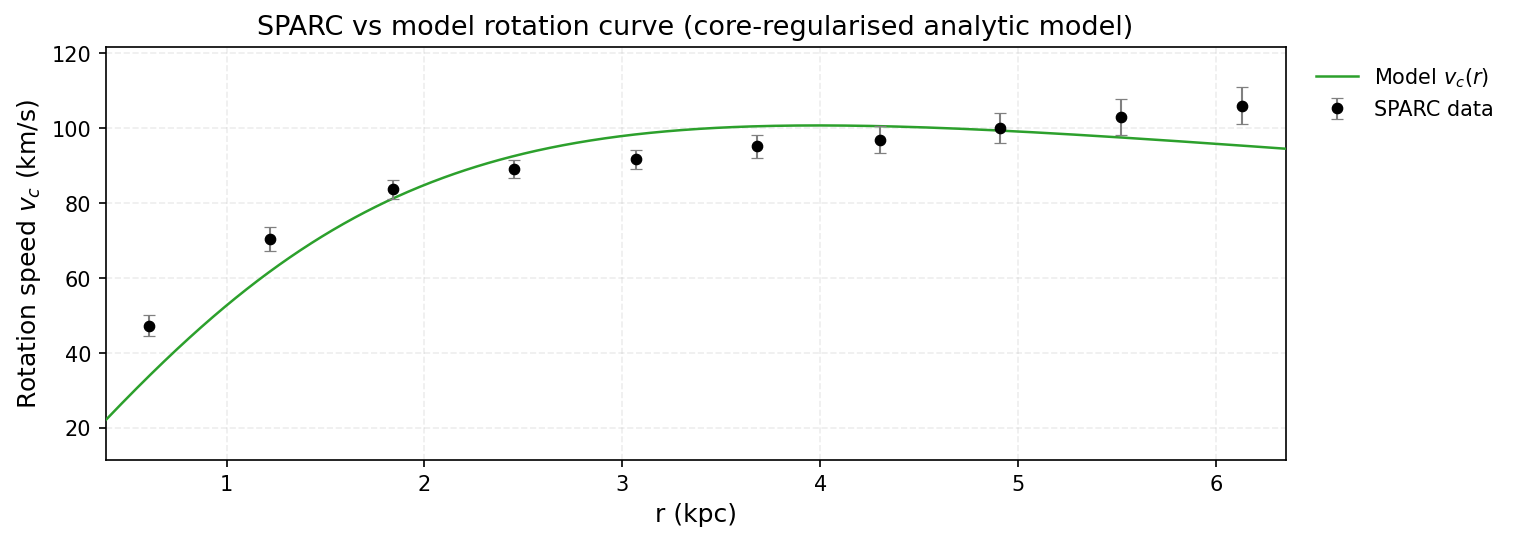}
\caption{The predicted rotation curves for the optimized SIDM
model of Eq. (\ref{ScaledependentEoSDM}), versus the SPARC
observational data for the galaxy UGC07399.} \label{UGC07399}
\end{figure}

In Table \ref{collUGC07399} we present the optimized values of
$K_0$ and $\rho_0$ for the analytic SIDM model of Eq.
(\ref{ScaledependentEoSDM}) for which the maximum compatibility
with the SPARC data is achieved.
\begin{table}[h!]
  \begin{center}
    \caption{SIDM Optimization Values for the galaxy UGC07399}
    \label{collUGC07399}
     \begin{tabular}{|r|r|}
     \hline
      \textbf{Parameter}   & \textbf{Optimization Values}
      \\  \hline
     $\rho_0 $  ($M_{\odot}/\mathrm{Kpc}^{3}$) & $1.84111\times 10^8$
\\  \hline $K_0$ ($M_{\odot} \,
\mathrm{Kpc}^{-3} \, (\mathrm{km/s})^{2}$)& 4391.78
\\  \hline
    \end{tabular}
  \end{center}
\end{table}
In Figs. \ref{UGC07399dens}, \ref{UGC07399} we present the density
of the analytic SIDM model, the predicted rotation curves for the
SIDM model (\ref{ScaledependentEoSDM}), versus the SPARC
observational data and the sound speed, as a function of the
radius respectively. As it can be seen, for this galaxy, the SIDM
model produces non-viable rotation curves which are incompatible
with the SPARC data.

\begin{figure}[h!]
\centering
\includegraphics[width=20pc]{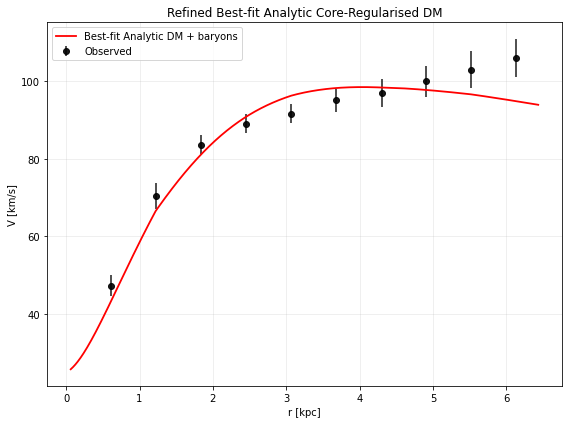}
\caption{The predicted rotation curves after using an optimization
for the SIDM model (\ref{ScaledependentEoSDM}), and the extended
SPARC data for the galaxy UGC07399. We included the rotation
curves of the gas, the disk velocities, the bulge (where present)
along with the SIDM model.} \label{extendedUGC07399}
\end{figure}

Now we shall include contributions to the rotation velocity from
the other components of the galaxy, namely the disk, the gas, and
the bulge if present. In Fig. \ref{extendedUGC07399} we present
the combined rotation curves including all the components of the
galaxy along with the SIDM. As it can be seen, the extended
collisional DM model is non-viable.
Also in Table \ref{evaluationextendedUGC07399} we present the
optimized values of the free parameters of the SIDM model for
which  we achieve the maximum compatibility with the SPARC data,
for the galaxy UGC07399, and also the resulting reduced
$\chi^2_{red}$ value.
\begin{table}[h!]
\centering \caption{Optimized Parameter Values of the Extended
SIDM model for the Galaxy UGC07399.}
\begin{tabular}{lc}
\hline
Parameter & Value  \\
\hline
$\rho_0 $ ($M_{\odot}/\mathrm{Kpc}^{3}$) & $1.34382\times 10^8$   \\
$K_0$ ($M_{\odot} \,
\mathrm{Kpc}^{-3} \, (\mathrm{km/s})^{2}$) & 3501.73   \\
$ml_{\text{disk}}$ & 1 \\
$ml_{\text{bulge}}$ & 0.9901 \\
$\alpha$ (Kpc) & 2.94555\\
$\chi^2_{red}$ & 2.80419 \\
\hline
\end{tabular}
\label{evaluationextendedUGC07399}
\end{table}

\subsection{The Galaxy UGC07524, Marginally Viable, Extended Viable}

For this galaxy, the optimization method we used, ensures maximum
compatibility of the analytic SIDM model of Eq.
(\ref{ScaledependentEoSDM}) with the SPARC data, if we choose
$\rho_0=2.48844\times 10^7$$M_{\odot}/\mathrm{Kpc}^{3}$ and
$K_0=2726.02
$$M_{\odot} \, \mathrm{Kpc}^{-3} \, (\mathrm{km/s})^{2}$, in which
case the reduced $\chi^2_{red}$ value is $\chi^2_{red}=1.23716$.
Also the parameter $\alpha$ in this case is $\alpha=6.0402 $Kpc.

\begin{figure}[h!]
\centering
\includegraphics[width=20pc]{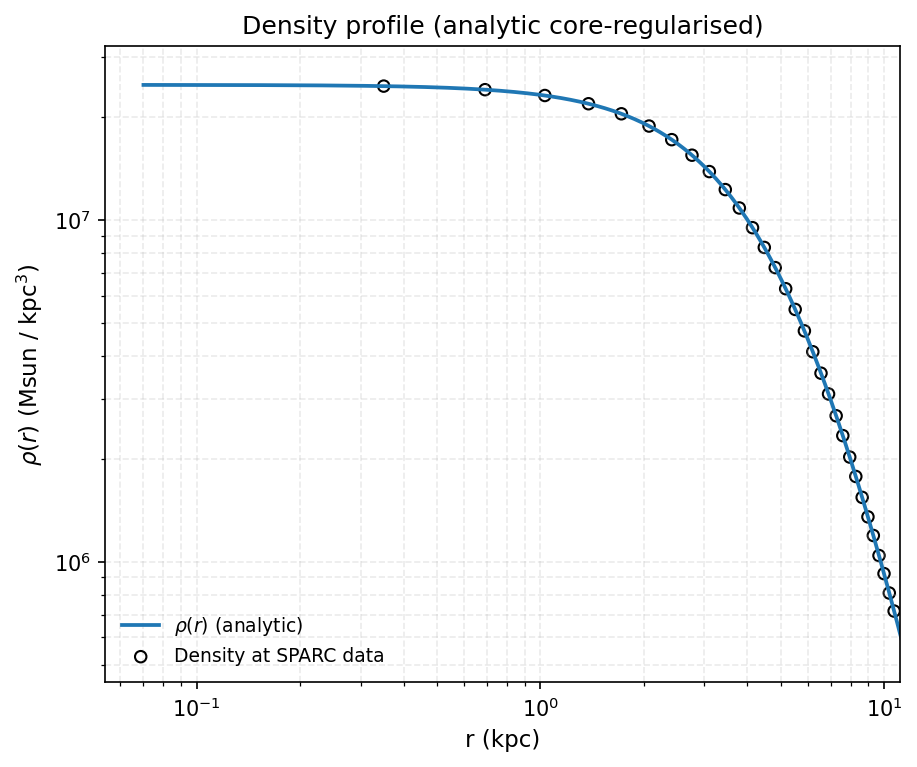}
\caption{The density of the SIDM model of Eq.
(\ref{ScaledependentEoSDM}) for the galaxy UGC07524, versus the
radius.} \label{UGC07524dens}
\end{figure}

\begin{figure}[h!]
\centering
\includegraphics[width=35pc]{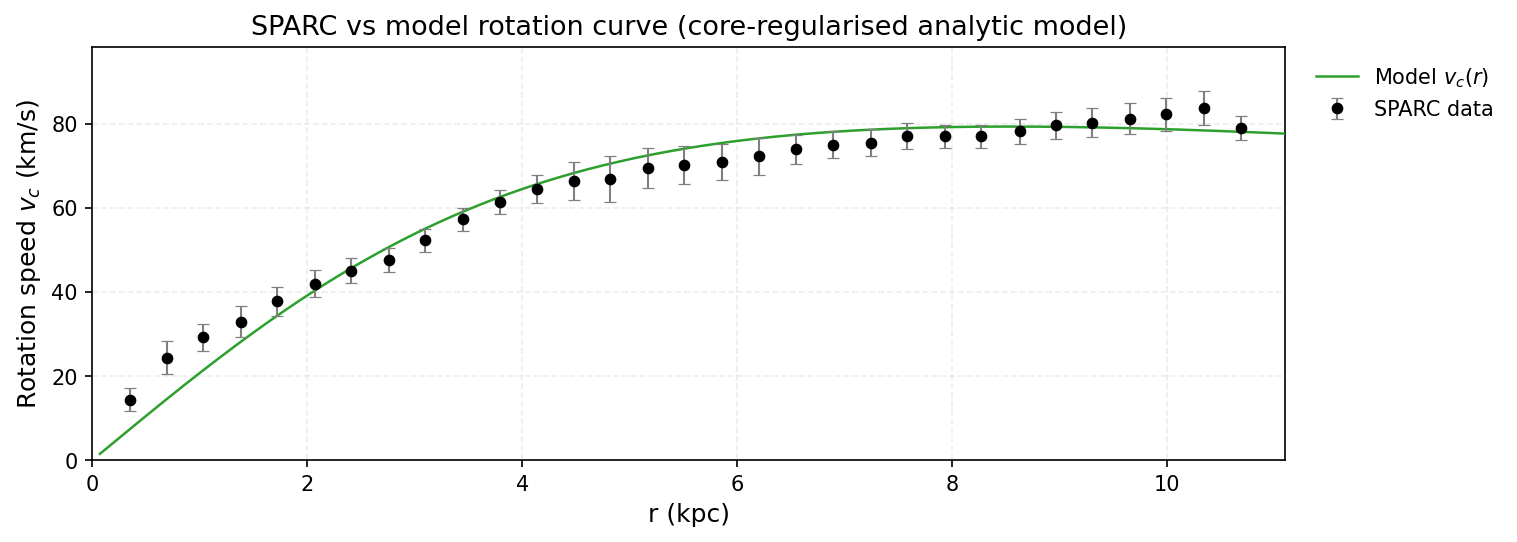}
\caption{The predicted rotation curves for the optimized SIDM
model of Eq. (\ref{ScaledependentEoSDM}), versus the SPARC
observational data for the galaxy UGC07524.} \label{UGC07524}
\end{figure}

In Table \ref{collUGC07524} we present the optimized values of
$K_0$ and $\rho_0$ for the analytic SIDM model of Eq.
(\ref{ScaledependentEoSDM}) for which the maximum compatibility
with the SPARC data is achieved.
\begin{table}[h!]
  \begin{center}
    \caption{SIDM Optimization Values for the galaxy UGC07524}
    \label{collUGC07524}
     \begin{tabular}{|r|r|}
     \hline
      \textbf{Parameter}   & \textbf{Optimization Values}
      \\  \hline
     $\rho_0 $  ($M_{\odot}/\mathrm{Kpc}^{3}$) & $2.48844\times 10^7$
\\  \hline $K_0$ ($M_{\odot} \,
\mathrm{Kpc}^{-3} \, (\mathrm{km/s})^{2}$)& 2726.02
\\  \hline
    \end{tabular}
  \end{center}
\end{table}
In Figs. \ref{UGC07524dens}, \ref{UGC07524}  we present the
density of the analytic SIDM model, the predicted rotation curves
for the SIDM model (\ref{ScaledependentEoSDM}), versus the SPARC
observational data and the sound speed, as a function of the
radius respectively. As it can be seen, for this galaxy, the SIDM
model produces marginally viable rotation curves which are
marginally compatible with the SPARC data.

\begin{figure}[h!]
\centering
\includegraphics[width=20pc]{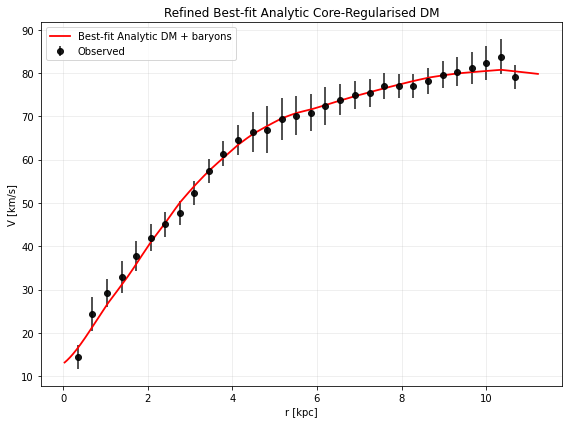}
\caption{The predicted rotation curves after using an optimization
for the SIDM model (\ref{ScaledependentEoSDM}), and the extended
SPARC data for the galaxy UGC07524. We included the rotation
curves of the gas, the disk velocities, the bulge (where present)
along with the SIDM model.} \label{extendedUGC07524}
\end{figure}

Now we shall include contributions to the rotation velocity from
the other components of the galaxy, namely the disk, the gas, and
the bulge if present. In Fig. \ref{extendedUGC07524} we present
the combined rotation curves including all the components of the
galaxy along with the SIDM. As it can be seen, the extended
collisional DM model is viable.
Also in Table \ref{evaluationextendedUGC07524} we present the
optimized values of the free parameters of the SIDM model for
which  we achieve the maximum compatibility with the SPARC data,
for the galaxy UGC07524, and also the resulting reduced
$\chi^2_{red}$ value.
\begin{table}[h!]
\centering \caption{Optimized Parameter Values of the Extended
SIDM model for the Galaxy UGC07524.}
\begin{tabular}{lc}
\hline
Parameter & Value  \\
\hline
$\rho_0 $ ($M_{\odot}/\mathrm{Kpc}^{3}$) & $1.57695\times 10^7$   \\
$K_0$ ($M_{\odot} \,
\mathrm{Kpc}^{-3} \, (\mathrm{km/s})^{2}$) & 1703.08  \\
$ml_{\text{disk}}$ & 1 \\
$ml_{\text{bulge}}$ & 0.1308 \\
$\alpha$ (Kpc) & 5.99659\\
$\chi^2_{red}$ & 0.201786 \\
\hline
\end{tabular}
\label{evaluationextendedUGC07524}
\end{table}

\subsection{The Galaxy UGC07559}

For this galaxy, the optimization method we used, ensures maximum
compatibility of the analytic SIDM model of Eq.
(\ref{ScaledependentEoSDM}) with the SPARC data, if we choose
$\rho_0=3.81394\times 10^7$$M_{\odot}/\mathrm{Kpc}^{3}$ and
$K_0=379.708
$$M_{\odot} \, \mathrm{Kpc}^{-3} \, (\mathrm{km/s})^{2}$, in which
case the reduced $\chi^2_{red}$ value is $\chi^2_{red}=0.265754$.
Also the parameter $\alpha$ in this case is $\alpha=1.82091 $Kpc.

\begin{figure}[h!]
\centering
\includegraphics[width=20pc]{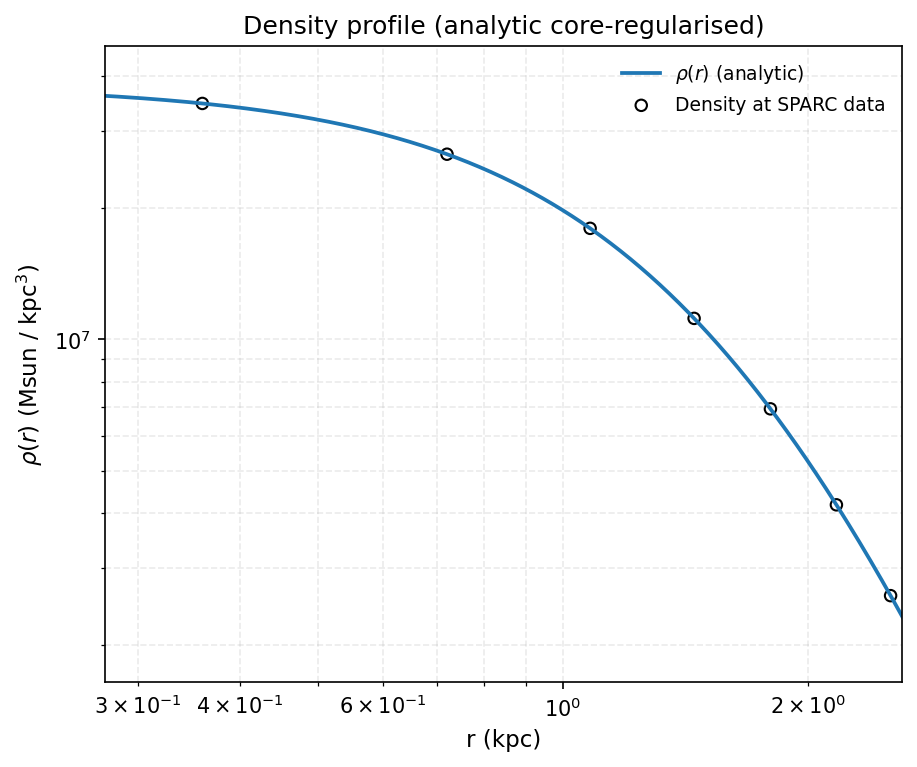}
\caption{The density of the SIDM model of Eq.
(\ref{ScaledependentEoSDM}) for the galaxy UGC07559, versus the
radius.} \label{UGC07559dens}
\end{figure}

\begin{figure}[h!]
\centering
\includegraphics[width=35pc]{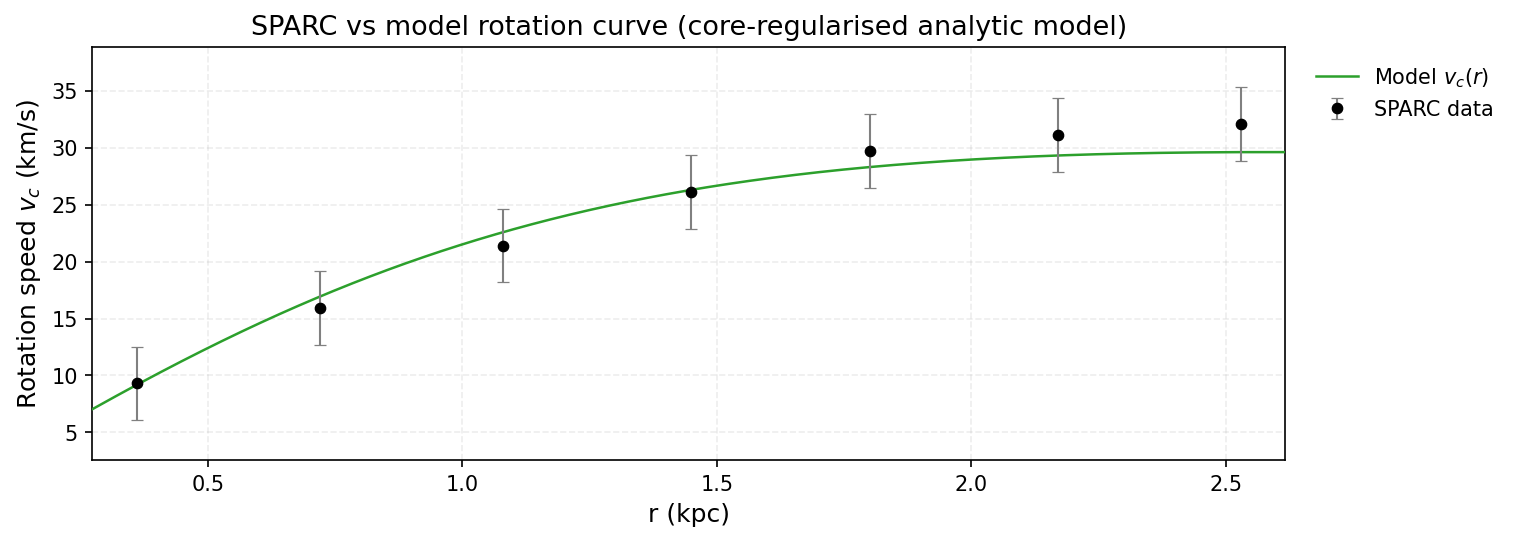}
\caption{The predicted rotation curves for the optimized SIDM
model of Eq. (\ref{ScaledependentEoSDM}), versus the SPARC
observational data for the galaxy UGC07559.} \label{UGC07559}
\end{figure}

In Table \ref{collUGC07559} we present the optimized values of
$K_0$ and $\rho_0$ for the analytic SIDM model of Eq.
(\ref{ScaledependentEoSDM}) for which the maximum compatibility
with the SPARC data is achieved.
\begin{table}[h!]
  \begin{center}
    \caption{SIDM Optimization Values for the galaxy UGC07559}
    \label{collUGC07559}
     \begin{tabular}{|r|r|}
     \hline
      \textbf{Parameter}   & \textbf{Optimization Values}
      \\  \hline
     $\rho_0 $  ($M_{\odot}/\mathrm{Kpc}^{3}$) & $3.81394\times 10^7$
\\  \hline $K_0$ ($M_{\odot} \,
\mathrm{Kpc}^{-3} \, (\mathrm{km/s})^{2}$)& 379.708
\\  \hline
    \end{tabular}
  \end{center}
\end{table}
In Figs. \ref{UGC07559dens}, \ref{UGC07559}  we present the
density of the analytic SIDM model, the predicted rotation curves
for the SIDM model (\ref{ScaledependentEoSDM}), versus the SPARC
observational data and the sound speed, as a function of the
radius respectively. As it can be seen, for this galaxy, the SIDM
model produces viable rotation curves which are compatible with
the SPARC data.

\subsection{The Galaxy UGC07577}

For this galaxy, the optimization method we used, ensures maximum
compatibility of the analytic SIDM model of Eq.
(\ref{ScaledependentEoSDM}) with the SPARC data, if we choose
$\rho_0=1.82893\times 10^7$$M_{\odot}/\mathrm{Kpc}^{3}$ and
$K_0=119.082
$$M_{\odot} \, \mathrm{Kpc}^{-3} \, (\mathrm{km/s})^{2}$, in which
case the reduced $\chi^2_{red}$ value is $\chi^2_{red}=0.38068$.
Also the parameter $\alpha$ in this case is $\alpha=1.47257 $Kpc.

\begin{figure}[h!]
\centering
\includegraphics[width=20pc]{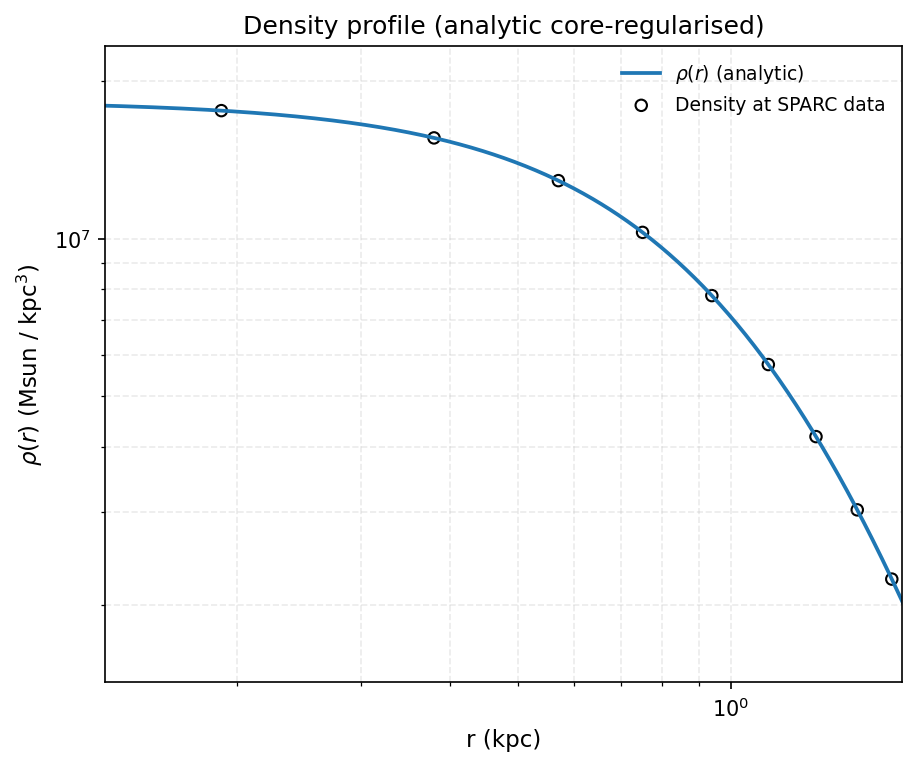}
\caption{The density of the SIDM model of Eq.
(\ref{ScaledependentEoSDM}) for the galaxy UGC07577, versus the
radius.} \label{UGC07577dens}
\end{figure}

\begin{figure}[h!]
\centering
\includegraphics[width=20pc]{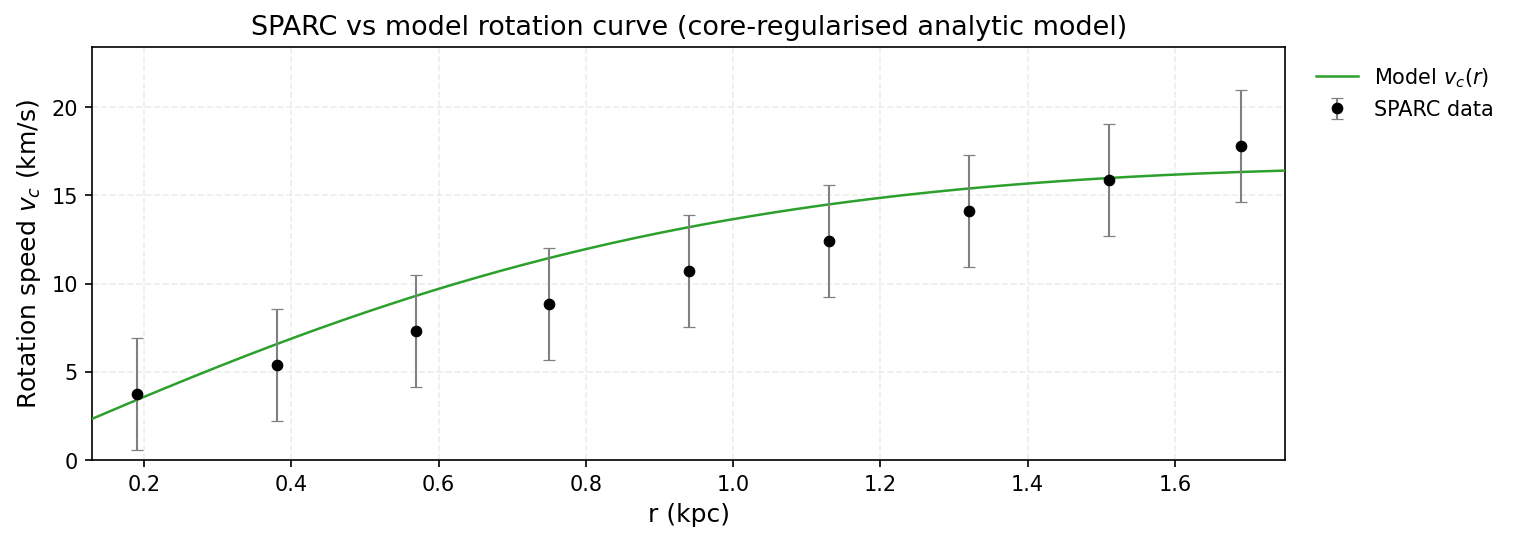}
\caption{The predicted rotation curves for the optimized SIDM
model of Eq. (\ref{ScaledependentEoSDM}), versus the SPARC
observational data for the galaxy UGC07577.} \label{UGC07577}
\end{figure}

In Table \ref{collUGC07577} we present the optimized values of
$K_0$ and $\rho_0$ for the analytic SIDM model of Eq.
(\ref{ScaledependentEoSDM}) for which the maximum compatibility
with the SPARC data is achieved.
\begin{table}[h!]
  \begin{center}
    \caption{SIDM Optimization Values for the galaxy UGC07577}
    \label{collUGC07577}
     \begin{tabular}{|r|r|}
     \hline
      \textbf{Parameter}   & \textbf{Optimization Values}
      \\  \hline
     $\rho_0 $  ($M_{\odot}/\mathrm{Kpc}^{3}$) & $1.82893\times 10^7$
\\  \hline $K_0$ ($M_{\odot} \,
\mathrm{Kpc}^{-3} \, (\mathrm{km/s})^{2}$)& 119.082
\\  \hline
    \end{tabular}
  \end{center}
\end{table}
In Figs. \ref{UGC07577dens}, \ref{UGC07577} we present the density
of the analytic SIDM model, the predicted rotation curves for the
SIDM model (\ref{ScaledependentEoSDM}), versus the SPARC
observational data and the sound speed, as a function of the
radius respectively. As it can be seen, for this galaxy, the SIDM
model produces viable rotation curves which are compatible with
the SPARC data.

\subsection{The Galaxy UGC07603}

For this galaxy, the optimization method we used, ensures maximum
compatibility of the analytic SIDM model of Eq.
(\ref{ScaledependentEoSDM}) with the SPARC data, if we choose
$\rho_0=1.39702\times 10^8$$M_{\odot}/\mathrm{Kpc}^{3}$ and
$K_0=1744.97
$$M_{\odot} \, \mathrm{Kpc}^{-3} \, (\mathrm{km/s})^{2}$, in which
case the reduced $\chi^2_{red}$ value is $\chi^2_{red}=0.537465$.
Also the parameter $\alpha$ in this case is $\alpha=2.03959 $Kpc.

\begin{figure}[h!]
\centering
\includegraphics[width=20pc]{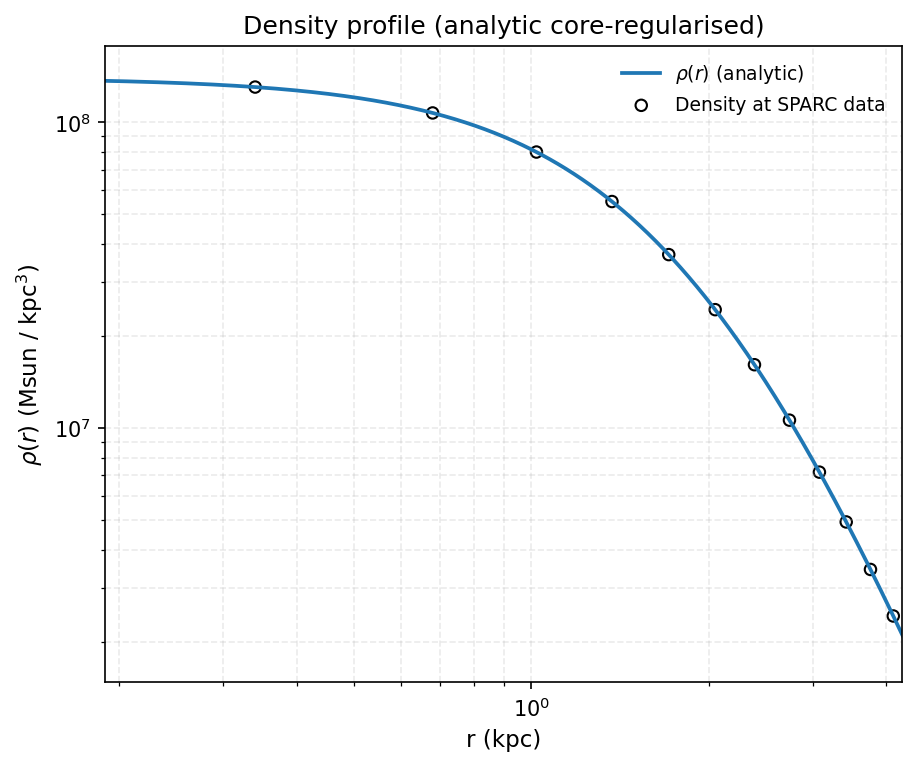}
\caption{The density of the SIDM model of Eq.
(\ref{ScaledependentEoSDM}) for the galaxy UGC07603, versus the
radius.} \label{UGC07603dens}
\end{figure}

\begin{figure}[h!]
\centering
\includegraphics[width=35pc]{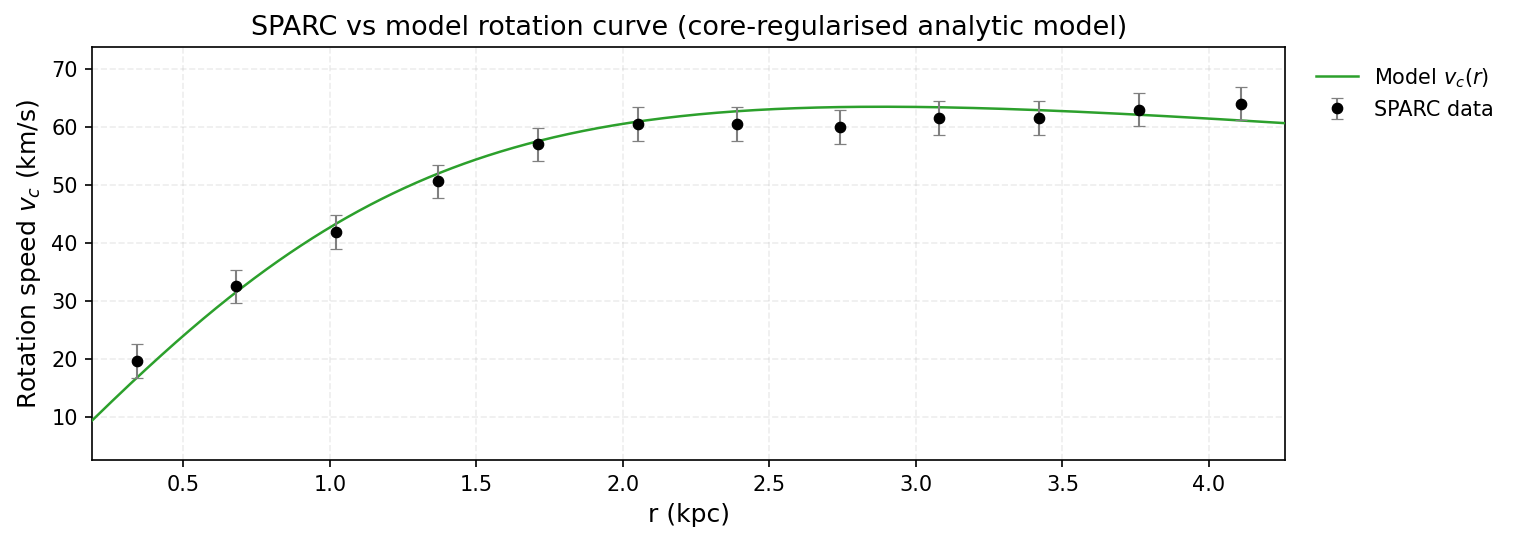}
\caption{The predicted rotation curves for the optimized SIDM
model of Eq. (\ref{ScaledependentEoSDM}), versus the SPARC
observational data for the galaxy UGC07603.} \label{UGC07603}
\end{figure}

In Table \ref{collUGC07603} we present the optimized values of
$K_0$ and $\rho_0$ for the analytic SIDM model of Eq.
(\ref{ScaledependentEoSDM}) for which the maximum compatibility
with the SPARC data is achieved.
\begin{table}[h!]
  \begin{center}
    \caption{SIDM Optimization Values for the galaxy UGC07603}
    \label{collUGC07603}
     \begin{tabular}{|r|r|}
     \hline
      \textbf{Parameter}   & \textbf{Optimization Values}
      \\  \hline
     $\rho_0 $  ($M_{\odot}/\mathrm{Kpc}^{3}$) & $1.39702\times 10^8$
\\  \hline $K_0$ ($M_{\odot} \,
\mathrm{Kpc}^{-3} \, (\mathrm{km/s})^{2}$)& 1744.97
\\  \hline
    \end{tabular}
  \end{center}
\end{table}
In Figs. \ref{UGC07603dens}, \ref{UGC07603} we present the density
of the analytic SIDM model, the predicted rotation curves for the
SIDM model (\ref{ScaledependentEoSDM}), versus the SPARC
observational data and the sound speed, as a function of the
radius respectively. As it can be seen, for this galaxy, the SIDM
model produces viable rotation curves which are compatible with
the SPARC data.

\begin{figure}[h!]
\centering
\includegraphics[width=20pc]{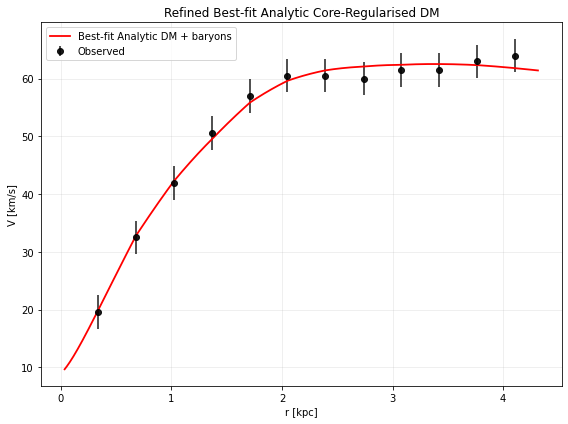}
\caption{The predicted rotation curves after using an optimization
for the SIDM model (\ref{ScaledependentEoSDM}), and the extended
SPARC data for the galaxy UGC07603. We included the rotation
curves of the gas, the disk velocities, the bulge (where present)
along with the SIDM model.} \label{extendedUGC07603}
\end{figure}

Now we shall include contributions to the rotation velocity from
the other components of the galaxy, namely the disk, the gas, and
the bulge if present. In Fig. \ref{extendedUGC07603} we present
the combined rotation curves including all the components of the
galaxy along with the SIDM. As it can be seen, the extended
collisional DM model is viable.
Also in Table \ref{evaluationextendedUGC07603} we present the
optimized values of the free parameters of the SIDM model for
which  we achieve the maximum compatibility with the SPARC data,
for the galaxy UGC07603, and also the resulting reduced
$\chi^2_{red}$ value.
\begin{table}[h!]
\centering \caption{Optimized Parameter Values of the Extended
SIDM model for the Galaxy UGC07603.}
\begin{tabular}{lc}
\hline
Parameter & Value  \\
\hline
$\rho_0 $ ($M_{\odot}/\mathrm{Kpc}^{3}$) & $8.03207\times 10^7$   \\
$K_0$ ($M_{\odot} \,
\mathrm{Kpc}^{-3} \, (\mathrm{km/s})^{2}$) & 1447.94  \\
$ml_{\text{disk}}$ & 0.7492 \\
$ml_{\text{bulge}}$ & 0.2680 \\
$\alpha$ (Kpc) & 2.44995\\
$\chi^2_{red}$ & 0.235451 \\
\hline
\end{tabular}
\label{evaluationextendedUGC07603}
\end{table}

\subsection{The Galaxy UGC07690}

For this galaxy, the optimization method we used, ensures maximum
compatibility of the analytic SIDM model of Eq.
(\ref{ScaledependentEoSDM}) with the SPARC data, if we choose
$\rho_0=4.34713\times 10^8$$M_{\odot}/\mathrm{Kpc}^{3}$ and
$K_0=1786.83
$$M_{\odot} \, \mathrm{Kpc}^{-3} \, (\mathrm{km/s})^{2}$, in which
case the reduced $\chi^2_{red}$ value is $\chi^2_{red}=0.48353$.
Also the parameter $\alpha$ in this case is $\alpha=1.17001 $Kpc.

\begin{figure}[h!]
\centering
\includegraphics[width=20pc]{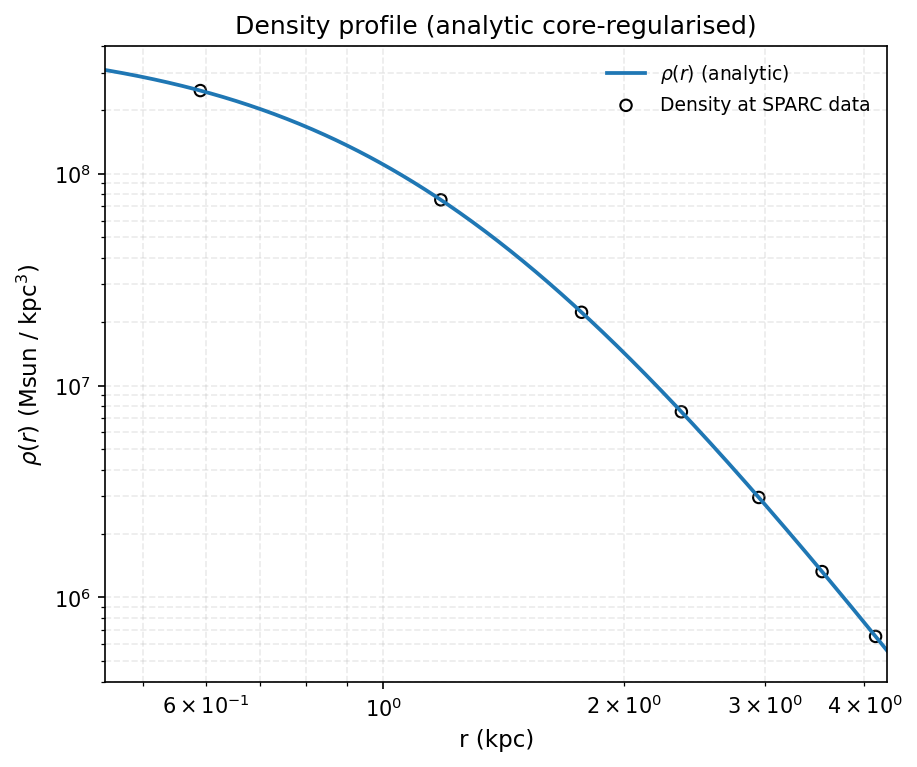}
\caption{The density of the SIDM model of Eq.
(\ref{ScaledependentEoSDM}) for the galaxy UGC07690, versus the
radius.} \label{UGC07690dens}
\end{figure}

\begin{figure}[h!]
\centering
\includegraphics[width=35pc]{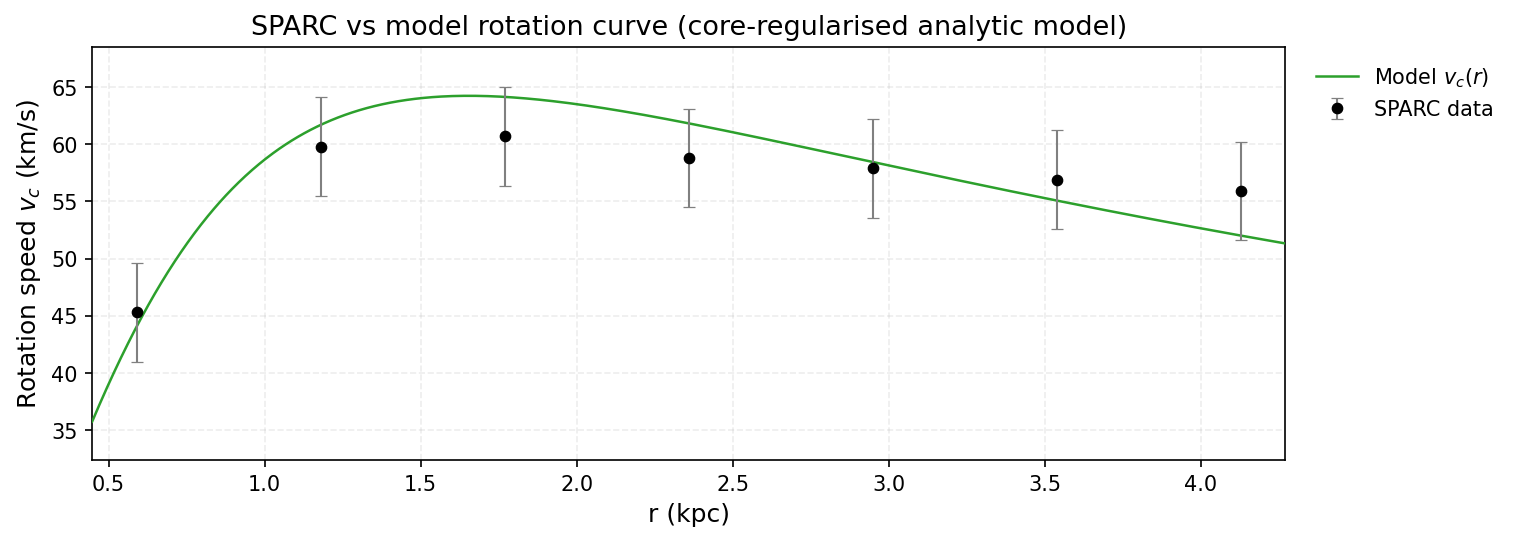}
\caption{The predicted rotation curves for the optimized SIDM
model of Eq. (\ref{ScaledependentEoSDM}), versus the SPARC
observational data for the galaxy UGC07690.} \label{UGC07690}
\end{figure}

In Table \ref{collUGC07690} we present the optimized values of
$K_0$ and $\rho_0$ for the analytic SIDM model of Eq.
(\ref{ScaledependentEoSDM}) for which the maximum compatibility
with the SPARC data is achieved.
\begin{table}[h!]
  \begin{center}
    \caption{SIDM Optimization Values for the galaxy UGC07690}
    \label{collUGC07690}
     \begin{tabular}{|r|r|}
     \hline
      \textbf{Parameter}   & \textbf{Optimization Values}
      \\  \hline
     $\rho_0 $  ($M_{\odot}/\mathrm{Kpc}^{3}$) & $4.34713\times 10^8$
\\  \hline $K_0$ ($M_{\odot} \,
\mathrm{Kpc}^{-3} \, (\mathrm{km/s})^{2}$)& 1786.83
\\  \hline
    \end{tabular}
  \end{center}
\end{table}
In Figs. \ref{UGC07690dens}, \ref{UGC07690} we present the density
of the analytic SIDM model, the predicted rotation curves for the
SIDM model (\ref{ScaledependentEoSDM}), versus the SPARC
observational data and the sound speed, as a function of the
radius respectively. As it can be seen, for this galaxy, the SIDM
model produces viable rotation curves which are compatible with
the SPARC data.

\subsection{The Galaxy UGC07866}

For this galaxy, the optimization method we used, ensures maximum
compatibility of the analytic SIDM model of Eq.
(\ref{ScaledependentEoSDM}) with the SPARC data, if we choose
$\rho_0=4.19666\times 10^7$$M_{\odot}/\mathrm{Kpc}^{3}$ and
$K_0=522.148
$$M_{\odot} \, \mathrm{Kpc}^{-3} \, (\mathrm{km/s})^{2}$, in which
case the reduced $\chi^2_{red}$ value is $\chi^2_{red}=0.463844$.
Also the parameter $\alpha$ in this case is $\alpha=2.03562 $Kpc.

\begin{figure}[h!]
\centering
\includegraphics[width=20pc]{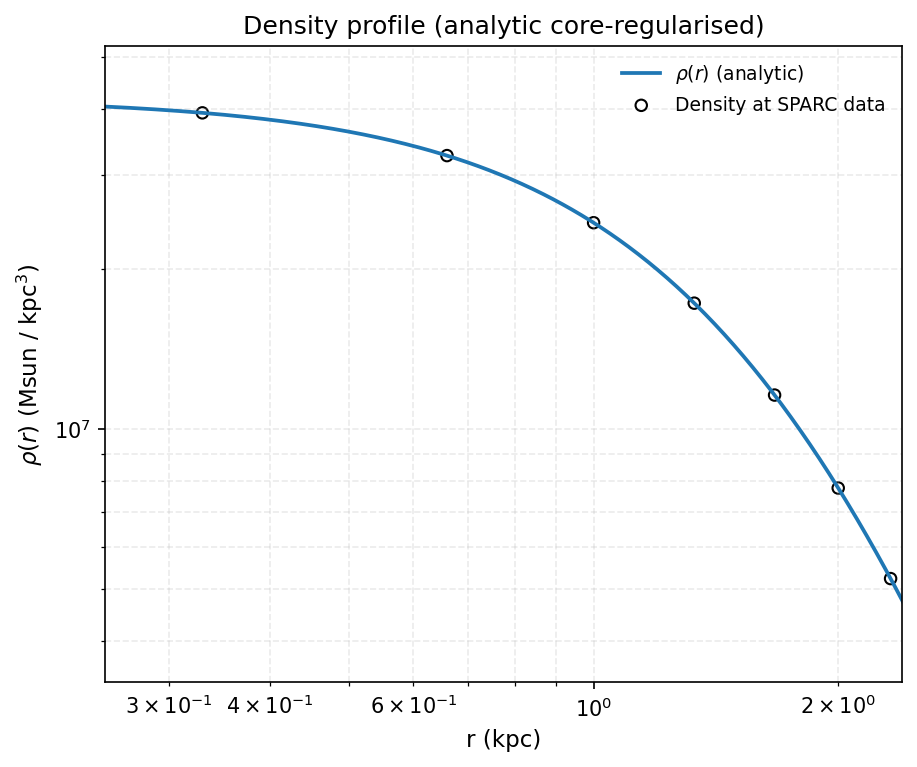}
\caption{The density of the SIDM model of Eq.
(\ref{ScaledependentEoSDM}) for the galaxy UGC07866, versus the
radius.} \label{UGC07866dens}
\end{figure}

\begin{figure}[h!]
\centering
\includegraphics[width=35pc]{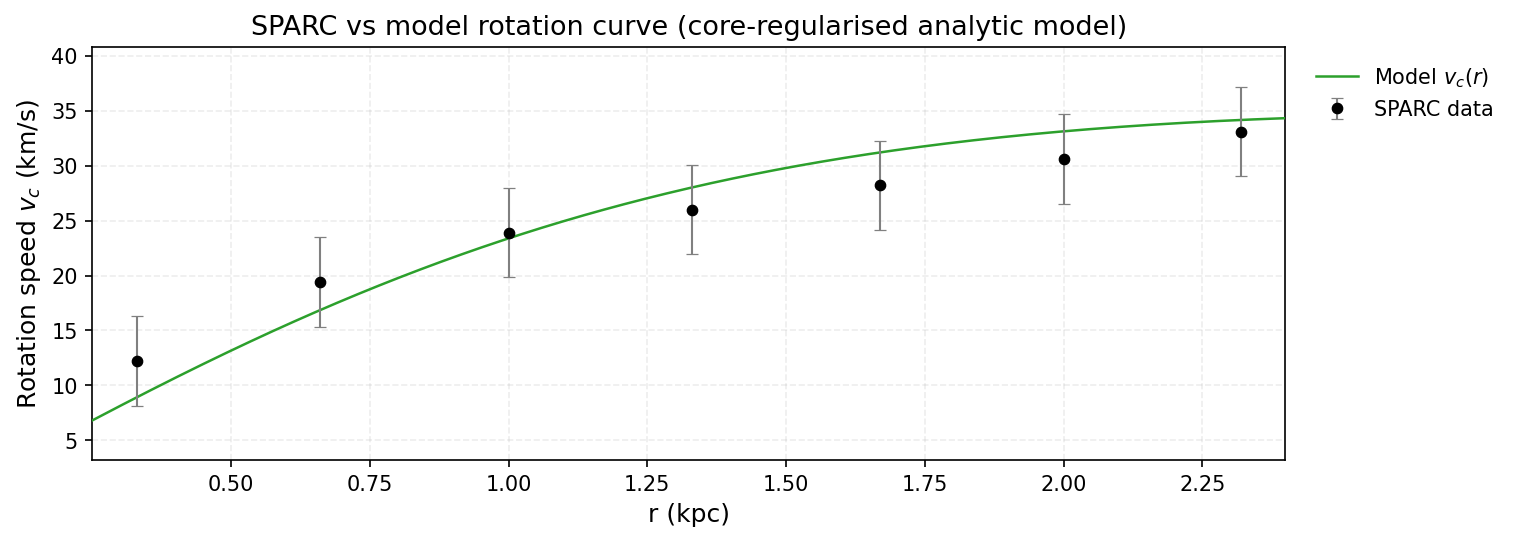}
\caption{The predicted rotation curves for the optimized SIDM
model of Eq. (\ref{ScaledependentEoSDM}), versus the SPARC
observational data for the galaxy UGC07866.} \label{UGC07866}
\end{figure}

In Table \ref{collUGC07866} we present the optimized values of
$K_0$ and $\rho_0$ for the analytic SIDM model of Eq.
(\ref{ScaledependentEoSDM}) for which the maximum compatibility
with the SPARC data is achieved.
\begin{table}[h!]
  \begin{center}
    \caption{SIDM Optimization Values for the galaxy UGC07866}
    \label{collUGC07866}
     \begin{tabular}{|r|r|}
     \hline
      \textbf{Parameter}   & \textbf{Optimization Values}
      \\  \hline
     $\rho_0 $  ($M_{\odot}/\mathrm{Kpc}^{3}$) & $4.19666\times 10^7$
\\  \hline $K_0$ ($M_{\odot} \,
\mathrm{Kpc}^{-3} \, (\mathrm{km/s})^{2}$)& 522.148
\\  \hline
    \end{tabular}
  \end{center}
\end{table}
In Figs. \ref{UGC07866dens}, \ref{UGC07866}  we present the
density of the analytic SIDM model, the predicted rotation curves
for the SIDM model (\ref{ScaledependentEoSDM}), versus the SPARC
observational data and the sound speed, as a function of the
radius respectively. As it can be seen, for this galaxy, the SIDM
model produces viable rotation curves which are compatible with
the SPARC data.

\subsection{The Galaxy UGC08286, Non-viable}

For this galaxy, the optimization method we used, ensures maximum
compatibility of the analytic SIDM model of Eq.
(\ref{ScaledependentEoSDM}) with the SPARC data, if we choose
$\rho_0=8.99622\times 10^7$$M_{\odot}/\mathrm{Kpc}^{3}$ and
$K_0=3052.87
$$M_{\odot} \, \mathrm{Kpc}^{-3} \, (\mathrm{km/s})^{2}$, in which
case the reduced $\chi^2_{red}$ value is $\chi^2_{red}=4.76634$.
Also the parameter $\alpha$ in this case is $\alpha=3.36182 $Kpc.

\begin{figure}[h!]
\centering
\includegraphics[width=20pc]{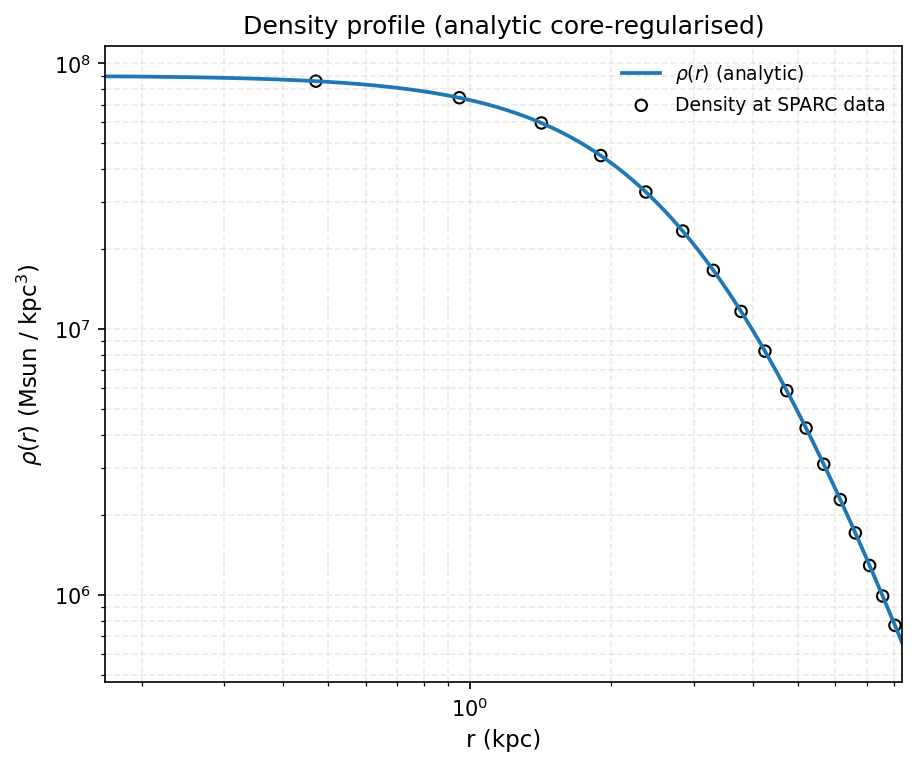}
\caption{The density of the SIDM model of Eq.
(\ref{ScaledependentEoSDM}) for the galaxy UGC08286, versus the
radius.} \label{UGC08286dens}
\end{figure}

\begin{figure}[h!]
\centering
\includegraphics[width=35pc]{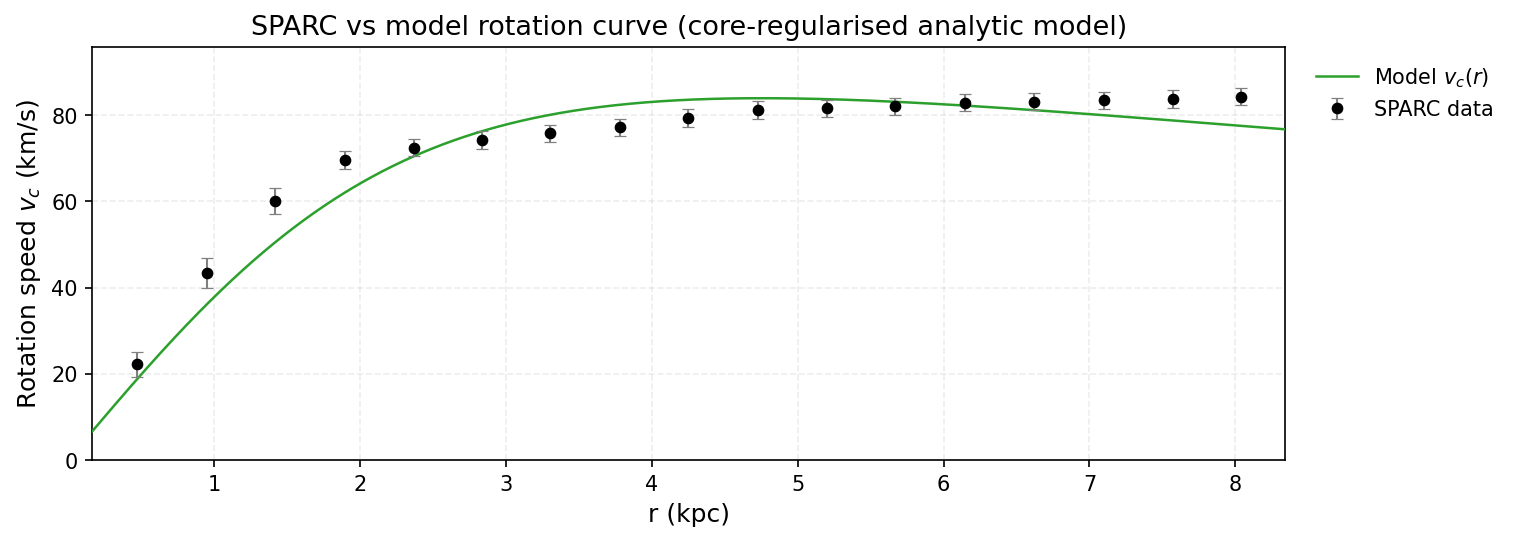}
\caption{The predicted rotation curves for the optimized SIDM
model of Eq. (\ref{ScaledependentEoSDM}), versus the SPARC
observational data for the galaxy UGC08286.} \label{UGC08286}
\end{figure}

In Table \ref{collUGC08286} we present the optimized values of
$K_0$ and $\rho_0$ for the analytic SIDM model of Eq.
(\ref{ScaledependentEoSDM}) for which the maximum compatibility
with the SPARC data is achieved.
\begin{table}[h!]
  \begin{center}
    \caption{SIDM Optimization Values for the galaxy UGC08286}
    \label{collUGC08286}
     \begin{tabular}{|r|r|}
     \hline
      \textbf{Parameter}   & \textbf{Optimization Values}
      \\  \hline
     $\rho_0 $  ($M_{\odot}/\mathrm{Kpc}^{3}$) & $8.99622\times 10^7$
\\  \hline $K_0$ ($M_{\odot} \,
\mathrm{Kpc}^{-3} \, (\mathrm{km/s})^{2}$)& =3052.87
\\  \hline
    \end{tabular}
  \end{center}
\end{table}
In Figs. \ref{UGC08286dens}, \ref{UGC08286}  we present the
density of the analytic SIDM model, the predicted rotation curves
for the SIDM model (\ref{ScaledependentEoSDM}), versus the SPARC
observational data and the sound speed, as a function of the
radius respectively. As it can be seen, for this galaxy, the SIDM
model produces non-viable rotation curves which are incompatible
with the SPARC data.

\begin{figure}[h!]
\centering
\includegraphics[width=20pc]{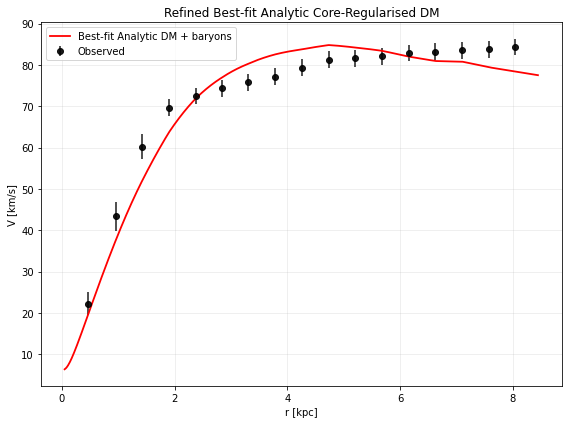}
\caption{The predicted rotation curves after using an optimization
for the SIDM model (\ref{ScaledependentEoSDM}), and the extended
SPARC data for the galaxy UGC08286. We included the rotation
curves of the gas, the disk velocities, the bulge (where present)
along with the SIDM model.} \label{extendedUGC08286}
\end{figure}

Now we shall include contributions to the rotation velocity from
the other components of the galaxy, namely the disk, the gas, and
the bulge if present. In Fig. \ref{extendedUGC08286} we present
the combined rotation curves including all the components of the
galaxy along with the SIDM. As it can be seen, the extended
collisional DM model is non-viable.
Also in Table \ref{evaluationextendedUGC08286} we present the
optimized values of the free parameters of the SIDM model for
which  we achieve the maximum compatibility with the SPARC data,
for the galaxy UGC08286, and also the resulting reduced
$\chi^2_{red}$ value.
\begin{table}[h!]
\centering \caption{Optimized Parameter Values of the Extended
SIDM model for the Galaxy UGC08286.}
\begin{tabular}{lc}
\hline
Parameter & Value  \\
\hline
$\rho_0 $ ($M_{\odot}/\mathrm{Kpc}^{3}$) & $9.81387\times 10^7$   \\
$K_0$ ($M_{\odot} \,
\mathrm{Kpc}^{-3} \, (\mathrm{km/s})^{2}$) & 2886.24   \\
$ml_{\text{disk}}$ & 0.77 \\
$ml_{\text{bulge}}$ & 0.2364 \\
$\alpha$ (Kpc) & 3.12927\\
$\chi^2_{red}$ & 4.54762 \\
\hline
\end{tabular}
\label{evaluationextendedUGC08286}
\end{table}

\subsection{The Galaxy UGC08490, Non-viable}

For this galaxy, the optimization method we used, ensures maximum
compatibility of the analytic SIDM model of Eq.
(\ref{ScaledependentEoSDM}) with the SPARC data, if we choose
$\rho_0=1.24736\times 10^8$$M_{\odot}/\mathrm{Kpc}^{3}$ and
$K_0=3195.87
$$M_{\odot} \, \mathrm{Kpc}^{-3} \, (\mathrm{km/s})^{2}$, in which
case the reduced $\chi^2_{red}$ value is $\chi^2_{red}=4.99288$.
Also the parameter $\alpha$ in this case is $\alpha=2.92112 $Kpc.

\begin{figure}[h!]
\centering
\includegraphics[width=20pc]{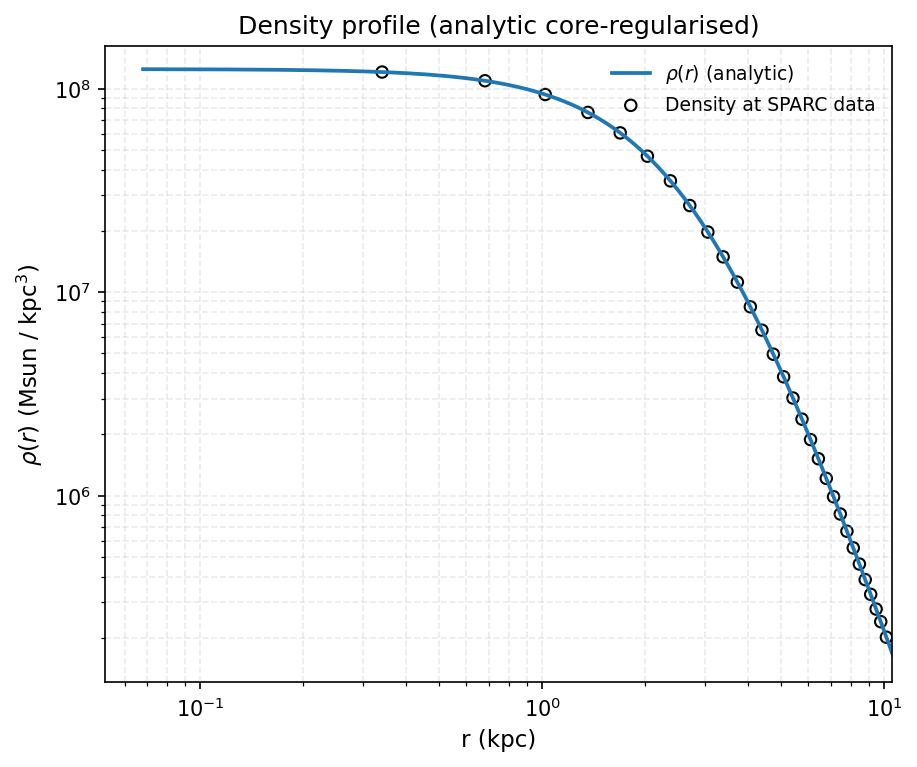}
\caption{The density of the SIDM model of Eq.
(\ref{ScaledependentEoSDM}) for the galaxy UGC08490, versus the
radius.} \label{UGC08490dens}
\end{figure}

\begin{figure}[h!]
\centering
\includegraphics[width=35pc]{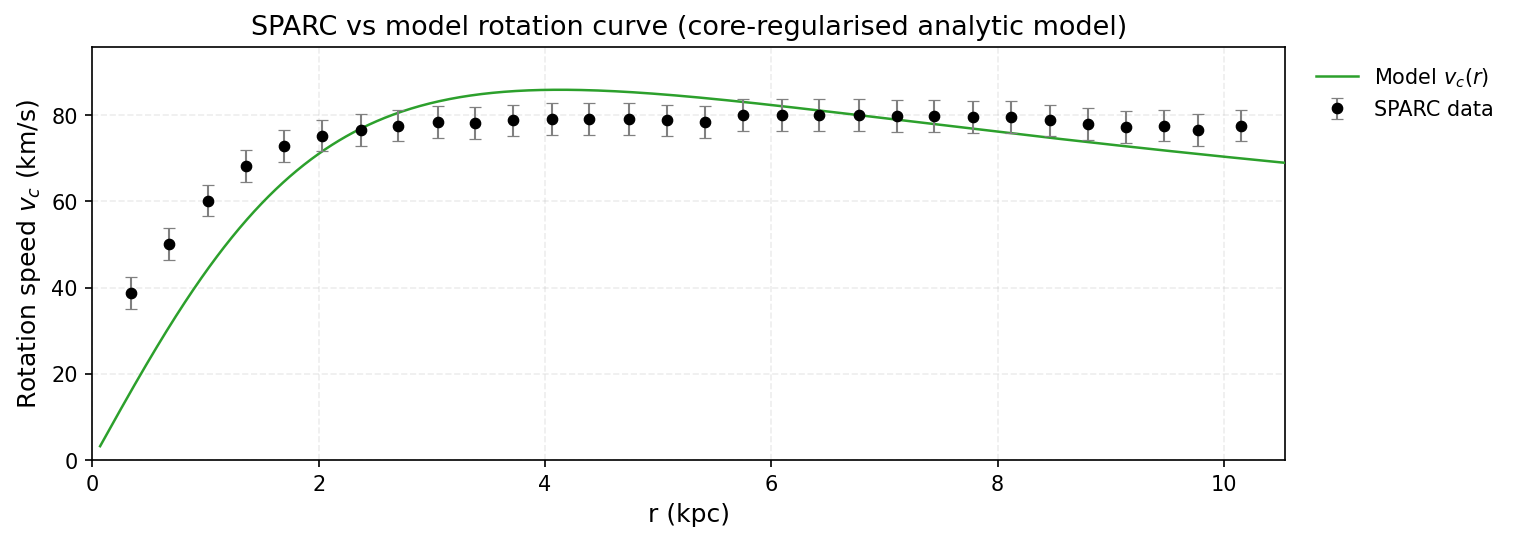}
\caption{The predicted rotation curves for the optimized SIDM
model of Eq. (\ref{ScaledependentEoSDM}), versus the SPARC
observational data for the galaxy UGC08490.} \label{UGC08490}
\end{figure}

In Table \ref{collUGC08490} we present the optimized values of
$K_0$ and $\rho_0$ for the analytic SIDM model of Eq.
(\ref{ScaledependentEoSDM}) for which the maximum compatibility
with the SPARC data is achieved.
\begin{table}[h!]
  \begin{center}
    \caption{SIDM Optimization Values for the galaxy UGC08490}
    \label{collUGC08490}
     \begin{tabular}{|r|r|}
     \hline
      \textbf{Parameter}   & \textbf{Optimization Values}
      \\  \hline
     $\rho_0 $  ($M_{\odot}/\mathrm{Kpc}^{3}$) & $1.24736\times 10^8$
\\  \hline $K_0$ ($M_{\odot} \,
\mathrm{Kpc}^{-3} \, (\mathrm{km/s})^{2}$)& 3195.87
\\  \hline
    \end{tabular}
  \end{center}
\end{table}
In Figs. \ref{UGC08490dens}, \ref{UGC08490}  we present the
density of the analytic SIDM model, the predicted rotation curves
for the SIDM model (\ref{ScaledependentEoSDM}), versus the SPARC
observational data and the sound speed, as a function of the
radius respectively. As it can be seen, for this galaxy, the SIDM
model produces non-viable rotation curves which are incompatible
with the SPARC data.

\begin{figure}[h!]
\centering
\includegraphics[width=20pc]{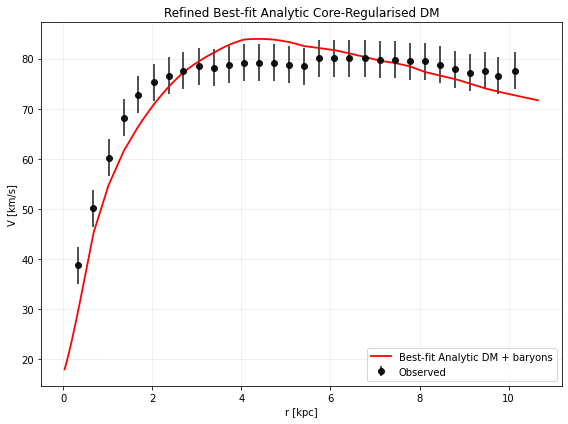}
\caption{The predicted rotation curves after using an optimization
for the SIDM model (\ref{ScaledependentEoSDM}), and the extended
SPARC data for the galaxy UGC08490. We included the rotation
curves of the gas, the disk velocities, the bulge (where present)
along with the SIDM model.} \label{extendedUGC08490}
\end{figure}

Now we shall include contributions to the rotation velocity from
the other components of the galaxy, namely the disk, the gas, and
the bulge if present. In Fig. \ref{extendedUGC08490} we present
the combined rotation curves including all the components of the
galaxy along with the SIDM. As it can be seen, the extended
collisional DM model is non-viable.
Also in Table \ref{evaluationextendedUGC08490} we present the
optimized values of the free parameters of the SIDM model for
which  we achieve the maximum compatibility with the SPARC data,
for the galaxy UGC08490, and also the resulting reduced
$\chi^2_{red}$ value.
\begin{table}[h!]
\centering \caption{Optimized Parameter Values of the Extended
SIDM model for the Galaxy UGC08490.}
\begin{tabular}{lc}
\hline
Parameter & Value  \\
\hline
$\rho_0 $ ($M_{\odot}/\mathrm{Kpc}^{3}$) & $5.38884\times 10^7$   \\
$K_0$ ($M_{\odot} \,
\mathrm{Kpc}^{-3} \, (\mathrm{km/s})^{2}$) & 2344.78   \\
$ml_{\text{disk}}$ & 1 \\
$ml_{\text{bulge}}$ & 0.4631 \\
$\alpha$ (Kpc) & 3.80627\\
$\chi^2_{red}$ & 1.254 \\
\hline
\end{tabular}
\label{evaluationextendedUGC08490}
\end{table}

\subsection{The Galaxy UGC08550, Non-viable}

For this galaxy, the optimization method we used, ensures maximum
compatibility of the analytic SIDM model of Eq.
(\ref{ScaledependentEoSDM}) with the SPARC data, if we choose
$\rho_0=9.10509\times 10^7$$M_{\odot}/\mathrm{Kpc}^{3}$ and
$K_0=1417.93
$$M_{\odot} \, \mathrm{Kpc}^{-3} \, (\mathrm{km/s})^{2}$, in which
case the reduced $\chi^2_{red}$ value is $\chi^2_{red}=5.10517$.
Also the parameter $\alpha$ in this case is $\alpha=2.27738 $Kpc.

\begin{figure}[h!]
\centering
\includegraphics[width=20pc]{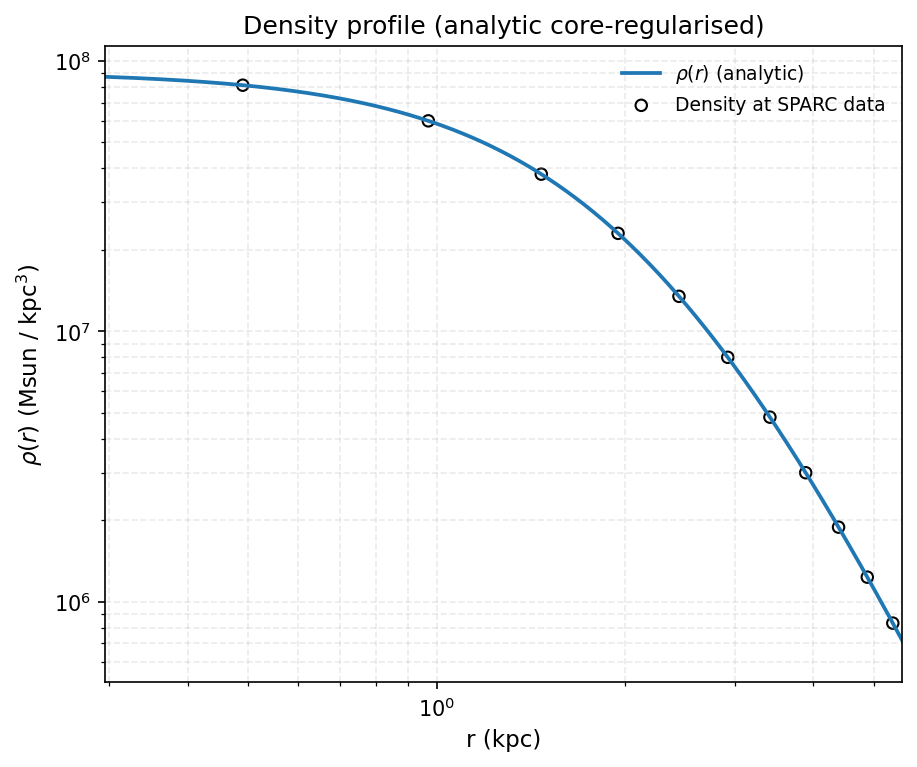}
\caption{The density of the SIDM model of Eq.
(\ref{ScaledependentEoSDM}) for the galaxy UGC08550, versus the
radius.} \label{UGC08550dens}
\end{figure}

\begin{figure}[h!]
\centering
\includegraphics[width=35pc]{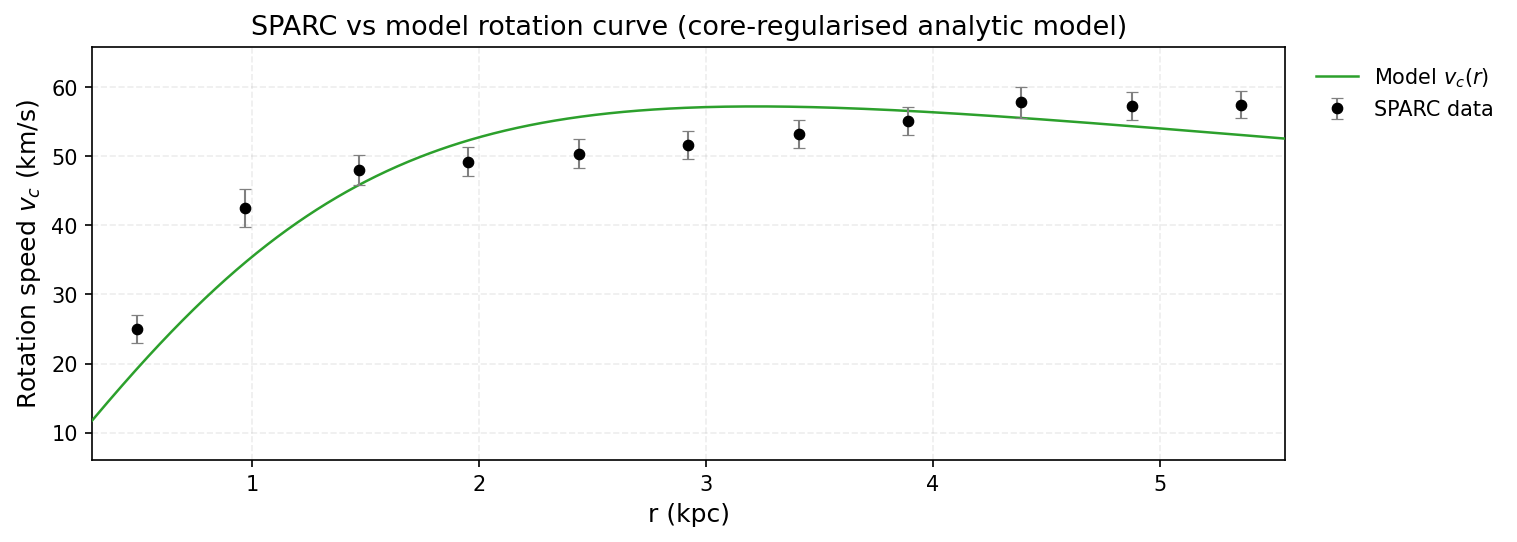}
\caption{The predicted rotation curves for the optimized SIDM
model of Eq. (\ref{ScaledependentEoSDM}), versus the SPARC
observational data for the galaxy UGC08550.} \label{UGC08550}
\end{figure}

In Table \ref{collUGC08550} we present the optimized values of
$K_0$ and $\rho_0$ for the analytic SIDM model of Eq.
(\ref{ScaledependentEoSDM}) for which the maximum compatibility
with the SPARC data is achieved.
\begin{table}[h!]
  \begin{center}
    \caption{SIDM Optimization Values for the galaxy UGC08550}
    \label{collUGC08550}
     \begin{tabular}{|r|r|}
     \hline
      \textbf{Parameter}   & \textbf{Optimization Values}
      \\  \hline
     $\rho_0 $  ($M_{\odot}/\mathrm{Kpc}^{3}$) & $9.10509\times 10^7$
\\  \hline $K_0$ ($M_{\odot} \,
\mathrm{Kpc}^{-3} \, (\mathrm{km/s})^{2}$)& 1417.93
\\  \hline
    \end{tabular}
  \end{center}
\end{table}
In Figs. \ref{UGC08550dens}, \ref{UGC08550}  we present the
density of the analytic SIDM model, the predicted rotation curves
for the SIDM model (\ref{ScaledependentEoSDM}), versus the SPARC
observational data and the sound speed, as a function of the
radius respectively. As it can be seen, for this galaxy, the SIDM
model produces non-viable rotation curves which are incompatible
with the SPARC data.

\begin{figure}[h!]
\centering
\includegraphics[width=20pc]{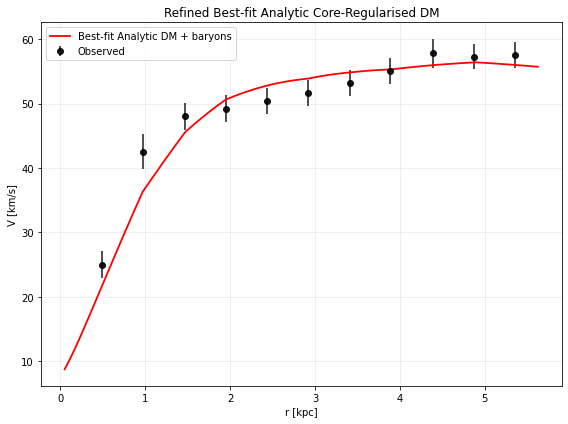}
\caption{The predicted rotation curves after using an optimization
for the SIDM model (\ref{ScaledependentEoSDM}), and the extended
SPARC data for the galaxy UGC08550. We included the rotation
curves of the gas, the disk velocities, the bulge (where present)
along with the SIDM model.} \label{extendedUGC08550}
\end{figure}

Now we shall include contributions to the rotation velocity from
the other components of the galaxy, namely the disk, the gas, and
the bulge if present. In Fig. \ref{extendedUGC08550} we present
the combined rotation curves including all the components of the
galaxy along with the SIDM. As it can be seen, the extended
collisional DM model is non-viable.
Also in Table \ref{evaluationextendedUGC08550} we present the
optimized values of the free parameters of the SIDM model for
which  we achieve the maximum compatibility with the SPARC data,
for the galaxy UGC08550, and also the resulting reduced
$\chi^2_{red}$ value.
\begin{table}[h!]
\centering \caption{Optimized Parameter Values of the Extended
SIDM model for the Galaxy UGC08550.}
\begin{tabular}{lc}
\hline
Parameter & Value  \\
\hline
$\rho_0 $ ($M_{\odot}/\mathrm{Kpc}^{3}$) & $4.04304\times 10^7$   \\
$K_0$ ($M_{\odot} \,
\mathrm{Kpc}^{-3} \, (\mathrm{km/s})^{2}$) & 1080.68   \\
$ml_{\text{disk}}$ & 1 \\
$ml_{\text{bulge}}$ & 0.3544 \\
$\alpha$ (Kpc) & 2.98326\\
$\chi^2_{red}$ & 2.04733 \\
\hline
\end{tabular}
\label{evaluationextendedUGC08550}
\end{table}

\subsection{The Galaxy UGC08699, Non-viable}

For this galaxy, the optimization method we used, ensures maximum
compatibility of the analytic SIDM model of Eq.
(\ref{ScaledependentEoSDM}) with the SPARC data, if we choose
$\rho_0=2.06633\times 10^8$$M_{\odot}/\mathrm{Kpc}^{3}$ and
$K_0=16391.8
$$M_{\odot} \, \mathrm{Kpc}^{-3} \, (\mathrm{km/s})^{2}$, in which
case the reduced $\chi^2_{red}$ value is $\chi^2_{red}=155.888$.
Also the parameter $\alpha$ in this case is $\alpha=5.14 $Kpc.

\begin{figure}[h!]
\centering
\includegraphics[width=20pc]{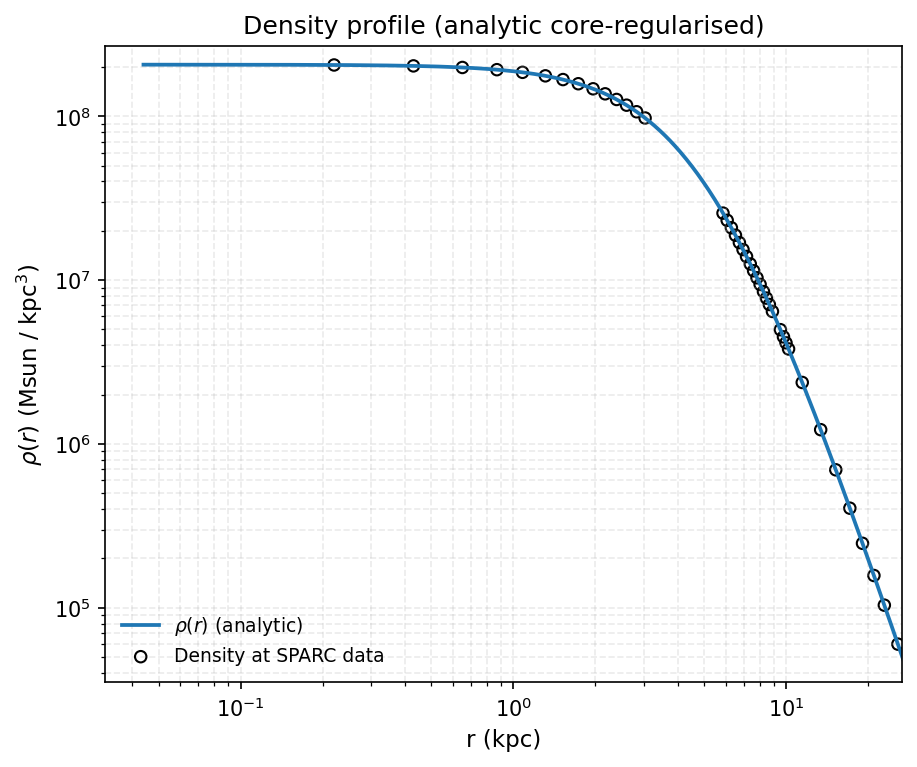}
\caption{The density of the SIDM model of Eq.
(\ref{ScaledependentEoSDM}) for the galaxy UGC08699, versus the
radius.} \label{UGC08699dens}
\end{figure}

\begin{figure}[h!]
\centering
\includegraphics[width=35pc]{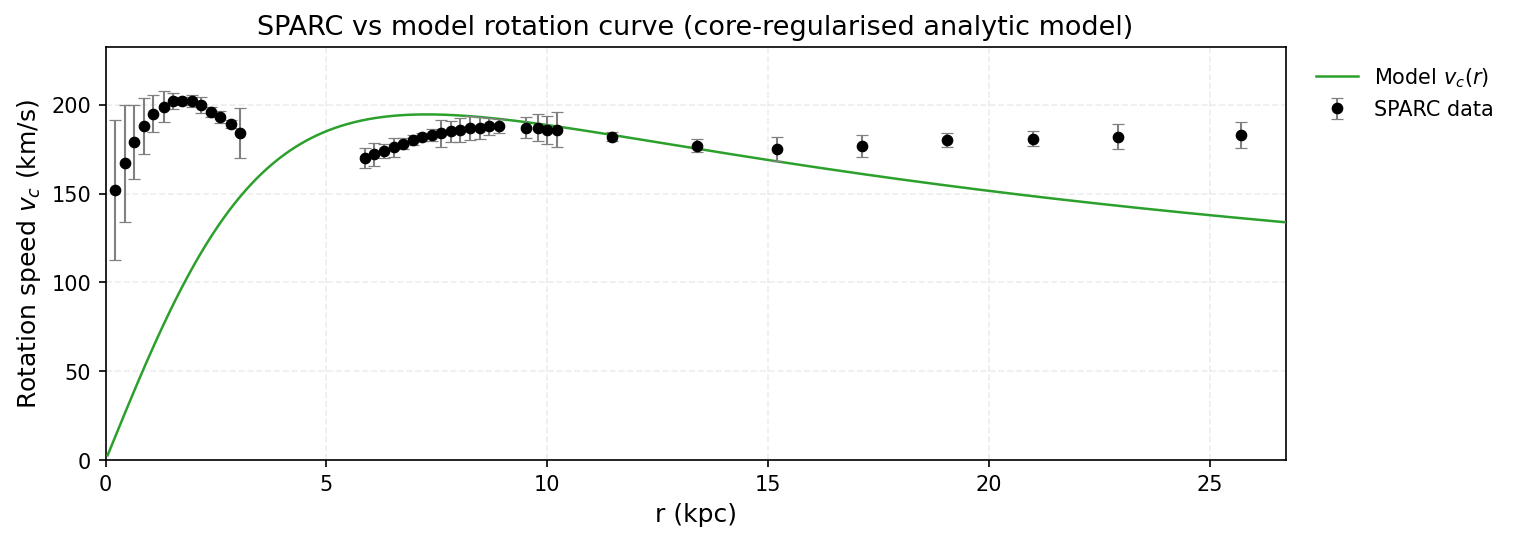}
\caption{The predicted rotation curves for the optimized SIDM
model of Eq. (\ref{ScaledependentEoSDM}), versus the SPARC
observational data for the galaxy UGC08699.} \label{UGC08699}
\end{figure}

In Table \ref{collUGC08699} we present the optimized values of
$K_0$ and $\rho_0$ for the analytic SIDM model of Eq.
(\ref{ScaledependentEoSDM}) for which the maximum compatibility
with the SPARC data is achieved.
\begin{table}[h!]
  \begin{center}
    \caption{SIDM Optimization Values for the galaxy UGC08699}
    \label{collUGC08699}
     \begin{tabular}{|r|r|}
     \hline
      \textbf{Parameter}   & \textbf{Optimization Values}
      \\  \hline
     $\rho_0 $  ($M_{\odot}/\mathrm{Kpc}^{3}$) & $2.06633\times 10^8$
\\  \hline $K_0$ ($M_{\odot} \,
\mathrm{Kpc}^{-3} \, (\mathrm{km/s})^{2}$)& 16391.8
\\  \hline
    \end{tabular}
  \end{center}
\end{table}
In Figs. \ref{UGC08699dens}, \ref{UGC08699}  we present the
density of the analytic SIDM model, the predicted rotation curves
for the SIDM model (\ref{ScaledependentEoSDM}), versus the SPARC
observational data and the sound speed, as a function of the
radius respectively. As it can be seen, for this galaxy, the SIDM
model produces non-viable rotation curves which are incompatible
with the SPARC data.

\begin{figure}[h!]
\centering
\includegraphics[width=20pc]{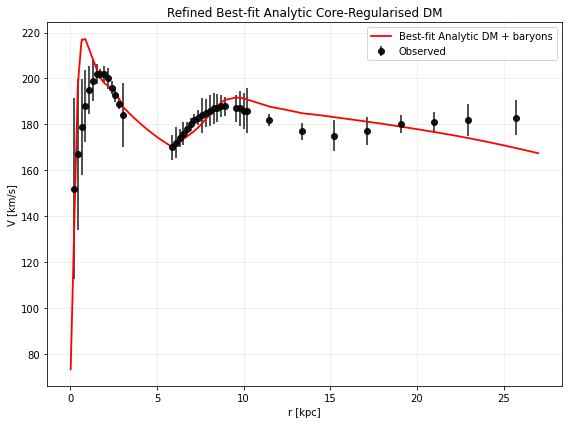}
\caption{The predicted rotation curves after using an optimization
for the SIDM model (\ref{ScaledependentEoSDM}), and the extended
SPARC data for the galaxy UGC08699. We included the rotation
curves of the gas, the disk velocities, the bulge (where present)
along with the SIDM model.} \label{extendedUGC08699}
\end{figure}

Now we shall include contributions to the rotation velocity from
the other components of the galaxy, namely the disk, the gas, and
the bulge if present. In Fig. \ref{extendedUGC08699} we present
the combined rotation curves including all the components of the
galaxy along with the SIDM. As it can be seen, the extended
collisional DM model is non-viable.
Also in Table \ref{evaluationextendedUGC08699} we present the
optimized values of the free parameters of the SIDM model for
which  we achieve the maximum compatibility with the SPARC data,
for the galaxy UGC08699, and also the resulting reduced
$\chi^2_{red}$ value.
\begin{table}[h!]
\centering \caption{Optimized Parameter Values of the Extended
SIDM model for the Galaxy UGC08699.}
\begin{tabular}{lc}
\hline
Parameter & Value  \\
\hline
$\rho_0 $ ($M_{\odot}/\mathrm{Kpc}^{3}$) & $1.15203\times 10^7$   \\
$K_0$ ($M_{\odot} \,
\mathrm{Kpc}^{-3} \, (\mathrm{km/s})^{2}$) & 8544.33   \\
$ml_{\text{disk}}$ & 1.0000 \\
$ml_{\text{bulge}}$ & 0.8248 \\
$\alpha$ (Kpc) & 15.7146\\
$\chi^2_{red}$ & 1.21621 \\
\hline
\end{tabular}
\label{evaluationextendedUGC08699}
\end{table}

\subsection{The Galaxy UGC08837}

For this galaxy, the optimization method we used, ensures maximum
compatibility of the analytic SIDM model of Eq.
(\ref{ScaledependentEoSDM}) with the SPARC data, if we choose
$\rho_0=1.25744\times 10^7$$M_{\odot}/\mathrm{Kpc}^{3}$ and
$K_0=1886.63
$$M_{\odot} \, \mathrm{Kpc}^{-3} \, (\mathrm{km/s})^{2}$, in which
case the reduced $\chi^2_{red}$ value is $\chi^2_{red}=0.408737$.
Also the parameter $\alpha$ in this case is $\alpha=7.06888 $Kpc.

\begin{figure}[h!]
\centering
\includegraphics[width=20pc]{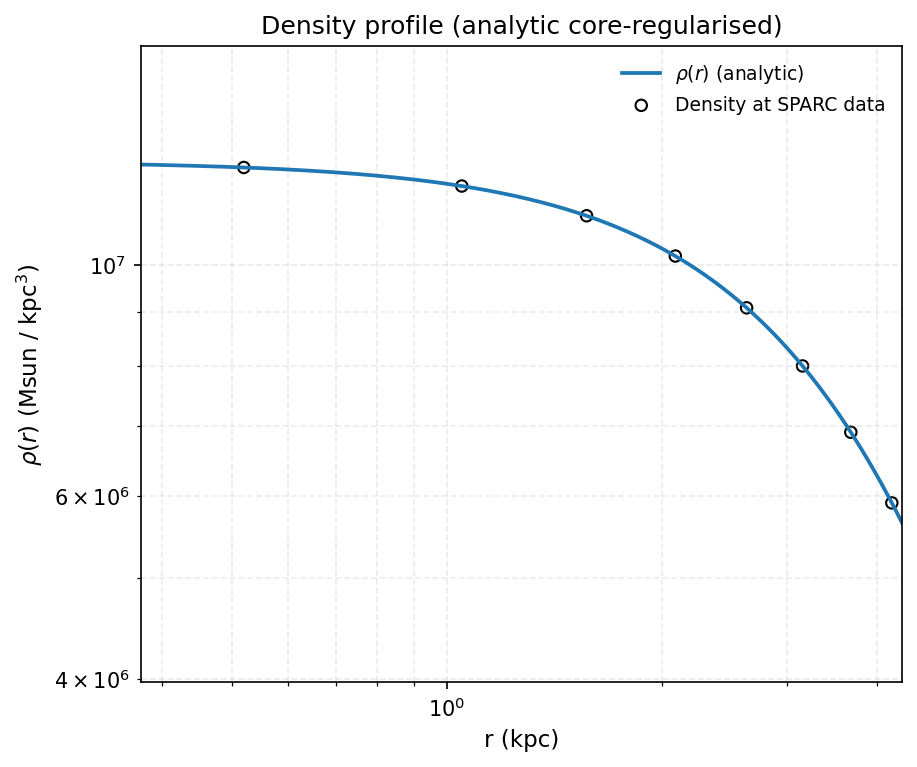}
\caption{The density of the SIDM model of Eq.
(\ref{ScaledependentEoSDM}) for the galaxy UGC08837, versus the
radius.} \label{UGC08837dens}
\end{figure}

\begin{figure}[h!]
\centering
\includegraphics[width=35pc]{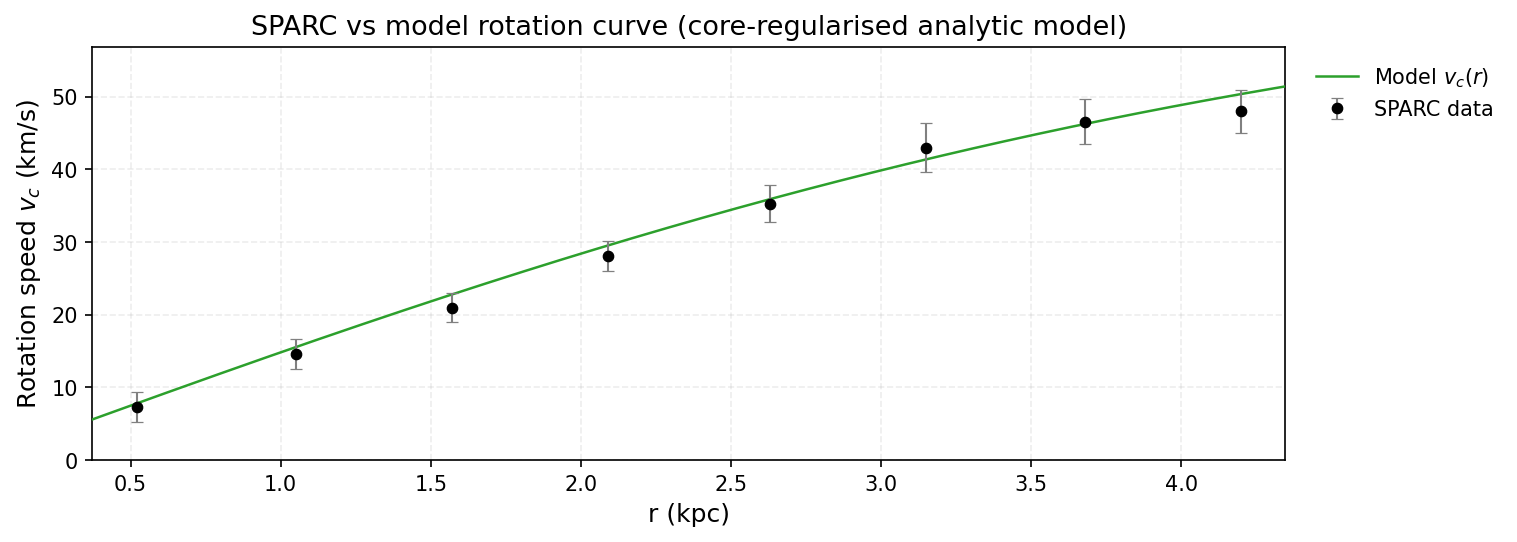}
\caption{The predicted rotation curves for the optimized SIDM
model of Eq. (\ref{ScaledependentEoSDM}), versus the SPARC
observational data for the galaxy UGC08837.} \label{UGC08837}
\end{figure}

In Table \ref{collUGC08837} we present the optimized values of
$K_0$ and $\rho_0$ for the analytic SIDM model of Eq.
(\ref{ScaledependentEoSDM}) for which the maximum compatibility
with the SPARC data is achieved.
\begin{table}[h!]
  \begin{center}
    \caption{SIDM Optimization Values for the galaxy UGC08837}
    \label{collUGC08837}
     \begin{tabular}{|r|r|}
     \hline
      \textbf{Parameter}   & \textbf{Optimization Values}
      \\  \hline
     $\rho_0 $  ($M_{\odot}/\mathrm{Kpc}^{3}$) & $1.25744\times 10^7$
\\  \hline $K_0$ ($M_{\odot} \,
\mathrm{Kpc}^{-3} \, (\mathrm{km/s})^{2}$)& 1886.63
\\  \hline
    \end{tabular}
  \end{center}
\end{table}
In Figs. \ref{UGC08837dens}, \ref{UGC08837}  we present the
density of the analytic SIDM model, the predicted rotation curves
for the SIDM model (\ref{ScaledependentEoSDM}), versus the SPARC
observational data and the sound speed, as a function of the
radius respectively. As it can be seen, for this galaxy, the SIDM
model produces viable rotation curves which are compatible with
the SPARC data.

\subsection{The Galaxy UGC09037, Non-viable, Extended Viable}

For this galaxy, the optimization method we used, ensures maximum
compatibility of the analytic SIDM model of Eq.
(\ref{ScaledependentEoSDM}) with the SPARC data, if we choose
$\rho_0=2.58578\times 10^7$$M_{\odot}/\mathrm{Kpc}^{3}$ and
$K_0=10243.5
$$M_{\odot} \, \mathrm{Kpc}^{-3} \, (\mathrm{km/s})^{2}$, in which
case the reduced $\chi^2_{red}$ value is $\chi^2_{red}=2.59911$.
Also the parameter $\alpha$ in this case is $\alpha=11.4863 $Kpc.

\begin{figure}[h!]
\centering
\includegraphics[width=20pc]{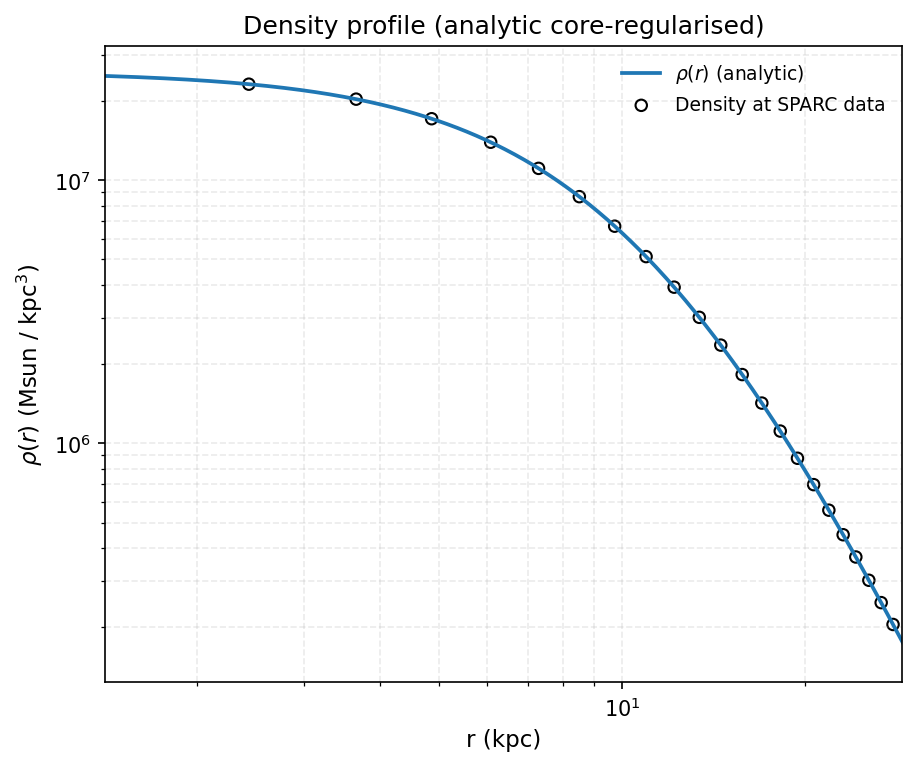}
\caption{The density of the SIDM model of Eq.
(\ref{ScaledependentEoSDM}) for the galaxy UGC09037, versus the
radius.} \label{UGC09037dens}
\end{figure}

\begin{figure}[h!]
\centering
\includegraphics[width=35pc]{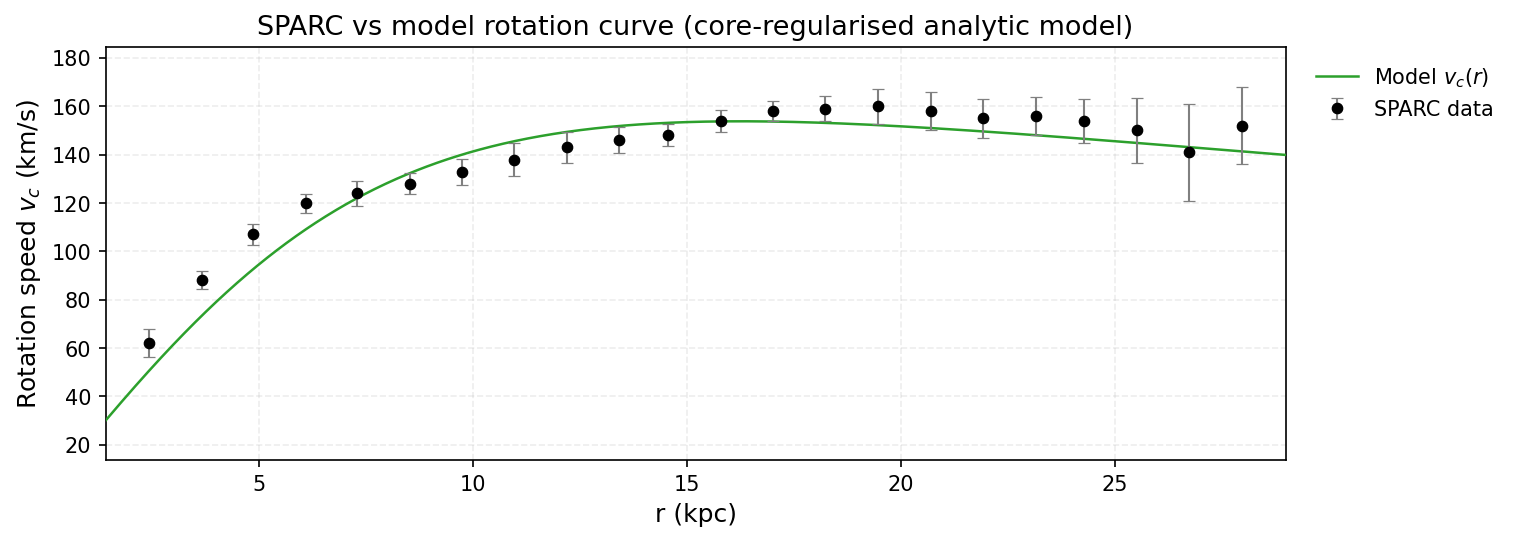}
\caption{The predicted rotation curves for the optimized SIDM
model of Eq. (\ref{ScaledependentEoSDM}), versus the SPARC
observational data for the galaxy UGC09037.} \label{UGC09037}
\end{figure}

In Table \ref{collUGC09037} we present the optimized values of
$K_0$ and $\rho_0$ for the analytic SIDM model of Eq.
(\ref{ScaledependentEoSDM}) for which the maximum compatibility
with the SPARC data is achieved.
\begin{table}[h!]
  \begin{center}
    \caption{SIDM Optimization Values for the galaxy UGC09037}
    \label{collUGC09037}
     \begin{tabular}{|r|r|}
     \hline
      \textbf{Parameter}   & \textbf{Optimization Values}
      \\  \hline
     $\rho_0 $  ($M_{\odot}/\mathrm{Kpc}^{3}$) & $2.58578\times 10^7$
\\  \hline $K_0$ ($M_{\odot} \,
\mathrm{Kpc}^{-3} \, (\mathrm{km/s})^{2}$)& 10243.5
\\  \hline
    \end{tabular}
  \end{center}
\end{table}
In Figs. \ref{UGC09037dens}, \ref{UGC09037}  we present the
density of the analytic SIDM model, the predicted rotation curves
for the SIDM model (\ref{ScaledependentEoSDM}), versus the SPARC
observational data and the sound speed, as a function of the
radius respectively. As it can be seen, for this galaxy, the SIDM
model produces non-viable rotation curves which are incompatible
with the SPARC data.

\begin{figure}[h!]
\centering
\includegraphics[width=20pc]{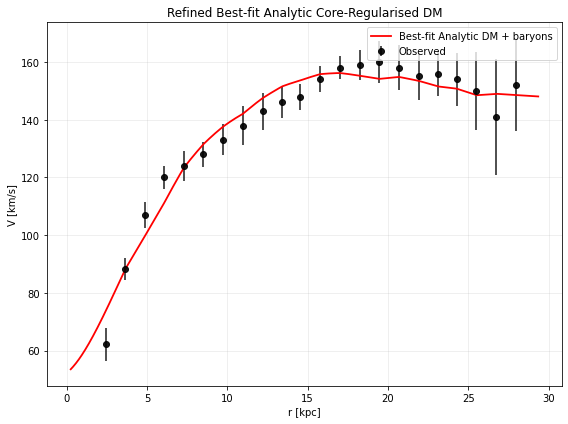}
\caption{The predicted rotation curves after using an optimization
for the SIDM model (\ref{ScaledependentEoSDM}), and the extended
SPARC data for the galaxy UGC09037. We included the rotation
curves of the gas, the disk velocities, the bulge (where present)
along with the SIDM model.} \label{extendedUGC09037}
\end{figure}

Now we shall include contributions to the rotation velocity from
the other components of the galaxy, namely the disk, the gas, and
the bulge if present. In Fig. \ref{extendedUGC09037} we present
the combined rotation curves including all the components of the
galaxy along with the SIDM. As it can be seen, the extended
collisional DM model is viable.
Also in Table \ref{evaluationextendedUGC09037} we present the
optimized values of the free parameters of the SIDM model for
which  we achieve the maximum compatibility with the SPARC data,
for the galaxy UGC09037, and also the resulting reduced
$\chi^2_{red}$ value.
\begin{table}[h!]
\centering \caption{Optimized Parameter Values of the Extended
SIDM model for the Galaxy UGC09037.}
\begin{tabular}{lc}
\hline
Parameter & Value  \\
\hline
$\rho_0 $ ($M_{\odot}/\mathrm{Kpc}^{3}$) & $1.37733\times 10^7$   \\
$K_0$ ($M_{\odot} \,
\mathrm{Kpc}^{-3} \, (\mathrm{km/s})^{2}$) & 7327.13   \\
$ml_{\text{disk}}$ & 0.4595 \\
$ml_{\text{bulge}}$ & 0.5412 \\
$\alpha$ (Kpc) & 13.309\\
$\chi^2_{red}$ & 1.05996 \\
\hline
\end{tabular}
\label{evaluationextendedUGC09037}
\end{table}

\subsection{The Galaxy UGC09133, Non-viable}

For this galaxy, the optimization method we used, ensures maximum
compatibility of the analytic SIDM model of Eq.
(\ref{ScaledependentEoSDM}) with the SPARC data, if we choose
$\rho_0=2.56523\times 10^7$$M_{\odot}/\mathrm{Kpc}^{3}$ and
$K_0=36142.8
$$M_{\odot} \, \mathrm{Kpc}^{-3} \, (\mathrm{km/s})^{2}$, in which
case the reduced $\chi^2_{red}$ value is $\chi^2_{red}=829.59$.
Also the parameter $\alpha$ in this case is $\alpha=21.662 $Kpc.

\begin{figure}[h!]
\centering
\includegraphics[width=20pc]{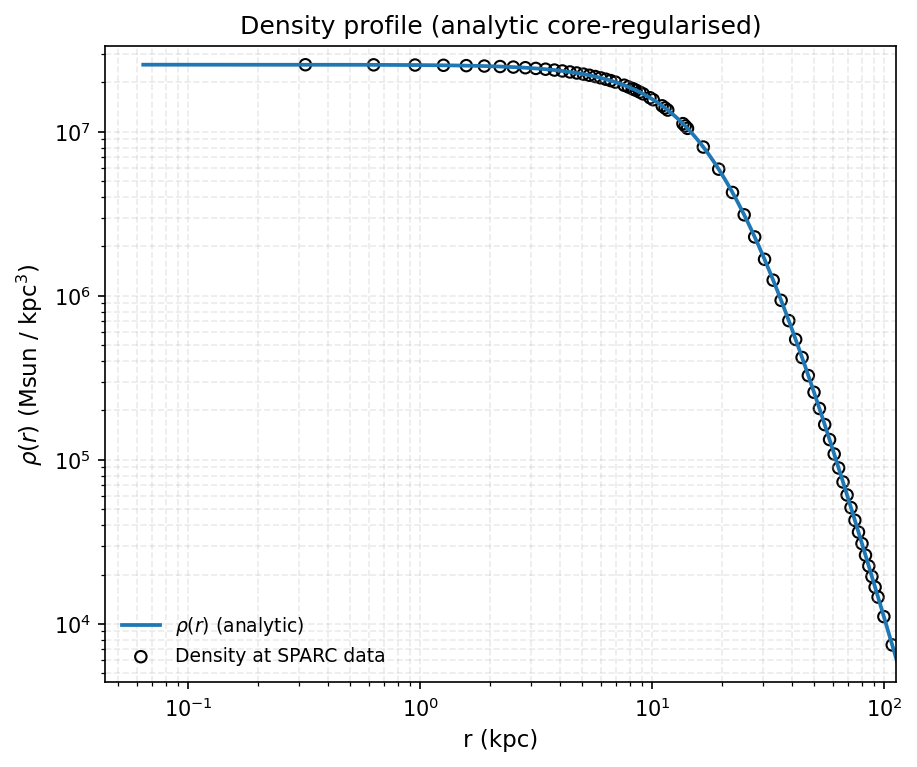}
\caption{The density of the SIDM model of Eq.
(\ref{ScaledependentEoSDM}) for the galaxy UGC09133, versus the
radius.} \label{UGC09133dens}
\end{figure}

\begin{figure}[h!]
\centering
\includegraphics[width=35pc]{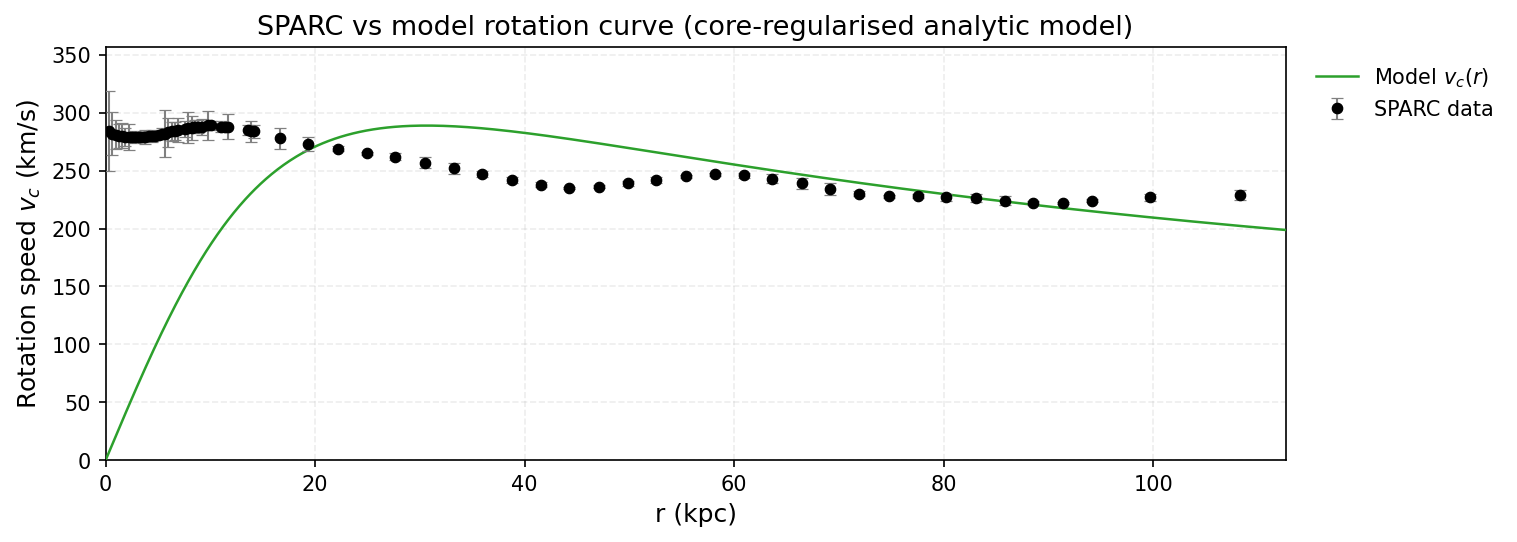}
\caption{The predicted rotation curves for the optimized SIDM
model of Eq. (\ref{ScaledependentEoSDM}), versus the SPARC
observational data for the galaxy UGC09133.} \label{UGC09133}
\end{figure}

In Table \ref{collUGC09133} we present the optimized values of
$K_0$ and $\rho_0$ for the analytic SIDM model of Eq.
(\ref{ScaledependentEoSDM}) for which the maximum compatibility
with the SPARC data is achieved.
\begin{table}[h!]
  \begin{center}
    \caption{SIDM Optimization Values for the galaxy UGC09133}
    \label{collUGC09133}
     \begin{tabular}{|r|r|}
     \hline
      \textbf{Parameter}   & \textbf{Optimization Values}
      \\  \hline
     $\rho_0 $  ($M_{\odot}/\mathrm{Kpc}^{3}$) & $2.56523\times 10^7$
\\  \hline $K_0$ ($M_{\odot} \,
\mathrm{Kpc}^{-3} \, (\mathrm{km/s})^{2}$)& 36142.8
\\  \hline
    \end{tabular}
  \end{center}
\end{table}
In Figs. \ref{UGC09133dens}, \ref{UGC09133}  we present the
density of the analytic SIDM model, the predicted rotation curves
for the SIDM model (\ref{ScaledependentEoSDM}), versus the SPARC
observational data and the sound speed, as a function of the
radius respectively. As it can be seen, for this galaxy, the SIDM
model produces non-viable rotation curves which are incompatible
with the SPARC data.

\begin{figure}[h!]
\centering
\includegraphics[width=20pc]{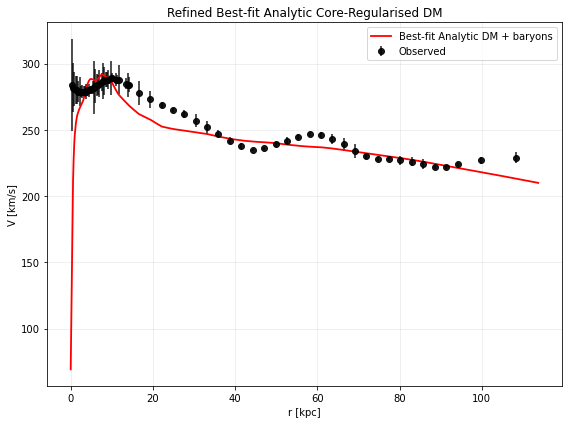}
\caption{The predicted rotation curves after using an optimization
for the SIDM model (\ref{ScaledependentEoSDM}), and the extended
SPARC data for the galaxy UGC09133. We included the rotation
curves of the gas, the disk velocities, the bulge (where present)
along with the SIDM model.} \label{extendedUGC09133}
\end{figure}

Now we shall include contributions to the rotation velocity from
the other components of the galaxy, namely the disk, the gas, and
the bulge if present. In Fig. \ref{extendedUGC09133} we present
the combined rotation curves including all the components of the
galaxy along with the SIDM. As it can be seen, the extended
collisional DM model is non-viable.
Also in Table \ref{evaluationextendedUGC09133} we present the
optimized values of the free parameters of the SIDM model for
which  we achieve the maximum compatibility with the SPARC data,
for the galaxy UGC09133, and also the resulting reduced
$\chi^2_{red}$ value.
\begin{table}[h!]
\centering \caption{Optimized Parameter Values of the Extended
SIDM model for the Galaxy UGC09133.}
\begin{tabular}{lc}
\hline
Parameter & Value  \\
\hline
$\rho_0 $ ($M_{\odot}/\mathrm{Kpc}^{3}$) & $1.90461\times 10^6$   \\
$K_0$ ($M_{\odot} \,
\mathrm{Kpc}^{-3} \, (\mathrm{km/s})^{2}$) & 15594.2   \\
$ml_{\text{disk}}$ & 1 \\
$ml_{\text{bulge}}$ & 0.7274 \\
$\alpha$ (Kpc) & 52.2126\\
$\chi^2_{red}$ & 9.34344 \\
\hline
\end{tabular}
\label{evaluationextendedUGC09133}
\end{table}

\subsection{The Galaxy UGC09992}

For this galaxy, the optimization method we used, ensures maximum
compatibility of the analytic SIDM model of Eq.
(\ref{ScaledependentEoSDM}) with the SPARC data, if we choose
$\rho_0=1.31564\times 10^8$$M_{\odot}/\mathrm{Kpc}^{3}$ and
$K_0=575.789
$$M_{\odot} \, \mathrm{Kpc}^{-3} \, (\mathrm{km/s})^{2}$, in which
case the reduced $\chi^2_{red}$ value is $\chi^2_{red}=0.484757$.
Also the parameter $\alpha$ in this case is $\alpha=1.2073 $Kpc.

\begin{figure}[h!]
\centering
\includegraphics[width=20pc]{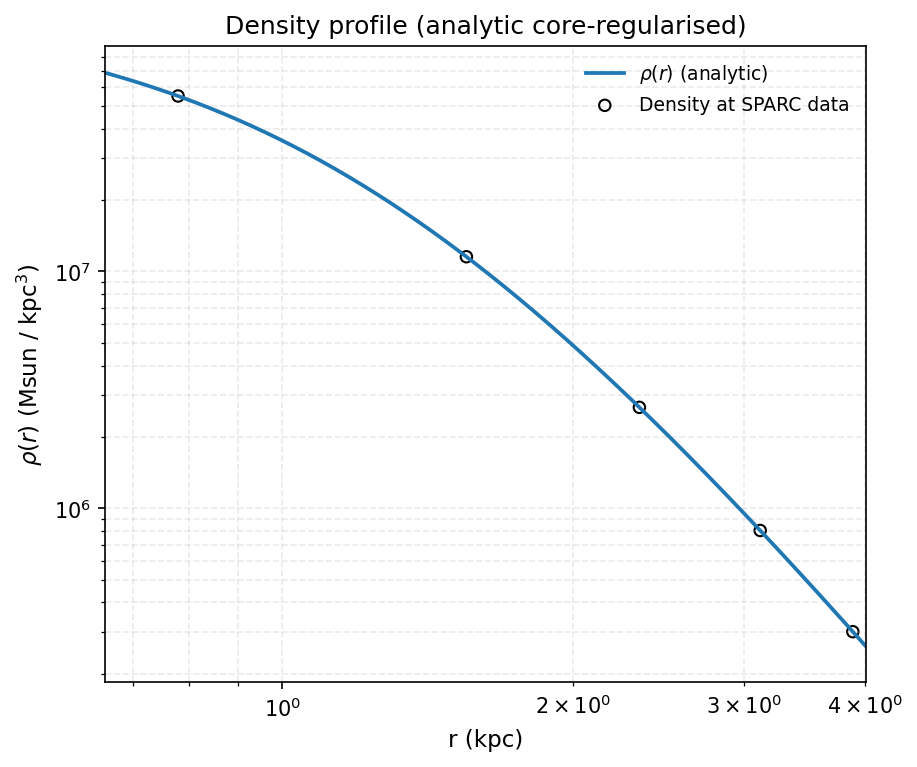}
\caption{The density of the SIDM model of Eq.
(\ref{ScaledependentEoSDM}) for the galaxy UGC09992, versus the
radius.} \label{UGC09992dens}
\end{figure}

\begin{figure}[h!]
\centering
\includegraphics[width=35pc]{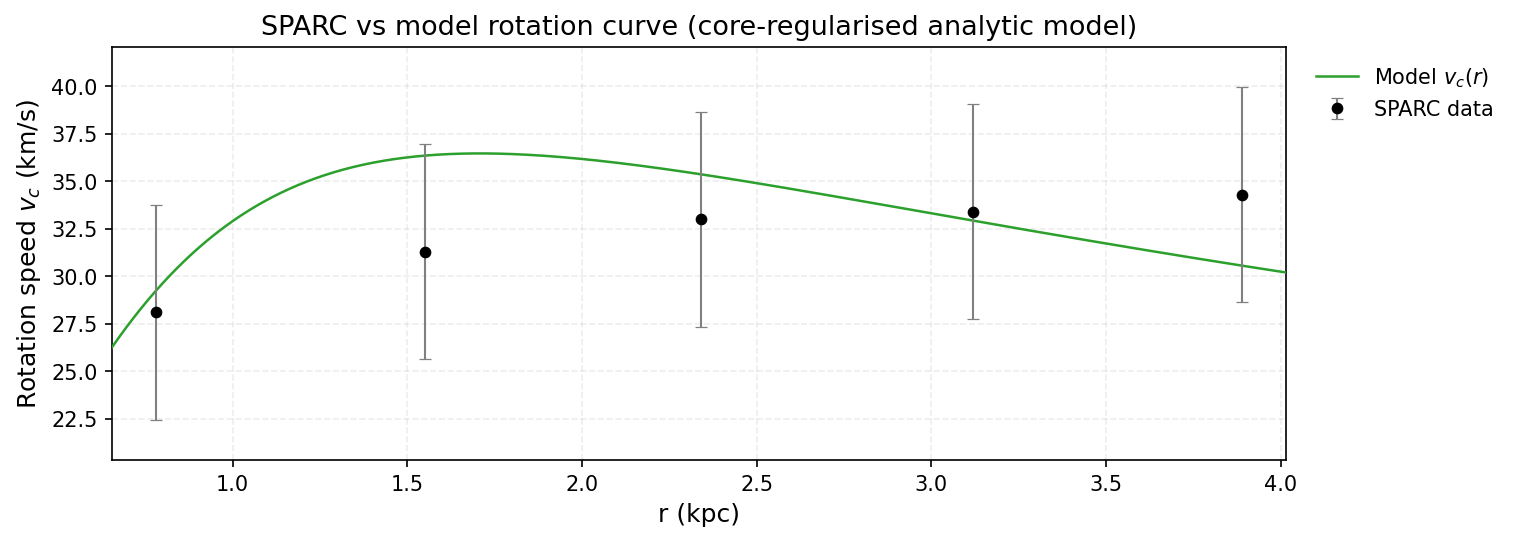}
\caption{The predicted rotation curves for the optimized SIDM
model of Eq. (\ref{ScaledependentEoSDM}), versus the SPARC
observational data for the galaxy UGC09992.} \label{UGC09992}
\end{figure}

In Table \ref{collUGC09992} we present the optimized values of
$K_0$ and $\rho_0$ for the analytic SIDM model of Eq.
(\ref{ScaledependentEoSDM}) for which the maximum compatibility
with the SPARC data is achieved.
\begin{table}[h!]
  \begin{center}
    \caption{SIDM Optimization Values for the galaxy UGC09992}
    \label{collUGC09992}
     \begin{tabular}{|r|r|}
     \hline
      \textbf{Parameter}   & \textbf{Optimization Values}
      \\  \hline
     $\rho_0 $  ($M_{\odot}/\mathrm{Kpc}^{3}$) & $1.31564\times 10^8$
\\  \hline $K_0$ ($M_{\odot} \,
\mathrm{Kpc}^{-3} \, (\mathrm{km/s})^{2}$)& 575.789
\\  \hline
    \end{tabular}
  \end{center}
\end{table}
In Figs. \ref{UGC09992dens}, \ref{UGC09992}  we present the
density of the analytic SIDM model, the predicted rotation curves
for the SIDM model (\ref{ScaledependentEoSDM}), versus the SPARC
observational data and the sound speed, as a function of the
radius respectively. As it can be seen, for this galaxy, the SIDM
model produces viable rotation curves which are compatible with
the SPARC data.

\subsection{The Galaxy UGC10310}

For this galaxy, the optimization method we used, ensures maximum
compatibility of the analytic SIDM model of Eq.
(\ref{ScaledependentEoSDM}) with the SPARC data, if we choose
$\rho_0=5.77999\times 10^7$$M_{\odot}/\mathrm{Kpc}^{3}$ and
$K_0=2353.19
$$M_{\odot} \, \mathrm{Kpc}^{-3} \, (\mathrm{km/s})^{2}$, in which
case the reduced $\chi^2_{red}$ value is $\chi^2_{red}=0.33138$.
Also the parameter $\alpha$ in this case is $\alpha=3.6822 $Kpc.

\begin{figure}[h!]
\centering
\includegraphics[width=20pc]{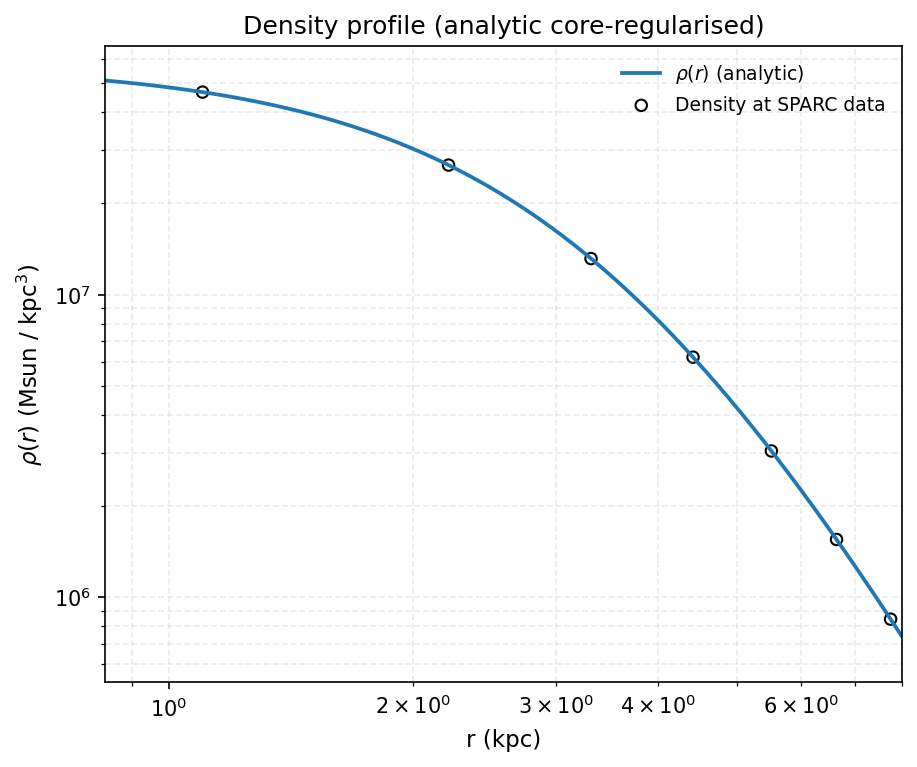}
\caption{The density of the SIDM model of Eq.
(\ref{ScaledependentEoSDM}) for the galaxy UGC10310, versus the
radius.} \label{UGC10310dens}
\end{figure}

\begin{figure}[h!]
\centering
\includegraphics[width=20pc]{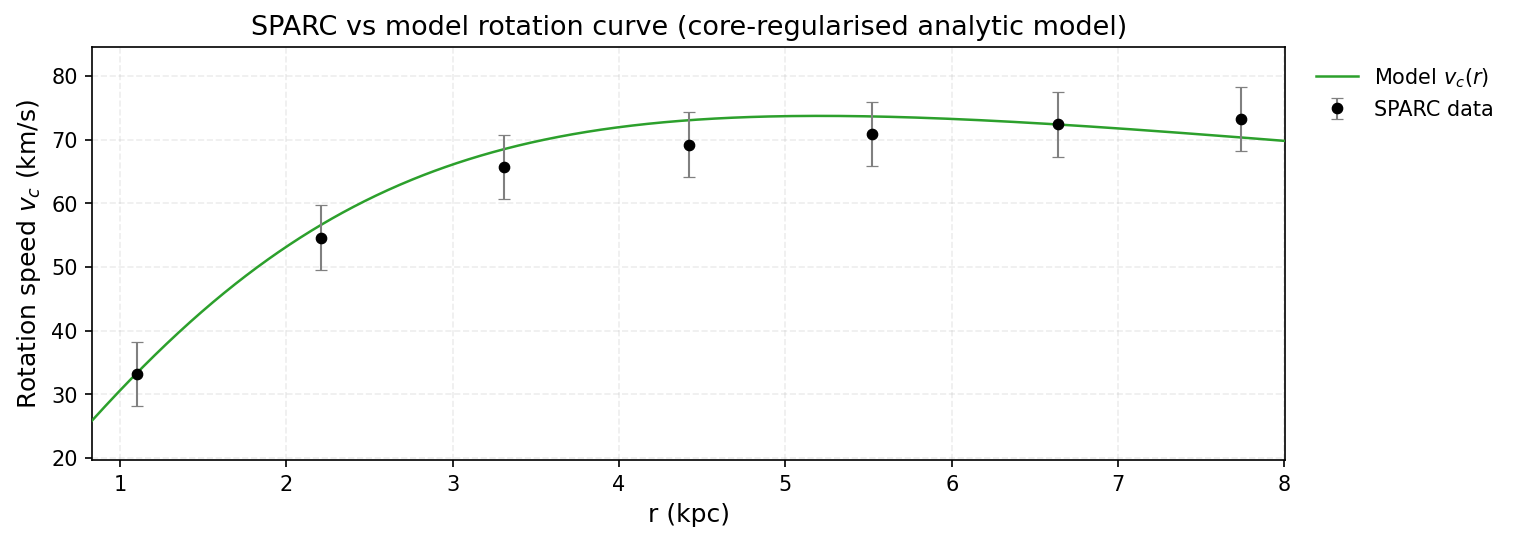}
\caption{The predicted rotation curves for the optimized SIDM
model of Eq. (\ref{ScaledependentEoSDM}), versus the SPARC
observational data for the galaxy UGC10310.} \label{UGC10310}
\end{figure}

In Table \ref{collUGC10310} we present the optimized values of
$K_0$ and $\rho_0$ for the analytic SIDM model of Eq.
(\ref{ScaledependentEoSDM}) for which the maximum compatibility
with the SPARC data is achieved.
\begin{table}[h!]
  \begin{center}
    \caption{SIDM Optimization Values for the galaxy UGC10310}
    \label{collUGC10310}
     \begin{tabular}{|r|r|}
     \hline
      \textbf{Parameter}   & \textbf{Optimization Values}
      \\  \hline
     $\rho_0 $  ($M_{\odot}/\mathrm{Kpc}^{3}$) & $5.77999\times 10^7$
\\  \hline $K_0$ ($M_{\odot} \,
\mathrm{Kpc}^{-3} \, (\mathrm{km/s})^{2}$)& 2353.19
\\  \hline
    \end{tabular}
  \end{center}
\end{table}
In Figs. \ref{UGC10310dens}, \ref{UGC10310} we present the density
of the analytic SIDM model, the predicted rotation curves for the
SIDM model (\ref{ScaledependentEoSDM}), versus the SPARC
observational data and the sound speed, as a function of the
radius respectively. As it can be seen, for this galaxy, the SIDM
model produces viable rotation curves which are compatible with
the SPARC data.

\subsection{The Galaxy UGC11455, Non-viable}

For this galaxy, the optimization method we used, ensures maximum
compatibility of the analytic SIDM model of Eq.
(\ref{ScaledependentEoSDM}) with the SPARC data, if we choose
$\rho_0=7.79297\times 10^7$$M_{\odot}/\mathrm{Kpc}^{3}$ and
$K_0=37694.7
$$M_{\odot} \, \mathrm{Kpc}^{-3} \, (\mathrm{km/s})^{2}$, in which
case the reduced $\chi^2_{red}$ value is $\chi^2_{red}=3.84583$.
Also the parameter $\alpha$ in this case is $\alpha=12.6923 $Kpc.

\begin{figure}[h!]
\centering
\includegraphics[width=20pc]{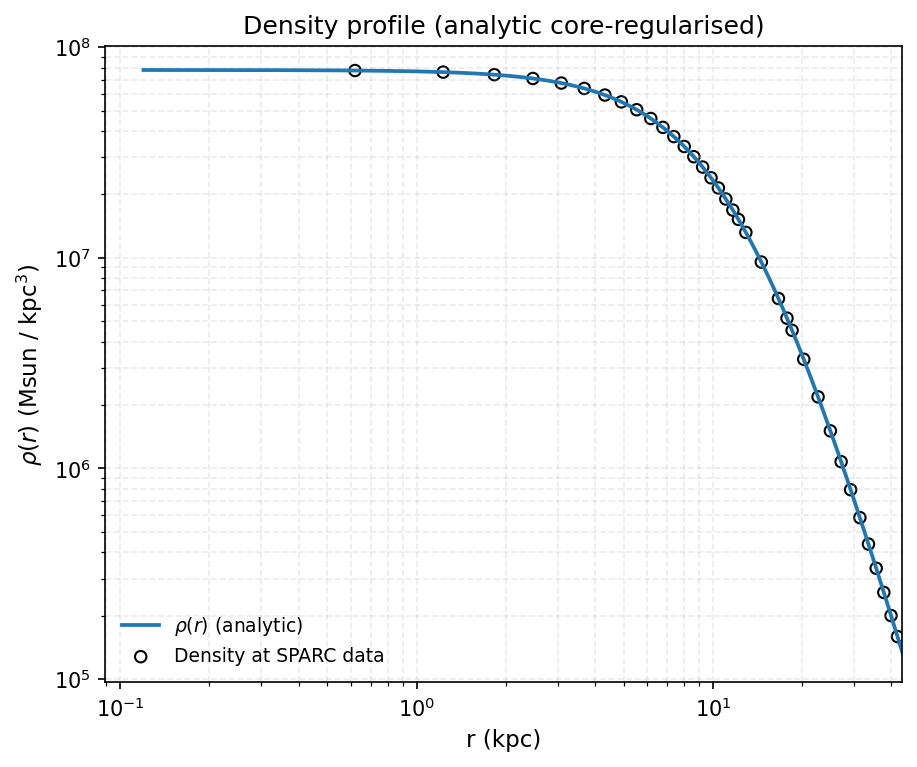}
\caption{The density of the SIDM model of Eq.
(\ref{ScaledependentEoSDM}) for the galaxy UGC11455, versus the
radius.} \label{UGC11455dens}
\end{figure}

\begin{figure}[h!]
\centering
\includegraphics[width=35pc]{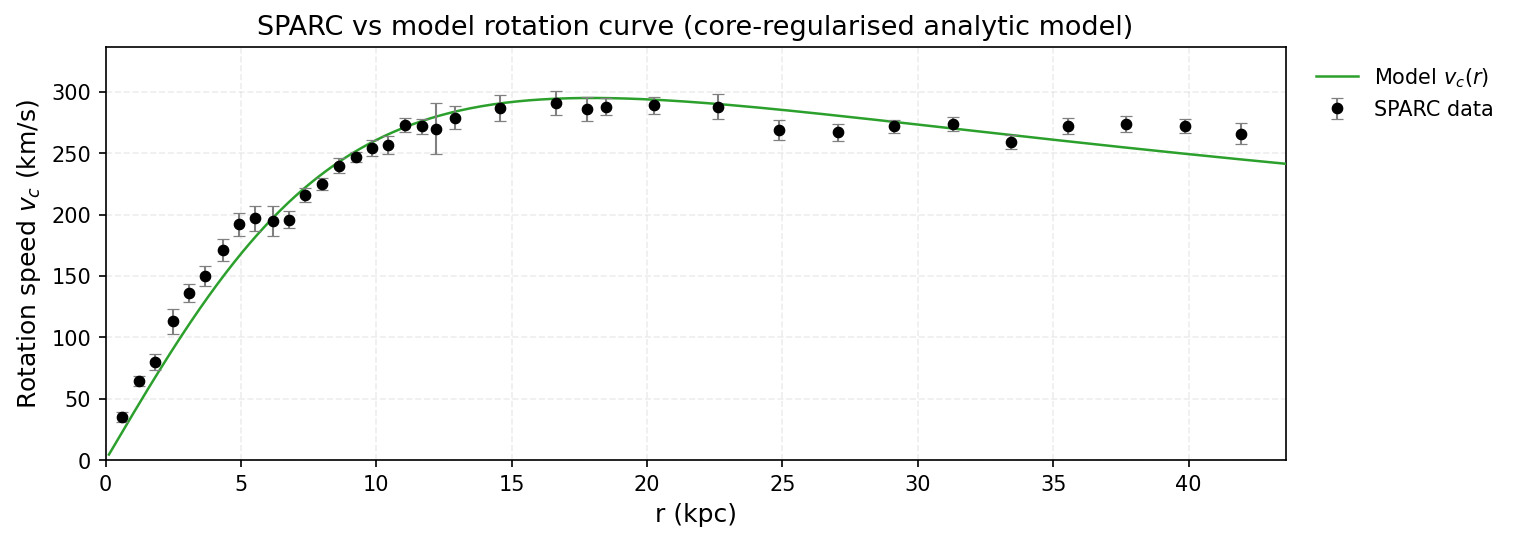}
\caption{The predicted rotation curves for the optimized SIDM
model of Eq. (\ref{ScaledependentEoSDM}), versus the SPARC
observational data for the galaxy UGC11455.} \label{UGC11455}
\end{figure}

In Table \ref{collUGC11455} we present the optimized values of
$K_0$ and $\rho_0$ for the analytic SIDM model of Eq.
(\ref{ScaledependentEoSDM}) for which the maximum compatibility
with the SPARC data is achieved.
\begin{table}[h!]
  \begin{center}
    \caption{SIDM Optimization Values for the galaxy UGC11455}
    \label{collUGC11455}
     \begin{tabular}{|r|r|}
     \hline
      \textbf{Parameter}   & \textbf{Optimization Values}
      \\  \hline
     $\rho_0 $  ($M_{\odot}/\mathrm{Kpc}^{3}$) & $7.79297\times 10^7$
\\  \hline $K_0$ ($M_{\odot} \,
\mathrm{Kpc}^{-3} \, (\mathrm{km/s})^{2}$)& 37694.7
\\  \hline
    \end{tabular}
  \end{center}
\end{table}
In Figs. \ref{UGC11455dens}, \ref{UGC11455}  we present the
density of the analytic SIDM model, the predicted rotation curves
for the SIDM model (\ref{ScaledependentEoSDM}), versus the SPARC
observational data and the sound speed, as a function of the
radius respectively. As it can be seen, for this galaxy, the SIDM
model produces non-viable rotation curves which are incompatible
with the SPARC data.

\begin{figure}[h!]
\centering
\includegraphics[width=20pc]{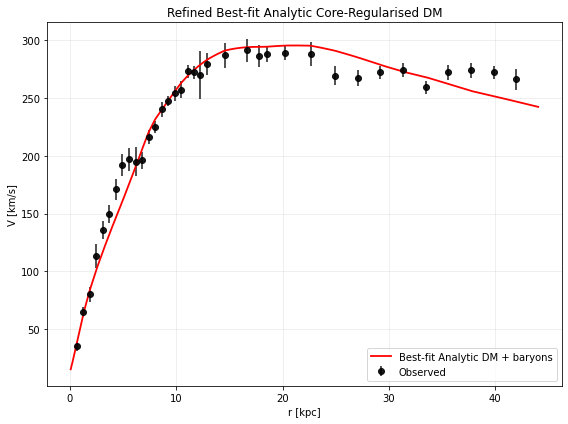}
\caption{The predicted rotation curves after using an optimization
for the SIDM model (\ref{ScaledependentEoSDM}), and the extended
SPARC data for the galaxy UGC11455. We included the rotation
curves of the gas, the disk velocities, the bulge (where present)
along with the SIDM model.} \label{extendedUGC11455}
\end{figure}

Now we shall include contributions to the rotation velocity from
the other components of the galaxy, namely the disk, the gas, and
the bulge if present. In Fig. \ref{extendedUGC11455} we present
the combined rotation curves including all the components of the
galaxy along with the SIDM. As it can be seen, the extended
collisional DM model is non-viable.
Also in Table \ref{evaluationextendedUGC11455} we present the
optimized values of the free parameters of the SIDM model for
which  we achieve the maximum compatibility with the SPARC data,
for the galaxy UGC11455, and also the resulting reduced
$\chi^2_{red}$ value.
\begin{table}[h!]
\centering \caption{Optimized Parameter Values of the Extended
SIDM model for the Galaxy UGC11455.}
\begin{tabular}{lc}
\hline
Parameter & Value  \\
\hline
$\rho_0 $ ($M_{\odot}/\mathrm{Kpc}^{3}$) & $4.73819\times 10^7$   \\
$K_0$ ($M_{\odot} \,
\mathrm{Kpc}^{-3} \, (\mathrm{km/s})^{2}$) & 28755   \\
$ml_{\text{disk}}$ & 0.4580 \\
$ml_{\text{bulge}}$ & 0.5154 \\
$\alpha$ (Kpc) & 14.215\\
$\chi^2_{red}$ & 2.87458 \\
\hline
\end{tabular}
\label{evaluationextendedUGC11455}
\end{table}

\subsection{The Galaxy UGC11557}

For this galaxy, the optimization method we used, ensures maximum
compatibility of the analytic SIDM model of Eq.
(\ref{ScaledependentEoSDM}) with the SPARC data, if we choose
$\rho_0=2.88776\times 10^7$$M_{\odot}/\mathrm{Kpc}^{3}$ and
$K_0=3089.47
$$M_{\odot} \, \mathrm{Kpc}^{-3} \, (\mathrm{km/s})^{2}$, in which
case the reduced $\chi^2_{red}$ value is $\chi^2_{red}=0.423089$.
Also the parameter $\alpha$ in this case is $\alpha=6.85794 $Kpc.

\begin{figure}[h!]
\centering
\includegraphics[width=20pc]{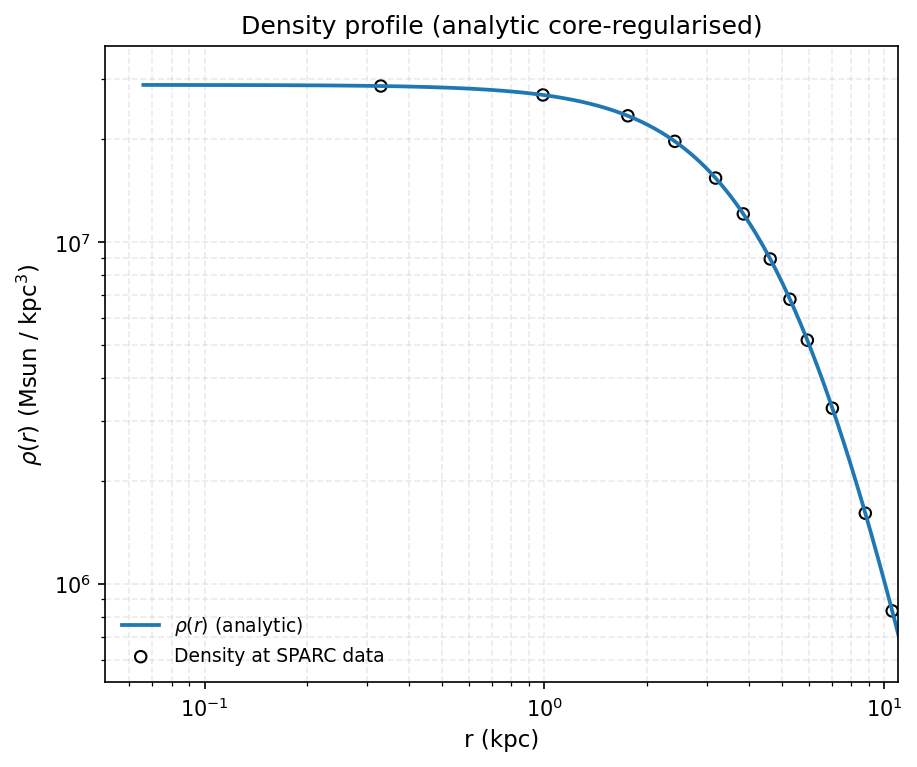}
\caption{The density of the SIDM model of Eq.
(\ref{ScaledependentEoSDM}) for the galaxy UGC11557, versus the
radius.} \label{UGC11557dens}
\end{figure}

\begin{figure}[h!]
\centering
\includegraphics[width=35pc]{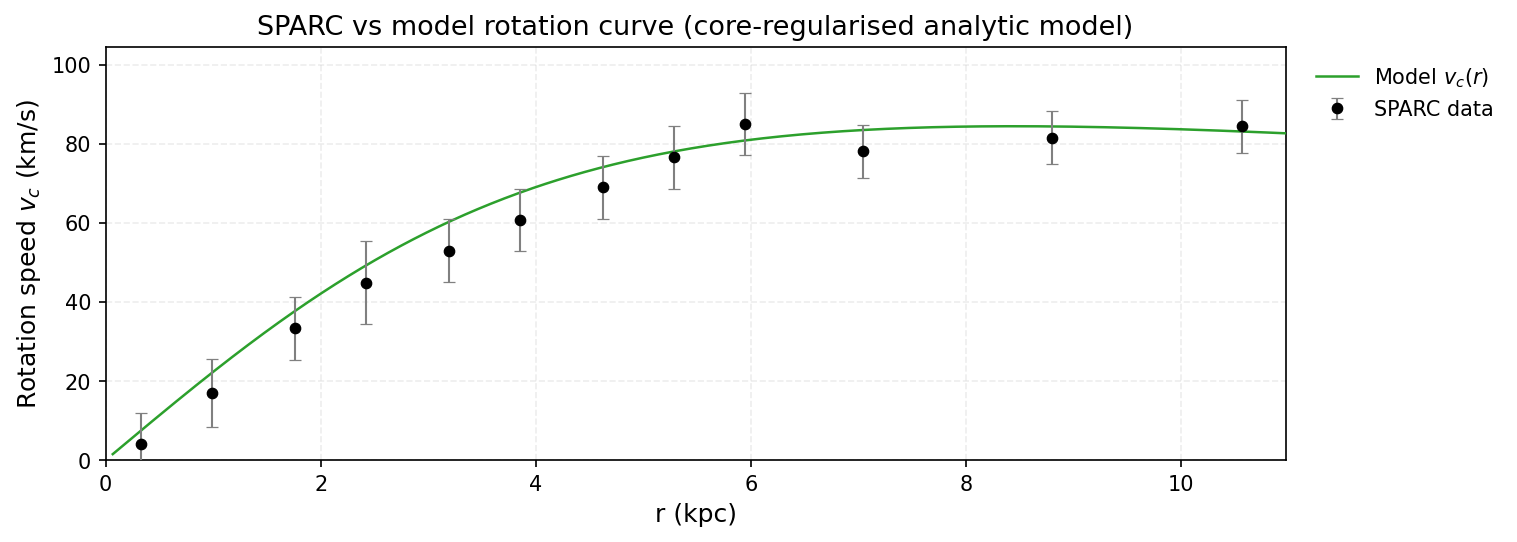}
\caption{The predicted rotation curves for the optimized SIDM
model of Eq. (\ref{ScaledependentEoSDM}), versus the SPARC
observational data for the galaxy UGC11557.} \label{UGC11557}
\end{figure}

In Table \ref{collUGC11557} we present the optimized values of
$K_0$ and $\rho_0$ for the analytic SIDM model of Eq.
(\ref{ScaledependentEoSDM}) for which the maximum compatibility
with the SPARC data is achieved.
\begin{table}[h!]
  \begin{center}
    \caption{SIDM Optimization Values for the galaxy UGC11557}
    \label{collUGC11557}
     \begin{tabular}{|r|r|}
     \hline
      \textbf{Parameter}   & \textbf{Optimization Values}
      \\  \hline
     $\rho_0 $  ($M_{\odot}/\mathrm{Kpc}^{3}$) & $2.88776\times 10^7$
\\  \hline $K_0$ ($M_{\odot} \,
\mathrm{Kpc}^{-3} \, (\mathrm{km/s})^{2}$)& 3089.47
\\  \hline
    \end{tabular}
  \end{center}
\end{table}
In Figs. \ref{UGC11557dens}, \ref{UGC11557}  we present the
density of the analytic SIDM model, the predicted rotation curves
for the SIDM model (\ref{ScaledependentEoSDM}), versus the SPARC
observational data and the sound speed, as a function of the
radius respectively. As it can be seen, for this galaxy, the SIDM
model produces viable rotation curves which are compatible with
the SPARC data.

\subsection{The Galaxy UGC11820, Non-viable}

For this galaxy, the optimization method we used, ensures maximum
compatibility of the analytic SIDM model of Eq.
(\ref{ScaledependentEoSDM}) with the SPARC data, if we choose
$\rho_0=9.84877\times 10^7$$M_{\odot}/\mathrm{Kpc}^{3}$ and
$K_0=3474.48
$$M_{\odot} \, \mathrm{Kpc}^{-3} \, (\mathrm{km/s})^{2}$, in which
case the reduced $\chi^2_{red}$ value is $\chi^2_{red}=188.997$.
Also the parameter $\alpha$ in this case is $\alpha=3.42772 $Kpc.

\begin{figure}[h!]
\centering
\includegraphics[width=20pc]{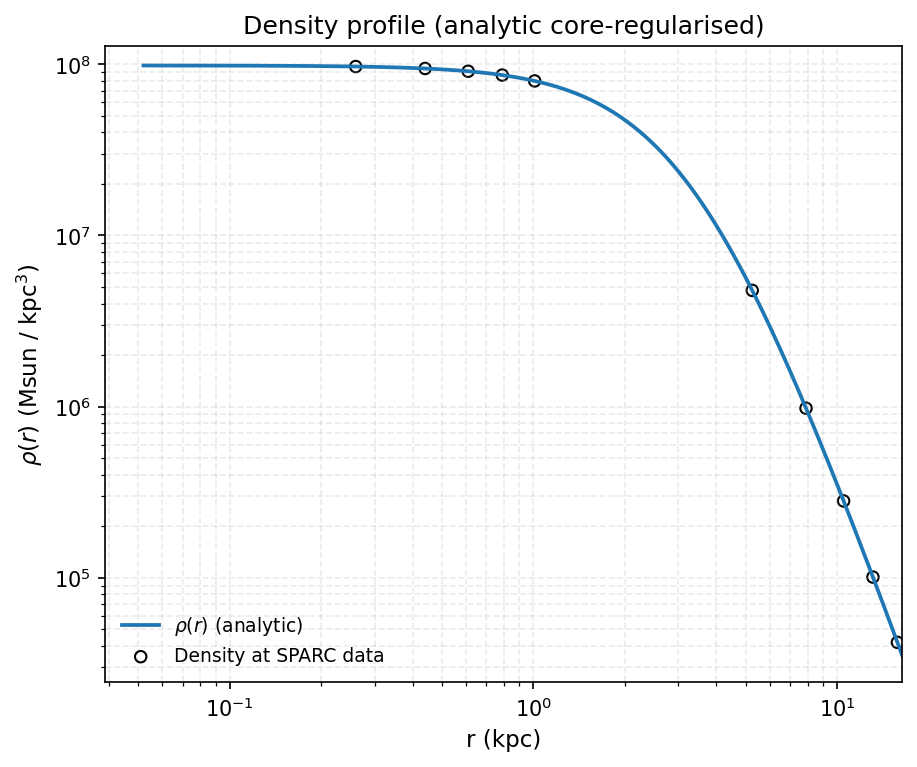}
\caption{The density of the SIDM model of Eq.
(\ref{ScaledependentEoSDM}) for the galaxy UGC11820, versus the
radius.} \label{UGC11820dens}
\end{figure}

\begin{figure}[h!]
\centering
\includegraphics[width=35pc]{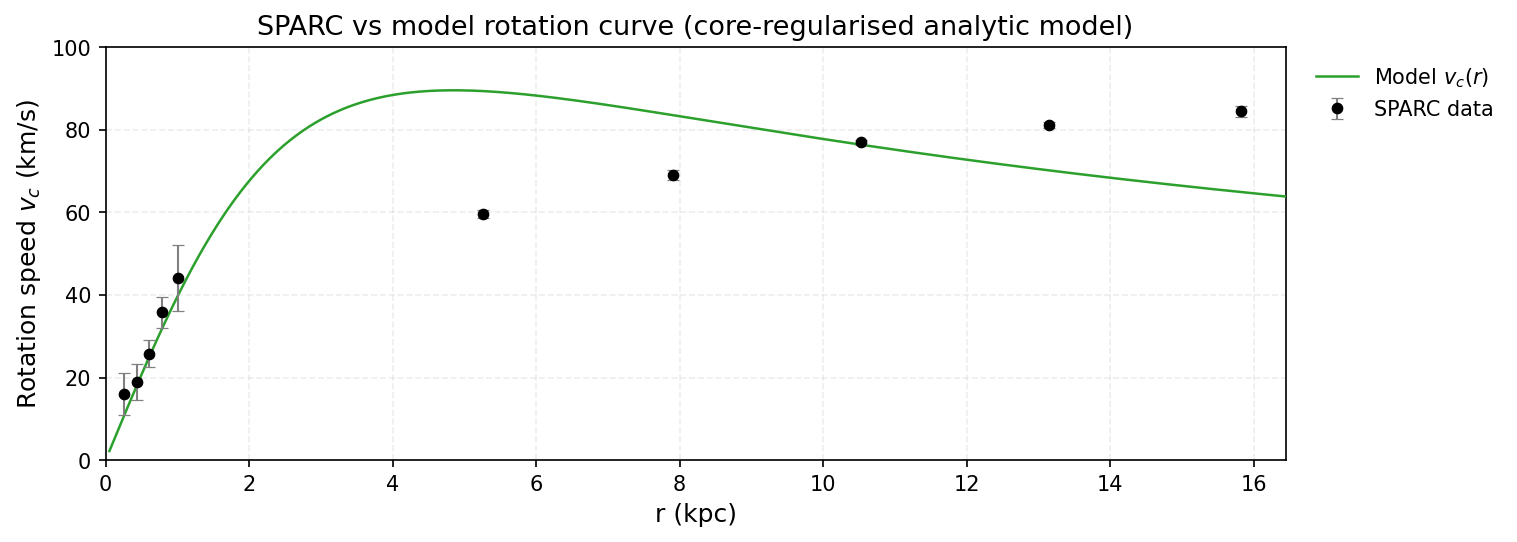}
\caption{The predicted rotation curves for the optimized SIDM
model of Eq. (\ref{ScaledependentEoSDM}), versus the SPARC
observational data for the galaxy UGC11820.} \label{UGC11820}
\end{figure}

In Table \ref{collUGC11820} we present the optimized values of
$K_0$ and $\rho_0$ for the analytic SIDM model of Eq.
(\ref{ScaledependentEoSDM}) for which the maximum compatibility
with the SPARC data is achieved.
\begin{table}[h!]
  \begin{center}
    \caption{SIDM Optimization Values for the galaxy UGC11820}
    \label{collUGC11820}
     \begin{tabular}{|r|r|}
     \hline
      \textbf{Parameter}   & \textbf{Optimization Values}
      \\  \hline
     $\rho_0 $  ($M_{\odot}/\mathrm{Kpc}^{3}$) & $5\times 10^7$
\\  \hline $K_0$ ($M_{\odot} \,
\mathrm{Kpc}^{-3} \, (\mathrm{km/s})^{2}$)& 1250
\\  \hline
    \end{tabular}
  \end{center}
\end{table}
In Figs. \ref{UGC11820dens}, \ref{UGC11820}  we present the
density of the analytic SIDM model, the predicted rotation curves
for the SIDM model (\ref{ScaledependentEoSDM}), versus the SPARC
observational data and the sound speed, as a function of the
radius respectively. As it can be seen, for this galaxy, the SIDM
model produces non-viable rotation curves which are incompatible
with the SPARC data.

\begin{figure}[h!]
\centering
\includegraphics[width=20pc]{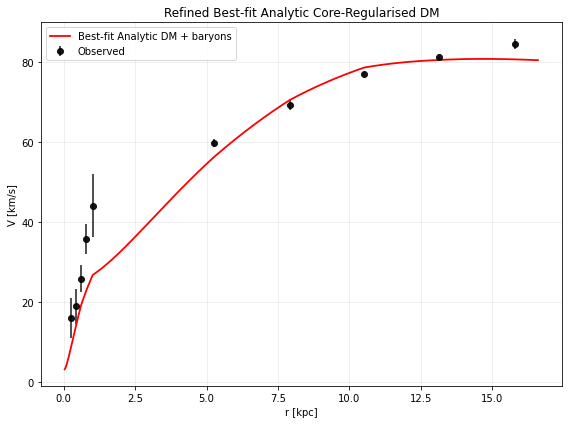}
\caption{The predicted rotation curves after using an optimization
for the SIDM model (\ref{ScaledependentEoSDM}), and the extended
SPARC data for the galaxy UGC11820. We included the rotation
curves of the gas, the disk velocities, the bulge (where present)
along with the SIDM model.} \label{extendedUGC11820}
\end{figure}

Now we shall include contributions to the rotation velocity from
the other components of the galaxy, namely the disk, the gas, and
the bulge if present. In Fig. \ref{extendedUGC11820} we present
the combined rotation curves including all the components of the
galaxy along with the SIDM. As it can be seen, the extended
collisional DM model is non-viable.
Also in Table \ref{evaluationextendedUGC11820} we present the
optimized values of the free parameters of the SIDM model for
which  we achieve the maximum compatibility with the SPARC data,
for the galaxy UGC11820, and also the resulting reduced
$\chi^2_{red}$ value.
\begin{table}[h!]
\centering \caption{Optimized Parameter Values of the Extended
SIDM model for the Galaxy UGC11820.}
\begin{tabular}{lc}
\hline
Parameter & Value  \\
\hline
$\rho_0 $ ($M_{\odot}/\mathrm{Kpc}^{3}$) & $5.88535\times 10^6$   \\
$K_0$ ($M_{\odot} \,
\mathrm{Kpc}^{-3} \, (\mathrm{km/s})^{2}$) & 2149.24  \\
$ml_{\text{disk}}$ & 1 \\
$ml_{\text{bulge}}$ & 0.2255 \\
$\alpha$ (Kpc) & 11.0269\\
$\chi^2_{red}$ & 9.15937 \\
\hline
\end{tabular}
\label{evaluationextendedUGC11820}
\end{table}

\subsection{The Galaxy UGC11914, Non-viable}

For this galaxy, the optimization method we used, ensures maximum
compatibility of the analytic SIDM model of Eq.
(\ref{ScaledependentEoSDM}) with the SPARC data, if we choose
$\rho_0=3.80519\times 10^9$$M_{\odot}/\mathrm{Kpc}^{3}$ and
$K_0=44161.3
$$M_{\odot} \, \mathrm{Kpc}^{-3} \, (\mathrm{km/s})^{2}$, in which
case the reduced $\chi^2_{red}$ value is $\chi^2_{red}=54.7408$.
Also the parameter $\alpha$ in this case is $\alpha=1.966 $Kpc.

\begin{figure}[h!]
\centering
\includegraphics[width=20pc]{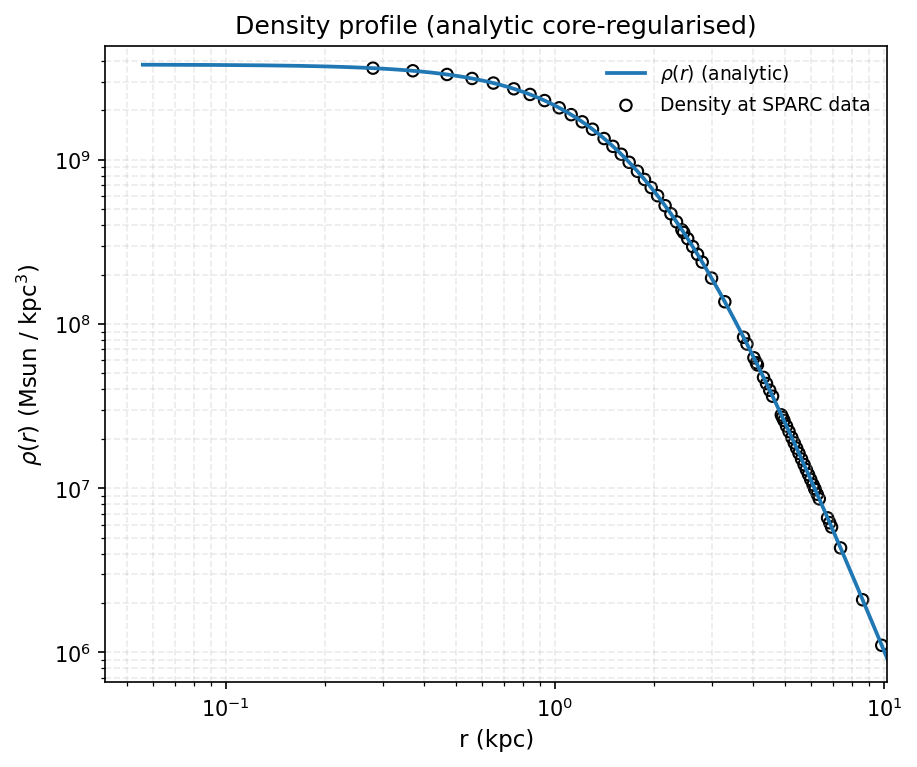}
\caption{The density of the SIDM model of Eq.
(\ref{ScaledependentEoSDM}) for the galaxy UGC11914, versus the
radius.} \label{UGC11914dens}
\end{figure}

\begin{figure}[h!]
\centering
\includegraphics[width=35pc]{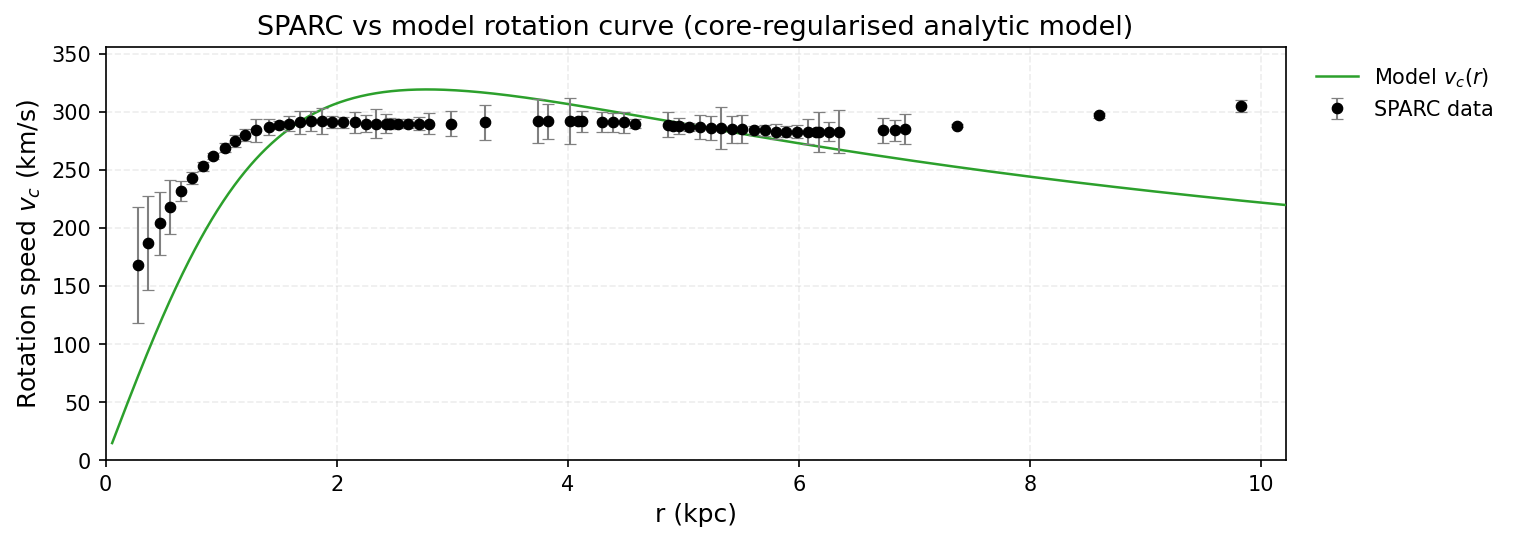}
\caption{The predicted rotation curves for the optimized SIDM
model of Eq. (\ref{ScaledependentEoSDM}), versus the SPARC
observational data for the galaxy UGC11914.} \label{UGC11914}
\end{figure}

In Table \ref{collUGC11914} we present the optimized values of
$K_0$ and $\rho_0$ for the analytic SIDM model of Eq.
(\ref{ScaledependentEoSDM}) for which the maximum compatibility
with the SPARC data is achieved.
\begin{table}[h!]
  \begin{center}
    \caption{SIDM Optimization Values for the galaxy UGC11914}
    \label{collUGC11914}
     \begin{tabular}{|r|r|}
     \hline
      \textbf{Parameter}   & \textbf{Optimization Values}
      \\  \hline
     $\rho_0 $  ($M_{\odot}/\mathrm{Kpc}^{3}$) & $3.80519\times 10^9$
\\  \hline $K_0$ ($M_{\odot} \,
\mathrm{Kpc}^{-3} \, (\mathrm{km/s})^{2}$)& 44161.3
\\  \hline
    \end{tabular}
  \end{center}
\end{table}
In Figs. \ref{UGC11914dens}, \ref{UGC11914} we present the density
of the analytic SIDM model, the predicted rotation curves for the
SIDM model (\ref{ScaledependentEoSDM}), versus the SPARC
observational data and the sound speed, as a function of the
radius respectively. As it can be seen, for this galaxy, the SIDM
model produces non-viable rotation curves which are incompatible
with the SPARC data.

\begin{figure}[h!]
\centering
\includegraphics[width=20pc]{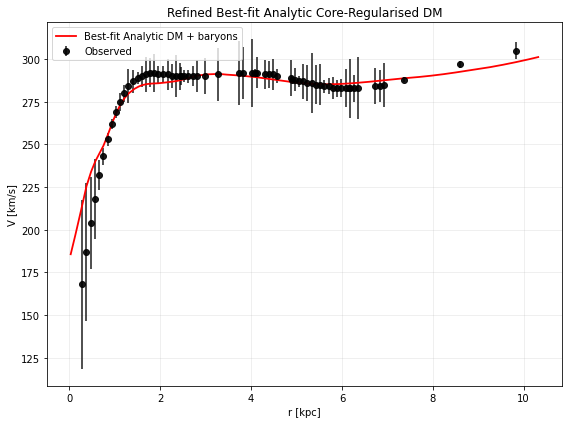}
\caption{The predicted rotation curves after using an optimization
for the SIDM model (\ref{ScaledependentEoSDM}), and the extended
SPARC data for the galaxy UGC11914. We included the rotation
curves of the gas, the disk velocities, the bulge (where present)
along with the SIDM model.} \label{extendedUGC11914}
\end{figure}

Now we shall include contributions to the rotation velocity from
the other components of the galaxy, namely the disk, the gas, and
the bulge if present. In Fig. \ref{extendedUGC11914} we present
the combined rotation curves including all the components of the
galaxy along with the SIDM. As it can be seen, the extended
collisional DM model is non-viable.
Also in Table \ref{evaluationextendedUGC11914} we present the
optimized values of the free parameters of the SIDM model for
which  we achieve the maximum compatibility with the SPARC data,
for the galaxy UGC11914, and also the resulting reduced
$\chi^2_{red}$ value.
\begin{table}[h!]
\centering \caption{Optimized Parameter Values of the Extended
SIDM model for the Galaxy UGC11914.}
\begin{tabular}{lc}
\hline
Parameter & Value  \\
\hline
$\rho_0 $ ($M_{\odot}/\mathrm{Kpc}^{3}$) & $2.37289\times 10^7$   \\
$K_0$ ($M_{\odot} \,
\mathrm{Kpc}^{-3} \, (\mathrm{km/s})^{2}$) & 125459   \\
$ml_{\text{disk}}$ & 0.7914 \\
$ml_{\text{bulge}}$ & 1 \\
$\alpha$ (Kpc) & 118.456\\
$\chi^2_{red}$ & 0.527133 \\
\hline
\end{tabular}
\label{evaluationextendedUGC11914}
\end{table}

\subsection{The Galaxy UGC12506, Non-viable}

For this galaxy, the optimization method we used, ensures maximum
compatibility of the analytic SIDM model of Eq.
(\ref{ScaledependentEoSDM}) with the SPARC data, if we choose
$\rho_0=6.11517\times 10^7$$M_{\odot}/\mathrm{Kpc}^{3}$ and
$K_0=30315.4
$$M_{\odot} \, \mathrm{Kpc}^{-3} \, (\mathrm{km/s})^{2}$, in which
case the reduced $\chi^2_{red}$ value is $\chi^2_{red}=2.65417$.
Also the parameter $\alpha$ in this case is $\alpha=12.8492 $Kpc.

\begin{figure}[h!]
\centering
\includegraphics[width=20pc]{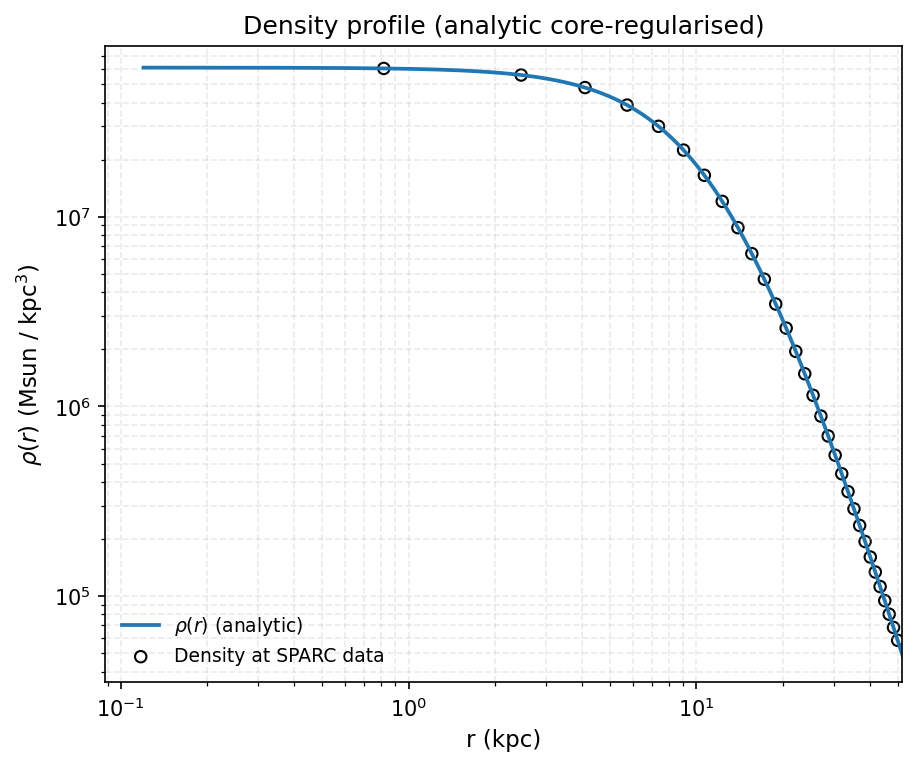}
\caption{The density of the SIDM model of Eq.
(\ref{ScaledependentEoSDM}) for the galaxy UGC12506, versus the
radius.} \label{UGC12506dens}
\end{figure}

\begin{figure}[h!]
\centering
\includegraphics[width=35pc]{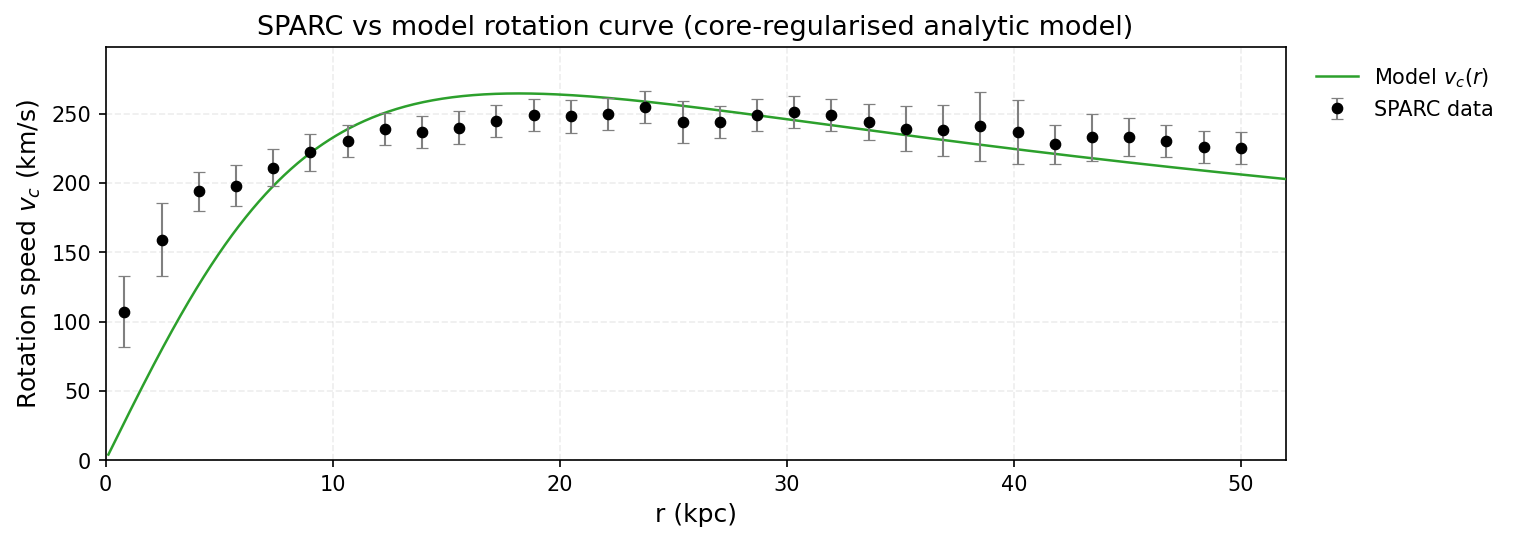}
\caption{The predicted rotation curves for the optimized SIDM
model of Eq. (\ref{ScaledependentEoSDM}), versus the SPARC
observational data for the galaxy UGC12506.} \label{UGC12506}
\end{figure}

In Table \ref{collUGC12506} we present the optimized values of
$K_0$ and $\rho_0$ for the analytic SIDM model of Eq.
(\ref{ScaledependentEoSDM}) for which the maximum compatibility
with the SPARC data is achieved.
\begin{table}[h!]
  \begin{center}
    \caption{SIDM Optimization Values for the galaxy UGC12506}
    \label{collUGC12506}
     \begin{tabular}{|r|r|}
     \hline
      \textbf{Parameter}   & \textbf{Optimization Values}
      \\  \hline
     $\rho_0 $  ($M_{\odot}/\mathrm{Kpc}^{3}$) & $6.11517\times 10^7$
\\  \hline $K_0$ ($M_{\odot} \,
\mathrm{Kpc}^{-3} \, (\mathrm{km/s})^{2}$)& 30315.4
\\  \hline
    \end{tabular}
  \end{center}
\end{table}
In Figs. \ref{UGC12506dens}, \ref{UGC12506}  we present the
density of the analytic SIDM model, the predicted rotation curves
for the SIDM model (\ref{ScaledependentEoSDM}), versus the SPARC
observational data and the sound speed, as a function of the
radius respectively. As it can be seen, for this galaxy, the SIDM
model produces non-viable rotation curves which are incompatible
with the SPARC data.

\begin{figure}[h!]
\centering
\includegraphics[width=20pc]{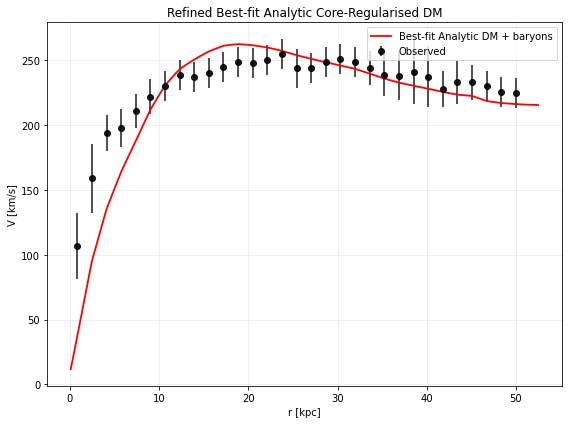}
\caption{The predicted rotation curves after using an optimization
for the SIDM model (\ref{ScaledependentEoSDM}), and the extended
SPARC data for the galaxy UGC12506. We included the rotation
curves of the gas, the disk velocities, the bulge (where present)
along with the SIDM model.} \label{extendedUGC12506}
\end{figure}

Now we shall include contributions to the rotation velocity from
the other components of the galaxy, namely the disk, the gas, and
the bulge if present. In Fig. \ref{extendedUGC12506} we present
the combined rotation curves including all the components of the
galaxy along with the SIDM. As it can be seen, the extended
collisional DM model is non-viable.
Also in Table \ref{evaluationextendedUGC12506} we present the
optimized values of the free parameters of the SIDM model for
which  we achieve the maximum compatibility with the SPARC data,
for the galaxy UGC12506, and also the resulting reduced
$\chi^2_{red}$ value.
\begin{table}[h!]
\centering \caption{Optimized Parameter Values of the Extended
SIDM model for the Galaxy UGC12506.}
\begin{tabular}{lc}
\hline
Parameter & Value  \\
\hline
$\rho_0 $ ($M_{\odot}/\mathrm{Kpc}^{3}$) & $2.08225\times 10^7$   \\
$K_0$ ($M_{\odot} \,
\mathrm{Kpc}^{-3} \, (\mathrm{km/s})^{2}$) & 15949.2   \\
$ml_{\text{disk}}$ & 1 \\
$ml_{\text{bulge}}$ & 1 \\
$\alpha$ (Kpc) & 15.9698\\
$\chi^2_{red}$ & 2.0016 \\
\hline
\end{tabular}
\label{evaluationextendedUGC12506}
\end{table}

\subsection{The Galaxy UGC12632}

For this galaxy, the optimization method we used, ensures maximum
compatibility of the analytic SIDM model of Eq.
(\ref{ScaledependentEoSDM}) with the SPARC data, if we choose
$\rho_0=2.59734\times 10^7$$M_{\odot}/\mathrm{Kpc}^{3}$ and
$K_0=2234.72
$$M_{\odot} \, \mathrm{Kpc}^{-3} \, (\mathrm{km/s})^{2}$, in which
case the reduced $\chi^2_{red}$ value is $\chi^2_{red}=0.984724$.
Also the parameter $\alpha$ in this case is $\alpha=5.35301 $Kpc.

\begin{figure}[h!]
\centering
\includegraphics[width=20pc]{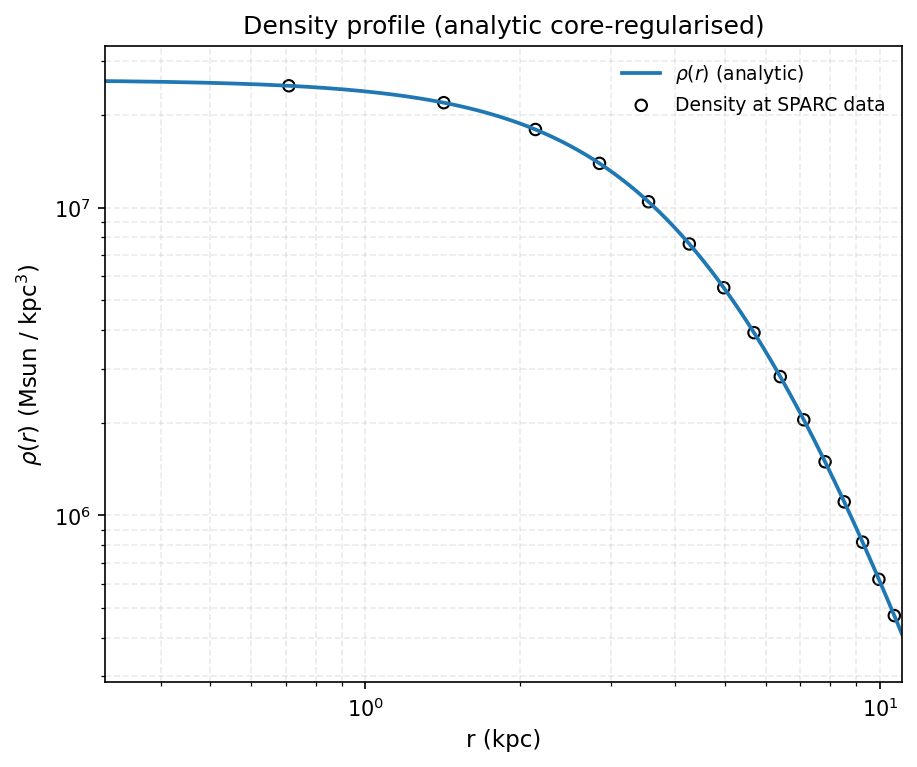}
\caption{The density of the SIDM model of Eq.
(\ref{ScaledependentEoSDM}) for the galaxy UGC12632, versus the
radius.} \label{UGC12632dens}
\end{figure}

\begin{figure}[h!]
\centering
\includegraphics[width=35pc]{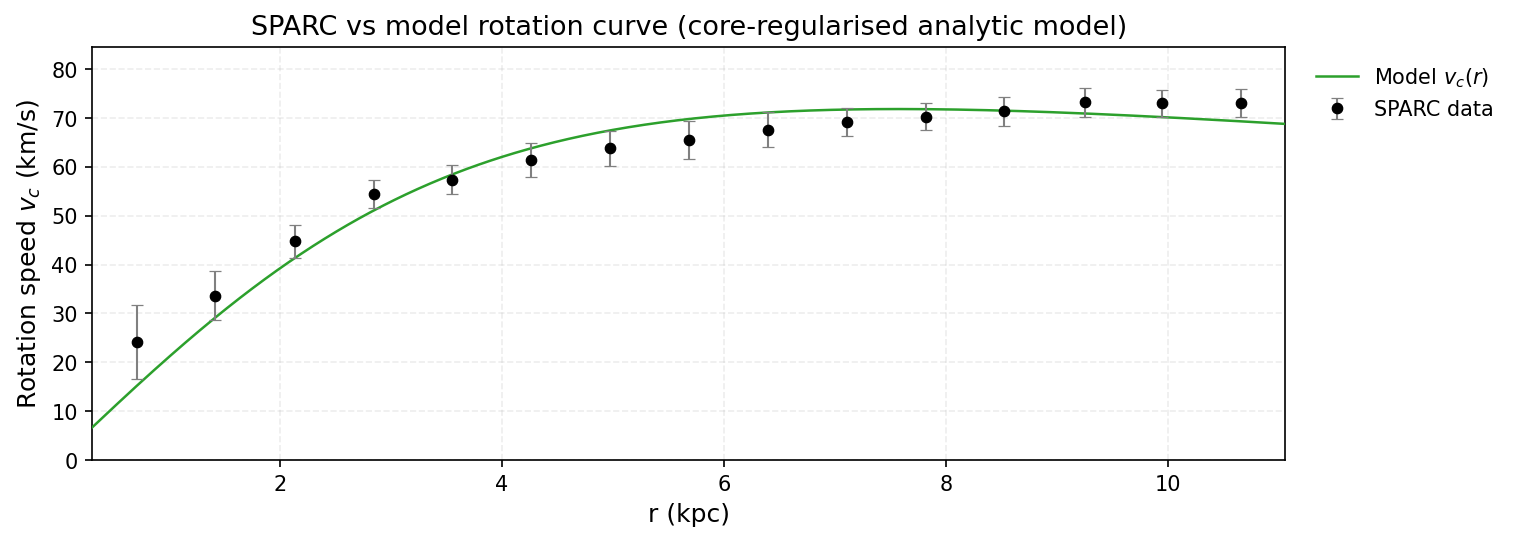}
\caption{The predicted rotation curves for the optimized SIDM
model of Eq. (\ref{ScaledependentEoSDM}), versus the SPARC
observational data for the galaxy UGC12632.} \label{UGC12632}
\end{figure}

In Table \ref{collUGC12632} we present the optimized values of
$K_0$ and $\rho_0$ for the analytic SIDM model of Eq.
(\ref{ScaledependentEoSDM}) for which the maximum compatibility
with the SPARC data is achieved.
\begin{table}[h!]
  \begin{center}
    \caption{SIDM Optimization Values for the galaxy UGC12632}
    \label{collUGC12632}
     \begin{tabular}{|r|r|}
     \hline
      \textbf{Parameter}   & \textbf{Optimization Values}
      \\  \hline
     $\rho_0 $  ($M_{\odot}/\mathrm{Kpc}^{3}$) & $2.59734\times 10^7$
\\  \hline $K_0$ ($M_{\odot} \,
\mathrm{Kpc}^{-3} \, (\mathrm{km/s})^{2}$)& 2234.72
\\  \hline
    \end{tabular}
  \end{center}
\end{table}
In Figs. \ref{UGC12632dens}, \ref{UGC12632}  we present the
density of the analytic SIDM model, the predicted rotation curves
for the SIDM model (\ref{ScaledependentEoSDM}), versus the SPARC
observational data and the sound speed, as a function of the
radius respectively. As it can be seen, for this galaxy, the SIDM
model produces viable rotation curves which are compatible with
the SPARC data.

\subsection{The Galaxy UGC12732, Non-viable}

For this galaxy, the optimization method we used, ensures maximum
compatibility of the analytic SIDM model of Eq.
(\ref{ScaledependentEoSDM}) with the SPARC data, if we choose
$\rho_0=1.98779\times 10^7$$M_{\odot}/\mathrm{Kpc}^{3}$ and
$K_0=3344.88
$$M_{\odot} \, \mathrm{Kpc}^{-3} \, (\mathrm{km/s})^{2}$, in which
case the reduced $\chi^2_{red}$ value is $\chi^2_{red}=6.14551$.
Also the parameter $\alpha$ in this case is $\alpha=7.48609 $Kpc.

\begin{figure}[h!]
\centering
\includegraphics[width=20pc]{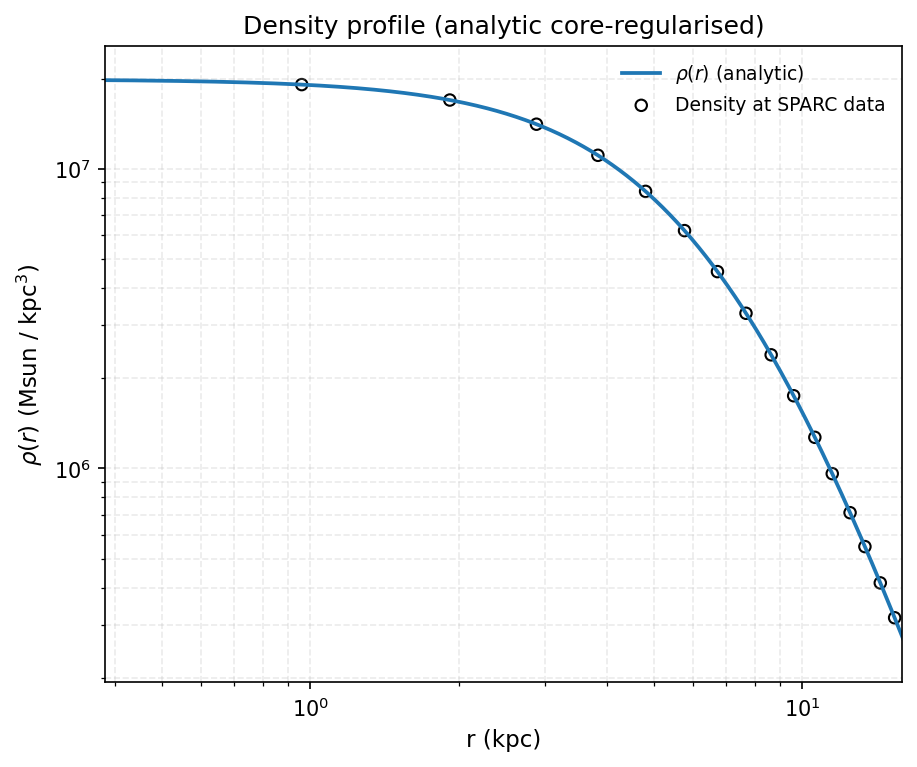}
\caption{The density of the SIDM model of Eq.
(\ref{ScaledependentEoSDM}) for the galaxy UGC12732, versus the
radius.} \label{UGC12732dens}
\end{figure}

\begin{figure}[h!]
\centering
\includegraphics[width=35pc]{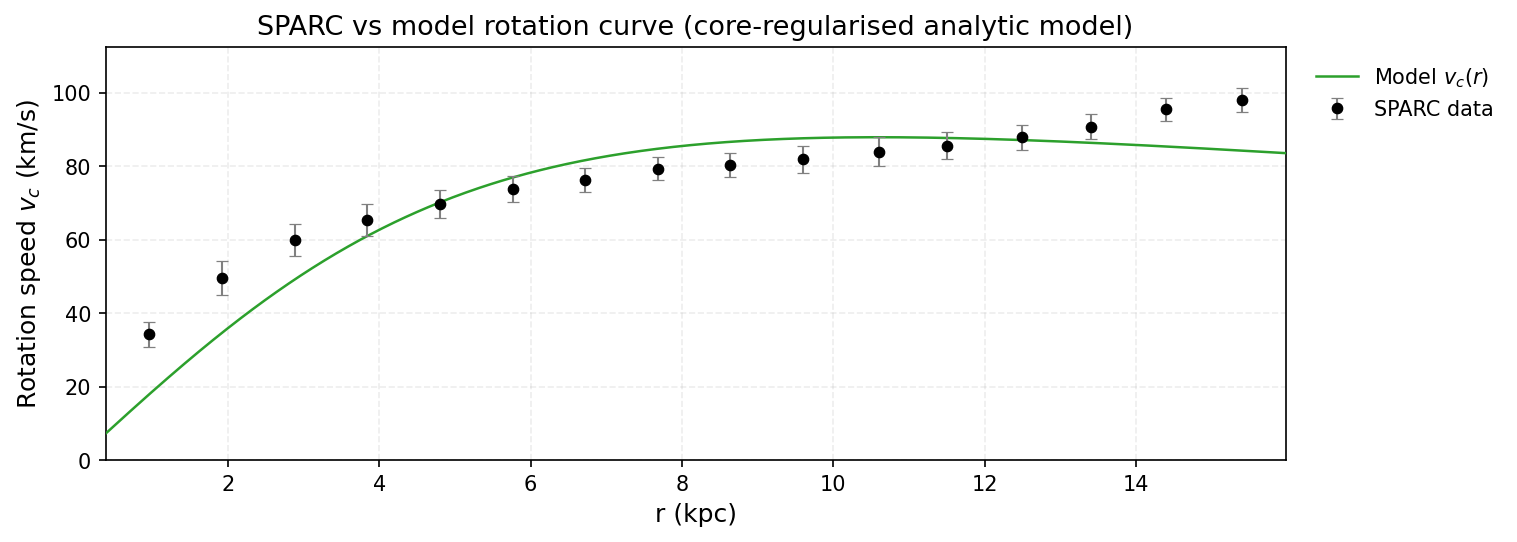}
\caption{The predicted rotation curves for the optimized SIDM
model of Eq. (\ref{ScaledependentEoSDM}), versus the SPARC
observational data for the galaxy UGC12732.} \label{UGC12732}
\end{figure}

In Table \ref{collUGC12732} we present the optimized values of
$K_0$ and $\rho_0$ for the analytic SIDM model of Eq.
(\ref{ScaledependentEoSDM}) for which the maximum compatibility
with the SPARC data is achieved.
\begin{table}[h!]
  \begin{center}
    \caption{SIDM Optimization Values for the galaxy UGC12732}
    \label{collUGC12732}
     \begin{tabular}{|r|r|}
     \hline
      \textbf{Parameter}   & \textbf{Optimization Values}
      \\  \hline
     $\rho_0 $  ($M_{\odot}/\mathrm{Kpc}^{3}$) & $1.98779\times 10^7$
\\  \hline $K_0$ ($M_{\odot} \,
\mathrm{Kpc}^{-3} \, (\mathrm{km/s})^{2}$)& 3344.88
\\  \hline
    \end{tabular}
  \end{center}
\end{table}
In Figs. \ref{UGC12732dens}, \ref{UGC12732}  we present the
density of the analytic SIDM model, the predicted rotation curves
for the SIDM model (\ref{ScaledependentEoSDM}), versus the SPARC
observational data and the sound speed, as a function of the
radius respectively. As it can be seen, for this galaxy, the SIDM
model produces non-viable rotation curves which are incompatible
with the SPARC data.

\begin{figure}[h!]
\centering
\includegraphics[width=20pc]{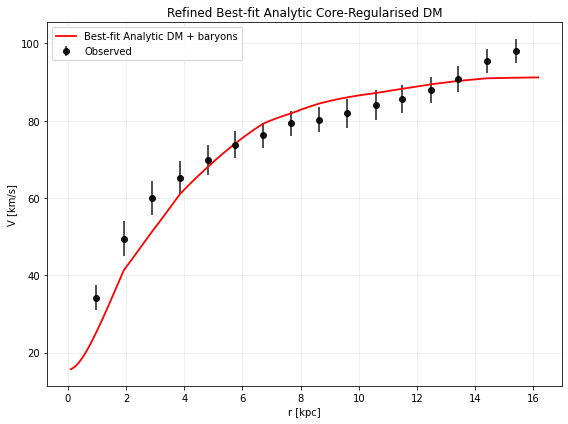}
\caption{The predicted rotation curves after using an optimization
for the SIDM model (\ref{ScaledependentEoSDM}), and the extended
SPARC data for the galaxy UGC12732. We included the rotation
curves of the gas, the disk velocities, the bulge (where present)
along with the SIDM model.} \label{extendedUGC12732}
\end{figure}

Now we shall include contributions to the rotation velocity from
the other components of the galaxy, namely the disk, the gas, and
the bulge if present. In Fig. \ref{extendedUGC12732} we present
the combined rotation curves including all the components of the
galaxy along with the SIDM. As it can be seen, the extended
collisional DM model is non-viable.
Also in Table \ref{evaluationextendedUGC12732} we present the
optimized values of the free parameters of the SIDM model for
which  we achieve the maximum compatibility with the SPARC data,
for the galaxy UGC12732, and also the resulting reduced
$\chi^2_{red}$ value.
\begin{table}[h!]
\centering \caption{Optimized Parameter Values of the Extended
SIDM model for the Galaxy UGC12732.}
\begin{tabular}{lc}
\hline
Parameter & Value  \\
\hline
$\rho_0 $ ($M_{\odot}/\mathrm{Kpc}^{3}$) & $1.18726\times 10^7$   \\
$K_0$ ($M_{\odot} \,
\mathrm{Kpc}^{-3} \, (\mathrm{km/s})^{2}$) & 2646.01   \\
$ml_{\text{disk}}$ & 1 \\
$ml_{\text{bulge}}$ & 0.5865 \\
$\alpha$ (Kpc) & 8.61428\\
$\chi^2_{red}$ & 2.38378 \\
\hline
\end{tabular}
\label{evaluationextendedUGC12732}
\end{table}

\subsection{The Galaxy UGCA442}

For this galaxy, the optimization method we used, ensures maximum
compatibility of the analytic SIDM model of Eq.
(\ref{ScaledependentEoSDM}) with the SPARC data, if we choose
$\rho_0=3.53164\times 10^7$$M_{\odot}/\mathrm{Kpc}^{3}$ and
$K_0=1389.69
$$M_{\odot} \, \mathrm{Kpc}^{-3} \, (\mathrm{km/s})^{2}$, in which
case the reduced $\chi^2_{red}$ value is $\chi^2_{red}=1.16858$.
Also the parameter $\alpha$ in this case is $\alpha=3.6201 $Kpc.

\begin{figure}[h!]
\centering
\includegraphics[width=20pc]{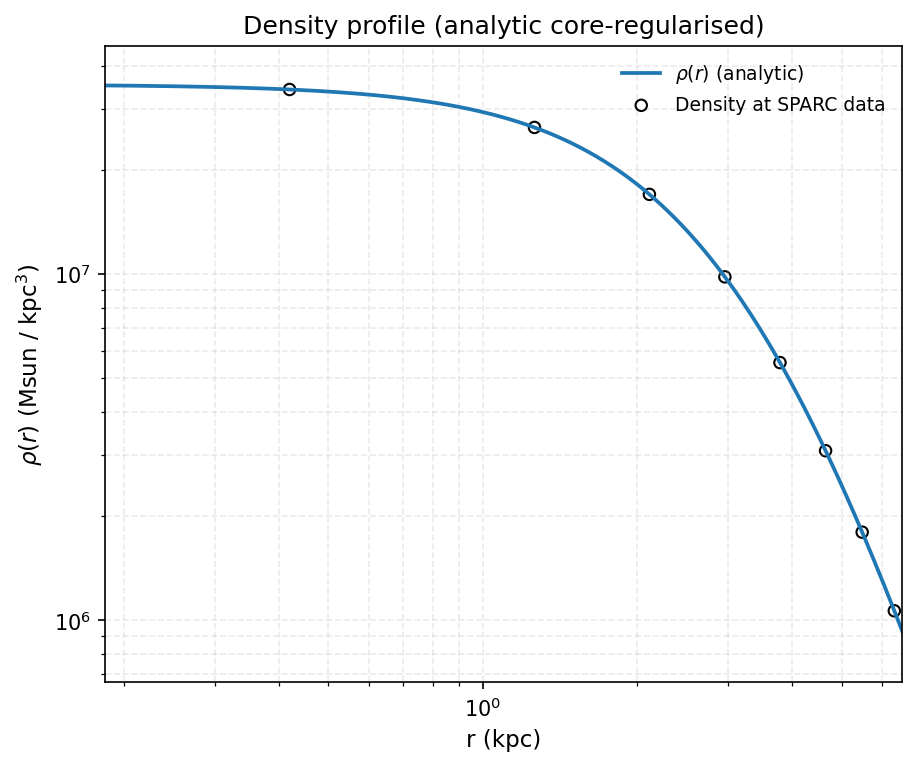}
\caption{The density of the SIDM model of Eq.
(\ref{ScaledependentEoSDM}) for the galaxy UGCA442, versus the
radius.} \label{UGCA442dens}
\end{figure}

\begin{figure}[h!]
\centering
\includegraphics[width=35pc]{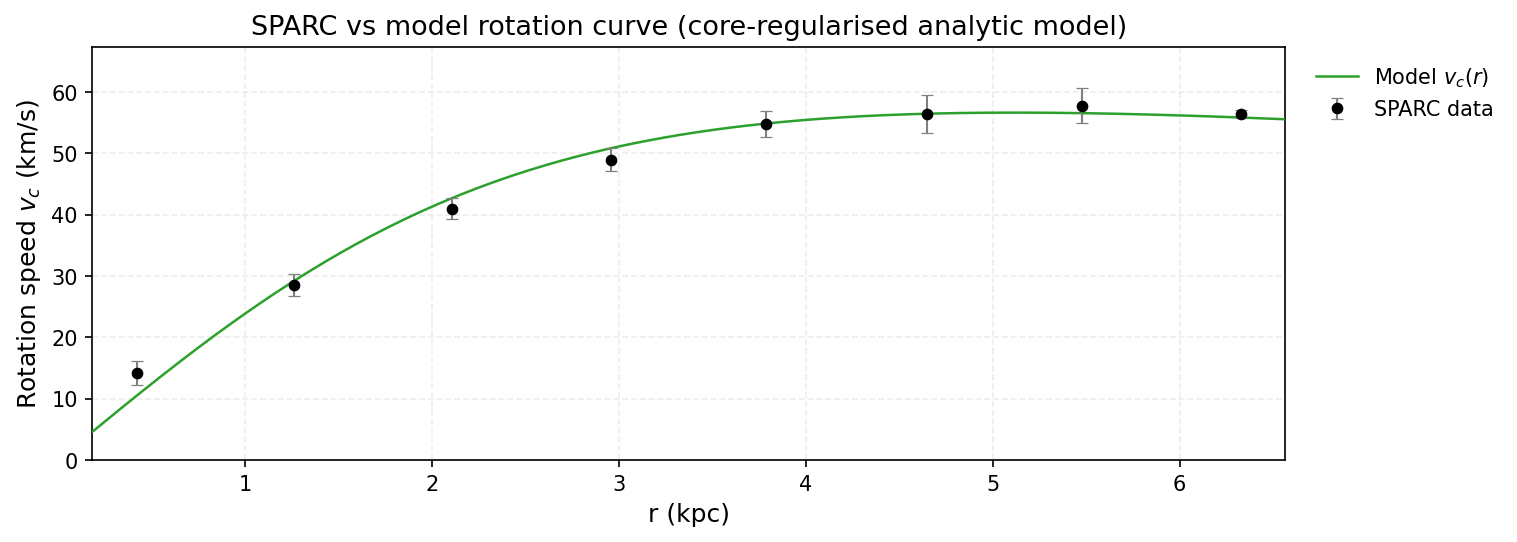}
\caption{The predicted rotation curves for the optimized SIDM
model of Eq. (\ref{ScaledependentEoSDM}), versus the SPARC
observational data for the galaxy UGCA442.} \label{UGCA442}
\end{figure}

In Table \ref{collUGCA442} we present the optimized values of
$K_0$ and $\rho_0$ for the analytic SIDM model of Eq.
(\ref{ScaledependentEoSDM}) for which the maximum compatibility
with the SPARC data is achieved.
\begin{table}[h!]
  \begin{center}
    \caption{SIDM Optimization Values for the galaxy UGCA442}
    \label{collUGCA442}
     \begin{tabular}{|r|r|}
     \hline
      \textbf{Parameter}   & \textbf{Optimization Values}
      \\  \hline
     $\rho_0 $  ($M_{\odot}/\mathrm{Kpc}^{3}$) & $3.53164\times 10^7$
\\  \hline $K_0$ ($M_{\odot} \,
\mathrm{Kpc}^{-3} \, (\mathrm{km/s})^{2}$)& 1389.69
\\  \hline
    \end{tabular}
  \end{center}
\end{table}
In Figs. \ref{UGCA442dens}, \ref{UGCA442} we present the density
of the analytic SIDM model, the predicted rotation curves for the
SIDM model (\ref{ScaledependentEoSDM}), versus the SPARC
observational data and the sound speed, as a function of the
radius respectively. As it can be seen, for this galaxy, the SIDM
model produces viable rotation curves which are compatible with
the SPARC data.

\subsection{The Galaxy UGCA444}

For this galaxy, the optimization method we used, ensures maximum
compatibility of the analytic SIDM model of Eq.
(\ref{ScaledependentEoSDM}) with the SPARC data, if we choose
$\rho_0=5.91282\times 10^7$$M_{\odot}/\mathrm{Kpc}^{3}$ and
$K_0=446.298
$$M_{\odot} \, \mathrm{Kpc}^{-3} \, (\mathrm{km/s})^{2}$, in which
case the reduced $\chi^2_{red}$ value is $\chi^2_{red}=0.35789$.
Also the parameter $\alpha$ in this case is $\alpha=1.5855 $Kpc.

\begin{figure}[h!]
\centering
\includegraphics[width=20pc]{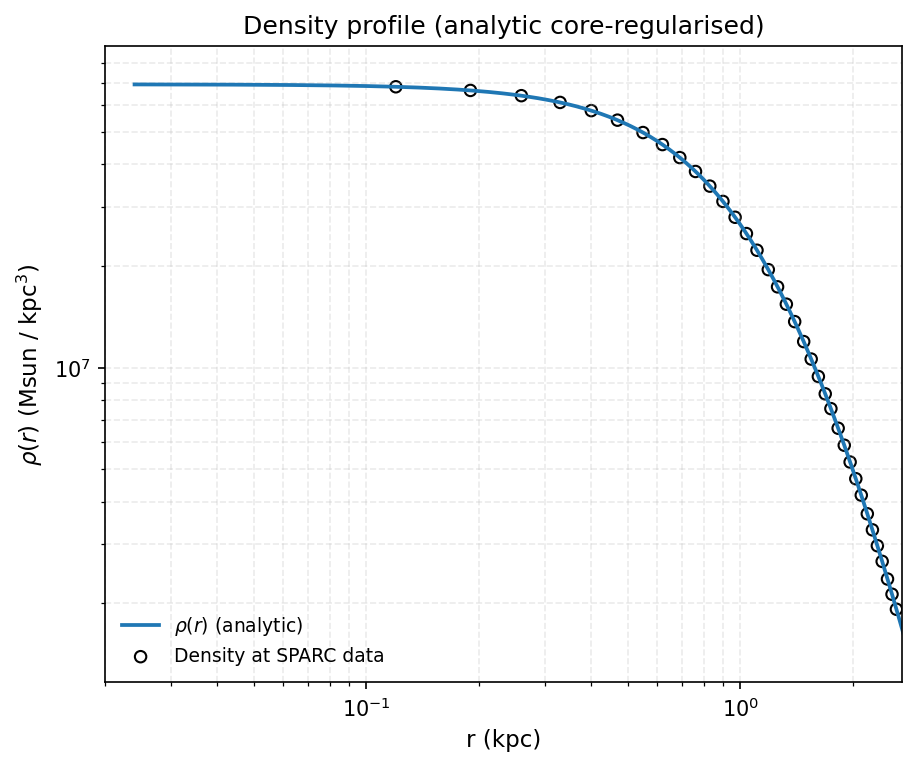}
\caption{The density of the SIDM model of Eq.
(\ref{ScaledependentEoSDM}) for the galaxy UGCA444, versus the
radius.} \label{UGCA444dens}
\end{figure}

\begin{figure}[h!]
\centering
\includegraphics[width=35pc]{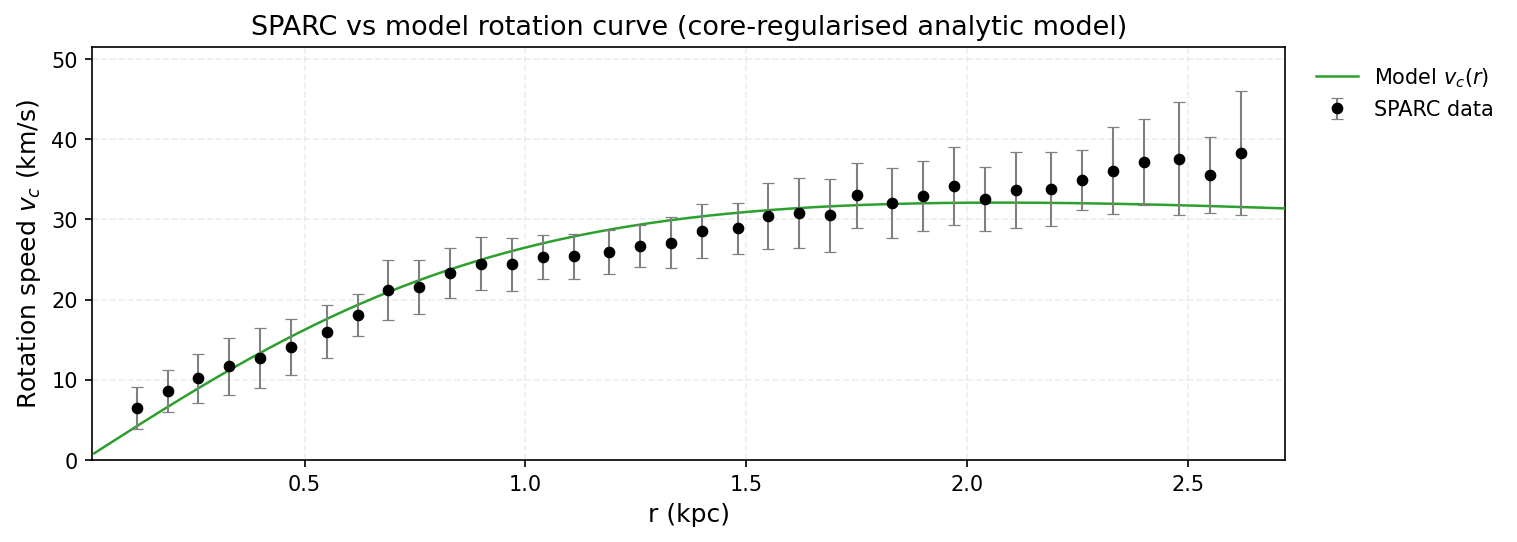}
\caption{The predicted rotation curves for the optimized SIDM
model of Eq. (\ref{ScaledependentEoSDM}), versus the SPARC
observational data for the galaxy UGCA444.} \label{UGCA444}
\end{figure}

In Table \ref{collUGCA444} we present the optimized values of
$K_0$ and $\rho_0$ for the analytic SIDM model of Eq.
(\ref{ScaledependentEoSDM}) for which the maximum compatibility
with the SPARC data is achieved.
\begin{table}[h!]
  \begin{center}
    \caption{SIDM Optimization Values for the galaxy UGCA444}
    \label{collUGCA444}
     \begin{tabular}{|r|r|}
     \hline
      \textbf{Parameter}   & \textbf{Optimization Values}
      \\  \hline
     $\rho_0 $  ($M_{\odot}/\mathrm{Kpc}^{3}$) & $5.91282\times 10^7$
\\  \hline $K_0$ ($M_{\odot} \,
\mathrm{Kpc}^{-3} \, (\mathrm{km/s})^{2}$)& 446.298
\\  \hline
    \end{tabular}
  \end{center}
\end{table}
In Figs. \ref{UGCA444dens}, \ref{UGCA444} we present the density
of the analytic SIDM model, the predicted rotation curves for the
SIDM model (\ref{ScaledependentEoSDM}), versus the SPARC
observational data and the sound speed, as a function of the
radius respectively. As it can be seen, for this galaxy, the SIDM
model produces viable rotation curves which are compatible with
the SPARC data.

\subsection{The Galaxy F561-1}

For this galaxy, the optimization method we used, ensures maximum
compatibility of the analytic SIDM model of Eq.
(\ref{ScaledependentEoSDM}) with the SPARC data, if we choose
$\rho_0=1.19454\times 10^7$$M_{\odot}/\mathrm{Kpc}^{3}$ and
$K_0=943.258
$$M_{\odot} \, \mathrm{Kpc}^{-3} \, (\mathrm{km/s})^{2}$, in which
case the reduced $\chi^2_{red}$ value is $\chi^2_{red}=0.631898$.
Also the parameter $\alpha$ in this case is $\alpha=5.12822 $Kpc.

\begin{figure}[h!]
\centering
\includegraphics[width=20pc]{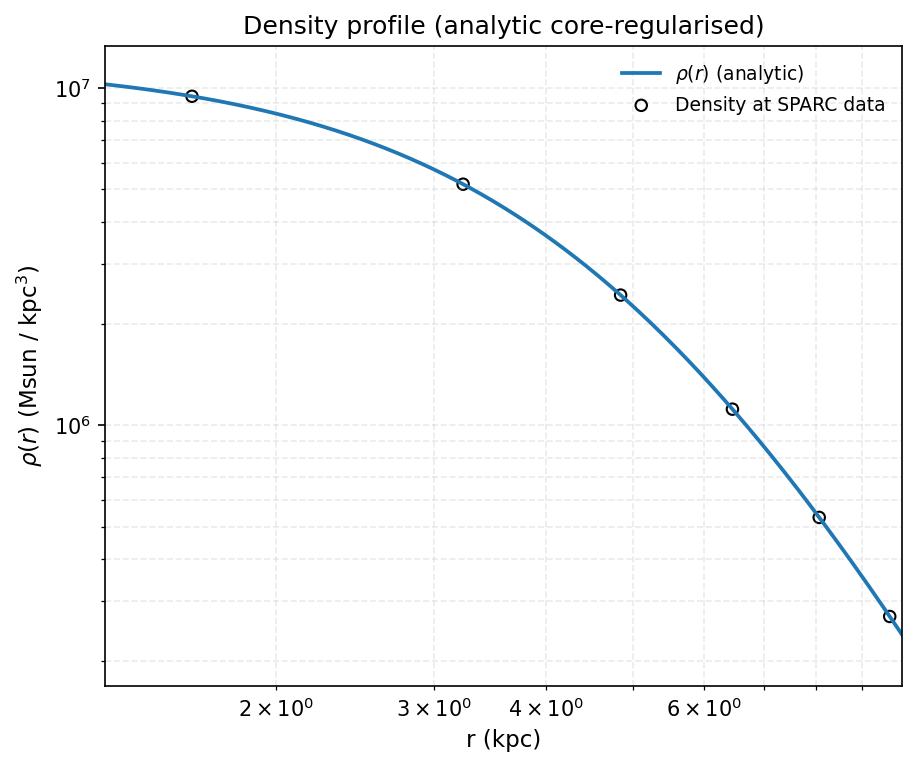}
\caption{The density of the SIDM model of Eq.
(\ref{ScaledependentEoSDM}) for the galaxy F561-1, versus the
radius.} \label{F561-1dens}
\end{figure}

\begin{figure}[h!]
\centering
\includegraphics[width=35pc]{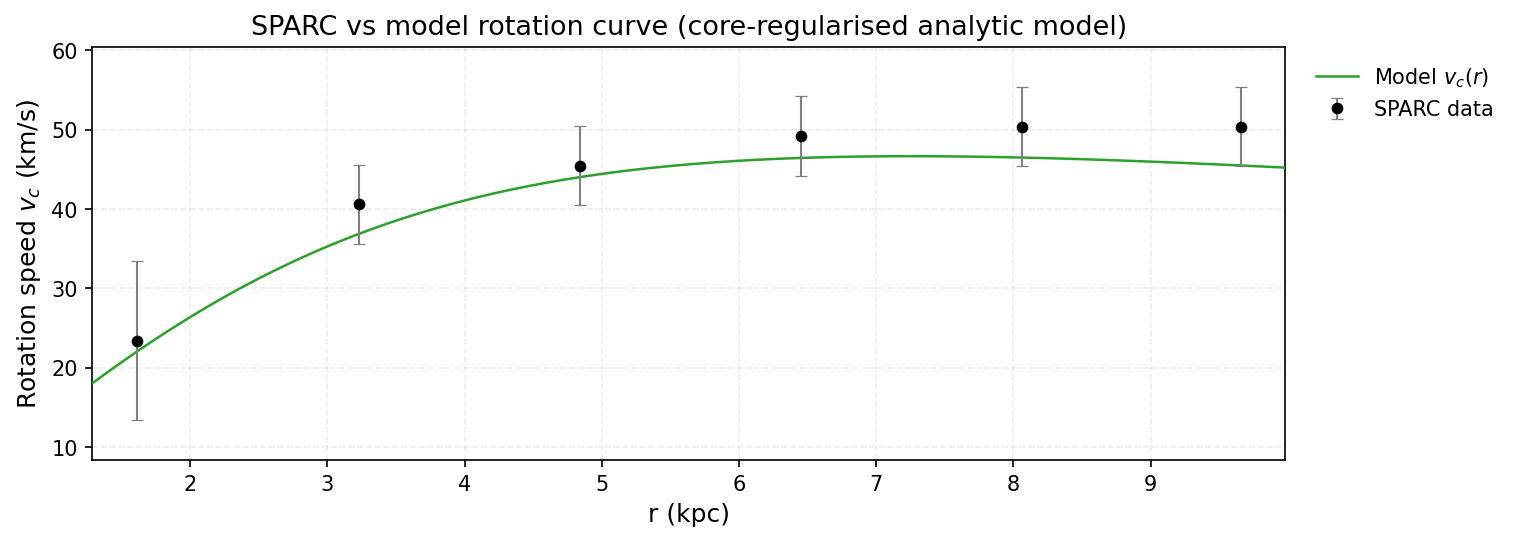}
\caption{The predicted rotation curves for the optimized SIDM
model of Eq. (\ref{ScaledependentEoSDM}), versus the SPARC
observational data for the galaxy F561-1.} \label{F561-1}
\end{figure}

In Table \ref{collF561-1} we present the optimized values of $K_0$
and $\rho_0$ for the analytic SIDM model of Eq.
(\ref{ScaledependentEoSDM}) for which the maximum compatibility
with the SPARC data is achieved.
\begin{table}[h!]
  \begin{center}
    \caption{SIDM Optimization Values for the galaxy F561-1}
    \label{collF561-1}
     \begin{tabular}{|r|r|}
     \hline
      \textbf{Parameter}   & \textbf{Optimization Values}
      \\  \hline
     $\rho_0 $  ($M_{\odot}/\mathrm{Kpc}^{3}$) & $1.19454\times 10^7$
\\  \hline $K_0$ ($M_{\odot} \,
\mathrm{Kpc}^{-3} \, (\mathrm{km/s})^{2}$)& 943.258
\\  \hline
    \end{tabular}
  \end{center}
\end{table}
In Figs. \ref{F561-1dens}, \ref{F561-1}  we present the density of
the analytic SIDM model, the predicted rotation curves for the
SIDM model (\ref{ScaledependentEoSDM}), versus the SPARC
observational data and the sound speed, as a function of the
radius respectively. As it can be seen, for this galaxy, the SIDM
model produces viable rotation curves which are compatible with
the SPARC data.

\subsection{The Galaxy F563-1}

For this galaxy, the optimization method we used, ensures maximum
compatibility of the analytic SIDM model of Eq.
(\ref{ScaledependentEoSDM}) with the SPARC data, if we choose
$\rho_0=3.47649\times 10^7$$M_{\odot}/\mathrm{Kpc}^{3}$ and
$K_0=5149.08
$$M_{\odot} \, \mathrm{Kpc}^{-3} \, (\mathrm{km/s})^{2}$, in which
case the reduced $\chi^2_{red}$ value is $\chi^2_{red}=1.05607$.
Also the parameter $\alpha$ in this case is $\alpha=7.02337 $Kpc.

\begin{figure}[h!]
\centering
\includegraphics[width=20pc]{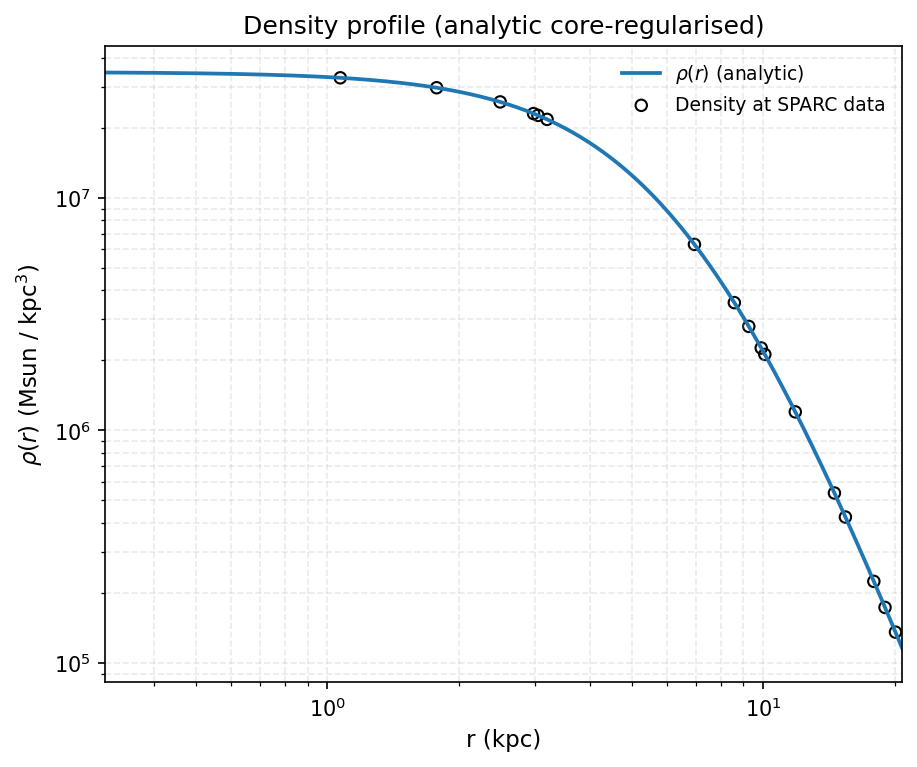}
\caption{The density of the SIDM model of Eq.
(\ref{ScaledependentEoSDM}) for the galaxy F563-1, versus the
radius.} \label{F563-1dens}
\end{figure}

\begin{figure}[h!]
\centering
\includegraphics[width=35pc]{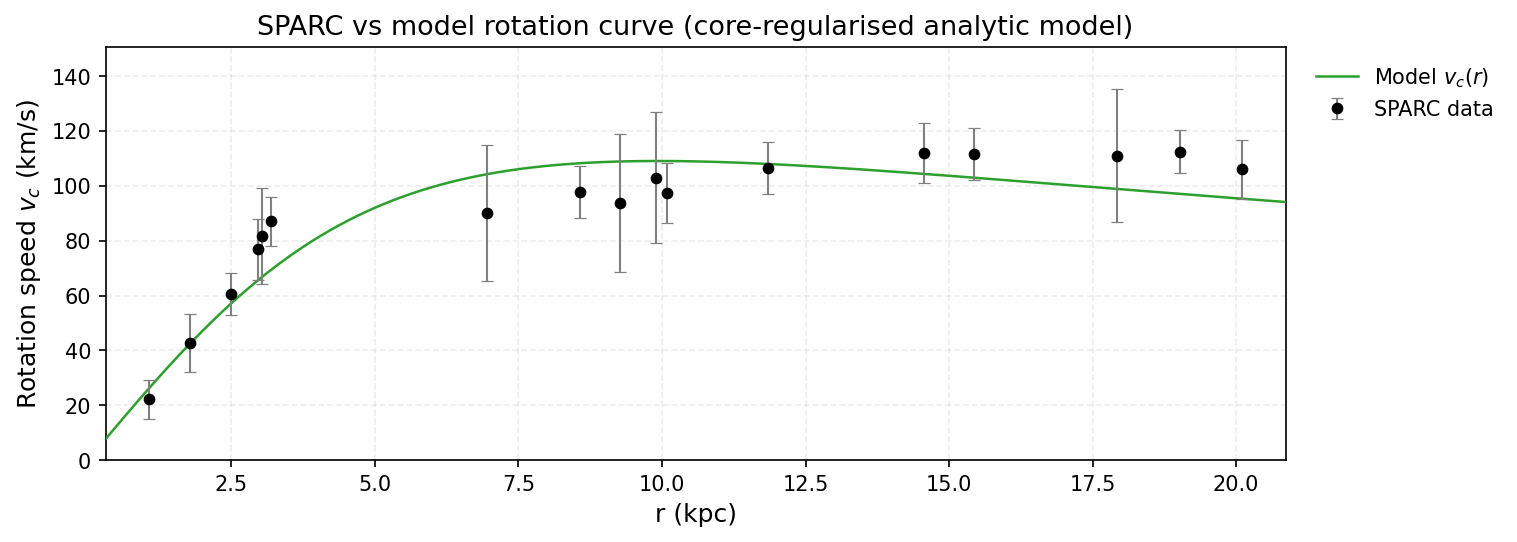}
\caption{The predicted rotation curves for the optimized SIDM
model of Eq. (\ref{ScaledependentEoSDM}), versus the SPARC
observational data for the galaxy F563-1.} \label{F563-1}
\end{figure}

In Table \ref{collF563-1} we present the optimized values of $K_0$
and $\rho_0$ for the analytic SIDM model of Eq.
(\ref{ScaledependentEoSDM}) for which the maximum compatibility
with the SPARC data is achieved.
\begin{table}[h!]
  \begin{center}
    \caption{SIDM Optimization Values for the galaxy F563-1}
    \label{collF563-1}
     \begin{tabular}{|r|r|}
     \hline
      \textbf{Parameter}   & \textbf{Optimization Values}
      \\  \hline
     $\rho_0 $  ($M_{\odot}/\mathrm{Kpc}^{3}$) & $3.47649\times 10^7$
\\  \hline $K_0$ ($M_{\odot} \,
\mathrm{Kpc}^{-3} \, (\mathrm{km/s})^{2}$)& 5149.08
\\  \hline
    \end{tabular}
  \end{center}
\end{table}
In Figs. \ref{F563-1dens}, \ref{F563-1}  we present the density of
the analytic SIDM model, the predicted rotation curves for the
SIDM model (\ref{ScaledependentEoSDM}), versus the SPARC
observational data and the sound speed, as a function of the
radius respectively. As it can be seen, for this galaxy, the SIDM
model produces viable rotation curves which are incompatible with
the SPARC data.


\subsection{The Galaxy F563-V1}

For this galaxy, the optimization method we used, ensures maximum
compatibility of the analytic SIDM model of Eq.
(\ref{ScaledependentEoSDM}) with the SPARC data, if we choose
$\rho_0=5.23175\times 10^6$$M_{\odot}/\mathrm{Kpc}^{3}$ and
$K_0=421.941
$$M_{\odot} \, \mathrm{Kpc}^{-3} \, (\mathrm{km/s})^{2}$, in which
case the reduced $\chi^2_{red}$ value is $\chi^2_{red}=0.235344$.
Also the parameter $\alpha$ in this case is $\alpha=5.18267 $Kpc.

\begin{figure}[h!]
\centering
\includegraphics[width=20pc]{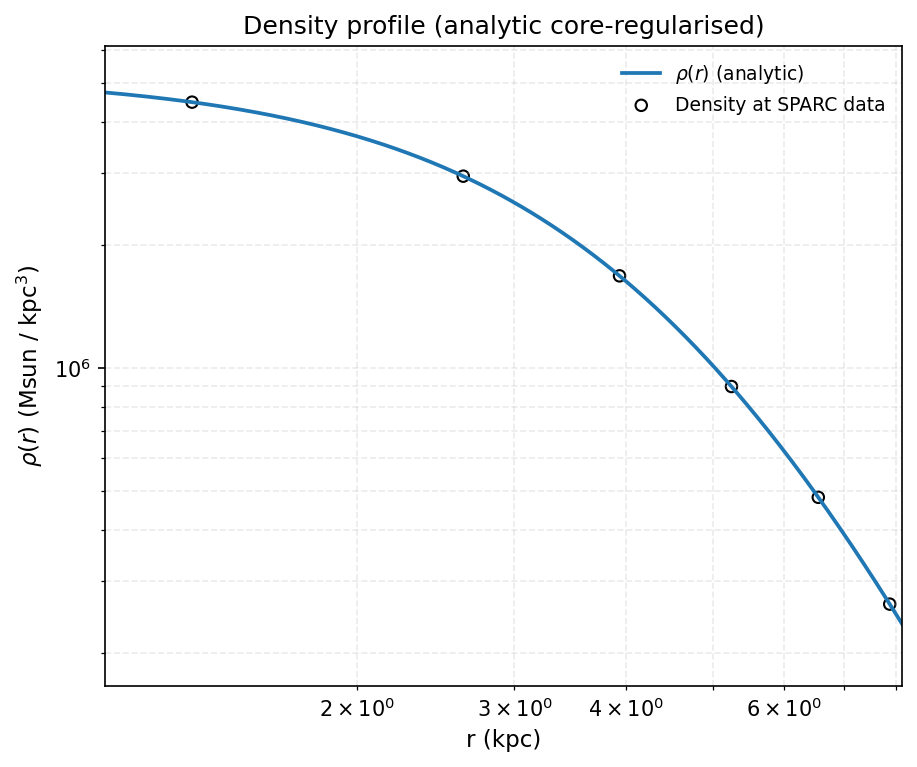}
\caption{The density of the SIDM model of Eq.
(\ref{ScaledependentEoSDM}) for the galaxy F563-V1, versus the
radius.} \label{F563-V1dens}
\end{figure}

\begin{figure}[h!]
\centering
\includegraphics[width=35pc]{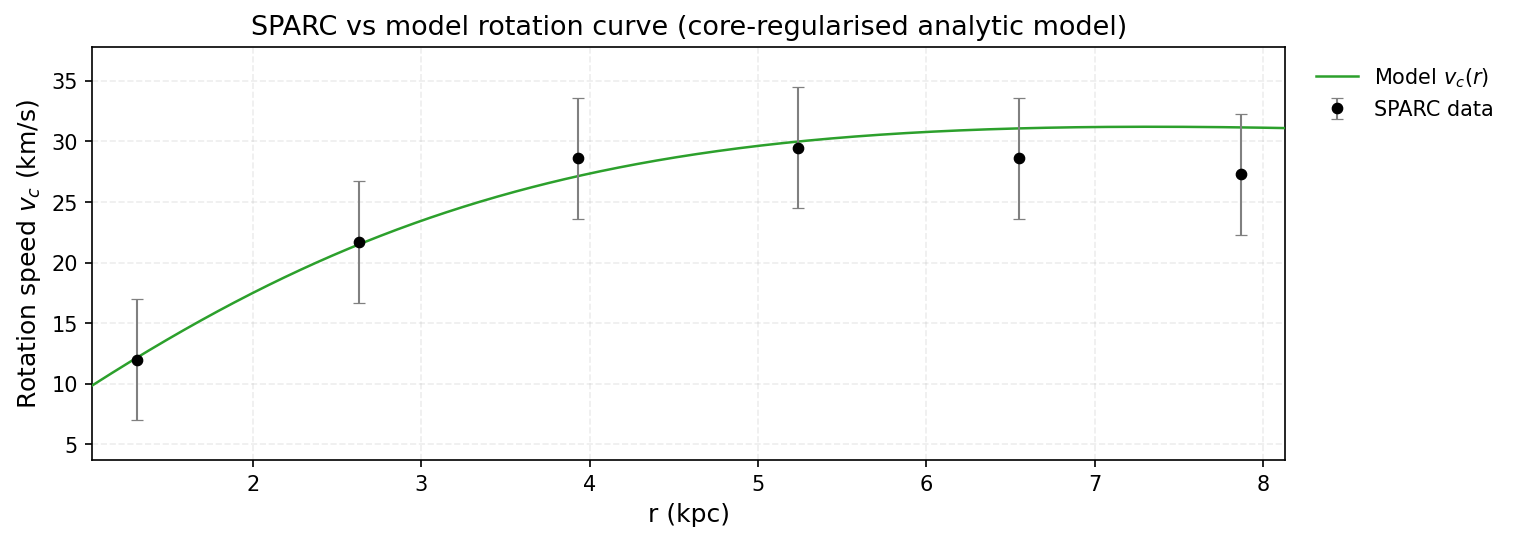}
\caption{The predicted rotation curves for the optimized SIDM
model of Eq. (\ref{ScaledependentEoSDM}), versus the SPARC
observational data for the galaxy F563-V1.} \label{F563-V1}
\end{figure}

In Table \ref{collF563-V1} we present the optimized values of
$K_0$ and $\rho_0$ for the analytic SIDM model of Eq.
(\ref{ScaledependentEoSDM}) for which the maximum compatibility
with the SPARC data is achieved.
\begin{table}[h!]
  \begin{center}
    \caption{SIDM Optimization Values for the galaxy F563-V1}
    \label{collF563-V1}
     \begin{tabular}{|r|r|}
     \hline
      \textbf{Parameter}   & \textbf{Optimization Values}
      \\  \hline
     $\rho_0 $  ($M_{\odot}/\mathrm{Kpc}^{3}$) & $5.23175\times 10^6$
\\  \hline $K_0$ ($M_{\odot} \,
\mathrm{Kpc}^{-3} \, (\mathrm{km/s})^{2}$)& 421.941
\\  \hline
    \end{tabular}
  \end{center}
\end{table}
In Figs. \ref{F563-V1dens}, \ref{F563-V1} we present the density
of the analytic SIDM model, the predicted rotation curves for the
SIDM model (\ref{ScaledependentEoSDM}), versus the SPARC
observational data and the sound speed, as a function of the
radius respectively. As it can be seen, for this galaxy, the SIDM
model produces viable rotation curves which are compatible with
the SPARC data.

\subsection{The Galaxy F563-V2}

For this galaxy, the optimization method we used, ensures maximum
compatibility of the analytic SIDM model of Eq.
(\ref{ScaledependentEoSDM}) with the SPARC data, if we choose
$\rho_0=7.91462\times 10^7$$M_{\odot}/\mathrm{Kpc}^{3}$ and
$K_0=5793.97
$$M_{\odot} \, \mathrm{Kpc}^{-3} \, (\mathrm{km/s})^{2}$, in which
case the reduced $\chi^2_{red}$ value is $\chi^2_{red}=0.336871$.
Also the parameter $\alpha$ in this case is $\alpha=4.93769 $Kpc.

\begin{figure}[h!]
\centering
\includegraphics[width=20pc]{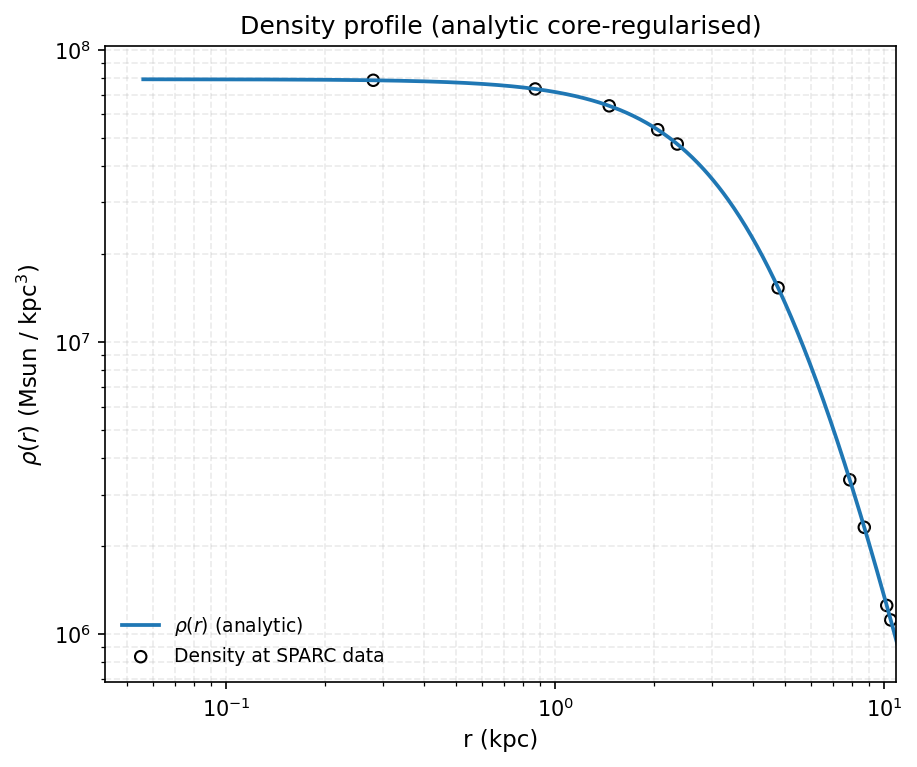}
\caption{The density of the SIDM model of Eq.
(\ref{ScaledependentEoSDM}) for the galaxy F563-V2, versus the
radius.} \label{F563-V2dens}
\end{figure}

\begin{figure}[h!]
\centering
\includegraphics[width=35pc]{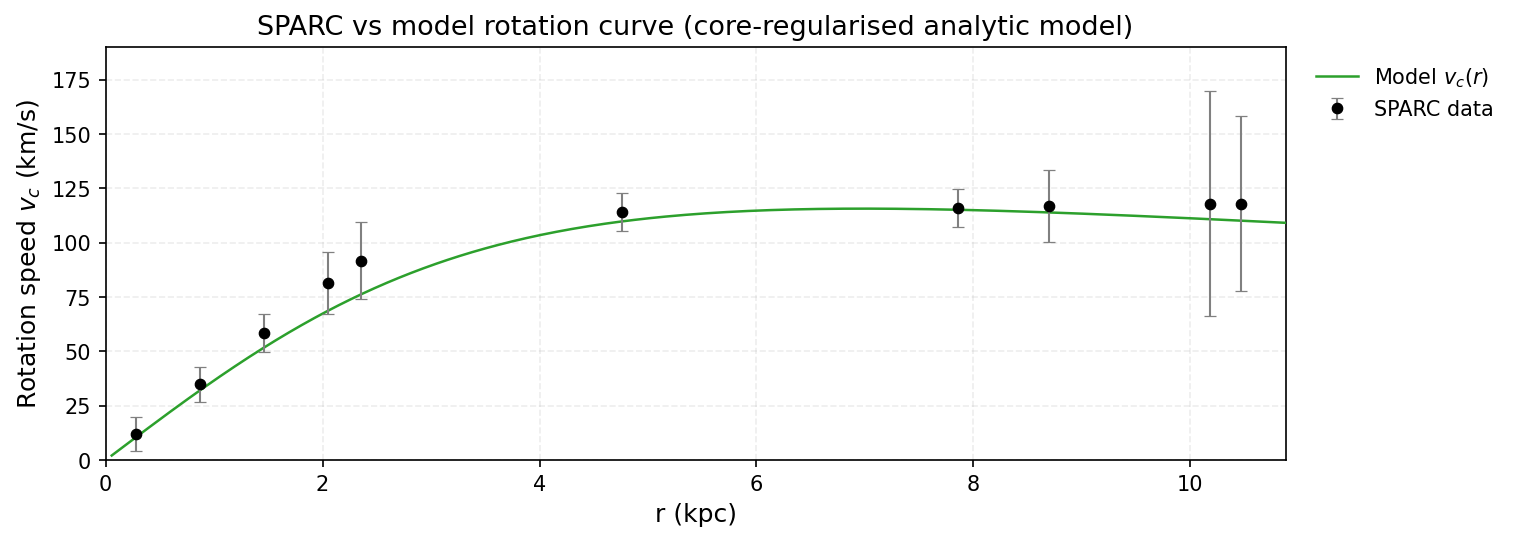}
\caption{The predicted rotation curves for the optimized SIDM
model of Eq. (\ref{ScaledependentEoSDM}), versus the SPARC
observational data for the galaxy F563-V2.} \label{F563-V2}
\end{figure}

In Table \ref{collF563-V2} we present the optimized values of
$K_0$ and $\rho_0$ for the analytic SIDM model of Eq.
(\ref{ScaledependentEoSDM}) for which the maximum compatibility
with the SPARC data is achieved.
\begin{table}[h!]
  \begin{center}
    \caption{SIDM Optimization Values for the galaxy F563-V2}
    \label{collF563-V2}
     \begin{tabular}{|r|r|}
     \hline
      \textbf{Parameter}   & \textbf{Optimization Values}
      \\  \hline
     $\rho_0 $  ($M_{\odot}/\mathrm{Kpc}^{3}$) & $7.91462\times 10^7$
\\  \hline $K_0$ ($M_{\odot} \,
\mathrm{Kpc}^{-3} \, (\mathrm{km/s})^{2}$)& 5793.97
\\  \hline
    \end{tabular}
  \end{center}
\end{table}
In Figs. \ref{F563-V2dens}, \ref{F563-V2} we present the density
of the analytic SIDM model, the predicted rotation curves for the
SIDM model (\ref{ScaledependentEoSDM}), versus the SPARC
observational data and the sound speed, as a function of the
radius respectively. As it can be seen, for this galaxy, the SIDM
model produces viable rotation curves which are compatible with
the SPARC data.

\subsection{The Galaxy F567-2}

For this galaxy, the optimization method we used, ensures maximum
compatibility of the analytic SIDM model of Eq.
(\ref{ScaledependentEoSDM}) with the SPARC data, if we choose
$\rho_0=1.4789\times 10^7$$M_{\odot}/\mathrm{Kpc}^{3}$ and
$K_0=1151.54
$$M_{\odot} \, \mathrm{Kpc}^{-3} \, (\mathrm{km/s})^{2}$, in which
case the reduced $\chi^2_{red}$ value is $\chi^2_{red}=0.27007$.
Also the parameter $\alpha$ in this case is $\alpha=5.09239 $Kpc.

\begin{figure}[h!]
\centering
\includegraphics[width=20pc]{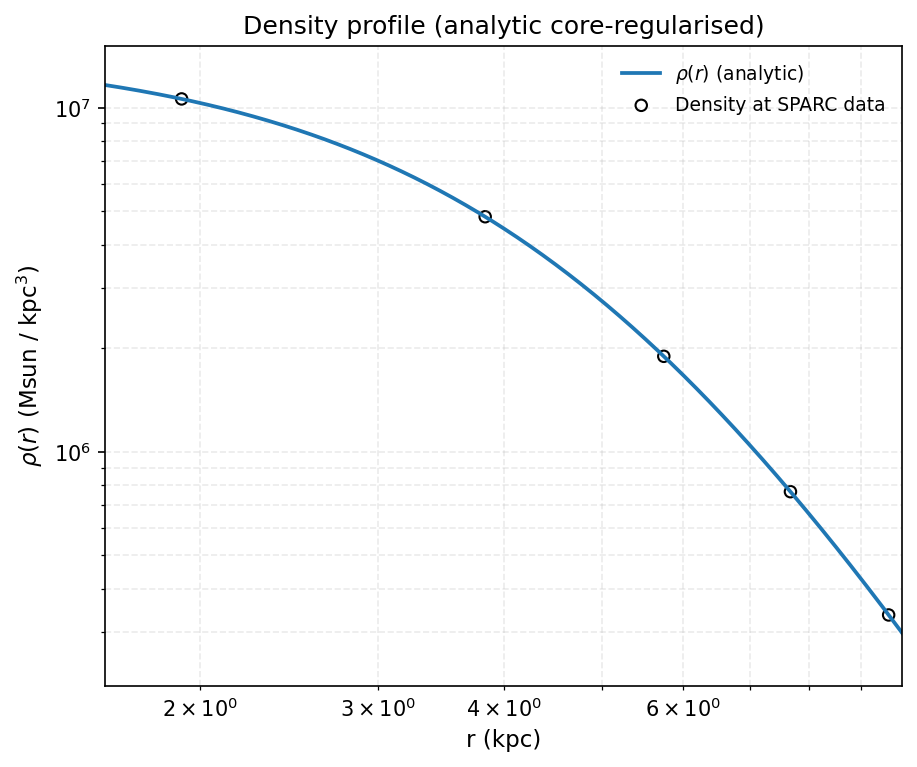}
\caption{The density of the SIDM model of Eq.
(\ref{ScaledependentEoSDM}) for the galaxy F567-2, versus the
radius.} \label{F567-2dens}
\end{figure}

\begin{figure}[h!]
\centering
\includegraphics[width=35pc]{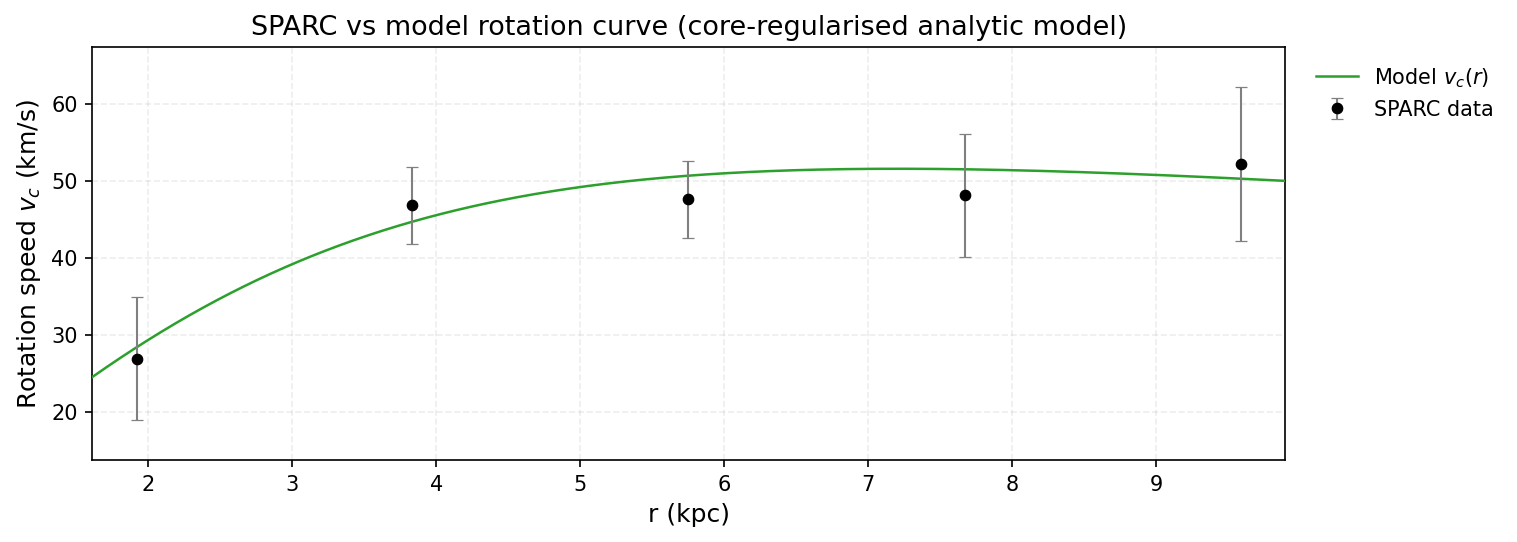}
\caption{The predicted rotation curves for the optimized SIDM
model of Eq. (\ref{ScaledependentEoSDM}), versus the SPARC
observational data for the galaxy F567-2.} \label{F567-2}
\end{figure}

In Table \ref{collF567-2} we present the optimized values of $K_0$
and $\rho_0$ for the analytic SIDM model of Eq.
(\ref{ScaledependentEoSDM}) for which the maximum compatibility
with the SPARC data is achieved.
\begin{table}[h!]
  \begin{center}
    \caption{SIDM Optimization Values for the galaxy F567-2}
    \label{collF567-2}
     \begin{tabular}{|r|r|}
     \hline
      \textbf{Parameter}   & \textbf{Optimization Values}
      \\  \hline
     $\rho_0 $  ($M_{\odot}/\mathrm{Kpc}^{3}$) & $1.4789\times 10^7$
\\  \hline $K_0$ ($M_{\odot} \,
\mathrm{Kpc}^{-3} \, (\mathrm{km/s})^{2}$)& 1151.54
\\  \hline
    \end{tabular}
  \end{center}
\end{table}
In Figs. \ref{F567-2dens}, \ref{F567-2} we present the density of
the analytic SIDM model, the predicted rotation curves for the
SIDM model (\ref{ScaledependentEoSDM}), versus the SPARC
observational data and the sound speed, as a function of the
radius respectively. As it can be seen, for this galaxy, the SIDM
model produces viable rotation curves which are compatible with
the SPARC data.

\subsection{The Galaxy F568-1}

For this galaxy, the optimization method we used, ensures maximum
compatibility of the analytic SIDM model of Eq.
(\ref{ScaledependentEoSDM}) with the SPARC data, if we choose
$\rho_0=5.59467\times 10^7$$M_{\odot}/\mathrm{Kpc}^{3}$ and
$K_0=7857.31
$$M_{\odot} \, \mathrm{Kpc}^{-3} \, (\mathrm{km/s})^{2}$, in which
case the reduced $\chi^2_{red}$ value is $\chi^2_{red}=0.256689$.
Also the parameter $\alpha$ in this case is $\alpha=6.83913 $Kpc.

\begin{figure}[h!]
\centering
\includegraphics[width=20pc]{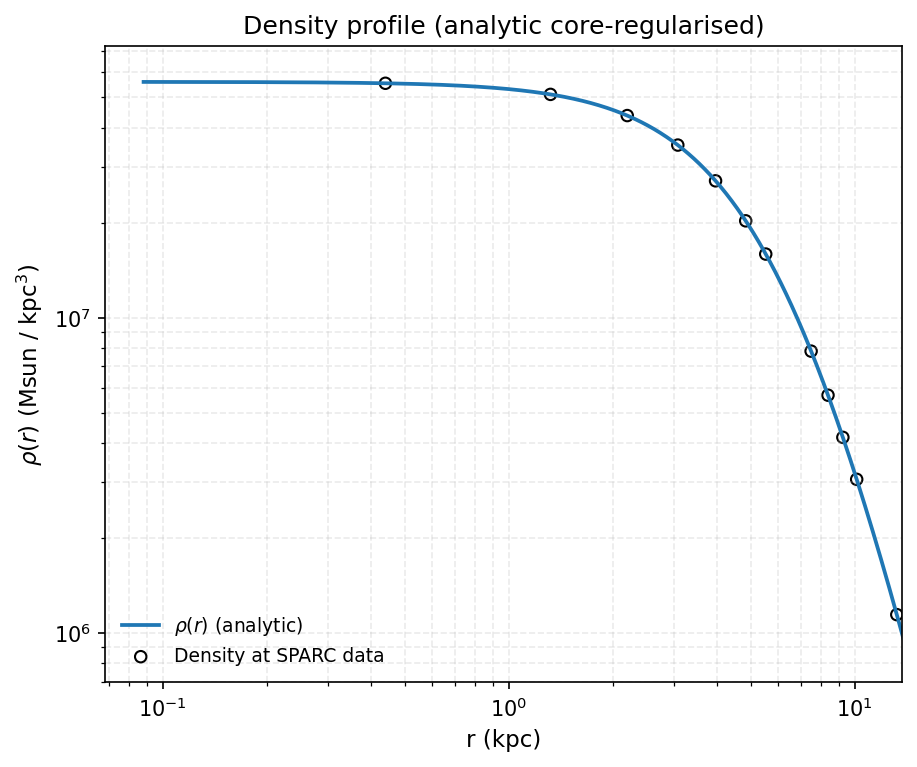}
\caption{The density of the SIDM model of Eq.
(\ref{ScaledependentEoSDM}) for the galaxy F568-1, versus the
radius.} \label{F568-1dens}
\end{figure}

\begin{figure}[h!]
\centering
\includegraphics[width=35pc]{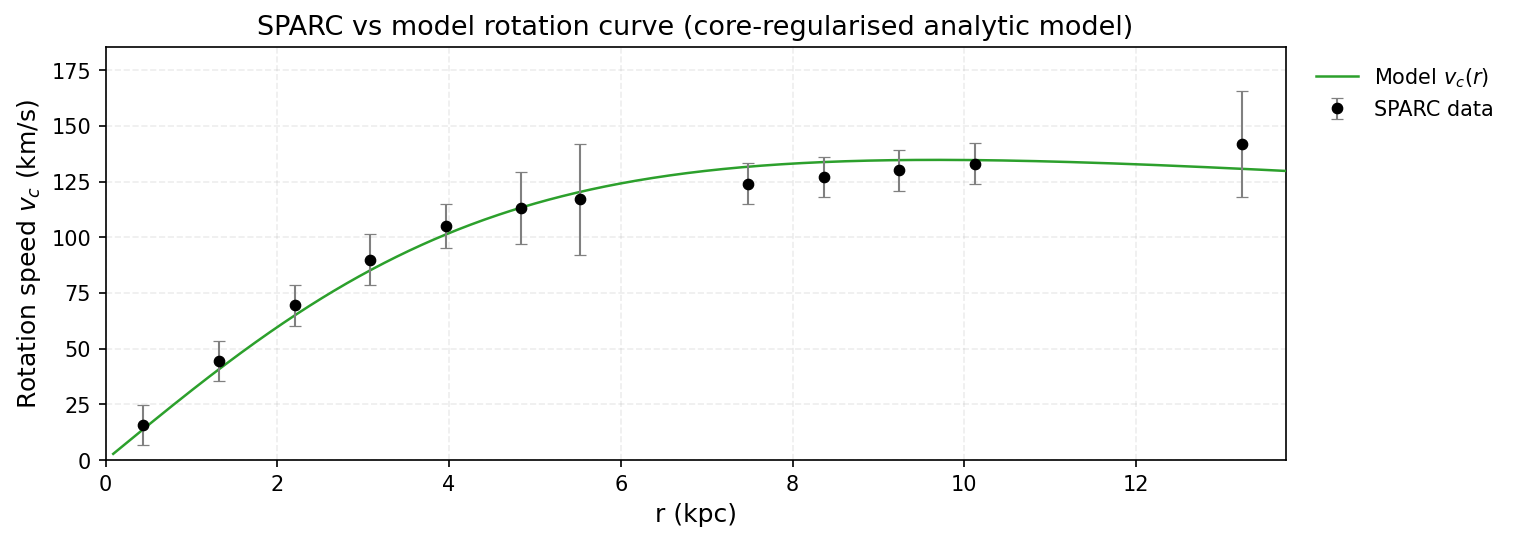}
\caption{The predicted rotation curves for the optimized SIDM
model of Eq. (\ref{ScaledependentEoSDM}), versus the SPARC
observational data for the galaxy F568-1.} \label{F568-1}
\end{figure}

In Table \ref{collF568-1} we present the optimized values of $K_0$
and $\rho_0$ for the analytic SIDM model of Eq.
(\ref{ScaledependentEoSDM}) for which the maximum compatibility
with the SPARC data is achieved.
\begin{table}[h!]
  \begin{center}
    \caption{SIDM Optimization Values for the galaxy F568-1}
    \label{collF568-1}
     \begin{tabular}{|r|r|}
     \hline
      \textbf{Parameter}   & \textbf{Optimization Values}
      \\  \hline
     $\rho_0 $  ($M_{\odot}/\mathrm{Kpc}^{3}$) & $5.59467\times 10^7$
\\  \hline $K_0$ ($M_{\odot} \,
\mathrm{Kpc}^{-3} \, (\mathrm{km/s})^{2}$)& 7857.31
\\  \hline
    \end{tabular}
  \end{center}
\end{table}
In Figs. \ref{F568-1dens}, \ref{F568-1}  we present the density of
the analytic SIDM model, the predicted rotation curves for the
SIDM model (\ref{ScaledependentEoSDM}), versus the SPARC
observational data and the sound speed, as a function of the
radius respectively. As it can be seen, for this galaxy, the SIDM
model produces viable rotation curves which are compatible with
the SPARC data.

\subsection{The Galaxy F574-2}

For this galaxy, the optimization method we used, ensures maximum
compatibility of the analytic SIDM model of Eq.
(\ref{ScaledependentEoSDM}) with the SPARC data, if we choose
$\rho_0=2.36801\times 10^6$$M_{\odot}/\mathrm{Kpc}^{3}$ and
$K_0=1131.68
$$M_{\odot} \, \mathrm{Kpc}^{-3} \, (\mathrm{km/s})^{2}$, in which
case the reduced $\chi^2_{red}$ value is $\chi^2_{red}=0.640246$.
Also the parameter $\alpha$ in this case is $\alpha=12.616 $Kpc.

\begin{figure}[h!]
\centering
\includegraphics[width=20pc]{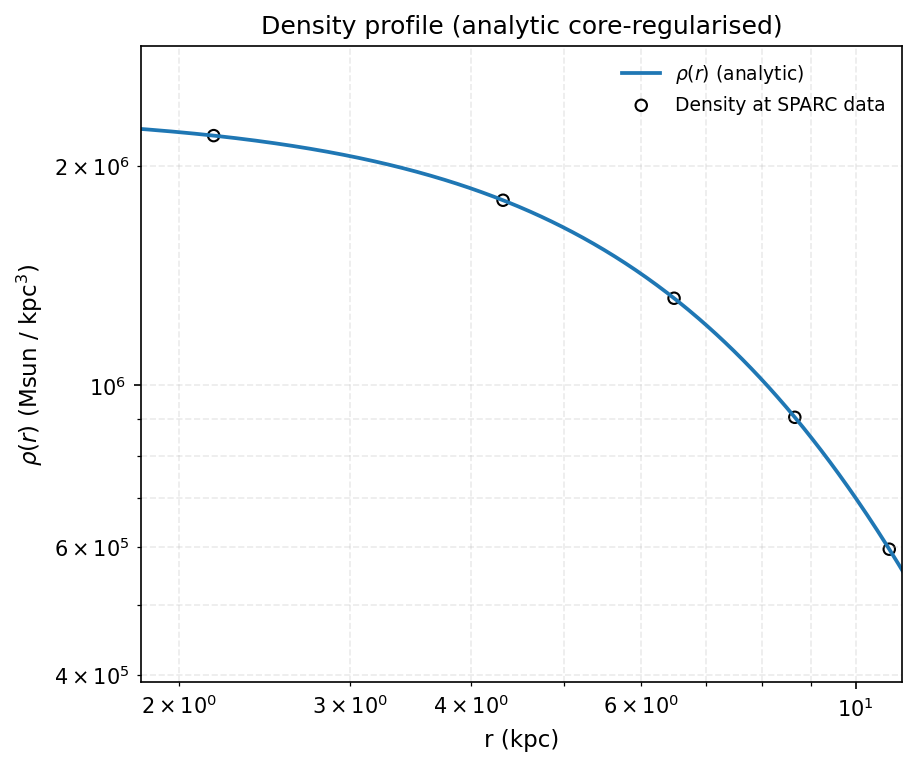}
\caption{The density of the SIDM model of Eq.
(\ref{ScaledependentEoSDM}) for the galaxy F574-2, versus the
radius.} \label{F574-2dens}
\end{figure}

\begin{figure}[h!]
\centering
\includegraphics[width=35pc]{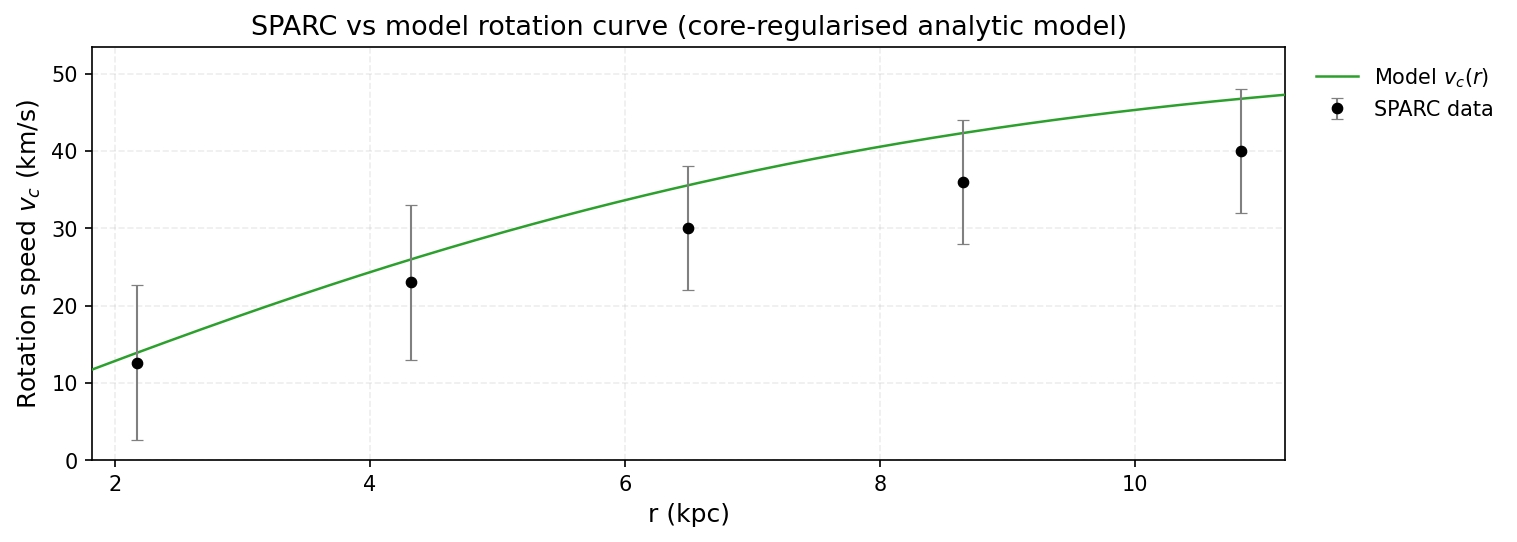}
\caption{The predicted rotation curves for the optimized SIDM
model of Eq. (\ref{ScaledependentEoSDM}), versus the SPARC
observational data for the galaxy F574-2.} \label{F574-2}
\end{figure}

In Table \ref{collF574-2} we present the optimized values of $K_0$
and $\rho_0$ for the analytic SIDM model of Eq.
(\ref{ScaledependentEoSDM}) for which the maximum compatibility
with the SPARC data is achieved.
\begin{table}[h!]
  \begin{center}
    \caption{SIDM Optimization Values for the galaxy F574-2}
    \label{collF574-2}
     \begin{tabular}{|r|r|}
     \hline
      \textbf{Parameter}   & \textbf{Optimization Values}
      \\  \hline
     $\rho_0 $  ($M_{\odot}/\mathrm{Kpc}^{3}$) & $2.36801\times 10^6$
\\  \hline $K_0$ ($M_{\odot} \,
\mathrm{Kpc}^{-3} \, (\mathrm{km/s})^{2}$)& 1131.68
\\  \hline
    \end{tabular}
  \end{center}
\end{table}
In Figs. \ref{F574-2dens}, \ref{F574-2}  we present the density of
the analytic SIDM model, the predicted rotation curves for the
SIDM model (\ref{ScaledependentEoSDM}), versus the SPARC
observational data and the sound speed, as a function of the
radius respectively. As it can be seen, for this galaxy, the SIDM
model produces viable rotation curves which are compatible with
the SPARC data.

\subsection{The Galaxy F579-V1, Non-viable, Extended Viable}

For this galaxy, the optimization method we used, ensures maximum
compatibility of the analytic SIDM model of Eq.
(\ref{ScaledependentEoSDM}) with the SPARC data, if we choose
$\rho_0=1.25165\times 10^8$$M_{\odot}/\mathrm{Kpc}^{3}$ and
$K_0=6219.26
$$M_{\odot} \, \mathrm{Kpc}^{-3} \, (\mathrm{km/s})^{2}$, in which
case the reduced $\chi^2_{red}$ value is $\chi^2_{red}=1.62409$.
Also the parameter $\alpha$ in this case is $\alpha=4.06798 $Kpc.

\begin{figure}[h!]
\centering
\includegraphics[width=20pc]{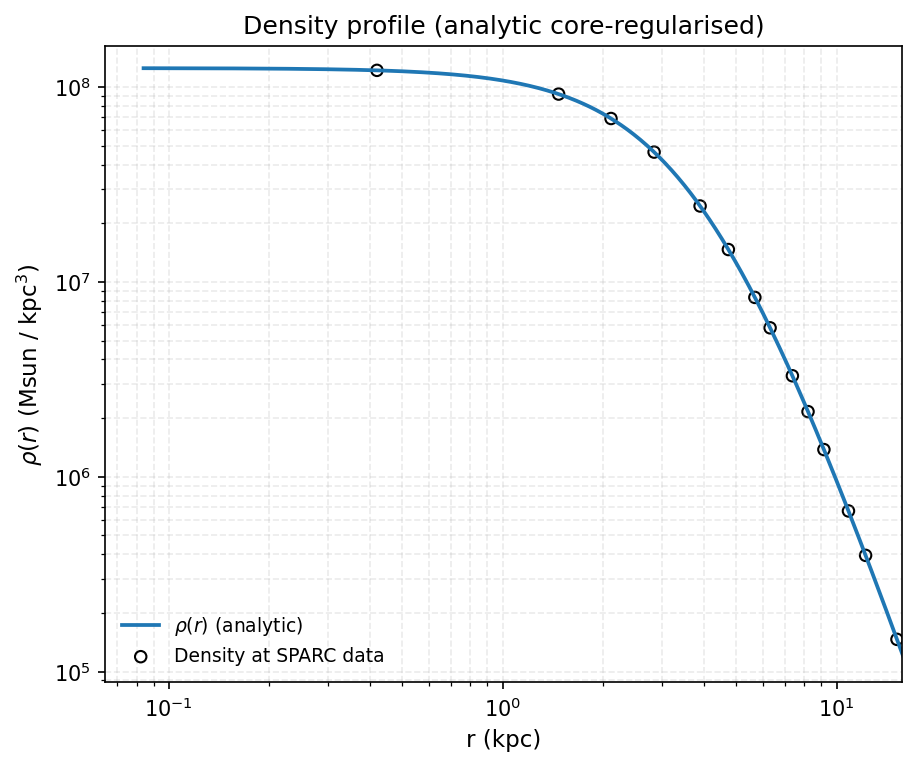}
\caption{The density of the SIDM model of Eq.
(\ref{ScaledependentEoSDM}) for the galaxy F579-V1, versus the
radius.} \label{F579-V1dens}
\end{figure}

\begin{figure}[h!]
\centering
\includegraphics[width=35pc]{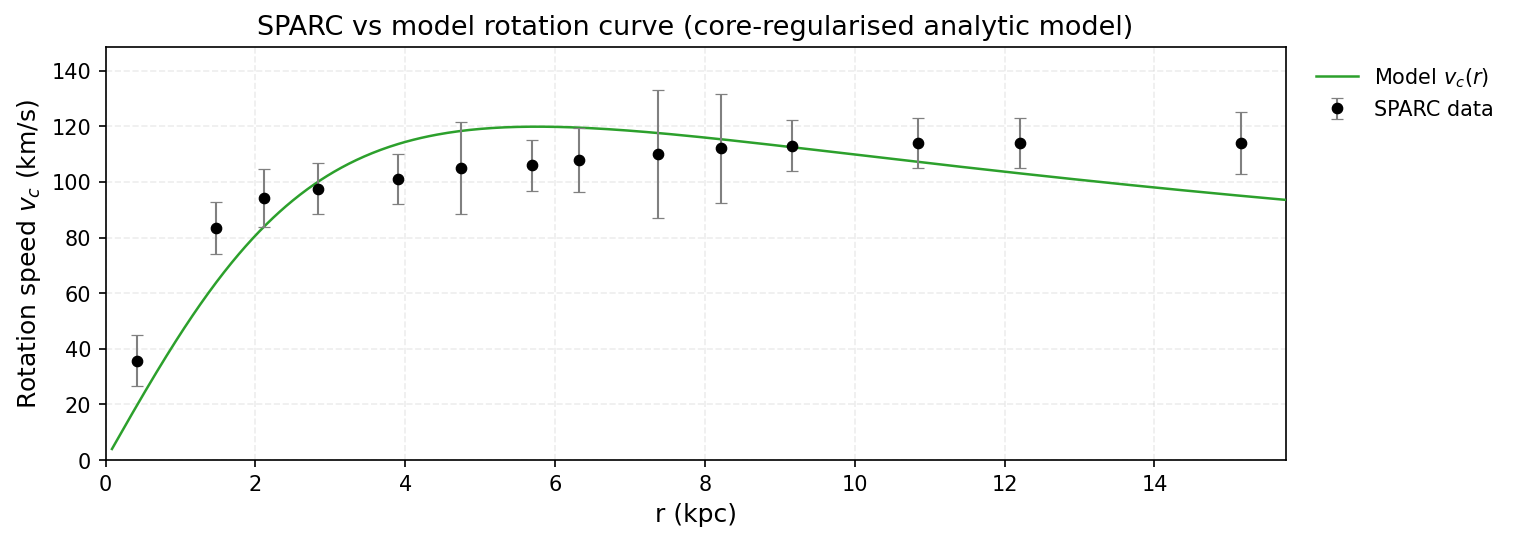}
\caption{The predicted rotation curves for the optimized SIDM
model of Eq. (\ref{ScaledependentEoSDM}), versus the SPARC
observational data for the galaxy F579-V1.} \label{F579-V1}
\end{figure}

In Table \ref{collF579-V1} we present the optimized values of
$K_0$ and $\rho_0$ for the analytic SIDM model of Eq.
(\ref{ScaledependentEoSDM}) for which the maximum compatibility
with the SPARC data is achieved.
\begin{table}[h!]
  \begin{center}
    \caption{SIDM Optimization Values for the galaxy F579-V1}
    \label{collF579-V1}
     \begin{tabular}{|r|r|}
     \hline
      \textbf{Parameter}   & \textbf{Optimization Values}
      \\  \hline
     $\rho_0 $  ($M_{\odot}/\mathrm{Kpc}^{3}$) & $1.25165\times 10^8$
\\  \hline $K_0$ ($M_{\odot} \,
\mathrm{Kpc}^{-3} \, (\mathrm{km/s})^{2}$)& 6219.26
\\  \hline
    \end{tabular}
  \end{center}
\end{table}
In Figs. \ref{F579-V1dens}, \ref{F579-V1} we present the density
of the analytic SIDM model, the predicted rotation curves for the
SIDM model (\ref{ScaledependentEoSDM}), versus the SPARC
observational data and the sound speed, as a function of the
radius respectively. As it can be seen, for this galaxy, the SIDM
model produces non-viable rotation curves which are incompatible
with the SPARC data.

\begin{figure}[h!]
\centering
\includegraphics[width=20pc]{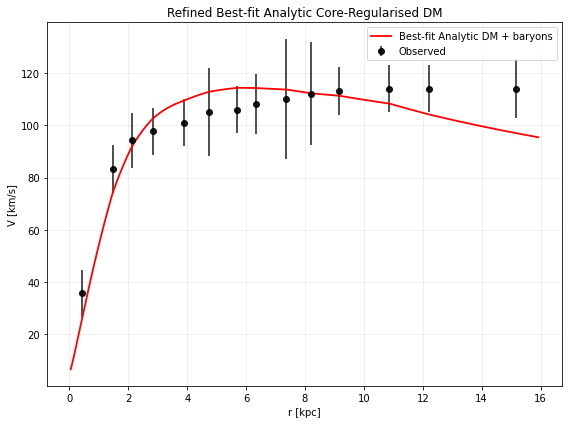}
\caption{The predicted rotation curves after using an optimization
for the SIDM model (\ref{ScaledependentEoSDM}), and the extended
SPARC data for the galaxy F579-V1. We included the rotation curves
of the gas, the disk velocities, the bulge (where present) along
with the SIDM model.} \label{extendedF579-V1}
\end{figure}

Now we shall include contributions to the rotation velocity from
the other components of the galaxy, namely the disk, the gas, and
the bulge if present. In Fig. \ref{extendedF579-V1} we present the
combined rotation curves including all the components of the
galaxy along with the SIDM. As it can be seen, the extended
collisional DM model is viable.
Also in Table \ref{evaluationextendedF579-V1} we present the
optimized values of the free parameters of the SIDM model for
which  we achieve the maximum compatibility with the SPARC data,
for the galaxy F579-V1, and also the resulting reduced
$\chi^2_{red}$ value.
\begin{table}[h!]
\centering \caption{Optimized Parameter Values of the Extended
SIDM model for the Galaxy F579-V1.}
\begin{tabular}{lc}
\hline
Parameter & Value  \\
\hline
$\rho_0 $ ($M_{\odot}/\mathrm{Kpc}^{3}$) & $1.34471\times 10^8$   \\
$K_0$ ($M_{\odot} \,
\mathrm{Kpc}^{-3} \, (\mathrm{km/s})^{2}$) & 3756.51   \\
$ml_{\text{disk}}$ & 1 \\
$ml_{\text{bulge}}$ & 0.4708 \\
$\alpha$ (Kpc) & 3.04982\\
$\chi^2_{red}$ & 0.870182 \\
\hline
\end{tabular}
\label{evaluationextendedF579-V1}
\end{table}

\subsection{The Galaxy NGC1705, Non-viable, Extended Viable}

For this galaxy, the optimization method we used, ensures maximum
compatibility of the analytic SIDM model of Eq.
(\ref{ScaledependentEoSDM}) with the SPARC data, if we choose
$\rho_0=4.15243\times 10^8$$M_{\odot}/\mathrm{Kpc}^{3}$ and
$K_0=2692.68
$$M_{\odot} \, \mathrm{Kpc}^{-3} \, (\mathrm{km/s})^{2}$, in which
case the reduced $\chi^2_{red}$ value is $\chi^2_{red}=3.00796$.
Also the parameter $\alpha$ in this case is $\alpha=1.46957 $Kpc.

\begin{figure}[h!]
\centering
\includegraphics[width=20pc]{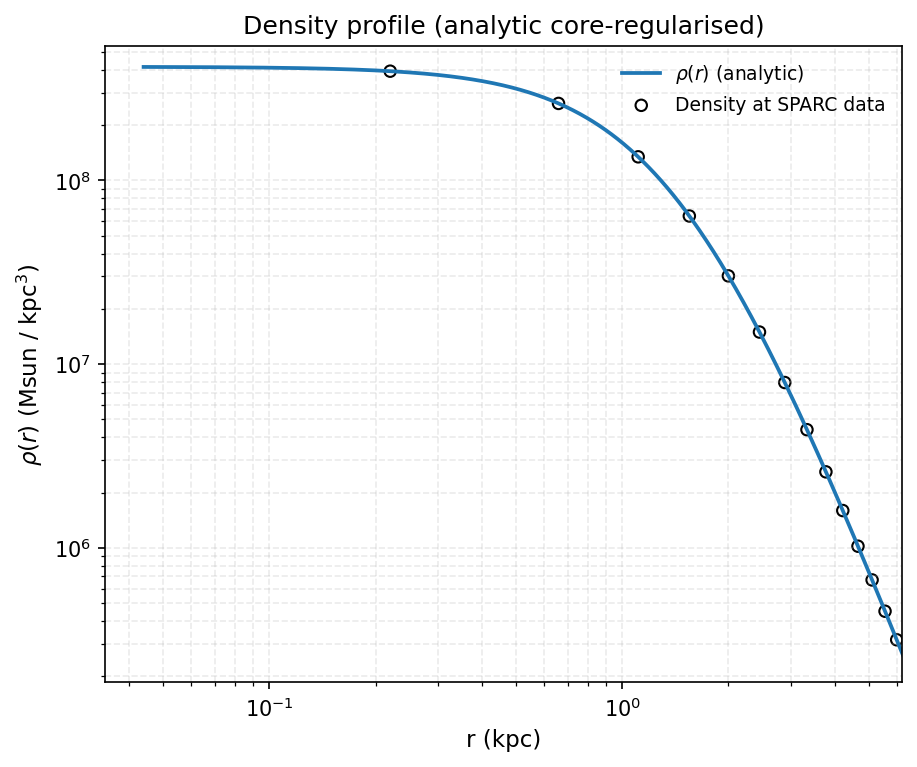}
\caption{The density of the SIDM model of Eq.
(\ref{ScaledependentEoSDM}) for the galaxy NGC1705, versus the
radius.} \label{NGC1705dens}
\end{figure}

\begin{figure}[h!]
\centering
\includegraphics[width=35pc]{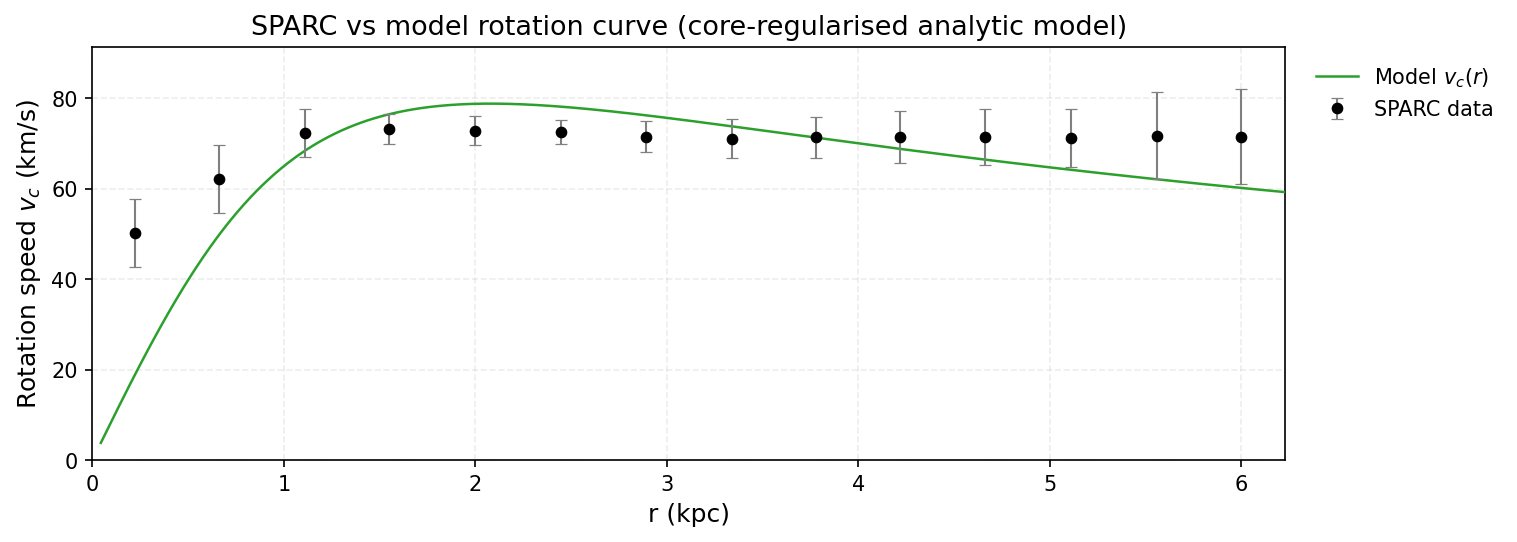}
\caption{The predicted rotation curves for the optimized SIDM
model of Eq. (\ref{ScaledependentEoSDM}), versus the SPARC
observational data for the galaxy NGC1705.} \label{NGC1705}
\end{figure}

In Table \ref{collNGC1705} we present the optimized values of
$K_0$ and $\rho_0$ for the analytic SIDM model of Eq.
(\ref{ScaledependentEoSDM}) for which the maximum compatibility
with the SPARC data is achieved.
\begin{table}[h!]
  \begin{center}
    \caption{SIDM Optimization Values for the galaxy NGC1705}
    \label{collNGC1705}
     \begin{tabular}{|r|r|}
     \hline
      \textbf{Parameter}   & \textbf{Optimization Values}
      \\  \hline
     $\rho_0 $  ($M_{\odot}/\mathrm{Kpc}^{3}$) & $4.15243\times 10^8$
\\  \hline $K_0$ ($M_{\odot} \,
\mathrm{Kpc}^{-3} \, (\mathrm{km/s})^{2}$)& 2692.68
\\  \hline
    \end{tabular}
  \end{center}
\end{table}
In Figs. \ref{NGC1705dens}, \ref{NGC1705} we present the density
of the analytic SIDM model, the predicted rotation curves for the
SIDM model (\ref{ScaledependentEoSDM}), versus the SPARC
observational data and the sound speed, as a function of the
radius respectively. As it can be seen, for this galaxy, the SIDM
model produces non-viable rotation curves which are incompatible
with the SPARC data.

\begin{figure}[h!]
\centering
\includegraphics[width=20pc]{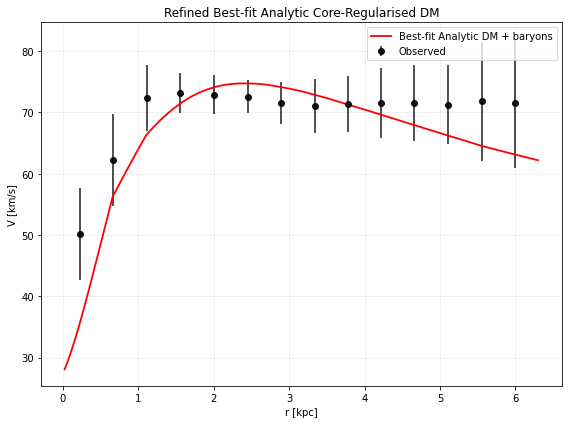}
\caption{The predicted rotation curves after using an optimization
for the SIDM model (\ref{ScaledependentEoSDM}), and the extended
SPARC data for the galaxy NGC1705. We included the rotation curves
of the gas, the disk velocities, the bulge (where present) along
with the SIDM model.} \label{extendedNGC1705}
\end{figure}

Now we shall include contributions to the rotation velocity from
the other components of the galaxy, namely the disk, the gas, and
the bulge if present. In Fig. \ref{extendedNGC1705} we present the
combined rotation curves including all the components of the
galaxy along with the SIDM. As it can be seen, the extended
collisional DM model is viable.
Also in Table \ref{evaluationextendedNGC1705} we present the
optimized values of the free parameters of the SIDM model for
which  we achieve the maximum compatibility with the SPARC data,
for the galaxy NGC1705, and also the resulting reduced
$\chi^2_{red}$ value.
\begin{table}[h!]
\centering \caption{Optimized Parameter Values of the Extended
SIDM model for the Galaxy NGC1705.}
\begin{tabular}{lc}
\hline
Parameter & Value  \\
\hline
$\rho_0 $ ($M_{\odot}/\mathrm{Kpc}^{3}$) & $1.42866\times 10^8$   \\
$K_0$ ($M_{\odot} \,
\mathrm{Kpc}^{-3} \, (\mathrm{km/s})^{2}$) & 1918.09  \\
$ml_{\text{disk}}$ & 1 \\
$ml_{\text{bulge}}$ & 0.34 \\
$\alpha$ (Kpc) & 2.1143\\
$\chi^2_{red}$ & 0.980005 \\
\hline
\end{tabular}
\label{evaluationextendedNGC1705}
\end{table}

\subsection{The Galaxy NGC2366}

For this galaxy, the optimization method we used, ensures maximum
compatibility of the analytic SIDM model of Eq.
(\ref{ScaledependentEoSDM}) with the SPARC data, if we choose
$\rho_0=3.31738\times 10^7$$M_{\odot}/\mathrm{Kpc}^{3}$ and
$K_0=1175.78
$$M_{\odot} \, \mathrm{Kpc}^{-3} \, (\mathrm{km/s})^{2}$, in which
case the reduced $\chi^2_{red}$ value is $\chi^2_{red}=0.260425$.
Also the parameter $\alpha$ in this case is $\alpha=3.43571 $Kpc.

\begin{figure}[h!]
\centering
\includegraphics[width=20pc]{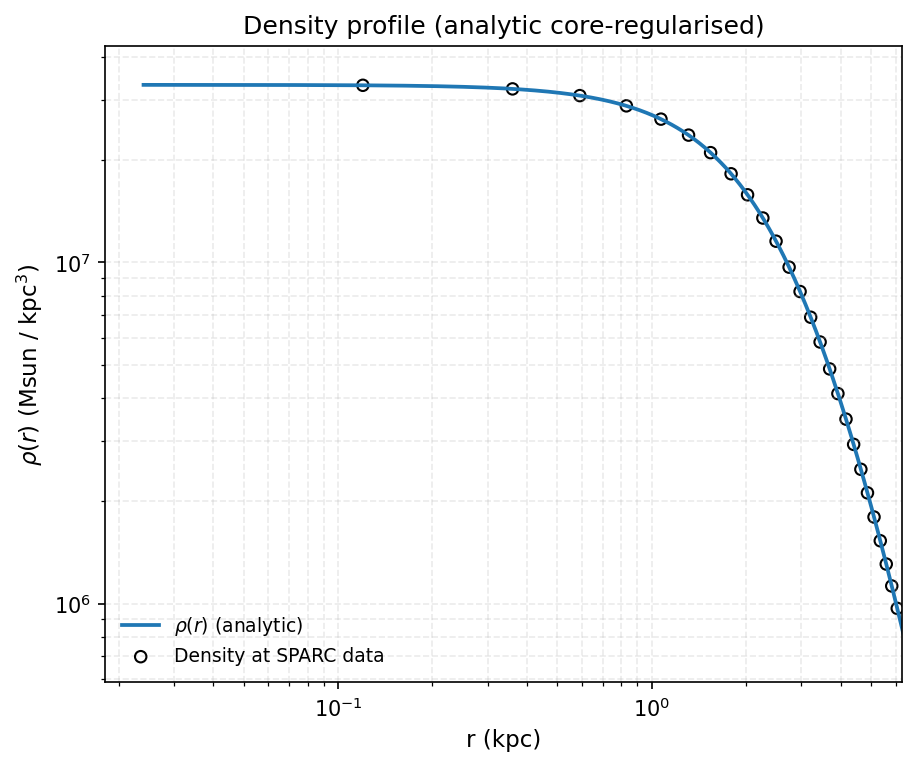}
\caption{The density of the SIDM model of Eq.
(\ref{ScaledependentEoSDM}) for the galaxy NGC2366, versus the
radius.} \label{NGC2366dens}
\end{figure}

\begin{figure}[h!]
\centering
\includegraphics[width=35pc]{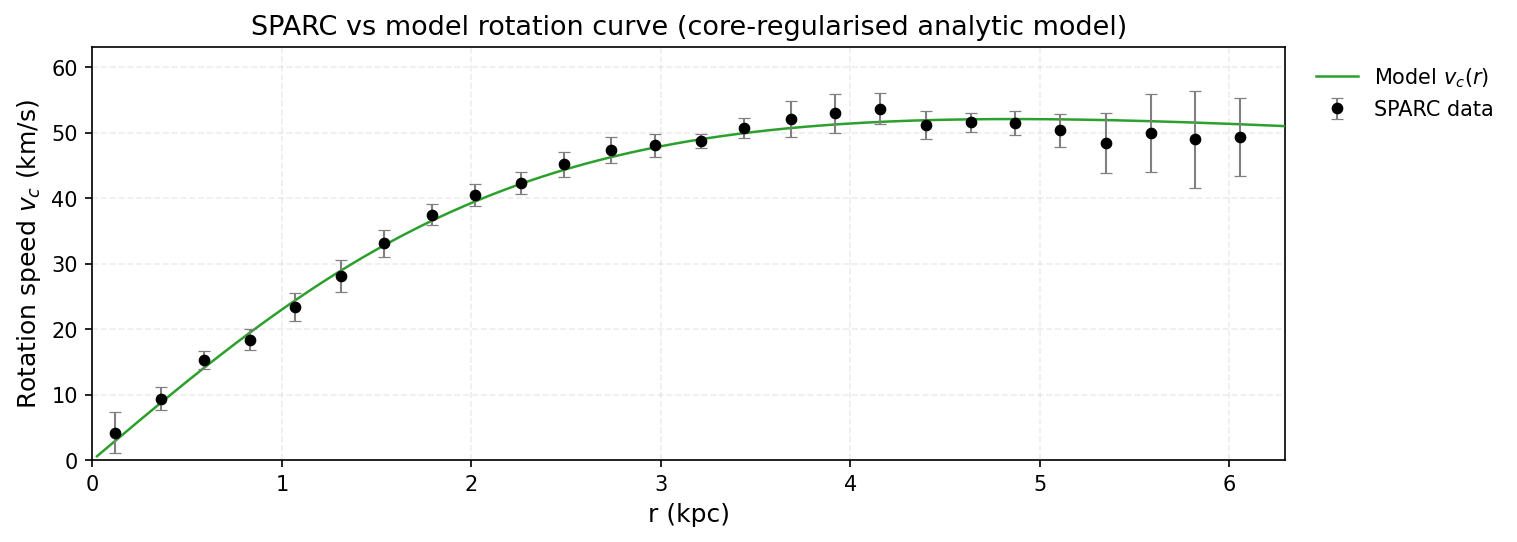}
\caption{The predicted rotation curves for the optimized SIDM
model of Eq. (\ref{ScaledependentEoSDM}), versus the SPARC
observational data for the galaxy NGC2366.} \label{NGC2366}
\end{figure}

In Table \ref{collNGC2366} we present the optimized values of
$K_0$ and $\rho_0$ for the analytic SIDM model of Eq.
(\ref{ScaledependentEoSDM}) for which the maximum compatibility
with the SPARC data is achieved.
\begin{table}[h!]
  \begin{center}
    \caption{SIDM Optimization Values for the galaxy NGC2366}
    \label{collNGC2366}
     \begin{tabular}{|r|r|}
     \hline
      \textbf{Parameter}   & \textbf{Optimization Values}
      \\  \hline
     $\rho_0 $  ($M_{\odot}/\mathrm{Kpc}^{3}$) & $3.31738\times 10^7$
\\  \hline $K_0$ ($M_{\odot} \,
\mathrm{Kpc}^{-3} \, (\mathrm{km/s})^{2}$)& 1175.78
\\  \hline
    \end{tabular}
  \end{center}
\end{table}
In Figs. \ref{NGC2366dens}, \ref{NGC2366} we present the density
of the analytic SIDM model, the predicted rotation curves for the
SIDM model (\ref{ScaledependentEoSDM}), versus the SPARC
observational data and the sound speed, as a function of the
radius respectively. As it can be seen, for this galaxy, the SIDM
model produces viable rotation curves which are compatible with
the SPARC data.

\subsection{The Galaxy NGC4214, Non-viable, Extended Viable}

For this galaxy, the optimization method we used, ensures maximum
compatibility of the analytic SIDM model of Eq.
(\ref{ScaledependentEoSDM}) with the SPARC data, if we choose
$\rho_0=3.18236\times 10^8$$M_{\odot}/\mathrm{Kpc}^{3}$ and
$K_0=3022.45
$$M_{\odot} \, \mathrm{Kpc}^{-3} \, (\mathrm{km/s})^{2}$, in which
case the reduced $\chi^2_{red}$ value is $\chi^2_{red}=4.23692$.
Also the parameter $\alpha$ in this case is $\alpha=1.77851 $Kpc.

\begin{figure}[h!]
\centering
\includegraphics[width=20pc]{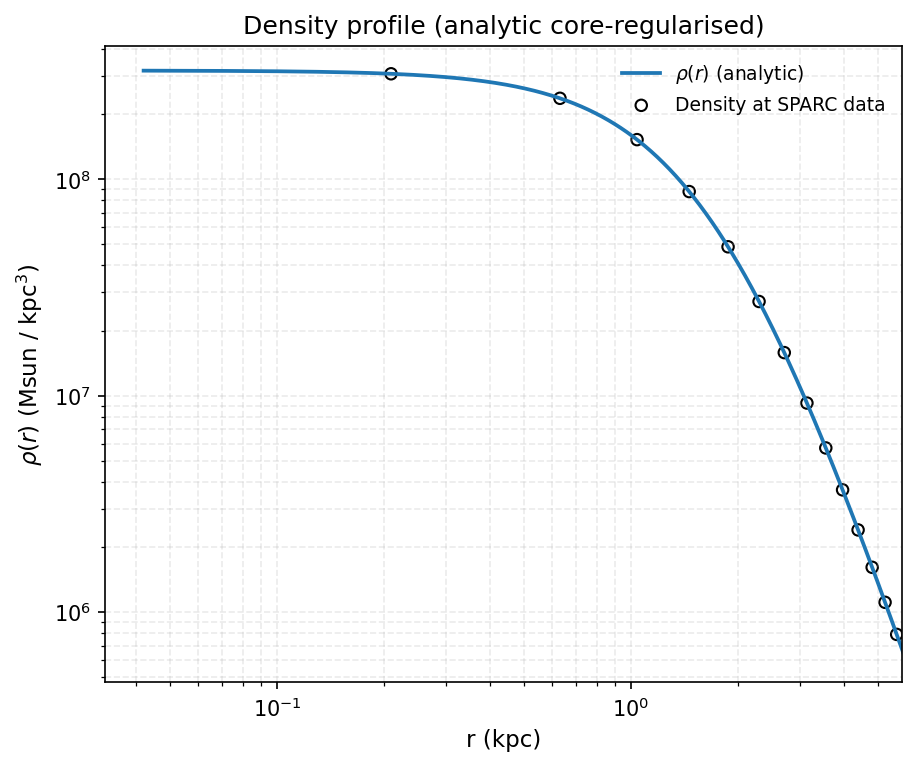}
\caption{The density of the SIDM model of Eq.
(\ref{ScaledependentEoSDM}) for the galaxy NGC4214, versus the
radius.} \label{NGC4214dens}
\end{figure}

\begin{figure}[h!]
\centering
\includegraphics[width=35pc]{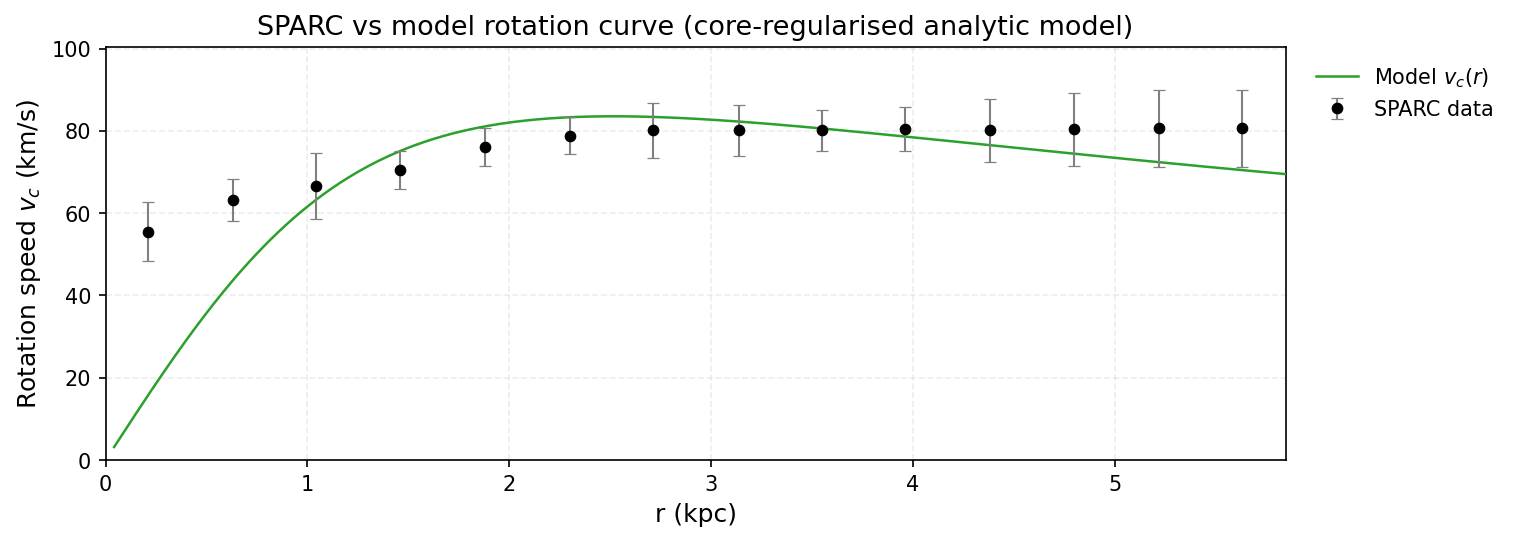}
\caption{The predicted rotation curves for the optimized SIDM
model of Eq. (\ref{ScaledependentEoSDM}), versus the SPARC
observational data for the galaxy NGC4214.} \label{NGC4214}
\end{figure}

In Table \ref{collNGC4214} we present the optimized values of
$K_0$ and $\rho_0$ for the analytic SIDM model of Eq.
(\ref{ScaledependentEoSDM}) for which the maximum compatibility
with the SPARC data is achieved.
\begin{table}[h!]
  \begin{center}
    \caption{SIDM Optimization Values for the galaxy NGC4214}
    \label{collNGC4214}
     \begin{tabular}{|r|r|}
     \hline
      \textbf{Parameter}   & \textbf{Optimization Values}
      \\  \hline
     $\rho_0 $  ($M_{\odot}/\mathrm{Kpc}^{3}$) & $3.18236\times 10^8$
\\  \hline $K_0$ ($M_{\odot} \,
\mathrm{Kpc}^{-3} \, (\mathrm{km/s})^{2}$)& 3022.45
\\  \hline
    \end{tabular}
  \end{center}
\end{table}
In Figs. \ref{NGC4214dens}, \ref{NGC4214} we present the density
of the analytic SIDM model, the predicted rotation curves for the
SIDM model (\ref{ScaledependentEoSDM}), versus the SPARC
observational data and the sound speed, as a function of the
radius respectively. As it can be seen, for this galaxy, the SIDM
model produces non-viable rotation curves which are incompatible
with the SPARC data.

\begin{figure}[h!]
\centering
\includegraphics[width=20pc]{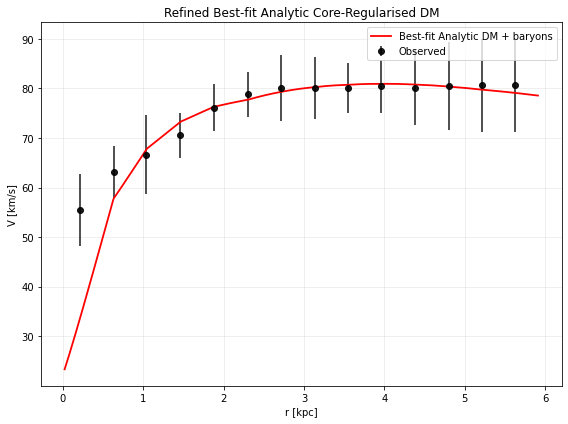}
\caption{The predicted rotation curves after using an optimization
for the SIDM model (\ref{ScaledependentEoSDM}), and the extended
SPARC data for the galaxy NGC4214. We included the rotation curves
of the gas, the disk velocities, the bulge (where present) along
with the SIDM model.} \label{extendedNGC4214}
\end{figure}

Now we shall include contributions to the rotation velocity from
the other components of the galaxy, namely the disk, the gas, and
the bulge if present. In Fig. \ref{extendedNGC4214} we present the
combined rotation curves including all the components of the
galaxy along with the SIDM. As it can be seen, the extended
collisional DM model is viable.
Also in Table \ref{evaluationextendedNGC4214} we present the
optimized values of the free parameters of the SIDM model for
which  we achieve the maximum compatibility with the SPARC data,
for the galaxy NGC4214, and also the resulting reduced
$\chi^2_{red}$ value.
\begin{table}[h!]
\centering \caption{Optimized Parameter Values of the Extended
SIDM model for the Galaxy NGC4214.}
\begin{tabular}{lc}
\hline
Parameter & Value  \\
\hline
$\rho_0 $ ($M_{\odot}/\mathrm{Kpc}^{3}$) & $6.0791\times 10^7$   \\
$K_0$ ($M_{\odot} \,
\mathrm{Kpc}^{-3} \, (\mathrm{km/s})^{2}$) & 2251.03   \\
$ml_{\text{disk}}$ & 1 \\
$ml_{\text{bulge}}$ & 0.2552 \\
$\alpha$ (Kpc) & 3.5113\\
$\chi^2_{red}$ & 1.09386 \\
\hline
\end{tabular}
\label{evaluationextendedNGC4214}
\end{table}

\subsection{The Galaxy NGC4389}

For this galaxy, the optimization method we used, ensures maximum
compatibility of the analytic SIDM model of Eq.
(\ref{ScaledependentEoSDM}) with the SPARC data, if we choose
$\rho_0=4.85533\times 10^7$$M_{\odot}/\mathrm{Kpc}^{3}$ and
$K_0=6639.95
$$M_{\odot} \, \mathrm{Kpc}^{-3} \, (\mathrm{km/s})^{2}$, in which
case the reduced $\chi^2_{red}$ value is $\chi^2_{red}=0.381669$.
Also the parameter $\alpha$ in this case is $\alpha=6.74875 $Kpc.

\begin{figure}[h!]
\centering
\includegraphics[width=20pc]{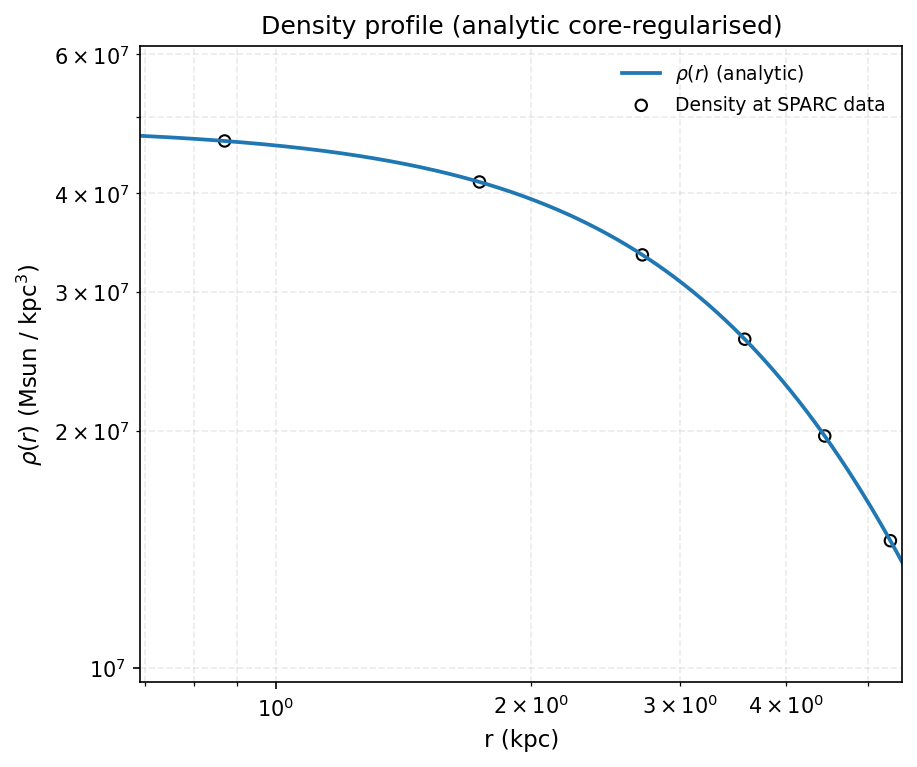}
\caption{The density of the SIDM model of Eq.
(\ref{ScaledependentEoSDM}) for the galaxy NGC4389, versus the
radius.} \label{NGC4389dens}
\end{figure}

\begin{figure}[h!]
\centering
\includegraphics[width=35pc]{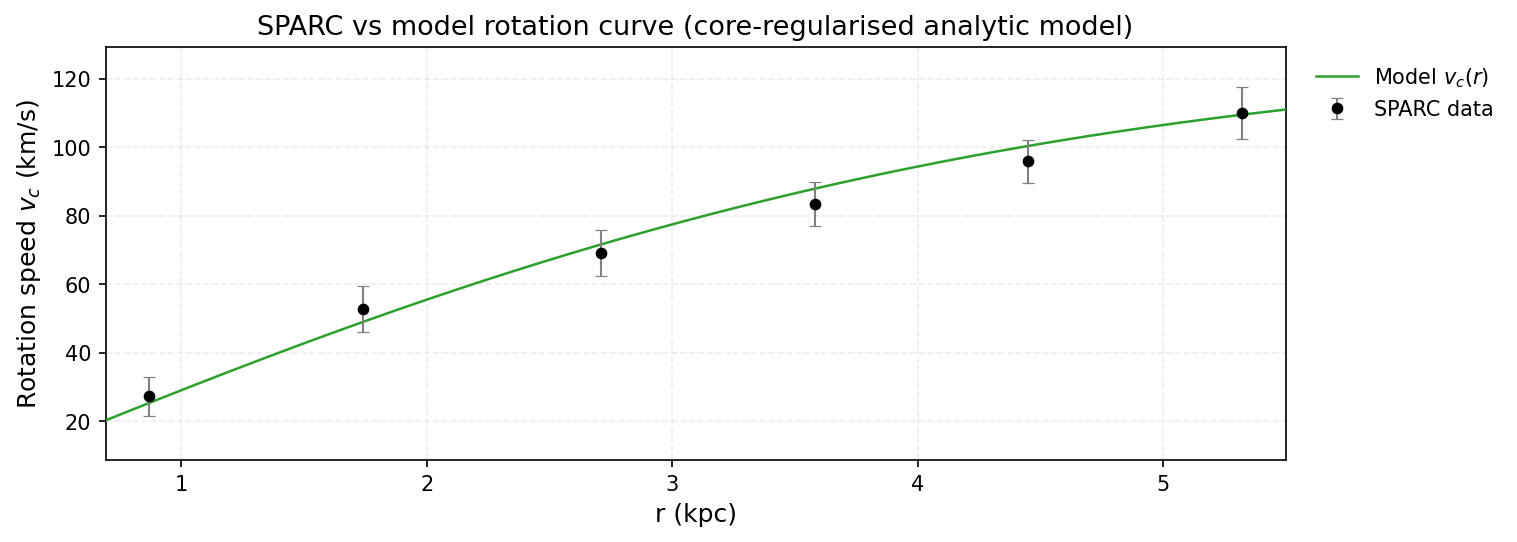}
\caption{The predicted rotation curves for the optimized SIDM
model of Eq. (\ref{ScaledependentEoSDM}), versus the SPARC
observational data for the galaxy NGC4389.} \label{NGC4389}
\end{figure}

In Table \ref{collNGC4389} we present the optimized values of
$K_0$ and $\rho_0$ for the analytic SIDM model of Eq.
(\ref{ScaledependentEoSDM}) for which the maximum compatibility
with the SPARC data is achieved.
\begin{table}[h!]
  \begin{center}
    \caption{SIDM Optimization Values for the galaxy NGC4389}
    \label{collNGC4389}
     \begin{tabular}{|r|r|}
     \hline
      \textbf{Parameter}   & \textbf{Optimization Values}
      \\  \hline
     $\rho_0 $  ($M_{\odot}/\mathrm{Kpc}^{3}$) & $4.85533\times 10^7$
\\  \hline $K_0$ ($M_{\odot} \,
\mathrm{Kpc}^{-3} \, (\mathrm{km/s})^{2}$)& 6639.95
\\  \hline
    \end{tabular}
  \end{center}
\end{table}
In Figs. \ref{NGC4389dens}, \ref{NGC4389} we present the density
of the analytic SIDM model, the predicted rotation curves for the
SIDM model (\ref{ScaledependentEoSDM}), versus the SPARC
observational data and the sound speed, as a function of the
radius respectively. As it can be seen, for this galaxy, the SIDM
model produces viable rotation curves which are compatible with
the SPARC data.

\subsection{The Galaxy NGC6946, Non-viable}

For this galaxy, the optimization method we used, ensures maximum
compatibility of the analytic SIDM model of Eq.
(\ref{ScaledependentEoSDM}) with the SPARC data, if we choose
$\rho_0=1.91358\times 10^8$$M_{\odot}/\mathrm{Kpc}^{3}$ and
$K_0=14582.9
$$M_{\odot} \, \mathrm{Kpc}^{-3} \, (\mathrm{km/s})^{2}$, in which
case the reduced $\chi^2_{red}$ value is $\chi^2_{red}=31.94$.
Also the parameter $\alpha$ in this case is $\alpha=5.0379 $Kpc.

\begin{figure}[h!]
\centering
\includegraphics[width=20pc]{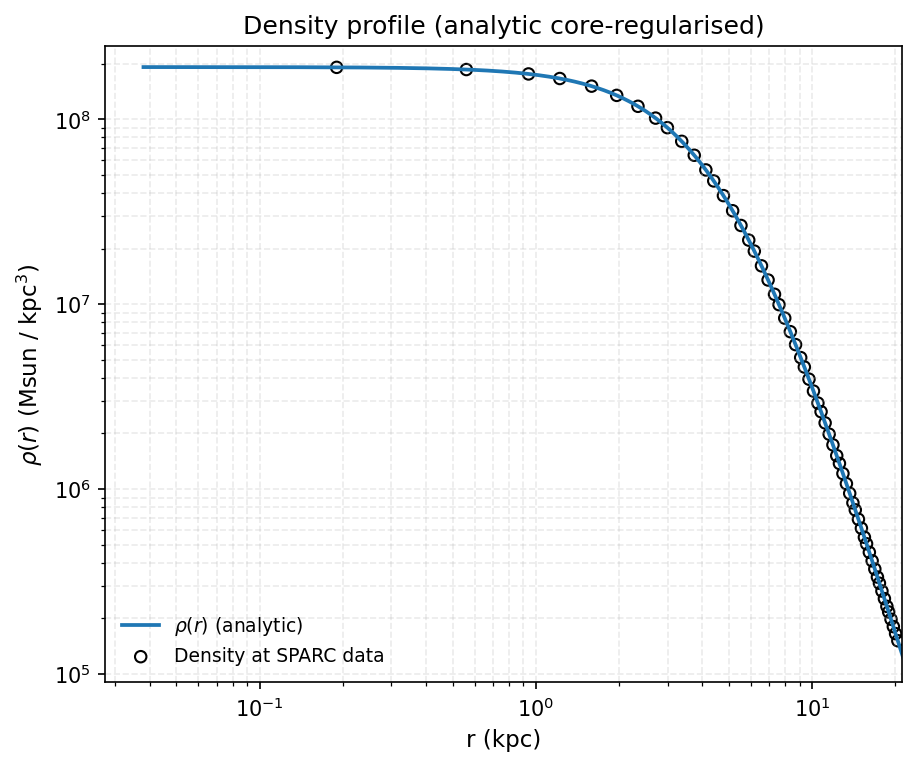}
\caption{The density of the SIDM model of Eq.
(\ref{ScaledependentEoSDM}) for the galaxy NGC6946, versus the
radius.} \label{NGC6946dens}
\end{figure}

\begin{figure}[h!]
\centering
\includegraphics[width=35pc]{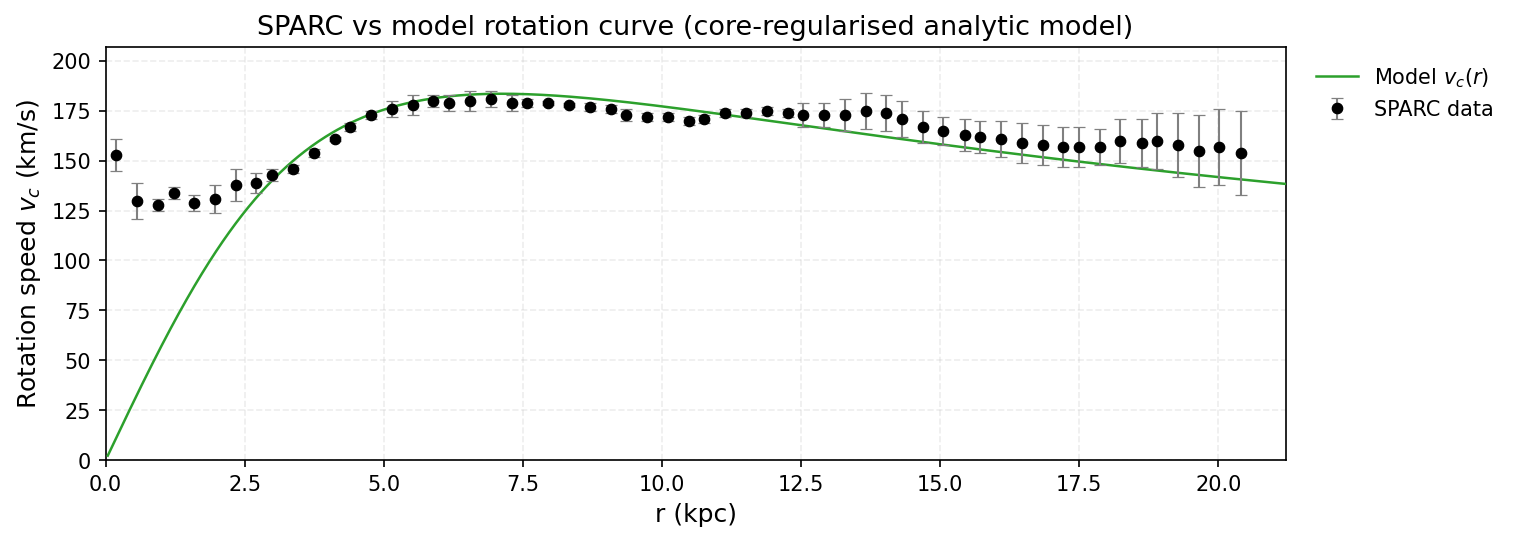}
\caption{The predicted rotation curves for the optimized SIDM
model of Eq. (\ref{ScaledependentEoSDM}), versus the SPARC
observational data for the galaxy NGC6946.} \label{NGC6946}
\end{figure}

In Table \ref{collNGC6946} we present the optimized values of
$K_0$ and $\rho_0$ for the analytic SIDM model of Eq.
(\ref{ScaledependentEoSDM}) for which the maximum compatibility
with the SPARC data is achieved.
\begin{table}[h!]
  \begin{center}
    \caption{SIDM Optimization Values for the galaxy NGC6946}
    \label{collNGC6946}
     \begin{tabular}{|r|r|}
     \hline
      \textbf{Parameter}   & \textbf{Optimization Values}
      \\  \hline
     $\rho_0 $  ($M_{\odot}/\mathrm{Kpc}^{3}$) & $5\times 10^7$
\\  \hline $K_0$ ($M_{\odot} \,
\mathrm{Kpc}^{-3} \, (\mathrm{km/s})^{2}$)& 1250
\\  \hline
    \end{tabular}
  \end{center}
\end{table}
In Figs. \ref{NGC6946dens}, \ref{NGC6946} we present the density
of the analytic SIDM model, the predicted rotation curves for the
SIDM model (\ref{ScaledependentEoSDM}), versus the SPARC
observational data and the sound speed, as a function of the
radius respectively. As it can be seen, for this galaxy, the SIDM
model produces non-viable rotation curves which are incompatible
with the SPARC data.

\begin{figure}[h!]
\centering
\includegraphics[width=20pc]{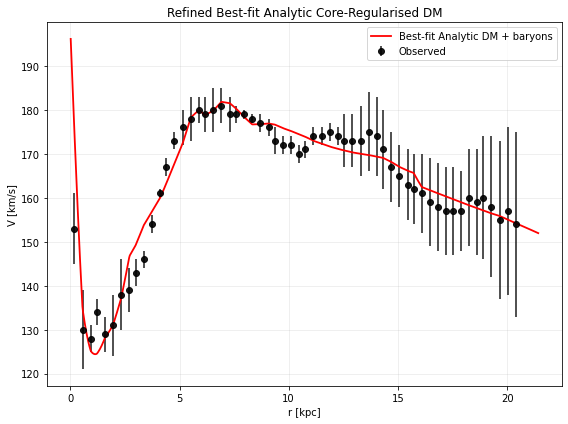}
\caption{The predicted rotation curves after using an optimization
for the SIDM model (\ref{ScaledependentEoSDM}), and the extended
SPARC data for the galaxy NGC6946. We included the rotation curves
of the gas, the disk velocities, the bulge (where present) along
with the SIDM model.} \label{extendedNGC6946}
\end{figure}

Now we shall include contributions to the rotation velocity from
the other components of the galaxy, namely the disk, the gas, and
the bulge if present. In Fig. \ref{extendedNGC6946} we present the
combined rotation curves including all the components of the
galaxy along with the SIDM. As it can be seen, the extended
collisional DM model is non-viable.
Also in Table \ref{evaluationextendedNGC6946} we present the
optimized values of the free parameters of the SIDM model for
which  we achieve the maximum compatibility with the SPARC data,
for the galaxy NGC6946, and also the resulting reduced
$\chi^2_{red}$ value.
\begin{table}[h!]
\centering \caption{Optimized Parameter Values of the Extended
SIDM model for the Galaxy NGC6946.}
\begin{tabular}{lc}
\hline
Parameter & Value  \\
\hline
$\rho_0 $ ($M_{\odot}/\mathrm{Kpc}^{3}$) & $1.06206\times 10^7$   \\
$K_0$ ($M_{\odot} \,
\mathrm{Kpc}^{-3} \, (\mathrm{km/s})^{2}$) & 5917.91   \\
$ml_{\text{disk}}$ & 0.7724 \\
$ml_{\text{bulge}}$ & 0.7454 \\
$\alpha$ (Kpc) & 13.6209\\
$\chi^2_{red}$ & 1.48679 \\
\hline
\end{tabular}
\label{evaluationextendedNGC6946}
\end{table}

\subsection{The Galaxy PGC51017, Non-viable, Extended Viable}

For this galaxy, the optimization method we used, ensures maximum
compatibility of the analytic SIDM model of Eq.
(\ref{ScaledependentEoSDM}) with the SPARC data, if we choose
$\rho_0=1.37511\times 10^8$$M_{\odot}/\mathrm{Kpc}^{3}$ and
$K_0=217.625
$$M_{\odot} \, \mathrm{Kpc}^{-3} \, (\mathrm{km/s})^{2}$, in which
case the reduced $\chi^2_{red}$ value is $\chi^2_{red}=1.72797$.
Also the parameter $\alpha$ in this case is $\alpha=0.726 $Kpc.

\begin{figure}[h!]
\centering
\includegraphics[width=20pc]{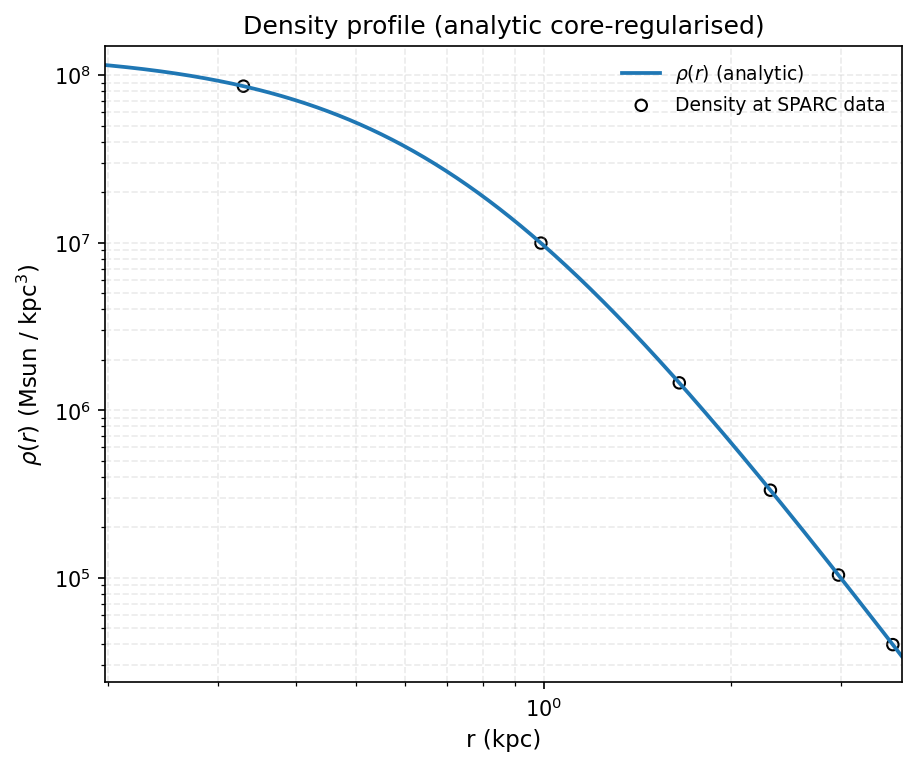}
\caption{The density of the SIDM model of Eq.
(\ref{ScaledependentEoSDM}) for the galaxy PGC51017, versus the
radius.} \label{PGC51017dens}
\end{figure}

\begin{figure}[h!]
\centering
\includegraphics[width=35pc]{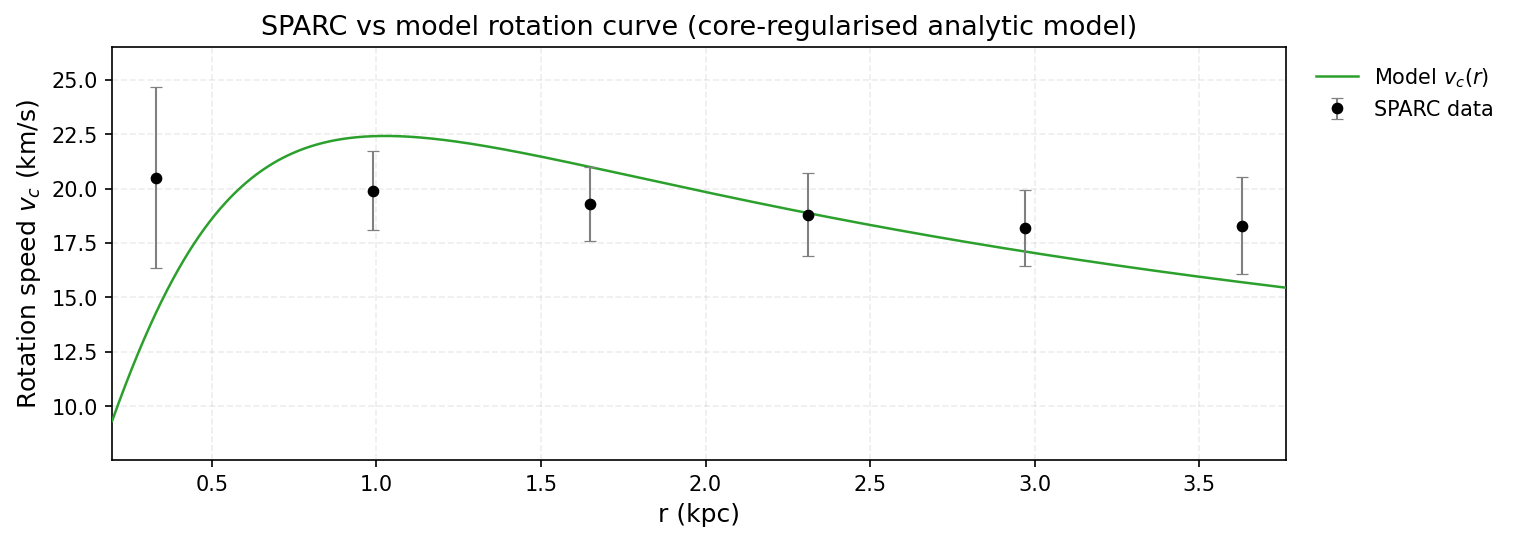}
\caption{The predicted rotation curves for the optimized SIDM
model of Eq. (\ref{ScaledependentEoSDM}), versus the SPARC
observational data for the galaxy PGC51017.} \label{PGC51017}
\end{figure}

In Table \ref{collPGC51017} we present the optimized values of
$K_0$ and $\rho_0$ for the analytic SIDM model of Eq.
(\ref{ScaledependentEoSDM}) for which the maximum compatibility
with the SPARC data is achieved.
\begin{table}[h!]
  \begin{center}
    \caption{SIDM Optimization Values for the galaxy PGC51017}
    \label{collPGC51017}
     \begin{tabular}{|r|r|}
     \hline
      \textbf{Parameter}   & \textbf{Optimization Values}
      \\  \hline
     $\rho_0 $  ($M_{\odot}/\mathrm{Kpc}^{3}$) & $1.37511\times 10^8$
\\  \hline $K_0$ ($M_{\odot} \,
\mathrm{Kpc}^{-3} \, (\mathrm{km/s})^{2}$)& 217.625
\\  \hline
    \end{tabular}
  \end{center}
\end{table}
In Figs. \ref{PGC51017dens}, \ref{PGC51017}  we present the
density of the analytic SIDM model, the predicted rotation curves
for the SIDM model (\ref{ScaledependentEoSDM}), versus the SPARC
observational data and the sound speed, as a function of the
radius respectively. As it can be seen, for this galaxy, the SIDM
model produces non-viable rotation curves which are incompatible
with the SPARC data.

\begin{figure}[h!]
\centering
\includegraphics[width=20pc]{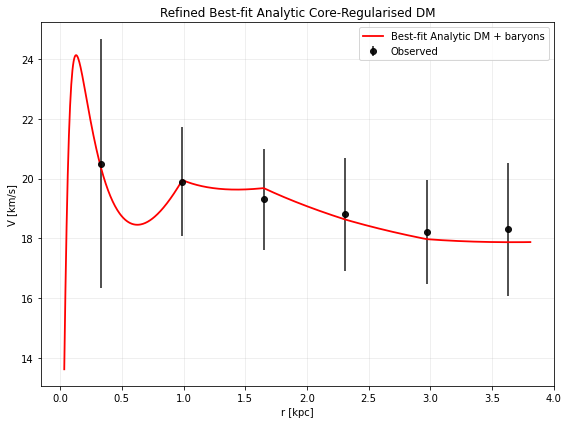}
\caption{The predicted rotation curves after using an optimization
for the SIDM model (\ref{ScaledependentEoSDM}), and the extended
SPARC data for the galaxy PGC51017. We included the rotation
curves of the gas, the disk velocities, the bulge (where present)
along with the SIDM model.} \label{extendedPGC51017}
\end{figure}

Now we shall include contributions to the rotation velocity from
the other components of the galaxy, namely the disk, the gas, and
the bulge if present. In Fig. \ref{extendedPGC51017} we present
the combined rotation curves including all the components of the
galaxy along with the SIDM. As it can be seen, the extended
collisional DM model is non-viable.
Also in Table \ref{evaluationextendedPGC51017} we present the
optimized values of the free parameters of the SIDM model for
which  we achieve the maximum compatibility with the SPARC data,
for the galaxy PGC51017, and also the resulting reduced
$\chi^2_{red}$ value.
\begin{table}[h!]
\centering \caption{Optimized Parameter Values of the Extended
SIDM model for the Galaxy PGC51017.}
\begin{tabular}{lc}
\hline
Parameter & Value  \\
\hline
$\rho_0 $ ($M_{\odot}/\mathrm{Kpc}^{3}$) & $1\times 10^9$   \\
$K_0$ ($M_{\odot} \,
\mathrm{Kpc}^{-3} \, (\mathrm{km/s})^{2}$) & 106.461   \\
$ml_{\text{disk}}$ & 0.6284  \\
$ml_{\text{bulge}}$ & 0.0628 \\
$\alpha$ (Kpc) & 2.67\\
$\chi^2_{red}$ & 0.303901 \\
\hline
\end{tabular}
\label{evaluationextendedPGC51017}
\end{table}

\subsection{The Galaxy UGC01230 Marginally Viable, Extended Viable}

For this galaxy, the optimization method we used, ensures maximum
compatibility of the analytic SIDM model of Eq.
(\ref{ScaledependentEoSDM}) with the SPARC data, if we choose
$\rho_0=2.89655\times 10^7$$M_{\odot}/\mathrm{Kpc}^{3}$ and
$K_0=6578.29
$$M_{\odot} \, \mathrm{Kpc}^{-3} \, (\mathrm{km/s})^{2}$, in which
case the reduced $\chi^2_{red}$ value is $\chi^2_{red}=0.686649$.
Also the parameter $\alpha$ in this case is $\alpha=8.69695 $Kpc.

\begin{figure}[h!]
\centering
\includegraphics[width=20pc]{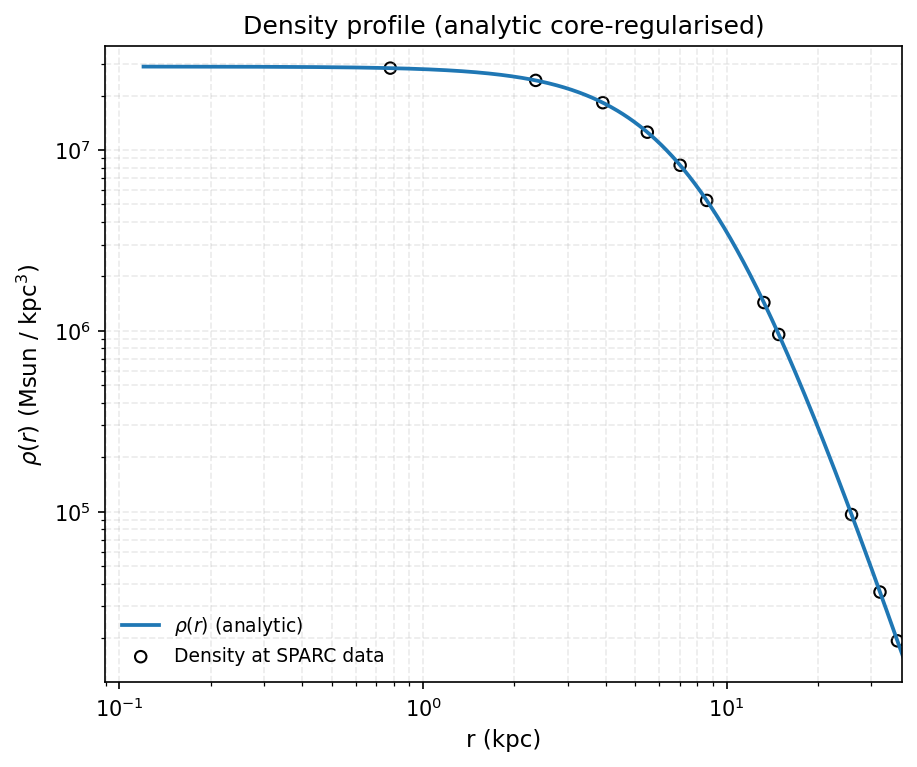}
\caption{The density of the SIDM model of Eq.
(\ref{ScaledependentEoSDM}) for the galaxy UGC01230, versus the
radius.} \label{UGC01230dens}
\end{figure}

\begin{figure}[h!]
\centering
\includegraphics[width=35pc]{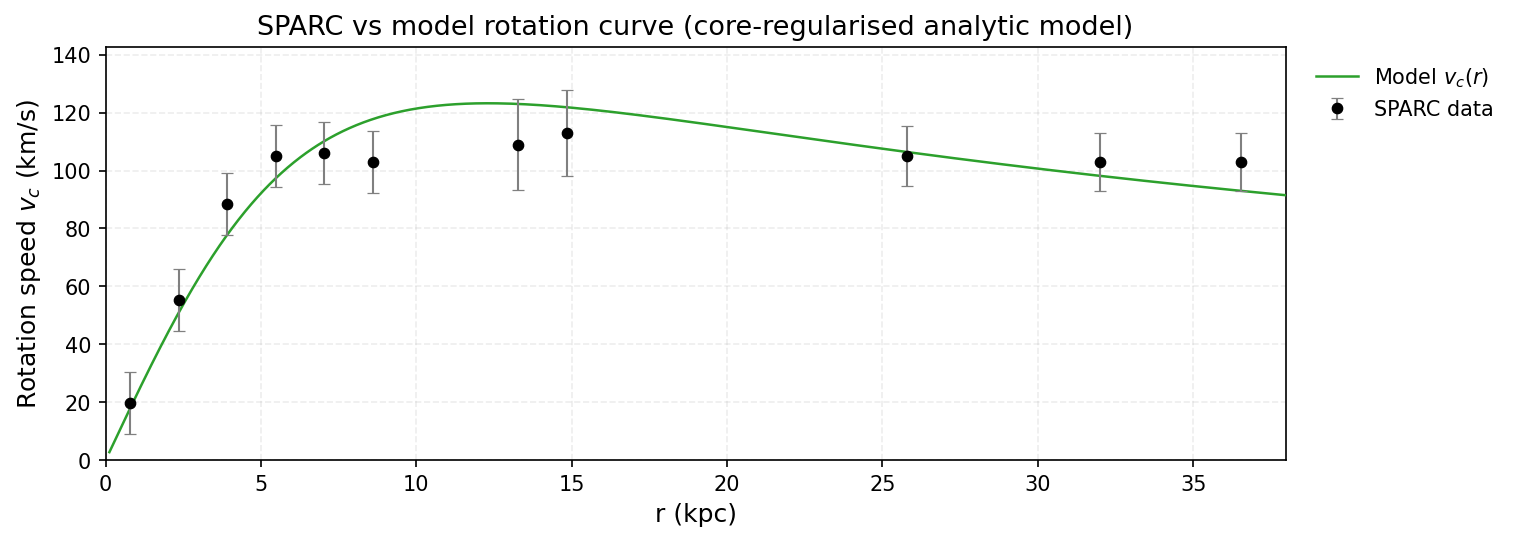}
\caption{The predicted rotation curves for the optimized SIDM
model of Eq. (\ref{ScaledependentEoSDM}), versus the SPARC
observational data for the galaxy UGC01230.} \label{UGC01230}
\end{figure}

In Table \ref{collUGC01230} we present the optimized values of
$K_0$ and $\rho_0$ for the analytic SIDM model of Eq.
(\ref{ScaledependentEoSDM}) for which the maximum compatibility
with the SPARC data is achieved.
\begin{table}[h!]
  \begin{center}
    \caption{SIDM Optimization Values for the galaxy UGC01230}
    \label{collUGC01230}
     \begin{tabular}{|r|r|}
     \hline
      \textbf{Parameter}   & \textbf{Optimization Values}
      \\  \hline
     $\rho_0 $  ($M_{\odot}/\mathrm{Kpc}^{3}$) & $2.89655\times 10^7$
\\  \hline $K_0$ ($M_{\odot} \,
\mathrm{Kpc}^{-3} \, (\mathrm{km/s})^{2}$)& 6578.29
\\  \hline
    \end{tabular}
  \end{center}
\end{table}
In Figs. \ref{UGC01230dens}, \ref{UGC01230}  we present the
density of the analytic SIDM model, the predicted rotation curves
for the SIDM model (\ref{ScaledependentEoSDM}), versus the SPARC
observational data and the sound speed, as a function of the
radius respectively. As it can be seen, for this galaxy, the SIDM
model produces marginally viable rotation curves which are
marginally compatible with the SPARC data.

\begin{figure}[h!]
\centering
\includegraphics[width=20pc]{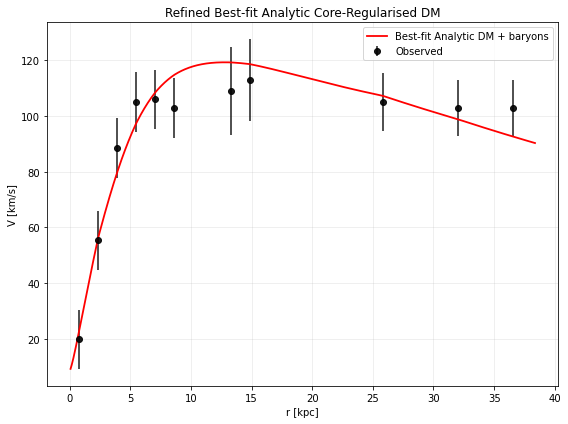}
\caption{The predicted rotation curves after using an optimization
for the SIDM model (\ref{ScaledependentEoSDM}), and the extended
SPARC data for the galaxy UGC01230. We included the rotation
curves of the gas, the disk velocities, the bulge (where present)
along with the SIDM model.} \label{extendedUGC01230}
\end{figure}

Now we shall include contributions to the rotation velocity from
the other components of the galaxy, namely the disk, the gas, and
the bulge if present. In Fig. \ref{extendedUGC01230} we present
the combined rotation curves including all the components of the
galaxy along with the SIDM. As it can be seen, the extended
collisional DM model is viable.
Also in Table \ref{evaluationextendedUGC01230} we present the
optimized values of the free parameters of the SIDM model for
which  we achieve the maximum compatibility with the SPARC data,
for the galaxy UGC01230, and also the resulting reduced
$\chi^2_{red}$ value.
\begin{table}[h!]
\centering \caption{Optimized Parameter Values of the Extended
SIDM model for the Galaxy UGC01230.}
\begin{tabular}{lc}
\hline
Parameter & Value  \\
\hline
$\rho_0 $ ($M_{\odot}/\mathrm{Kpc}^{3}$) & $2.4609\times 10^7$   \\
$K_0$ ($M_{\odot} \,
\mathrm{Kpc}^{-3} \, (\mathrm{km/s})^{2}$) & 5118.51   \\
$ml_{\text{disk}}$ & 0.7409 \\
$ml_{\text{bulge}}$ & 0.2610 \\
$\alpha$ (Kpc) & 8.32188\\
$\chi^2_{red}$ & 0.62717 \\
\hline
\end{tabular}
\label{evaluationextendedUGC01230}
\end{table}

\subsection{The Galaxy UGC02023}

For this galaxy, the optimization method we used, ensures maximum
compatibility of the analytic SIDM model of Eq.
(\ref{ScaledependentEoSDM}) with the SPARC data, if we choose
$\rho_0=1.5771\times 10^7$$M_{\odot}/\mathrm{Kpc}^{3}$ and
$K_0=1802.32
$$M_{\odot} \, \mathrm{Kpc}^{-3} \, (\mathrm{km/s})^{2}$, in which
case the reduced $\chi^2_{red}$ value is $\chi^2_{red}=0.536421$.
Also the parameter $\alpha$ in this case is $\alpha=6.16932 $Kpc.

\begin{figure}[h!]
\centering
\includegraphics[width=20pc]{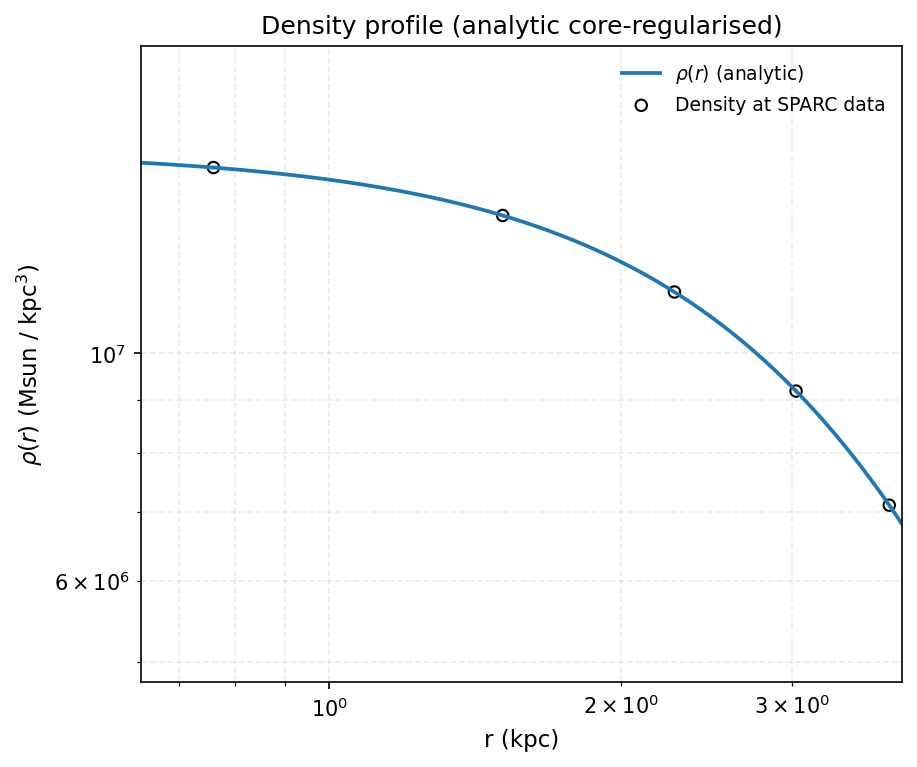}
\caption{The density of the SIDM model of Eq.
(\ref{ScaledependentEoSDM}) for the galaxy UGC02023, versus the
radius.} \label{UGC02023dens}
\end{figure}

\begin{figure}[h!]
\centering
\includegraphics[width=35pc]{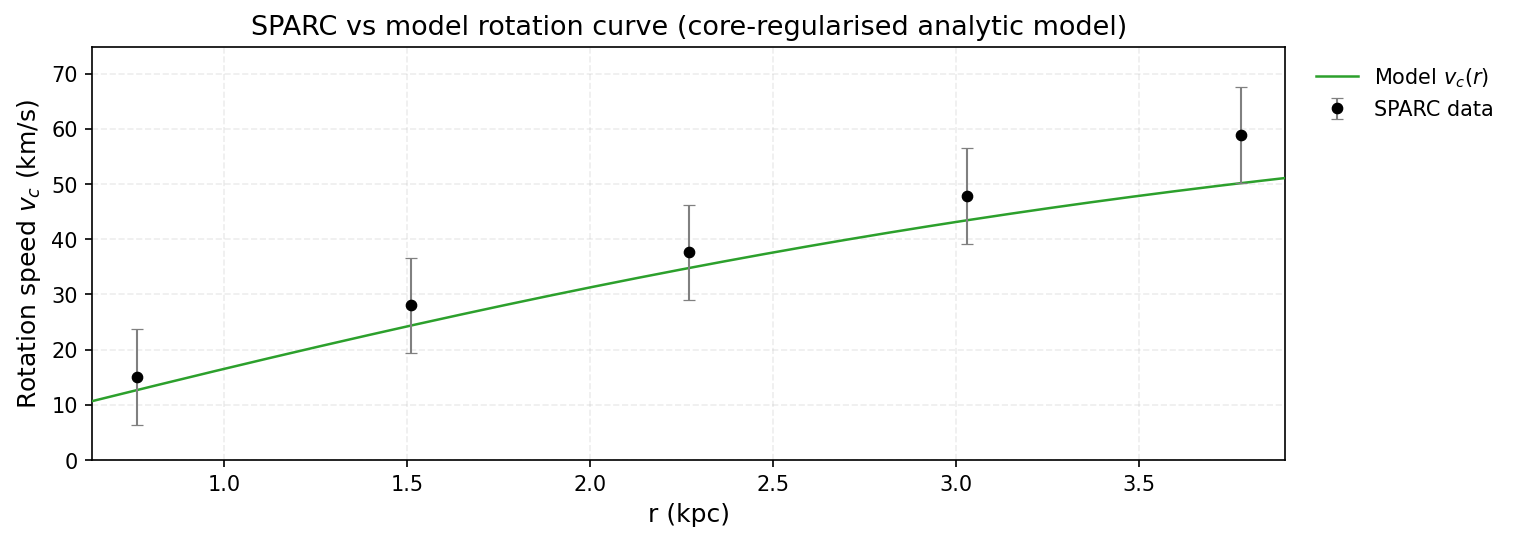}
\caption{The predicted rotation curves for the optimized SIDM
model of Eq. (\ref{ScaledependentEoSDM}), versus the SPARC
observational data for the galaxy UGC02023.} \label{UGC02023}
\end{figure}

In Table \ref{collUGC02023} we present the optimized values of
$K_0$ and $\rho_0$ for the analytic SIDM model of Eq.
(\ref{ScaledependentEoSDM}) for which the maximum compatibility
with the SPARC data is achieved.
\begin{table}[h!]
  \begin{center}
    \caption{SIDM Optimization Values for the galaxy UGC02023}
    \label{collUGC02023}
     \begin{tabular}{|r|r|}
     \hline
      \textbf{Parameter}   & \textbf{Optimization Values}
      \\  \hline
     $\rho_0 $  ($M_{\odot}/\mathrm{Kpc}^{3}$) & $1.5771\times 10^7$
\\  \hline $K_0$ ($M_{\odot} \,
\mathrm{Kpc}^{-3} \, (\mathrm{km/s})^{2}$)& 1802.32
\\  \hline
    \end{tabular}
  \end{center}
\end{table}
In Figs. \ref{UGC02023dens}, \ref{UGC02023}  we present the
density of the analytic SIDM model, the predicted rotation curves
for the SIDM model (\ref{ScaledependentEoSDM}), versus the SPARC
observational data and the sound speed, as a function of the
radius respectively. As it can be seen, for this galaxy, the SIDM
model produces viable rotation curves which are compatible with
the SPARC data.

\subsection{The Galaxy UGC04305}

For this galaxy, the optimization method we used, ensures maximum
compatibility of the analytic SIDM model of Eq.
(\ref{ScaledependentEoSDM}) with the SPARC data, if we choose
$\rho_0=4.51145\times 10^7$$M_{\odot}/\mathrm{Kpc}^{3}$ and
$K_0=537.831
$$M_{\odot} \, \mathrm{Kpc}^{-3} \, (\mathrm{km/s})^{2}$, in which
case the reduced $\chi^2_{red}$ value is $\chi^2_{red}=0.657603$.
Also the parameter $\alpha$ in this case is $\alpha=1.99258 $Kpc.

\begin{figure}[h!]
\centering
\includegraphics[width=20pc]{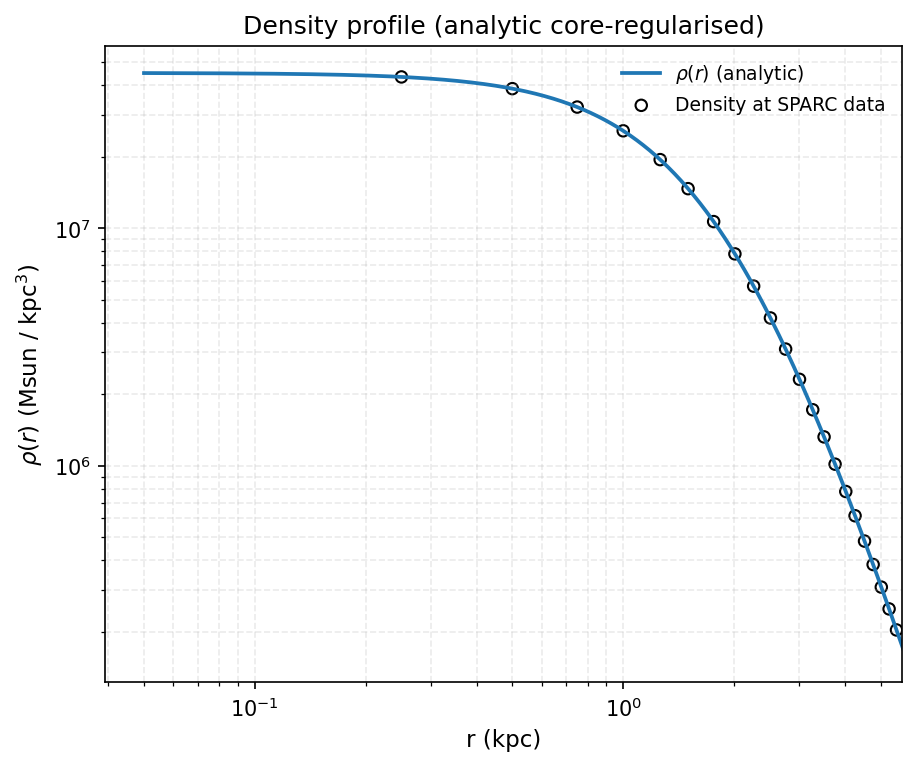}
\caption{The density of the SIDM model of Eq.
(\ref{ScaledependentEoSDM}) for the galaxy UGC04305, versus the
radius.} \label{UGC04305dens}
\end{figure}

\begin{figure}[h!]
\centering
\includegraphics[width=35pc]{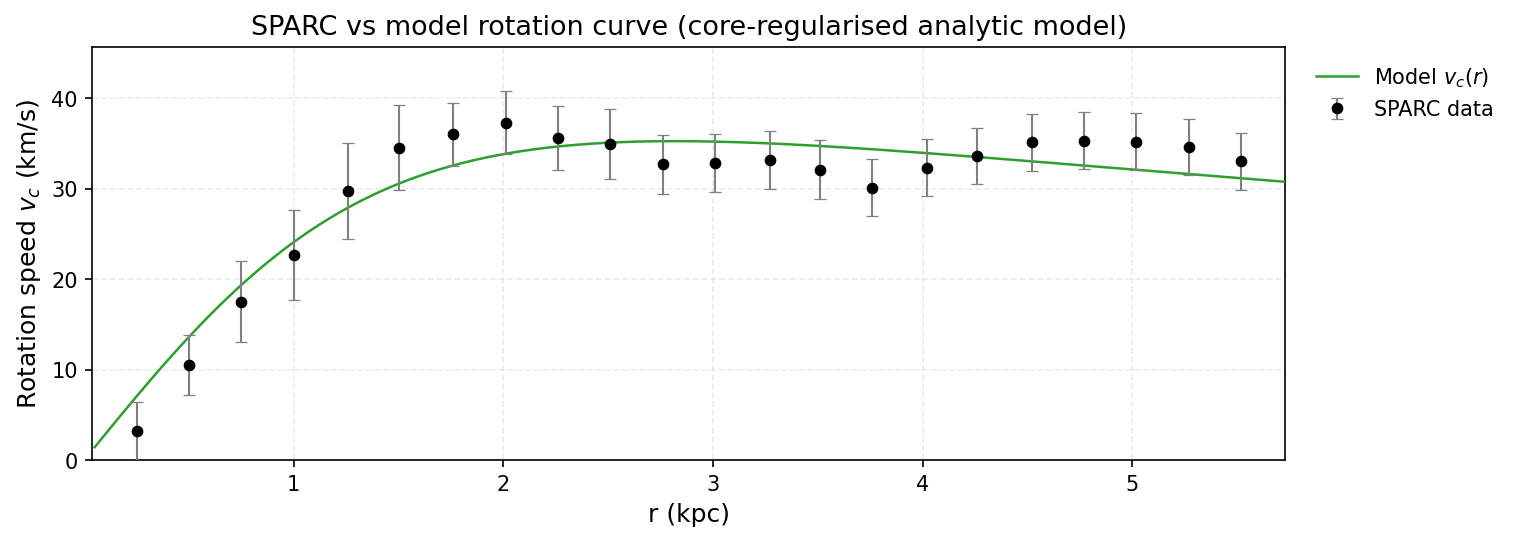}
\caption{The predicted rotation curves for the optimized SIDM
model of Eq. (\ref{ScaledependentEoSDM}), versus the SPARC
observational data for the galaxy UGC04305.} \label{UGC04305}
\end{figure}

In Table \ref{collUGC04305} we present the optimized values of
$K_0$ and $\rho_0$ for the analytic SIDM model of Eq.
(\ref{ScaledependentEoSDM}) for which the maximum compatibility
with the SPARC data is achieved.
\begin{table}[h!]
  \begin{center}
    \caption{SIDM Optimization Values for the galaxy UGC04305}
    \label{collUGC04305}
     \begin{tabular}{|r|r|}
     \hline
      \textbf{Parameter}   & \textbf{Optimization Values}
      \\  \hline
     $\rho_0 $  ($M_{\odot}/\mathrm{Kpc}^{3}$) & $4.51145\times 10^7$
\\  \hline $K_0$ ($M_{\odot} \,
\mathrm{Kpc}^{-3} \, (\mathrm{km/s})^{2}$)& 537.831
\\  \hline
    \end{tabular}
  \end{center}
\end{table}
In Figs. \ref{UGC04305dens}, \ref{UGC04305}  we present the
density of the analytic SIDM model, the predicted rotation curves
for the SIDM model (\ref{ScaledependentEoSDM}), versus the SPARC
observational data and the sound speed, as a function of the
radius respectively. As it can be seen, for this galaxy, the SIDM
model produces marginally viable rotation curves which are
compatible with the SPARC data.


\subsection{The Galaxy UGC05999}

For this galaxy, the optimization method we used, ensures maximum
compatibility of the analytic SIDM model of Eq.
(\ref{ScaledependentEoSDM}) with the SPARC data, if we choose
$\rho_0=1.1002\times 10^7$$M_{\odot}/\mathrm{Kpc}^{3}$ and
$K_0=4076.88
$$M_{\odot} \, \mathrm{Kpc}^{-3} \, (\mathrm{km/s})^{2}$, in which
case the reduced $\chi^2_{red}$ value is $\chi^2_{red}=0.448089$.
Also the parameter $\alpha$ in this case is $\alpha=11.1091 $Kpc.

\begin{figure}[h!]
\centering
\includegraphics[width=20pc]{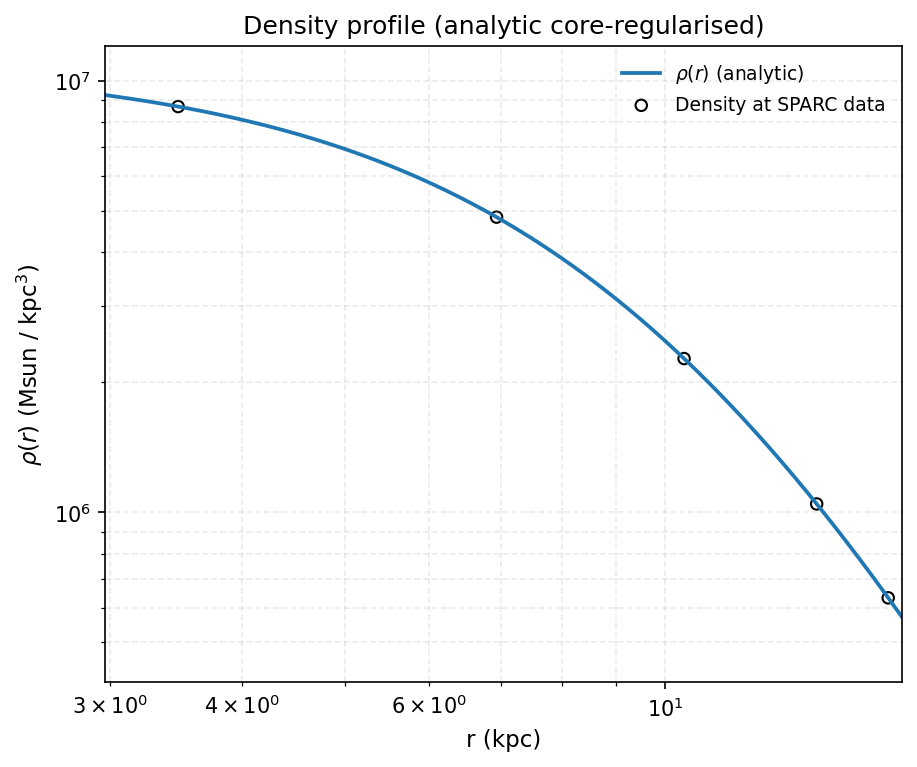}
\caption{The density of the SIDM model of Eq.
(\ref{ScaledependentEoSDM}) for the galaxy UGC05999, versus the
radius.} \label{UGC05999dens}
\end{figure}

\begin{figure}[h!]
\centering
\includegraphics[width=35pc]{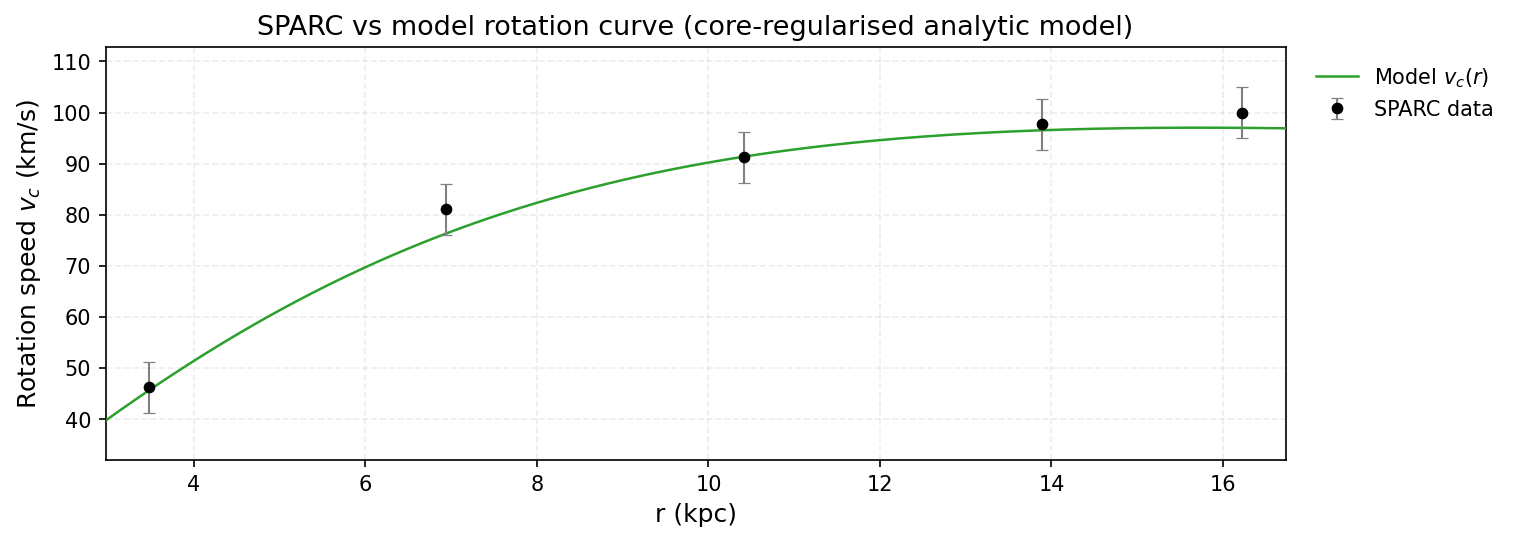}
\caption{The predicted rotation curves for the optimized SIDM
model of Eq. (\ref{ScaledependentEoSDM}), versus the SPARC
observational data for the galaxy UGC05999.} \label{UGC05999}
\end{figure}

In Table \ref{collUGC05999} we present the optimized values of
$K_0$ and $\rho_0$ for the analytic SIDM model of Eq.
(\ref{ScaledependentEoSDM}) for which the maximum compatibility
with the SPARC data is achieved.
\begin{table}[h!]
  \begin{center}
    \caption{SIDM Optimization Values for the galaxy UGC05999}
    \label{collUGC05999}
     \begin{tabular}{|r|r|}
     \hline
      \textbf{Parameter}   & \textbf{Optimization Values}
      \\  \hline
     $\rho_0 $  ($M_{\odot}/\mathrm{Kpc}^{3}$) & $1.1002\times 10^7$
\\  \hline $K_0$ ($M_{\odot} \,
\mathrm{Kpc}^{-3} \, (\mathrm{km/s})^{2}$)& 4076.88
\\  \hline
    \end{tabular}
  \end{center}
\end{table}
In Figs. \ref{UGC05999dens}, \ref{UGC05999} we present the density
of the analytic SIDM model, the predicted rotation curves for the
SIDM model (\ref{ScaledependentEoSDM}), versus the SPARC
observational data and the sound speed, as a function of the
radius respectively. As it can be seen, for this galaxy, the SIDM
model produces viable rotation curves which are compatible with
the SPARC data.

\subsection{The Galaxy UGC06628}

For this galaxy, the optimization method we used, ensures maximum
compatibility of the analytic SIDM model of Eq.
(\ref{ScaledependentEoSDM}) with the SPARC data, if we choose
$\rho_0=2.14819\times 10^7$$M_{\odot}/\mathrm{Kpc}^{3}$ and
$K_0=1032.319
$$M_{\odot} \, \mathrm{Kpc}^{-3} \, (\mathrm{km/s})^{2}$, in which
case the reduced $\chi^2_{red}$ value is $\chi^2_{red}=0.603718$.
Also the parameter $\alpha$ in this case is $\alpha=3.23535 $Kpc.

\begin{figure}[h!]
\centering
\includegraphics[width=20pc]{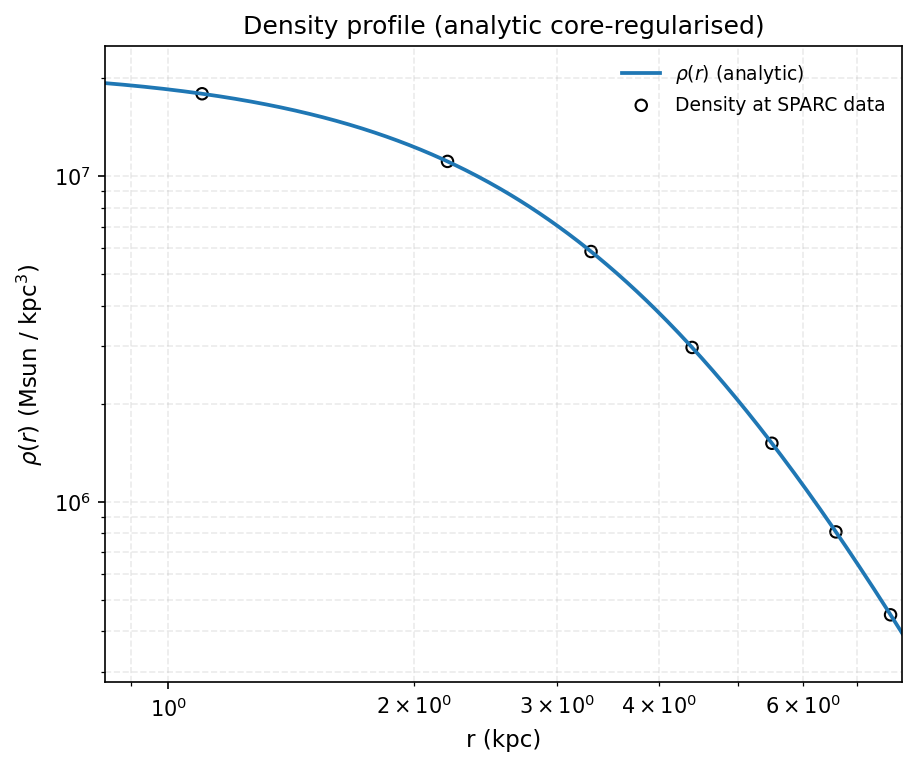}
\caption{The density of the SIDM model of Eq.
(\ref{ScaledependentEoSDM}) for the galaxy UGC06628, versus the
radius.} \label{UGC06628dens}
\end{figure}

\begin{figure}[h!]
\centering
\includegraphics[width=35pc]{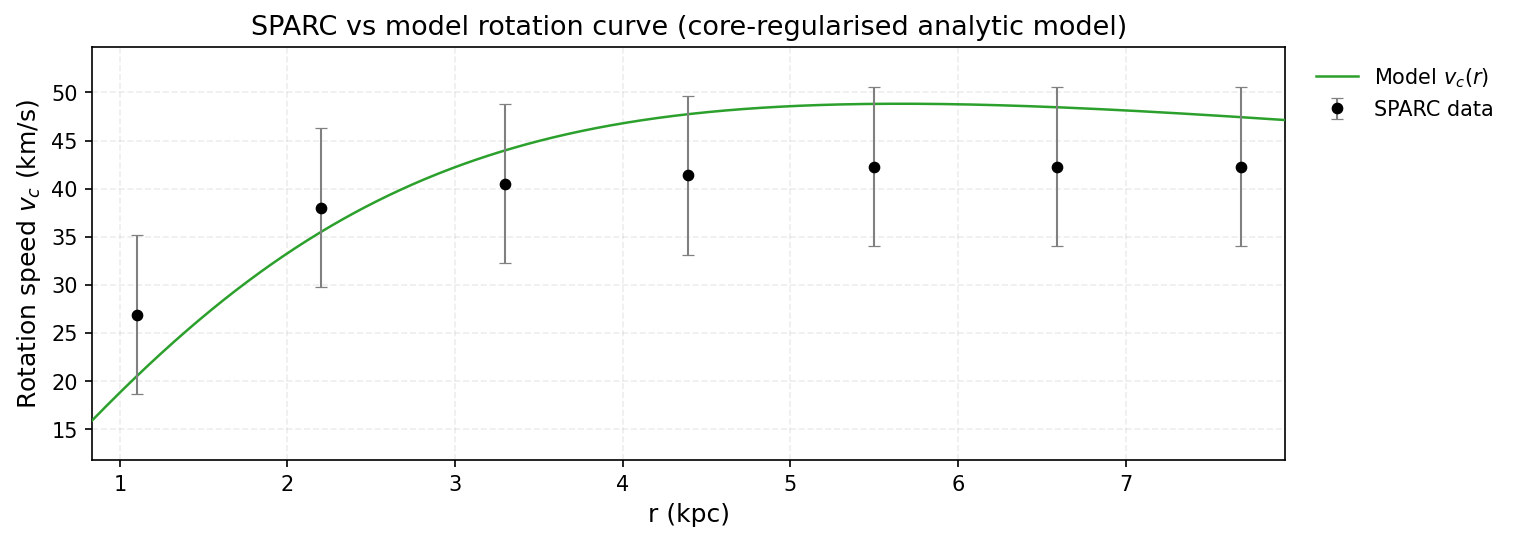}
\caption{The predicted rotation curves for the optimized SIDM
model of Eq. (\ref{ScaledependentEoSDM}), versus the SPARC
observational data for the galaxy UGC06628.} \label{UGC06628}
\end{figure}

In Table \ref{collUGC06628} we present the optimized values of
$K_0$ and $\rho_0$ for the analytic SIDM model of Eq.
(\ref{ScaledependentEoSDM}) for which the maximum compatibility
with the SPARC data is achieved.
\begin{table}[h!]
  \begin{center}
    \caption{SIDM Optimization Values for the galaxy UGC06628}
    \label{collUGC06628}
     \begin{tabular}{|r|r|}
     \hline
      \textbf{Parameter}   & \textbf{Optimization Values}
      \\  \hline
     $\rho_0 $  ($M_{\odot}/\mathrm{Kpc}^{3}$) & $2.14819\times 10^7$
\\  \hline $K_0$ ($M_{\odot} \,
\mathrm{Kpc}^{-3} \, (\mathrm{km/s})^{2}$)& 1032.319
\\  \hline
    \end{tabular}
  \end{center}
\end{table}
In Figs. \ref{UGC06628dens}, \ref{UGC06628} we present the density
of the analytic SIDM model, the predicted rotation curves for the
SIDM model (\ref{ScaledependentEoSDM}), versus the SPARC
observational data and the sound speed, as a function of the
radius respectively. As it can be seen, for this galaxy, the SIDM
model produces viable rotation curves which are compatible with
the SPARC data.

\subsection{The Galaxy UGC06973, Non-viable}

For this galaxy, the optimization method we used, ensures maximum
compatibility of the analytic SIDM model of Eq.
(\ref{ScaledependentEoSDM}) with the SPARC data, if we choose
$\rho_0=5.93142\times 10^8$$M_{\odot}/\mathrm{Kpc}^{3}$ and
$K_0=14203.9
$$M_{\odot} \, \mathrm{Kpc}^{-3} \, (\mathrm{km/s})^{2}$, in which
case the reduced $\chi^2_{red}$ value is $\chi^2_{red}=9.32271$.
Also the parameter $\alpha$ in this case is $\alpha=2.82407 $Kpc.

\begin{figure}[h!]
\centering
\includegraphics[width=20pc]{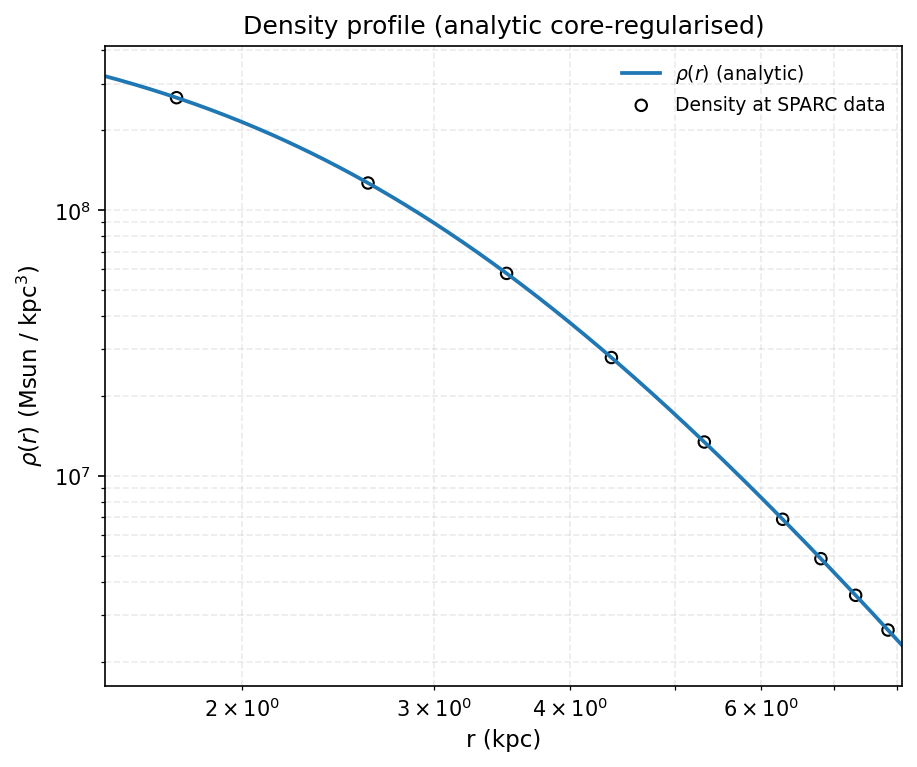}
\caption{The density of the SIDM model of Eq.
(\ref{ScaledependentEoSDM}) for the galaxy UGC06973, versus the
radius.} \label{UGC06973dens}
\end{figure}

\begin{figure}[h!]
\centering
\includegraphics[width=35pc]{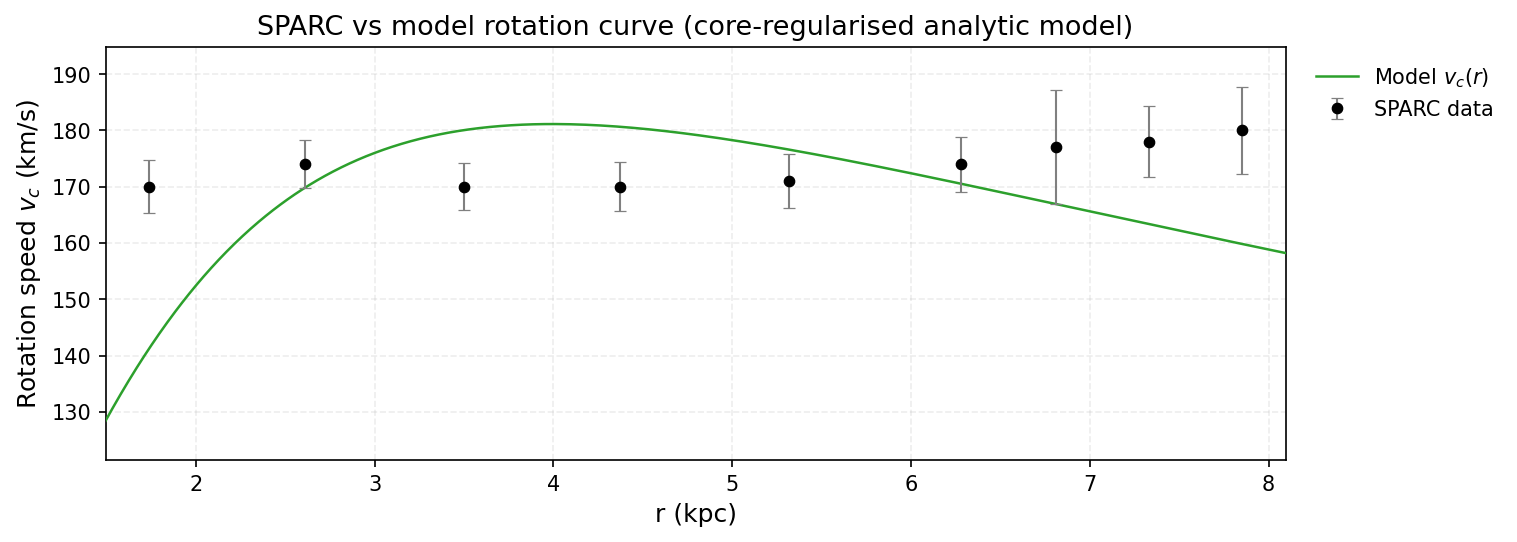}
\caption{The predicted rotation curves for the optimized SIDM
model of Eq. (\ref{ScaledependentEoSDM}), versus the SPARC
observational data for the galaxy UGC06973.} \label{UGC06973}
\end{figure}

In Table \ref{collUGC06973} we present the optimized values of
$K_0$ and $\rho_0$ for the analytic SIDM model of Eq.
(\ref{ScaledependentEoSDM}) for which the maximum compatibility
with the SPARC data is achieved.
\begin{table}[h!]
  \begin{center}
    \caption{SIDM Optimization Values for the galaxy UGC06973}
    \label{collUGC06973}
     \begin{tabular}{|r|r|}
     \hline
      \textbf{Parameter}   & \textbf{Optimization Values}
      \\  \hline
     $\rho_0 $  ($M_{\odot}/\mathrm{Kpc}^{3}$) & $5.93142\times 10^7$
\\  \hline $K_0$ ($M_{\odot} \,
\mathrm{Kpc}^{-3} \, (\mathrm{km/s})^{2}$)& 14203.9
\\  \hline
    \end{tabular}
  \end{center}
\end{table}
In Figs. \ref{UGC06973dens}, \ref{UGC06973}  we present the
density of the analytic SIDM model, the predicted rotation curves
for the SIDM model (\ref{ScaledependentEoSDM}), versus the SPARC
observational data and the sound speed, as a function of the
radius respectively. As it can be seen, for this galaxy, the SIDM
model produces non-viable rotation curves which are incompatible
with the SPARC data.

\begin{figure}[h!]
\centering
\includegraphics[width=20pc]{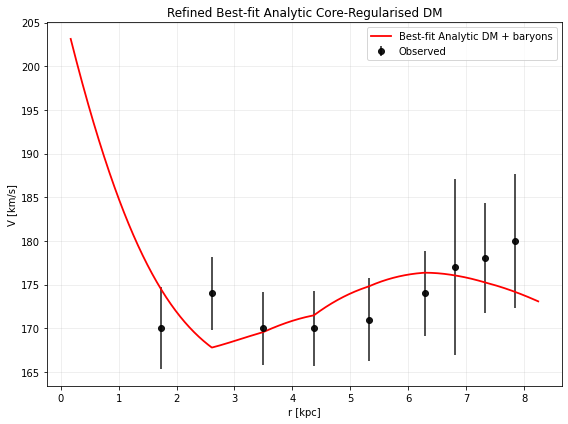}
\caption{The predicted rotation curves after using an optimization
for the SIDM model (\ref{ScaledependentEoSDM}), and the extended
SPARC data for the galaxy UGC06973. We included the rotation
curves of the gas, the disk velocities, the bulge (where present)
along with the SIDM model.} \label{extendedUGC06973}
\end{figure}

Now we shall include contributions to the rotation velocity from
the other components of the galaxy, namely the disk, the gas, and
the bulge if present. In Fig. \ref{extendedUGC06973} we present
the combined rotation curves including all the components of the
galaxy along with the SIDM. As it can be seen, the extended
collisional DM model is non-viable.
Also in Table \ref{evaluationextendedUGC06973} we present the
optimized values of the free parameters of the SIDM model for
which  we achieve the maximum compatibility with the SPARC data,
for the galaxy UGC06973, and also the resulting reduced
$\chi^2_{red}$ value.
\begin{table}[h!]
\centering \caption{Optimized Parameter Values of the Extended
SIDM model for the Galaxy UGC06973.}
\begin{tabular}{lc}
\hline
Parameter & Value  \\
\hline
$\rho_0 $ ($M_{\odot}/\mathrm{Kpc}^{3}$) & $9.611\times 10^7$   \\
$K_0$ ($M_{\odot} \,
\mathrm{Kpc}^{-3} \, (\mathrm{km/s})^{2}$) & 9991.29   \\
$ml_{\text{disk}}$ & 0.4852 \\
$ml_{\text{bulge}}$ & 0 \\
$\alpha$ (Kpc) & 5.88332\\
$\chi^2_{red}$ & 0.965879 \\
\hline
\end{tabular}
\label{evaluationextendedUGC06973}
\end{table}

\subsection{The Galaxy UGC07608}

For this galaxy, the optimization method we used, ensures maximum
compatibility of the analytic SIDM model of Eq.
(\ref{ScaledependentEoSDM}) with the SPARC data, if we choose
$\rho_0=7.199838\times 10^7$$M_{\odot}/\mathrm{Kpc}^{3}$ and
$K_0=1752.32
$$M_{\odot} \, \mathrm{Kpc}^{-3} \, (\mathrm{km/s})^{2}$, in which
case the reduced $\chi^2_{red}$ value is $\chi^2_{red}=0.916443$.
Also the parameter $\alpha$ in this case is $\alpha=3.92396 $Kpc.

\begin{figure}[h!]
\centering
\includegraphics[width=20pc]{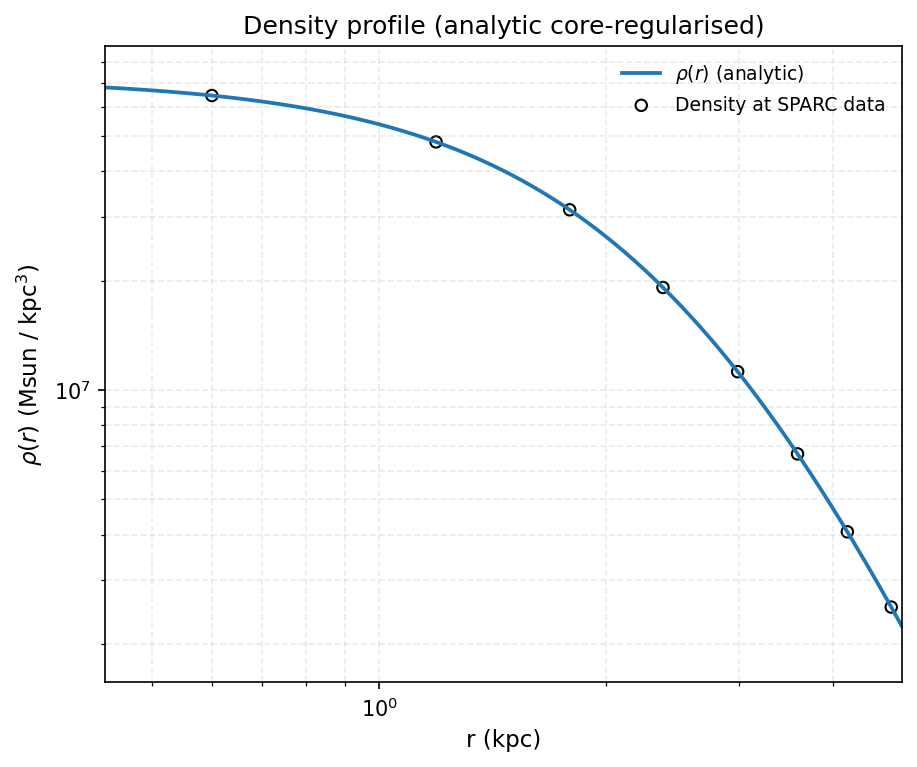}
\caption{The density of the SIDM model of Eq.
(\ref{ScaledependentEoSDM}) for the galaxy UGC07608, versus the
radius.} \label{UGC07608dens}
\end{figure}

\begin{figure}[h!]
\centering
\includegraphics[width=35pc]{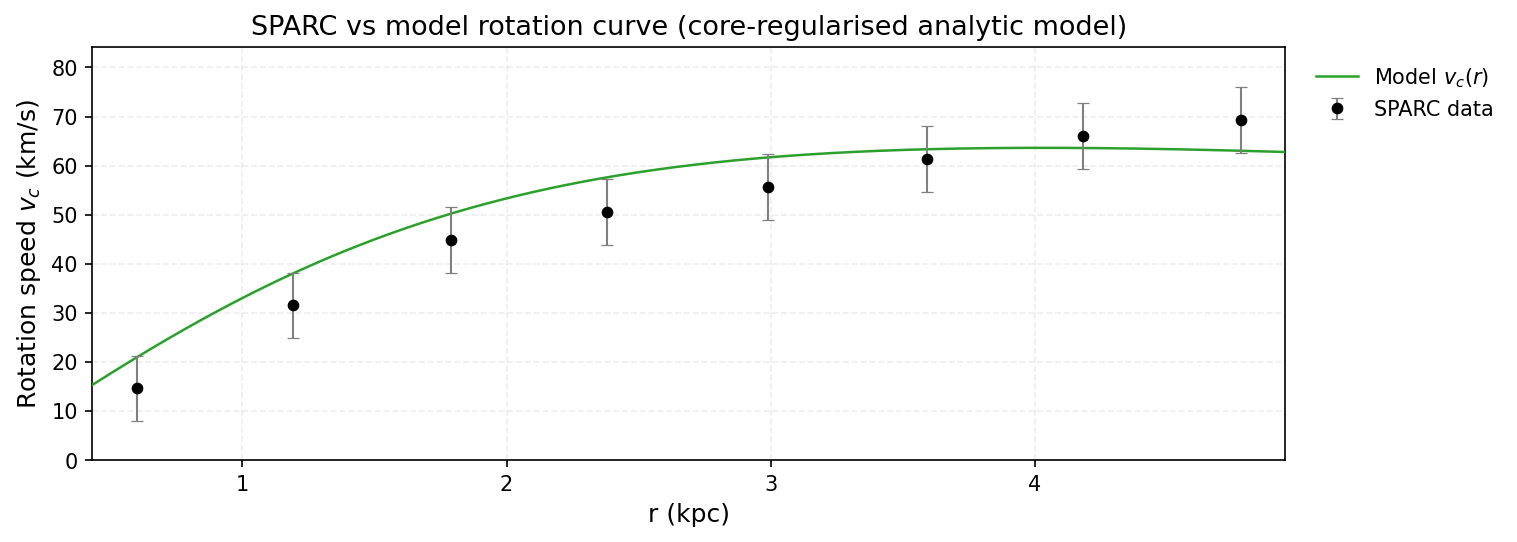}
\caption{The predicted rotation curves for the optimized SIDM
model of Eq. (\ref{ScaledependentEoSDM}), versus the SPARC
observational data for the galaxy UGC07608.} \label{UGC07608}
\end{figure}

In Table \ref{collUGC07608} we present the optimized values of
$K_0$ and $\rho_0$ for the analytic SIDM model of Eq.
(\ref{ScaledependentEoSDM}) for which the maximum compatibility
with the SPARC data is achieved.
\begin{table}[h!]
  \begin{center}
    \caption{SIDM Optimization Values for the galaxy UGC07608}
    \label{collUGC07608}
     \begin{tabular}{|r|r|}
     \hline
      \textbf{Parameter}   & \textbf{Optimization Values}
      \\  \hline
     $\rho_0 $  ($M_{\odot}/\mathrm{Kpc}^{3}$) & $7.199838\times 10^7$
\\  \hline $K_0$ ($M_{\odot} \,
\mathrm{Kpc}^{-3} \, (\mathrm{km/s})^{2}$)& 1752.32
\\  \hline
    \end{tabular}
  \end{center}
\end{table}
In Figs. \ref{UGC07608dens}, \ref{UGC07608}  we present the
density of the analytic SIDM model, the predicted rotation curves
for the SIDM model (\ref{ScaledependentEoSDM}), versus the SPARC
observational data and the sound speed, as a function of the
radius respectively. As it can be seen, for this galaxy, the SIDM
model produces viable rotation curves which are compatible with
the SPARC data.

\subsection{The Galaxy UGCA281}

For this galaxy, the optimization method we used, ensures maximum
compatibility of the analytic SIDM model of Eq.
(\ref{ScaledependentEoSDM}) with the SPARC data, if we choose
$\rho_0=2.12792\times 10^8$$M_{\odot}/\mathrm{Kpc}^{3}$ and
$K_0=379.708
$$M_{\odot} \, \mathrm{Kpc}^{-3} \, (\mathrm{km/s})^{2}$, in which
case the reduced $\chi^2_{red}$ value is $\chi^2_{red}=0.75153$.
Also the parameter $\alpha$ in this case is $\alpha=0.7709 $Kpc.

\begin{figure}[h!]
\centering
\includegraphics[width=20pc]{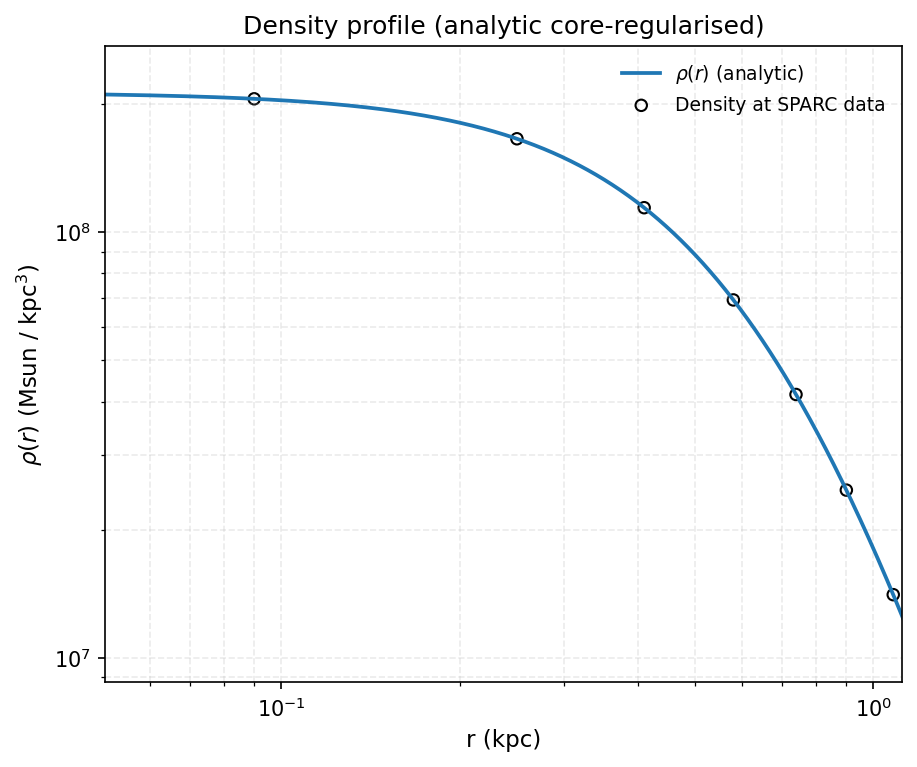}
\caption{The density of the SIDM model of Eq.
(\ref{ScaledependentEoSDM}) for the galaxy UGCA281, versus the
radius.} \label{UGCA281dens}
\end{figure}

\begin{figure}[h!]
\centering
\includegraphics[width=35pc]{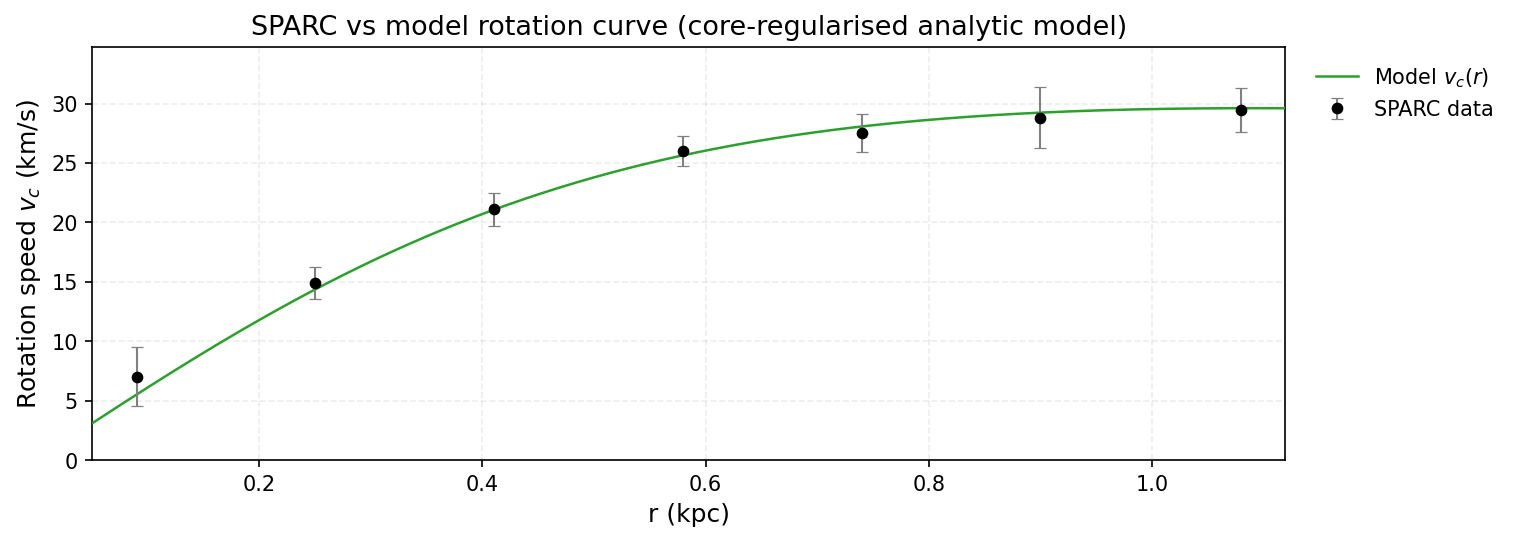}
\caption{The predicted rotation curves for the optimized SIDM
model of Eq. (\ref{ScaledependentEoSDM}), versus the SPARC
observational data for the galaxy UGCA281.} \label{UGCA281}
\end{figure}

In Table \ref{collUGCA281} we present the optimized values of
$K_0$ and $\rho_0$ for the analytic SIDM model of Eq.
(\ref{ScaledependentEoSDM}) for which the maximum compatibility
with the SPARC data is achieved.
\begin{table}[h!]
  \begin{center}
    \caption{SIDM Optimization Values for the galaxy UGCA281}
    \label{collUGCA281}
     \begin{tabular}{|r|r|}
     \hline
      \textbf{Parameter}   & \textbf{Optimization Values}
      \\  \hline
     $\rho_0 $  ($M_{\odot}/\mathrm{Kpc}^{3}$) & $2.12792\times 10^8$
\\  \hline $K_0$ ($M_{\odot} \,
\mathrm{Kpc}^{-3} \, (\mathrm{km/s})^{2}$)& 379.708
\\  \hline
    \end{tabular}
  \end{center}
\end{table}
In Figs. \ref{UGCA281dens}, \ref{UGCA281} we present the density
of the analytic SIDM model, the predicted rotation curves for the
SIDM model (\ref{ScaledependentEoSDM}), versus the SPARC
observational data and the sound speed, as a function of the
radius respectively. As it can be seen, for this galaxy, the SIDM
model produces viable rotation curves which are compatible with
the SPARC data.


\end{document}